\documentclass[a4paper,12pt]{article}
\pdfoutput=1 % if your are submitting a pdflatex (i.e. if you have
\usepackage{jheppub} % for details on the use of the package, please
\usepackage{hyperref}
\hypersetup{
    bookmarks=true,         % show bookmarks bar?
    unicode=false,          % non-Latin characters in AcrobatÃÂÃÂÃÂÃÂ¢ÃÂÃÂÃÂÃÂÃÂÃÂÃÂÃÂs bookmarks
    pdftoolbar=true,        % show AcrobatÃÂÃÂÃÂÃÂ¢ÃÂÃÂÃÂÃÂÃÂÃÂÃÂÃÂs toolbar?
    pdfmenubar=true,        % show AcrobatÃÂÃÂÃÂÃÂ¢ÃÂÃÂÃÂÃÂÃÂÃÂÃÂÃÂs menu?
    pdffitwindow=false,     % window fit to page when opened
    pdfstartview={FitH},    % fits the width of the page to the window
    pdftitle={My title},    % title
    pdfauthor={Author},     % author
    pdfsubject={Subject},   % subject of the document
    pdfcreator={Creator},   % creator of the document
    pdfproducer={Producer}, % producer of the document
    pdfkeywords={keyword1} {key2} {key3}, % list of keywords
    pdfnewwindow=true,      % links in new window
    colorlinks=true,       % false: boxed links; true: colored links
    linkcolor=red,          % color of internal links
    citecolor=purple,        % color of links to bibliography
    filecolor=magenta,      % color of file links
    urlcolor=blue,           % color of external links
    linktocpage=true
}
\usepackage{graphics}
\usepackage{rotating}
\usepackage{subfigure}
\usepackage{pstricks}
\usepackage{amsfonts}
\usepackage{mathrsfs,amsmath}
\usepackage{epsfig}
\usepackage{amsmath,amssymb,amsthm,graphicx,latexsym}
\usepackage{float}
\usepackage{textcomp}
\usepackage{array}
\usepackage{booktabs}

\usepackage{ulem}

\newcommand{\be}{\begin{equation}}
\newcommand{\ee}{\end{equation}}
\newcommand{\bc}{\begin{center}}
\newcommand{\ec}{\end{center}}
\newcommand{\bea}{\begin{eqnarray}}
\newcommand{\eea}{\end{eqnarray}}
\newcommand{\bml}{\begin{subequations}}
\newcommand{\eml}{\end{subequations}}
\newcommand{\bfig}{\begin{figure}}
\newcommand{\efig}{\end{figure}}

\newcommand{\ag}{\alpha}
\newcommand{\bg}{\beta}

\newcommand{\del}{\delta}

\newcommand{\thg}{\theta}

\newcommand{\lb}{\lambda}

\newcommand{\og}{\omega}

\newcommand{\Zstroke}{%
  \text{\ooalign{\hidewidth \raisebox{0.2ex}{--}\hidewidth\cr$Z$\cr}}%
  }
\newcommand{\Del}{\Delta}
\newcommand{\ta}{\tau}

\newcommand{\pl}{\partial}

\newcommand{\slashed}{\hspace{-1.1ex}/}

\newcommand{\bmat}{\begin{pmatrix}}
\newcommand{\emat}{\end{pmatrix}}
\begin{document}

%$~~~~~~~~~~~~~~~~~~~~~~~~~~~~~~~~~~~~~~~~~~~~~~~~~~~~~~~~~~~~~~~~~~~~~~~~~~~~~~~~~~~~$\textcolor{red}{\Large\bf TIFR/TH/17-30}
	\title{\textsc{\sffamily \bfseries \textcolor{violet}{Quantum Out-of-Equilibrium Cosmology}}}
	
	%Quantum chaos in the Sky : A new cosmological toolkit for quantifying stochastic randomness}}}

\author[a]{Sayantan Choudhury,
		\footnote{\textcolor{purple}{\bf Alternative
				E-mail: sayanphysicsisi@gmail.com}. ${}^{}$}}
	\affiliation[a]{Quantum Gravity and Unified Theory and Theoretical Cosmology Group, Max Planck Institute for Gravitational Physics (Albert Einstein Institute),
	   Am M$\ddot{u}$hlenberg 1,
	   14476 Potsdam-Golm, Germany.}
\author[b]{Arkaprava Mukherjee,}	   
\author[c]{Prashali Chauhan,}			
\author[d]{Sandipan Bhattacherjee}
\affiliation[b]{Department of Physical Sciences, Indian Institute of Science Education and Research Kolkata,
Mohanpur, West Bengal 741246, India}
\affiliation[c]{Ashoka University, Sonepat - 131029, India}		
\affiliation[d]{Department of Physics, Birla Institute of Technology, Mesra, Ranchi - 835215, India}
\emailAdd{sayantan.choudhury@aei.mpg.de,chauhan.prashali@gmail.com, www.farada@gmail.com, arka262016@gmail.com }

\abstract{In this work, our prime focus is to study the one to one correspondence between the conduction phenomena in electrical wires with impurity and the scattering events responsible for particle production during stochastic inflation and reheating implemented under a closed quantum mechanical system in early universe cosmology. In this connection, we also present a derivation of quantum corrected version of the Fokker Planck equation without dissipation and its fourth order corrected analytical solution for the probability distribution profile responsible for studying the dynamical features of the particle creation events in the stochastic inflation and reheating stage of the universe.  It is explicitly shown from our computation that quantum corrected Fokker Planck equation describe the particle creation phenomena better for Dirac delta type of scatterer. In this connection, we additionally discuss It$\hat{o}$, Stratonovich prescription and the explicit role of finite temperature effective potential for solving the probability distribution profile. Furthermore, we extend our discussion of particle production phenomena to describe the quantum description of randomness involved in the dynamics. We also present computation to derive the expression for the measure of the stochastic non-linearity (randomness or chaos) arising in the stochastic inflation and reheating epoch of the universe, often described by {\it  Lyapunov Exponent}. Apart from that, we quantify the quantum chaos arising in a closed system by a more strong measure, commonly known as {\it Spectral Form Factor} using the principles of Random Matrix Theory (RMT). Additionally, we discuss the role of out of time order correlation function (OTOC) to describe quantum chaos in the present non-equilibrium field theoretic setup and its consequences in early universe cosmology (stochastic inflation and reheating). Finally, for completeness, we also provide a bound on the measure of quantum chaos ( i.e. on {\it  Lyapunov Exponent} and {\it Spectral Form Factor}) arising due to the presence of stochastic non-linear dynamical interactions into the closed quantum system of the early universe in a completely model-independent way.}


\maketitle
\flushbottom
\newpage
\section{Introduction}

 Quantum fields in an inflationary background \cite{Baumann:2009ds,Baumann:2018muz,Baumann:2014nda,Cheung:2007st,Weinberg:2008hq,Delacretaz:2016nhw,Senatore:2016aui,Delacretaz:2015edn,LopezNacir:2011kk,Senatore:2010wk,Choudhury:2011sq,Choudhury:2011jt,Choudhury:2012yh,Choudhury:2013zna,Choudhury:2013jya,Choudhury:2013iaa,Choudhury:2014sxa,Choudhury:2014kma,Choudhury:2014sua,Choudhury:2014hja,Choudhury:2015hvr,Choudhury:2017cos,Naskar:2017ekm,Choudhury:2017glj,Choudhury:2015pqa} or during reheating \cite{ Kofman:1994rk,Shtanov:1994ce,Amin:2014eta,Ozsoy:2015rna,Giblin:2017qjp,Kofman:2005yz,Choudhury:2011rz} gives rise to the burst of particle production, which has been extensively studied in ref.~\cite{Gibbons:1977mu,SchuTzhold:2013fga,Winitzki:2005rw}. This has been studied to a  great extent in the background of the inflationary scenario of the universe in ref.~ \cite{Birrell:1982ix,Parker:1968mv,Fulling:1989nb}. Such phenomena has been compared to that of the scattering problem in quantum mechanics with a specific effective potential arising due to the impurity in the conduction wire, which can approximately be solved using the well known WKB technique \cite{SchuTzhold:2013fga,Birrell:1982ix}~\footnote{In the context of cosmology conformal time dependent effective mass profile exactly mimics the role of impurity potential in electrical conduction wire. Due to such one to one correspondence the time evolution equation (i.e. Klien Gordon equation) of the Fourier modes corresponding to the quantum fluctuation in the context of primordial cosmology can be described in terms of the Schrodinger equation in electrical conduction wire with specific impurity potential.  We have investigated this possibility in detail in this paper. Additionally, it is important to note that such time dependent effective mass profiles are also important to study the role of quantum critical quench and eigen state thermalization \cite{} during the reheating epoch of universe.}. It is important to note that such particle production events are completely random (or chaotic) when the evolution is non-adiabatic or tachyonic in nature.
\begin{figure}[htb]
	\includegraphics[width=18cm,height=8cm]{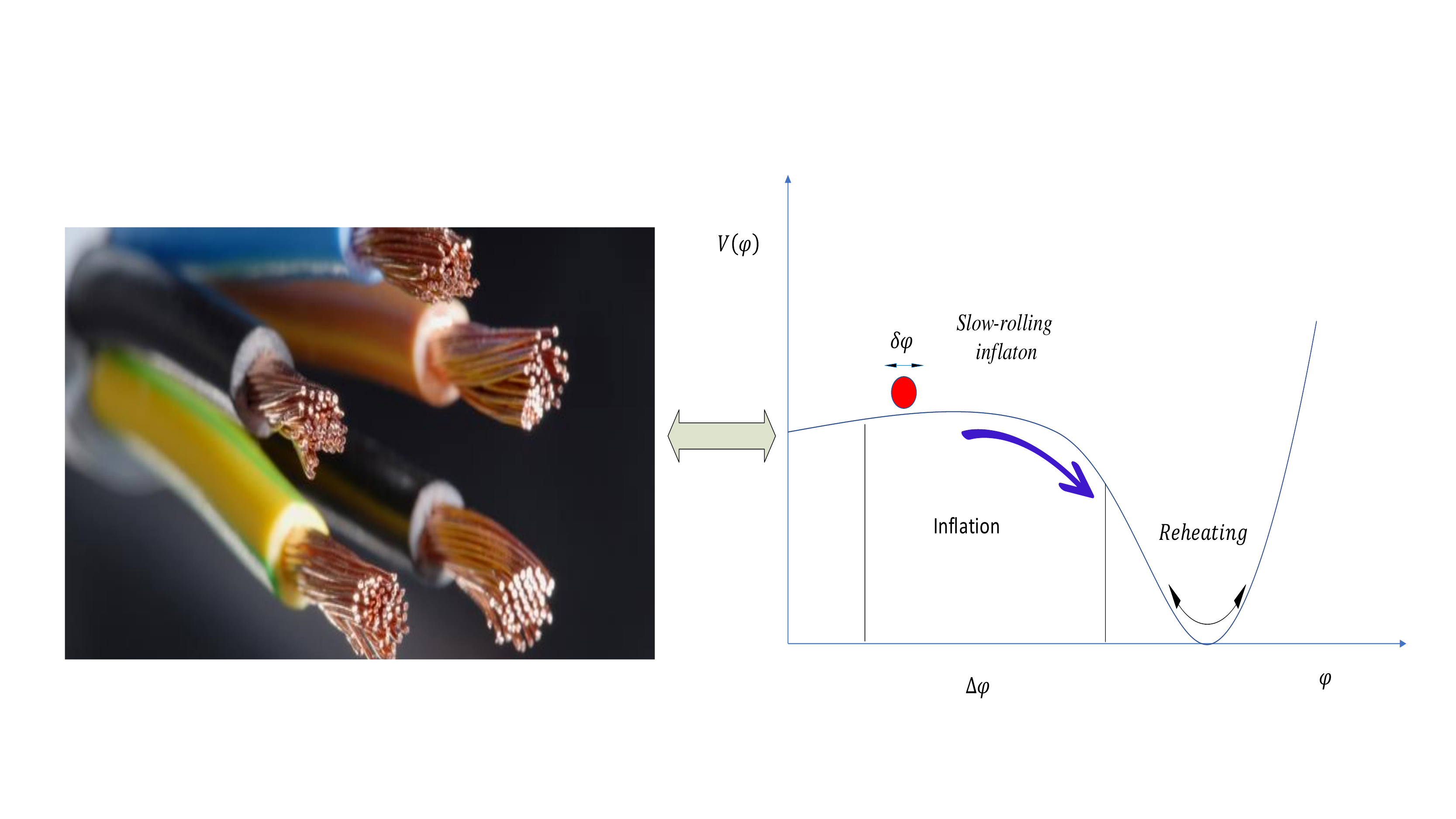}
	\caption{This schematic diagram shows the correspondence between the conduction phenomena in electrical wires with impurity to that of the cosmological random particle-creation events during the non-adiabatic stage of the early universe.}
	\label{chaos 2}
\end{figure}

A non-adiabatic change in the time dependent effective mass profiles of the fields (which is actually coming from integrating out the heavy degrees of freedom from the UV complete theory and after path integration finally one gets the time dependent effective coupling parameters between fields) as the background evolution of the fields passes through special points in field space produces these burst of particle creation in (quasi) de Sitter space time. There lies a physical and mathematical equivalence between such cosmological events to that of the stochastic random phenomena occurring in mesoscopic systems where fluctuations in physical quantities play a significant role of producing stochastic randomness in the system under consideration. We also discuss the cosmological systems which have been considered to be rather non-linear and dissipative due to the significant amounts of quantum fluctuations in the effective coupling terms (or in the time dependent effective mass profile) of the interactions between the fields. Important reviews on the non-linear and dissipative effects arising in the context of cosmology were put forward in the refs.~ \cite{Finelli:2011gd,Starobinsky:1994bd,Starobinsky:1986fx,LopezNacir:2011kk}. In this paper we explicitly discuss bout the various non-linear and dissipative effects in cosmological set up that arises in (quasi) de-Sitter space with $m^2>0$, where the term $m^2$ represents the effective mass squared of the created particle in  
(quasi) de-Sitter background. In this connection it is important to note that, the massless scalar field gets "thermalize" due to the effective time dependent interaction in the (quasi) de-Sitter background.
The cosmological events that we talk about in this paper are identified with those of the particle production stochastic random events. In this paper, we
present the dynamical features of inherent chaos (stochastic randomness) in the physical system and its connection with the quantum mechanics in detail. The model is exactly similar to that of "massless scalar field"
interacting with a scatterer in the background which are treated to be the heavy fields and are mainly responsible for cosmological particle
production in (qusi) de-Sitter space (see\cite{Birrell:1982ix}). In this context, when the free massless scalar field interacts
with the the heavy field in the background space time, it mimics the role of  thermalization phenomena of the field which occurs during 
the epoch of reheating of the universe.

\begin{figure}[htb]
	\includegraphics[width=17cm,height=8cm]{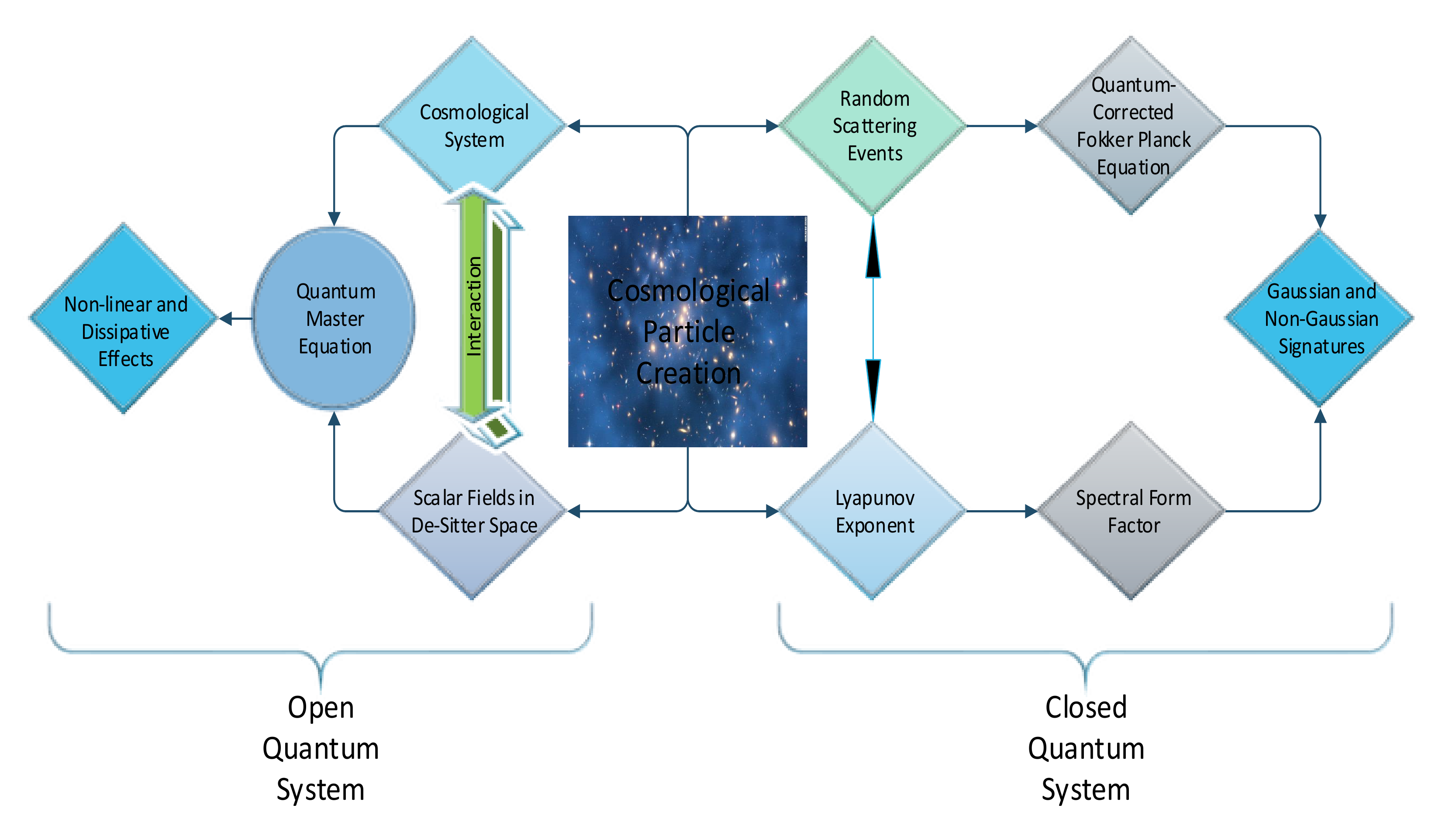}
	\caption{Overview of the computational strategy of the whole paper and how different parts are inter-related.}
	\label{chaos1}
\end{figure}

The specific problem we will discuss here is similar to one presented in ref.~\cite{Amin:2015ftc}. This problem is similar to that of a scattering problem in presence of impurity in quantum mechanics where the Schr{\"o}dinger equation yields approximate solutions to the wave-function of the particle which encounters a effective impurity potential barrier $V(x)$ of a given strength. The similarity in the following model is drawn between the current carrying electrons responsible for conduction in electrical wires to that of the particle creation in cosmology as a result of the non-adiabatic random events occurring in the early (inflation and reheating) stage of the universe. In this present problem for the sake of simplicity we consider an one dimensional conducting electrical wire, which implies that the current carrying electrons in the electrical wire has only a single propagating degree of freedom. As mentioned earlier, this has been considered to reduce clutter in our computation. But the similar problem can be generalized to more complicated situation~\footnote{For an example, one can generalize the same prescription in three space dimensions.}. Since, a current carrying wire consists of a large number of impurities, these act like the potential barriers $V(x)$, which are randomly distributed across the wire. Therefore, the motion of the electrons while confronting these scatterers gets hindered due to the presence of these randomly placed scatterers. One of the most important outcome of such an event is known as {\it Anderson Localization} as appearing in the context of condensed matter systems. Usually this is characterized by probability density of the localized wave-function:
\be \boxed{|\psi(x)|^2 \sim \exp{\left(|x|/\xi\right)}}~~,\ee with $ \xi $ being the localization length of the quantum mechanical wave-function $\psi(x)$. This phenomena of {\it Anderson Localization} usually occurs due to the interference of the waves scattered from the impurities present in the conduction wire. By formulating cosmological particle production as a random scattering problem, it has been shown in \cite{Amin:2015ftc} that {\it Anderson localization} maps to a problem of estimating exponential particle production, as given by:
\be \boxed{|\phi_{k}(\tau)|^2 \sim \exp{\left(\mu_{k}\tau\right)}}~~,\ee 
where $\mu_{k}$ is the mean particle production rate which is characterized by the conformal time dependent scalar field $\phi_{k}(\tau)$. A striking similarity has been observed between such scattering problems in conducting wires to that of the burst of particle production in cosmological random events shown in ref.~\cite{Amin:2015ftc}. In such cases, it has been observed that the solving a scattering problem in quantum mechanics using Schr{\"o}dinger equation is similar to solving a Klein-Gordon equation for a massless scalar field
in presence of a conformal time-dependent effective mass squared coupling parameter $ m^2(\tau)$. In this context the scalar field with time-dependent mass $ m^2(\tau) $ mimics the role of coupling strength parameter which characterizes the 
scattering to the massless scalar field in (quasi) de Sitter background. For more details see refs.~\cite{Bassett:1997gb,Zanchin:1997gf,Zanchin:1998fj}. Moreover, such stochasticity in a cosmological set up arises
due to the stochastic time evolution of Hubble parameter $ H(t) $, so that the inflaton (or the field participating in reheating) evolves with time stochastically due to the quantum fluctuations in the FLRW background. In the similar context the role of interacting scalar field has been studied to a great deal in ref.~\cite{Bassett:1997gb}.

 In this context,we have presented the  the amount by which the quantum mechanical system deviates with respect to the initial conditions. This means that more the value of this exponent, more is the chaos or stochastic randomness in the system under consideration in this paper. This exponent plays a significant role in our scenario as the number of particles produces in a given scattering event per unit time is random in nature. In a system of randomly spaced scatterers chaos emerges out of the random scattering events that an electron encounters while drifting across the wire with some drift velocity $\textit{v}$ within the conducting wire. The number that quantifies this increase in stochastic
       randomness or {\it chaos} in the system is the {\it Lyapunov Exponent}.  In refs.~\cite{Bassett:1997gb,Amin:2015ftc}, particle production phenomena in cosmological non-adiabatic events has been exclusively studied which yields the fact that the particle occupation number depends on Floquet indices $\mu_{k}$, which finally control the number of produced particles with the following number density:
        \bea \label{Occu1} n_{k}(\tau)= \int_{0}^{\infty} dk~k^2\exp{\left[2m(\tau)\mu_{k}\tau\right]}~~,\eea
         as well as the variances in the field fluctuation. The quantum fluctuations in the inflationary state of the universe results in the randomization of these bursts of particle production. The number density has been a random variable which is rendered stochastic due to the scattering events in the context of early universe cosmology. Our main objective in this paper to quantify this characteristic number for the massless scalar field having a conformal time-dependent mass coupling with it.  One of the prime reasons for finding a signature of chaos in such a system is the well known {\it thermalization} phenomena, which means that the FLRW background which embeds the massless scalar field into it is being thermalized by the massive field in interaction with the FLRW set up, which constantly being giving rise to a burst of particle production in the context of early universe cosmology. The scalar fields that we considering in our paper are said to be massive or heavy fields ($m\geq H$) which mimics the role of the scatterers in the Schr{\"o}dinger problem in quantum mechanics where the strength of the effective potential or the scatterer is given by the probability distribution function of the effective potential function. We draw a picturesque landscape by considering three distinct mass profiles:
 \bea \begin{array}{lll}\label{eq17a}
		\displaystyle   m^{2} (\tau)=\left\{\begin{array}{lll}
			\displaystyle  
			\frac{m_{0}^{2}}{2}\left[1 -\tanh (\rho \tau)\right]\,,~~~~~~~~~~~~ &
			\mbox{\small  \textcolor{red}{\bf  {Profile~ I }}}  \\ 
			\displaystyle  
			m_{0}^{2}~ {\rm sech}^2(\rho \tau)\,,~~~~~~~~~~~~ &
			\mbox{\small  \textcolor{red}{\bf  {Profile~ II }}}  \\ 
			\displaystyle   
			m_{0}^{2}~\Theta (-\tau).\,~~~~~~~~~~~~ &
			\mbox{\small \textcolor{red}{\bf  {Profile~ III }}}  
		\end{array}
		\right.
	\end{array},\eea
which exactly mimics the role of cosmological scatterers in early universe. We thereby investigate the momentum scale dependent behaviour of the Lyapunov exponent. In this context, the incoming momenta of the mode functions of the quantized massless scalar field having random interactions with the scatterer. In the following class of model, the Bogoliubov coefficients arise due to the interaction between massless scalar field with the heavy field. These Bogoliubov coefficients gives the information about the transmission coefficient viz.a.viz in similar problem to that of a scattering problem in quantum mechanics, that we solve using the well known WKB approximation technique~\footnote{To find approximate solution of the Schr{\"o}dinger equation (or in other words the Klein-Gordon field equation) in presence of an arbitrary impurity effective potential, (or the conformal time dependent mass coupling parameter) WKB approximation method plays crucial role \cite{Choudhury:2016cso,Choudhury:2016pfr}.}. These WKB solutions are extremely useful as it tell us the dynamical feature of the particle production in cosmological scattering events.

In continuation with this, we  discuss about the epoch of reheating which occurs after the end of inflationary stage of the universe which finally results in the stochastic random burst of particle production. The dynamics of these stochastic random
bursts of particle production can be well understood by using a {\it Fokker-Planck} equation, which gives us a statistical interpretation of the
number density of particles created per scattering event. Since, the number of particles created in a given non-adiabatic event is not discrete in nature but rather its random, which means that there must be a probability distribution function associated with the particle number. The various dynamical features of this type of probability distribution and its physical consequences has been studied in ref.~\cite{Amin:2015ftc}. It has been phenomenologically proposed in ref.~\cite{Amin:2015ftc} that such  probability density function would necessarily is Gaussian one. The occupation number of the produced particles, $n_{k}$, executes a drifting Brownian motion and a Fokker-Planck (FP) equation that evolves the probability distribution, $P(n_{k};\tau)$, emerging out of this Brownian motion has been studied in ref.~\cite{Amin:2015ftc}. We further compute the analytical expressions for the mean, variance and other higher order moments which are commonly known as, skewness and kurtosis and such additional statistical higher order moments are very useful to study the exact mathematical form and asymptotic limits of the probability distribution function. The evolution of mean, variance, skewness and kurtosis finally gives a coarse-grained analysis of the {\it Fokker-Planck} dynamics to more corrected orders of magnitude in quantum regime. We show in this paper explicitly that though Gaussianity is an inherent part of the probability density function, but the consideration of the higher order moments in the {\it Fokker-Planck} equation tells us that the density function may not be a Gaussian one but with some higher-order corrections entailed into it due to the quantum mechanical origin. Therefore, to a greater extent we extend the more corrected quantum version of the {\it Fokker-Planck} equation used to describe the dynamics of the probability distribution function used in ref.~\cite{Amin:2015ftc} that tells us the dynamics of the bursts of particle production in these random scattering events. The more quantum corrected version tells us that the probability amplitude of the particle production in the scattering events is more than a Log normal distribution. The distribution profile of the probability distribution function depends largely on the profile of the scatterer, i.e., the effective potential $V(x)$ in the Schr{\"o}dinger-like equation. While calculating the {\it Fokker-Planck} dynamics we observe that the skewness gives us a clue about the rate at which the particle production occurs meaning that longer the trailing part of the profile more is the number density of particles in the scattering event for a given time in the frame of the observer, whereas, kurtosis tells us the width of the probability distribution function which is essentially the amplitude with which the particle production phenomena occurs, which more suggestively tells us about the standard deviation of the density function from Gaussianity. This may be a signature of non-Gaussianity that arises in various models in early universe cosmology. 

In this connection it is important to note that, such stochastic approaches to the early universe scenario have been studied in details in  \cite{Linde:2002yd,Koks:1997rk}, where the authors give an account of how chaos arises  in the context of eternal inflation. As any rapidly oscillating classical field looses
its energy by creating pairs of elementary particles, these particles interact with each other and comes to a state of showing thermal behaviour at some temperature $T$. This implies that we must eliminate the necessary assumption of the universe being in thermal equilibrium.  This means that the inflating universe is rather thermal in the sense that the particle creation events that occurs during the quantum fluctuation in the randomly distributed scalar fields $\phi$ which results in a chaotic model of the inflationary scenario of the universe 
thereby leading to a generation of stochastic idea of the particle creation events during the {\it thermalization} of the quantum states of the field randomly distributed over the space-time. These particle creation events are more phenomenologically associated with one of the fundamental ideas in out-of-equilibrium statistical mechanics known as {\it Fokker-Planck equation} which gives the rate of the particle production during theses random events in stochastically emerging space-time along with the distribution function that this rate charts out. In
ref.~\cite{Amin:2015ftc}, such a phenomenon of particle creation events by the randomly spaced scatterers in due context of {\it cosmology} has been shown where the the statistics of the produced particles as a function of time which is the probability distribution function $P(n_{k},\tau)$ has been predicted to be following a Log-Normal distribution. The entire process have been carried out with the delta-scatterers which are localized in space-time. 

Following ref.~\cite{Amin:2015ftc}, in this paper we give a more improved quantum corrected version of the same approach to the probability density function of the particle production events and our prediction from the results show that the higher order quantum correction terms being included into the {\it Fokker-planck} equation introduces an approximation to the theory. This tells us that the number of particles produced in a given non-adiabatic event during the reheating stage of the universe is quantized, which would mean that the rate of particle production in a given event gives rise to a discrete set of  occupation number $n_{k}$. Furthermore, the quantum corrected terms obtained by deriving the  {\it Fokker-Planck} equation takes the general form, which is linear in $n_{k}$ being the first order in $\tau$. Using this information we calculate further the leading order, second and third order terms in the {\it Fokker-Planck} equation. Hence, we derive the analytic expression of the quantum corrected version of {\it Fokker-Planck} equation. We also calculate the various higher moments in order to get an overview of the nature of the solution of the quantum corrected {\it Fokker-Planck} equation which are - standard deviation, skewness and kurtosis which gives the hint of how the probability density function deviates from its Gaussian nature when the higher order quantum corrections are taken into account in the computation. This in turn may will be another indirect signature of the primordial non-Gaussianity in cosmology other than obtaining the signatures provided by the 3-point functions from scalar fluctuations.

Apart from that, we discuss about spectral form factor (SFF), which
measures the random distribution of eigen values of the energy hamiltonian of a chaotic system. For this computation of SFF we use the principles of random matrix theory (RMT) in this paper.  In the present context an upper bound on SFF denotes the saturation of eigen value distribution hence supports the ref.~\cite{Maldacena:2015waa} for quantum chaotic system. Within the framework of quantum physics, chaotic systems can be characterized using only some additional constraints. This theoretical approach is discussed in refs.~\cite{Dyson:1962es,Dyson:1970tza,Dyson:1962a,Dyson:II,Dyson:IV} and the authors  use the theory of random matrices to characterize quantum mechanical system. In this method, any arbitrarily complicated many-body Hamiltonian can be replaced by matrix of random numbers drawn from a Gaussian statistical ensemble. This random matrix approach towards quantum mechanics help to characterize and understand the underlying features of the chaotic random system. After studying the behaviour of SFF with time one can further comment that whether it is valid for a cosmological particle production event (semi-classical) or not. For our purpose we discuss generalized version of  SFF for different even order polynomial structure of random potential and then extend that result to describe the cosmological particle production events \cite{Maldacena:2016hyu,Garcia-Garcia:2016mno}. For any random potential we can use this method of SFF and we can deal with scatterer of any arbitrary type. For any such scatterer we can get a bound on randomness in the chaotic system characterised by SFF.  Also using specific transfer matrix for different conformal time dependent effective mass profiles which are precisely known in this paper, we can finally compute {\it Lyapunov exponent} which also measure stochastic randomness.

Also it is important to note that in ref.~\cite{Amin:2015ftc}, the scatterers were considered to be some localized potential functions in space-time. On the contrary the choice of our specific time dependent mass profiles mimics the role of thermalized fields or effective potential functions, which are playing the role of scatterers in this context. We see that the choice of these time dependent mass profiles leads to particle production which is chaotic in nature and therefore, to determine the rise of chaos in such a system we quantify as well as analyse chaos by a well known quantities known as the , {\it Lyapunov exponent}\cite{Mandal:2015kxi} and Spectral Form Factor (SFF)     \cite{Gaikwad:2017odv}. Here fusing the principles of random matrix theory (RMT) we provide a generalized bound on randomness (or stochasticity) for any general random scatterer whose potential can be expressed in terms of an even polynomial. More precisely, we provide a possible method to compute the degree of randomness in a chaotic system and from that one can check the bound on chaos. 

The plan of the paper is as follows - In \underline{\textcolor{purple}{section} {\ref{model}}} we discuss about the model which is responsible for the quantum description of chaos during the cosmological particle production and have similarities with the quantum mechanical problem of electrical conducting wire with impurities. In \underline{\textcolor{purple} {section} \ref{Calspecmass}}, we have presented the analytical expressions for the Bogoliubov coefficients, transmisson and reflection coefficients, Lyapunov exponent, conductance, and  resistance for different time dependent mass profile. We have discussed the correspondence between In \underline{\textcolor{purple}{section} \ref{SpecForm}} the specific role of Spectral Form Factor (SFF) to quantify chaos in the context of particle production rate is discussed. In \underline{\textcolor{purple}{section} \ref{QuamCorrected}} the particle production event with quantum corrected {\it Fokker-Planck} equation is discussed by taking contribution upto fourth order and also different higher order  moments from the quantum corrected probability density function are explicitly computed. Finally, in \underline{\textcolor{purple}{section} \ref{conc}} we conclude with the future prospect and physical impacts of our work.

Additionally it is important to note that, throughout this paper, we use natural system of units, $ \hbar = c = 1 $.

%%%%%%%%%%%5
%%%%%%%%%%%%%
%%%%%%%%%%%%%
%%%%%%%
%%%%%
%%5

\section{Modelling randomness in cosmology}	
\label{model}

The background model which we consider in this section to quantify quantum chaos in cosmology consists of a massless scalar field interacting with coupled with a background
scalar field with conformal time dependent mass profile which in principle have heavier or comparable to the Hubble scale ($m\geq H$) \cite{Arkani-Hamed:2015bza,Maldacena:2015bha,Choudhury:2016cso}.  It is important to note that such heavy mass profiles play significant role in finding various cosmological correlation functions and also can be treated as an additional probe to break the degeneracy between various models of inflation from the perspective of implementing cosmological perturbation theory in (quasi) de Sitter background. We know that in usual set up of primordial cosmological perturbation such heavy fields are not appearing in the low energy effective field theory action. For that case in the simplest situation we actually start with an one field set up where the kinetic term is canonical in nature and the field is minimally coupled with the background gravity which is treated to be classical usually. Also such field has an effective structure of the interaction potential which play crucial role to study the time dynamics in FLRW cosmological background. Here specifically the field is treated to be massless compared to the Hubble scale ($m<<H$). However, this is not the complete story yet. To explain this let us start with a Ultra Violet (UV) complete set up of quantum field theory (QFT) such as string theory in higher dimensions. There are various examples of string theory from which one can start the computation, which are - Type II A, Type II B, Heterotic, M - theory etc. Also the low energy extension of such theories (supergravity) are also useful for the computation in the context of cosmology. Here it is important to note such all such theories contain massive ($m>>H$), intermediate mass ($m\approx H$) and  massless ($m<<H$) fields in the matter multiplet. To write down an effective field theory (EFT) one need to integrate out all such heavy degrees of freedom from the UV complete version of the action~\footnote{\underline{\textcolor{red}{\bf Important notes:}}\\$~~~~~~~$Here we note the following points which are very useful to study the consequences from EFT set up:
\begin{enumerate}
\item  In this context, one can construct an EFT by utilizing the underlying symmetries appearing in the field theoretic set up. In such a generalized description where EFT is constructed by following the top down approach, we really don't care about the exact UV completion of the parent theory i.e. detailed quantum field theory origin at high energy scale of such effective constructions are not important in this case. See ref.~\cite{Cheung:2007st,Choudhury:2017glj} for more technical details.

\item In a more generalized prescription of EFT one can construct the set up which requires to correctly account for all relevant self interactions of adiabatic modes around and after the cosmological horizon crossing.  Specifically the adiabatic mode contains all types of EFT relevant operators, including transient reductions in the effective sound speed $c_S$ each time the background field undertakes non-geodesic motion in background target space. In an EFT framework with single field setting, where heavy directions are such that the mass of the field under consideration is heavy compared to the Hubble scale i.e. $m>>H$, one gets transient drops in the effective sound speed $c_S$  during slow roll if the potential is such that the field traverses a bend even if the parent theory consists of canonically normalized scalar fields. So for general consideration one can allow many more possibilities without following any restriction to time dependent mass profile. However, these three specific types of time dependent mass profiles are very popular in the context of the study of quantum critical quench in a analytical fashion. We have considered these profiles particularly as our future objective is to apply the idea of quantum quench in the context of De Sitter space to quantify randomness during reheating phase. It is important to note that, using a simple field redefinition at the level of quantum fluctuation to the Mukhanov-Sasaki variable which results in a time dependent mass for the rescaled variable appearing with additional contributions of the mathematical form, $\dot{c_S}/c_S\sim SH$. Here $S=\dot{c_S}/Hc_S$ is the associated slow roll parameter with the effective time dependent sound speed $c_S$ and $H$ is the Hubble parameter and it is associated with the changes in the radius of curvature of the inflaton trajectory. In this case the effective sound speed is given by, $c^{-2}_S\equiv 1+4\dot{\phi}^2/\kappa^2 M^2$, where $\kappa$ is the radius of curvature of the background inflaton ($\phi$) trajectory and $M$ is the effective cut-off scale of the EFT at high energy (UV) regime. Equivalently, it refers to the degree to which effective sound speed $c_S$ is reduced, which actually quantify the distance from the adiabatic minimum of the potential in the background inflaton trajectory is forced by its evolution. Each of these possibilities has different applications in the low energy limiting region of EFT.  When the effective sound speed $c_S<<1$ and $\dot{c}_S\sim 0$ is fixed over few e-folds of expansion then it is extremely difficult to maintain a meaningful derivative expansion without considering other types of special symmetries appearing in the set up. However, as mentioned earlier, within certain limits one can consider an adiabatic region where the effective sound speed $c_S<<1$ and $\dot{c}_S\sim c_S H$ and $S=\dot{c_S}/Hc_S\sim 1$ is fixed over a very small e-fold of expansion and this in turn generate all possible consistent transient strong coupling parameters without violating perturbative uniterity and these terms are explicitly appearing in the derivative expansion in the EFT. Consequently, the nature of these two types of features in the effective sound speed $c_S$ give rise to distinctive contributions to the physical observables studied in the EFT set up. The positive detection of these physical observables in different experiments allow to extract the underlying non-trivial physics from the EFT set up. In the technical ground the adiabatic mode is identified with the Goldstone boson, which is appearing due to spontaneously broken time translational symmetry prior to the path integration of the heavy fields. In this context, the invariance of the parent theory completely fixes the entire non-perturbative structure of all possible Wilsonian EFT operators and the associated coupling parameters can be expressed entirely by the effective sound speed $c_S$ of adiabatic perturbations, where the adiabaticity conditions $c_S<<1$ and $\dot{c}_S\sim c_S H$ are respected. In principle, $c_S$ can be computed in terms of the parameters of the parent theory. Thus the additional contributions appearing in the adiabatic limit 
$c_S<<1$ and $\dot{c}_S\sim c_S H$ directly justifies the validity of our treatment in this paper. For further technical details of this EFT set up see ref.~\cite{Achucarro:2010da, Achucarro:2012sm,Chluba:2015bqa}.

\end{enumerate}

}. After doing dimensional reduction along with applying various compactification techniques one can derive various types of UV complete effective field theories at cosmological scale where the effective couplings of various relevant and irrelevant Wilsonian operators have time dependent profile in FLRW background and in such a case from the relevant quadratic operator one can also get the time dependent effective mass which is in general heavy ($m\geq H$). It is further important to mention here that, such heavy fields can give rise to non vanishing one point function for scalar (curvature) perturbation in cosmology, which carries the signature of Bell's inequality violation in primordial universe \cite{Maldacena:2015bha, Choudhury:2016cso, Kanno:2017dci,Kanno:2016qcc,Kanno:2014bma,Kanno:2014ifa,Kanno:2014lma}. Also it is important to note that such Bell violating set up can be explained using the theory of quantum entanglement in (quasi) de Sitter background and can give rise to non-vanishing quantum information theoretic measure i.e. Von Neumann entropy, R$\acute{e}$nyi entropy, quantum discord, logarithmic entangled negativity \cite{Choudhury:2017bou,Choudhury:2017qyl,Maldacena:2012xp,Albrecht:2018prr} etc. Additionally, one can get correct expression for two point function and also the three point function from scalar (curvature) perturbation, which will show significant effect in estimating primordial non-Gaussianity from single field models of inflation. Apart from this one can consider a simplest situation in four space-time dimensions where the cosmological dynamics is explained in terms of two interacting scalar fields. The light field ($m<<H$) is participating in inflation and the other heavy field ($m>>H$) is participating to explain the dynamics of reheating. If we path integrate out the reheating degrees of freedom then we get an effective field theory of inflation which is exactly same as we have explained earlier. But here one can consider the other possibility as well in which one can path integrate out the light inflaton degrees of freedom and write down an effective field theory to describe reheating in terms of the heavy fields ($m\geq H$). In such a description this reheating field have mass and in the effective field theory description one can write down some time dependent coupling in terms of the integrated inflaton degrees of freedom and the mass of the reheating field appearing  in the coefficient of the relevant quadratic operator. In this description such time dependent coupling is  treated to be the time dependent effective mass parameter profile which is considered in the present discussion. So it is evident from this discussion that using both the effective field theory of inflation and reheating one can actually explain the origin of such time dependent effective mass profiles in four dimensions. However, in this paper since our objective is to study the cosmological particle production phenomena, we will mostly focus on the reheating epoch of the universe.

The dynamics of this fluctuating scalar field~\footnote{Here it is important to note that, for inflation this scalar field is actually massless and in the effective field theory description one can construct the time dependent effective mass profile. On the other hand, in the context of reheating the scalar field is massive and in the effective field theory description one can construct time dependent effective mass in terms of the original mass of the reheating field and other degrees of freedom which are integrated out from the original theory.} in FLRW cosmological background with a time-dependent
coupling obeys the following Klein-Gordon equation~\footnote{Here we have assumed that the effective sound speed parameter, $c_S=1$, which indirectly implies the fact that for background time evolution we are considering a single scalar field with canonical kinetic term minimally coupled to the gravity. Effective mass of the scalar field is $m(\tau)$, which has time dependent profile.  However, one can generalize this prescription for any general non-canonical single field (i.e.$P(X,\phi)$ theory) theoretic framework where the effective sound speed parameter $c_S\neq 1$. }:
\be
\boxed{\left[\frac{d^2 }{d\tau^2} + \left(k^2  +  m^2(\tau) \right)\right]\phi_{k}(\tau) = 0}~~,
\ee
where ${m^{2}(\tau)}$  is the time dependent mass of the scalar field with which is originating from the effective field theory (EFT) of massless scalar field coupled with other heavy degrees of freedom by following the two possibilities:
\begin{enumerate}
\item In EFT time dependent couplings are appearing after path integrating out the massive degrees of freedom. This prescription is usually used to construct a most generic \underline{\textcolor{red}{\bf EFT of inflation}}.
\item In EFT time dependent couplings are appearing after path integrating out the massless degrees of freedom. This prescription is usually used to construct a most generic \underline{\textcolor{red}{\bf EFT of reheating}}.
\end{enumerate}  
Here $\phi_{k}(\tau)$ is the associated Fourier mode of the fluctuating scalar field with momentum $k$, where it plays the role of wave number in the present context. 

In this paper,  our prime objective is to find a precise equivalence between the dynamics of this scalar field resulting in
stochastic particle production in cosmological events during reheating and the similarity with the dynamics of the electron transport
in conduction wires. To establish this equivalence we start with the fact that the above mentioned Klein-Gordon equation for the fluctuating scalar field in (quasi) de Sitter background shows a striking similarity with the time-independent one dimensional Schr{\"o}dinger equation appearing in the context of quantum mechanical system which describes the space evolution of electron inside a wire in presence of impurity as given by:
 \bea \left[\frac{d^2}{dx^2} +E-  V(x)\right] \psi(x)=0,
\eea
\begin{table}[htb]
\begin{tabular}{||c||c||}
\hline
\textcolor{red}{\bf Scattering in conduction wire} \hspace{0.66 cm} & \textcolor{red}{\bf Cosmological particle creation} \hspace{1.89 cm} \\
\hline
\end{tabular}
\begin{tabular}{||c|c|c|c||}
\hline
\textcolor{blue}{\bf Symbol} & \textcolor{blue}{\bf Physical interpretation} & \textcolor{blue}{\bf Symbol} & \textcolor{blue}{\bf Physical interpretation} \\
\hline
x & Distance  & $\tau$ & Conformal time \\
\hline
V(x) & Potential & -$m^{2}(\tau)$ & Time Dependent mass parameter \\
\hline
$\Psi(x)$ & Wave Function & $\phi_{k}(\tau)$  & Mode function in Fourier space\\
\hline
$N_{s}$ & No. of Scatterers & $N_{s}$  & No. of non-adiabatic events\\
\hline
$\Del$x & Distance between scatterers & $\Del\tau$  & Time between non-adiabatic events\\
\hline
$\xi $ & Localization length & $\mu_{k}$  &  Mean particle production rate \\
\hline
$\rho(x)$ & Resistance & $n_{k}(\tau)$  & Particle occupation number \\
\hline
$E$ & Energy eigen value & $k^2$  & Wave number of Fourier modes\\
\hline
$N_c$ & Number of channels & $N_f$  & Number of fields\\
\hline
\end{tabular}
\caption{A brief overview of the connection between the scattering problem in quantum mechanics to that of
	        cosmological particle creation events.}
\label{chaoswire}
\label{ghg}
\end{table}
where, $V(x)$ corresponds to the time-dependent potential which is appearing as an outcome of impurity in the electrical wire and plays the similar role of negative of the square of time-dependent mass profile as appearing in the context of cosmology i.e. $-m^2(\tau)$. Also $E$ represents the energy eigen value which mimics the role of the wave number squared  i.e. $k^2$. Finally, $\psi(x)$ represents the wave function of the quantum mechanical system under consideration which is similar to the Fourier modes of the time dependent fluctuating scalar field in the context of cosmology i.e. $\phi_{k}(\tau)$.
The above set up can be 
re-expressed in terms of solving a transfer matrix problem since the scatterers can be thought as potential profiles in Schr{\"o}dinger problem
in quantum mechanics with the incoming and outgoing modes of the scalar field related to each other with the
Bogoliubov coefficients. A complete overview of the connection between the variables that quantify the scattering problem in the context of
quantum mechanics to the one in the cosmological particle production problem has been shown in table~(\ref{ghg}).

It is very well known fact that the conductance of the electrical wire is related to the transmission probability of electrons across 
the wire and this can be obtained by explicitly solving the time-independent Schr{\"o}dinger equation (see Eq~(\ref{schr})) in the presence of the impurities. Before going to the further details of the computation here we begin by reviewing the scattering problem by a single impurity localized at the position  $ x = x_{j}$. To the left (L) and the right (R) of the impurity potential, the wave-function can be written as a linear combination of right-propagating waves ($\exp\left(ikx\right)$) and the left-propagating waves
($\exp\left(-ikx\right)$) as:
 \bea \label{schr}
\psi_{\Delta}(x) = \beta_{\Delta} \exp\left(ikx\right)+ \alpha_{\Delta} \exp\left(-ikx\right)~~~~~~~~{\rm where}~\Delta=L,R,
\eea
This is essentially a scattering problem in the context of quantum mechanics in which the impurities act as interaction potentials or
scatterers across which the electrons get transmitted within the conduction wire. The map between the Bogoliubov coefficients $(\beta_{R},\alpha_{R})$ from the right (R) side and the Bogoliubov coefficients $(\beta_{L},\alpha_{L})$ from the left (L) side can be expressed in terms of the following Bogoliubov transformation equation as:
 \bea {\cal B}_{R}={\cal M}_{j}~{\cal B}_{L},\eea
where we define:
  \bea {\cal B}_{\Delta}=\begin{pmatrix} \beta_{\Delta} \\ \alpha_{\Delta} \end{pmatrix}~~~~~~~~~{\rm where}~\Delta=L,R,
\eea
and in this context the {\it transfer matrix  } for the $j$-th scatterer $ {\cal M}_{j}$ is given by the following expression:
\bea {\cal M}_{j} =\large \begin{pmatrix} \large~~ \frac{1}{t^{*}_{j}}~~~~ & \frac{-r^{*}_{j}}{t^{*}_{j}}~~ \\ ~~\frac{-r_{j}}{t_{j}}~~~~ & \frac{1}{t_{j}}~~ \end{pmatrix},\eea
 which is essentially an unitary matrix related the incoming and the outgoing wave functions and their normalization coefficients.
 
Ultimately, using this methodology our objective is to connect several impurities together. This is particularly very easy to describe in terms of the {\it transfer matrix}
approach, since the total {\it transfer matrix} across $N_s$ number of scatterers is simply given by the simple matrix multiplication of the individual
transfer matrices as given by the following expression:
\bea {\cal M}\equiv{\cal M}(N_{s}) = \prod^{N_s}_{j=1}{\cal M}_{j}= {\cal M}_{N_{s}} \otimes {\cal M}_{N_{s}-1}\otimes......\otimes {\cal M}_{3}\otimes {\cal M}_{2}\otimes {\cal M}_{1}. \eea
For our choice of convenience of symbols we will drop the term $N_{s}$ for the $N_s$ number of scatterers and hence we 
will be considering this to be equal to ${\cal M}$.
\begin{figure}[htb]
	\includegraphics[width=16cm,height=8cm]{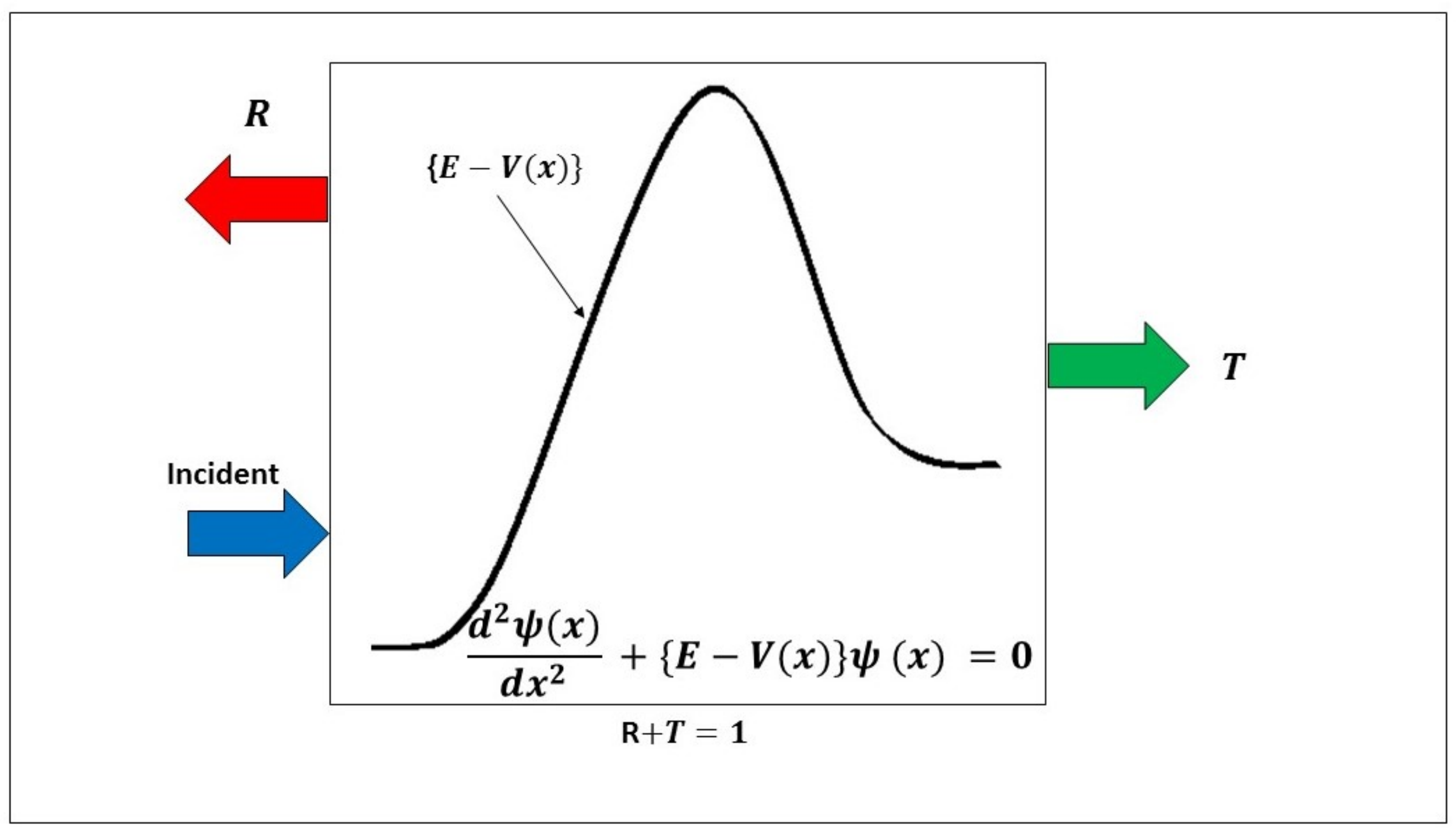}
	\caption{This diagram shows that incoming wave of electron encounter a scatterer and it partially passes through it with T(transmisson probability) and partially reflected back with R(reflection probability).}
	\label{scat2}
\end{figure}
In Fig:\ref{scat2} we show the electron(wave) encounter a potential(impurity or scatterer).It transmit and reflect through it.From simalirty of Klein-Gordon equation and the time-independent one dimensional Schr{\"o}dinger equation we calculate R and T for particle production event.
Further,  let us consider the simplest possibility of having two ($N_s=2$) scatterers across which the transmission probability can be written as:
\bea  T = \frac{T_{1}T_{2}}{| 1 - \sqrt{R_{1}R_{2}}e^{i\theta} |^2},
\eea
where the transmission and reflection coefficients for the $j-th$ scatterer can be expressed as: 
\bea T_{j} = t^*_{j}t_{j},~~~R_{j} = r^*_{j}r_{j}~~~~~\forall j=1,2.\eea  
and additionally $e^{i\theta}$ is the overall phase factor which describes the shift in phase 
between the reflecting waves across the scatterers due to the presence of impurities. If the distance between the two impurities is random in nature and uniformly distributed
over a region with the assumption, $ k\Delta x >> 1$ (where $\Delta x=x_2-x_1$ is the distance between the scatterers),  then the phase $\theta$ is also uniformly distributed over the interval $ 0 <\theta< 2\pi $. Using this fact explicitly we take
logarithm on both sides of the above equation and further doing average over the phase within the interval $ 0 <\theta< 2\pi $ we finally get~\footnote{Following this discussion, one can generalize this statement for $N_s$ number of scatterers as:
\bea \langle\log{T} \rangle_{\theta} =\log\left(\prod^{N_s}_{j=1}T_{j}\right)=\sum^{N_s}_{j=1}\log T_j.
\eea}:
\bea \langle\log{T} \rangle_{\theta} = \log{T_{1}} + \log{T_{2}}+\underbrace{2\langle\log \left|1-\sqrt{R_1R_2}e^{i\theta}\right|\rangle_{\theta}}_{=0}=\log\left(\prod^{2}_{j=1}T_{j}\right)=\sum^{2}_{j=1}\log T_j
\eea
The phase-averaged logarithm of the total transmission probability across $N_{s}$ number of scatterers then further simply can be written as:
\bea \langle\log{T} \rangle_{\theta} =\log\left(\prod^{N_s}_{j=1}T_{j}\right)=\sum^{N_s}_{j=1}\log T_j = -N_{s}\gamma,
\eea
where $\gamma$ is known as the {\it Lyapunov exponent}, which is defined as:
 \bea \label{connection2}
 \gamma = -N^{-1}_{s}\sum_{j=1}^{N_{s}}\log{T_{j}}= -N^{-1}_{s}\log\left(\prod^{N_s}_{j=1}T_{j}\right)=-N^{-1}_{s}\langle\log{T} \rangle_{\theta}.\eea  
 This actually 
determines the rise of chaos in the system. Using this information the {\it typical transmission probability} is defined as:
\bea T_{typ} \equiv \exp\left(\langle \log{T} \rangle_{\theta}\right)=\prod^{N_s}_{j=1}T_{j}=\exp\left(-N_s\gamma\right)=\exp\left(L/\xi\right), \eea 
which corresponds to the {\it most probable} transmission probability in the ensemble
of random potentials. 
Also it is important to note that,
\bea L \equiv N_{s}\Delta x=N_s\left(x_{N_s}-x_1\right),\eea represents the total length of the conduction wire. Here the {\it localization length} is defined as:
 \bea 
\label{localen} \xi \equiv \frac{\Delta x}{\gamma}=-L\left(\sum_{j=1}^{N_{s}}\log{T_{j}}\right)^{-1}=-L\left(\log\left(\prod^{N_s}_{j=1}T_{j}\right)\right)^{-1}=-L\left(\langle\log{T} \rangle_{\theta}\right)^{-1}. \eea
In one spacial dimension, the {\it localization length} is of the same order as the transport mean free path as pointed in ref.~\cite{Rolf Landauer (2006) Electrical resistance of disordered one-dimensional lattices,Localization distance and mean free path in one-dimensional disordered systems}. If the mean distance between scatterers, $\Delta x$, and the average logarithm of the transmission probability
per scattering, $\gamma$, are fixed, then the total transmission probability decays exponentially with the length $L$ of the conduction wire~\footnote{ Equivalently, here one can say that it exponentially decays with the number of scatterers.}. This is commonly known as {\it Anderson localization} \cite{Absence of Diffusion in Certain Random Lattices}.

Naturally,  it is well known that the resistance of the conduction wire scales inversely with the total transmission probability. At zero temperature, all one-dimensional conduction wire are therefore can be treated as an insulator, which is independent of the strength of the impurities appearing in the wire. However, the mathematical structure of the total transmission probability $T$ is preserved for $N_s$ number of such scatterers and this can be shown as:
\be
\boxed{{\cal M}=\large 
\begin{pmatrix} ~~\frac{1}{t^{*}_{N_s}} ~~~& \frac{-r^{*}_{N_s}}{t^{*}_{N_s}}~~ \\ ~~\frac{-r_{N_s}}{t_{N_s}}~~~ & \frac{1}{t_{N_s}}~~ \end{pmatrix}\otimes\cdots\otimes\begin{pmatrix} ~~\frac{1}{t^{*}_{3}}~~~ & \frac{-r^{*}_{3}}{t^{*}_{3}}~~ \\ ~~\frac{-r_{3}}{t_{3}}~~~ & \frac{1}{t_{3}}~~ \end{pmatrix}\otimes
\begin{pmatrix} ~~\frac{1}{t^{*}_{2}}~~~ & \frac{-r^{*}_{2}}{t^{*}_{2}}~~ \\ ~~\frac{-r_{2}}{t_{2}}~~~ & \frac{1}{t_{2}}~~ \end{pmatrix}\otimes
\begin{pmatrix}~~ \frac{1}{t^{*}_{1}}~~~ & \frac{-r^{*}_{1}}{t^{*}_{1}}~~ \\ ~~\frac{-r_{1}}{t_{1}}~~~ & \frac{1}{t_{1}} ~~\end{pmatrix}}~~.
\ee

For a one-dimensional non relativistic electron in conduction wire under the influence of certain potential $V(x)$ evolution of the wave function $\psi(x,t)$ is given by: 
\bea \left[\frac{d^2}{dx^2} +E-  V(x)\right] \psi(x)=0,
\eea
with the Hamiltonian for this particle is given by:
\bea H=\frac{p^{2}}{2m}+V(x)\eea
If we consider the particle is initially prepared in presence of potential $V(x)$ wave-packet take the specific form of $\psi(x,t)$. The final stationary density distribution $|\psi(x,t)|^{2}$ at long time carries important information both in their average and fluctuations. The quantum mechanical wave can tunnel through potential hills and reflect for by small fluctuations. So the initial wave packet split on each potential fluctuations into a transmitted and a reflected part. After huge number of scattering instances this reduce to a random walk problem and on average the motion at long times will have the diffusion constant in it. This is exactly the case of electron is propagating in a conduction wire. At long times average dynamics \cite{1005.0915} of the wave packet freeze and it takes the shape as given by the  equation
\bea |\psi(z,t)|^{2} \propto exp\left(-\frac{|z|}{\xi}\right),
\eea
Here, $\xi$ is the localization length as discussed in Eq:-\ref{localen}.
An electron in random potential is normally studied using statistical ensemble of random one-electron matrix Hamiltonians. Using {\it Tight Binding} approximation in orthogonalized lattice-site basis representation. The diagonal matrix elements are chosen from a flat probability distribution of width W,the strength of disorder. The off-diagonal hopping matrix elements for every pair of nearest-neighbour sites and represented by $2$x$2$ matrix where potential take the form,
\bea V=t^{0}.{\bf I}+i\mu {\bf t}.{\bf \sigma}=t^{0}.{\bf I}+i\mu(t^{x}.\sigma_{x}+t^{y}.\sigma_{y}+t^{z}.\sigma_{z})
\eea
where, ${\bf I}$, $\sigma_{x},\sigma_{y},\sigma_{z}$ are identity and Pauli spin matrices forming the complete basis set. Here $\mu$ is the random spin-orbit coupling strength, $t^{x},t^{y},t^{z}$ are independent random variables taken from uniform distribution on interval $[-1/2,1/2]$. The metal-insulator transformation occur at specific values. Below that mobility edges appear in band separating localised states near edges from extended states near band center.

The tight-binding random matrix ensembles (TBME) classified scheme is possible on symmetry. Orthogonal Ensemble in random potential..Localisation and mobility occur in all 3-D tight-binding ensembles and in 2-D for symplectic and unitary classes. From this distinction one found striking similarities with symmetry classification of Gaussian random matrix ensemble. The Gaussian ensemble belongs to high dimensionality limit of TBME and always metallic. So  the metallic phase is well approximated by Gaussian random matrix theory. From our discussion on RMT we use the Nearest Neighbour Spacing Distribution function [P($\omega$)see-Eq\ref{connect1}] and measure it in units of mean level spacing $\Delta$. Around the mobility edge and intermediate law from $P(\omega)$ can be obtained. Anderson localization in this context mimics quantum chaotic transition. The fluctuations for the density of states are partially responsible for the conductance fluctuations. Although average density of states is insensitive to Anderson transition its higher order moments are sensitive to it.
On an other approach we can relate {\it Anderson localization} to RMT using {\it Lyapunov Exponents}. Equation \ref{connection2} and \ref{localen} relates {\it Anderson localization} to {\it Lyapunov exponents}. Now statistical property of the {\it Lyapunov spectrum} with large number of degrees of freedom can be described universally by RMT. \cite{1702.06935} As described in \cite{1702.06935}, the spectrum of {\it Lyapunov exponents} is well approximated by the following expression:
\bea \rho(\lb,t)=\frac{3}{4\lb_{max}^{\frac{3}{2}}}\sqrt{\lb_{max}-|\lb|}
\eea
Here $\lb_{max}$ is the time independent parameter which approximately equals to bound of {\it Lyapunov exponent}. This equation shows striking similarity with Wigner law [Eq:-\ref{gaussian}]. In this approach we can also show the connection between {\it Anderson localization} and RMT. But there is a striking difference too. Random matrix theory takes all its entries from Gaussian random variables but for electronic models [Scattering matrix theory] matrix ensemble have short-ranged and sparse random matrix with most of the matrix elements having main diagonal non-zero.

\section{Randomness from conduction wire to cosmology: Dynamical study with time dependent protocols}
\label{Calspecmass}
In this section, our objective is to explain the various features from the time dependent effective mass profiles which are related to the quantum mechanical scattering problem in conduction wire as mentioned earlier. These features are appended bellow: 
\begin{enumerate}
\item \underline{\textcolor{red}{\bf Lyapunov exponent}:} ~It actually quantify the amount of chaos appearing in the quantum mechanical systems that we are studying in the context of early universe cosmology.
In our discussion it tells us the degree of randomness in the stochastic particle production. In our case, the chaos emerges due to the random scattering events which are non adiabatic and we call these as cosmological scattering events leading to particle production. 
In this section, we discuss about {\it Lyapunov exponent } and try to discuss their behaviour for the different time dependent mass profiles.
  In thus context,  {\it Lyapunov exponent } is defined as \cite{Amin:2015ftc,Localization1}:
  \bea \lambda = - {\rm log}~ T,
  \eea 
  where, $T$ is the transmission coefficient given by the following expression:
 \bea  T = t^{*}t=|t|^2,\eea   
 with $t$  and $t^{*}$  being the transmission amplitude of the incoming and the outgoing wave.
  In the present discussion, the transmission coefficient can be expressed as:
 \bea  \label{ew1}T = |t|^2=\frac{1}{|\alpha|^{2}},\eea
 where $\beta$ and $\alpha$ are the Bogoliubov coefficients. Also, it is important to note that, in the present context one can define the reflection coefficient as:
  \bea  \label{ew2} R=\tilde{ r}^{*}\tilde{ r}=|\tilde{r}|^2=\frac{|\beta|^2}{|\alpha|^{2}},\eea
  where $\tilde{ r}$  and $\tilde{ r}^{*}$  being the reflection amplitude of the incoming and the outgoing wave. Finally from Eq~(\ref{ew1}) and Eq~(\ref{ew2}), we get the following conservation equation:
 \bea R + T=|\tilde{r}|^2 + |t|^2=\frac{1+|\beta|^2}{|\alpha|^{2}}=1,\eea
  where we have used the following normalization condition for the Bogoliubov coefficients, as given by:
  \bea |\alpha|^2-|\beta|^2=1.\eea

\item \underline{\textcolor{red}{\bf Conductance}:} ~It quantify the degree of support of the flow of electron inside an electrical conduction wire. this is exactly reciprocal of resistance. In the present context, conductance refers to the ability of the massless scalar fields to transmit through the massive fields which are the
 specific heavy mass profiles that we have discussed above. This may be more suggestive in telling us about the interaction
 of the massless scalar field with the massive fields.
 More value of conductance refers to the larger transmitivity of the background fields through the scatterers and 
 vice-versa. Thus, conductance also carries a valuable information about the transmission coefficient of the scalar field
 interacting with the scatterer. In this context, the conductance can be expressed as:
  \bea G = \exp\left(-2\lb\right)=T^2=|t|^4=\frac{1}{|\alpha|^{4}}~~,
 \eea
 where $\lb$ is the {\it Lyapunov exponent}, $T$ is the transmission coefficient, $|t|$ is the transmission amplitude of the incoming/outgoing wave and $\beta,~\alpha$ are the Bogoliubov coefficients as mentioned above.
 \item \underline{\textcolor{red}{\bf Resistance}:} ~It quantify the degree of oppose of the flow of electrons inside an electrical conduction wire.
It is the property by the virtue of which the scatterers (which are the time dependent mass profiles in our case) resist the 
 massless scalar field to tunnel through them. In other words, it is the same Schr{\"o}dinger formulation in quantum mechanics 
 where the incoming wave interacts with a potential barrier and the strength of the barrier is the measurement of resistance to 
 the tunneling of the incoming particle through it. This means that more the resistance to the incoming wave, more is the 
 lower is the transmission probability across the barrier.
 Resistance is defined as the reciprocal of conductance $G(k)$, which gives:
  \bea r(k) = \frac{1}{G(k)}=\exp\left(2\lb\right)=\frac{1}{T^2}=\frac{1}{|t|^4}=|\alpha|^{4}~~.
 \eea
\end{enumerate}
We will discuss details of these features for three different mass profiles as mentioned in Eq~(\ref{eq17a}). All of these mass profiles that we choose here mimics the role of scatterers inside the conduction wire. 
Such scatterers provide the way for scattering events to occur resulting in random particle production in cosmological space-time. 

To study the cosmological particle creation problem during early epoch of universe (specifically during reheating) we use the analogy with the quantum mechanical scattering problem inside an electrical conduction wire in presence of time dependent effective mass profile we will perform the computation in (quasi) de Sitter space using FLRW spatially flat metric.

Here we consider a massive free scalar field with time-dependent mass ~\footnote{Here it is important to note that our approach is similar to 
that of used in refs.~\cite{Mandal:2015kxi, Das:2014hqa} to explain the time dynamics of quantum quench. }:
 \bea
S&=& -\frac12 \int d^4 x \sqrt{-g}(g^{\mu\nu}\partial_\mu \chi~ \partial_\nu \chi - m^2(\tau)\chi^2)\nonumber\\
&=& \frac12 \int d^3 x~d\tau~ a^2(\tau)\left[\left(\frac{\partial\chi({\bf x},\tau)}{\partial\tau}\right)^2-a^2(\tau)\left\{({\bf \nabla}\chi({\bf x},\tau))^2 +m^2(\tau)(\chi({\bf x},\tau))^2\right\}\right]\nonumber\\
&=& \frac12 \int \frac{d^3k}{(2\pi)^3}d\tau ~a^2(\tau)\left[\left|\frac{d\chi_{k}(\tau)}{d\tau}\right|^2 - a^2(\tau)(k^2+ m^2(\tau))|\chi_{k}(\tau)|^2\right], 
\label{scalar-action}
\eea
where the scalar field satisfies the following constraint:
 \bea \chi(-k,\tau)= \chi^*(k,\tau),\eea
and the Fourier transform of the field is defined as:
 \bea \chi({\bf x},\tau)= \int \frac{d^3k}{(2\pi)^3}\ \chi_{k}(\tau)\ e^{i{\bf k}. {\bf x}}.\eea
Also in the (quasi) de Sitter background the scale factor $a(\tau)$ can be expressed in terms of conformal time as~\footnote{In de Sitter and quasi de Sitter space one can compute the relation between the conformal time ($\tau$) and the physical time ($t$) as given by the following expressions: \bea\begin{array}{lll}\label{eqfcc}
		\displaystyle   \tau=\int \frac{dt}{a}=\left\{\begin{array}{lll}
			\displaystyle  
			-\frac{1}{Ha}=-\frac{1}{H}\exp(-Ht)\,,~~~~~~~~~~~~ &
			\mbox{\small  \textcolor{red}{\bf  {De Sitter}}}  \\ 
			\displaystyle  
			-\frac{1}{Ha}(1+\epsilon)=-\frac{1}{H}(1+\epsilon)\exp(-Ht)\,,~~~~~~~~~~~~ &
			\mbox{\small  \textcolor{red}{\bf  {Quasi De Sitter }}}  
		\end{array}
		\right.
	\end{array}~~,\eea}:
 \bea\begin{array}{lll}\label{eqf}
		\displaystyle   a(\tau)=\left\{\begin{array}{lll}
			\displaystyle  
			-\frac{1}{H\tau}\,,~~~~~~~~~~~~ &
			\mbox{\small  \textcolor{red}{\bf  {De Sitter}}}  \\ 
			\displaystyle  
			-\frac{1}{H\tau}(1+\epsilon)\,,~~~~~~~~~~~~ &
			\mbox{\small  \textcolor{red}{\bf  {Quasi De Sitter }}}  
		\end{array}
		\right.
	\end{array},\eea
	where $\epsilon$ is the slow-roll parameter in quasi de Sitter space, which is defined as:
	 \bea \epsilon=-\frac{1}{H^2}\frac{dH}{dt}=-\frac{1}{a(\tau)H^2}\frac{dH}{d\tau}\approx \frac{\bar{\epsilon}}{a(\tau)}=-H\bar{\epsilon}\tau. \eea
	Here, we define a new slow-roll parameter with respect to the conformal time:
	 \bea\bar{\epsilon}=-\frac{1}{H^2}\frac{dH}{d\tau}.\eea
Now we use the following field redefinition in Fourier space:
 \bea \phi_{k}(\tau)\equiv a(\tau)~\chi_{k}(\tau).\eea
Consequently, the scalar field action as stated in Eq~(\ref{scalar-action}) can be recast in terms of the newly defined field $\phi_{k}(\tau)$ as:
 \bea S
= \frac12 \int \frac{d^3k}{(2\pi)^3}d\tau ~\left(\left|\frac{d\phi_{k}(\tau)}{d\tau}-\frac{1}{a(\tau)}\frac{da(\tau)}{d\tau}\phi_{k}(\tau)\right|^2 - (k^2+ m^2(\tau))|\phi_{k}(\tau)|^2\right) 
\label{scalar-action},
\eea
Further, varying the above action with respect to the redefined field $\phi^{*}_{k}(\tau)$ we get the following equation of motion:
 \bea \label{p1}\left[\frac{d^2}{d\tau^2}+\frac{1}{a(\tau)}\frac{da(\tau)}{d\tau}\frac{d}{d\tau}+\left(k^2+m^2(\tau)-\left(\frac{1}{a(\tau)}\frac{da(\tau)}{d\tau}\right)^2\right)\right]\phi_{k}(\tau)=0.\eea
Further, Eq~(\ref{p1}) can be simplified for de Sitter and quasi de Sitter space as:
 \bea \label{pv1}\underline{\textcolor{red}{\bf De~ Sitter:}}\nonumber~~~~~~~~~~~~~~~~~~~~~~~~~~~~~~~~~~~~~~~~~\quad\quad\quad~~~~~~~~~~~~~~~~~~~~~~~~~~~~~~~~~~~~~~\\
 \left[\frac{d^2}{d\tau^2}-\frac{1}{\tau}\frac{d}{d\tau}+\left(k^2+m^2(\tau)-\frac{1}{\tau^2}\right)\right]\phi_{k}(\tau)=0.~~~~~~~~~~\\
\label{pv1}\underline{\textcolor{red}{\bf Quasi~De~ Sitter:}}\nonumber~~~~~~~~~~~~~~~~~~~~~~~~~~~~~~~~~~~~~~~~~\quad\quad\quad~~~~~~~~~~~~~~~~~~~~~~~~~~~~~~~~~~~~~~\\
\left[\frac{d^2}{d\tau^2}-\frac{1}{\tau}\left(1-\frac{2\epsilon^2}{1+\epsilon}\right)\frac{d}{d\tau}+\left(k^2+m^2(\tau)-\frac{1}{\tau^2}\left(1-\frac{2\epsilon^2}{1+\epsilon}\right)^2\right)\right]\phi_{k}(\tau)=0.~~~~~~~~~~\eea
It is important to note that, the
main contribution to particle production is originating from the excitations of the field  with $k/a >> m>>H$, at the stage of oscillations. Therefore, in the first approximation
we can neglect the expansion of the Universe, taking the scale factor $a(\tau)$ as a constant during reheating. We call it \underline{\textcolor{red}{\bf reheating approximation}}~\footnote{\underline{\textcolor{red}{\bf Important note:}}~In the present context, the analysis is perfectly valid for the highly localized particle production events after neglecting the cosmological expansion during \underline{\textcolor{red}{\bf reheating approximation}}. But this approximation fail for the events that are sufficiently spaced out. If we don't neglect the cosmological expansion in this computation then the conformal time dependent mass term of the form $2\tau^2$ is restored from the background cosmological background. This actually implies that the scattering problem is being performed  on a conformal time dependent potential of the form $1/r^2$ (inverse square), which makes the analytic computations of the Bogoliubov coefficients and all the other derived physical quantities to quantify quantum randomness from the present set up extremely difficult. Here it is important to note that, for long wavelength cosmological observables particle production  appears more than an e-fold apart and consequently the corrections appearing due to the cosmological expansion seem certainly relevant in the computation as the incoming and outgoing wave functions depart from plane waves. Although for localized particle production events, the \underline{\textcolor{red}{\bf reheating approximation}} considered in this paper perfectly holds good. In the present context this approximation breaks when we consider the particle production events for a sustained period of time or may be separated by times approaching an e-fold expansion.
} Consequently, one can approximately write Eq~(\ref{p1}) in the following simplified form~\footnote{Here it is important to note that, since the scale factor $a(\tau)$ is approximately a constant during reheating (\underline{\textcolor{red}{\bf reheating approximation}}), then conformal time ($\tau$) and the physical time ($t$) is related through the following coordinate rescaling transformation:
 \bea \tau=\int \frac{dt}{a}=\frac{t}{a}.\eea
}:
  \bea \label{p21}\left[\frac{d^2}{d\tau^2}+\left(k^2+m^2(\tau)\right)\right]\phi_{k}(\tau)=0.\eea
 The  Fourier modes of the scalar field follow the equation of motion in as stated in Eq~\ref{p21}, with every Fourier mode 
satisfying the Schr{\"o}dinger equation where $-m^2(\tau)$ playing the role of a potential.
\begin{figure}[htb]
	\includegraphics[width=16cm,height=8cm]{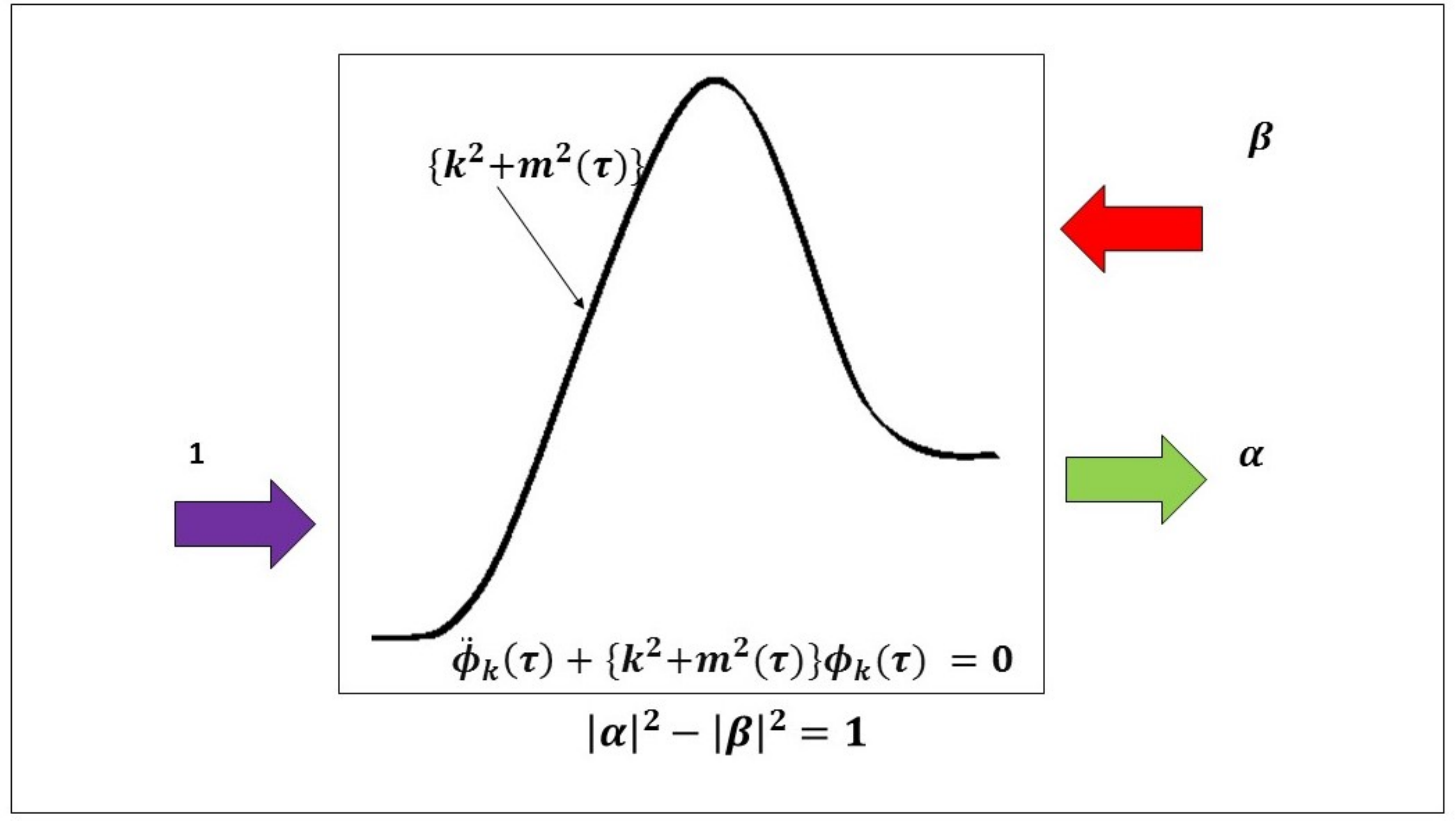}
	\caption{This diagram shows that ground state fluctuations from the past can in future be amplified which can be measured by the coefficient $\alpha$ whereas particle excitation from ground state can be measured by $\beta$.}
	\label{scat3}
\end{figure}
In Fig:\ref{scat3} the particle produced show fluctuation from ground state and from calculating the Bogoliubov coefficients we predicted all its properties.
For the solution we refer to ref.~\cite{Birrell:1982ix},
 for the field $\phi_{k}(\tau)$ can be expressed in two distinctive ways, as given by:
 \bea \phi_{k}(\tau)&= a_{in}(k) u_{in}(k,\tau) +  a^\dagger_{in}(-k) u^*_{in}(-k,\tau) \nonumber\\
&= a_{out}(k) u_{out}(k,\tau) + a^\dagger_{out}(-k) u^*_{out}(-k,\tau), 
\label{in-out}
\eea
where $u_{in,in}(k,\tau)$ and $u_{in,out}(k,\tau)$ are the `ingoing' and `outgoing' wave-functions.  Also, the in- and out- oscillators are related to
each other through the Bogoliubov coefficients $\ag(k)$ and $\bg(k)$
 \bea
a_{in}(k)&= \alpha^*(k) a_{out}(k) - \beta^*(k) a^\dagger_{out}(-k), \nonumber\\
a_{out}(k)&= \alpha(k) a_{in}(k) + \beta^*(k) a^\dagger_{in}(-k),   
\label{bogo}
\eea
Now, we calculate the various electrical properties and also the expression for the {\it Lyapunov exponent} to quantify quantum chaos for the various time dependent effective mass profiles which are equivalent to the impurity potential term in the time Independent Schr{\"o}dinger Equation describing a scattering problem inside a conduction wire.

 \subsection{Protocol I:~$m^{2} (\tau) = m^2_{0}(1 -\tanh (\rho \tau))/2$}
 \begin{figure}[htb]
\centering
\subfigure[$m^2(\tau)$ vs $\tau$ profile.]{	
	\includegraphics[width=7.8cm,height=8cm] {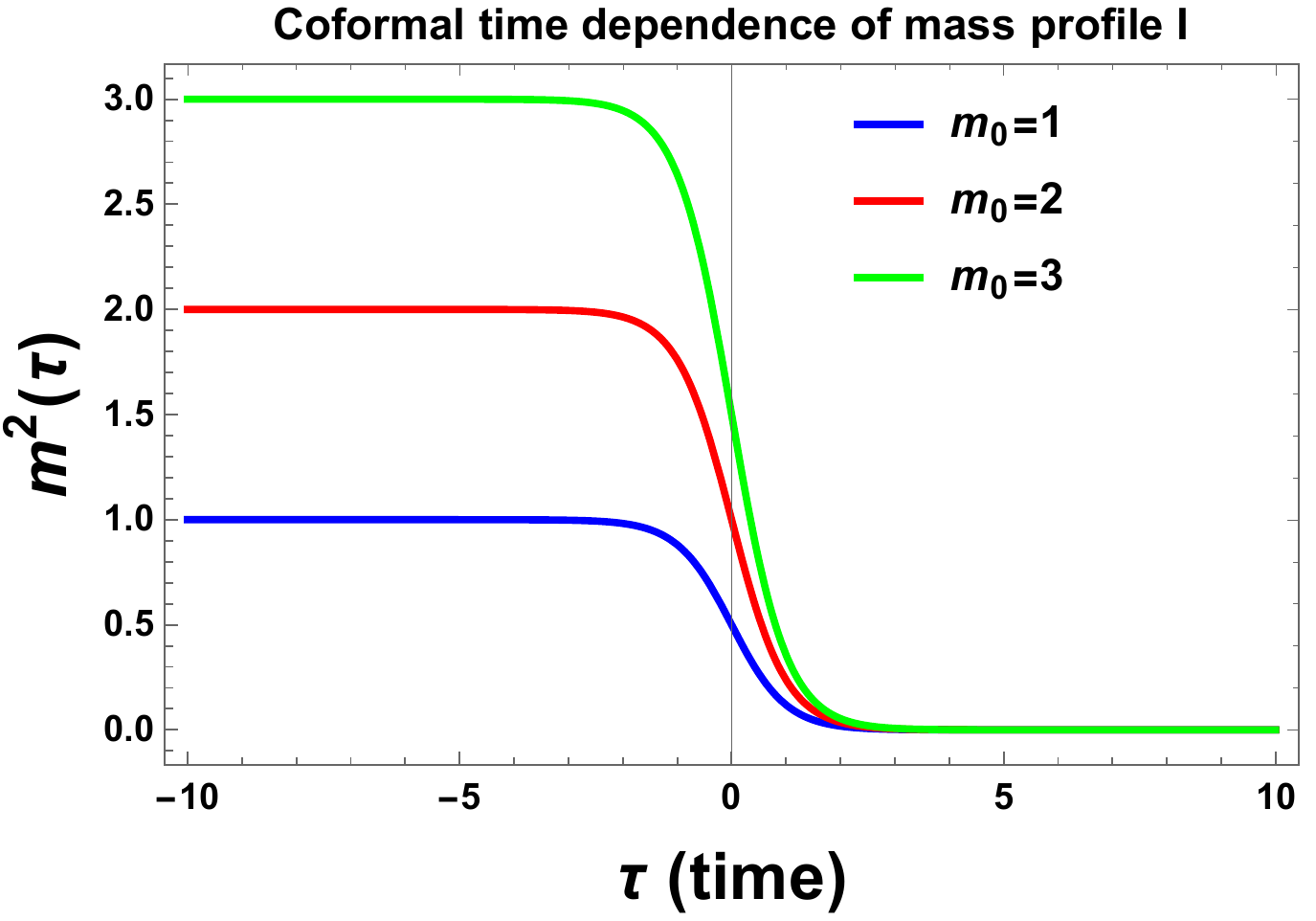}
	\label{gq1}
}
\subfigure[$V(\tau)$ vs $\tau$ profile.]{	
	\includegraphics[width=7.8cm,height=8cm] {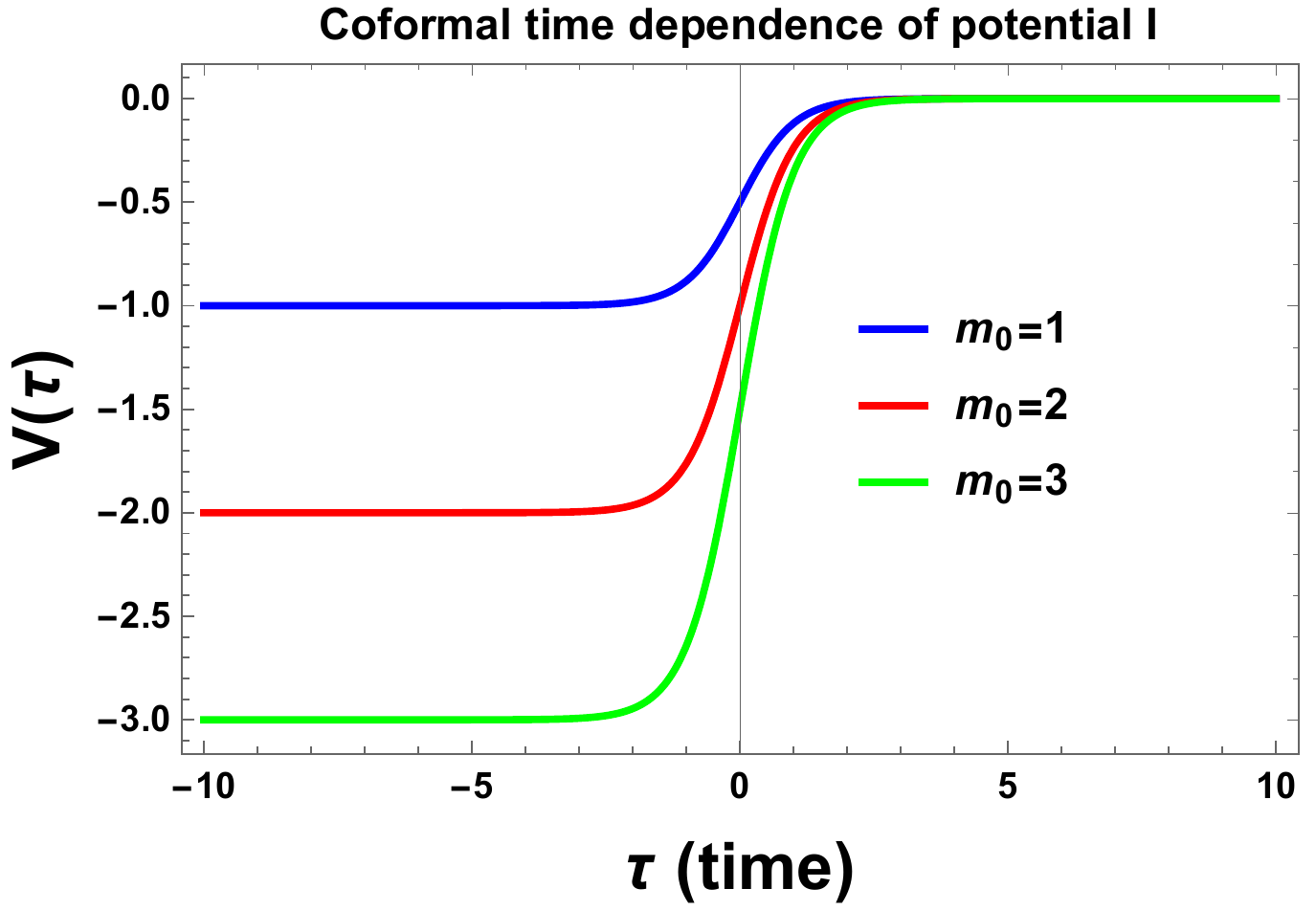}
	\label{gq2}
}
\caption{Conformal time dependent behaviour of the mass profile I and its corresponding potential used in Schr{\"o}dinger scattering problem is explicitly shown here. Here we fix $\rho=1$.}
\end{figure}
Here we start with the following mass profile: 
 \bea m^{2} (\tau) = m^2_{0}(1 -\tanh (\rho \tau))/2.\eea
  The corresponding Schr{\"o}dinger problem for this potential function can be solved by using the potential function as given bellow:
  \bea V(\tau)=-m^2(\tau)= -m^2_{0}(1 -\tanh (\rho \tau))/2.\eea
  In fig.~(\ref{gq1}) and fig.~(\ref{gq2}), we have explicitly shown the conformal time dependent behaviour of the mass profile under consideration and also the corresponding potential used in Schr{\"o}dinger scattering problem.
  
  We can find the following explicit solutions for
$u_{in}(k,t)$ and $u_{out}(k,t)$, as given by:
 \bea u_{in}(k,\tau) = \frac{e^{-i\omega_{in}\tau}}{\sqrt{2 \omega_{in}}} \; _2F_1\left(\frac{i\omega_-}{\rho },-\frac{i\omega_+}{\rho };1-\frac{i\omega_{in}}{\rho };-e^{2\rho \tau}\right), 
\\
u_{out}(k,\tau) =\frac{e^{-i\omega_{out}\tau}}{\sqrt{2 \omega_{out}}} \; _2F_1\left(\frac{i\omega_-}{\rho}, \frac{i\omega_+}{\rho};\frac{i\omega_{out}}{\rho}+1;-e^{-2\rho \tau}\right),
\eea
where we define $\omega_{\pm}$, $\omega_{in}$ and $\omega_{out}$ in the following:
 \bea\omega_{in}=\sqrt{k^2+m_0^2},\quad \omega_{out}=|k|, 
\quad \omega_{\pm}=\frac{1}{2}(\omega_{out}\pm\omega_{in}).
\eea
 \subsubsection{Bogoliubov coefficients}
For this specific mass profile the Bogoliubov coefficients can be expressed as:
 \bea\alpha(k) =\sqrt{\frac{\omega_{out}}{\omega_{in}}}\; \frac{\Gamma \left(-\frac{i \og_{out}}{\rho }\right) \Gamma \left(1-\frac{i\og_{in}}{\rho }\right)}{\Gamma \left(-\frac{i \og_+}{2 \rho }\right) \Gamma \left(1-\frac{i \og_+}{2 \rho }\right)}, ~~~
\beta(k) =\sqrt{\frac{\omega_{out}}{\omega_{in}}}\;  \frac{\Gamma \left(\frac{i \og_{out}}{\rho }\right) \Gamma \left(1-\frac{i \og_{in}}{\rho }\right)}{\Gamma \left(\frac{i \og_-}{2 \rho }\right)\Gamma \left(1+ \frac{i \og_-}{2 \rho }\right)}.~~~~\eea

%The Bogoliubov Coefficients for the above may be given as:
%\be
%\alpha (k)=\frac{\sqrt{\frac{k}{\sqrt{k^{2}+m_{0}^{2}}}} \Gamma (-i k) \Gamma (1-\sqrt{-k^{2}-m_{0}^{2}})}
%{\Gamma (-\frac{i k}{4}-\frac{1}{4} \sqrt{-k^{2}-m_{0}^{2}}) \Gamma (-\frac{i k}{4}-\frac{1}{4} \sqrt{-k^{2}-m_{0}^{2}}+1)}
%\ee
%and,
%\be
%\beta(k) = \frac{\sqrt{\frac{k}{\sqrt{k^2+m_{0}^2}}} \Gamma (i k) \Gamma (1-\sqrt{-k^2-m_{0}^2})}{\Gamma (\frac{i k}{4}-\frac{1}{4} \sqrt{-k^2-m_{0}^2}) \Gamma (\frac{i k}{4}-\frac{1}{4} \sqrt{-k^2-m_{0}^2}+1)}
%\ee
\begin{figure}[htb]
\centering
\subfigure[$\ag$ vs $k$ profile.]{	
	\includegraphics[width=7.8cm,height=8cm] {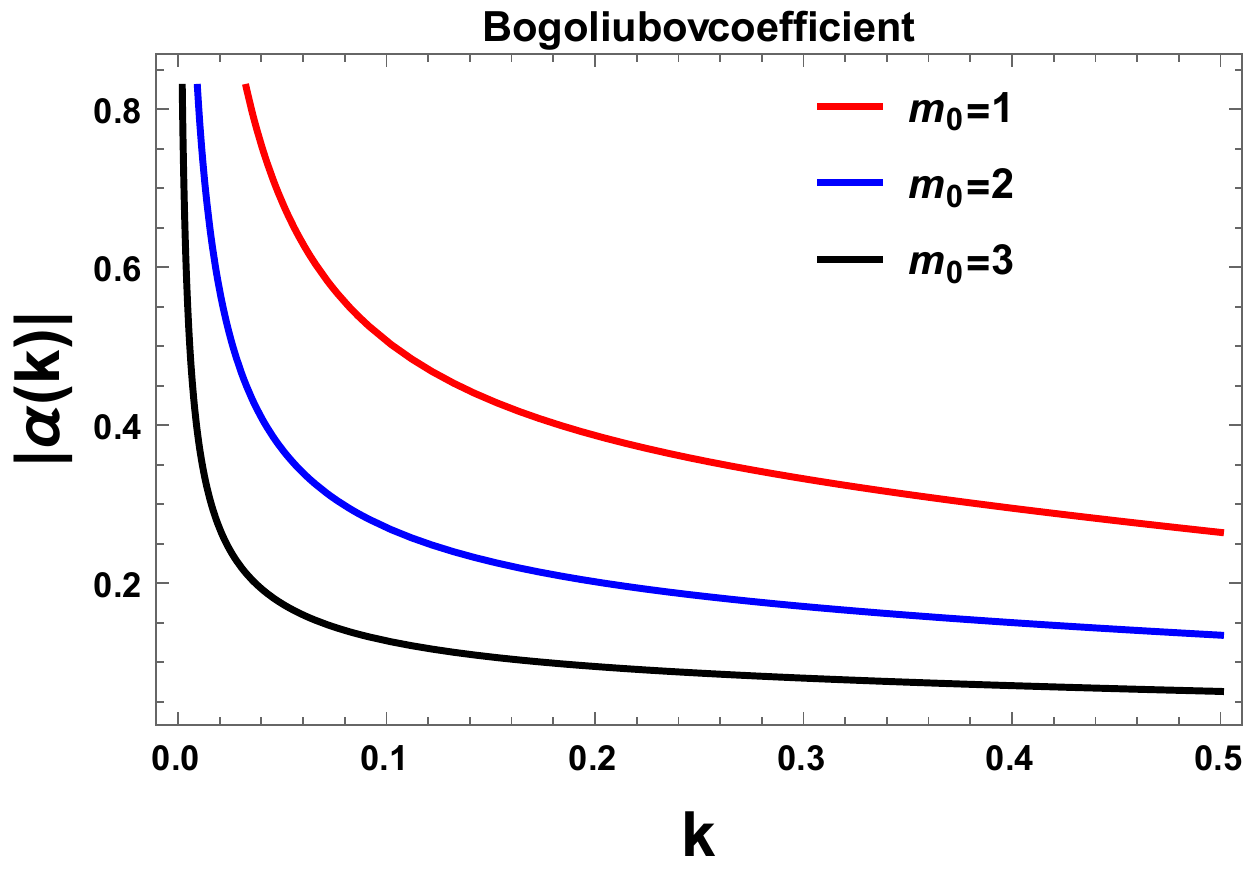}
	\label{BCA1}
}
\subfigure[$\bg$ vs $k$ profile.]{	
	\includegraphics[width=7.8cm,height=8cm] {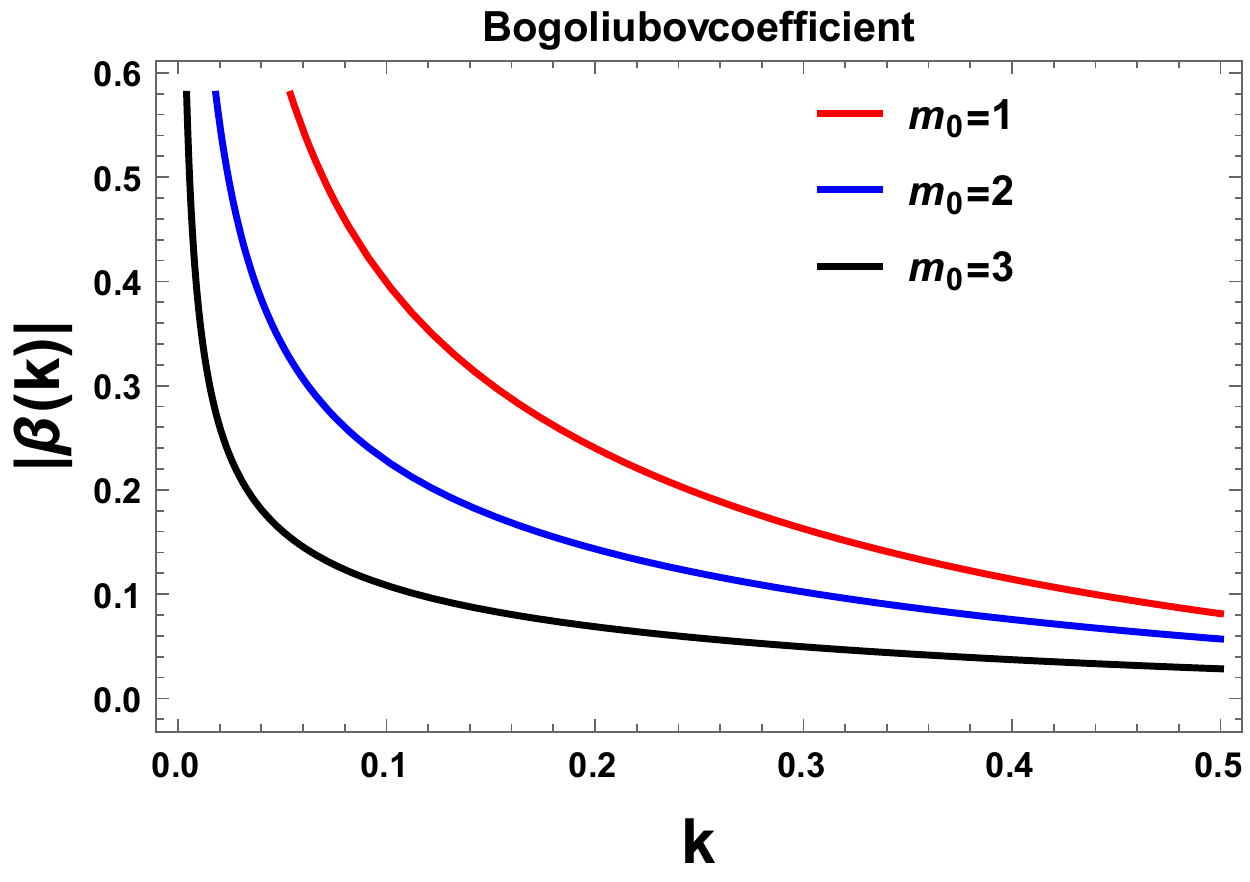}
	\label{BCB1}
}
\caption{Wave number dependence of the Bogoliubov coefficients from the mass profile I is shown here. Here we fix $\rho=1$.}
\end{figure}

In fig.~(\ref{BCA1}) and fig.~(\ref{BCB1}), we have shown the variation of the Bogoliubov Coefficients with wave number $k$. 

 \subsubsection{Optical properties: Reflection and transmission coefficients}
For this specific mass profile the  transmission and reflection coefficients can be expressed as:
 \bea &&T =\frac{1}{ | \alpha (k) |^2}=\frac{\omega_{in}}{\omega_{out}}\frac{|\Gamma \left(-\frac{i \og_+}{2 \rho }\right) \Gamma \left(1-\frac{i \og_+}{2 \rho }\right)|^2}{|\Gamma \left(-\frac{i \og_{out}}{\rho }\right) \Gamma \left(1-\frac{i\og_{in}}{\rho }\right)|^2},\\
&&R = \frac{| \beta (k) |^2}{| \alpha (k) |^2}=\frac{|\Gamma \left(\frac{i \og_{out}}{\rho }\right)|^2}{|\Gamma \left(-\frac{i \og_{out}}{\rho }\right) |^2}\frac{|\Gamma \left(-\frac{i \og_+}{2 \rho }\right) \Gamma \left(1-\frac{i \og_+}{2 \rho }\right)|^2}{|\Gamma \left(\frac{i \og_-}{2 \rho }\right)\Gamma \left(1+ \frac{i \og_-}{2 \rho }\right)|^2}.\eea
\begin{figure}[htb]
\centering
\subfigure[$T(k)$ vs $k$ plot.]{
	\includegraphics[width=7.8cm,height=8cm] {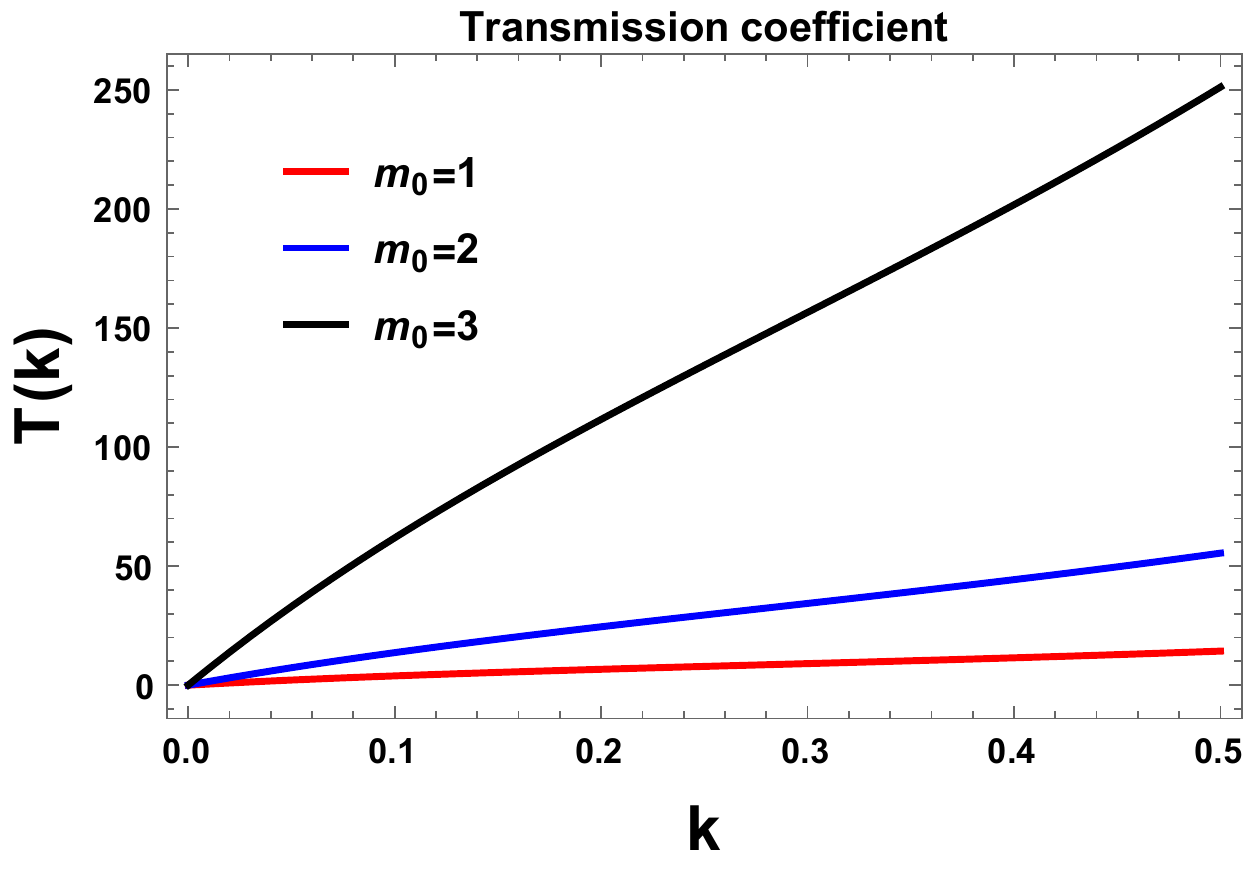}
	\label{TRC1}
}
\subfigure[$R(k)$ vs $k$ plot.]{
	\includegraphics[width=7.8cm,height=8cm] {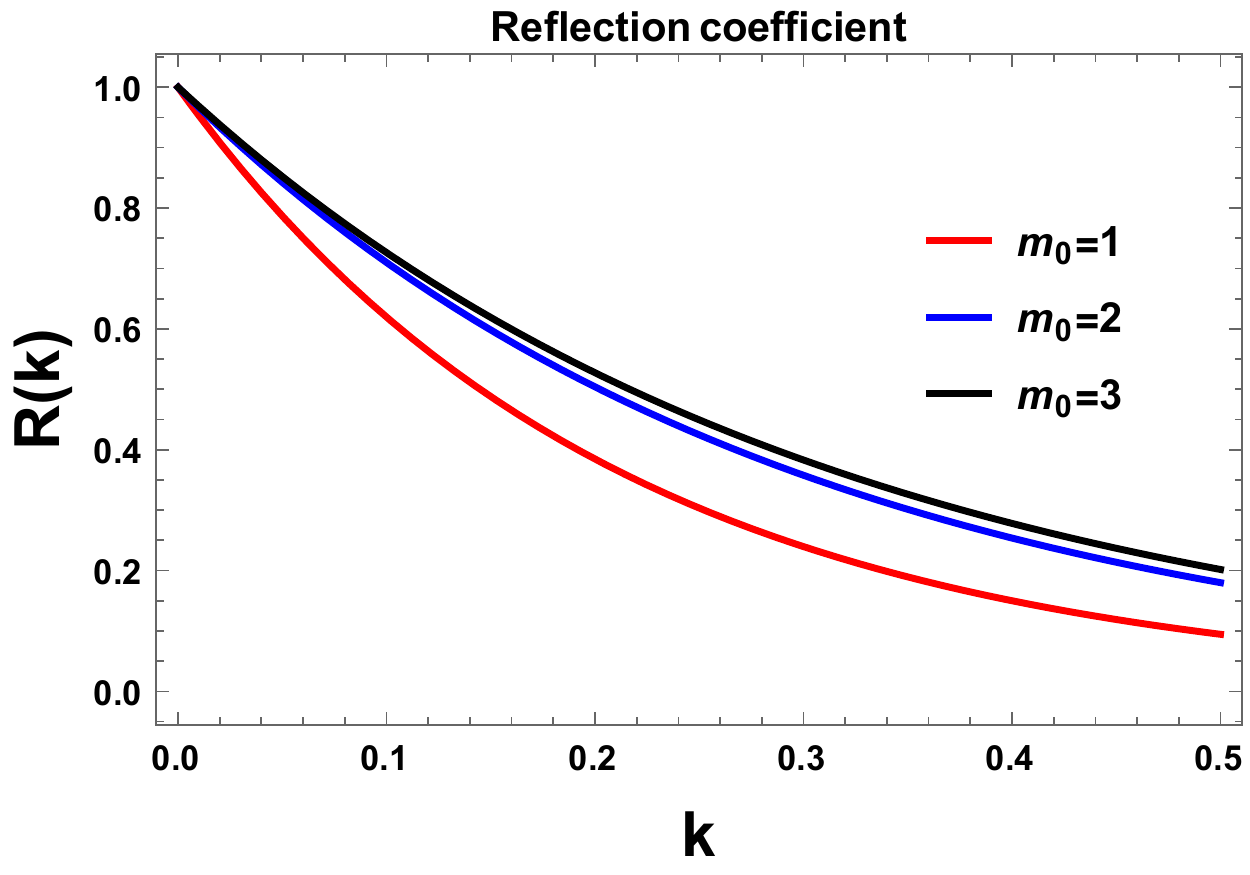}
	\label{REC1}
}
\label{transref1}
\caption{Wave number dependence of transmission and reflection  coefficients for the mass profile I is shown here. Here we fix $\rho=1$. }
\end{figure}
In fig.~(\ref{TRC1}) and fig.~(\ref{REC1}), we have shown the variation of the transmission and reflection coefficients with wave number $k$. 
 \subsubsection{Chaotic property: Lyapunov exponent}

 For this specific mass profile the {\it  Lyapunov exponent} can be expressed as:
 \bea\lb(k) = -\log~T=2\log| \alpha (k)|=2\log\left| \sqrt{\frac{\omega_{out}}{\omega_{in}}}\; \frac{\Gamma \left(-\frac{i \og_{out}}{\rho }\right) \Gamma \left(1-\frac{i\og_{in}}{\rho }\right)}{\Gamma \left(-\frac{i \og_+}{2 \rho }\right) \Gamma \left(1-\frac{i \og_+}{2 \rho }\right)}\right|.
\eea

\begin{figure}[htb]
\centering
{
    \includegraphics[width=13cm,height=8cm]{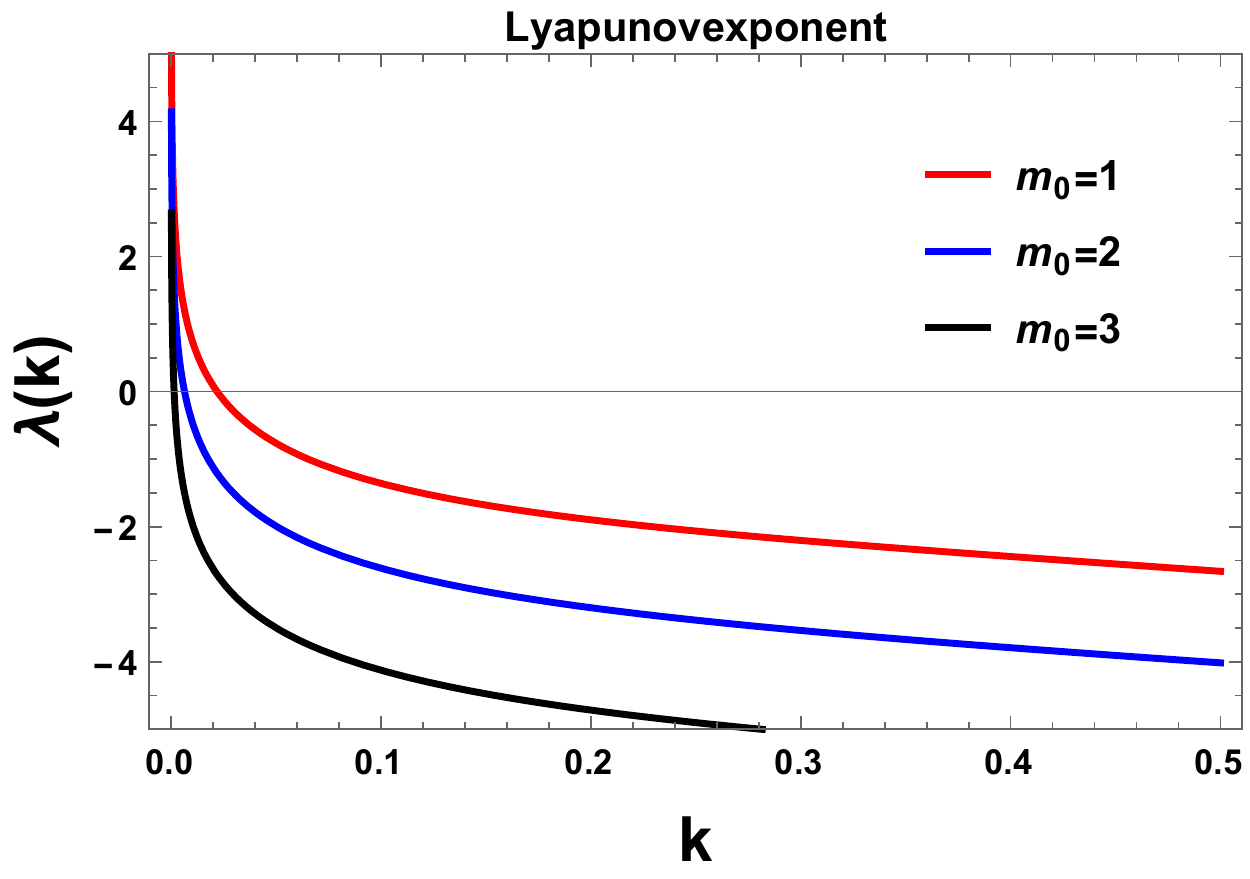}
}
\caption{Wave number dependence of {\it Lyapunov exponent} is shown for mass profile I. Here we fix $\rho=1$.}
\label{LAL1}
\end{figure}

In  fig.~\ref{LAL1}, we observe that with increase in wave number $k$ the {\it Lyapunov exponent} decreases. This shows that
  the {\it Lyapunov exponent} is dependent on the momenta values of the fields interacting with the massive field acting as
  a scatterer. Furthermore, we discover that for the mass profile I, the chaos in the event reduces with increase in 
  the wave number. This suggests that lesser the number of fields interacting with the massive field more is the chaos in 
  the quantum system considered in this paper. Since, a negative value of Lyapunov Exponent pulls a system out of chaos, this further tells us that the {\it Lyapunov
  exponent} is inversely related to the number of background fields interacting with the scatterer or the massive field. This may be interpreted in the following way in the context of Schr{\"o}dinger problem in quantum mechanics that a higher value of wave
  number $k$ of the incoming wave would be able to cross a potential barrier of a given strength and would be able to
  get transmitted through the barrier and the pulse won't damp easily than that of a wave with lower $k$ value. This
  means that the scatterer acts as a definitive medium which allows only certain wave numbers to pass through thus
  reducing the chaos in the system.

 \subsubsection{Conduction properties: Conductance and Resistance}
 For the given mass profile the expression for conductance and resistance can be expressed as:
 \bea~G(k) =\exp(-2\lb(k))=2\log\left| \sqrt{\frac{\omega_{in}}{\omega_{out}}}\; \frac{\Gamma \left(-\frac{i \og_+}{2 \rho }\right) \Gamma \left(1-\frac{i \og_+}{2 \rho }\right)}{\Gamma \left(-\frac{i \og_{out}}{\rho }\right) \Gamma \left(1-\frac{i\og_{in}}{\rho }\right)}\right| \\
 r(k) = \exp(2\lb(k))=2\log\left| \sqrt{\frac{\omega_{out}}{\omega_{in}}}\; \frac{\Gamma \left(-\frac{i \og_{out}}{\rho }\right) \Gamma \left(1-\frac{i\og_{in}}{\rho }\right)}{\Gamma \left(-\frac{i \og_+}{2 \rho }\right) \Gamma \left(1-\frac{i \og_+}{2 \rho }\right)}\right| 
 \eea
 %%%%
%%%%%%
%%%%%%
%%%%%%%%
\begin{figure}[htb]
\centering
\subfigure[$G(k)$ vs $k$ plot.]{
	\includegraphics[width=7.8cm,height=8cm] {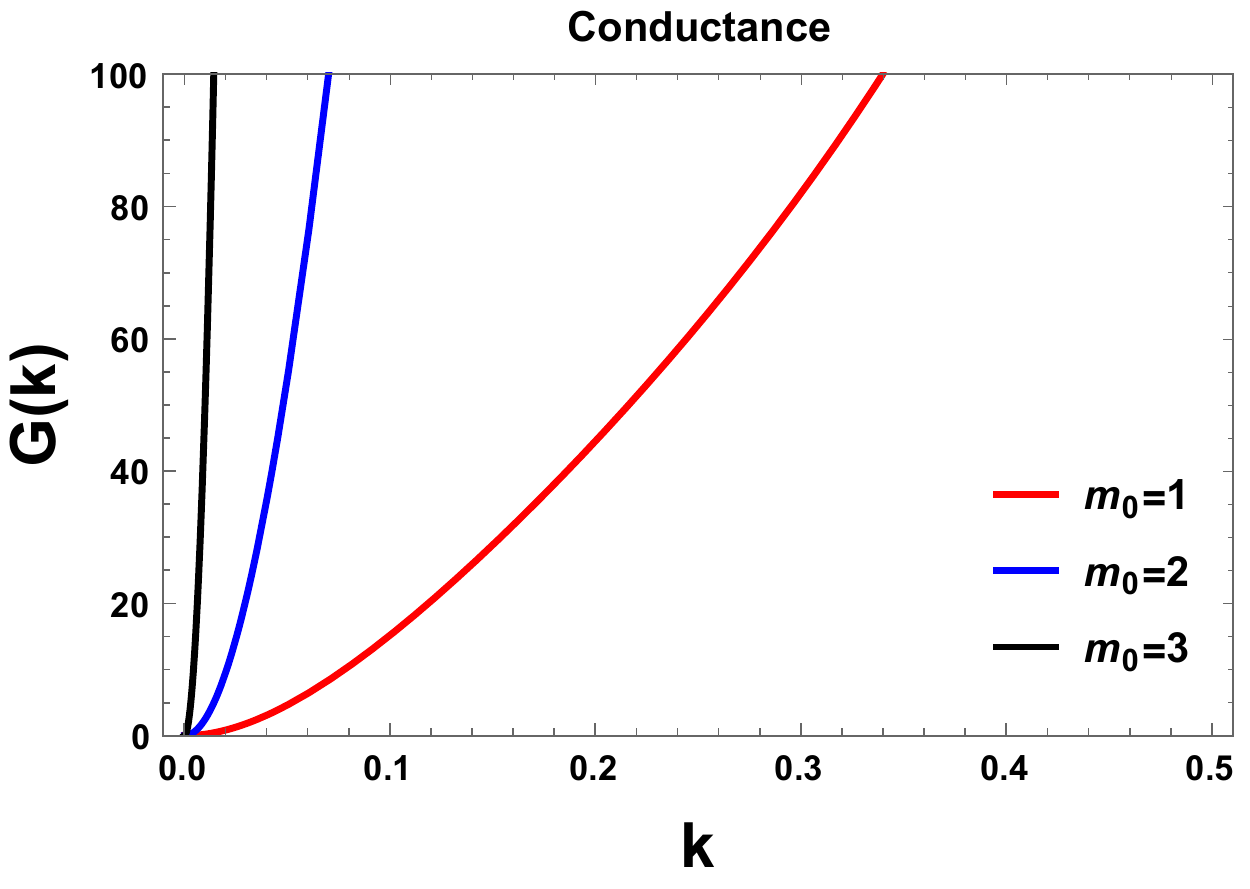}
	\label{MC1}
}
\subfigure[$r(k)$ vs $k$ plot.]{
	\includegraphics[width=7.8cm,height=8cm] {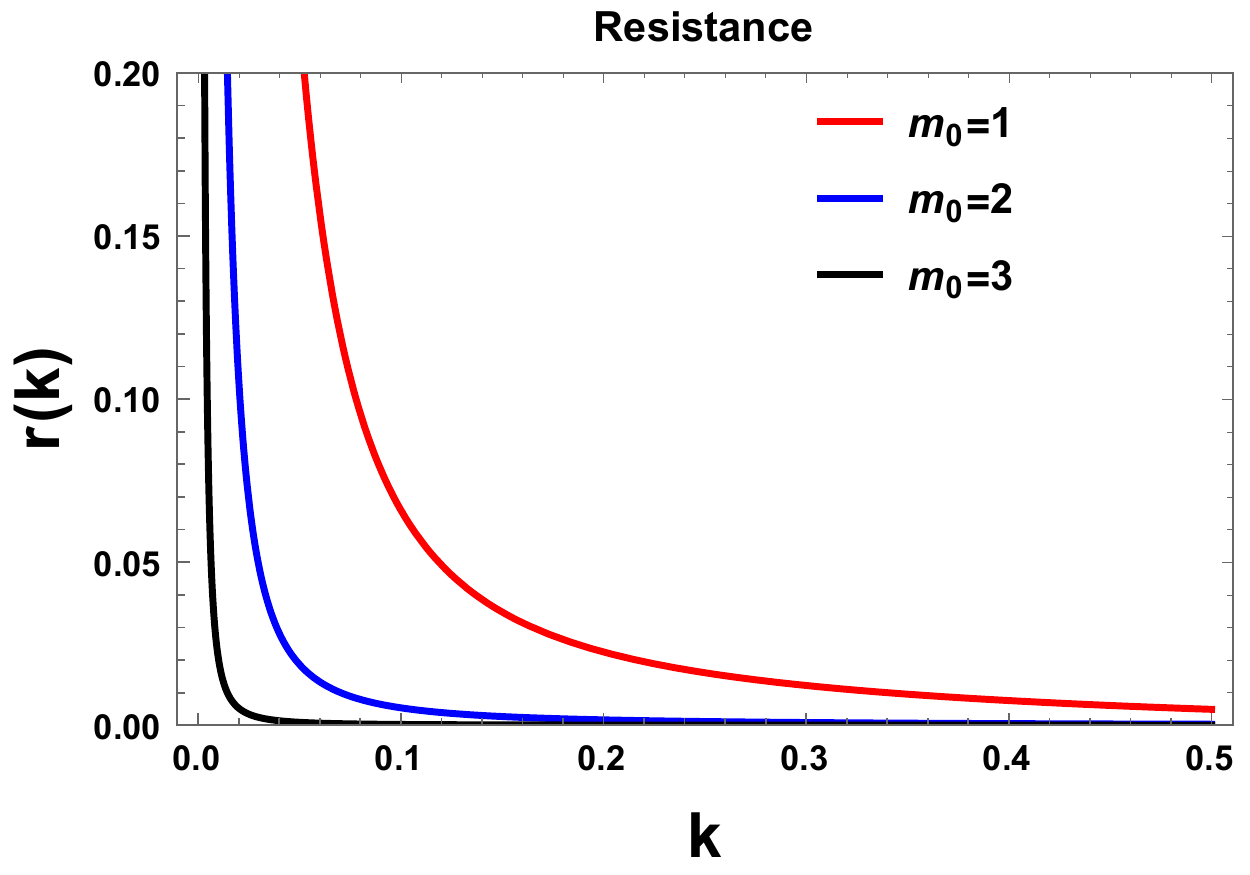}
	\label{MR1}
}
\caption{Wave number dependence of conductance and resistance for the mass profile I is shown here. Here we fix $\rho=1$. }
\end{figure}

 In fig.~\ref{MC1} we have shown the variation of conductance  with wave number $k$. This figure shows that with increase in the momenta value 
 of the massless scalar field, the conductance also increases. Now, accounting for $m_{0}$ values, we see that for $m_{0} = 1$
 the conductance shoots up at a much lower $k$ value than that of $m_{0} = 2$ and $m_{0} = 3$. This suggests that for
 $m_{0} = 1$ the field has a much higher transmission probability than that of $m_{0} = 2$ and $m_{0} = 3$. An increase in 
 transmission probability gives a direct evidence of the conductance value. Therefore, we conclude that for $m_{0} = 1$ the
 field has more conductance value in comparison to $m_{0} = 2$ and $m_{0} = 3$. We also conclude that larger the momenta value,
 more is the transmission coefficient and thereby shoots up the conductance of the system. This means that an incoming wave
 with large momenta value would eventually cross a barrier potential field thereby increasing the conductance of the system as
 the transmission probability would be much higher than an incoming wave with lower momenta value.

 In fig.~\ref{MR1} we have shown the variation of resistance with wave number. We observe that with an increase
 in the value of $k$ the resistance starts decreasing which suggests that with an increase in momenta value the
 transmission probability across the scatterer. This may be viewed in accordance with the potential barrier in the
 Schr{\"o}dinger equation in quantum mechanics also starts increasing thereby allowing the incoming wave to tunnel 
 through the barrier thereby increasing the transmission probability and hence,reducing the resistance. We also observe that with an increase in $k$ value the resistance reduces less rapidly for $m_{0} = 1$ than that
 of  $m_{0} = 3$ and  $m_{0} = 2$. Whereas, it reduces more rapidly for  $m_{0} = 3$ suggesting that higher the value of the constant $m_{0}$ lower is the value of resistance offered.
\subsection{Protocol II:~$m^{2} (\tau) = m_{0}^{2}~ {\rm sech}^2(\rho \tau)$}

 \begin{figure}[htb]
\centering
\subfigure[$m^2(\tau)$ vs $\tau$ profile.]{	
	\includegraphics[width=7.8cm,height=8cm] {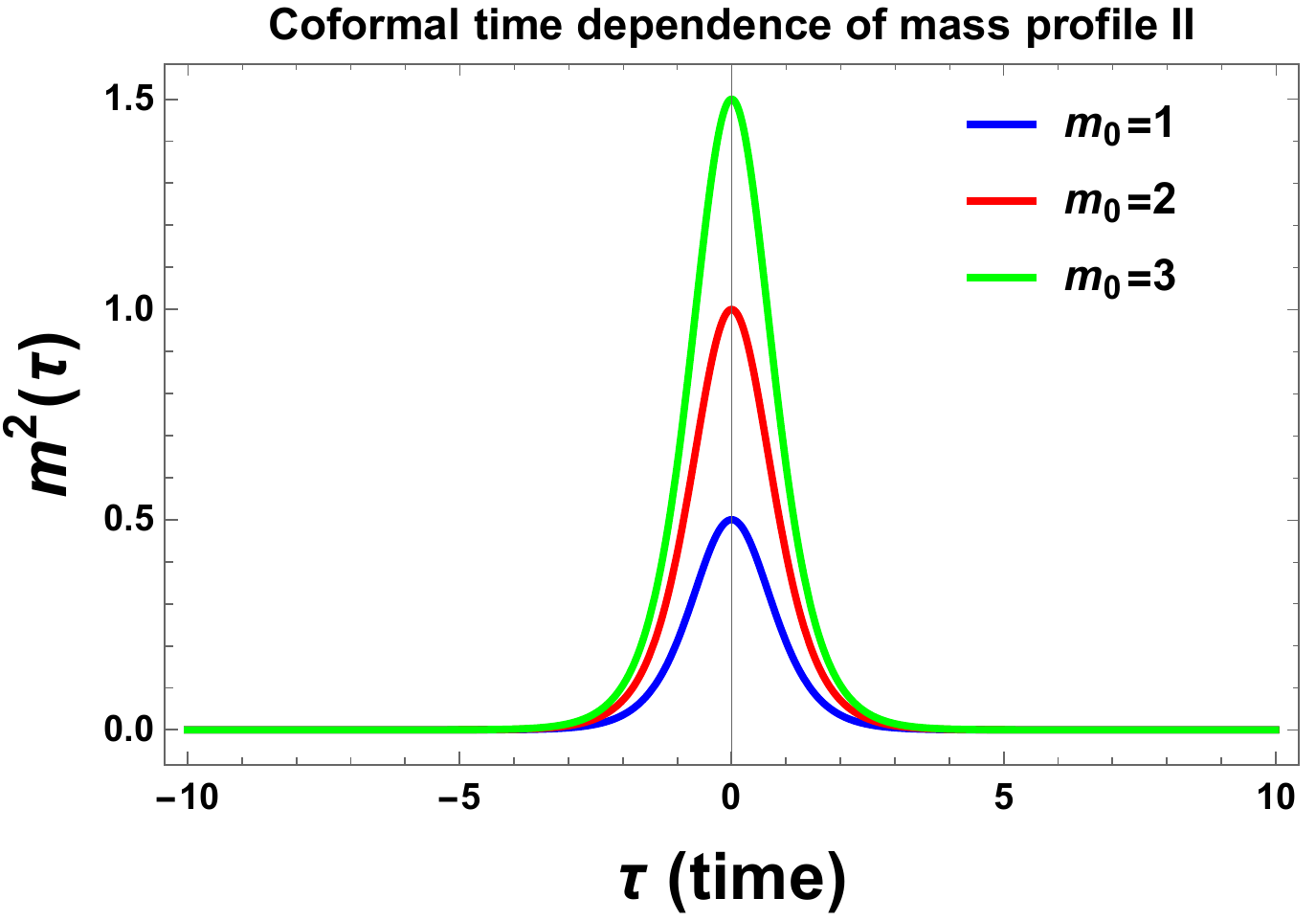}
	\label{gq3}
}
\subfigure[$V(\tau)$ vs $\tau$ profile.]{	
	\includegraphics[width=7.8cm,height=8cm] {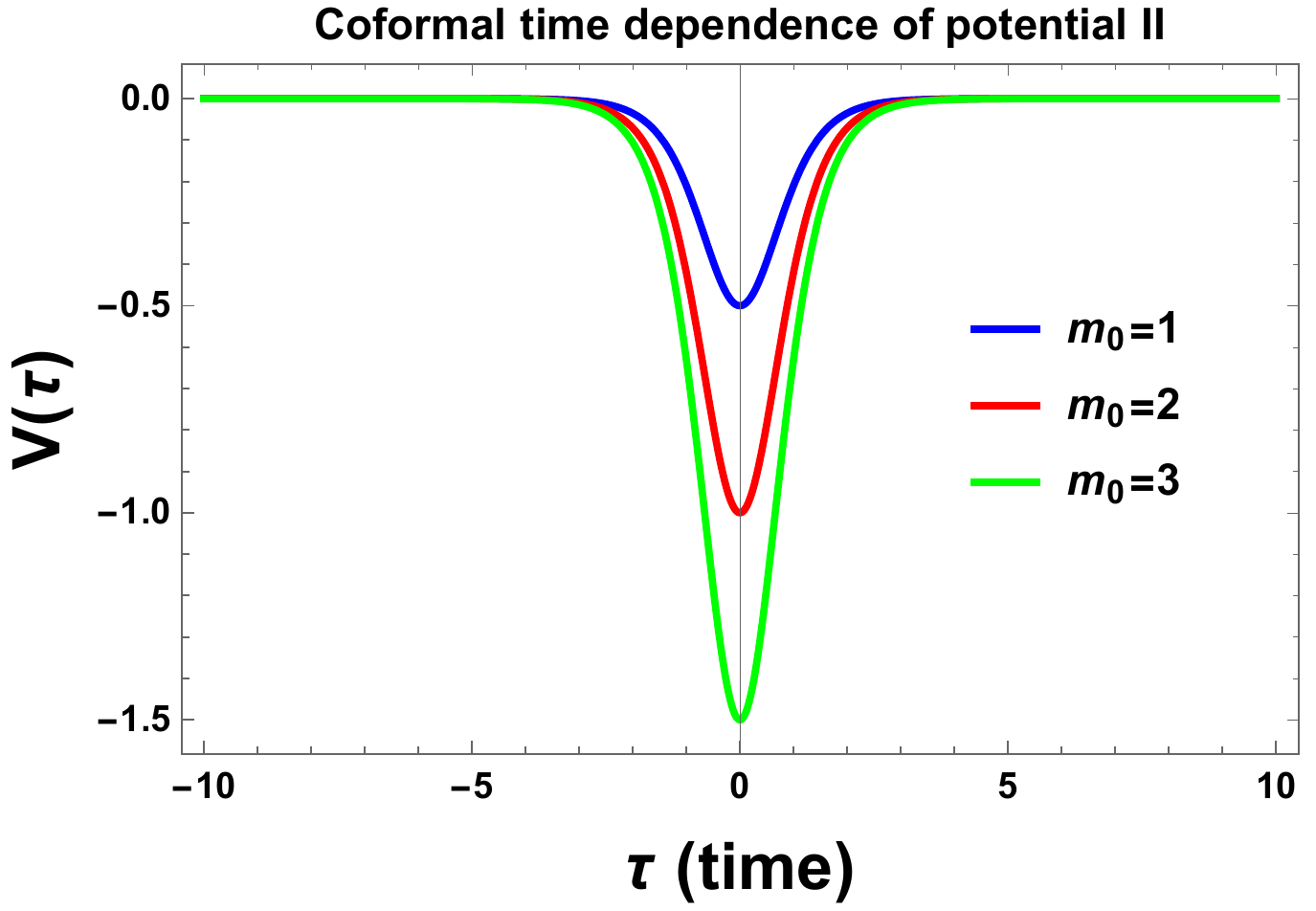}
	\label{gq4}
}
\caption{Conformal time dependent behaviour of the mass profile II and its corresponding potential used in Schr{\"o}dinger scattering problem is explicitly shown here. Here we fix $\rho=1$.}
\end{figure}
Here we consider the following mass profile: 
 \bea~m^{2} (\tau) = m_{0}^{2}~ {\rm sech}^2(\rho \tau).
\eea
  The corresponding Schr{\"o}dinger problem for this potential function can be solved by using the potential function as given bellow:
  \bea V(\tau)=-m^2(\tau)= -m_{0}^{2}~ {\rm sech}^2(\rho \tau).\eea
  In fig.~(\ref{gq3}) and fig.~(\ref{gq4}), we have explicitly shown the conformal time dependent behaviour of the mass profile under consideration and also the corresponding potential used in Schr{\"o}dinger scattering problem.
  
Now using the coordinate transformation $y=e^{2\rho \tau}$  \cite{Mandal:2015kxi,Das:2014hqa} one can recast the equation
of motion, analogous to the time independent Schr{\"o}dinger equation takes the following form:
 \bea\phi''_{k}(y) + \frac{\phi'_{k}(y)}{y} + \left(\frac{k^2}{4\rho^2y} + \frac{m_0^2}{\rho^2(1 + y)^2}\right) \phi_{k}(y) = 0.
\eea
The solution of this equation is given by:
 \bea\begin{array}{lll}\footnotesize
u(k,\tau)=e^{-ik\tau} (1 + e^{2\rho \tau})^{\alpha}\left[C_1\ e^{2ik\tau}\ {}_2F_1\left(\alpha , \frac{ik}{\rho} + \alpha, 1 + \frac{ik}{\rho}, -e^{2\rho \tau}\right) \right.\\
\left.~~~~~+  C_2\, {}_2F_1\left(\alpha , -\frac{ik}{\rho} +\alpha , 1 - \frac{ik}{\rho}, -e^{2\rho t}\right)\right],
\end{array}\eea
where we define a parameter $\alpha$ as:
 \bea \alpha=\frac{1}{2} + \frac{1}{\rho}\sqrt{4 m_0^2 + \rho^2}\eea
 \subsubsection{Bogoliubov coefficients}
Now we fix $C_1=1$ and $C_2=0$, which  gives the incoming solution $u_{in}(k)$. Further taking the $t\to +\infty$
limit and using Bogoliubov transformation we can express incoming solution in terms of the outgoing solution as given by:
 \bea
u_{in}(k)=\alpha(k)u_{out}(k) + \beta(k) u^*_{out}(k),
\eea 
where $\alpha(k)$ and $\beta(k)$ are the Bogoliubov coefficients, which are defined as:
 \bea\alpha(k)=\frac{\Gamma(\frac{i k}{\rho} +1) \Gamma (\frac{i k}{\rho})}{\Gamma(\frac{i k}{\rho}-\alpha +1)\Gamma(\frac{i k}{\rho} +\alpha)},~~~ \; \beta(k)=i \sin (\pi  \alpha ) \text{cosech}\left(\frac{\pi  k}{\rho}\right).
\eea

\begin{figure}[htb]
\centering
\subfigure[$\ag$ vs $k$ profile.]{
	\includegraphics[width=7.8cm,height=8cm] {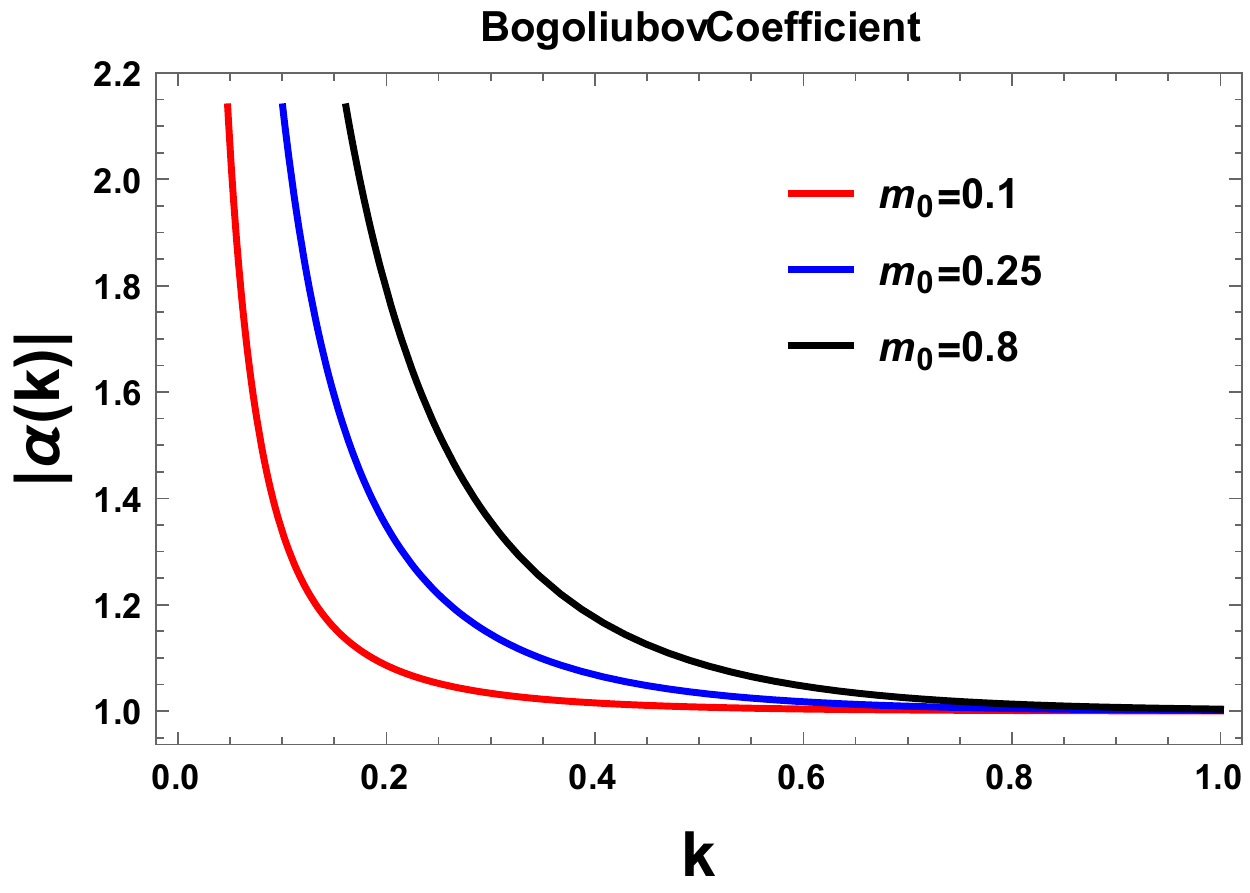}
	\label{MP2BA}
	}
\subfigure[$\bg$ vs $k$ profile.]{
	\includegraphics[width=7.8cm,height=8cm] {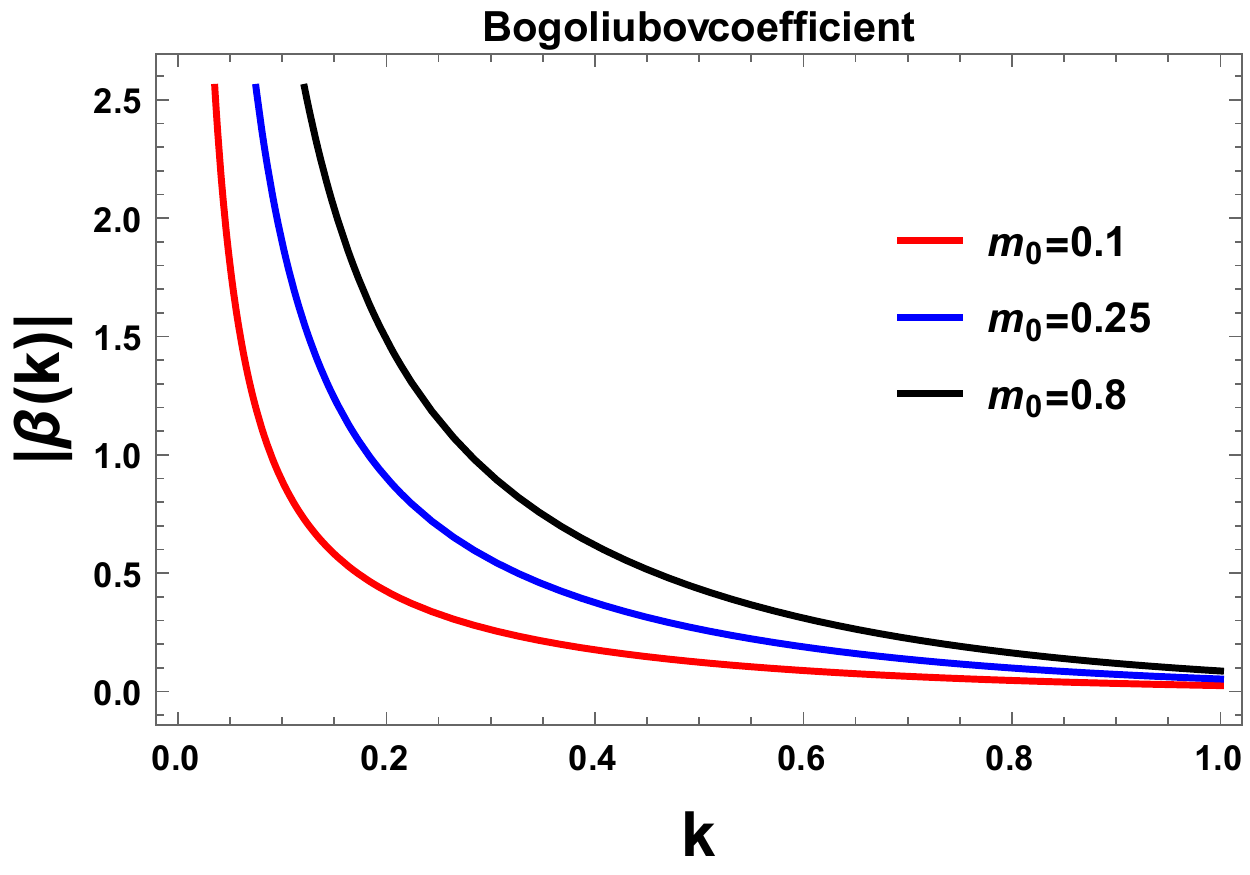}
	\label{MP2BB}
	}
	\caption{Wave number dependence of the Bogoliubov coefficients are shown here for mass profile II. Here we fix $\rho=1$. }
\end{figure}
In fig.~(\ref{MP2BA}) and fig.~(\ref{MP2BB}), we have shown the variation of the Bogoliubov coefficients with wave number $k$. 
\subsubsection{Optical properties: Reflection and transmission coefficients}
For this specific mass profile the transmission and the reflection coefficients can be expressed as:
 \bea T(K) = \frac{1}{|\frac{\Gamma (\frac{i k}{\rho} +1) \Gamma (\frac{i k}{\rho})}{\Gamma (\frac{i k}{\rho}-\alpha  +1)\Gamma (\frac{i k}{\rho} +\alpha)^2}|^2},~~~~ \; R(k) = \frac{|i \sin (\pi  \alpha ) \text{cosech}\left(\frac{\pi  k}{\rho}\right)|^2}{|\frac{\Gamma (\frac{i k}{\rho} +1)\Gamma(\frac{i k}{\rho})}{\Gamma(\frac{i k}{\rho}-\alpha  +1)\Gamma(\frac{i k}{\rho} +\alpha)}|^2}.
\eea

%%%%%%%
%%%%
\begin{figure}[htb]
\centering
\subfigure[$T(k)$ vs $k$ plot.]{
	\includegraphics[width=7.8cm,height=8cm] {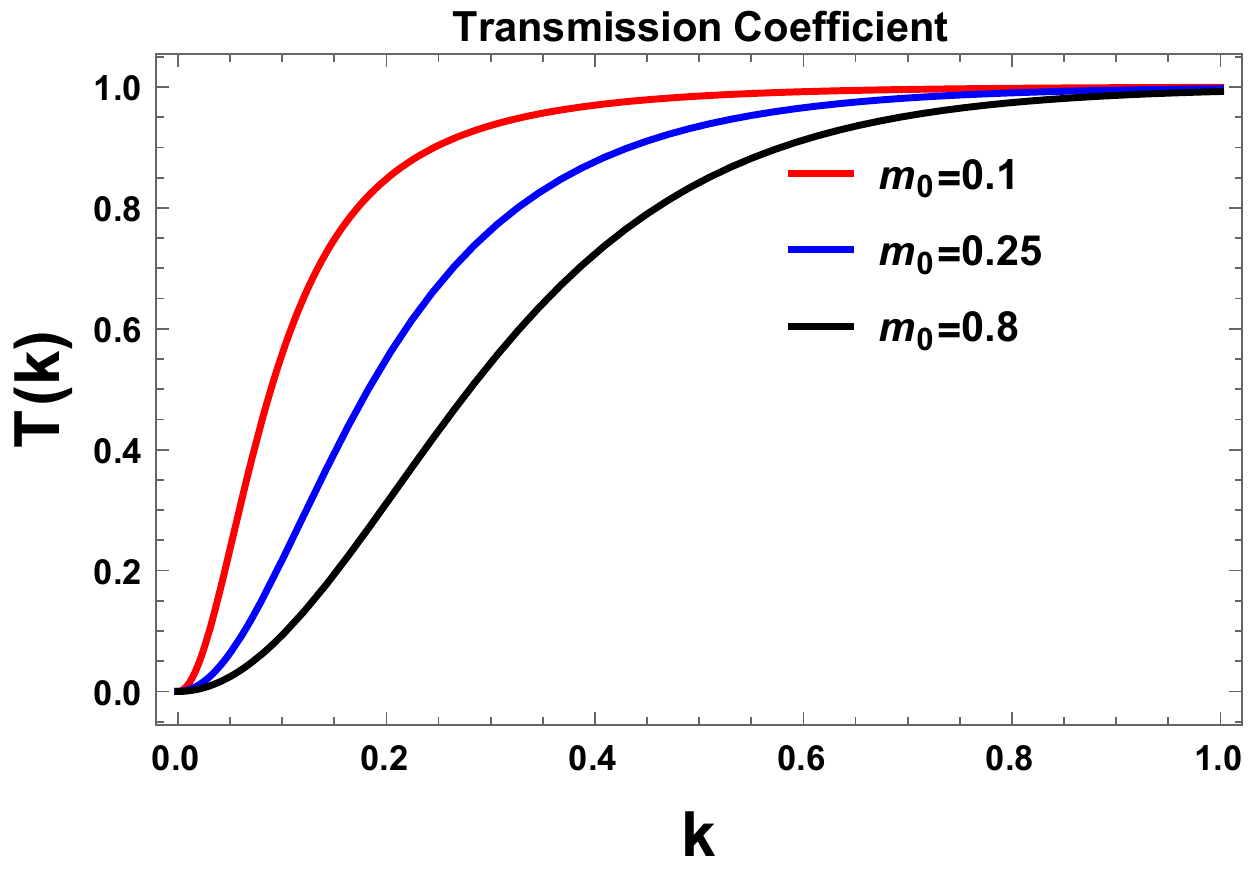}
	\label{MP2TC}
}
\subfigure[$R(k)$ vs $k$ plot.]{
	 \includegraphics[width=7.8cm,height=8cm] {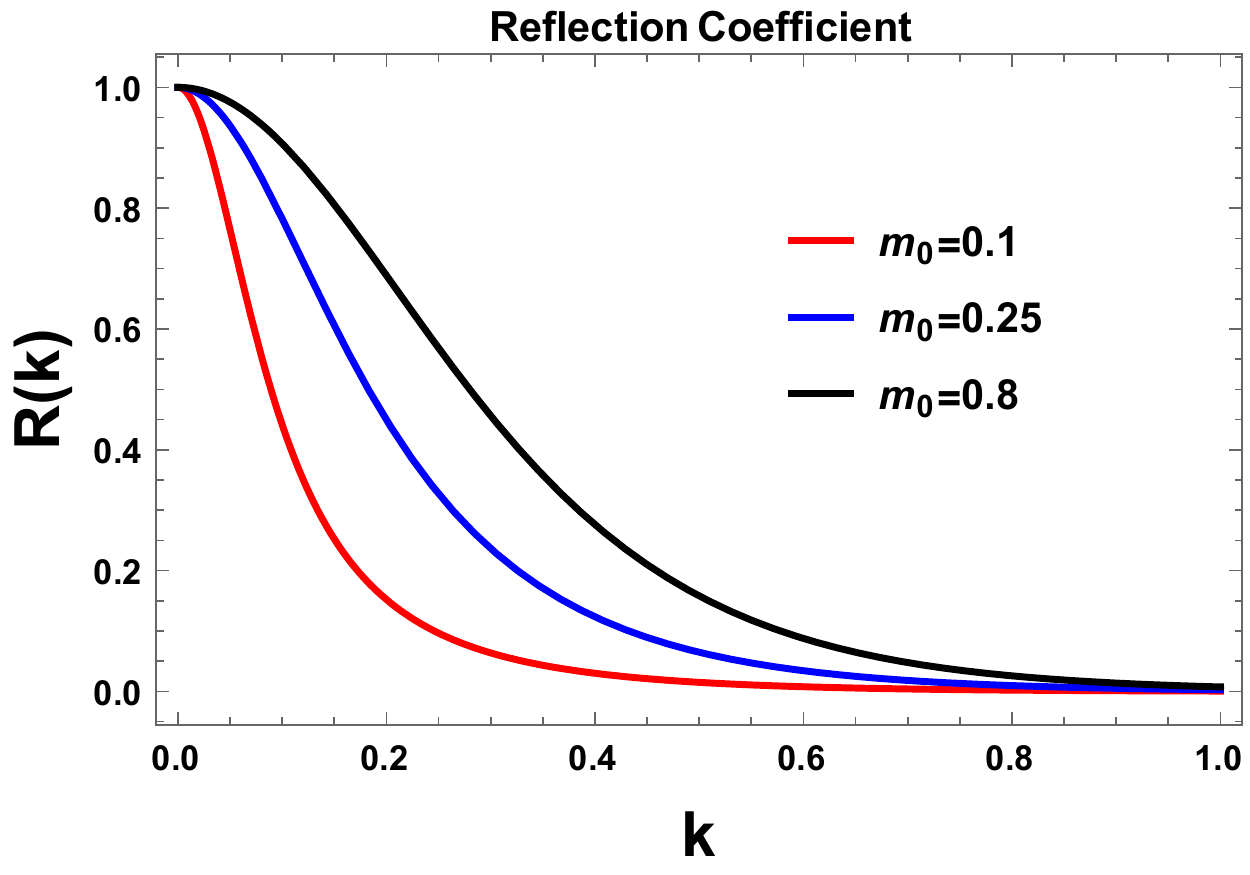}
	\label{M2ReflecCoeff}
}
\label{transref2}
\caption{Transmission and Reflection  Coefficient for mass profile $m^{2} (t) = m_{0}^{2} sech^2(\rho \tau)$}
\end{figure}
In fig.~(\ref{MP2TC}) and fig.~(\ref{M2ReflecCoeff}), we have shown the variation of the transmission and reflection coefficients with wave number $k$. 
\subsubsection{Chaotic property: Lyapunov exponent}

The Lyapunov exponent for this case may be given as:\\
 \bea\lb = -\log~T=2\log|\alpha (k)|=2 \log\left|\frac{\Gamma (\frac{i k}{\rho} +1) \Gamma (\frac{i k}{\rho})}{\Gamma (\frac{i k}{\rho}-\alpha  +1)\Gamma (\frac{i k}{\rho} +\alpha)^2}\right|.
\eea

\begin{figure}[htb]
\centering{
    \includegraphics[width=13cm,height=8cm]{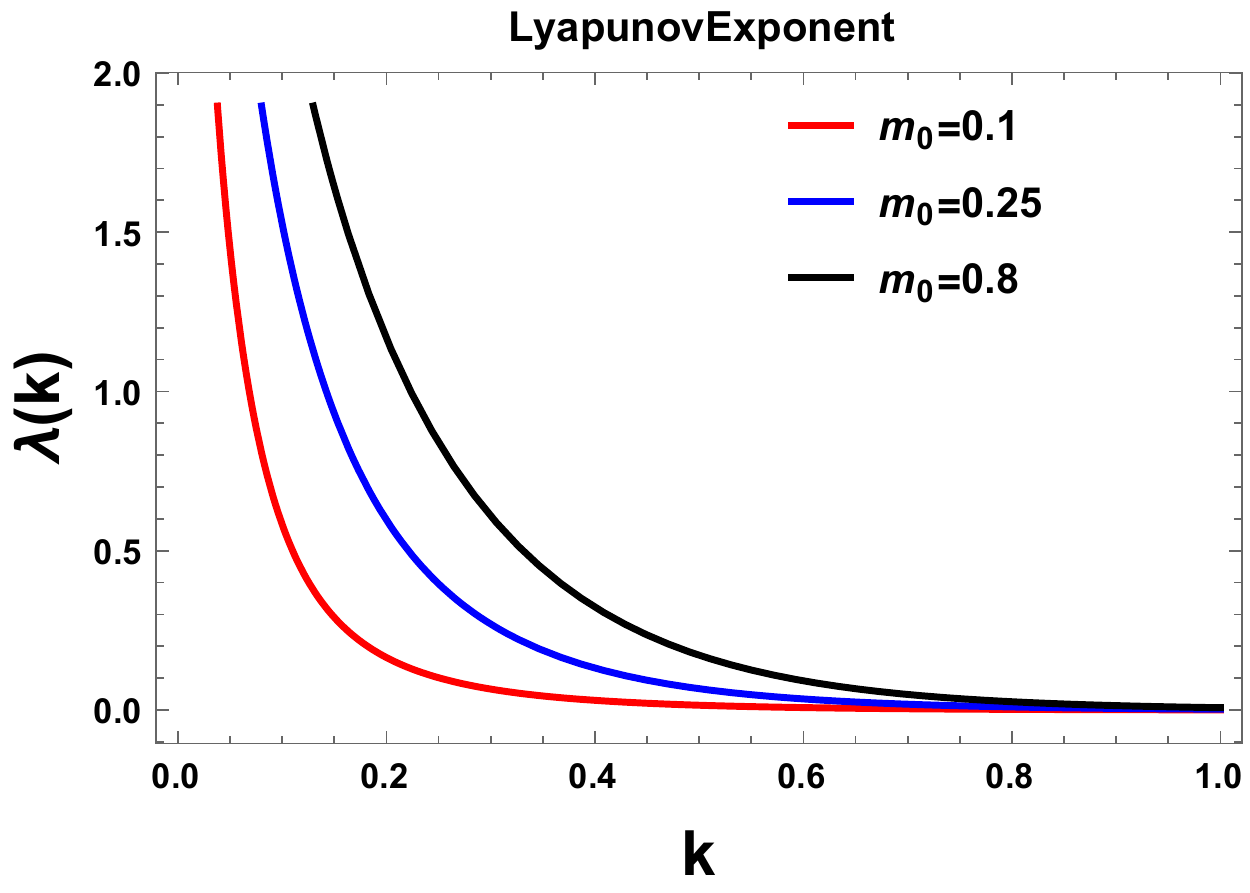}
}
\caption{ This shows the variation of Lyapunov exponent with momenta values
	for mass profiles with $m_{0} = 1$, $m_{0} = 2$ and  $m_{0} = 3$}
\label{L2L}
\end{figure}
 In fig.~\ref{L2L}, we have shown the wave number dependence of {\it Lyapunov exponent}. Here we observe that with increase in $k$ value the {\it Lyapunov exponent} decreases. This implies that
 the {\it Lyapunov exponent} is dependent on the momenta values of the fields interacting with the massive field acting as
 a scatterer. Furthermore, we also discover that for this mass profile II, the chaos in the event reduces with increase in 
 the $k$ value.

\subsubsection{Conduction properties: Conductance and Resistance}
For this specific mass profile the expression for the conductance and resistance can be computed as:
 \bea G(k) = \exp(-2\lambda(k))=\left|\frac{\Gamma (\frac{i k}{\rho} +1) \Gamma (\frac{i k}{\rho})}{\Gamma (\frac{i k}{\rho}-\alpha  +1)\Gamma (\frac{i k}{\rho} +\alpha)^2}\right|^4,\\
 r(k) = \exp(2\lambda(k))=\left|\frac{\Gamma (\frac{i k}{\rho}-\alpha  +1)\Gamma (\frac{i k}{\rho} +\alpha)^2}{\Gamma (\frac{i k}{\rho} +1) \Gamma (\frac{i k}{\rho})}\right|^4.
 \eea

%%%%%%%%
\begin{figure}[htb]
\centering
\subfigure[$G(k)$ vs $k$ plot.]{
	\includegraphics[width=7.8cm,height=8cm] {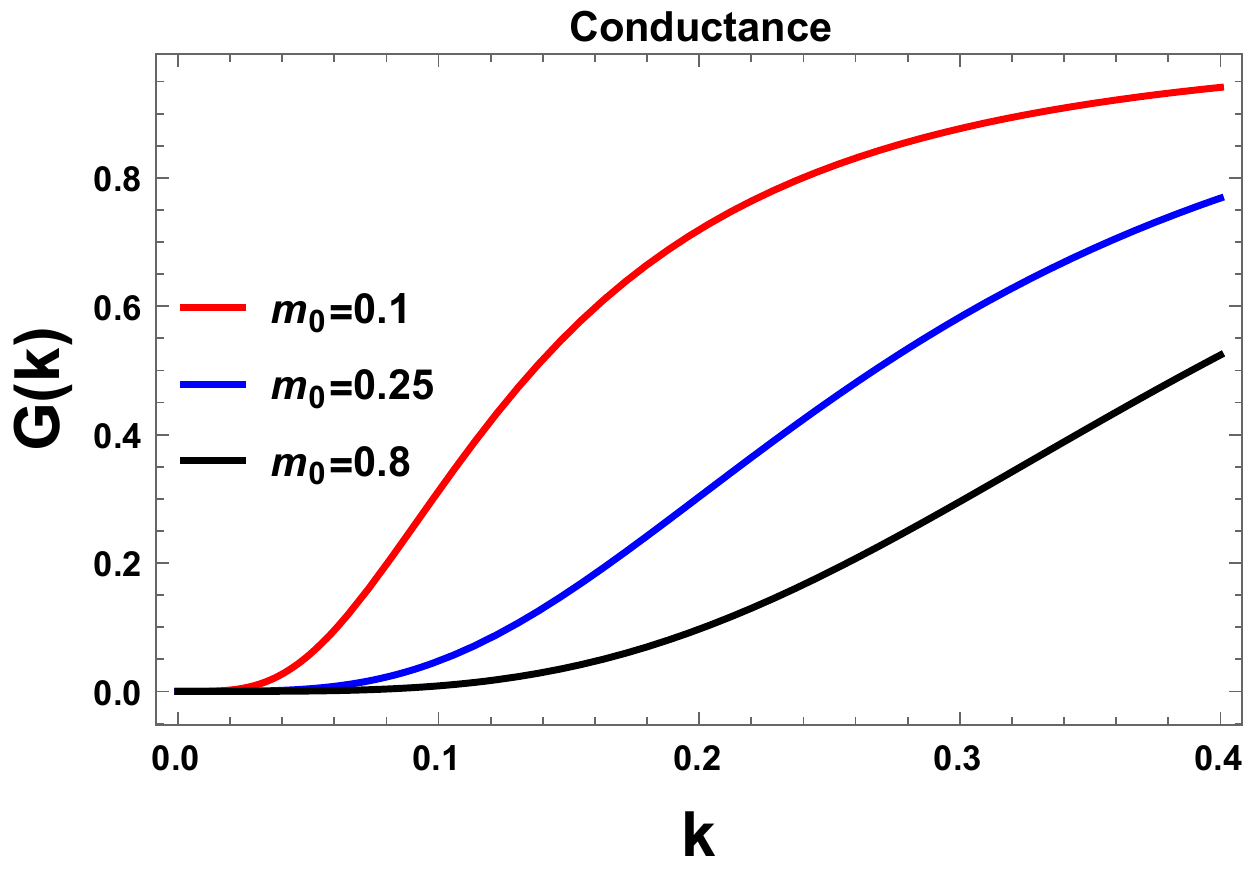}
	\label{MC2}
}
\subfigure[$r(k)$ vs $k$ plot.]{
	\includegraphics[width=7.8cm,height=8cm] {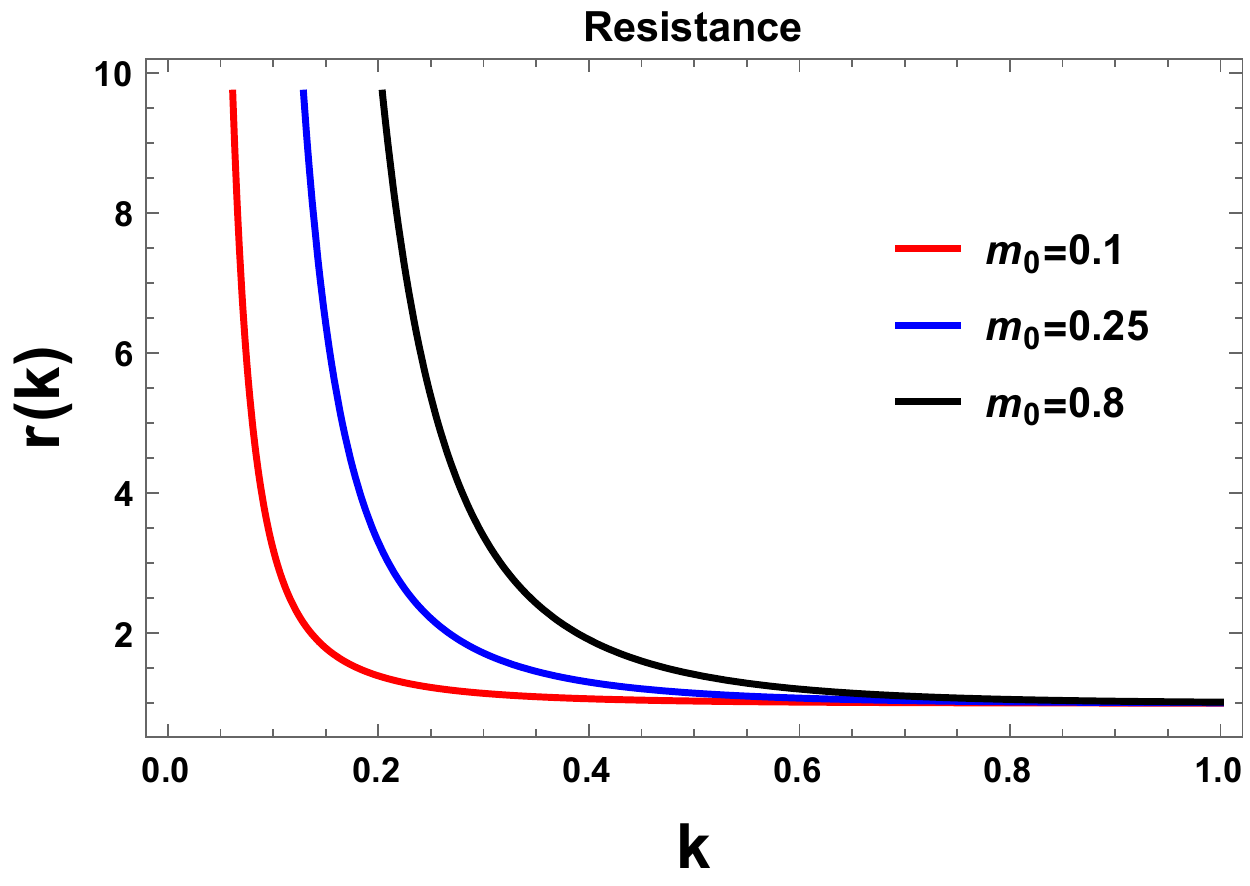}
	\label{MR2}
}
\caption{Wave number dependence of conductance and resistance for the mass profile II is shown here. Here we fix $\rho=1$. }
\end{figure}
 %%%%
%%%%%%
%%%%%%
%%%%%%%%
 
 In  fig.~\ref{MC2} we have shown the wave number dependence of conductance. We observe that for $m_{0} = 1$ conductance starts increasing at
 a larger value of $k$ than that of  $m_{0} =2$ and $m_{0} = 3$. But, in contrast to the variation of conductance
 with momenta $k$ in the above figure, here the conductance starts increasing rapidly for $m_{0} = 3$ than that
 for $m_{0} = 1$ which suggests that the transmission probability for $m_{0} = 3$ is much higher than $m_{0} = 1$
 and $m_{0} =2$, thereby making it more conductive than the other two.

 In fig.~\ref{MR2}, we have shown the wave number dependence of resistance. Here like the first mass profile the resistance for $m_{0} = 3$ falls more rapidly than that of
 $m_{0} = 2$ and $m_{0} = 1$. This suggests that for the given mass profile II, as the value of $m_{0}$ increases, the
 value of resistance also decreases. But unlike the first mass profile, the resistance for $m_{0} = 3$ falls more
 rapidly suggesting that for $m_{0} = 3$ this specific mass profile offers more resistance than the first one. Therefore,
 we conclude that for the same values of $m_{0}$ this mass profile offers less resistance in comparison to the first
 mass profile.
\subsection{Protocol III:~$m^{2} (\tau) = m_{0}^{2}~\Theta (-\tau)$}
 \begin{figure}[htb]
\centering
\subfigure[$m^2(\tau)$ vs $\tau$ profile.]{	
	\includegraphics[width=7.8cm,height=8cm] {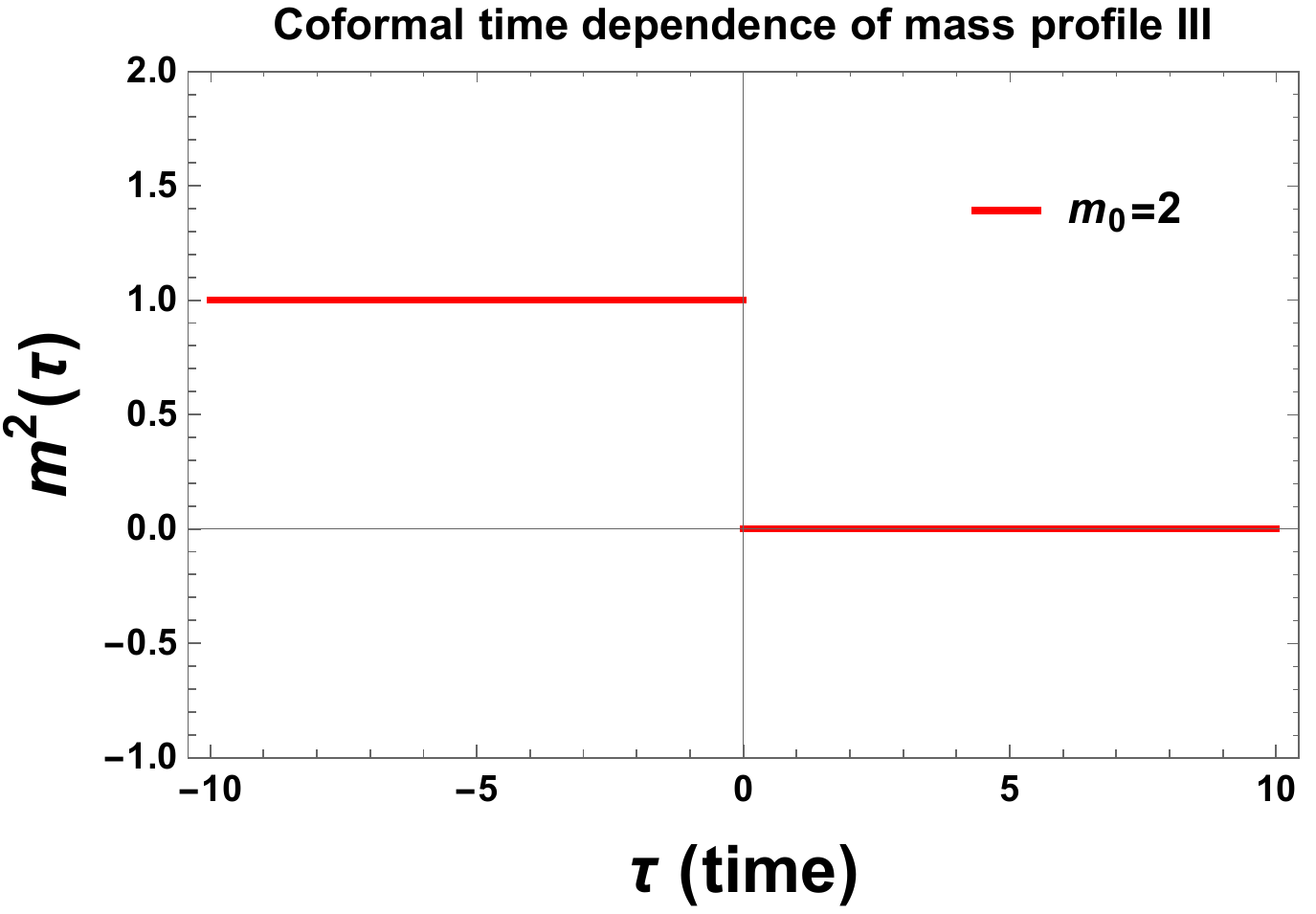}
	\label{gq5}
}
\subfigure[$V(\tau)$ vs $\tau$ profile.]{	
	\includegraphics[width=7.8cm,height=8cm] {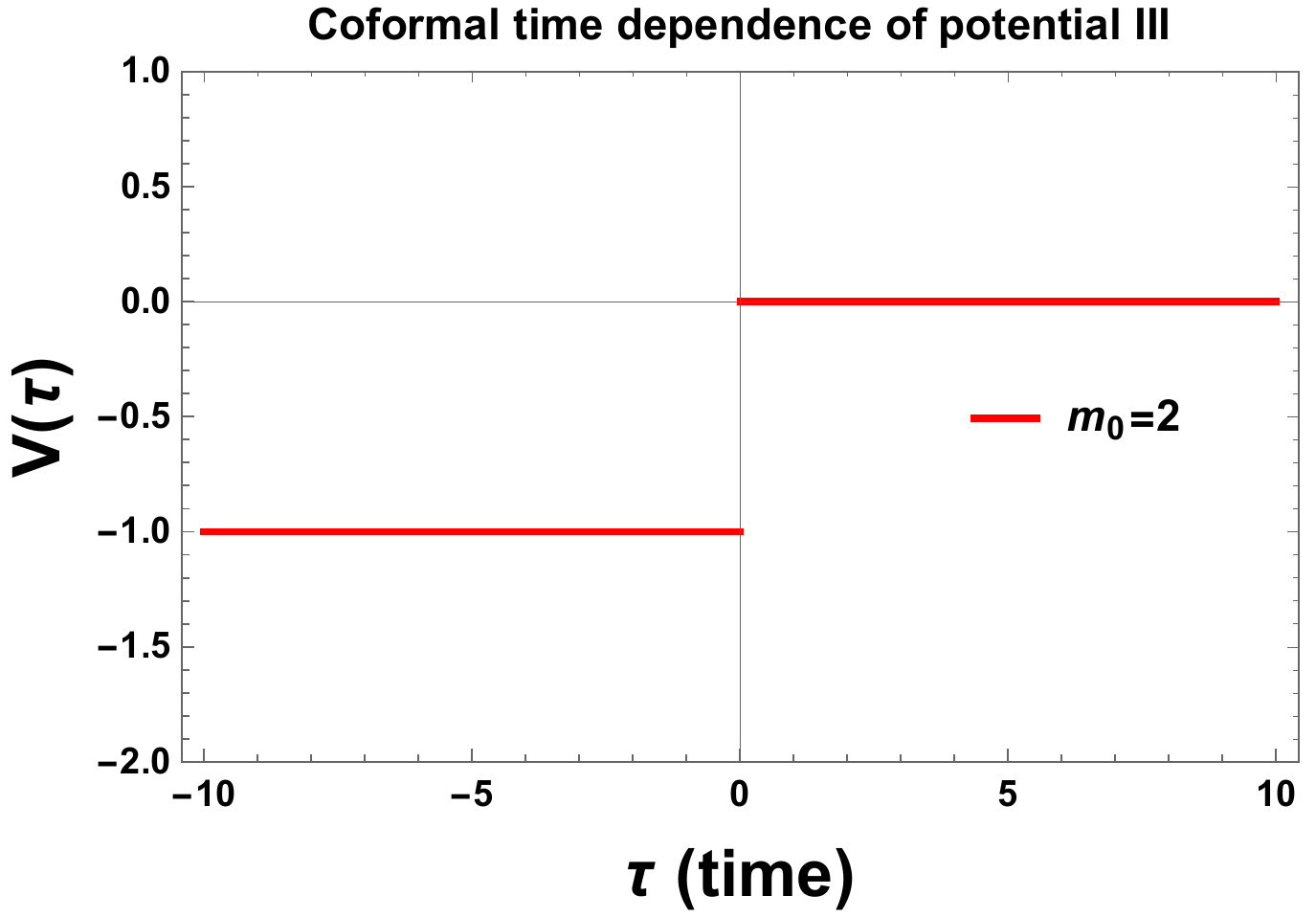}
	\label{gq6}
}
\caption{Conformal time dependent behaviour of the mass profile II and its corresponding potential used in Schr{\"o}dinger scattering problem is explicitly shown here. }
\end{figure}
Here we consider the following time dependent mass profile:
 \bea m^{2} (\tau) = m_{0}^{2}~\Theta (-\tau).
\eea
This $\Theta$  function in $\tau$ makes the mass profile a quenched one.

The corresponding Schr{\"o}dinger problem for this potential function can be solved by using the potential function as given bellow:
  \bea V(\tau)=-m^2(\tau)= -m_{0}^{2}~\Theta (-\tau).\eea
   In fig.~(\ref{gq5}) and fig.~(\ref{gq6}), we have explicitly shown the conformal time dependent behaviour of the mass profile under consideration and also the corresponding potential used in Schr{\"o}dinger scattering problem.
\subsubsection{Bogoliubov coefficients}
For this specific mass profile the Bogoliubov coefficients can be expressed as:
 \bea\alpha(k)=\frac{1}{2}\frac{|k|+\og_{in}}{\sqrt{|k|\og_{in}}},~~~~ \; \beta(k)=\frac{1}{2}\frac{|k|-\og_{in}}{\sqrt{|k|\og_{in}}}.
\eea
with the solution of the incoming and the outgoing waves are given by the following expressions:
 \bea u_{in}(k,t)= \frac{e^{-i\omega_{in}t}}{\sqrt{2 \omega_{in}}},~~~
u_{out}(k,t) =\frac{e^{-i\omega_{out}t}}{\sqrt{2 \omega_{out}}}. 
\label{wave-sudden}
\eea
\begin{figure}[htb]
\centering
\subfigure[$\alpha$ vs $k$ profile.]{
    \includegraphics[width=7.8cm,height=8cm]{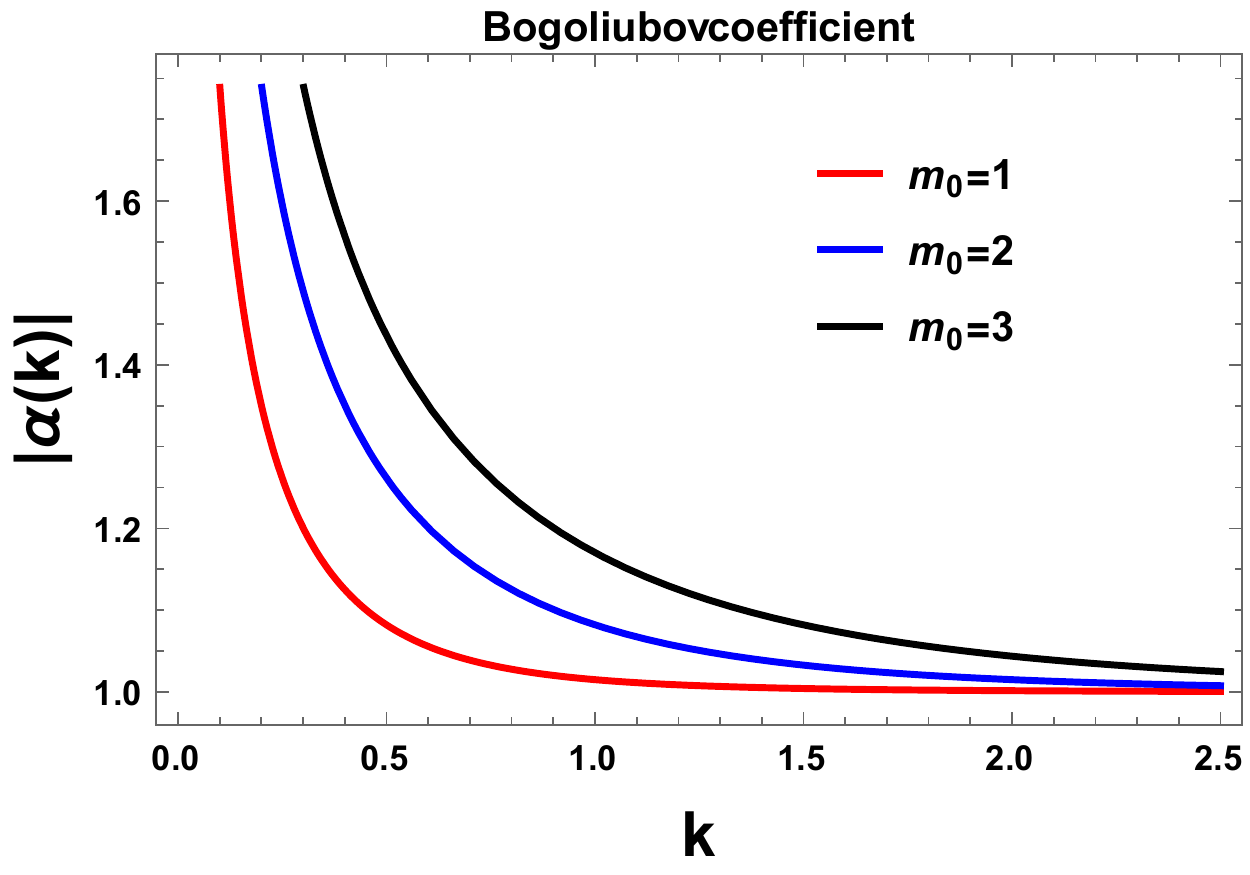}
    \label{B3B}
}
\subfigure[$\beta$ vs $k$ profile.]{
    \includegraphics[width=7.8cm,height=8cm]{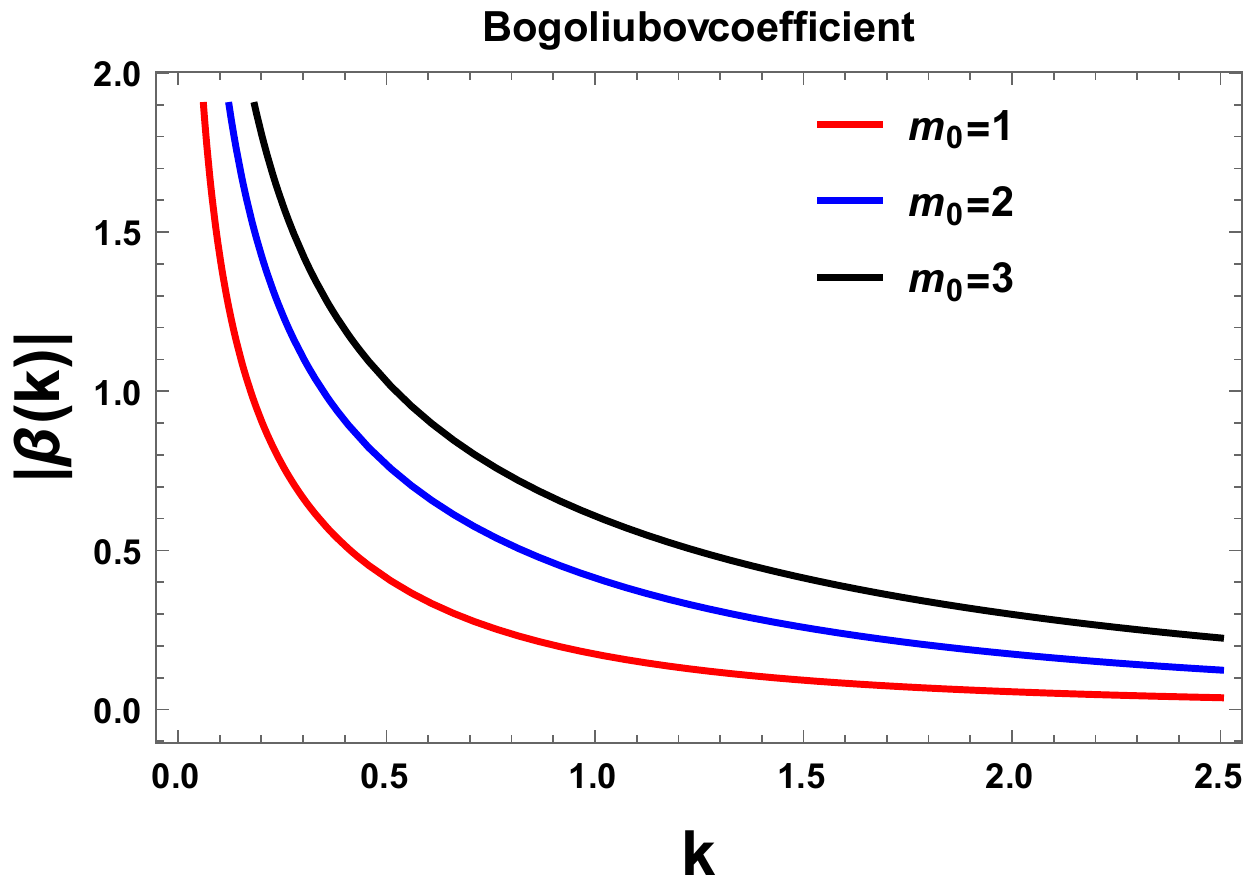}
    \label{B3A}
}
\caption{Wave number dependence of Bogoliubov coefficients for mass profile II is shown here.}
\end{figure}
In fig.~(\ref{B3B}) and fig.~(\ref{B3A}), we have shown the variation of the transmission and reflection coefficients with wave number $k$. 
\subsubsection{Optical properties: Refeclection and transmission coefficients}
For this specific mass profile the transmission and the reflection coefficients can be computed as:
 \bea T(k) = \left|\frac{2\sqrt{|k|\og_{in}}}{|k|+\og_{in}}\right|^2,~~~ R(k) = \left|\frac{2\sqrt{|k|\og_{in}}}{|k|-\og_{in}}\right|^2.
\eea

\begin{figure}[htb]
\centering
\subfigure[$T(k)$ vs $k$ plot.]{
    \includegraphics[width=7.8cm,height=8cm]{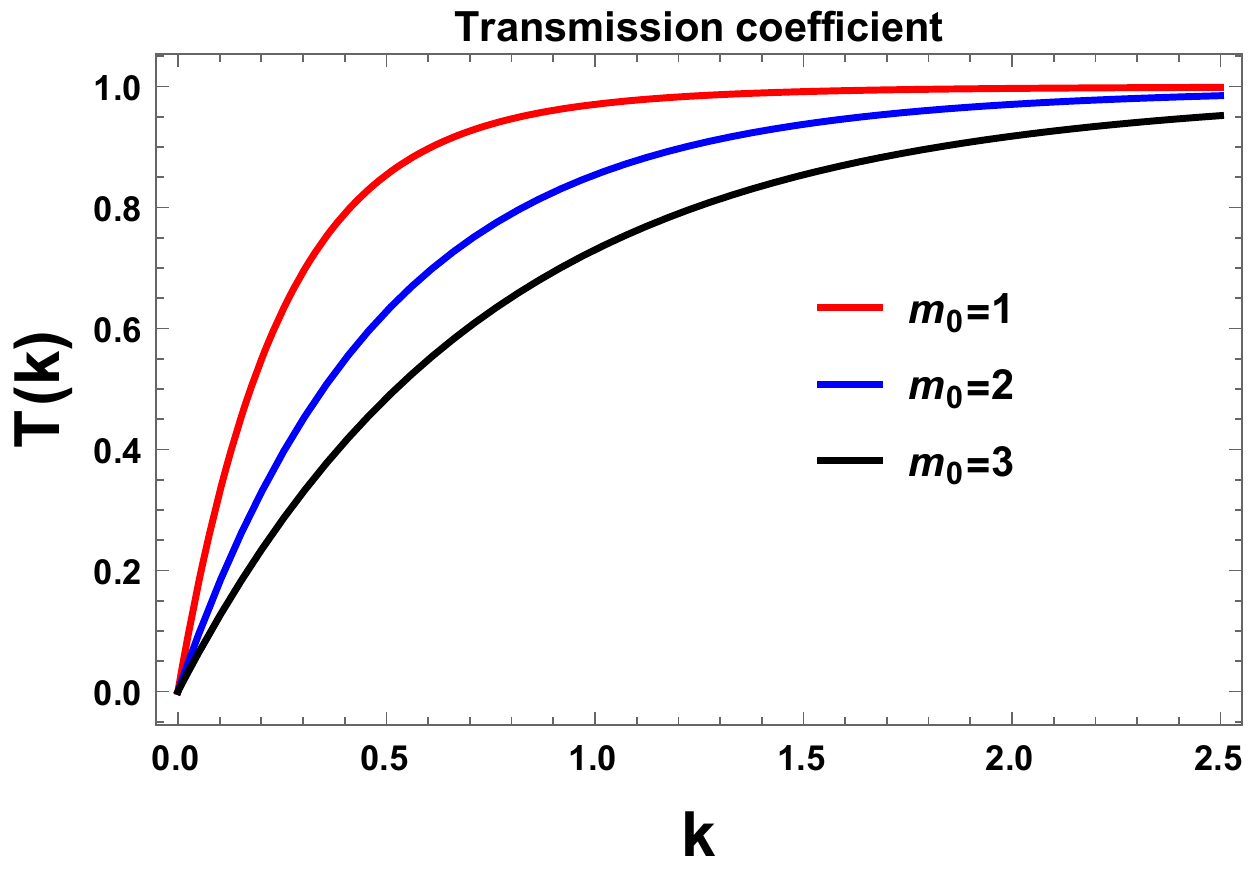}
    \label{B3B}
}
\subfigure[$R(k)$ vs $k$ plot.]{
    \includegraphics[width=7.8cm,height=8cm]{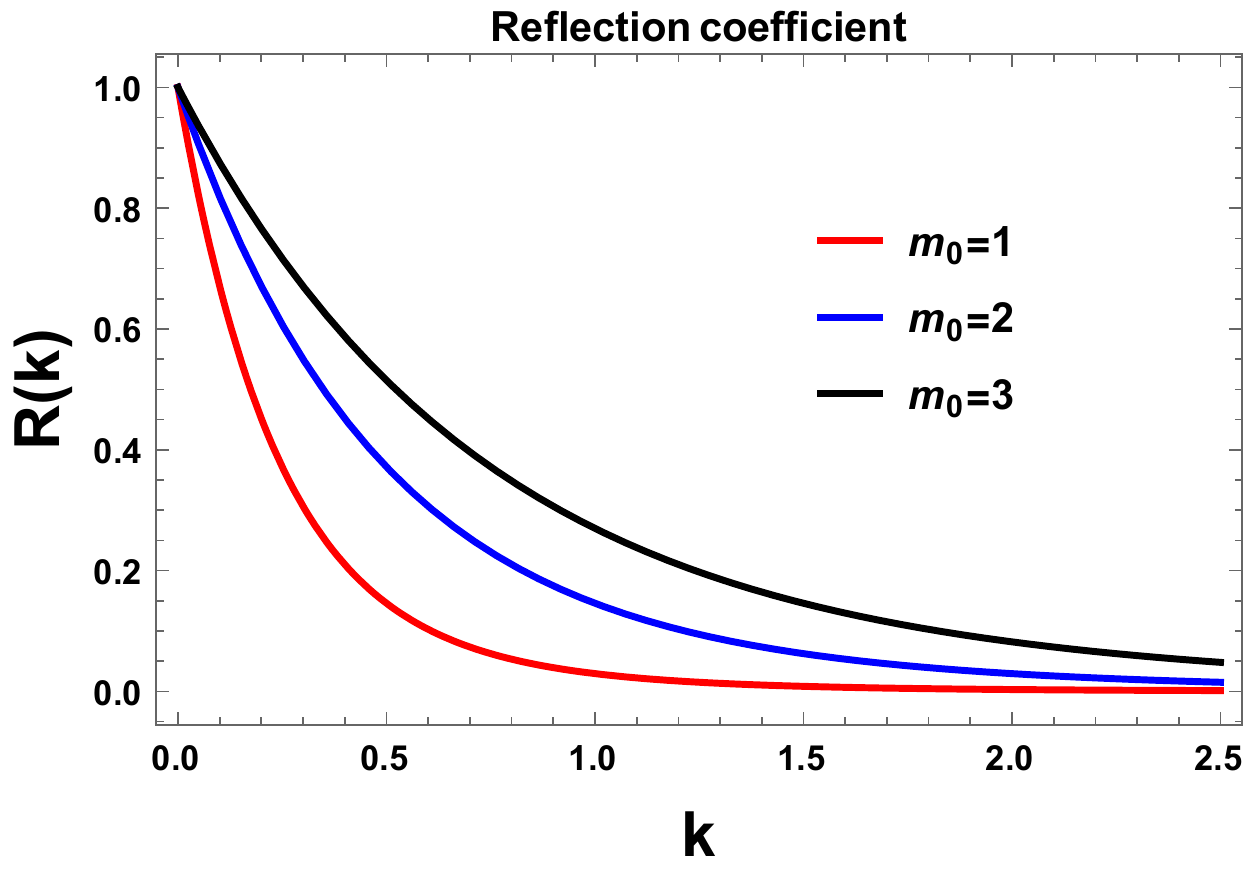}
    \label{R3R}
}
\caption{Wave number dependence of the transmission and reflection coefficients for mass profile III is shown here. }
\end{figure}
In fig.~(\ref{B3B}) and fig.~(\ref{R3R}), we have shown the variation of the transmission and reflection coefficients with wave number $k$. 
\subsubsection{Chaotic property: Lyapunov exponent}
 
 The Lyapunov in this case is written as:
  \bea\lb =-2\log~T=2\log|\alpha(k)|= 2 \log\left|\frac{1}{2}\frac{|k|+\og_{in}}{\sqrt{|k|\og_{in}}}\right|.
 \eea
 
\begin{figure}[H]
\centering
 {
 \includegraphics[width=13cm,height=8cm] {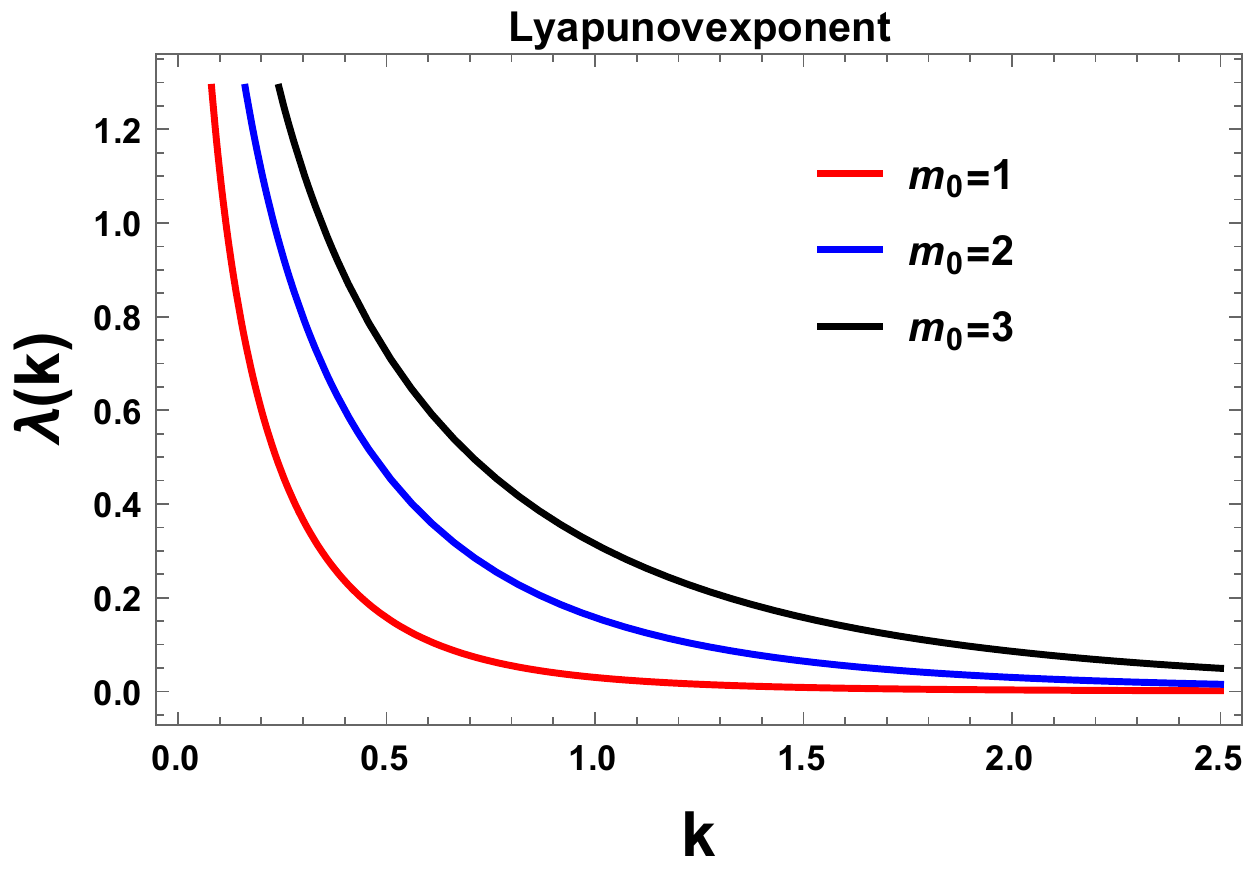}
 }
\caption{ Variation of {\it Lyapunov exponent} is shown with respect to the wave number.}
\label{L3Lz}
\end{figure}
From fig.~\ref{L3Lz} we observe that with increase in wave number the {\it Lyapunov exponent} decreases more like a
 rectangular hyperbolic fashion. In comparison to the other two mass profiles where the reduction in the value of the 
 {\it Lyapunov exponent } is much less rapid in comparison to this mass profile discussed here. This suggests that since, the mass profile
 is a heavy side theta function, which is a quenched mass protocol, the {\it Lyapunov exponent} also gives a similar like profile.
 This shows that the {\it Lyapunov exponent} is dependent on the wave number of the fields interacting with the massive
 field acting as a scatterer. Furthermore, we discover that for this given mass profile, the chaos in the event reduces
 with increase in the $k$ value. So, in this case the {\it Lyapunov exponent} decays much rapidly than the first two mass profiles.
Next, we will try to find an upper bound of Lyapunov exponent using the definition of \cite{RMT}

\subsubsection{Conduction properties: Conductance and Resistance}
For this specific mass profile the expression for the conductance and resistance can be written as:
 \bea G(k)=\exp(-2\lambda(k))=\left|\frac{2\sqrt{|k|\og_{in}}}{|k|+\og_{in}}\right|^4,
 \\ 
r(k)=\exp(2\lambda(k))=\left|\frac{1}{2}\frac{|k|+\og_{in}}{\sqrt{|k|\og_{in}}}\right|^4,
 \eea 
 
%%%%%%%%
\begin{figure}[H]
\centering
\subfigure[$G(k)$ vs $k$ plot.]{
	\includegraphics[width=7.8cm,height=8cm] {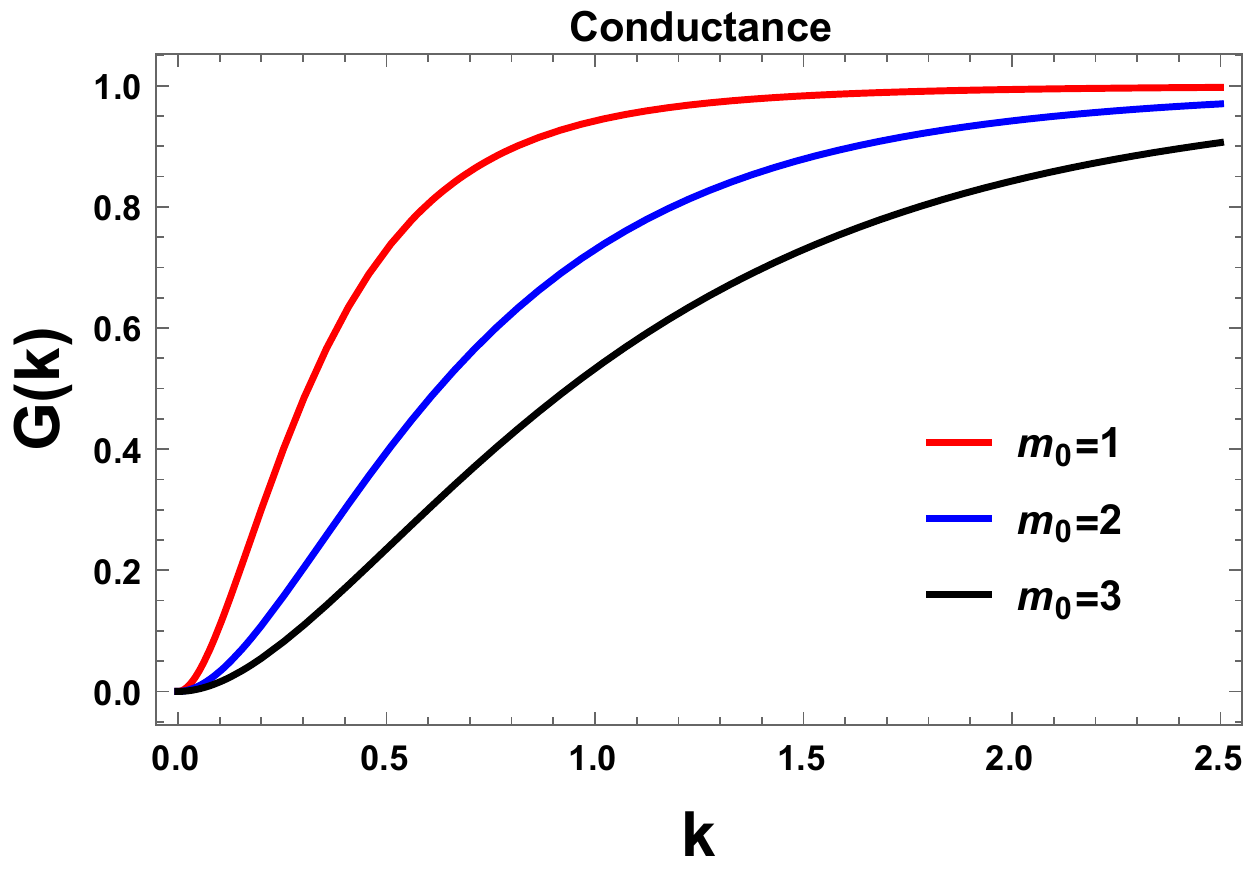}
	\label{MC3}
}
\subfigure[$r(k)$ vs $k$ plot.]{
	\includegraphics[width=7.8cm,height=8cm] {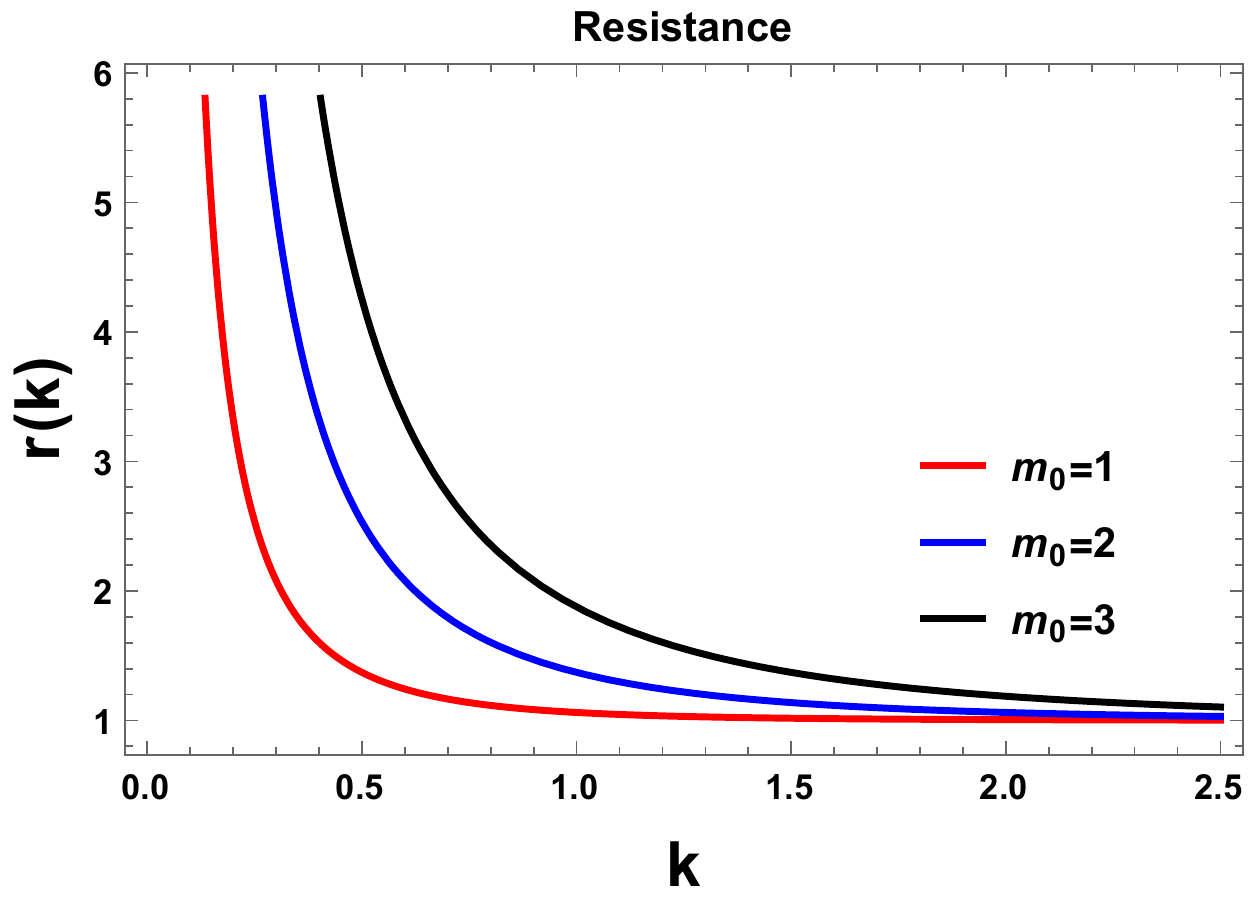}
	\label{MR3}
}
\caption{Wave number dependence of conductance and resistance for the mass profile III is shown here. }
\end{figure}
 %%%%
%%%%%%
%%%%%%
%%%%%%%%
 %%%%
%%%%%%
%%%%%%

In fig.~\ref{MC3}, we have shown the wave number dependence of conductance. This figure shows that with increase in the wave number of the massless scalar field, the conductance also increases. Now, accounting for $m_{0}$ values, we see that for $m_{0} = 1$
 the conductance shoots up at a much lower $k$ value than that of $m_{0} = 2$ and $m_{0} = 3$. This suggests that for
 $m_{0} = 1$ the field has a much higher transmission probability than that of $m_{0} = 2$ and $m_{0} = 3$. An increase in 
 transmission probability gives a direct evidence of the conductance value. Therefore, we conclude that for $m_{0} = 1$ the
 field has more conductance value in comparison to $m_{0} = 2$ and $m_{0} = 3$.

 In fig.~\ref{MR3}, unlike the mass profile I the resistance for $m_{0} = 1$ falls more rapidly than that of
 $m_{0} = 2$ and $m_{0} = 3$. This suggest that for the given mass profile, as the value of $m_{0}$ increases, the
 value of resistance also increases suggesting that heavier the field gets lesser is the transmission probability
 of the incoming wave to tunnel through it thereby reducing the value of conductance for this specific mass profile.

\section{Quantum chaos from out of time ordered correlators (OTOC)}
\subsection{Chaos bound in out-of-equilibrium quantum field theory (OEQFT) and its application to cosmology}
We know that in the context of quantum field theory it is possible to achieve the following universal bound on the {\it Lyapunov exponent}~\cite{Maldacena:2015waa}~\footnote{For this specific discussion only we keep the Planck's constant $\hbar$ and the Boltzmann constant $k_B$ in our computation. But for the rest of the paper we fix $\hbar=1$ and $k_B=1$ for which the parameter $\beta$ can be written as, $\beta=1/T$. In such a situation chaos bound is given by, $\lambda<2\pi/\beta$.}~\footnote{In the context of weakly coupled gauge theory one can introduce 't Hooft coupling $\lambda_T$ which is independent of $N$ and in such a theory the {\it Lyapunov exponent} is given by the following expression:
 \bea
{\textcolor{blue}{\bf \underline{Lyapunov~exponent~in~gauge~theory:}}}~~~~~~~~\lambda_{G} = \frac{\lambda_T}{\beta}=\lambda_T k_B T=\frac{\hbar\lambda_T}{2\pi}\frac{2\pi k_B T}{\hbar}<\lambda
\eea }:
 \bea
{\textcolor{blue}{\bf \underline{Universal~chaos ~bound~in~OEQFT:}}}~~~~~~~~\lambda \leq \frac{2 \pi  k_B T}{\hbar }=\frac{2\pi}{\hbar\beta},
\eea
where $k_B$ is the Boltzmann constant and $T$ is the temperature associated with the dynamical system. This upper bound of the {\it Lyapunov exponent} is treated as the saturation bound of chaos~\footnote{Considering the bulk contribution weakly coupled with string theory with large radius of curvature one can show that the perturbative stringy correction to the Einstein gravity computation of the {\it scrambling } can give rise to the following first order corrected expression for the {\it Lyapunov exponent} \cite{}:
 \bea\textcolor{blue}{\bf Stringy~correction:}~~~~\lambda=\frac{2\pi}{\beta}\left[1-\underbrace{\frac{\mu^2}{2}L_s^2+\cdots}_{\textcolor{red}{\bf Stringy~correction}}\right],\eea
where $L_s$ is the stringy length scale and $\mu^2$ is a specific constant which is appearing in the shock wave equation propagating along the horizon.
 }. Our aim is to establish this bound in the context of cosmology and study its further consequences. This bound was first proposed in the context of quantum information theory of black hole \cite{chaosblackhole,Lowe:2017ehz,Polchinski:2016hrw,Polchinski:2015cea}. Additionally it is important that, the bound on the {\it Lyapunov exponent} saturates in the context of Sachdev-Ye-Kitaev (SYK) model \cite{Maldacena:2016hyu,Choudhury:2017tax,Mandal:2017thl,Gross:2017hcz,Das:2017pif,Garcia-Garcia:2016mno,Garcia-Garcia:2017pzl,Witten:2016iux,Nishinaka:2016nxg}, which describes the quantum features of Majorana fermions in presence of infinitely long range disorder. Saturation of the {\it Lyapunov exponent} implies that SYK model mimics a quantum description of black hole via AdS/CFT correspondence. In the strict classical limit $\hbar\rightarrow 0$  the {\it Lyapunov exponent} take any values, which is consistent with the requirement. 
 
 To give an explicit derivation of the chaos bound on {\it Lyapunov exponent} in the context of cosmology let us follow the steps appended below:
 \begin{enumerate}
 \item Let us start with a completely mathematical problem described by a time dependent function $g(\tau)$, which satisfy the following set of properties:
 \begin{enumerate}
 \item In the complex plane $g(\tau+iT)$ is analytic in the half strip described within $\tau>0$ and $-\frac{\beta}{4}\leq T \leq \frac{\beta}{4}$. In this context, $\tau$ and $T$ represent the real and imaginary part of the complex number $\tau+iT$ after analytical continuation in complex plane.
 \item The function $g(\tau)$ is completely real at $T=0$. 
 \item After analytical continuation the function in the complex plane satisfy the following constraint:
 \be \label{poq} |g(\tau+iT)|\leq 1,\ee
 which is perfectly valid in the complete half strip.
 \end{enumerate}
 \item Next, we actually conformally map the entire half strip to a unit thermal circle in the complex plane, which can be done using the following {\it M$\ddot{o}$bius transformation}:
  \bea\textcolor{blue}{\bf \underline{M\ddot{o}bius~transformation:}}~~~~
 z=\frac{1-\Delta_{\beta}(\tau+iT)}{1+\Delta_{\beta}(\tau+iT)},\eea
 where $\Delta_{\beta}(\tau+iT)$ is the temperature dependent function in the complex plane, described by the following expression:
  \bea \Delta_{\beta}(\tau+iT):=\sinh \left(\frac{2\pi}{\beta}(\tau+iT)\right).\eea
  \begin{figure}[htb]
\centering
\subfigure[$|z|$ with $\beta=1$.]{	
	\includegraphics[width=7.8cm,height=8cm] {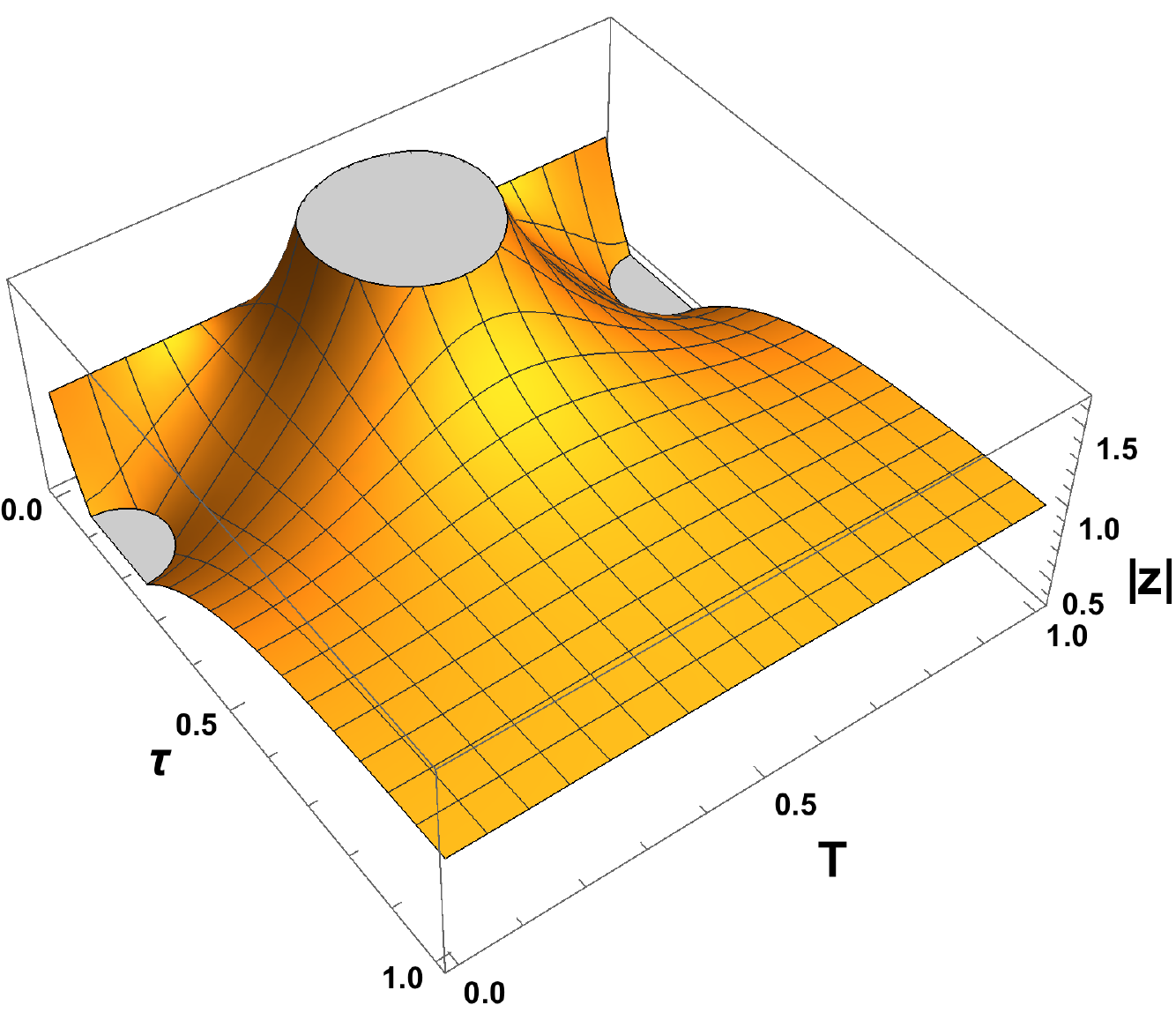}
	\label{CON5}
}
\subfigure[z with $\beta=1$ and $T=0$.]{	
	\includegraphics[width=7.8cm,height=8cm] {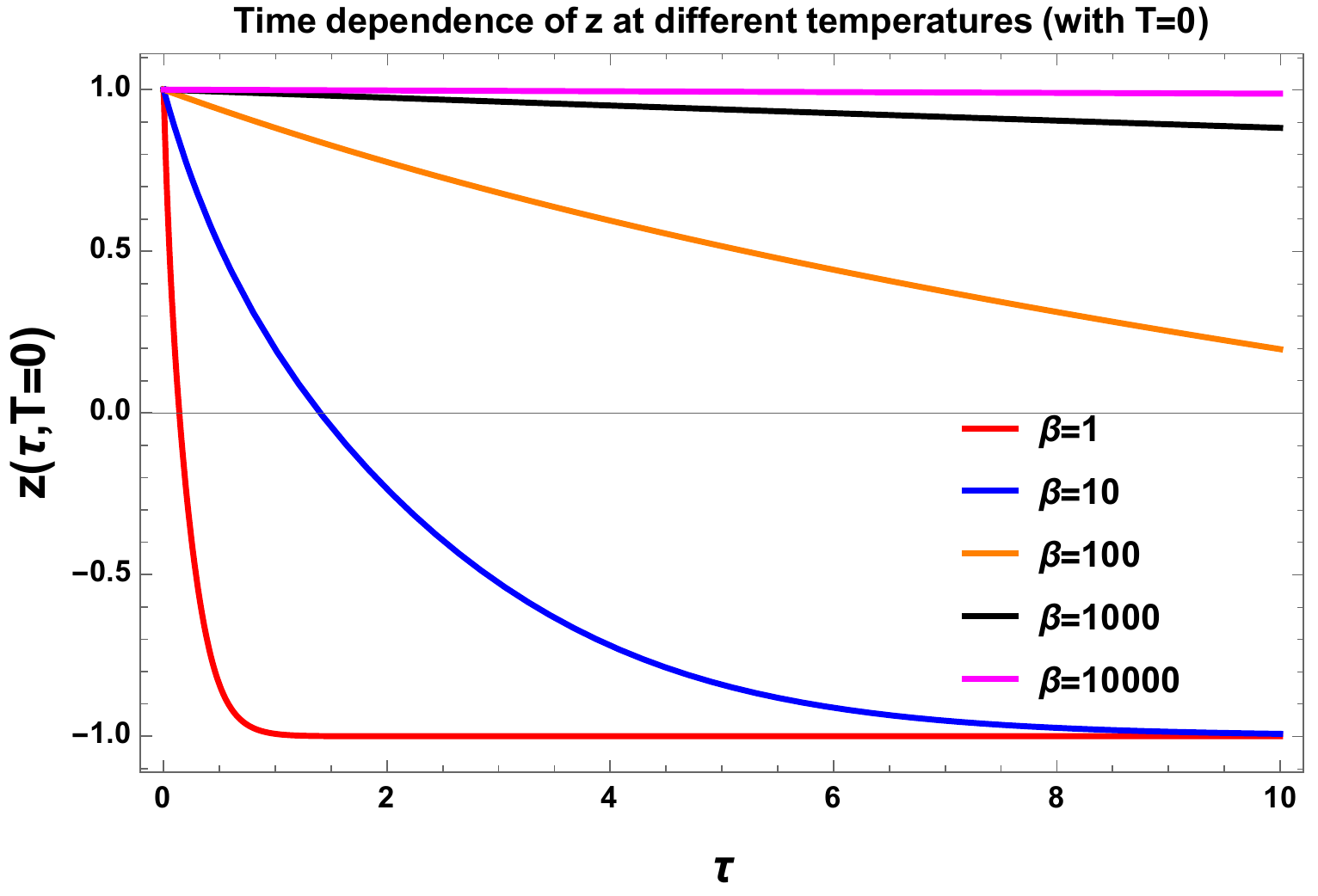}
	\label{CON6}
}
\caption{Parametric dependence $|z|$ and $z(T=0)$ at different temperatures.}
\end{figure}
In fig.~(\ref{CON5}), we have shown the behaviour of the amplitude of the complex number $z$ with respect to the parameters ($\tau,T$) in 3D plot. Finally, to check the consistency with {\it Schwarz-Pick inequality} we have plotted the complex number $z$ at $T=0$ in fig.~(\ref{CON6}).
 \item Further using Eq~(\ref{poq}) one can further say that the complex function $g(z)$ is an analytic function from one to one conformal map from unit disk to unit disk.
 \item  A variant of the {\it Schwarz lemma} can be represented as a invariant contribution under analytic automorphisms on the unit disk, which implies the bijective holomorphic mappings of the unit disc to itself. This specific variant is known as the {\it Schwarz–Pick theorem}.

 \item Now the hyperbolic metric in complex plane is defined as:
  \bea ds^2=4\frac{dz d\bar{z}}{\left(1-|z|^2\right)^2}=\left(\frac{2|dz|}{\left(1-|z|^2\right)}\right)^2.\eea
 Further using this metric and applying {\it Schwarz–Pick theorem} one can write:
 \bea \label{ineq} \textcolor{blue}{\bf \underline{Schwarz–Pick~inequality:}}~~~~\frac{|dg|}{\left(1-|g(z)|^2\right)}\leq ds=\frac{2|dz|}{\left(1-|z|^2\right)}.\eea
 
 \item Further applying the fact that the function $g(\tau)$ is real at $T=0$ and using Eq~(\ref{ineq}) we get the following simplified result:
  \bea \label{dfq1e} \frac{1}{1-g^{2}(\tau)}\left|\frac{dg(\tau)}{d\tau}\right|\leq \left[\frac{1}{1-|z|^2}\left|\frac{dz}{d\tau}\right|\right]_{T=0}=\frac{\pi}{\beta}\coth \left(\frac{2\pi\tau}{\beta}\right).\eea
 \item Further, rearranging Eq~(\ref{dfq1e}) we get the following final result:
   \bea \label{dfq2ex} \frac{1}{(1-g(\tau))}\left|\frac{dg(\tau)}{d\tau}\right|\leq \frac{1}{2}(1+g(\tau))\frac{2\pi}{\beta}\coth \left(\frac{2\pi\tau}{\beta}\right),\eea
   which is the outcome of {\it Schwarz-Pick inequality} and very very useful to prove the universal chaos bound in OEQFT.
  
 Now it is important to note that in this context,
   \bea  \frac{1}{2}(1+g(\tau))\coth \left(\frac{2\pi\tau}{\beta}\right)\leq 1+\frac{\beta}{2\pi}{\cal O}\left(\exp\left(-\frac{4\pi\tau}{\beta}\right)\right).~\eea
  This further implies that:
   \bea \label{dfq2edd} \frac{1}{(1-g(\tau))}\left|\frac{dg(\tau)}{d\tau}\right|\leq \frac{2\pi}{\beta}+{\cal O}\left(\exp\left(-\frac{4\pi\tau}{\beta}\right)\right).\eea
  Now at very large time scale ($\tau\rightarrow \infty$) or at very high temperature ($\beta=1/T\rightarrow 0$) one can neglect the contribution from the second sub-leading term. As a result we get the following inequality:
    \bea\label{dfq2edd} \frac{1}{(1-g(\tau))}\left|\frac{dg(\tau)}{d\tau}\right|\leq \frac{2\pi}{\beta},\eea
   
 \item    Further, we take the following phenomenological function:
   \bea g(\tau)=1-k\exp[\hbar\lambda \tau],\eea
   where $k$ is constant and $\lambda$ is the {\it Lyapunov exponent}. This function satisfy all the requirements that we have mentioned earlier explicitly. Further substituting this function in the result obtained in Eq~(\ref{dfq2ex}) we get the following simplified result~\footnote{Henceforth we set $\hbar=1$ for which the bound is translated to $\lambda\leq \frac{2\pi}{\beta}$, which we will use for the further application purposes.}:
  \bea\lambda \leq \frac{2\pi}{\hbar\beta},\eea
  which proves the {\it Universal chaos bound} in OEQFT. 
  \item This bound on the {\it Lyapunov exponent} is an unique feature of all classes of OEQFT set up. It has a very strong impact in the context of early universe cosmology, specifically during reheating epoch. By knowing specific time dependent couplings in the context of effective field theory (EFT) it is possible to give an estimate of {\it Lyapunov exponent} in such OEQFT set up. We will show this feature in the next section for three known model of interactions appearing in EFT. In such a situation one can give an estimate of the upper bound on reheating temperature using this bound, which is again obviously an universal bound itself. The earlier study in the context of reheating actually predicts a very crude estimate of reheating temperature which is based on the assumption that reheating is extremely model dependent. It actually means that to write an EFT of reheating we need to know the all interacting relativistic degrees of freedom in a specific model. In this framework the total energy density during reheating can be expressed in terms of total number of relativistic degrees of freedom by the following expression:
   \bea \rho_{\rm reh}=\frac{\pi^2}{30}g_{*}(T_{\rm reh})T^4_{\rm reh}.\eea
  Using this expression of energy density during reheating epoch one can able to express the reheating temperature as:
 \bea T_{\rm reh}=\left(\frac{30}{\pi^2 g_{*}(T_{\rm reh})}\right)^{1/4}\rho^{1/4}_{\rm reh}\approx \left(\frac{30}{\pi^2 g_{*}(T_{\rm reh})}\right)^{1/4}V^{1/4}_{\rm reh},\eea
  where $g_{*}(T_{\rm reh})$ is the effective number of total relativistic degrees of freedom present in the thermal bath at temperature $T=T_{\rm reh}$ and $V_{reh}$ is the scale of reheating which can be obtained by fixing the field value at $\phi=\phi_{\rm reh}$ for a specific model. Counting all the degrees of freedom in the particle physics model one can fix $g_{*}(T_{\rm reh})$ in the present context. To find the reheating constraint from the prescribed set up let us further introduce the number of e-foldings at the epoch of reheating, which is defined as:
  \bea {\cal N}_{\rm reh}=\int^{t_{e}}_{t_{\rm reh}}H~dt={\cal N}_{\rm total}-\widetilde{\Delta{\cal N}}\approx -\frac{1}{M^2_p}\int^{\phi_e}_{\phi_{\rm reh}}\frac{V(\phi)}{V^{'}(\phi)}~d\phi,\eea
  where ${\cal N}_{\rm total}$ is the total number of e-foldings which is defined as:
  \be {\cal N}_{\rm total}=\int^{t_{e}}_{t_{i}}H~dt -\frac{1}{M^2_p}\int^{\phi_e}_{\phi_{i}}\frac{V(\phi)}{V^{'}(\phi)}~d\phi\sim \underbrace{{\cal O}(60-70)}_{\textcolor{red}{\bf From ~Planck~observation}}~.\ee
  Here $t_{e}$, $t_{i}$ and $t_{\rm reh}$ are the representative time to specify end of inflation, starting of inflation and time scale at the end of reheating respectively. Similarly $\phi_{e}$ and $\phi_{\rm reh}$ are the field values at the end of inflation and reheating respectively, which can be computed for a given known model of inflation. Also it is important to note that in this context, $\widetilde{\Delta{\cal N}}$ is defined as:
  \bea \widetilde{\Delta{\cal N}}={\cal N}_{\rm total}-{\cal N}_{\rm reh}=\Delta{\cal N}-\left({\cal N}_{\rm reh}-{\cal N}_{\rm cmb}\right)\Longrightarrow \Delta{\cal N}-\widetilde{\Delta{\cal N}}=\left({\cal N}_{\rm reh}-{\cal N}_{\rm cmb}\right).\eea
  Here $\Delta{\cal N}$ is defined as:
  \bea \Delta{\cal N}={\cal N}_{\rm total}-{\cal N}_{\rm cmb}.\eea
  From different models of inflation one can explicitly compute e-foldings at horizon exit, which is given by the following expression:
   \bea {\cal N}_{\rm cmb}=\int^{t_e}_{t_{\rm cmb}}H~dt\approx -\frac{1}{M^2_p}\int^{\phi_e}_{\phi_{\rm cmb}}\frac{V(\phi)}{V^{'}(\phi)}~d\phi\sim \underbrace{{\cal O}(8-10)}_{\textcolor{red}{\bf From ~Planck~observation}}~.\eea
  Consequently, the value of $\Delta{\cal N}$ from observation can be estimated as:
  \bea\Delta{\cal N}\sim {\cal O}(52-60).\eea
  Now, to give a numerical estimate of the reheating temperature let us consider the following simplest monomial model:
  \bea e V(\phi)=V_0 \left(\frac{\phi}{M_p}\right)^p,\eea
  where $V_0$ fix the overall scale of the potential and $p$ is the degree of the monomial which depends on the characteristic of the model. For this model the field value during reheating can be expressed as:
   \bea \phi_{\rm reh}=\sqrt{2p{\cal N}_{\rm reh}+\left(\frac{\phi_e}{M_p}\right)^2}~M_p.\eea
  The reheating scale is quantified in terms of the number of e-foldings as: 
  \bea V(\phi_{\rm reh})=V_0 \left(\frac{\phi_{\rm reh}}{M_p}\right)^p=V_0~\left[2p{\cal N}_{\rm reh}+\left(\frac{\phi_e}{M_p}\right)^2\right]^{p/2}.\eea
  Consequently, for the monomial model the reheating temperature can be quantified as:
   \bea &&\textcolor{blue}{\bf \underline{Reheating~bound~from~model:}}\\
  &&T_{\rm reh}=\left(\frac{30}{\pi^2 g_{*}(T_{\rm reh})}\right)^{1/4}V^{1/4}_0\left[2p{\cal N}_{\rm reh}+\left(\frac{\phi_e}{M_p}\right)^2\right]^{p/8}<V^{1/4}_{\rm inf}.\nonumber\eea
  Here $V_{\rm inf}$ is the scale of inflation which is quantified by the following expression:
 \bea \textcolor{blue}{\bf \underline{Upper~bound~on~inflationary~scale:}}~~~~V^{1/4}_{\rm \inf}\leq 1.67\times 10^{16}{\rm GeV}\left(\frac{r(k_*)}{0.064}\right)^{1/4}.\eea
  As a result, we get the following bound on the reheating temperature:
   \bea &\textcolor{blue}{\bf \underline{Upper-bound~on~reheating ~temperature~from~inflation:}}\nonumber\\
  &~~~~~~~~~~~~~~~~~~~~T_{\rm reh}\leq 1.67\times 10^{16}{\rm GeV}\left(\frac{r(k_*)}{0.064}\right)^{1/4},\eea
  which is true for any models of inflation. From the {\rm Planck 2018+BICEP2/Keck Array BK14 data} the upper bound on the tensor-to-scalar ratio (primordial gravitational waves) is restricted to:
   \bea r(k_*)<0.064,\eea
  where $k_*\sim 0.05 {\rm Mpc}^{-1}$ is the pivot scale of momentum. This implies that the upper bound of reheating temperature from the {\rm Planck 2018+BICEP2/Keck Array BK14 data} is given by:
   \bea T_{\rm reh}\leq 1.67\times 10^{16}{\rm GeV}.\eea
  
  Here to writing down this expression for reheating temperature it is important to consider the following assumption:
  \begin{enumerate}
  \item Contribution from the kinetic term of the field which is mainly responsible for reheating is neglected.
  
  \item We also assume that reheating is described by scalar field.
  \end{enumerate}
  This further implies that depending on the  background particle physics model reheating temperature actually varies in a wide range and one cannot able to determine exactly its value as there is no such universal bound available earlier in this context. This is the main shortcoming of the phenomenological prediction of reheating temperature in the context of early universe cosmology.
  
  On the other hand, just only considering the dynamical details of quantum chaos one can express the reheating temperature in terms of the {\it Lyapunov exponent}:
  \bea \textcolor{blue}{\bf \underline{Universal~lower-bound`on~reheating~temperature:}} ~~~~T_{\rm reh}\geq \frac{\lambda}{2\pi},\eea
  which is an universal lower bound on reheating temperature in the present context of discussion as it is not involve any model dependence from the background theory. This implies that the universal bound on quantum chaos in OEQFT restrict us to fix an universal model independent lower bound on reheating temperature. Combing the obtained bound in this paper and the upper bound obtained from inflation one can restrict the reheating temperature within a specified range. Additionally, the present analysis helps us put an unique upper bound on the {\it Lyapunov exponent} in terms of the scale of inflation (or tensor-to-scalar ratio) as:
  \bea \lambda \leq V^{1/4}_{\rm inf}=1.67\times 10^{16}{\rm GeV}\left(\frac{r(k_*)}{0.064}\right)^{1/4}.\eea

 \end{enumerate}
 
 \subsection{Out of time ordered correlators (OTOC) in OEQFT }
 \subsubsection{What is OTOC?}
 Now it is important to note that the universal bound on quantum chaos can be achieved by computing the out of time ordered correlators (OTOC), which in general can be expressed in terms of commutators. In the study of quantum chaos, specifically in the context of {\it Butterfly effect} one can introduce two time dependent operators $W(\tau)$ and $V(\tau^{'})$ from which one can define a commutator, $\left[W(\tau),V(0)\right]$, where the operators are in general Hermitian in nature and they have introduced with time separation $\Delta \tau= \tau-\tau^{'}=\tau$ with $\tau^{'}=0$. This commutator actually captures the effect of perturbation by the operator $V(0)$ on the later time measurement on the operator $W(\tau)$ and the converse statement is also true. The time dependence of the operator $W(\tau)$ in this context of discussion can be expressed in the Heisenberg representation as:
 \bea W(\tau)=\exp\left[iH\tau\right]~W(0)~\exp\left[-iH\tau\right].\eea
The strength of such chaotic effect is characterised by the following measure:
 \bea\textcolor{blue}{\bf \underline{Quantum~OTOC:}}~~~~~~~{\cal C}(\tau):=-\langle \underbrace{\left[W(\tau),V(0)\right]^2}_{\textcolor{red}{\bf Four~point~quantum~operator}}\rangle, \eea
where the expectation value is in general the thermal averaged~\footnote{Thermal averaging is a very important concept in the context of AdS/CFT correspondence as the dual description of the quantum field theory of balck holes can be treated as a thermal bath which have Hawking temperature.}, which is defined as:
 \bea {\cal C}(\tau)=-\langle \left[W(\tau),V(0)\right]^2\rangle=-\frac{1}{Z}{\rm Tr} \left\{\exp(-\beta H) \left[W(\tau),V(0)\right]\right\}.\eea
Here $Z$ is the partition function which is defined as:
 \bea Z={\rm Tr}\left\{\exp[-\beta H]\right\},\eea
and $H$ is the Hamiltonian of the chaotic system under consideration. Here it is important to note that to construct the chaotic OTOC measure instead of using the two point operator, $\left[W(\tau),V(0)\right]$ (commutator), here we have actually used the four point quantum operator, $\left[W(\tau),V(0)\right]^2$ (square of the commutator). The specific reason for such choice is following. To describe this let us first assume that we replace the commutator bracket by the Poisson bracket by considering the semi-classical limiting situation. In such a case the Poisson bracket shows typically an exponential growth, $\exp[\lambda \tau]$, where $\lambda$ is the {\it Lyapunov exponent}. But the signature of its coefficient can be anything, either positive or negative. Now further if we take the thermal averaging over this two point operator then both the contributions are cancelled each other in the semi-classical limit and will not contribute to describe chaos. From the quantum mechanical perspective, the two point thermal averaged operator, $\langle \left[W(\tau),V(0)\right] \rangle $ actually captures the description of correlation between the quantum Hermitian operators $W(\tau)$ and $V(0)$., which decays in the large time limit ($\tau\rightarrow \infty$) and cannot describe the chaotic behaviour. On the other hand, the four point quantum operator after transforming it to the Poisson bracket in the semi-classical picture don't show any ambiguity in the signature of the co-efficient as it takes only positive value. After taking thermal average we get non vanishing result using which one can describe quantum chaos. In the quantum mechanical picture the four point thermal averaged operator, $\langle \left[W(\tau),V(0)\right]^2 \rangle $ not decays exponentially at the leading order in the large time limit ($\tau\rightarrow \infty$). 

Now, in the quantum mechanical description of the {\it Butterfly effect} predicts the following result:
 \bea  {\cal C}(\tau)\sim 2\langle V(0) V(0) W(\tau) W(\tau) \rangle= 2\langle V(0) V(0) \rangle \langle W(\tau) W(\tau) \rangle ~~~{\rm for}~ \tau\rightarrow \infty, \eea
for any mathematical structure of the operators $V(0)$ and $W(t)$. Here it is important to note that, $V(0)W(\tau)W(\tau)V(0)$ contribution is not directly effected by the quantum chaos.

Also it is important to note that, in the present context for the sake of simplicity we additionally assume that:
 \bea \langle V(0) \rangle &=0,\\
\langle W(\tau) \rangle &= 0,\eea
i.e. both the one point function or the thermal averaged expectation values of these operators vanishes.

\subsubsection{Estimation of scrambling and dissipation time scales from OTOC}
In the context of quantum chaos two important time scales are associated: 
\begin{enumerate}
\item \textcolor{red}{\bf \underline{Scrambing time:}}\\
Here the associated time scale where the operator ${\cal C}(\tau)$ is relevant is known as the {\it scrambling time scale} $\tau_{*}$. Sometimes in literature this is known as the {\it Ehrenfest time scale}. A possible distinction between the classical and quantum description of chaos can be described by the {\it Ehrenfest time scale} in which the previously mentioned OTOC don't grow with respect to the associated time scale and saturates at the same scale. In the next section we have provided a alternative chaos bound on OTOC (i.e. SFF in our case) from which we have further give an estimate of the bound on the {\it Ehrenfest time scale}. 
\item \textcolor{red}{\bf \underline{Dissipation time:}}\\
Another time scale for chaos is the exponential decay time scale $\tau_d$ in which the two point thermal correlation function behaves like $\langle V(0)V(\tau)\rangle$. Sometimes in this literature it is known as the {\it dissipation time scale} or the {\it collision time scale}. In the context of strongly coupled quantum field theories at finite temperature it is expected that the {\it dissipation time scale} $\tau_d\sim \beta$. It is also expected that for large time limit the more general form of the OTOC during this time scaled as:
 \bea \langle V(0)V(0)W(\tau)W(\tau)\rangle \sim \langle V(0)V(0) \rangle \langle W(\tau)W(\tau) \rangle+{\cal O}(\exp[-\tau/\tau_d])+\cdots, \eea
where $\cdots$ represent higher order terms which are more suppressed by the {\it dissipation time scale} $\tau_d$.
\end{enumerate}
In the present context, additionally one can predict the connection between the quantum mechanical operator ${\cal C}(\tau)$ and quantum chaos by considering the semi-classical limit of a chaotic system which involves a single particle. To demonstrate this argument one can consider  semi-classical billiards as a toy example. In the semi-classical limit one can take,
 $V(0)=p(0),~W(\tau)=q(\tau),$
where $p$ and $q$ is the generalized momentum and coordinate respectively. As a result in the semi-classical limiting approximation one can map the previously defined commutator bracket to the Poisson bracket, as given by:
 \bea\left[q(\tau),p(0)\right]\Longrightarrow i\hbar \left\{q(\tau),p(0)\right\}_{\bf PB}=i\hbar \frac{\pl q(\tau)}{\pl q(0)},\eea
which can be treated as the classical analogous version of the quantum mechanical {\it Butterfly effect}. It is also expected that for such a system the nearby dynamical trajectories scale as,
$q(\tau)\sim q(0)\exp\left[\lambda \tau\right],$
where $\lambda$ is the {\it Lyapunov exponent}. It is in principle divergent in nature for large time limiting situation. Now at the {\it dissipation time scale} $\tau_d$ it is also expected that,
$\tau_d \sim 1/\lambda,$
for which the nearby trajectory is convergent and is of the order of $e$. On the other hand, the prescribed OTOC can approximately expressed in semi-classical limit as~\footnote{\textcolor{blue}{\bf \underline{Classical~result:}}~~~Here one can perform the exact classical computation of OTOC to check whether the quantum and classical descriptions give the same result or not. In the case of billiards, the Poisson bracket is given by,
$\left\{q(\tau),p(0)\right\}_{\bf PB}\sim \exp[\lambda\tau].$
One can explicitly show that in this context the {\it Lyapunov exponent} can be expressed as, 
$ \lambda \sim \frac{v}{\sqrt{A}}=\frac{p(0)}{\sqrt{A}},$
where $A=\pi R^2+4aR$ is the area of the stadium and $v$ is the velocity of the particle. Then the classical OTOC can be expressed as:
\bea {\cal C}(\tau)&=&\frac{1}{Z_{\rm cl}}\int \frac{d^2q}{2\pi}\frac{d^2p(0)}{2\pi}~\exp\left[-\beta p^2(0)+\frac{2p(0)}{\sqrt{A}}\right]\nonumber\\
&=&\frac{1}{Z_{\rm cl}}\int^{\infty}_{0}\frac{dp}{2\pi}~p~\exp\left[-\beta\left(p(0)-\frac{\tau}{\beta\sqrt{A}}\right)^2+\frac{\tau^2}{\beta^2 A}\right]\nonumber\\
&=& \left\{1+ \frac{\sqrt{\pi } \tau }{ \sqrt{A\beta }}\exp\left[\frac{\tau ^2}{A \beta ^2}\right]\left(\text{erf}\left(\frac{\tau }{2 \sqrt{A\beta }}\right)+1\right)\right\}, \eea
where $Z_{\rm cl}$ is the classical partition function defined as:
\be Z_{\rm cl}=\int \frac{d^2q}{2\pi}\frac{d^2p(0)}{2\pi}~e^{-\beta p^2(0)}=\int^{\infty}_{0}\frac{dp}{2\pi}~p~\exp\left[-\beta\left(p(0)\right)^2\right]=\frac{1}{4\pi \beta}.\ee
Further taking $A=1$ for simplicity we get:
\bea \textcolor{blue}{\bf \underline{Classical~OTOC:}}~~~{\cal C}(\tau)&=&\left\{1+ \frac{\sqrt{\pi } \tau }{2 \sqrt{\beta }}\exp\left[\frac{\tau ^2}{4  \beta ^2}\right]\left(\text{erf}\left(\frac{\tau }{2 \sqrt{\beta }}\right)+1\right)\right\} \eea
Further taking the limit $t>>\sqrt{\beta}$ we get the following simplified answer for classical OTOC for billiards:
 \bea\textcolor{blue}{\bf \underline{Classical~OTOC:}}~~~{\cal C}(\tau)&=&\frac{\sqrt{\pi } \tau }{\sqrt{\beta }}\exp\left[\frac{\tau ^2}{4  \beta ^2}\right]~~~~~~{\rm for}~~t>>\sqrt{\beta}~. \eea
This result implies that in classical OTOC and in semi-classical (or quantum) OTOC the time dependence is completely different. In the case of classical OTOC it shows faster growth with respect to the result obtained for quantum OTOC.}:
 \bea\textcolor{blue}{\bf \underline{Semi-classical~OTOC:}}~~~~~{\cal C}(\tau)\sim\hbar^2\left(\frac{\pl q(\tau)}{\pl q(0)}\right)^2=\hbar^2\exp\left[2\lambda \tau\right]. \eea
\begin{figure}[H]
\centering
\subfigure[Semi-classical ~OTOC]{	
	\includegraphics[width=7.8cm,height=8cm] {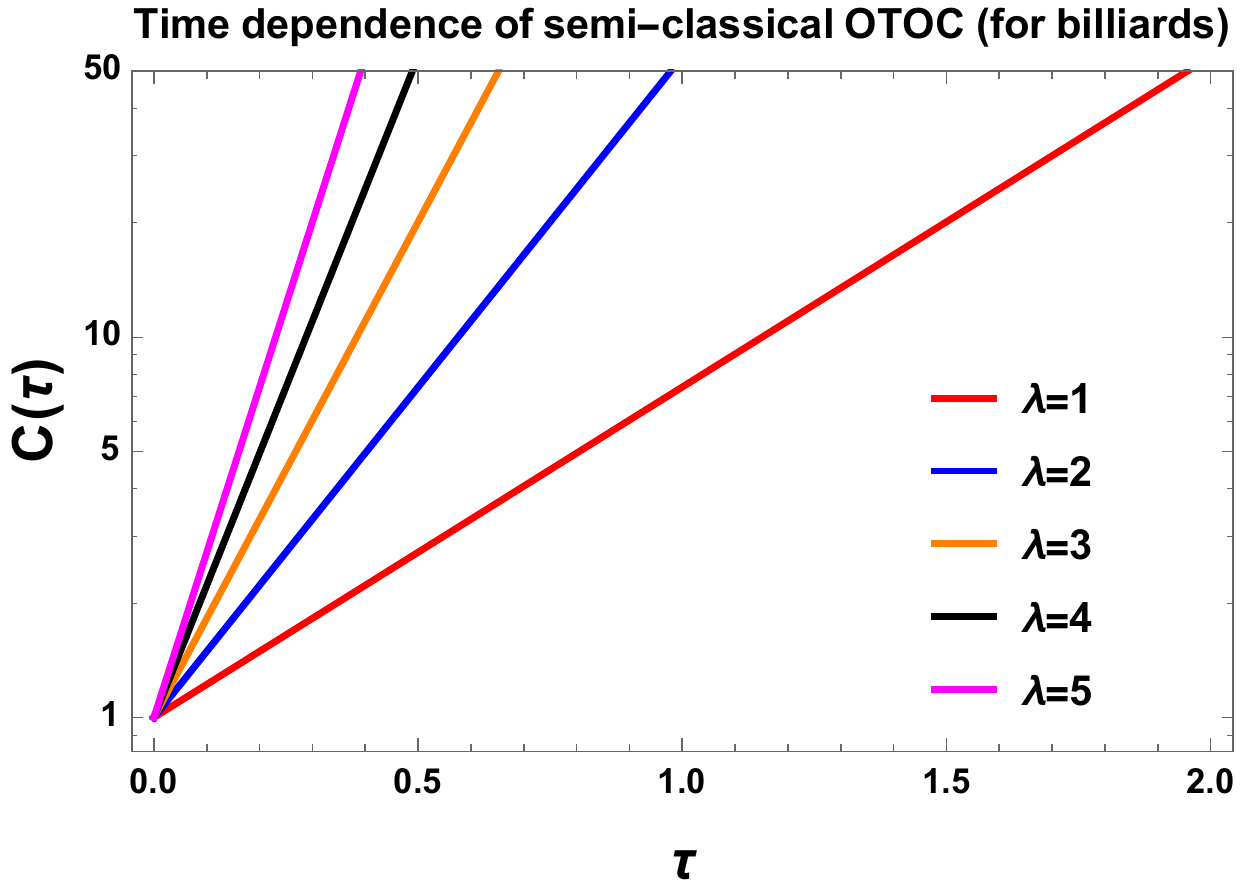}
	\label{OTOC1}
}
\subfigure[Classical ~OTOC]{	
	\includegraphics[width=7.8cm,height=8cm] {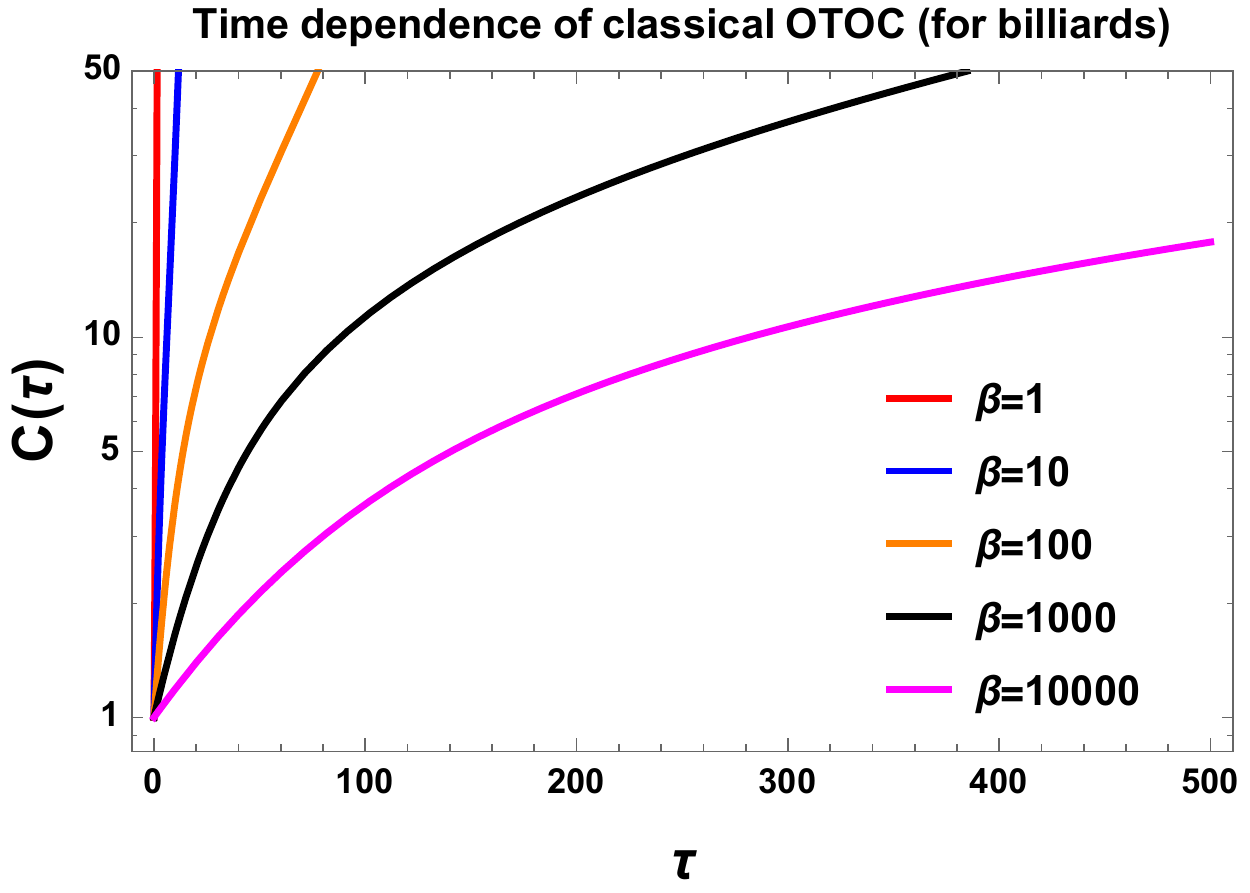}
	\label{OTOC2}
}
\caption{Time dependent behaviour of semi-classical and classical OTOC for billiards.}
\end{figure}

In fig.~(\ref{OTOC1}) and fig.~(\ref{OTOC2}), we have shown the variation of the time dependent behaviour of semi-classical and classical OTOC for billiards, which show they are different in both the cases.

Now at the {\it scrambling time scale}, $\tau_*$ and {\it dissipation time scale}, $\tau_d$ the OTOC approximately in the semi-classical limit scaled as:
\bea {\cal C}(\tau_*)\sim 1,~~~~~{\cal C}(\tau_d)\sim \hbar^2 e^2,\eea
from which the {\it scrambling time scale}, $\tau_*$ can be estimated as:
\bea \tau_{*}&\sim & \frac{1}{\lambda}\ln \frac{1}{\hbar}.\eea
This further implies that, in the semi-classical limit  the {\it scrambling time scale}, $\tau_*$ and {\it dissipation time scale}, $\tau_d$ are related by the following expression:
 \bea \tau_{*}&\sim \tau_d\ln \frac{1}{\hbar},\eea
which explicitly shows that both the time scales for quantum chaos is different from each other and the fractional difference is given by the following expression:
 \bea \frac{\tau_d - \tau_*}{\tau_d}&=&1-\ln\frac{1}{\hbar}=\ln \hbar, \eea
which is actually a large amount of hierarchy at the semi-classical limit as $\hbar\rightarrow 0$.

Now, the OTOC in the present context actually quantify the temporal growth of the Hermitian quantum mechanical operator $W(\tau)$ is is introduced earlier in this section. In the general prescriptions of quantum field theory (QFT) such OTOC can be expressed in terms of the addition of simple type of operators, which span the quantum basis. Now, if the OTOC is large~\footnote{In the present context large OTOC (${\cal C}(\tau)$ implies that the quantum operator for chaos $W(\tau)$ completely destroy the effect of the initial factor $\exp[iH\tau]$ and the final factor $\exp[-iH\tau]$ to cancel their contribution in the definition of the operator $W(\tau)$.} then in such a situation with non-local interactions the {\it scrambling time scale}, $\tau_*$ can be estimated as:
 \bea t_*\sim \ln {\cal N}_{\rm bit}~~~~~~{\rm for}~~{\cal C}(\tau)\rightarrow \infty,\eea 
where ${\cal N}_{\rm bit}$ is the number of qubits. Similarly for local interactions he {\it scrambling time scale}, $\tau_*$ can be estimated by computing the separation between the quantum operators $W$ and $V$. 
Additionally it is important to note that, quantization of a classical chaotic system may accommodate positive {\it Lyapunov exponent} from the OTOC mentioned earlier. To quantify quantum chaos also the nearest neighbour distribution (NDD) for the spectrum of the energy is alternatively used~\footnote{In the context of integrable and non-integrable quantum mechanical system nearest neighbour distribution (NDD) is described by {\it Poisson} and {\it Wigner functional}.}. Except {\it Lyapunov exponent}, in the present context of discussion OTOC (in our discussion it is SFF) play crucial role to quantify quantum chaos to explain dynamical features in the early epoch of universe.

\section{Quantum chaos from RMT: An alternative treatment in cosmology}
\label{SpecForm}
In this section, we will try to generalize spectral form factor (SFF) for any order of even polynomial potential. To serve this purpose, one can create such ensemble, such that all possible interaction between energy levels of many-body Hamiltonian would be accounted for by various matrices in the ensemble. If the Hamiltonian is time-reversal symmetric the required distribution will be invariant under orthogonal transformation. Else, it is invariant under unitary transformation. 

In the thermodynamic limit ($N \rightarrow \infty$) eigen value of density of random matrices showed a universal behaviour characterised by {\it Wigner's Semicircle law}. The results seemed to be applicable to a varied class of quantum system displaying chaotic behaviour. Chaos was also a hallmark of a few-body Hamiltonian ($N$ finite), but better diagnostic for quantum systems was devised in which nearest neighbour spacing distribution (NNSD) of eigenvalues of the system will be chaotic if distribution is {\it Wigner Dyson} type:
 \bea\label{connect1}
P(\og = E_{n+1}-E_{n})=A_{\bg} \og^{\beta} e^{(-\bg \og)},
\eea
Here it is important to note that, here $\beta$ is fixed at, $\bg =1$ for Gaussian orthogonal ensemble and $\bg=2$ for Gaussian unitary ensemble. In the present context of discussion Spectral Form Factor (SFF) is a tool for characterising spectrum ( i.e. discreteness of energy spectrum) of quantum system under consideration and defined by the following expression:
 \bea{\bf SFF}=|Z(\bg + i \tau)|^{2}=\sum_{m,n} e^{-\bg (E_{m}+E_{n})} e^{-it(E_{m}-E_{n})}.
\eea
Here $Z(\bg)$ is the partition function of the quantum system and $\bg=1/T$.  For $\bg=0$, the expression pick out contribution only form the difference between nearest neighbour energy eigenvalues at very late times. SFF when averaged over Gaussian random matrices, has very particular behaviour at large N with initial decay followed by a linear rise and then after a critical point saturation.This approach can relate a saturation limit for large N which can be treated as {\it bound on chaos}.
 Additionally, it is important to note that quantifying chaos through finding SFF is very useful when one cannot have a specific time dependent mass profile during cosmological particle production. In terms of scattering problem in the conduction wire if we don't know precisely the structure of interaction potential, then one can quantify chaos in terms of SFF rather than using {\it Lyapunov exponent}, as we have used in the previous section. Here we will discuss general approach to find SFF to quantify chaos for various even polynomial potential.
\subsection{Quantifying chaos using RMT}
Gaussian matrix ensemble is a collection of large number of matrices which are filled with random numbers picked arbitrarily from a Gaussian probability distribution. See refs.~\cite{Eynard:2015aea,1997PhRvE..55.4067B} for more details.

\begin{table}[H]
\centering
\footnotesize
\begin{tabular}{|||c||c||c|||}
\hline\hline\hline
\textcolor{red}{\bf Element of matrix} & \textcolor{blue}{\bf Type of ensemble} & \textcolor{purple}{Relation} \\
\hline\hline
Elements are real &  Gaussian Orthogonal Ensemble  &  time reversal symmetric Hamiltonian \\
\hline\hline
Elements are complex& Gaussian Unitary Ensemble& broken time reversal symmetric Hamiltonian \\
\hline\hline
Elements are quaternion& Gaussian Sympletic Ensemble &- \\
\hline\hline\hline
\end{tabular}
\caption{Properties of Gaussian matrix ensemble in Random Matrix Theory (RMT).}
\label{vgf1}
\end{table}
In table~(\ref{vgf1}), we have explicitly mentioned the properties of the each elements of the Gaussian matrix ensemble in Random Matrix Theory (RMT).

Further, the joint probability distribution of such random matrix, which is characterized by the Gaussian potential is given by the following expression:
 \bea P(M)dM=\exp\left(-\frac{1}{2}trM^{2}\right)dM=\exp\left(-\frac{1}{2}\sum_{i=1}^{N}x_{ii}^{2}\right) \exp\left(-\sum_{i\neq j}^{N}x_{ii}^{2}\right)\prod_{i\eqslantless j=1}^{N} dx_{ij},
\eea
where $N$ represents the rank of the matrix $M$. If we consider any ensemble of matrices to keep this  measure invariant under similarity transformation: \be M\rightarrow U^{-1}MU,\ee such that it satisfies the following constraint:
 \be P(U^{-1}MU)=P(M).\ee 
 Here $U$ being an orthogonal or unitary matrix. Then for most generalized ensemble one can implement the concepts of time independent Random Matrix Theory \cite{RMT1} in the present context of discussion.  Now here integrating over the random matrix measure one can construct the following expression for the partition function for the  
 Gaussian matrix ensemble, as given by:
 \bea
\Zstroke=\int dM~ e^{-{\rm Tr}(V(M)}.
\eea
Further, using similarity transformation one can diagonalize the random matrix $M$ as:
\be M=U^{-1}DU. \ee
On the other hand,  ensemble in basis of eigenvalues of the matrix the partition function can be written as:
 \bea \Zstroke=                                                                                                                                                                                                                                                                                                                                                                                                                                                          \prod_{i=1}^{N}\int d \lb_{i}~e^{-N^{2}S(\lb_{1},....\lb_{N})}
\eea
where the action $S(\lb_{i})$ is defined as~\footnote{This formalism is very useful when we can't able specify the particle interaction in the effective action. More precisely, in this situation when we really don't have any information about the particle interaction one can't able to define the action in terms of the usual language. Additionally it is important to note that, in our computation we consider that gravitational background is classical and non dynamical. However it will not explicitly appearing in the action for the distribution of eigen values of random matrices. Also during reheating since one can neglect the contribution from the expansion of our universe, then considering only the representative action for random distribution is sufficient enough for our discussion when we don't have any knowledge about the particle interactions at the level of action. In such a situation gravitational background is treated to be not evolving with time during reheating.}:
 \bea S(\lb_{1},....\lb_{N})=\frac{1}{N}\sum_{i=1}^{N} V(\lb_{i})+\bg \sum_{i< j}^{N}\log|\lb_{i}-\lb_{j}|.
 \eea                                                                                                                                         
 Here we fix $\bg=1$ for GOE and $\bg=2$ for GUE. The overall $1/N$ come from scaling of eigenvalues by factor $\sqrt{N}$. To find a solution we need to extremize the action w.r.t  $\lb_{i}$, such that we get:
  \bea\frac{d S}{d \lb_{i}}=0 \iff V'(\lb_{i})=\frac{2}{N}\sum_{j \neq i} \frac{1}{\lb_{i}-\lb_{j}}.
 \eea
 Now we need the method of resolvents to derive the expression for the partition function ($Z(\bg)$) in the present context.
 In continum limit of eigenvalues we can use density of states (eigen values) $\rho(\lb)$, which gives the number of eigen values lying in between $\lb$ and $\lb+d\lb$. Therefore, saddle point of $V'(\lb_{i})$ is given by the following expression: 
  \bea\label{V'}
 V'(\lb_{i})=2 {\rm Pr}\left(\int d u \frac{\rho(u)}{\lb -u}\right)\eea
 Here {\rm Pr} represents the principal part of the integral. Solution of the principal part of the integral Eq~(\ref{V'}) gives the eigen value density $\rho(u)$ at large N limit.
 
 Now, we can define resolvent as given by: 
  \bea\label{pw10}
\og(x)=\frac{1}{N}\sum_{i=1}^{N} \frac{1}{x-\lb_{i}}\eea
 further, using Eq~(\ref{pw10}) we compute the following function:
  \bea\label{oid1}
 \og^2(x)+\frac{1}{N}\og'(x)&=&\frac{1}{N^{2}}\left[\sum_{i=1}^{N}\frac{1}{x-\lb_{i}}\right]^{2}-\frac{1}{N^{2}}\sum_{i=1}^{N}\frac{1}{(x-\lb_{i})^{2}}\nonumber\\
&=&\frac{1}{N^{2}}\left[\sum_{i=1}^{N}\frac{1}{x-\lb_{i}} \sum_{i \neq j=1}^{N} \frac{\lb_{j}-\lb{i}}{(x-\lb_{i})(x-\lb_{j})}\right]\nonumber\\
&=&\frac{1}{N^{2}}\left[\sum_{i=1}^{N}\frac{1}{x-\lb_{i}} \sum_{i \neq j=1}^{N} \frac{1}{(\lb_{i}-\lb_{j})}\right]\eea
 Next, we use the following resolvent identities for our computation
 performed in this paper:
 \bea
  R(Z;A)-R(\og;A) = (Z-\og)R(Z;A)R(\og;A),\\
 R(Z;A)-R(Z;B) = R(Z;A)(B-A)R(Z;B).\eea
 Here $R$ denotes the resolvent and $A,B$  both defined over same linear space. Consequently, Eq~(\ref{oid1}) can be recast into the following simplified form: 
 \bea\label{res}
\og(x)^{2}+\frac{1}{N}\og'(x)=\frac{1}{N}\sum_{i=1}^{N} \frac{V'(\lb_{i})}{x-\lb_{i}}+\frac{1}{N}\sum_{i=1}^{N} \frac{V'(x)-V'(\lb_{i})}{x-\lb_{i}}
=V'(x)\og(x)-\rho(x).
\eea
Here we define:
 \bea\label{ba1}  \rho(x)=\sum_{i=1}^{N} \frac{V'(x)-V'(\lb_{i})}{x-\lb_{i}},~~~\og = \frac{1}{N}\sum_{i=1}^{N} \frac{1}{x-\lb_{i}}=\frac{1}{N}\frac{\Psi'}{\Psi},\eea
which implies that, here $\og'$ can be expressed as:
\be \og'=\frac{1}{N}\left(\frac{\Psi''}{\Psi}-\frac{\Psi'}{\Psi^{2}}\right).\ee
Finally, in terms of newly defined function $\Psi$ as stated in Eq~(\ref{ba1}), one can further recast Eq~(\ref{res}) as:
 \bea\frac{1}{N^{2}}\frac{\Psi'^{2}}{\Psi^{2}}+\frac{1}{N^{2}}\left(\frac{\Psi''}{\Psi}-\frac{\Psi'}{\Psi^{2}}\right)=V'(x) \frac{1}{N}\frac{\Psi'}{\Psi} -\rho(x).
 \eea
 Further, comparing the two equivalent definition of $\og(x)$ we get the following differential equation for $\Psi$ in terms of the eigen values of the random matrix, as given by:
 \bea\frac{\Psi'}{\Psi}=\sum_{i=1}^{N} \frac{1}{x-\lb_{i}}. 
 \eea
 Therefore , the solution for $\Psi(x) $ is given by the following characterestic polynomial :
  \bea\Psi(x)=\prod_{i=1}^{N}(x-\lb_{i})={\rm det}(x~{\rm I}-{\rm M}).
 \eea
Here it is important to note that, the solution obtained in large N limit can be compared with the solution obtained using WKB approximation in Schr{\"o}dinger equation. Then we can neglect the term $\frac{1}{N}\og'(x)$ in Eq~\ref{res} and write down the following approximated algebraic equation of $\og(x)$, given by:
 \bea\overline{\og}^{2}(x)-V'(x)\overline{\og}(x)+\overline{\rho}(x)=0\eea
where we have introduced two new quantities $\overline{\og}(x)$ and $\overline{\rho}(x)$, which are defined as:
\bea \overline{\og}(x)=\lim_{N\to\infty}\og(x),\\ \overline{\rho}(x)=\lim_{N\to\infty}\rho(x).\eea
Then solution of $\overline\og(x)$ is given by the following expression:
 \bea\overline\og(x)\equiv\overline\og_{\pm}(x)=\frac{1}{2}\left[V'(x)\pm \sqrt{(V'(x))^2-4\overline{\rho}(x)}\right].
\eea
Here for our discussion $\overline\og_{+}(x)$ is redundant and only acceptable solution for our purpose is given by the following expression:
 \bea\label{sq1a}
\overline\og(x)\equiv\overline\og_{-}(x)=\frac{1}{2}\left[V'(x)- \sqrt{(V'(x))^2-4\overline{\rho}(x)}\right].
\eea
Additionally, it important to mention that in large $N$  limit we can write,
$\overline{\rho}(x)=\rho(x)=V''(x),$ 
where $\rho(x)$ is the density of eigen values from {\it Wigner's semi-circle law}. Consequently, the solution obtained in Eq~(\ref{sq1a}) can be recast in the following simplified form in the large $N$ limit as:
 \bea\label{sq1b}
\hat{\og}(x)\equiv\lim _{N\rightarrow \infty}\overline\og(x)\equiv\lim _{N\rightarrow \infty}\overline\og_{-}(x)=\frac{1}{2}\left[V'(x)- \sqrt{(V'(x))^2-4V''(x)}\right].
\eea
This implies that, just by knowing the even polynomial structure of the potential $V(x)$ one can able to find out the solution for the distribution of $\overline\og(x)$ in terms of the random variable $x$. In this context, which further implies that the {Wigner's semicircle law} is defined as the probability density function of eigen values of many random matrices is a semi-circle as $N \rightarrow \infty$. On the other hand, for finite $N$, Schr{\"o}dinger equation gives the corrections comparing with calculated result obtained in Eq~(\ref{sq1b}), which is given by the following expression:
 \bea\label{sq1c}
\overline\og(x)\equiv\overline\og_{-}(x)&=\frac{1}{2\sqrt{2}}\sqrt{4\hat{\og}(x)+1}\left[1-\sqrt{\frac{16((\hat{\og}(x))^2+V^{''}(x))}{(4\hat{\og}(x)+1)}}\right]^{\frac{1}{2}}\nonumber\\&~~~~\times\left[1- \sqrt{1-\frac{4\overline{\rho}(x)}{(4\hat{\og}(x)+1)\left[1-\sqrt{\frac{16((\hat{\og}(x))^2+V^{''}(x))}{(4\hat{\og}(x)+1)}}\right]}}\right].
\eea
\begin{figure}[htb]
\centering
    \includegraphics[width=14cm,height=7cm] {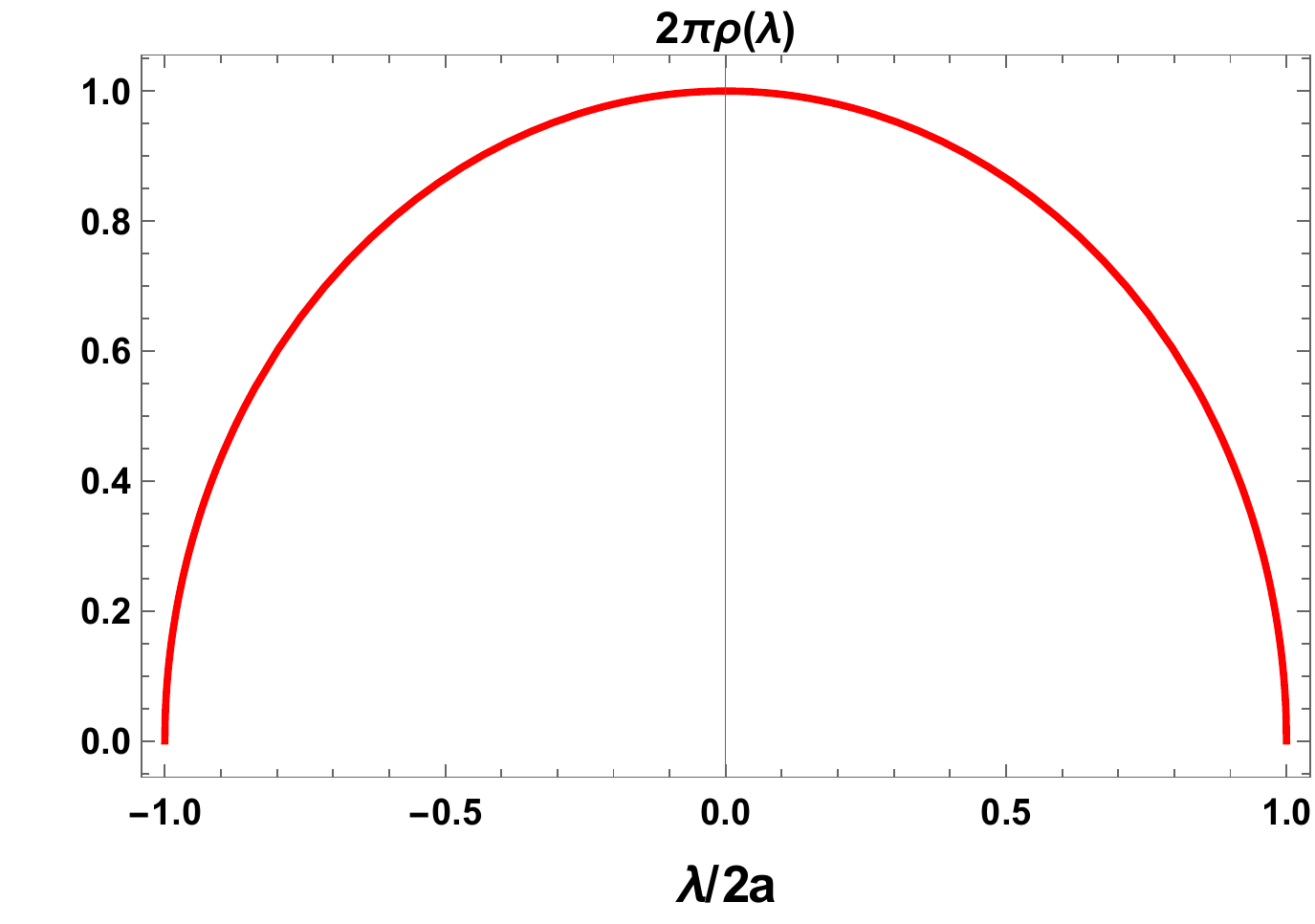}
\caption{Schematic representation of Wigner semicircle law for Gaussian random matrices. }
\label{wignerlaw}
\end{figure}
In fig~\ref{wignerlaw} density function $\rho(\lb)$ for quadratic or Gaussian potential is plotted against $\lb$ with scaling factor $\frac{1}{2a}$.The semicircle nature predicted from Eq.~(\ref{gaussian}).

Consequently, one can write:
 \bea\label{xxcq}
S[\overline\rho]=\int_{\bf R} dx~ \overline\rho(x)V(x)-\int_{\bf R^{2}} dx~ dx'~ \overline\rho(x)~\overline\rho(x')~ \log|x-x'|+L\left(1-\int_{\bf R} dx~ \overline\rho(x)\right),
\eea
 where, $L$ is the Lagrange multiplier and $1$ denotes the total density.
 
 Now, we can generalize it to normal matrix model whose eigen value belongs to $V_{i}$ (union of contours). To characterize this here we introduce filling functions, which are described by the symbol $\epsilon_{i}$ and consider the contours as:
  \be \label{plq1a}\gamma^{-n^{-1}}=\prod_{i=1}^{d} \epsilon_{i}^{n_{i}}.\ee Here
   \be \sum_{i=1}^{d}=N,\ee where $d=$dimension and $n_{i}$ eigen values are integrated over $\gamma_{i}$.
   
 Further, we define \be \epsilon_{i}=\frac{n_{i}}{N}.\ee
 Consequently, from Eq~(\ref{plq1a}) one can write:
  \bea\gamma_{i}=\sum_{i,j} C_{i,j} \gamma_{j} \iff \epsilon_{i}=\sum_{i,j} C_{ij}\epsilon_{j}.
 \eea
 \bea\Zstroke\left(\sum_{n} C_{n} \gamma^{n^{-1}}\right)=\sum_{n}C_{n}\Zstroke(\gamma^{n^{-1}}) ~~~~~{\rm where}~~ C_{n} \in {\bf C},
 \eea
 which will be helpful for further computation.
 
 Now, for a contour, which is represented by:
  \be \gamma=\sum_{i=1}^{d} C_{i}~\gamma_{i}~\epsilon~ H_{1}(e^{-V(\lb)}d \lb)\ee
  one can write:
 \be
 \frac{1}{N!}\Zstroke(\gamma^{N})=\sum_{n}\frac{\prod_{i=1}^{d} C_{i}^{n_{i}}}{\prod_{i=1}^{d} n_{i}!}\Zstroke(\gamma^{-n^{-1}}).
 \ee
 Consequently, Eq~(\ref{xxcq}) can be recast into the following simplified form:
 \bea\label{ed}
S[\overline \rho]=\int_{\gamma} dx~ \overline \rho(x)~V(x)-\int_{\gamma^{2}} dx~ dx'~ \overline\rho(x)~ \overline\rho(x)~ \log|x-x'|+\sum_{i} C_{i}~\left(\epsilon_{i}-\int_{\gamma_{i}}dx\overline\rho(x)\right).
\eea
Now the Fourier transform of the density function $\rho(x)$ can be written as:
 \be \widetilde \rho(k)=\int_{\bf R} dx~ e^{ikx}~ \overline \rho(x),\ee
 using which the second term of Eq~(\ref{ed}) can be written in Fourier space as:
\be
-\int_{\bf R \times R} dx~ dx' ~\overline\rho(x)~ \overline\rho(x')~\log|x-x'|=\int_{\bf R}\frac{dk}{|k|}~\widetilde\rho(k)~\widetilde\rho(-k)=\frac{1}{2}\int_{0}^{\infty} \frac{dk}{k}~|\widetilde\rho(k)|^{2}
\ee
Now we know  that the saddle points can be computed by imposing the following condition:
  \bea\frac{\del S}{\del \widetilde\rho(x)}=0.\eea 
During this computation one can further define the effective random potential, which is given by the following expression:
 \bea V_{\rm eff}(x)=L=V(x)-2 \int_{\bf R} dx'~\overline\rho(x')~\log|x-x'|. \eea
 Then one can recast Eq~(\ref{ed}) in terms of the effective potential as:
  \bea\label{ed1}
S[\overline \rho]=\int_{\bf R} dx~ \overline \rho(x)~V_{\rm eff}(x)+\sum_{i} C_{i}~\left(\epsilon_{i}-\int_{\gamma_{i}}dx\overline\rho(x)\right).
\eea
Further imposing the saddle point condition we get:
 \bea V'(x)=2\int_{R} \frac{dx'}{x-x'}~\overline\rho(x'),
\eea
which can be further written in terms of the eigen values of the random matrices as:
 \bea
V'(\lb_{i})=\sum_{j}\frac{1}{\lb_{i}-\lb_{j}}.
\eea
Therefore within supp of $\overline\rho\overline\rho$ one can write:
\bea \label{vvx1}\overline\og(x)=\int_{\rm supp~ \overline\rho} \frac{dx'}{x-x'}~\overline\rho(x')\\ 
\label{vvx2} V'(x)=\overline\og(x+i0)+\overline\og(x-i0).\eea
On the other hand outside the supp of $\overline\rho$ since $\overline\rho(x)\rightarrow0$, then in the large $N$ limit one can write: 
 \bea\hat{\og}(x)\equiv\lim_{N\to\infty} \overline\og(x)=\frac{1}{x}+O\left(\frac{1}{x^{2}}\right).\eea
 Therefore jump (discontinuity) on real line along the support $\overline\rho(x)$ is given by the following expression:
  \bea\label{df}
\Delta \overline\og(x)=\hat{\og}(x)- \overline\og(x)=\frac{1}{x}+O\left(\frac{1}{x^{2}}\right)-\int_{supp \overline\rho} \frac{dx'}{x-x'}\overline\rho(x'). 
 \eea
 Then using Eq~(\ref{df}), we get the following simplified expression for the jump (discontinuity)
 \be
 \boxed{\overline\og(x+i0)-\overline\og(x-i0)=\frac{1}{(x+i0)}+O\left(\frac{1}{(x+i0)^{2}}\right)-\int_{{\rm supp}~  \overline\rho} \frac{dx'}{x-i0-x'}~\overline\rho(x')=2\pi i~ \overline\rho(x-i0)}~~.
 \ee
 Now one can introduce a new function $P(x)$ of random variable $x$ as:
 \be P(x)=V'(x)\overline\og(x)-\overline\og(x)^{2}\ee which is analytic on {\bf C} as it gives zero value of the jump. This is explicitly shown in the following:
 \bea
 P(x+i0)-P(x-i0)&=&V'(x+i0)\overline\og(x+i0)-\overline\og(x+i0)^{2}-V'(x-i0)\overline\og(x-i0)+\overline\og(x-i0)^{2}\nonumber\\
 &=&V'(x)[\overline\og(x+i0)-\overline\og(x-i0)]\nonumber\\
 &&~~~~~~~~~-[\overline\og(x+i0)-\overline\og(x-i0)][\overline\og(x+i0)+\overline\og(x-i0)]\nonumber\\
 &=&0~~~~~~~~~~~~~~{\rm  on support of}~ \overline\rho
 \eea
 Additionally, it is important to note that, using the previous results we get:
\bea
\og(\lb+i0)=\frac{1}{2}V'(\lb)-i\pi \overline\rho(\lb),\\
  \og(\lb-i0)=\frac{1}{2}V'(\lb)+ii \pi \overline\rho(\lb)
\eea
 Here the most general solution for the density function is given by the following expression~\footnote{Additionally, it is important to note that the the density function satisfy the following normalization condition:
 \be \int_{{\rm supp }~\mu}d \mu~ \rho(\mu) =1\ee}: 
  \bea \rho(\lb)=\frac{1}{2\pi}M(\lb)\sqrt{-\sigma(\lb)},
 \eea
 where both $M(\lb)$ and $\sigma(\lb)$ are polynomial in $\lb$ are defined as:
  \bea M(\lb)=\sum _{k=1}^{\infty} a_{n-k} \lambda ^{2 (n-k)},~~~~\sigma(\lb)=\prod_{i=1}^{n}(\lb-a_{2i-1})(\lb-a_{2i}),
 \eea
Here we consider $n$ number of intervals on which $\rho(\lb)$ is supported and $a_{2i-1}$ and $a_{2i}$ are the end point. 
 
 Further we consider a general case where instead of the specific form of the mass profile we only know the polynomial structure of interaction random potential $V(M)$ which is characterized in terms of the random matrix $M$. For our purpose we take it to be even polynomial potential written in the following general form:
 \bea V(M)=\sum _{i=1}^{\infty} C_{2 i} M^{2 i}=C_{2} M^{2}+C_{4} M^{4}+C_6 M^6+.....
 \eea
 Here after diagonalizing the random matrix $M$ we get its eigen values $\lb_{1},\lb_{2},.............\lb_{N}$, from which we can compute the distribution of this eigen values for large $N$ limit and it turns out to be w be the density function $\rho(\lb)$, which is already introduced earlier. 
 
 Now let us consider that the degree of the polynomial $P$, $\sigma$ and $M$  are:
  \be {\rm deg}(P)=2k,~~~ {\rm deg}(\sigma)=2n,~~~ {\rm deg}(M)=2k-n-1.\ee 
Now considering $n=1$ and $n=2$ we get:
  \bea {\rm For~ n=1:}~~~{\rm deg}(P)=2k,~~~ {\rm deg}(\sigma)=2,~~~ {\rm deg}(M)=2k-2,\\
  {\rm For~ n=2:}~~~{\rm deg}(P)=2k,~~~ {\rm deg}(\sigma)=4,~~~ {\rm deg}(M)=2k-3.\eea
For $n=1$ we also  get the following simplified expressions for the polynomial $M(\lambda)$ and $\sigma(\lb)$:
\bea M(\lb)&=&\sum _{k=1}^{\infty} a_{1-k} \lambda ^{2 (1-k)},\\
\sigma (\lambda )&=&\lambda ^2-4 a^2.\eea
Consequently, for $n=1$ we get the following expression for the density function on semi-circle:
 \bea\rho(\lb)=\frac{1}{\pi}\sqrt{4 a^2-\lambda ^2} ~\sum _{k=1}^{\infty} a_{1-k} \lambda ^{2 (1-k)}.
 \eea
 Now we use this $\rho(\lb)$ in $\og(\lb+i0)$ and Taylor expand in the limit $\lb\rightarrow\infty$ we get:
  \bea\og(\lb\rightarrow\infty)=\frac{1}{\lb}+O\left(\lb^{2}\right),
 \eea 
 which implies that all coefficients of $\lb^{r}$ for $r>0$ is zero and this gives $n$ number of equations. This finally gives the full equation of $M(\lb)$ in terms of the coefficients $C_{2i}$. Solving these equations we get:
 \bea\label{cc1}
\frac{1}{2}\left(-2\lb+\frac{4a^{2}}{\lb}+\frac{4a^{4}}{\lb^{2}}+\frac{8a^{6}}{\lb^{5}}+O\left(\frac{1}{\lb}\right)^{6}\right) \sum _{k=1}^\infty a_{n-k} \lambda ^{2 (n-k)}+\sum _{i=1}^\infty 2 i~ C_{2 i} ~\lambda ^{2 i-1}=\frac{1}{\lb}.
\eea
Further equating the coefficients on both sides of the Eq~(\ref{cc1})  we get:
 \bea
&&2nC_{2n}-2a_{n-1}=0,\\
&&4a^{2}a_{n-1}-2a_{n-2}+2(n-1)C_{2n-2}=0\\
&&4a^{4}a_{n-1}+4a^{2}a_{n-2}-2a_{n-3}+2(n-2)C_{2n-4}=0,
 \eea
 and it will continue upto term by term giving all $a_{n}$ and we get the unique polynomial $M(\lb)$. We will verify this generalization for $n=1,2,3,4,5$ and check their SFF in this work accordingly.
For more general discussions see ref.~ \cite{BHANOT1990388,Mandal1} also.

\subsection{OTOC in Random Matrix Theory (RMT)}
In earlier section we have introduced OTOC and its application to cosmology. In this subsection, we will discuss about OTOC appearing in the context of RMT. 
\subsubsection{Two point OTOC}
For this purpose, we start with two point correlation functions for the GUE which is described by the following equation:
 \bea \langle {\cal O}_{1}(0){\cal O}_{2}(\tau)\rangle_{\rm GUE}\equiv 
\int dH ~\langle {\cal O}_{1}(0){\cal O}_{2}(\tau)\rangle,\eea
where the operator ${\cal O}_{2}(\tau)$ in Heisenberg picture can be expressed as:
\be {\cal O}_{2}(\tau)=\exp[-iH\tau]{\cal O}_{2}(0)\exp[iH\tau].\ee
Here it is important to note that the GUE measure $ dH$ is represented by the Hamiltonian $H$., which is invariant under the following unitary conjugation operation, which is described by:
\be dH=d(UHU^{\dagger})~~~~~\forall ~U.\ee
Here $U$ is the unitary matrix. Consequently, the GUE two point correlation function can be further expressed as:
\be \langle {\cal O}_{1}(0){\cal O}_{2}(\tau)\rangle_{\rm GUE}=\int \int dH~dU~\langle {\cal O}_{1}U\exp[-iH\tau]U^{\dagger}{\cal O}_{2}U\exp[iH\tau]U^{\dagger}\rangle,\ee
where $dU$ is the Haar measure appearing in this context. After integrating over the Haar measure we get the following expression for the  GUE two point correlation function:
 \bea \langle {\cal O}_{1}(0){\cal O}_{2}(\tau)\rangle_{\rm GUE}=\langle {\cal O}_{1}\rangle \langle {\cal O}_{2}\rangle+\frac{{\bf SFF}(\tau)-1}{{\cal I}^2-1}\langle \langle {\cal O}_{1}{\cal O}_{2}\rangle \rangle_C, \eea
where the connected two point correlation function $\langle \langle {\cal O}_{1}{\cal O}_{2}\rangle \rangle$ is defined as:
\be \langle \langle {\cal O}_{1}{\cal O}_{2}\rangle \rangle_C=\langle {\cal O}_{1}{\cal O}_{2}\rangle-\langle {\cal O}_{1}\rangle\langle {\cal O}_{2}\rangle.\ee
Now we consider a special case where ${\cal O}_{1}$ and ${\cal O}_{2}$ are described Pauli operators. In such a situation, the GUE two point correlation function can be expressed as:
 \bea\begin{array}{lll}\label{qaaq1}
		\displaystyle   \langle {\cal O}_{1}(0){\cal O}_{2}(\tau)\rangle_{\rm GUE}=\left\{\begin{array}{lll}
			\displaystyle  
			\frac{{\bf SFF}(\tau)-1}{{\cal I}^2-1}\,,~~~~~~~~~~~~ &
			\mbox{\small  \textcolor{red}{\bf  {${\cal O}_1={\cal O}_2$}}}  \\ 
			\displaystyle  
			0\,,~~~~~~~~~~~~ &
			\mbox{\small  \textcolor{red}{\bf  {${\cal O}_1\neq {\cal O}_2$}}}
		\end{array}
		\right.
	\end{array},\eea
where ${\bf SFF}(\tau)$ is the two point {\it Spectral Form Factor} (SFF) which we have defined explicitly earlier. Further, one can consider the situation where ${\bf SFF}(\tau)>>1$ and ${\cal O}_2(\tau)={\cal O}^{\dagger}_1(\tau)$. For this case the GUE two point correlation function is simplified to the following expression:
\be \langle {\cal O}_{1}(0){\cal O}_{2}(\tau)\rangle_{\rm GUE}\sim \frac{{\bf SFF}(\tau)}{{\cal I}^2}.\ee
Here ${\cal I}$ represents the $2^n$ dimensional Hilbert space in the present computation. To derive this above mentioned expression we have not assumed any additional assumption expect the fact that the Haar measure of GUE $dH$ is invariant. This is a very useful information to study the physical characteristics of chaotic Hamiltonian at macroscopic scales.
\subsubsection{Four point OTOC}
Now we discuss about the four point OTOC for the GUE prescription. Here the fourth point OTOC can be expressed in terms of fourth Haar moment:
\bea \langle {\cal O}_1(0){\cal O}_2(\tau){\cal O}_3(0){\cal O}_4(\tau)\rangle_{\rm GUE}&&=\int \int dH~dU~\langle {\cal O}_1U\exp[-iH\tau]U^{\dagger}{\cal O}_2U\nonumber\\
&&~~~~~~~~~~~~~\exp[iH\tau]U^{\dagger}{\cal O}_3U\exp[-iH\tau]U^{\dagger}{\cal O}_4U\exp[iH\tau]U^{\dagger}\rangle,\nonumber\\
&&\eea
where we consider $(4!)^2=576$ terms in this expression for four point OTOC. Now we consider a special situation, where all these operators appearing in the expression for the four point OTOC for GUE are described by Pauli operators. In such a case, the four point OTOC for GUE can be simplified as:
 \bea \langle {\cal O}_1(0){\cal O}_2(\tau){\cal O}_3(0){\cal O}_4(\tau)\rangle_{\rm GUE}&\simeq \langle {\cal O}_1{\cal O}_2{\cal O}_3{\cal O}_4\rangle \times \frac{{\bf SFF}_4(\tau)}{{\cal I}^4},\eea
where ${\bf SFF}_4(\tau)$ is the four point SFF for GUE, which is defined by the following expression:
\bea {\bf SFF}_4(\tau)&\equiv& \langle Z(\tau)Z(\tau)Z^{*}(\tau)Z^{*}(\tau)\rangle_{\rm GUE}\nonumber\\
&=&\int D\lambda \sum_{i,j,k,l}\exp[i(\lambda_i+\lambda_j+\lambda_k+\lambda_m)\tau]\nonumber\\
&=&{\cal I}^4\frac{J^4_1(2\tau)}{\tau^4}+\frac{\tau}{2}(\tau-2)\nonumber\\
&\sim & \frac{{\cal I}^6}{\pi^2\tau^6}+\frac{\tau}{2}(\tau-2),~~~~~\eea
and this is derived only by considering the leading order behaviour of four point SFF. Here additionally it is important to note that if we fix:
\be \langle {\cal O}_1{\cal O}_2{\cal O}_3{\cal O}_4\rangle ={\bf I}.\ee
This will give rise to non-zero expression for the four point OTOC for GUE. For other situations, where 
\be \langle {\cal O}_1{\cal O}_2{\cal O}_3{\cal O}_4\rangle =0,\ee
we get zero contribution to the four point OTOC for GUE. 

One can further generalise this statement for any arbitrary $2p$ point OTOC for GUE, which is given by the following expression:
 \bea \langle {\cal O}_1(0){\cal Q}_1(\tau)\cdots{\cal O}_p(0){\cal Q}_p(\tau)\rangle_{\rm GUE}\simeq \langle {\cal O}_1{\cal Q}_1\cdots{\cal O}_p{\cal Q}_p\rangle\times \frac{{\bf SFF}_{2p}(\tau)}{{\cal I}^{2p}}.\eea
Generalizing the previous argument one can conclude that the final result for the $2p$ point OTOC for GUE is non zero when we have the following constraint:
\bea \langle {\cal O}_1{\cal Q}_1\cdots{\cal O}_p{\cal Q}_p\rangle= {\bf I}.\eea
Here one can further show that for the GUE we get:
\be \langle {\cal O}_1(0){\cal O}_2(\tau){\cal O}_3(2\tau){\cal O}_4(\tau)\rangle_{\rm GUE}\simeq \langle {\cal O}_1(0){\cal O}_2(\tau){\cal O}_3(0){\cal O}_4(\tau)\rangle_{\rm GUE}\simeq \langle {\cal O}_1{\cal O}_2{\cal O}_3{\cal O}_4\rangle \times \frac{{\bf SFF}_4(\tau)}{{\cal I}^4},\ee
which indirectly implies that GUE is not sensitive to the fact that the operators as appearing in this context are out-of-time ordered or something else. Additionally, it is important to note that, if we compute the expression for the OTOC correlation function for a specified class of Hamiltonian operators, which are in general invariant under the operation of conjugation on the unitary matrix $U$. In such a situation from OTOC one can further express the OTOC in terms of SFF. This is a very well known technique in the study of many -body QFT systems, where particularly to study the underlying physics of thermalization and quantum quench \cite{}. In the next subsection we will provide an  analytical proof of the equivalence of the two point SFF and the two point OTOC, which can be further generalized to any arbitrary  $2p$ point correlation functions.

\subsection{Spectral Form Factor (SFF) from OTOC}
From the traditional perspective the idea of quantum chaos is used in the context of study of spectral aspects of statistical field theory. Recent developments are made in the context of black hole theory and quantum information theory where using OTOC one can quantify quantum chaos. However in this paper our one of the prime objective to apply the concept of quantum chaos to study early universe cosmology, which is obviously another new direction of future research area. In this subsection, our air is to give a formal proof which establish the connection between Spectral Form Factor (SFF) and OTOC in OEQFT. First of all we consider a limit where $\beta=1/T=0$ in which distribution of quantum operator insertions around a thermal circular path is very straightforward.

Let us consider a quantum mechanical Hamiltonian operator $H$ operating on an ${\cal I}=2^n$ dimensional Hilbert space and consists of $n$ number of quantum bits (qbits). Next, we consider the two point correlation function $\langle {\cal O}(0){\cal O}^{\dagger}(\tau)\rangle$ using which one define the following averaged two point correlation function:
 \bea \label{xcg2} \int d{\cal O}~\langle {\cal O}(0){\cal O}^{\dagger}(\tau)\rangle:&\equiv& \frac{1}{{\cal I}}\int d{\cal O}~{\rm Tr}\left({\cal O}\exp[-iH\tau]{\cal O}^{\dagger}\exp[iH\tau]\right)\nonumber\\
&=&\frac{1}{{\cal I}^3}\sum^{{\cal I}^2}_{k=1}{\rm Tr}\left({\cal O}_k\exp[-H\tau]{\cal O}^{\dagger}_{k}\exp[iH\tau]\right).\eea
Here we assume that ${\cal O}$ is the Unitary operator which is integrated over a Haar measure on ${\cal U}(2^n)$. Also it is important to note that the integral over the Haar measure can be translated in terms of the Pauli operators ${\cal O}_k$ and ${\cal I}^2=2^{2n}=4^n$ represents the total number of Pauli operators for this quantum $n$ qubit system.

Further, it is important to note that, to derive the expression for SFF from the present context additionally we need the first moment of the Haar ensemble, which is defined as:
\bea \int d{\cal O}~{\cal O}{\cal D}{\cal O}^{\dagger}=\frac{1}{{\cal I}}{\rm Tr}({\cal D})~{\bf I},\eea
which can be be equivalently expressed in terms of the language of Pauli operator as:
\bea \label{xcg1} \int d{\cal O}~{\cal O}^{k}_{m}{\cal O}^{l}_{n}=\frac{1}{{\cal I}}\delta^{k}_{n}\delta^{l}_{m}.\eea
Next using Eq~(\ref{xcg1}) in Eq~(\ref{xcg2}), we get the following simplified result:
 \bea \label{xcg3} \textcolor{red}{\bf Quantum~averaged~OTOC}&=&\int d{\cal O}~\langle {\cal O}(0){\cal O}^{\dagger}(\tau)\rangle\nonumber
\\&=&\frac{1}{{\cal I}^2}|{\rm Tr}(\exp[-iH\tau])|^2\nonumber\\
&=&\frac{1}{{\cal I}^2}{\bf SFF}(\tau)\propto \textcolor{red}{\bf Two~point~ SFF}.,\eea
where the two point SFF at infinite temperature is defined in terms of the quantum Hamiltonian $H$ as:
\be {\bf SFF}=|\exp[-iH\tau]|^2.\ee
Here the result obtained in Eq~(\ref{xcg3}) implies that the quantum averaged OTOC is proportional to the two point SFF at infinite temperature
of the present context.

This prescription can be further generalised to make the connection between any arbitrary $2p$ point quantum OTOC and $2p$ point SFF in this context. To establish this connection let us consider a $2p$ point quantum OTOC, which is described by:
\be \langle {\cal O}_1(0){\cal Q}_1(\tau)\cdots{\cal O}_p(0){\cal Q}_p(\tau\rangle~~~~~~~~{\rm with}~~~{\cal O}_1{\cal Q}_1\cdots{\cal O}_p{\cal Q}_p={\bf I}.\ee
Now taking the average over such $2p$ point OTOC we get:
\bea \int d{\cal O}_1d{\cal Q}_1\cdots d{\cal O}_{k-1}d{\cal Q}_k~ \langle {\cal O}_1(0){\cal Q}_1(\tau)\cdots{\cal O}_p(0){\cal Q}_p(\tau)\rangle &=&\frac{1}{{\cal I}^{2p}}|{\rm Tr}(\exp[-iH\tau])|^{2p}\nonumber\\
&=&\frac{1}{{\cal I}^{2p}}{\bf SFF}_{2p}(\tau),\eea
where ${\cal Q}_{p}$ is defined as:
\be {\cal Q}_p={\cal O}^{\dagger}_p\cdots{\cal Q}^{\dagger}_1{\cal O}^{\dagger}_1. \ee
Here one can consider a special case where \be {\cal Q}_p={\cal O}^{\dagger}_p~~~\forall ~p.\ee
Consequently, the average over such $2p$ point OTOC can be further simplified to the following form:
\be \int d{\cal O}_1d{\cal O}_2\cdots d{\cal O}_{k-1}d{\cal O}_k~ \langle {\cal O}_1(0){\cal O}^{\dagger}_1(\tau)\cdots{\cal O}_p(0){\cal O}^{\dagger}_p(\tau)\rangle =\frac{1}{{\cal I}^{p+1}}\underbrace{{\rm Tr}(\exp[-iH\tau])^p{\rm Tr}(\exp[ipH\tau])}.\ee
Here it is important to note that the terms appearing in the $\underbrace{}$ are not symmetric because the operator ${\cal O}_1(0){\cal O}^{\dagger}_1(\tau)\cdots{\cal O}_p(0){\cal O}^{\dagger}_p(\tau)$ is an non-Hermitian quantum operator. This result establishes a direct connection between the spectral physics in statistical field theory and other physical observables. Apart from theoretical perspective one can use two point {\bf SFF} as a good estimator for experimental measure. For this purpose one can consider the following standard deviation (or experimental estimation error) of the unitary operator ${\cal O}$ given by:
 \bea \sigma_{{\cal O}}=\sqrt{{\rm Var}({\cal O})}=\sqrt{\int d{\cal O}|{\cal O}(0){\cal O}^{\dagger}(\tau)|^2-\left|\int d{\cal O}{\cal O}(0){\cal O}^{\dagger}(\tau)\right|^2}={\cal O}\left(\frac{1}{{\cal I}}\right).\eea
By choosing the Haar unitary operator ${\cal O}$ as a random Clifford operator one can find a good estimator of two point {\bf SFF}.

To give the similar proof at finite temperature let us consider the energy eigenvalue representation of OTOC, which is given by the following expression:
\be {\cal C}(\tau)=\frac{1}{|Z(\beta)|^2}\sum_{n,m}c_{n,m}(\tau)\exp[-\beta (E_n+E_m)],\ee
where the time dependent expansion coefficient can be expressed as:
\be c_{n,m}(\tau)=-\langle n|[e^{-iH\tau},x]^2|m\rangle=\exp\left[-i(E_n-E_m)\tau\right].\ee
Here we have used the fact that, $H|n\rangle=E_n|n\rangle$. Consequently we get:
 \bea \textcolor{red}{\bf Quantum ~OTOC}~~~~~{\cal C}(\tau)&=\frac{1}{|Z(\beta)|^2}\sum_{n,m}\exp[-\beta (E_n+E_m)]\exp\left[-i(E_n-E_m)\tau\right]\nonumber\\
&=\frac{|Z(\beta+i\tau)|^2}{|Z(\beta)|^2}=\textcolor{red}{\bf Two~point~SFF}.\eea
This establishes the connection between OTOC and two point SFF at finite temperature

 \subsection{Two point SFF and thermal Green's function in RMT}
 In this subsection our prime objective is to explicitly compute the
  expression for SFF for different even polynomial potential of random matrices. This is very useful to quantify chaos when we have no information about the interaction or time dependent effective mass profile which will finally give rise to scattering in conduction wire in presence of impurity or cosmological particle creation during reheating.
  
  Let us now consider a Thermofield Double State (TDS) associated with canonical quantum mechanical state at finite temperature. The time evolution of the TDS can be expressed as:
  \be |\Psi(\beta,\tau)\rangle_{\bf TDS}=\frac{1}{\sqrt{Z(\beta)}}\sum_{n}\exp\left[-\frac{\beta}{2}H\right]\exp[iH\tau]. \ee 
  Using this information one can define Spectral Form Factor (SFF) as:
 \bea {\bf SFF}=|{}_{\bf TDS}\langle \Psi(\beta,0|\Psi(\beta,\tau)\rangle_{\bf TDS}|^2=\frac{1}{|Z(\beta)|^2}\sum_{m,n} e^{-\bg (E_{m}+E_{n})} e^{-i\tau(E_{m}-E_{n})}=\frac{|Z(\bg + i \tau)|^{2}}{|Z(\beta)|^2}.
 \eea
 Here $E_n$ and $E_m$ correspond to the $n$ -th and $m$ -th level of the quantum system under consideration. Here the Boltzmann factor $\beta=1/T$, where $T$ is the temperature associated to the system. Apart from temperature dependent Boltzmann factor the definition of SFF also involves involves conformal time $\tau$, which we have define in earlier section of this paper and during reheating $\tau\propto t$. Here $t$ is the physical time scale and the proportionality factor is constant in space time.
 
 Now at very high temperature ($\beta=1/T\rightarrow 0$) and low temperature ($\beta=1/T\rightarrow \infty$) we get the following limiting behaviour of SFF, as given by:
  \bea\begin{array}{lll}\label{eqxxx}
		\displaystyle   {\bf SFF}=\left\{\begin{array}{lll}
			\displaystyle  
			\sum_{m,n} e^{-i\tau(E_{m}-E_{n})}\,,~~~~~~~~~~~~ &
			\mbox{\small  \textcolor{red}{\bf  {$\beta=1/T\rightarrow 0$ }}}  \\ 
			\displaystyle  
			0\,,~~~~~~~~~~~~ &
			\mbox{\small  \textcolor{red}{\bf  {$\beta=1/T\rightarrow \infty$}}}  
		\end{array}
		\right.
	\end{array},\eea
 It is also observed that in $\tau\rightarrow\infty$ limiting situation the nearest neighbour energy spacings contribute only to the quantification of SFF. This implies that the concept of SFF also helps in understanding the time dynamics of the quantum system under consideration and also very useful tool to analyze the discreteness in energy spectrum. Chaotic system satisfy {\it Wigner's formula} which makes SFF  a good observable for quantifying chaos.
 
 In usual prescriptions, SFF is averaged over an statistical ensemble of random matrix.  This is a very particular feature of SFF and can be directly linked to the quantification of quantum chaos. Before going to discuss further here it is important to note that, all distribution for eigenvalues are different from each other but quite similar at small scales. This is a very crucial information for the computation of SFF to quantify chaos.
 
 Now in the present context we define a new function $G(\bg,\tau)$, which is represented by the following expression:
 \bea G(\bg,\tau)=\frac{\langle|Z(\bg+i\tau)|^{2}\rangle_{\rm GUE}}{\langle Z(\bg)\rangle^{2}_{\rm GUE}}=\frac{\int_{{\rm supp}~ \overline\rho} d\lb~ d\mu~ e^{-\bg(\lb+\mu)}~e^{-i\tau(\lb-\mu)}\langle D(\lb)D(\mu)\rangle_{\rm GUE}}{\int_{{\rm supp }~ \overline\rho} d\lb~ d\mu~ e^{-\bg(\lb+\mu)}\langle D(\lb)\rangle \langle D(\mu)\rangle_{\rm GUE}}.
 \eea
 Here, $D(\lb)=\overline\rho(\lb)=$eigen value density. In the present context, $G(\bg,\tau)$ characterize the two point correlation function which measures SFF.
 
 Now, one can divide the total Green's function $G$ in two parts (connected and disconnected part of the Green's function), as given by:
 \be G(\beta,\tau)=G_{dc}(\beta,\tau)+G_{c}(\beta,\tau),\ee
  where disconnected part of the Green's function $G_{dc}$ and connected part of the Green's function $G_{c}$  can be expressed as:
 \be
G_{dc}(\beta,\tau)=\left[\frac{\langle Z(\bg+i\tau)\rangle \langle Z(\bg-i\tau)\rangle}{\langle Z(\bg)\rangle^{2}}\right]=\frac{\int d\lb~ d\mu~ e^{-\beta(\lb+\mu)}~e^{-i\tau(\lb-\mu)}~\langle D(\lb)\rangle \langle D(\mu)\rangle}{\int d\lb~ d\mu ~ e^{-\beta(\lb+\mu)}~\langle D(\lb)\rangle \langle D(\mu)\rangle}~~.
 \ee
 \bea G_{c}(\beta,\tau)=G(\beta,\tau)-G_{dc}(\beta,\tau)&=&\left[\frac{\langle|Z(\bg+i\tau)|^{2}\rangle_{\rm GUE}}{\langle Z(\bg)\rangle^{2}_{\rm GUE}}\right]-\left[\frac{\langle Z(\bg+i\tau)\rangle \langle Z(\bg-i\tau)\rangle}{\langle Z(\bg)\rangle^{2}}\right]\nonumber\\
 &=&\frac{\int d\lb~ d\mu~ e^{-\beta(\lb+\mu)}~ e^{-i\tau(\lb-\mu)}~\langle D(\lb)D(\mu)\rangle_{\bf c}}{\int d\lb~ d\mu~ e^{-\beta(\lb+\mu)}~ \langle D(\lb)\rangle \langle D(\mu)\rangle}~~.
\eea
 
 Now, for further analysis we consider the high temperature limit ($\beta=1/T\rightarrow 0$) and also can divide the total Green's function $G$ in two parts (connected and disconnected part of the Green's function), as given by:
 \be G(\beta\rightarrow 0,\tau)=G(\tau)=G_{dc}(\tau)+G_{c}(\tau),\ee
  where disconnected part of the Green's function $G_{dc}$ and connected part of the Green's function $G_{c}$  can be expressed as:
 \be
G_{dc}(\tau)=\left[\frac{\langle Z(\bg+i\tau)\rangle \langle Z(\bg-i\tau)\rangle}{\langle Z(\bg)\rangle^{2}}\right]_{\bg=0}=\frac{\int d\lb~ d\mu~ e^{-i\tau(\lb-\mu)}~\langle D(\lb)\rangle \langle D(\mu)\rangle}{\int d\lb~ d\mu ~\langle D(\lb)\rangle \langle D(\mu)\rangle}~~.
 \ee
 \bea G_{c}(\tau)=G(\tau)-G_{dc}(\tau)&=&\left[\frac{\langle|Z(\bg+i\tau)|^{2}\rangle_{\rm GUE}}{\langle Z(\bg)\rangle^{2}_{\rm GUE}}\right]_{\beta=0}-\left[\frac{\langle Z(\bg+i\tau)\rangle \langle Z(\bg-i\tau)\rangle}{\langle Z(\bg)\rangle^{2}}\right]_{\bg=0}\nonumber\\
 &=&\frac{\int d\lb~ d\mu~ e^{-i\tau(\lb-\mu)}~\langle D(\lb)D(\mu)\rangle_{\bf c}}{\int d\lb~ d\mu~ \langle D(\lb)\rangle \langle D(\mu)\rangle}~~.
\eea
Here we define the connected two-point correlation function, which is given by the following expression:
  \bea\langle D(\lb)D(\mu)\rangle_{\bf c}\equiv (\langle D(\lb)D(\mu)\rangle-\langle D(\lb)\rangle \langle D(\mu)\rangle).\eea 
 To quantify this explicitly one can define the eigen value density function $D(\lb)$ in the neighbourhood of extremum of level density ($\overline\rho(\lb)$) as:
 \be D(\lb)=\overline D (\lb)+\del D(\lb),\ee 
 where $\overline D (\lb)$ is the average of the eigen value density function over the statistical ensemble of eigen values of the random matrices and $\del D(\lb)$ represents the quantum fluctuation on $\overline D (\lb)$. 
 
Consequently,  using this fact the two point correlation function reduced to the following form:
 \bea
\langle D(\lb)D(\mu)\rangle_{\bf c}= \langle \del D(\lb) \del D(\mu)\rangle
\eea
and using this connected part of the Green's function $G_c$ can be further simplified as:
 \bea\label{ccon}
G_{c}(\tau)=G(\tau)-G_{dc}(\tau)=\frac{\int d\lb~ d\mu~ e^{-i\tau(\lb-\mu)}~\langle \del D(\lb) \del D(\mu)\rangle}{\int d\lb~ d\mu~ \langle D(\lb)\rangle \langle D(\mu)\rangle}.
\eea
 Additionally, it is important to note that, the mean level density can be normalised in a semi circle using the following two conditions:
 \bea \int_{-2a}^{2a} d\lb~D(\lb) &=&N,\\
\int_{-2a}^{2a} d\lb~\rho(\lb) &=&1.\eea
Here $D(\lb)$ actually represents the number of eigen values lying between the small interval ($\lb,\lb+d\lb$) and in the present context it is proportional to $~O(\sqrt{N})$. On the other hand, $\rho(\lb)$ is the density which we get by extremising the action and treated to be free from all factor of $N$ and all eigen values which are just $O(1)$. In this context, the two variables $\lb$ and $\sigma$ are related by the follwing expression:
\be \lb=\sqrt{N}\sigma.\ee

To compute the $G_{dc}$ and $G_{c}$ part of SFF explicitly let us first start with the one point function on the semi-circle as given by:
 \bea\label{xnmx1}
\langle Z(\bg\pm i\tau)\rangle_{\rm nGUE}=\int d \lb~ e^{\mp i\tau\lb}~e^{-\beta\lambda}~\langle \rho(\lb)\rangle_{\rm nGUE}=\int_{-2a}^{2a} d \lb~ e^{\mp i\tau\lb}~e^{-\beta\lambda}~ \rho(\lb).
\eea
At high temperature ($\beta=1/T\rightarrow 0$) this result can be simplified as::
 \bea\label{xnmx}
\left[\langle Z(\bg\pm i\tau)\rangle_{\rm nGUE}\right]_{\bg=0}=\int d \lb~ e^{\mp i\tau\lb}~\langle \rho(\lb)\rangle_{\rm nGUE}=\int_{-2a}^{2a} d \lb~ e^{\mp i\tau\lb}~ \rho(\lb).
\eea
On the other hand at very low temperature limit ($\beta=1/T\rightarrow \infty$) we get:
simplified as::
 \bea\label{xnmx2}
\left[\langle Z(\bg+i\tau)\rangle_{\rm nGUE}\right]_{\bg\rightarrow \infty}\rightarrow 0.
\eea
Here it is important to note that, for different polynomial random potential we will get different expressions for the integral measure. Now we need to find the specific point after which properties of SFF drastically changes. We define this points as {\it critical points}. For general even order polynomial potential one can write down the following expression for the density function of the eigenvalues of the random matrices:
 \bea\label{ssq2}
\rho(\lb)=\frac{1}{\pi}\sqrt{4 a^2-\lambda ^2}~ \sum _{k=1}^n a_{n-k} \lambda ^{2 (n-k)}~~~~~~\forall~ {\rm even~n}
\eea
Further substituting Eq~(\ref{ssq2}) in Eq~(\ref{xnmx1}) we get the following simplified expression for the one point function on the semi-circle:
\bea \langle Z(\bg\pm i\tau)\rangle_{\rm nGUE}&=&\frac{1}{\pi}\int_{-2a}^{2a} d \lb~ e^{\mp i\tau\lb}~e^{-\beta\lambda}~ \sqrt{4 a^2-\lambda ^2}~ \sum _{k=1}^n a_{n-k} \lambda ^{2 (n-k)}~~~~~~\forall~ {\rm even~n}\nonumber\\ &=& \sum^{n}_{k=1}a_{n-k}~ a^2 \left(-a^2\right)^{-2 k} 4^{n-k} \left[\left(e^{2 i \pi  k}+e^{2 i \pi  n}\right) a^{2 (k+n)} \Gamma \left(-k+n+\frac{1}{2}\right)\right. \nonumber\\ && \left.~~~~~~~~~~~~~~\times  \, _1\tilde{F}_2\left(-k+n+\frac{1}{2};\frac{1}{2},-k+n+2;a^2 (\beta\pm i \tau)^2\right)\right. \nonumber\\ && \left.~~~~~~~~~~~~~~+a (\beta\pm i \tau) \left(a^{2 k} (-a)^{2 n}-(-a)^{2 k} a^{2 n}\right) \Gamma (-k+n+1)\right. \nonumber\\ && \left.~~~~~~~~~~~~~~\times   \, _1\tilde{F}_2\left(-k+n+1;\frac{3}{2},-k+n+\frac{5}{2};a^2 (\beta\pm i \tau)^2\right)\right]~~~~~~\forall~ {\rm even~n}.\nonumber\\ &&\eea
where $\, _1\tilde{F}_2\left(A;B,C;D\right)$ is the regularized Hypergeometric function.

Repeating the same calculation in high temperature ($\beta=1/T\rightarrow 0$) limit we get:
\bea
\left[\langle Z(\bg\pm i\tau)\rangle_{\rm nGUE}\right]_{\bg=0}&=&\frac{1}{\pi}\int_{-2a}^{2a} d \lb~ e^{\mp i\tau\lb}~ \sqrt{4 a^2-\lambda ^2}~ \sum _{k=1}^n a_{n-k} \lambda ^{2 (n-k)}~~~~~~\forall~ {\rm even~n}\nonumber\\
&=&\sum^{n}_{k=1}a_{n-k}~\frac{ e^{-2 i \pi  k} 4^{n-k} a^{-2 k+2 n+2}}{\sqrt{\pi }\Gamma (-k+n+2) \Gamma \left(-k+n+\frac{5}{2}\right)} \nonumber\\
&&~~\times\left[\left\{(-1)^{2 k}+(-1)^{2 n}\right\} \Gamma \left(-k+n+\frac{1}{2}\right) \Gamma \left(-k+n+\frac{5}{2}\right)\right. \nonumber\\ && \left.~~~~~~~~~~~~~~\times \, _1F_2\left(-k+n+\frac{1}{2};\frac{1}{2},-k+n+2;-a^2 \tau^2\right)\right. \nonumber\\ && \left.~~~~~~~~\mp 2 i a \tau \left\{(-1)^{2 k}+(-1)^{2 n+1}\right\} \Gamma (-k+n+1) \Gamma (-k+n+2)\right. \nonumber\\ && \left.~~~~~~~~~~~~~~\times  \, _1F_2\left(-k+n+1;\frac{3}{2},-k+n+\frac{5}{2};-a^2 \tau^2\right)\right]~~~\forall~ {\rm even~n}.~~~~~~~~~~~
\eea
For different polynomial potentials we can actually calculate the expansion coefficients $a_{n-k}$ and get the exact form of $Z(\bg\pm i\tau)$.

At finite temperature the disconnected part of the Green's function ($G_{dc}(\beta,\tau)$) can be expressed as:
\bea G_{dc}(\beta,\tau)&=&\frac{\langle Z(\bg+i\tau)\rangle \langle Z(\bg-i\tau)\rangle}{\langle Z(\bg)\rangle^{2}}=\left\{ \sum^{n}_{q=1}a_{n-q}~\left(-a^2\right)^{-2 q} 4^{-q} \left[\left(e^{2 i \pi  q}+e^{2 i \pi  n}\right) a^{2 (q+n)} \Gamma \left(-q+n+\frac{1}{2}\right)\right. \right.\nonumber\\ && \left.\left.\times  \, _1\tilde{F}_2\left(-q+n+\frac{1}{2};\frac{1}{2},-q+n+2;a^2 \beta^2\right)+a \beta \left(a^{2 q} (-a)^{2 n}-(-a)^{2 q} a^{2 n}\right) \Gamma (-q+n+1)\right.\right. \nonumber\\ && \left.\left.~~~~~~~~~~~~~~\times   \, _1\tilde{F}_2\left(-q+n+1;\frac{3}{2},-q+n+\frac{5}{2};a^2 \beta^2\right)\right]\right\}^{-2}\nonumber\\
&& \times \left\{ \sum^{n}_{k=1}a_{n-k}~ \left(-a^2\right)^{-2 k} 4^{-k} \left[\left(e^{2 i \pi  k}+e^{2 i \pi  n}\right) a^{2 (k+n)} \Gamma \left(-k+n+\frac{1}{2}\right)\right. \right.\nonumber\\ && \left.\left.  \, _1\tilde{F}_2\left(-k+n+\frac{1}{2};\frac{1}{2},-k+n+2;a^2 (\beta+ i \tau)^2\right)\right.\right. \nonumber\\ && \left.\left.+a (\beta+ i \tau) \left(a^{2 k} (-a)^{2 n}-(-a)^{2 k} a^{2 n}\right) \Gamma (-k+n+1) \right. \right.\nonumber\\ && \left.\left.  \, _1\tilde{F}_2\left(-k+n+1;\frac{3}{2},-k+n+\frac{5}{2};a^2 (\beta+ i \tau)^2\right)\right]\right\}\nonumber\\
&&\times\left\{ \sum^{n}_{m=1}a_{n-m}~  \left(-a^2\right)^{-2 m} 4^{-m} \left[\left(e^{2 i \pi  m}+e^{2 i \pi  n}\right) a^{2 (m+n)} \Gamma \left(-m+n+\frac{1}{2}\right)\right. \right.\nonumber\\ && \left.\left.~~~~~~~~~~~~~~\times  \, _1\tilde{F}_2\left(-m+n+\frac{1}{2};\frac{1}{2},-m+n+2;a^2 (\beta- i \tau)^2\right)\right.\right. \nonumber\\ && \left.\left.~~~~~~~~~~~~~~+a (\beta- i \tau) \left(a^{2 m} (-a)^{2 n}-(-a)^{2 m} a^{2 n}\right) \Gamma (-m+n+1)\right.\right. \nonumber\\ && \left.\left.~~~~~~~~~~~~~~\times   \, _1\tilde{F}_2\left(-m+n+1;\frac{3}{2},-m+n+\frac{5}{2};a^2 (\beta- i \tau)^2\right)\right]\right\}~~~\forall~ {\rm even~n,m}.~~~~~~~~~~~.\eea
Further taking high temperature limit we get the following simplified expression for SFF as given by:
\bea G_{dc}(\tau)&=&\left[\frac{\langle Z(\bg+i\tau)\rangle \langle Z(\bg-i\tau)\rangle}{\langle Z(\bg)\rangle^{2}}\right]_{\beta=0}\nonumber\\
&=&\frac{1}{N^2}\left\{\sum^{n}_{k=1}a_{n-k}~\frac{ e^{-2 i \pi  k} 4^{n-k} a^{-2 k+2 n+2}}{\sqrt{\pi }\Gamma (-k+n+2) \Gamma \left(-k+n+\frac{5}{2}\right)} \right.\nonumber\\ 
&&\left.~~\times\left[\left\{(-1)^{2 k}+(-1)^{2 n}\right\} \Gamma \left(-k+n+\frac{1}{2}\right) \Gamma \left(-k+n+\frac{5}{2}\right)\right. \right.\nonumber\\ && \left.\left.~~~~~~~~~~~~~~\times \, _1F_2\left(-k+n+\frac{1}{2};\frac{1}{2},-k+n+2;-a^2 \tau^2\right)\right.\right. \nonumber\\ && \left.\left.~~~~~~~~- 2 i a \tau \left\{(-1)^{2 k}+(-1)^{2 n+1}\right\} \Gamma (-k+n+1) \Gamma (-k+n+2)\right.\right. \nonumber\\ && \left.\left.~~~~~~~~~~~~~~\times  \, _1F_2\left(-k+n+1;\frac{3}{2},-k+n+\frac{5}{2};-a^2 \tau^2\right)\right]\right\}\nonumber\\
&&\times \left\{\sum^{n}_{m=1}a_{n-k}~\frac{ e^{-2 i \pi  m} 4^{n-m} a^{-2 m+2 n+2}}{\sqrt{\pi }\Gamma (-m+n+2) \Gamma \left(-m+n+\frac{5}{2}\right)} \right.\nonumber\\ 
&&\left.~~\times\left[\left\{(-1)^{2 m}+(-1)^{2 n}\right\} \Gamma \left(-m+n+\frac{1}{2}\right) \Gamma \left(-m+n+\frac{5}{2}\right)\right. \right.\nonumber\\ && \left.\left.~~~~~~~~~~~~~~\times \, _1F_2\left(-m+n+\frac{1}{2};\frac{1}{2},-m+n+2;-a^2 \tau^2\right)\right.\right. \nonumber\\ && \left.\left.~~~~~~~~+ 2 i a \tau \left\{(-1)^{2 m}+(-1)^{2 n+1}\right\} \Gamma (-m+n+1) \Gamma (-m+n+2)\right.\right. \nonumber\\ && \left.\left.~~~~~~~~~~~~~~\times  \, _1F_2\left(-m+n+1;\frac{3}{2},-m+n+\frac{5}{2};-a^2 \tau^2\right)\right]\right\}\eea
Next, we will consider late time limiting behaviour of the one point function, which can be expressed as:
\be\lim_{\tau \to\infty}\langle Z(\bg\pm i\tau)\rangle_{nGUE}\equiv \langle Z(\beta\pm i\infty)\rangle_{nGUE}\ee
and at the high temperature ($\beta=1/T\rightarrow 0$) limit we get:
\be\lim_{\tau \to\infty}\left[\langle Z(\bg\pm i\tau)\rangle_{nGUE}\right]_{\bg=0}\equiv \langle Z(0\pm i\infty)\rangle_{nGUE}\ee
Now, it is important to note from the previous discussion on SFF that, the connected part of the Green's function $G_{c}$ part of SFF depends on the two point correlation function $\langle \del D(\lb) \del D(\mu)\rangle$ and from RMT the exact from of this two-point function near the centre of spectrum (mean) of the eigen values  is known and can be expressed in the following form:
 \bea\label{con1}
\langle \del D(\lb) \del D(\mu)\rangle=-\frac{\sin^{2}[N(\lb-\mu)]}{(\pi N(\lb-\mu))^{2}}+\frac{1}{\pi N}\del(\lb-\mu)
\eea
which can be derived using the method of orthogonal polynomials for Gaussian ensembles. This is true for any polynomial potential measure whose matrix (operator) is of single trace. Various polynomial potentials change only the eigen value distribution near edges of the distribution.
There are two parts and they give different measures:
\begin{enumerate}
\item $1/N^{2}$ part with sine squared function gives the ramp and have subdominant contribution.
\item $1/N$ part with Delta function gives the plateau and dominant.
\end{enumerate}
Next, using Eq~(\ref{con1}) in Eq~(\ref{ccon}) we get the following simplified expression for the connected part of the Green's function $G_c$ as given by:
 \bea\label{ccong}
G_{c}(\tau)=G(\tau)-G_{dc}(\tau)=\frac{1}{N^2}\int d\lb~ d\mu~ e^{-i\tau(\lb-\mu)}~\left[-\frac{\sin^{2}[N(\lb-\mu)]}{(\pi N(\lb-\mu))^{2}}+\frac{1}{\pi N}\del(\lb-\mu)\right].
\eea
where we have used the fact that:
\be \int d\lb~ d\mu~ \langle D(\lb)\rangle \langle D(\mu)\rangle=N^2.\ee
To perform the integral present in the expression for $G_{c}$ we further substitute,  
$\lb+\mu=E,~~~
\lb-\mu=\og.$ Consequently, the measure can be expressed as, 
$d\lb~ d\mu=dE~ d\og.$ Then at high temperature using this fact Eq~(\ref{ccong}) can be recast as:
 \bea\label{ccongg}
G_{c}(\tau)=G(\tau)-G_{dc}(\tau)=\frac{1}{N^2}\int^{\infty}_{-\infty}\int^{\infty}_{-\infty} dE~ d\og~ e^{-i\tau\og}~\left[-\frac{1}{\pi^2}\frac{\sin^{2}[N\og]}{(N\og)^{2}}+\frac{1}{\pi N}\del(\og)\right].
\eea
Then, at finite temperature the connected part of the Green's function can be written as:
\bea\label{ccongg}
G_{c}(\beta,\tau)&=&G(\beta,\tau)-G_{dc}(\beta,\tau)\nonumber\\
&=&\frac{1}{N^2}\int^{\infty}_{-\infty}\int^{\infty}_{-\infty} dE~ d\og~ e^{-\beta E}e^{-i\tau\og}~\left[-\frac{1}{\pi^2}\frac{\sin^{2}[N\og]}{(N\og)^{2}}+\frac{1}{\pi N}\del(\og)\right]\nonumber\\
&=&\frac{2\pi}{N^2}\delta(\beta)\int^{\infty}_{-\infty}~ d\og~ e^{-i\tau\og}~\left[-\frac{1}{\pi^2}\frac{\sin^{2}[N\og]}{(N\og)^{2}}+\frac{1}{\pi N}\del(\og)\right],
\eea
where $\delta(\beta)$ is the Dirac Delta Function, which is defined as:
\be \delta(\beta)=\frac{1}{2\pi}\int^{\infty}_{-\infty}~dE~e^{-\beta E}.\ee
Since the integral over $E$ gives trivial Diarc Delta function we choose our working region for which $E=0$ (at hight temperature limit). Then the remaining integrand is only over $\og$ and it finally gives:
 \bea\label{cconggx}
S(\tau)=N^2G_c(\tau)=\int^{\infty}_{-\infty}~ d\og~ e^{-i\tau\og}~\left[-\frac{1}{\pi^2}\frac{\sin^{2}[N\og]}{(N\og)^{2}}+\frac{1}{\pi N}\del(\og)\right].
\eea
which gives us finally the following simplified expression:
  \bea\begin{array}{lll}\label{eqxxxgx}
		\displaystyle   S(\tau)=\left\{\begin{array}{lll}
			\displaystyle  
			\frac{\tau}{(2\pi N)^{2}}-\frac{1}{N}+\frac{1}{(\pi N)}\,,~~~~~~~~~~~~ &
			\mbox{\small  \textcolor{red}{\bf  {$\tau<2\pi N$ }}}  \\ 
			\displaystyle  
			\frac{1}{\pi N}\,,~~~~~~~~~~~~ &
			\mbox{\small  \textcolor{red}{\bf  {$\tau>2\pi N$}}}  
		\end{array}
		\right.
	\end{array},\eea
From the obtained result it is clearly observed that we get the linear growth in the region $\tau<2\pi N$ and the constant plateau type behaviour in the region $\tau>2\pi N$. Also it is important to note that change in behaviour from  region $\tau<2\pi N$ to region $\tau>2\pi N$  is abrupt. To show the behaviour of SFF explicitly we define argument of $\sin$ function as:
 \be x\equiv N(\lb-\mu)=N\og={\rm constant} \ee as we 
 choose $N\rightarrow\infty$ and $\og\rightarrow 0$. In this limiting situation we get the following results:
 \begin{enumerate}
 \item For large $x(>>1)$, $\frac{\sin x}{x}\rightarrow 0$ and only the Dirac Delta function remains intact. So in this specific limit the vanishing of $\sin$ term implies that the oscillatory fluctuations don't contribute in the final expression for SFF. This limiting situation is called {\it  spectral rigidity}.
 
 \item For small $x(<<1)$, $\frac{\sin x}{x}\rightarrow 1$. In this limiting situation the integral gets maximum contribution from the $\og=0$ region.  And this part contributes in ramp region.
 \end{enumerate}
We can also measure dip-time and it will give the change of decay behaviour exactly at the critical point.
A direct relation between fall-off behaviour of the SFF and the edge behaviour of level density, at critical points can be established using {\it Paley-Wiener Theorem} \cite{Gaikwad:2017odv}.

Now we consider a function $g(\zeta)$ which is defined on a compact spatial support and its Fourier transform $F(\eta)$ has a lower bound on the rate of decay is given by the following expression:
\be
\boxed{|F(\eta)|\leqslant(1+\eta)^{-N}\gamma_{N}}
\ee
Here $N$ is a rational number and $\gamma_{N}$ is a real constant.
 A direct relation between the decay of the SFF and the edge effect of mean level density is given by the following expression:
 \bea|\langle Z(\pm i\tau)\rangle|\leqslant\frac{1}{(\pm \tau)^{n}}(4a)~\left|\int_{-2a}^{2a}\frac{d^{n}}{d\lb^{n}}(\overline\rho(\lb))~d\lb\right|
 \eea
  For the proof of this statement see ref.~\cite{Gaikwad:2017odv}. For decay behaviour of SFF at late time we use asymptotic behaivour of the solution appearing in the ref.~\cite{Dyer:2016pou}.
  
 Now to compute  SFF we need to add both connected and disconnected part of the Green's function $G$($=G_{c}+G_{dc}$). Therefore, for different even polynomial potential we get finally the following expression for SFF at finite temperature:
 \be\boxed{\begin{array}{lll}\label{wq1}
		\displaystyle  {\bf SFF}(\beta,\tau)\equiv G(\beta,\tau)=\left\{\begin{array}{lll}
			\displaystyle  
			G_{dc}(\beta,\tau)+\frac{\tau}{(2\pi N)^{2}}-\frac{1}{N}+\frac{1}{(\pi N)}\,,~~~~~~~~~~~~ &
			\mbox{\small  \textcolor{red}{\bf  {$\tau<2\pi N$ }}}  \\ 
			\displaystyle  
			G_{dc}(\beta,\tau)+\frac{1}{\pi N}\,,~~~~~~~~~~~~ &
			\mbox{\small  \textcolor{red}{\bf  {$\tau>2\pi N$}}}  
		\end{array}
		\right.
	\end{array}}~~,\ee
where ${\bf SFF}(\tau)$ is defined with proper normalization. 

After substituting the expression for $G_{dc}(\beta,\tau)$ we get the following expression for the SFF at finite temperature:
\bea\boxed{\begin{array}{lll}\label{wsq1}
		\small\displaystyle  {\bf SFF}(\beta,\tau)&\equiv& \displaystyle
		\left\{ \sum^{n}_{q=1}a_{n-q}~\left(-a^2\right)^{-2 q} 4^{-q} \left[\left(e^{2 i \pi  q}+e^{2 i \pi  n}\right) a^{2 (q+n)} \Gamma \left(-q+n+\frac{1}{2}\right)\right. \right.\\ && \left.\left.\displaystyle\times  \, _1\tilde{F}_2\left(-q+n+\frac{1}{2};\frac{1}{2},-q+n+2;a^2 \beta^2\right)\right. \right.\\ && \left.\left.~~~~~~~~~~~~~~\displaystyle+a \beta \left(a^{2 q} (-a)^{2 n}-(-a)^{2 q} a^{2 n}\right) \Gamma (-q+n+1)\right.\right. \\ && \left.\left.~~~~~~~~~~~~~~\displaystyle\times   \, _1\tilde{F}_2\left(-q+n+1;\frac{3}{2},-q+n+\frac{5}{2};a^2 \beta^2\right)\right]\right\}^{-2}\\
&& 	\displaystyle \times \left\{ \sum^{n}_{k=1}a_{n-k}~ \left(-a^2\right)^{-2 k} 4^{-k} \left[\left(e^{2 i \pi  k}+e^{2 i \pi  n}\right) a^{2 (k+n)} \Gamma \left(-k+n+\frac{1}{2}\right)\right. \right.\\ && \left.\left. \displaystyle \, _1\tilde{F}_2\left(-k+n+\frac{1}{2};\frac{1}{2},-k+n+2;a^2 (\beta+ i \tau)^2\right)\right.\right. \nonumber\\ && \left.\left.+a (\beta+ i \tau) \left(a^{2 k} (-a)^{2 n}-(-a)^{2 k} a^{2 n}\right) \Gamma (-k+n+1) \right. \right.\\ && \left.\left. \displaystyle \, _1\tilde{F}_2\left(-k+n+1;\frac{3}{2},-k+n+\frac{5}{2};a^2 (\beta+ i \tau)^2\right)\right]\right\}\\
&&	\displaystyle \times\left\{ \sum^{n}_{m=1}a_{n-m}~  \left(-a^2\right)^{-2 m} 4^{-m} \left[\left(e^{2 i \pi  m}+e^{2 i \pi  n}\right) a^{2 (m+n)} \Gamma \left(-m+n+\frac{1}{2}\right)\right. \right.\\ && \left.\left.~~~~~~~~~~~~~~\displaystyle\times  \, _1\tilde{F}_2\left(-m+n+\frac{1}{2};\frac{1}{2},-m+n+2;a^2 (\beta- i \tau)^2\right)\right.\right. \\ && \left.\left.~~~~~~~~~~~~~~+a (\beta- i \tau) \left(a^{2 m} (-a)^{2 n}-(-a)^{2 m} a^{2 n}\right) \Gamma (-m+n+1)\right.\right. \\ && \left.\left.~~~~~~~~~~~~~~\displaystyle\times   \, _1\tilde{F}_2\left(-m+n+1;\frac{3}{2},-m+n+\frac{5}{2};a^2 (\beta- i \tau)^2\right)\right]\right\}\\ 
&&+\left\{\begin{array}{lll}
			\displaystyle  
			\frac{\tau}{(2\pi N)^{2}}-\frac{1}{N}+\frac{1}{(\pi N)}\,,~~~~~~~~~~~~ &
			\mbox{\small  \textcolor{red}{\bf  {$\tau<2\pi N$ }}}  \\  
			\displaystyle  
			\frac{1}{\pi N}\,,~~~~~~~~~~~~ &
			\mbox{\small  \textcolor{red}{\bf  {$\tau>2\pi N$}}}  
		\end{array}
		\right.
	\end{array}}~~\\
	&& \label{bound1}\eea
	Further taking the high temperature limit we get the following simplified expression for SFF as given by:
\bea\boxed{\begin{array}{lll}\label{wsq4}
		\displaystyle  {\bf SFF}(\tau)&\equiv& \displaystyle
		\frac{1}{N^2}\left\{\sum^{n}_{k=1}a_{n-k}~\frac{ e^{-2 i \pi  k} 4^{n-k} a^{-2 k+2 n+2}}{\sqrt{\pi }\Gamma (-k+n+2) \Gamma \left(-k+n+\frac{5}{2}\right)} \right.\\ 
&&\left.~~\displaystyle\times\left[\left\{(-1)^{2 k}+(-1)^{2 n}\right\} \Gamma \left(-k+n+\frac{1}{2}\right) \Gamma \left(-k+n+\frac{5}{2}\right)\right. \right.\\ && \left.\left.~~~~~~~~~~~~~~\displaystyle  \times \, _1F_2\left(-k+n+\frac{1}{2};\frac{1}{2},-k+n+2;-a^2 \tau^2\right)\right.\right. \\ && \left.\left.~~~~~~~~- 2 i a \tau \left\{(-1)^{2 k}+(-1)^{2 n+1}\right\} \Gamma (-k+n+1) \Gamma (-k+n+2)\right.\right. \\ && \left.\left.~~~~~~~~~~~~~~\displaystyle  \times  \, _1F_2\left(-k+n+1;\frac{3}{2},-k+n+\frac{5}{2};-a^2 \tau^2\right)\right]\right\}\\
&&\displaystyle\times \left\{\sum^{n}_{m=1}a_{n-m}~\frac{ e^{-2 i \pi  m} 4^{n-m} a^{-2 m+2 n+2}}{\sqrt{\pi }\Gamma (-m+n+2) \Gamma \left(-m+n+\frac{5}{2}\right)} \right.\\ 
&&\left.~~\displaystyle  \times\left[\left\{(-1)^{2 m}+(-1)^{2 n}\right\} \Gamma \left(-m+n+\frac{1}{2}\right) \Gamma \left(-m+n+\frac{5}{2}\right)\right. \right.\\ && \left.\left.~~~~~~~~~~~~~~\displaystyle  \times \, _1F_2\left(-m+n+\frac{1}{2};\frac{1}{2},-m+n+2;-a^2 \tau^2\right)\right.\right. \\ && \left.\left.~~~~~~~~+ 2 i a \tau \left\{(-1)^{2 m}+(-1)^{2 n+1}\right\} \Gamma (-m+n+1) \Gamma (-m+n+2)\right.\right.\\ && \left.\left.~~~~~~~~~~~~~~\displaystyle  \times  \, _1F_2\left(-m+n+1;\frac{3}{2},-m+n+\frac{5}{2};-a^2 \tau^2\right)\right]\right\}\nonumber\\ \\
&&+\left\{\begin{array}{lll}
			\displaystyle  
			\frac{\tau}{(2\pi N)^{2}}-\frac{1}{N}+\frac{1}{(\pi N)}\,,~~~~~~~~~~~~ &
			\mbox{\small  \textcolor{red}{\bf  {$\tau<2\pi N$ }}}  \\  \\
			\displaystyle  
			\frac{1}{\pi N}\,,~~~~~~~~~~~~ &
			\mbox{\small  \textcolor{red}{\bf  {$\tau>2\pi N$}}}  
		\end{array}
		\right.
	\end{array}}~~\\
	&& \label{bound2}\eea	
  \subsection{SFF for even polynomial random potentials}
  \subsubsection{For Gaussian random potential}
  Let us start our discussion with Gaussian random potential given by:
   \bea V(M)=\frac{1}{2}M^{2}.
  \eea
  Now for a single interval ($n=1$) with end points $-2a$ and $2a$ (semi-circle) we get:
  \bea\og(\lb+i0)=\frac{\lambda }{2}+i a_0 \sqrt{4 a^2-\lambda ^2}.
  \eea
  and we get the following expression for density function for eigen value of the random matrix $M$ as given by:
 \bea\rho(\lb)=\frac{1}{\pi}\sqrt{4 a^2-\lambda ^2} ~a_0.
\eea
  Further, Taylor expanding $\og(\lb+i0)$ we get the following expression:
  \be
  \frac{4 a_0 a^6}{\lambda ^5}+\frac{2 a_0 a^4}{\lambda ^3}+\frac{2 a_0 a^2}{\lambda }+\left(\frac{1}{2}-a_0\right) \lambda +O\left(\left(\frac{1}{\lb}\right)^{6}\right)=\frac{1}{\lb}.
  \ee
  Then comparing the both the sides of above expression we get:
  \bea a_{0}&=&1/2, \\ a&=&\textpm 1. \eea 
  Then the density function in terms of the eigen value of random matrix $M$ is given by the following expression:
   \bea \label{gaussian} \rho(\lb)=\frac{1}{2\pi}\sqrt{4a^{2}-\lb^{2}}.\eea
  and one point function of the partition function in presence of the Gaussian random potential can be expressed as:
  \be
\langle Z(\bg\pm i\tau)\rangle= \frac{1}{2\pi}\int_{-2 a}^{2 a}d\lb~\sqrt{4 a^2-\lambda ^2} ~e^{\mp \text{i$\tau $} \lambda }~ e^{-\beta\lb}= a^2 \, _0\tilde{F}_1\left(2;a^2 (\beta\pm i \tau)^2\right),
\ee
where $ \, _0\tilde{F}_1\left(A;B\right)$ is the regularized Hypergeometric function. Now here substituting $\tau=0$ we get:
 \be
\langle Z(\bg)\rangle= \frac{1}{2\pi}\int_{-2 a}^{2 a}d\lb~\sqrt{4 a^2-\lambda ^2} ~ e^{-\beta\lb}= a^2 \, _0\tilde{F}_1\left(2;a^2 \beta^2\right).
\ee

Further taking high temperature limit we get the following simplified expression for the one point function:
\be
\left[\langle Z(\bg\pm i\tau)\rangle\right]_{\bg=0}= \frac{1}{2\pi}\int_{-2 a}^{2 a}d\lb~\sqrt{4 a^2-\lambda ^2} ~e^{\mp \text{i$\tau $} \lambda }=\pm \frac{a J_1(\pm 2 a \text{$\tau $})}{\text{$\tau $}},
\ee
which can be further simplified by taking the limit ${\cal T}=\sqrt{N}\tau\rightarrow\infty$ as:
\bea \label{tgaussian}
\left[\langle Z(\bg\pm i{\cal T})\rangle\right]_{\bg=0}=-\frac{1}{\sqrt{\pi}}a^2 \left(\pm \frac{1}{a {\cal T}  }\right)^{3/2} \cos \left(\frac{1}{4} (\pm 8 a {\cal T} +\pi )\right).
\eea
Now for the quadratic random potential disconnected part of the Green's function can be computed at finite temperature as:
\bea G_{dc}(\beta,\tau)&=&\frac{\langle Z(\bg+i\tau)\rangle \langle Z(\bg-i\tau)\rangle}{\langle Z(\bg)\rangle^{2}}= \frac{ \, _0\tilde{F}_1\left(2;a^2 (\beta+ i \tau)^2\right)\, _0\tilde{F}_1\left(2;a^2 (\beta+ i \tau)^2\right)}{ \left(\, _0\tilde{F}_1\left(2;a^2 \beta^2\right)\right)^2},~~~~~~\eea
which can be further simplified in the high temperature limiting situation as:
\bea  G_{dc}(\tau)&=&\left[\frac{\langle Z(\bg+i\tau)\rangle \langle Z(\bg-i\tau)\rangle}{\langle Z(\bg)\rangle^{2}}\right]_{\beta=0}=-\frac{a^2}{\tau^4}\frac{ J_1( 2 a \text{$\tau $})J_1(- 2 a \text{$\tau $})}{ N^2}.~~~~~~\eea
Further taking the limit ${\cal T}=\sqrt{N}\tau\rightarrow\infty$ we get the following simplified result:
\bea G_{dc}({\cal T})&=&\left[\frac{\langle Z(\bg+i{\cal T})\rangle \langle Z(\bg-i{\cal T})\rangle}{\langle Z(\bg)\rangle^{2}}\right]_{\beta=0}=(-1)^{3/2}\frac{a^4}{2N^2\pi}\left(\frac{1}{a {\cal T}  }\right)^{3} \cos \left(\frac{1}{2} ( 8 a {\cal T} +\pi )\right).~~~~~~~~~~\eea
Now to compute  SFF we need to add both connected and disconnected part of the Green's function $G$($=G_{c}+G_{dc}$). Therefore, for quadratic polynomial potential we get finally the following expression for SFF at finite temp:
\bea \begin{array}{lll}\label{cqf1}
		\displaystyle   {\bf SFF}(\beta,\tau)\equiv \left\{\begin{array}{lll}
			\displaystyle  
			\frac{ \, _0\tilde{F}_1\left(2;a^2 (\beta+ i \tau)^2\right)\, _0\tilde{F}_1\left(2;a^2 (\beta+ i \tau)^2\right)}{ \left(\, _0\tilde{F}_1\left(2;a^2 \beta^2\right)\right)^2}\\
			\displaystyle~~~~~~~~~+\frac{\tau}{(2\pi N)^{2}}-\frac{1}{N}+\frac{1}{(\pi N)}\,,~~~~~~~~~~~~ &
			\mbox{\small  \textcolor{red}{\bf  {$\tau<2\pi N$ }}}  \\  \\
			\displaystyle  
			\frac{ \, _0\tilde{F}_1\left(2;a^2 (\beta+ i \tau)^2\right)\, _0\tilde{F}_1\left(2;a^2 (\beta+ i \tau)^2\right)}{ \left(\, _0\tilde{F}_1\left(2;a^2 \beta^2\right)\right)^2}\\
			\displaystyle~~~~~~~~~+\frac{1}{\pi N}\,,~~~~~~~~~~~~ &
			\mbox{\small  \textcolor{red}{\bf  {$\tau>2\pi N$}}}  
		\end{array}
		\right.
	\end{array},\eea
where ${\bf SFF}(\beta,\tau)$ is defined with proper normalization and in our prescription it gives the total Green's function as mentioned above. 
Further simplifying the result for high temperature limit we get the following expression for SFF, as given by:
\be\boxed{\begin{array}{lll}\label{cqf2}
		\footnotesize\displaystyle   {\bf SFF}(\tau)\equiv \left\{\begin{array}{lll}
			\displaystyle  
			-\frac{a^2}{\tau^4}\frac{ J_1( 2 a \text{$\tau $})J_1(- 2 a \text{$\tau $})}{ N^2}+\frac{\tau}{(2\pi N)^{2}}-\frac{1}{N}+\frac{1}{(\pi N)}\,,~~~~~~~~~~~~ &
			\mbox{\small  \textcolor{red}{\bf  {$\tau<2\pi N$ }}}  \\ 
			\displaystyle  
			-\frac{a^2}{\tau^4}\frac{ J_1( 2 a \text{$\tau $})J_1(- 2 a \text{$\tau $})}{ N^2}+\frac{1}{\pi N}\,,~~~~~~~~~~~~ &
			\mbox{\small  \textcolor{red}{\bf  {$\tau>2\pi N$}}}  
		\end{array}
		\right.
	\end{array}}~~,\ee
	Further taking the limit ${\cal T}=\sqrt{N}\tau\rightarrow\infty$ we get the following simplified result for SFF:
\be\boxed{\begin{array}{lll}\label{cqf3}
		\footnotesize\displaystyle   {\bf SFF}({\cal T})\equiv \left\{\begin{array}{lll}
			\displaystyle  
			(-1)^{3/2}\frac{a^4}{2N^2\pi}\left(\frac{1}{a {\cal T}  }\right)^{3} \cos \left(\frac{1}{2} ( 8 a {\cal T} +\pi )\right)+\frac{{\cal T}}{(2\pi)^2 N^{5/2}}-\frac{1}{N}+\frac{1}{(\pi N)}\,,~~~~~~~~~~~~ &
			\mbox{\small  \textcolor{red}{\bf  {${\cal T}<2\pi N^{3/2}$ }}}  \\ 
			\displaystyle  
			(-1)^{3/2}\frac{a^4}{2N^2\pi}\left(\frac{1}{a {\cal T}  }\right)^{3} \cos \left(\frac{1}{2} ( 8 a {\cal T} +\pi )\right)+\frac{1}{\pi N}\,,~~~~~~~~~~~~ &
			\mbox{\small  \textcolor{red}{\bf  {${\cal T}>2\pi N^{3/2}$}}}  
		\end{array}
		\right.
	\end{array}}~~.\ee	
\begin{figure}[H]
\centering
\subfigure[SFF for gaussian potential at $\bg=10$.]{
    \includegraphics[width=7.8cm,height=8.7cm] {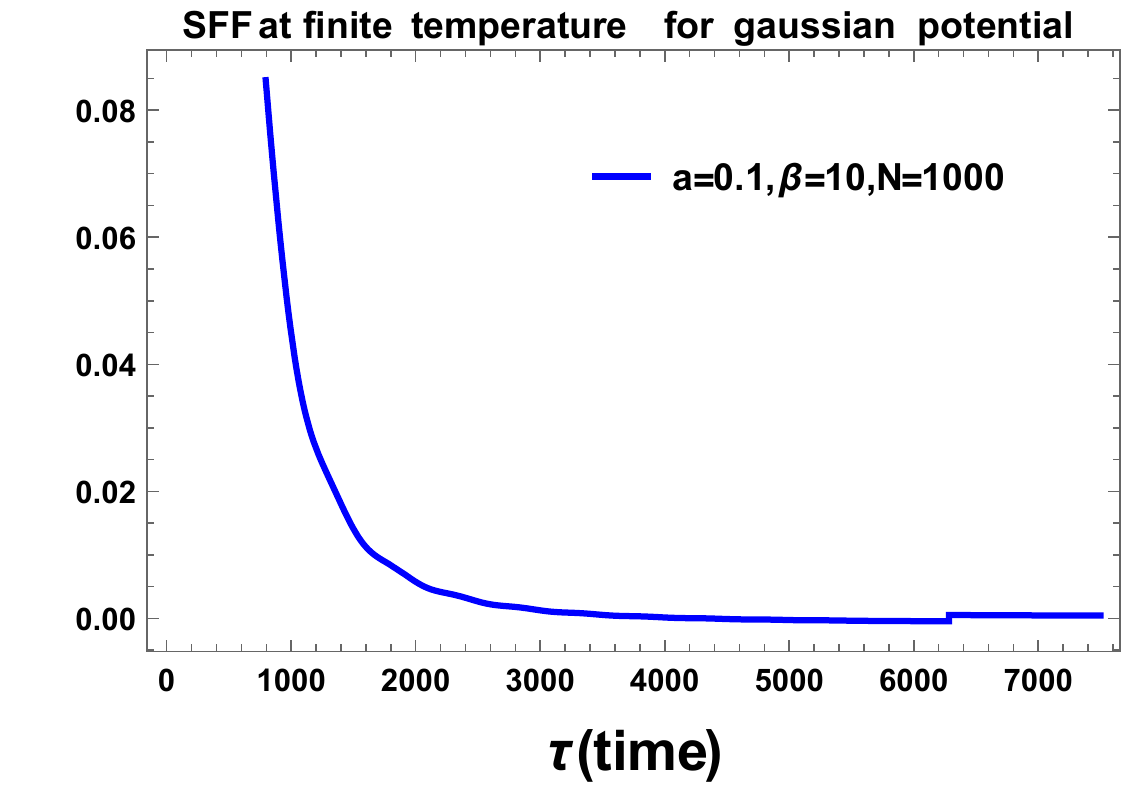}
    \label{gua1}
}
\subfigure[SFF for gaussian potential at $\bg=100$.]{
    \includegraphics[width=7.8cm,height=8.7cm] {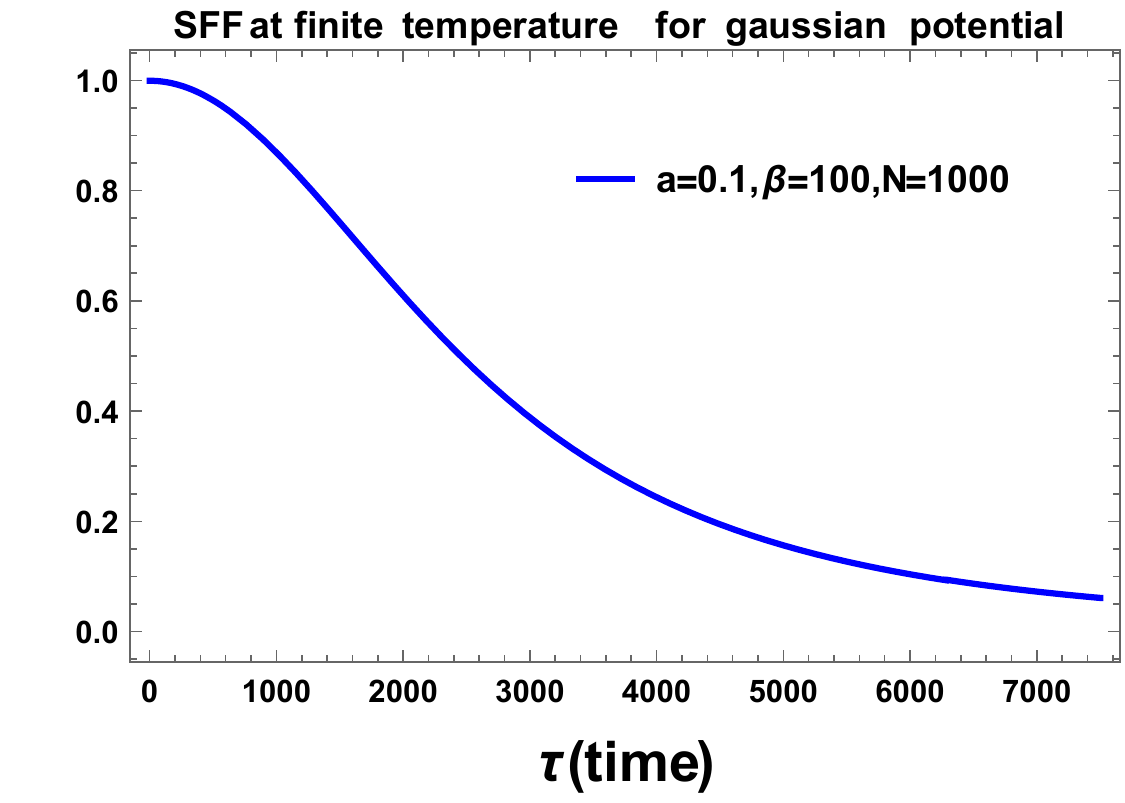}
    \label{gua2}
}
\subfigure[SFF for gaussian potential at $\bg=200$.]{
    \includegraphics[width=7.8cm,height=8.7cm] {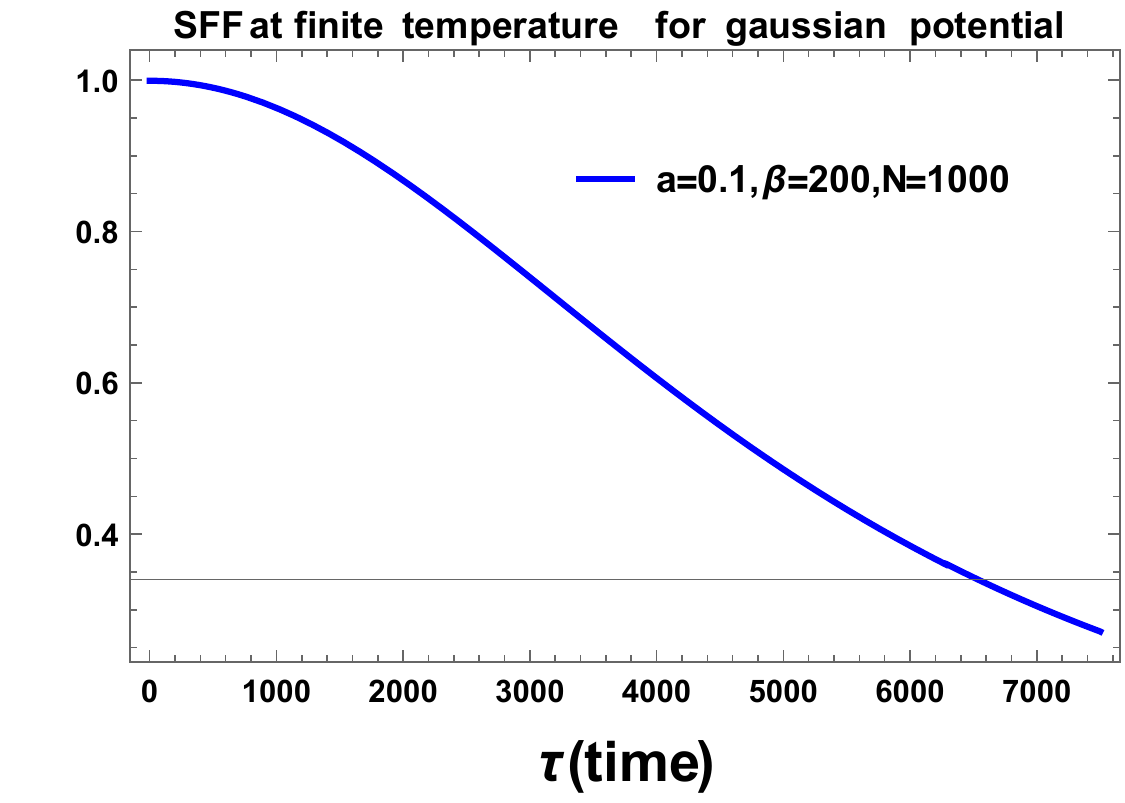}
    \label{gua3}
}
\subfigure[SFF for gaussian potential at $\bg=1000$.]{
    \includegraphics[width=7.8cm,height=8.7cm] {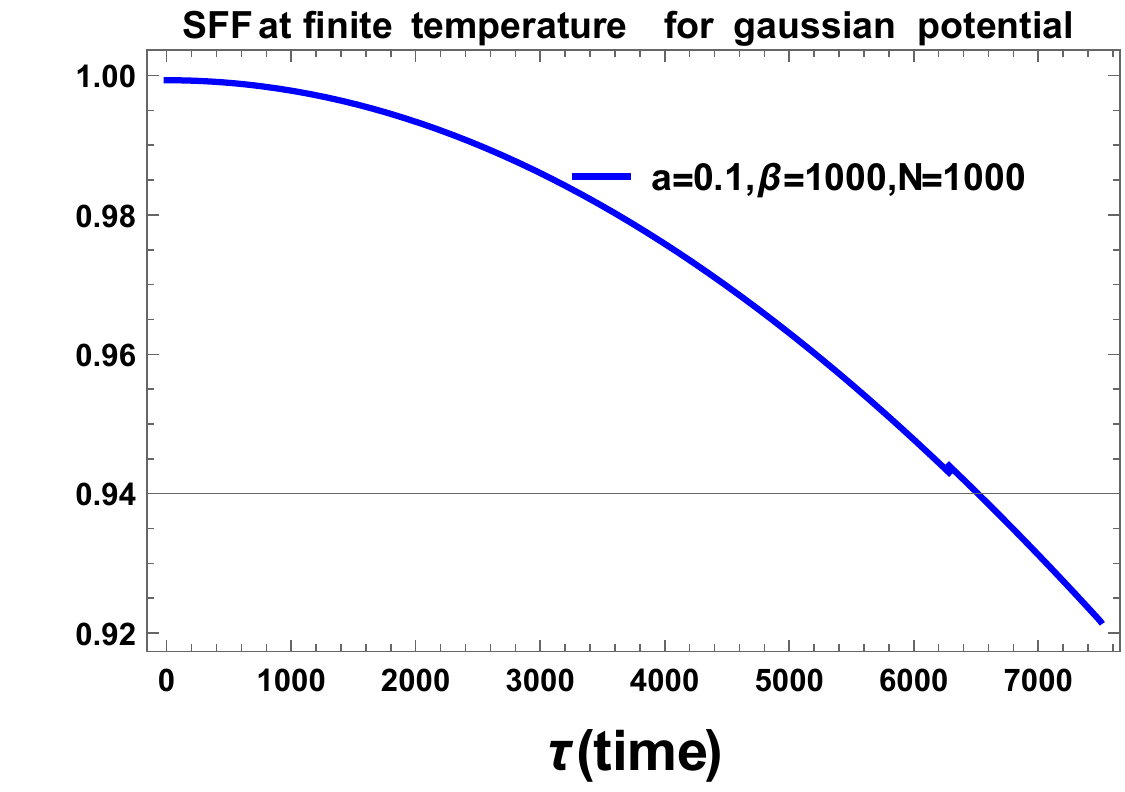}
    \label{gua4}
}
\caption{Spectral Form Factor for Gaussian  potential at different finite temperature[$\bg$] with $N=1000$ and $a=0.1$ }
\end{figure}
From fig.~\ref{gua1}, fig.~\ref{gua2},  fig.~\ref{gua3} and fig.~\ref{gua4} we see that SFF at finite temperature decays with increasing $\tau$ and reach zero. But with changing $\bg$, SFF values remains almost same initially (for higher $\bg$ or lower temperature).

%\begin{figure}[H]
%\centering
%\subfigure[SFF for gaussian potential at $N=10$.]{
 %   \includegraphics[width=7.8cm,height=8.5cm] {SFF_G_N10.pdf}
 %   \label{gua1a}
%}
%\subfigure[SFF for gaussian potential at $N=100$.]{
 %   \includegraphics[width=7.8cm,height=8.5cm] {SFF_G_N100.pdf}
   % \label{gua2a}
%}
%\subfigure[SFF for gaussian potential at $N=1000$.]{
  %  \includegraphics[width=7.8cm,height=8.5cm] {SFF_G_N1000.pdf}
  %  \label{gua3a}
%}
%\subfigure[SFF for gaussian potential at $N=10000$.]{
 %   \includegraphics[width=7.8cm,height=8.5cm] {SFF_G_N10000.pdf}
   % \label{gua4a}
%}
%\caption{Spectral Form Factor for gaussian  potential varying with different N at finite temperature[$\bg=100 $] and $a=0.1$ }
%\end{figure}

%From fig.~\ref{gua1a}, fig.~\ref{gua2a},  fig.~\ref{gua3a} and fig.~\ref{gua4a} we see that SFF at finite temperature decays with increasing $\tau$ and reach zero. But at different values of $N$ SFF values remains same initially.

For both the plots  we have shown that SFF decays to zero for finite temperature.
\begin{figure}[htb]
\centering
\subfigure[SFF for gaussian for $a=.1,N=100$]{
    \includegraphics[width=7.8cm,height=8cm] {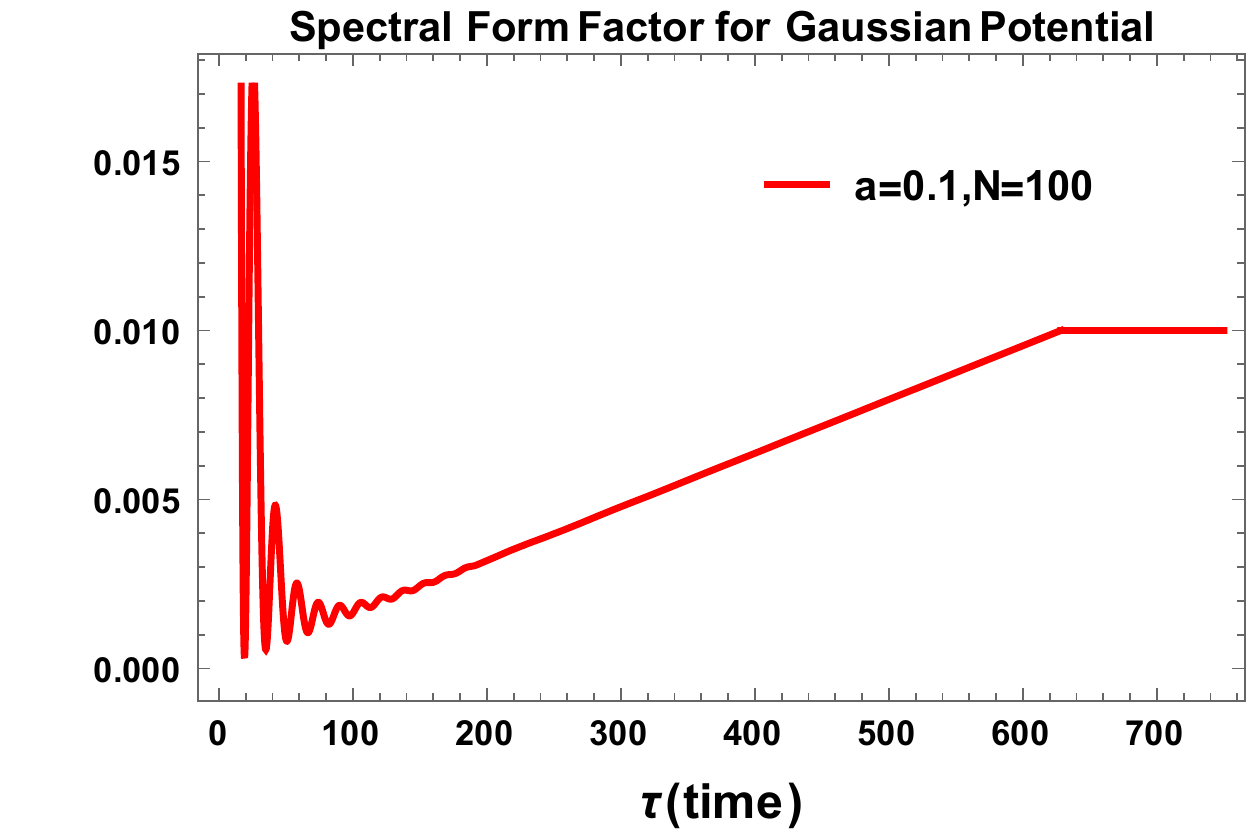}
    \label{SFFG21}
}
\subfigure[SFF for gaussian for $a=.1,N=1000$]{
    \includegraphics[width=7.8cm,height=8cm] {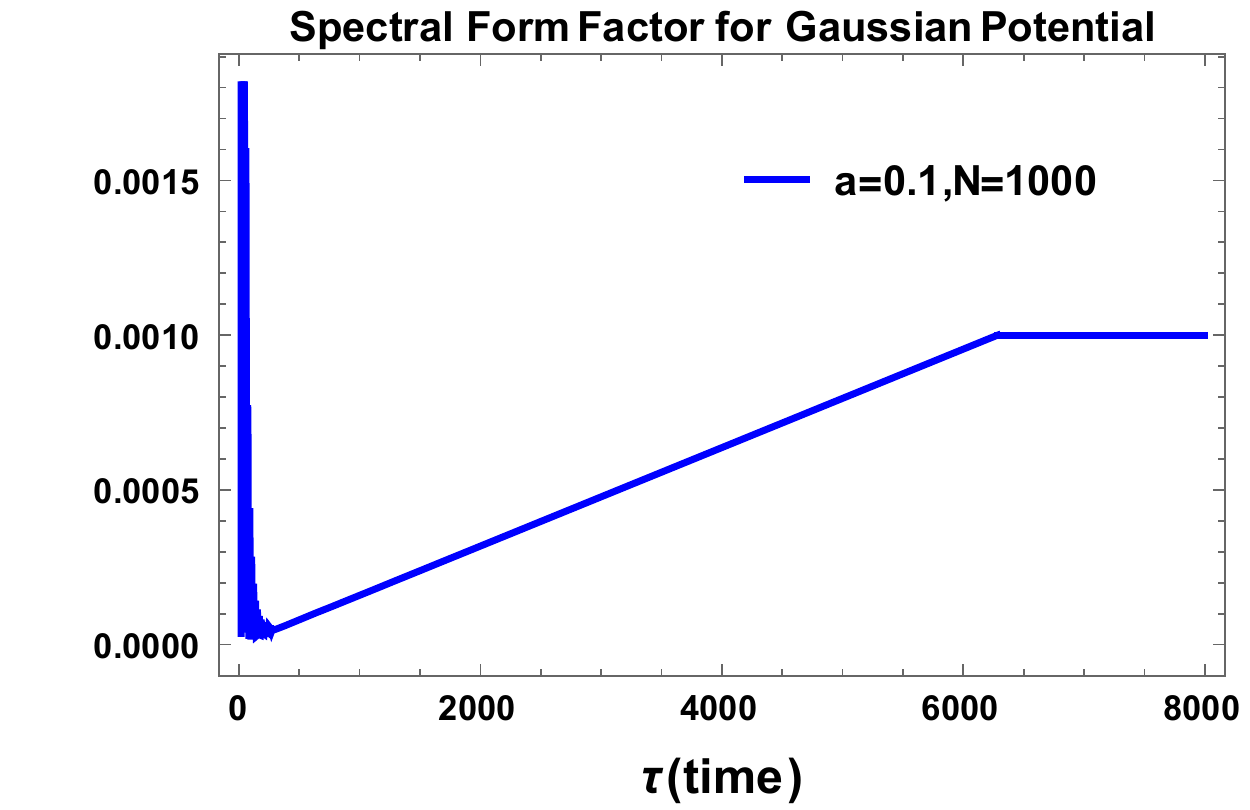}
    \label{SFFG23}
}
\subfigure[SFF for gaussian for $a=.1,N=10000$]{
    \includegraphics[width=10.8cm,height=8cm] {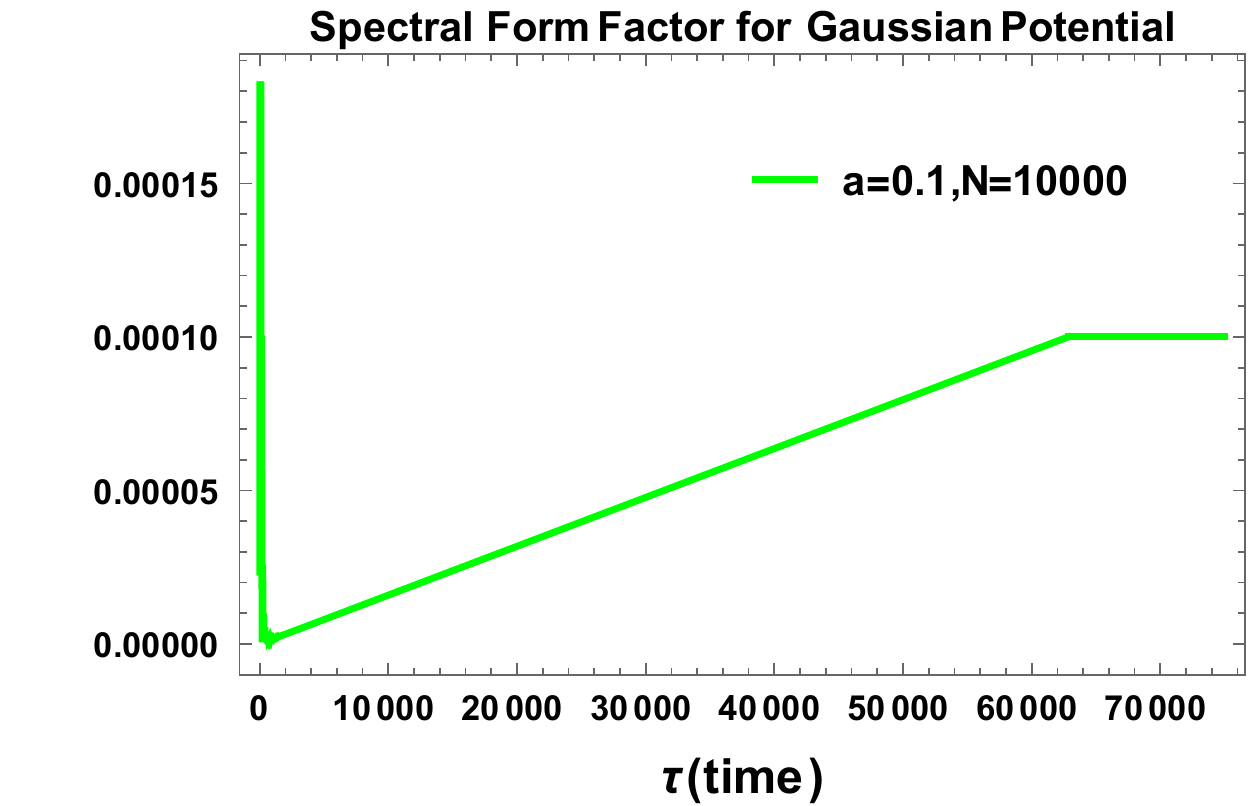}
    \label{SFFG22}
}
\caption{Time variation of SFF for different N at $\bg=0$. Here we used a scale factor $SFF+0.01137$ }
\label{SFF_N_variation}
\end{figure} 
In fig.~\ref{SFFG21}, fig.~\ref{SFFG23},  fig.~\ref{SFFG22} it is observed that SFF with variation in $N$ get saturated at different value of $\tau$. But with increasing $N$ the value of the saturation point, will decrease.
Subtracting the change of axis[$SFF|_{\tau=0}$] we get the predicted bound of SFF.

\subsubsection{For Quartic random potential}
Here we consider quartic random potential which can be written as:
\bea V(M)=\frac{1}{2}M^{2}+g M^{4}.
\eea
For a single interval ($n=1$) with end points -2a and 2a (semi-circle) we get the following expression for density function for eigen value of the random matrix $M$ as given by:
\bea \rho(\lb)=\frac{1}{\pi}\sqrt{4 a^2-\lambda ^2} ~\left(a_1 \lambda ^2+a_0\right).
\eea
Now, for the quartic random potential $\og(\lb+i0)$ can be expressed as:
\bea \og(\lb+i0)=\frac{1}{2} \left(2 C_2 \lambda +4 g \lambda ^3\right)+i \sqrt{4 a^2-\lambda ^2} \left(a_1 \lambda ^2+a_0\right).
\eea
Now Taylor expanding $\og(\lb+i0)$  near $\lb\rightarrow\infty$ gives the following expression:
\bea
&&\left(2 a_1 a^2-a_0+\frac{1}{2}\right) \lambda+\lambda ^3 \left(2 g-a_1\right)+\frac{10 a_1 a^8+4 a_0 a^6}{\lambda ^5}\nonumber\\
&&~~~~~~~~~~~~~~~~~~~~~~+\frac{4 a_1 a^6+2 a_0 a^4}{\lambda ^3}+\frac{2 a_1 a^4+2 a_0 a^2}{\lambda }+O\left(\left(\frac{1}{\lb}\right)^{6}\right)=\frac{1}{\lb}.
\eea
Therefore equating both sides of the above equation gives:
\bea a_{1}=2g,\\ a_{0}=4a^{2}g+\frac{1}{2},\eea
along with the following  constraint condition:
\be 12 g a^4+ a^{2}=1.\ee
 Then the density function in terms of the eigen value of random matrix $M$ is given by the following expression: 
\bea \label{quart1}
\rho(\lb)=\frac{1}{\pi}\left(\frac{1}{2}+4ga^{2}+2g\lb^{2}\right)~\sqrt{4a^{2}-\lb^{2}}.
\eea
Further solving the constraint we get:
\bea a^{2}=\frac{\sqrt{48 g+1}-1}{24 g}.\eea and 
Here $a^{2}$ has imaginary value for $g\leqslant -\frac{1}{48}$ and the critical value is given by:
\be \label{nonexsist1} \boxed{g_{c}=-\frac{1}{48}}~.\ee
\begin{figure}[htb]
\centering
\subfigure[$\rho(\lb)$ for quartic potential for different $g$.]{
    \includegraphics[width=7.8cm,height=9cm] {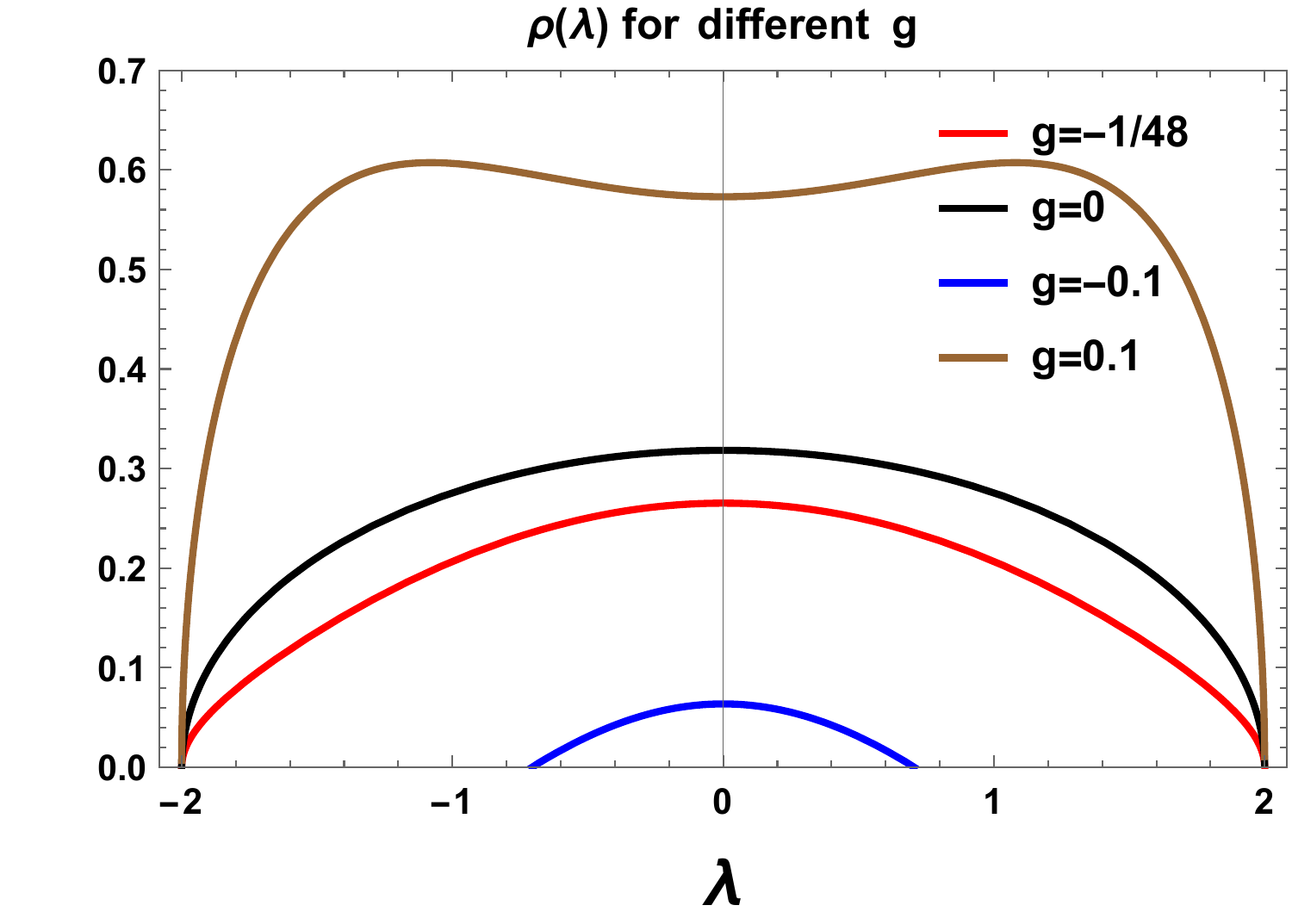}
    \label{Rho21}
}
\subfigure[$\rho(\lb)$ for quartic potential for different $g$.]{
    \includegraphics[width=7.8cm,height=9cm] {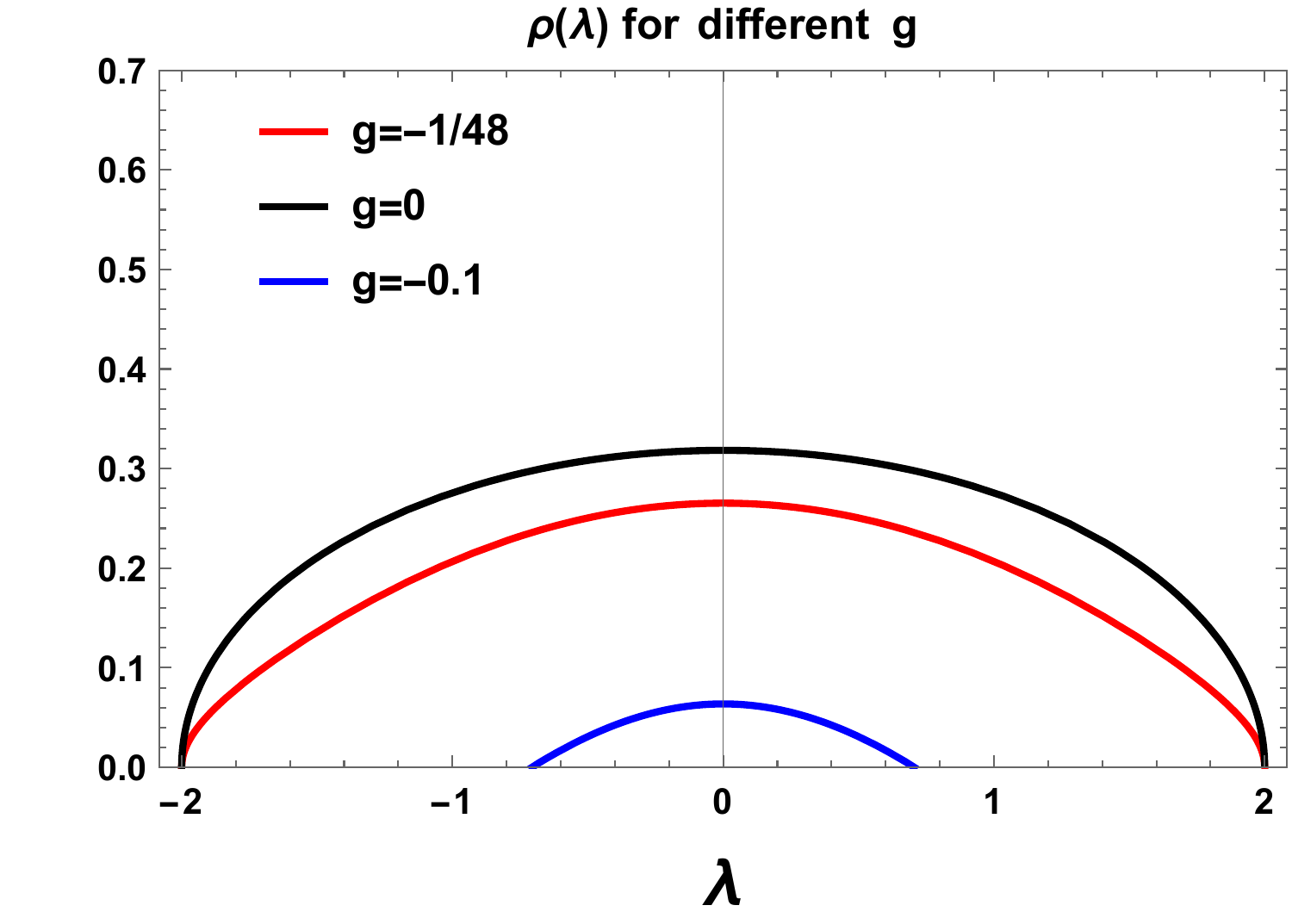}
    \label{Rho22}
}
\caption{Eigen value distribution curve of density function for quartic  potential for different parameter values. Here we fix $a=1$. }
\end{figure}
In fig.~\ref{Rho21} and fig.~\ref{Rho22} density function $\rho(\lb)$ for quartic potential is plotted with $a=1$. The curve follows from Eq.~(\ref{quart1}). When $g=0$ it matches with {\it Wigner law}. For $g>0$ the curve shows a plateau region whereas for $g<0$ it preserve the semicircular nature with minor deviation.The plateau region denotes the deviation from {\it Wigner law} even at very less effect of quartic term (as $g$ is chosen to be small). The plateau region though converge with semicircle at end point. At $g_{c}=-\frac{1}{48}$ the curve deviates but converge to semicircle at end points where as for $g<g_{c}$ the curve never converge to semicircle one supporting its non-existence. (See eq.~(\ref{nonexsist1}) for details.).

Now we will calculate the one point function of the partition function for quartic random potential, which is given by the following expression:
  \bea
\langle Z(\bg\pm i\tau)\rangle &=& \frac{1}{\pi}\int_{-2 a}^{2 a}d\lb~\left(\frac{1}{2}+4ga^{2}+2g\lb^{2}\right)~\sqrt{4 a^2-\lambda ^2} ~e^{\mp \text{i$\tau $} \lambda }~ e^{-\beta\lb}\nonumber\\
&=&\frac{a}{(\beta\pm i \tau)^2} \left[\left(24 a^2 g+1\right) (\beta\pm i \tau) I_1(2 a (\beta\pm i \tau))-24 a g I_2(2 a (\beta\pm i \tau))\right],~~~~~~~~~~
\eea
where $I_{n}(x)$ is the modified Bessel function of first kind with order $n$.

Further taking the high temperature limit we get the following simplified expression for the one point function:
\bea
\left[\langle Z(\bg\pm i\tau)\rangle\right]_{\bg=0}&=& \frac{1}{\pi}\int_{-2 a}^{2 a}d\lb~\left(\frac{1}{2}+4ga^{2}+2g\lb^{2}\right)~\sqrt{4 a^2-\lambda ^2} ~e^{\mp\text{i$\tau $} \lambda }\nonumber\\
&=&\frac{a }{\text{$\tau $}^2}\left[\pm \left(24 a^2 g+1\right) \text{$\tau $} I_1(\pm 2 a \text{$\tau $})-24 a g I_2(\pm 2 a \text{$\tau $})\right].
\eea
Therefore the first term vanishes exactly at the critical point $g_{c}=-\frac{1}{48}$ which gives: \be \boxed{a^{2}=\frac{1}{24g_{c}}=2}~.\ee
Now taking the limit ${\cal T}=\sqrt{N}t\rightarrow\infty$ we get finally the following simplified result for the one point function:
\bea\label{tquartic}
\left[\langle Z(\bg\pm i{\cal T})\rangle\right]_{\bg=0}&=&-\frac{1}{(\pm {\cal T})^{3/2}}\sqrt{\frac{a}{\pi}}\left[\left(24 a^2 g+1\right) \cos \left(\pm 2 a {\cal T} +\frac{\pi }{4}\right)\pm \frac{24 g \sin \left(\pm 2 a {\cal T}+ \frac{\pi}{4}\right)}{{\cal T}}\right]\nonumber\\
&&~~~~~~~~~~~~~~~~~~~~~~~~~~~~~~~~~~~~~~~
~~~~~~~~~~~~~~~~~~~~~~~~~     +O\left(\frac{1}{(\pm {\cal T})^{\frac{7}{2}}}\right)~.~~~~~~~~~~~~~
\eea

%Now if we don't choose $\bg\rightarrow 0$ we can get the $G_{dc}$
%\be\begin{array}{lllll}
%\displaystyle G_{dc}=a^3 \beta ^2 [(24 a^2 g+1) \, _0\tilde{F}_1(;2;a^2 (\beta -i \tau )^2)  -24 a^2 g \, _0\tilde{F}_1(;3;a^2 (\beta -i \tau )^2)) ((24 a^2 g+1) \, _0\tilde{F}_1(;2;a^2 (\beta +i \tau )^2)\\
% \displaystyle \hspace{1cm}-24 a^2 g \, _0\tilde{F}_1(;3;a^2 (\beta +i \tau )^2))]/(I_1(2 a \beta ) (24 a^2 \beta  g+\beta)-24 a g I_2(2 a \beta )
%\end{array}\ee

Now for the quartic random potential disconnected part of the Green's function can be computed at finite temperature as:
\bea G_{dc}(\beta,\tau)&=&\frac{\langle Z(\bg+i\tau)\rangle \langle Z(\bg-i\tau)\rangle}{\langle Z(\bg)\rangle^{2}}\nonumber\\
&=&\frac{\beta^4}{(\beta^2+ \tau^2)^2} \frac{1}{\left[\left(24 a^2 g+1\right) \beta I_1(2 a \beta)-24 a g I_2(2 a \beta)\right]^2}\nonumber\\
&&~~~~~~~~~~\times \left[\left(24 a^2 g+1\right) (\beta+ i \tau) I_1(2 a (\beta+ i \tau))-24 a g I_2(2 a (\beta+ i \tau))\right]\nonumber\\
&&~~~~~~~~~~\times \left[\left(24 a^2 g+1\right) (\beta- i \tau) I_1(2 a (\beta- i \tau))-24 a g I_2(2 a (\beta- i \tau))\right],~~~~~~~~~~\eea
which can be further simplified in the high temperature limiting situation as:
\bea G_{dc}(\tau)&=&\left[\frac{\langle Z(\bg+i\tau)\rangle \langle Z(\bg-i\tau)\rangle}{\langle Z(\bg)\rangle^{2}}\right]_{\beta=0}\nonumber\\
&=&\frac{a^2}{N^2\tau^4} \left[\left(24 a^2 g+1\right) ( i \tau) I_1(2 a ( i \tau))-24 a g I_2(2 a ( i \tau))\right]\nonumber\\
&&~~~~~~~~~~\times \left[\left(24 a^2 g+1\right) (- i \tau) I_1(2 a (- i \tau))-24 a g I_2(2 a (- i \tau))\right],~~~~~~~~~~.~~~~~~\eea
Further taking the limit ${\cal T}=\sqrt{N}\tau\rightarrow\infty$ we get the following simplified result:
\bea G_{dc}({\cal T})&=&\left[\frac{\langle Z(\bg+i{\cal T})\rangle \langle Z(\bg-i{\cal T})\rangle}{\langle Z(\bg)\rangle^{2}}\right]_{\beta=0}\nonumber\\
&=&\frac{i}{{\cal T}^{3}}{\frac{a}{N^2\pi}}\left\{\left[\left(24 a^2 g+1\right) \cos \left( 2 a {\cal T} +\frac{\pi }{4}\right)+\frac{24 g \sin \left( 2 a {\cal T}+ \frac{\pi}{4}\right)}{{\cal T}}\right]+O\left(\frac{1}{( {\cal T})^{\frac{7}{2}}}\right)\right\}\nonumber\\
&&~~~~\times \left\{\left[\left(24 a^2 g+1\right) \cos \left(2 a {\cal T} -\frac{\pi }{4}\right)-\frac{24 g \sin \left(2 a {\cal T}- \frac{\pi}{4}\right)}{{\cal T}}\right]+O\left(\frac{1}{(- {\cal T})^{\frac{7}{2}}}\right)\right\}.~~~~~~~~~~\eea
Now to compute  SFF we need to add both connected and disconnected part of the Green's function $G$($=G_{c}+G_{dc}$). Therefore, for quartic polynomial potential we get finally the following expression for SFF at finite temp:
 \be\boxed{\begin{array}{lll}\label{cqs1}
		\displaystyle   {\bf SFF}(\beta,\tau)\equiv \left\{\begin{array}{lll}
			\displaystyle  
			\frac{\beta^4}{(\beta^2+ \tau^2)^2} \frac{1}{\left[\left(24 a^2 g+1\right) \beta I_1(2 a \beta)-24 a g I_2(2 a \beta)\right]^2}\\\displaystyle
~~~~~~~~~~\times \left[\left(24 a^2 g+1\right) (\beta+ i \tau) I_1(2 a (\beta+ i \tau))-24 a g I_2(2 a (\beta+ i \tau))\right]\\\displaystyle
~~~~~~~~~~\times \left[\left(24 a^2 g+1\right) (\beta- i \tau) I_1(2 a (\beta- i \tau))-24 a g I_2(2 a (\beta- i \tau))\right] \\ \displaystyle
~~~~~~~~~~+\frac{\tau}{(2\pi N)^{2}}-\frac{1}{N}+\frac{1}{(\pi N)}\,,~~~~~~~~~~~~ &
			\mbox{\small  \textcolor{red}{\bf  {$\tau<2\pi N$ }}}  \\  \\
			\displaystyle  
			\frac{\beta^4}{(\beta^2+ \tau^2)^2} \frac{1}{\left[\left(24 a^2 g+1\right) \beta I_1(2 a \beta)-24 a g I_2(2 a \beta)\right]^2} \\ \displaystyle
~~~~~~~~~~\times \left[\left(24 a^2 g+1\right) (\beta+ i \tau) I_1(2 a (\beta+ i \tau))-24 a g I_2(2 a (\beta+ i \tau))\right]\\\displaystyle
~~~~~~~~~~\times \left[\left(24 a^2 g+1\right) (\beta- i \tau) I_1(2 a (\beta- i \tau))-24 a g I_2(2 a (\beta- i \tau))\right]\\ \displaystyle
~~~~~~~~~~+\frac{1}{\pi N}\,,~~~~~~~~~~~~ &
			\mbox{\small  \textcolor{red}{\bf  {$\tau>2\pi N$}}}  
		\end{array}
		\right.
	\end{array}}~~,\ee
where ${\bf SFF}(\beta,\tau)$ is defined with proper normalization and in our prescription it gives the total Green's function as mentioned above. 
Further simplifying the result for high temperature limit we get the following expression for SFF, as given by:
\be\boxed{\begin{array}{lll}\label{cqs2}
		\displaystyle   {\bf SFF}(\tau)\equiv \left\{\begin{array}{lll}
			\displaystyle  
			\frac{a^2}{N^2\tau^4} \left[\left(24 a^2 g+1\right) ( i \tau) I_1(2 a ( i \tau))-24 a g I_2(2 a ( i \tau))\right]\\ \displaystyle 
~~~~~~~~~~\times \left[\left(24 a^2 g+1\right) (- i \tau) I_1(2 a (- i \tau))-24 a g I_2(2 a (- i \tau))\right]\\ \displaystyle 
~~~~~~~~~~+\frac{\tau}{(2\pi N)^{2}}-\frac{1}{N}+\frac{1}{(\pi N)}\,,~~~~~~~~~~~~ &
			\mbox{\small  \textcolor{red}{\bf  {$\tau<2\pi N$ }}}  \\ \\
			\displaystyle  
			\frac{a^2}{N^2\tau^4} \left[\left(24 a^2 g+1\right) ( i \tau) I_1(2 a ( i \tau))-24 a g I_2(2 a ( i \tau))\right]\\ \displaystyle 
~~~~~~~~~~\times \left[\left(24 a^2 g+1\right) (- i \tau) I_1(2 a (- i \tau))-24 a g I_2(2 a (- i \tau))\right]\\ \displaystyle 
~~~~~~~~~~+\frac{1}{\pi N}\,,~~~~~~~~~~~~ &
			\mbox{\small  \textcolor{red}{\bf  {$\tau>2\pi N$}}}  
		\end{array}
		\right.
	\end{array}}~~,\ee
	
	Further taking the limit ${\cal T}=\sqrt{N}\tau\rightarrow\infty$ we get the following simplified result for SFF:
\bea\boxed{\begin{array}{lll}\label{cqs3}
		\small\displaystyle   {\bf SFF}({\cal T})\equiv \left\{\begin{array}{lll}
			\displaystyle  
			\frac{i}{{\cal T}^{3}}{\frac{a}{N^2\pi}}\left\{\left[\left(24 a^2 g+1\right) \cos \left( 2 a {\cal T} +\frac{\pi }{4}\right)+\frac{24 g \sin \left( 2 a {\cal T}+ \frac{\pi}{4}\right)}{{\cal T}}\right]+O\left(\frac{1}{( {\cal T})^{\frac{7}{2}}}\right)\right\}\\
\displaystyle~~~~\times \left\{\left[\left(24 a^2 g+1\right) \cos \left(2 a {\cal T} -\frac{\pi }{4}\right)-\frac{24 g \sin \left(2 a {\cal T}- \frac{\pi}{4}\right)}{{\cal T}}\right]+O\left(\frac{1}{(- {\cal T})^{\frac{7}{2}}}\right)\right\}\\
\displaystyle~~~~~~~~~ +\frac{{\cal T}}{(2\pi)^2 N^{5/2}}-\frac{1}{N}+\frac{1}{(\pi N)}\,,~~~~~~~~~~~~ &
			\mbox{\small  \textcolor{red}{\bf  {${\cal T}<2\pi N^{3/2}$ }}}  \\  \\ \\
			\displaystyle  
			\frac{i}{{\cal T}^{3}}{\frac{a}{N^2\pi}}\left\{\left[\left(24 a^2 g+1\right) \cos \left( 2 a {\cal T} +\frac{\pi }{4}\right)+\frac{24 g \sin \left( 2 a {\cal T}+ \frac{\pi}{4}\right)}{{\cal T}}\right]+O\left(\frac{1}{( {\cal T})^{\frac{7}{2}}}\right)\right\}\\
\displaystyle~~~~\times \left\{\left[\left(24 a^2 g+1\right) \cos \left(2 a {\cal T} -\frac{\pi }{4}\right)-\frac{24 g \sin \left(2 a {\cal T}- \frac{\pi}{4}\right)}{{\cal T}}\right]+O\left(\frac{1}{(- {\cal T})^{\frac{7}{2}}}\right)\right\}\\
\displaystyle~~~~~~~~~ +\frac{1}{\pi N}\,,~~~~~~~~~~~~ &
			\mbox{\small  \textcolor{red}{\bf  {${\cal T}>2\pi N^{3/2}$}}}  
		\end{array}
		\right.
	\end{array}}\nonumber\\
	&& \eea

Further simplifying the result for high temperature limit we get the following expression for SFF, as given by:
\be\boxed{\begin{array}{lll}\label{cqs2}
		\displaystyle   {\bf SFF}(\tau)\equiv \left\{\begin{array}{lll}
			\displaystyle  
			\frac{a^2}{N^2\tau^4} \left[\left(24 a^2 g+1\right) ( i \tau) I_1(2 a ( i \tau))-24 a g I_2(2 a ( i \tau))\right]\\ \displaystyle 
~~~~~~~~~~\times \left[\left(24 a^2 g+1\right) (- i \tau) I_1(2 a (- i \tau))-24 a g I_2(2 a (- i \tau))\right]\\ \displaystyle 
~~~~~~~~~~+\frac{\tau}{(2\pi N)^{2}}-\frac{1}{N}+\frac{1}{(\pi N)}\,,~~~~~~~~~~~~ &
			\mbox{\small  \textcolor{red}{\bf  {$\tau<2\pi N$ }}}  \\  \\ \\
			\displaystyle  
			\frac{a^2}{N^2\tau^4} \left[\left(24 a^2 g+1\right) ( i \tau) I_1(2 a ( i \tau))-24 a g I_2(2 a ( i \tau))\right]\\ \displaystyle 
~~~~~~~~~~\times \left[\left(24 a^2 g+1\right) (- i \tau) I_1(2 a (- i \tau))-24 a g I_2(2 a (- i \tau))\right]\\ \displaystyle 
~~~~~~~~~~+\frac{1}{\pi N}\,,~~~~~~~~~~~~ &
			\mbox{\small  \textcolor{red}{\bf  {$\tau>2\pi N$}}}  
		\end{array}
		\right.
	\end{array}}~~,\ee

	\begin{figure}[htb]
\centering
\subfigure[SFF for quartic potential at $\bg=10$.]{
    \includegraphics[width=7.8cm,height=8cm] {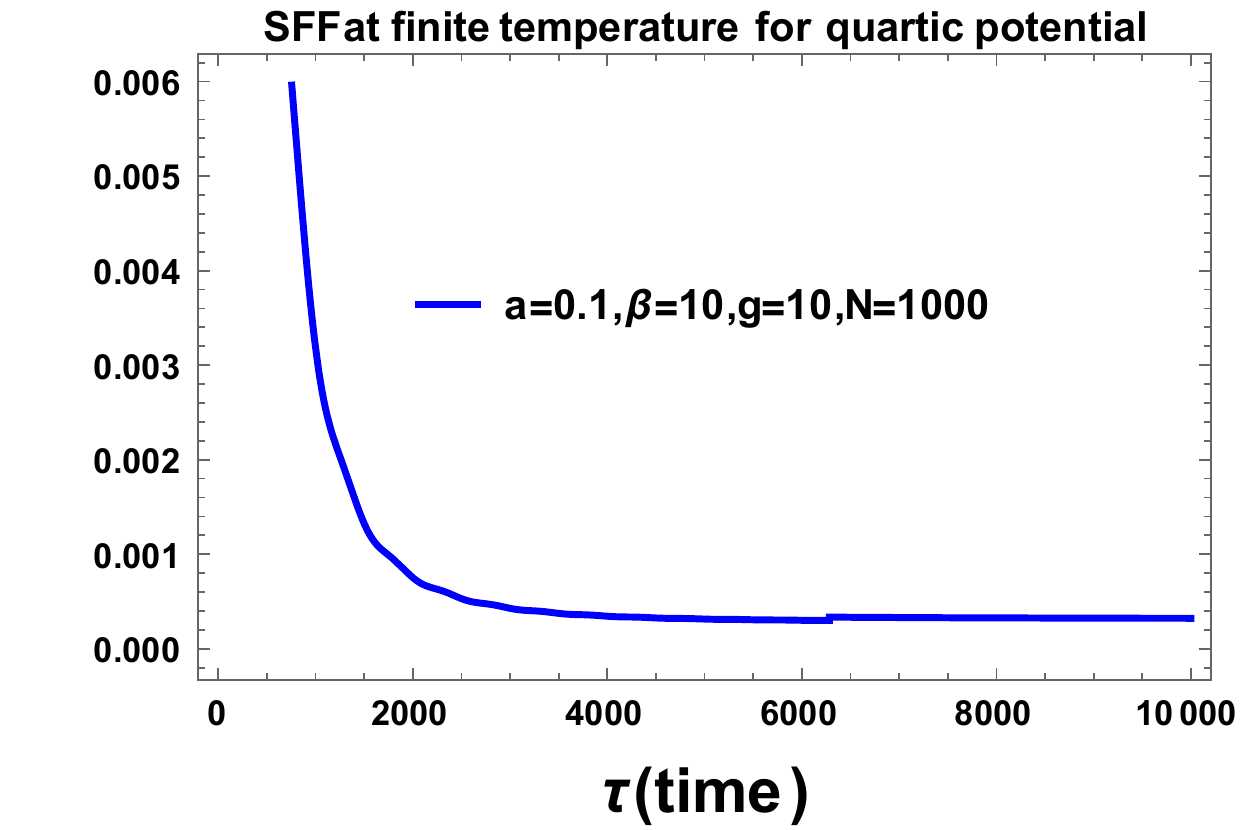}
    \label{quaf1}
}
\subfigure[SFF for quartic potential at $\bg=100$.]{
    \includegraphics[width=7.8cm,height=8cm] {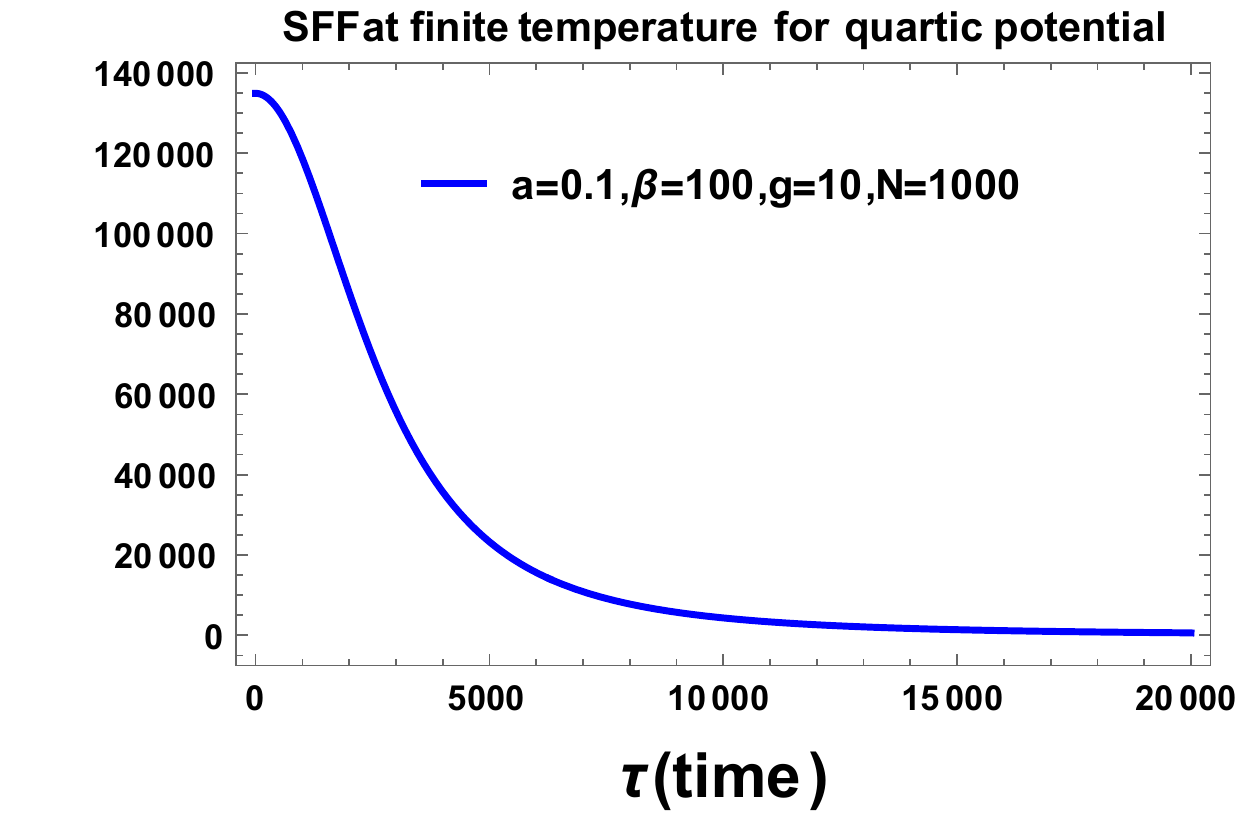}
    \label{quaf2}
}
%\subfigure[SFF for quartic potential at $\bg=200$.]{
%    \includegraphics[width=7.8cm,height=9.2cm] {SFF_B200.pdf}
 %   \label{quaf3}
%}
%\subfigure[SFF for quartic potential at $\bg=1000$.]{
%    \includegraphics[width=7.8cm,height=9.2cm] {SFF_B1000.pdf}
  %  \label{quaf4}
%}
\caption{Spectral Form Factor for quartic  potential at different finite temperature[$\bg$] with $N=1000$ and $a=0.1$ }
\end{figure}
From fig.~\ref{quaf1} and fig.~\ref{quaf2}, we see that SFF at finite temperature decays with increasing $\tau$ and reach zero. But with changing $\bg$ SFF values remains almost same initially (for higher $\bg$ or lower value of temperature).

%\begin{figure}[H]
%\centering
%\subfigure[SFF for quartic potential at $N=10$.]{
%    \includegraphics[width=7.8cm,height=7.8cm] {SFF_Q_N10.pdf}
%    \label{quax1}
%}
%\subfigure[SFF for quartic potential at $N=100$.]{
 %   \includegraphics[width=7.8cm,height=7.8cm] {SFF_Q_N100.pdf}
 %   \label{quax2}
%}
%\subfigure[SFF for quartic potential at $N=1000$.]{
 %   \includegraphics[width=7.8cm,height=7.8cm] {SFF_Q_N1000.pdf}
 %   \label{quax3}
%}
%\subfigure[SFF for quartic potential at $N=10000$.]{
  %  \includegraphics[width=7.8cm,height=7.8cm] {SFF_Q_N10000.pdf}
 %   \label{quax4}
%}
%\caption{Spectral Form Factor for quartic  potential varying with different N at finite temperature[$\bg=100 $] and $a=0.1$ }
%\end{figure}
%From fig.~\ref{quax1}, fig.~\ref{quax2}, fig.~\ref{quax3} and fig.~\ref{quax4},  we see that SFF at finite temperature decays with increasing $\tau$ and reach zero. But at different $N$ SFF values remains same initially.

From both the figures we have shown that SFF decays to zero for finite temperature.
\begin{figure}[htb]
\centering
\subfigure[SFF for gaussian for $a=.1,N=100$ with $SFF|_{\tau=0}= 0.0068169$ as origin]{
    \includegraphics[width=7.8cm,height=8cm] {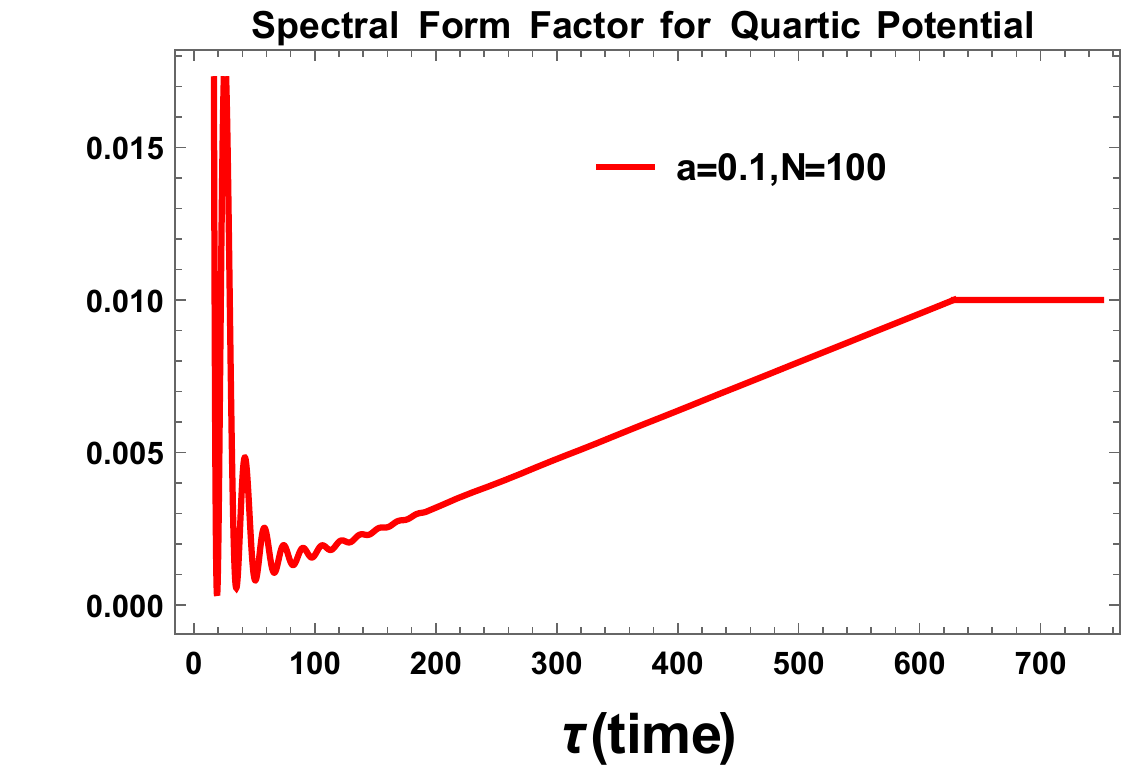}
    \label{SFFG31a}
}
\subfigure[SFF for gaussian for $a=.1,N=1000$ with  $SFF|_{\tau=0}=0.00068169$ as origin]{
    \includegraphics[width=7.8cm,height=8cm] {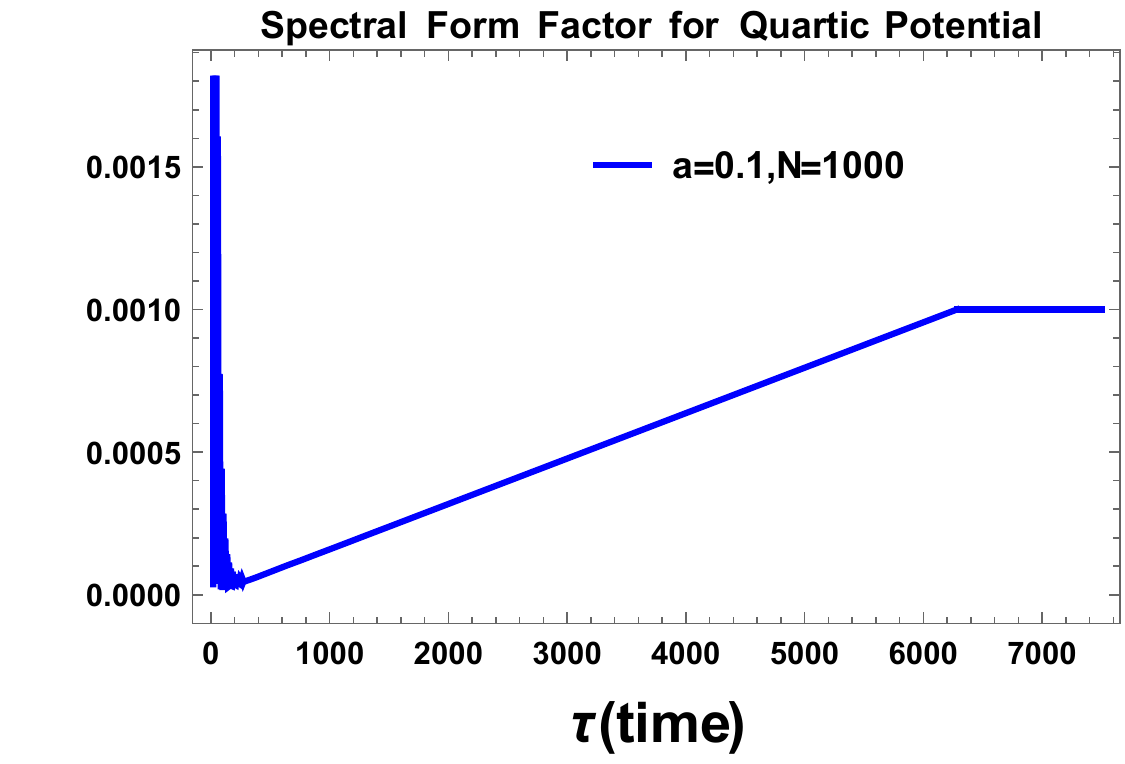}
    \label{SFFG33a}
}
\subfigure[SFF for gaussian for $a=.1,N=10000$ with  $SFF|_{\tau=0}=0.000068169$ as origin]{
    \includegraphics[width=10.8cm,height=8cm] {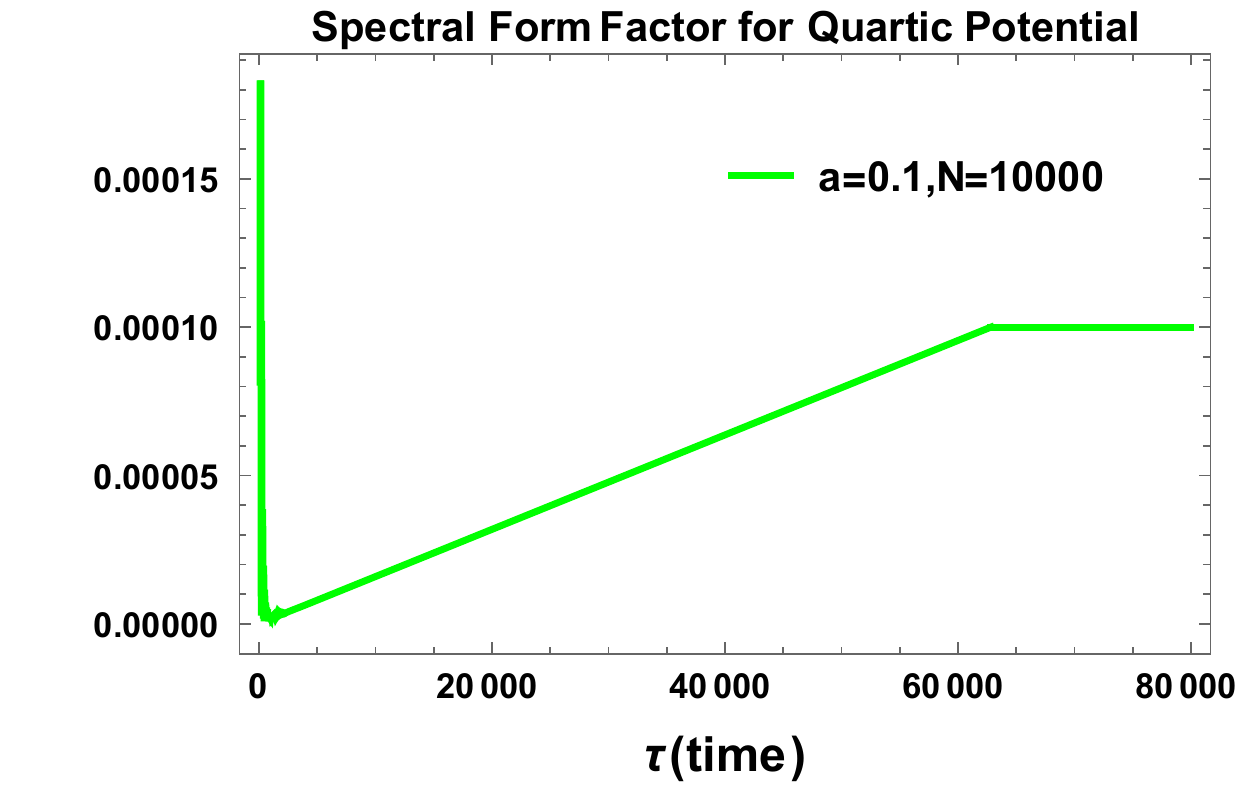}
    \label{SFFG32a}
}
\caption{Time variation of SFF for different N. Here we  shift  reference axis[SFF]  to $SFF|_{\tau=0}$ }
\label{quar223}
\end{figure} 
In fig.~\ref{SFFG31a}, fig.~\ref{SFFG33a} and fig.~\ref{SFFG32a}, it is observed that SFF with variation in $N$ get saturated at different value of $\tau$. But with increasing $N$ the value of the saturation point, will decrease.
Subtracting the change of axis[$SFF|_{\tau=0}$] we get the predicted bound of SFF.
\subsubsection{For Sextic random potential}
In this subsection we consider sextic random potential, as given by the following expression:
\bea V(M)=\frac{1}{2}M^{2}+g M^{4}+h M^{6}.
\eea
For a single interval ($n=1$) with end points -2a and 2a (semi-circle) we get the following expression for the density function in terms of the eigen value of random matrix $M$:
\bea \rho(\lb)=\frac{1}{\pi}~\sqrt{4 a^2-\lambda ^2}~ \left(a_2 \lambda ^4+a_1 \lambda ^2+a_0\right).
\eea
Also for sextic potential $\og(\lb+i0)$ can be expressed as:
\bea \og(\lb+i0)=\frac{1}{2} \left(4 g \lambda ^3+6 h \lambda ^5+\lambda \right)+i \sqrt{4 a^2-\lambda ^2} \left(a_2 \lambda ^4+a_1 \lambda ^2+a_0\right).
\eea
\begin{figure}[htb]
\centering
\subfigure[$\rho(\lb)$ for sextic potential for different $g, h$.]{
    \includegraphics[width=7.8cm,height=9cm] {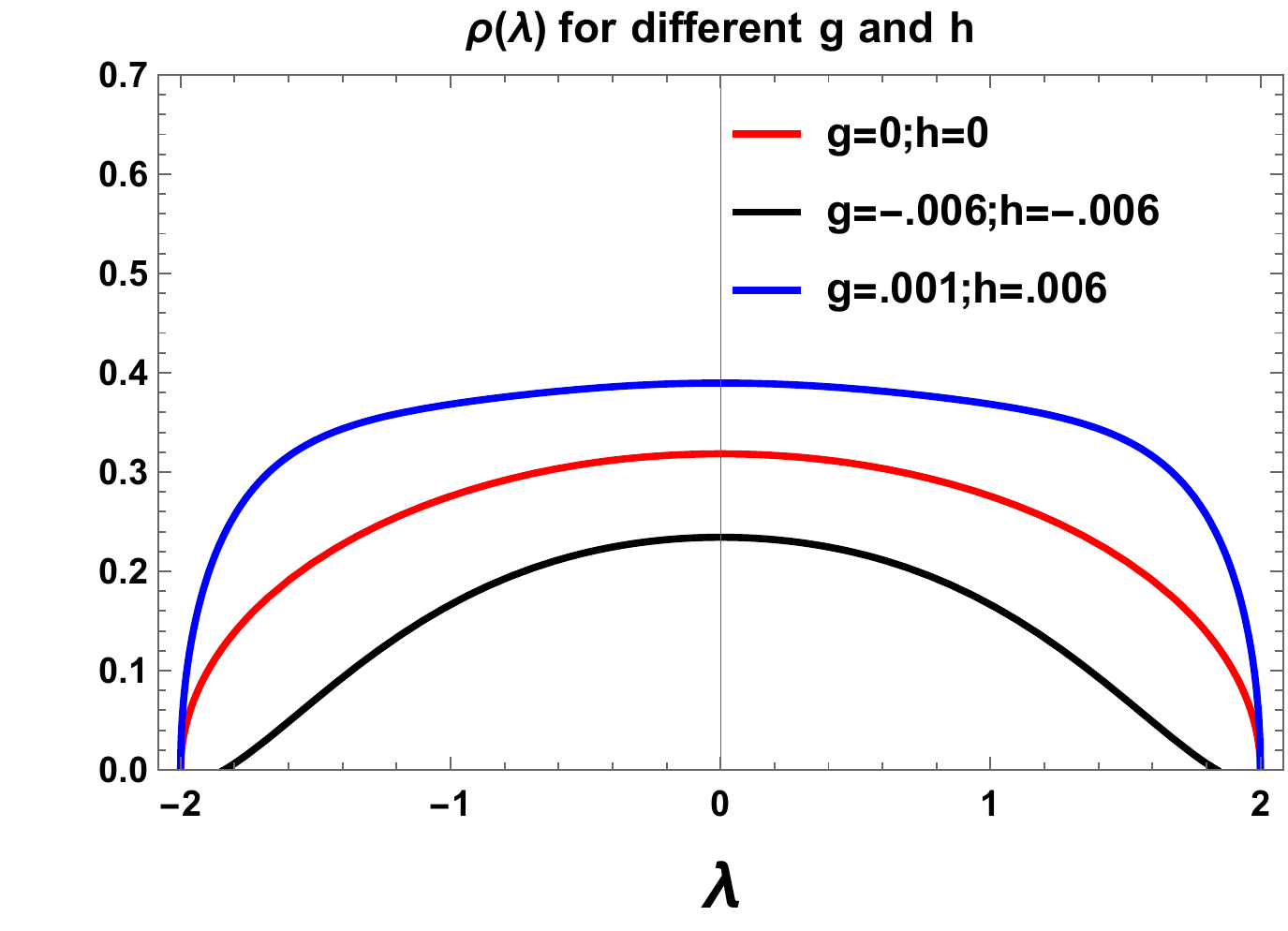}
    \label{Rho32}
}
\subfigure[$\rho(\lb)$ for sextic potential for different $g, h$.]{
    \includegraphics[width=7.8cm,height=9cm] {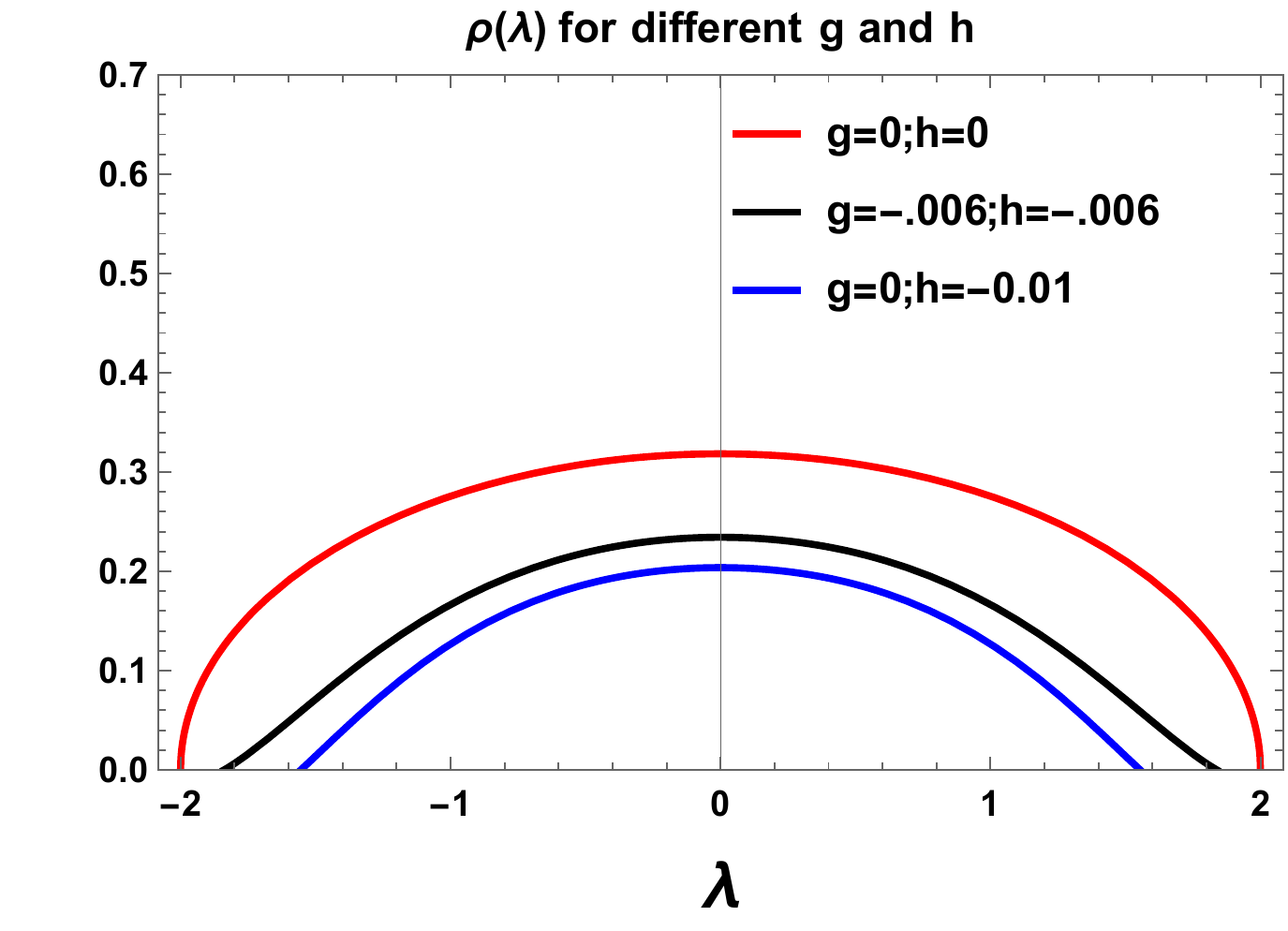}
    \label{Rho33}
}
\caption{Eigen value distribution curve of density function for  sextic potential for different parameter values. Here we fix $a=1$. }
\end{figure}
In fig.~\ref{Rho32} and fig.~\ref{Rho33} for sextic potential behaviour of density function $\rho(\lb)$ is shown. The curve follows from Eq.~(\ref{sextic11}). Again choosing $g=h=0$ will produce the {\it Wigner law}. Deviating $g$ and $h$ by small amount shows deviation from {\it Wigner semicircle law}. For $g>0,h>0$ the curve shows plateau region though merge with semicircle at end points. But choosing $g<0,h<0$ and $g=0$ and $h<0$ show deviation from semicircle and don't converge even at end points.

Further expanding $\og(\lb+i0)$ near $\lb\rightarrow\infty$ we get:
\bea
\lambda ^3 \left(2 a_2 a^2-a_1+2 g\right)&&+\left(2 a_2 a^4+2 a_1 a^2-a_0+\frac{1}{2}\right) \lambda\nonumber\\
&&~~~~~~~+\frac{4 a_2 a^6+2 a_1 a^4+2 a_0 a^2}{\lambda }+\lambda ^5 \left(3 h-a_2\right)+O\left(\left(\frac{1}{\lb}\right)^{3}\right)=\frac{1}{\lb}.~~~~~~~~~~~
\eea
Therefore, equating both the sides of the above equation we get:
\bea a_{2}&=&3h,\\
a_{1}&=&2g+6a^{2}h,\\
a_{0}&=&18a^{4}h+4a^{2}g+\frac{1}{2}.\eea
along with we get one additional constraint condition, as given by:
\bc
 $60a^{6}h+12 g a^4+ a^{2}=1$ 
\ec
Then, for the sextic random potential we get the following simplified expression for the density function in terms of the eigen value of the random matrix $M$, as given by:
\bea  \label{sextic11}
\rho(\lb)=\frac{1}{\pi}\sqrt{4 a^2-\lambda ^2}~ \left(18 a^4 h+\lambda ^2 \left(6 a^2 h+2 g\right)+4 a^2 g+3 h \lambda ^4+\frac{1}{2}\right).
\eea
Solving the constraint condition we get, $ a^{2}$ in terms of $g$ and $h$. The real root for $a^2$ is given by the following expression:
\bea a^2=\frac{{\cal F}(g,h)}{30 h}-\frac{180 h-144 g^2}{1080 h {\cal F}(g,h)}-\frac{g}{15 h}.
\eea
where we define the function ${\cal F}(g,h)$  as:
\bea {\cal F}(g,h)=\sqrt[3]{-8 g^3+5 \sqrt{-144 g^3 h^2-3 g^2 h^2+270 g h^3+2025 h^4+5 h^3}+15 g h+225 h^2}.~~~~~~~~~~\eea
Here we can check that putting $h=0$ the constraint condition reduces to the following simplified form:
\be
 12 g a^4+ a^{2}=1 
\ee
and the solution of this equation is given by the following expression:
\be  \boxed{a^{2}=\frac{\sqrt{48 g+1}-1}{24 g}}.\ee 
Here the critical value with $h=0$ is given by:
\be \boxed{g_{c}=-\frac{1}{48}}~,\ee 
which is exactly same result as obtained for quartic potential in the previous subsection.

Now the expression for the one point function for partition function at finite temperature can be computed as:
\bea
\langle Z(\bg\pm i\tau)\rangle &=&\frac{1}{\pi}\int_{-2 a}^{2 a} d\lb~\sqrt{4 a^2-\lambda ^2} \left(18 a^4 h+\lambda ^2 \left(6 a^2 h+2 g\right)+4 a^2 g+3 h \lambda ^4+\frac{1}{2}\right)~e^{\mp i\tau \lambda}~e^{-\beta\lb}\nonumber\\
&=&\frac{a }{(\beta\pm i \tau)^4}\left[(\beta\pm i \tau) I_1(2 a (\beta\pm i \tau))\right.\nonumber \\&& \left. \left(360 a^2 h+\beta^2 \left(180 a^4 h+24 a^2 g+1\right)\pm 2 i \beta \tau \left(180 a^4 h+24 a^2 g+1\right)\right.\right.\nonumber \\&& \left.\left.-\tau^2 \left(180 a^4 h+24 a^2 g+1\right)\right)\right.\nonumber \\&& \left.-24 a I_2(2 a (\beta\pm i \tau)) \left(30 h+(\beta\pm i \tau)^2 \left(15 a^2 h+g\right)\right)\right]
\eea
Further in the high temperature limit the one point function for partition function can be simplified as:
\bea
\left[\langle Z(\bg\pm it)\rangle\right]_{\bg=0}&=&\frac{1}{\pi}\int_{-2 a}^{2 a} d\lb~\sqrt{4 a^2-\lambda ^2} \left(18 a^4 h+\lambda ^2 \left(6 a^2 h+2 g\right)+4 a^2 g+3 h \lambda ^4+\frac{1}{2}\right)~e^{\mp i\tau \lambda}\nonumber\\
&=&\frac{a}{\tau ^4} \left[(J_1(\pm 2 a \tau )((\pm\tau) ^3(180 a^4 h+24 a^2 g+1)\mp 360 a^2 h \tau)\right. \nonumber\\&&\left.
\displaystyle~~~~~~~~~~~~~~~~~~~~~~ -24  J_2(\pm 2 a \tau ) (\tau ^2 (15 a^2 h+g)-30 h))\right]
\eea
Next, simplifying the result for one point function in the limit ${\cal T}=\sqrt{N}\tau\rightarrow\infty$ we get:
\bea\label{tsextic}\left[\langle Z(\bg\pm i{\cal T})\rangle\right]_{\bg=0}&=&\sqrt{\frac{a}{\pi}}\frac{1}{(\pm {\cal T})^{\frac{3}{2}}}\left[-\left(1+24 a^{2}g+180 a^{4}h\right)\cos\left(\frac{\pi}{4}\pm 2a{\cal T}\right)\right. \nonumber\\&& \left.~~~~~~~~\pm 24a\frac{g+15a^{2}h}{{\cal T}}\sin\left(\frac{\pi}{4}\pm 2a{\cal T}\right)\right. \nonumber\\&& \left.~~~~~~~~
+\frac{360a^{2}h}{{\cal T}^{2}}\cos\left(\frac{\pi}{4}\pm 2a{\cal T}\right)+O\left(\frac{1}{{\cal T}^{4}}\right)\right]\eea
Now for the quadratic random potential disconnected part of the Green's function can be computed at finite temperature as:
\bea G_{dc}(\beta,\tau)&=&\frac{\langle Z(\bg+i\tau)\rangle \langle Z(\bg-i\tau)\rangle}{\langle Z(\bg)\rangle^{2}}\nonumber\\
&=&\frac{\beta^8 }{(\beta^2+\tau^2)^4}\left[\beta I_1(2 a \beta)\left(360 a^2 h+\beta^2 \left(180 a^4 h+24 a^2 g+1\right)\right)\right.\nonumber \\&& \left.-24 a I_2(2 a \beta) \left(30 h+\beta^2 \left(15 a^2 h+g\right)\right)\right]^{-2}\nonumber\\
&& \times \left[(\beta+ i \tau) I_1(2 a (\beta+ i \tau))\right.\nonumber \\&& \left. \left(360 a^2 h+\beta^2 \left(180 a^4 h+24 a^2 g+1\right)+ 2 i \beta \tau \left(180 a^4 h+24 a^2 g+1\right)\right.\right.\nonumber \\&& \left.\left.-\tau^2 \left(180 a^4 h+24 a^2 g+1\right)\right)\right.\nonumber \\&& \left.-24 a I_2(2 a (\beta+ i \tau)) \left(30 h+(\beta+ i \tau)^2 \left(15 a^2 h+g\right)\right)\right]\nonumber\\
&&~~~~\times \left[(\beta- i \tau) I_1(2 a (\beta- i \tau))\right.\nonumber \\&& \left. \left(360 a^2 h+\beta^2 \left(180 a^4 h+24 a^2 g+1\right)- 2 i \beta \tau \left(180 a^4 h+24 a^2 g+1\right)\right.\right.\nonumber \\&& \left.\left.-\tau^2 \left(180 a^4 h+24 a^2 g+1\right)\right)\right.\nonumber \\&& \left.-24 a I_2(2 a (\beta- i \tau)) \left(30 h+(\beta- i \tau)^2 \left(15 a^2 h+g\right)\right)\right],~~~~~~\eea
which can be further simplified in the high temperature limiting situation as:
\bea G_{dc}(\tau)&=&\left[\frac{\langle Z(\bg+i\tau)\rangle \langle Z(\bg-i\tau)\rangle}{\langle Z(\bg)\rangle^{2}}\right]_{\beta=0}\nonumber\\
&=&\frac{a^2}{N^2\tau^{8}}\left[(J_1( 2 a \tau )(\tau ^3(180 a^4 h+24 a^2 g+1)- 360 a^2 h \tau)\right. \nonumber\\&&\left.
\displaystyle~~~~~~~~~~~~~~~~~~~~~~ -24  J_2( 2 a \tau ) (\tau ^2 (15 a^2 h+g)-30 h))\right]\nonumber\\
&&\times \left[(J_1(- 2 a \tau )((-\tau) ^3(180 a^4 h+24 a^2 g+1)+ 360 a^2 h \tau)\right. \nonumber\\&&\left.
\displaystyle~~~~~~~~~~~~~~~~~~~~~~ -24  J_2(- 2 a \tau ) (\tau ^2 (15 a^2 h+g)-30 h))\right].~~~~~~\eea
Further taking the limit ${\cal T}=\sqrt{N}\tau\rightarrow\infty$ we get the following simplified result:
\bea G_{dc}({\cal T})&=&\left[\frac{\langle Z(\bg+i{\cal T})\rangle \langle Z(\bg-i{\cal T})\rangle}{\langle Z(\bg)\rangle^{2}}\right]_{\beta=0}\nonumber\\
&=&\frac{i}{{\cal T}^3}\frac{a}{N^2\pi}\left[-\left(1+24 a^{2}g+180 a^{4}h\right)\cos\left(\frac{\pi}{4}+ 2a{\cal T}\right)\right. \nonumber\\&& \left.~~~~~~~~+ 24a\frac{g+15a^{2}h}{{\cal T}}\sin\left(\frac{\pi}{4}+ 2a{\cal T}\right)
+\frac{360a^{2}h}{{\cal T}^{2}}\cos\left(\frac{\pi}{4}+ 2a{\cal T}\right)+O\left(\frac{1}{{\cal T}^{4}}\right)\right]\nonumber\\
&&\times \left[-\left(1+24 a^{2}g+180 a^{4}h\right)\cos\left(\frac{\pi}{4}- 2a{\cal T}\right)\right. \nonumber\\&& \left.~~~~~~~~- 24a\frac{g+15a^{2}h}{{\cal T}}\sin\left(\frac{\pi}{4}- 2a{\cal T}\right)
+\frac{360a^{2}h}{{\cal T}^{2}}\cos\left(\frac{\pi}{4}- 2a{\cal T}\right)+O\left(\frac{1}{{\cal T}^{4}}\right)\right].~~~~~~~~~~\eea
Now to compute  SFF we need to add both connected and disconnected part of the Green's function $G$($=G_{c}+G_{dc}$). Therefore, for sextic polynomial potential we get finally the following expression for SFF at finite temp:
 \bea\boxed{\begin{array}{lll}\label{cqv1}
		  \footnotesize {\bf SFF}(\beta,\tau)&\equiv &\displaystyle\frac{\beta^8 }{(\beta^2+\tau^2)^4}\left[\beta I_1(2 a \beta)\left(360 a^2 h+\beta^2 \left(180 a^4 h+24 a^2 g+1\right)\right)\right. \\&& \left.-24 a I_2(2 a \beta) \left(30 h+\beta^2 \left(15 a^2 h+g\right)\right)\right]^{-2}\left[(\beta+ i \tau) I_1(2 a (\beta+ i \tau))\right. \\&& \left. \left(360 a^2 h+\beta^2 \left(180 a^4 h+24 a^2 g+1\right)+ 2 i \beta \tau \left(180 a^4 h+24 a^2 g+1\right)\right.\right. \\&& \left.\left.-\tau^2 \left(180 a^4 h+24 a^2 g+1\right)\right)-24 a I_2(2 a (\beta+ i \tau)) \left(30 h+(\beta+ i \tau)^2 \left(15 a^2 h+g\right)\right)\right]\\
&&~~~~\times \left[(\beta- i \tau) I_1(2 a (\beta- i \tau))\right. \\&& \left. \left(360 a^2 h+\beta^2 \left(180 a^4 h+24 a^2 g+1\right)- 2 i \beta \tau \left(180 a^4 h+24 a^2 g+1\right)\right.\right. \\&& \left.\left.-\tau^2 \left(180 a^4 h+24 a^2 g+1\right)\right)-24 a I_2(2 a (\beta- i \tau)) \left(30 h+(\beta- i \tau)^2 \left(15 a^2 h+g\right)\right)\right]\\
&&~~~+ \left\{\begin{array}{lll}
			\displaystyle  
			\frac{\tau}{(2\pi N)^{2}}-\frac{1}{N}+\frac{1}{(\pi N)}\,,~~~~~~~~~~~~ &
			\mbox{\small  \textcolor{red}{\bf  {$\tau<2\pi N$ }}}  \\ 
			\displaystyle  
			\frac{1}{\pi N}\,,~~~~~~~~~~~~ &
			\mbox{\small  \textcolor{red}{\bf  {$\tau>2\pi N$}}}  
		\end{array}
		\right.
	\end{array}}~~~~~~~\eea
where ${\bf SFF}(\beta,\tau)$ is defined with proper normalization and in our prescription it gives the total Green's function as mentioned above.

Further simplifying the result for high temperature limit we get the following expression for SFF, as given by:
\bea\boxed{\begin{array}{lll}\label{cqw2}
		   {\bf SFF}(\tau)&\equiv&\displaystyle \frac{a^2}{N^2\tau^{8}}\left[(J_1( 2 a \tau )(\tau ^3(180 a^4 h+24 a^2 g+1)- 360 a^2 h \tau)\right. \\&&\left.
\displaystyle~~~~~~~~~~~~~~~~~~~~~~ -24  J_2( 2 a \tau ) (\tau ^2 (15 a^2 h+g)-30 h))\right]\\
&&\times \left[(J_1(- 2 a \tau )((-\tau) ^3(180 a^4 h+24 a^2 g+1)+ 360 a^2 h \tau)\right. \\&&\left.
\displaystyle~~~~~~~~~~~~~~~~~~~~~~ -24  J_2(- 2 a \tau ) (\tau ^2 (15 a^2 h+g)-30 h))\right]\\
&&+\left\{\begin{array}{lll}
			\displaystyle  
			\frac{\tau}{(2\pi N)^{2}}-\frac{1}{N}+\frac{1}{(\pi N)}\,,~~~~~~~~~~~~ &
			\mbox{\small  \textcolor{red}{\bf  {$\tau<2\pi N$ }}}  \\ 
			\displaystyle  
			\frac{1}{\pi N}\,,~~~~~~~~~~~~ &
			\mbox{\small  \textcolor{red}{\bf  {$\tau>2\pi N$}}}  
		\end{array}
		\right.
	\end{array}}~~~~~~~~\eea
	Further taking the limit ${\cal T}=\sqrt{N}\tau\rightarrow\infty$ we get the following simplified result for SFF:
\bea\boxed{\begin{array}{lll}\label{cqw3}
	   {\bf SFF}({\cal T})&\equiv & \displaystyle \frac{i}{{\cal T}^3}\frac{a}{N^2\pi}\left[-\left(1+24 a^{2}g+180 a^{4}h\right)\cos\left(\frac{\pi}{4}+ 2a{\cal T}\right)\right. \\&& \left.~~~~~~~~+ 24a\frac{g+15a^{2}h}{{\cal T}}\sin\left(\frac{\pi}{4}+ 2a{\cal T}\right)
+\frac{360a^{2}h}{{\cal T}^{2}}\cos\left(\frac{\pi}{4}+ 2a{\cal T}\right)+O\left(\frac{1}{{\cal T}^{4}}\right)\right]\\
&&\times \left[-\left(1+24 a^{2}g+180 a^{4}h\right)\cos\left(\frac{\pi}{4}- 2a{\cal T}\right)\right. \\&& \left.~~~~~~~~- 24a\frac{g+15a^{2}h}{{\cal T}}\sin\left(\frac{\pi}{4}- 2a{\cal T}\right)
+\frac{360a^{2}h}{{\cal T}^{2}}\cos\left(\frac{\pi}{4}- 2a{\cal T}\right)+O\left(\frac{1}{{\cal T}^{4}}\right)\right]\\
&&+\left\{\begin{array}{lll}
			\displaystyle  
			\frac{{\cal T}}{(2\pi)^2 N^{5/2}}-\frac{1}{N}+\frac{1}{(\pi N)}\,,~~~~~~~~~~~~ &
			\mbox{\small  \textcolor{red}{\bf  {${\cal T}<2\pi N^{3/2}$ }}}  \\ 
			\displaystyle  
			\frac{1}{\pi N}\,,~~~~~~~~~~~~ &
			\mbox{\small  \textcolor{red}{\bf  {${\cal T}>2\pi N^{3/2}$}}}  
		\end{array}
		\right.
	\end{array}}~~~~~\label{boundsextic3}\eea
	\begin{figure}[htb]
\centering
\subfigure[SFF for sextic potential at $\bg=10$.]{
    \includegraphics[width=7.8cm,height=8cm] {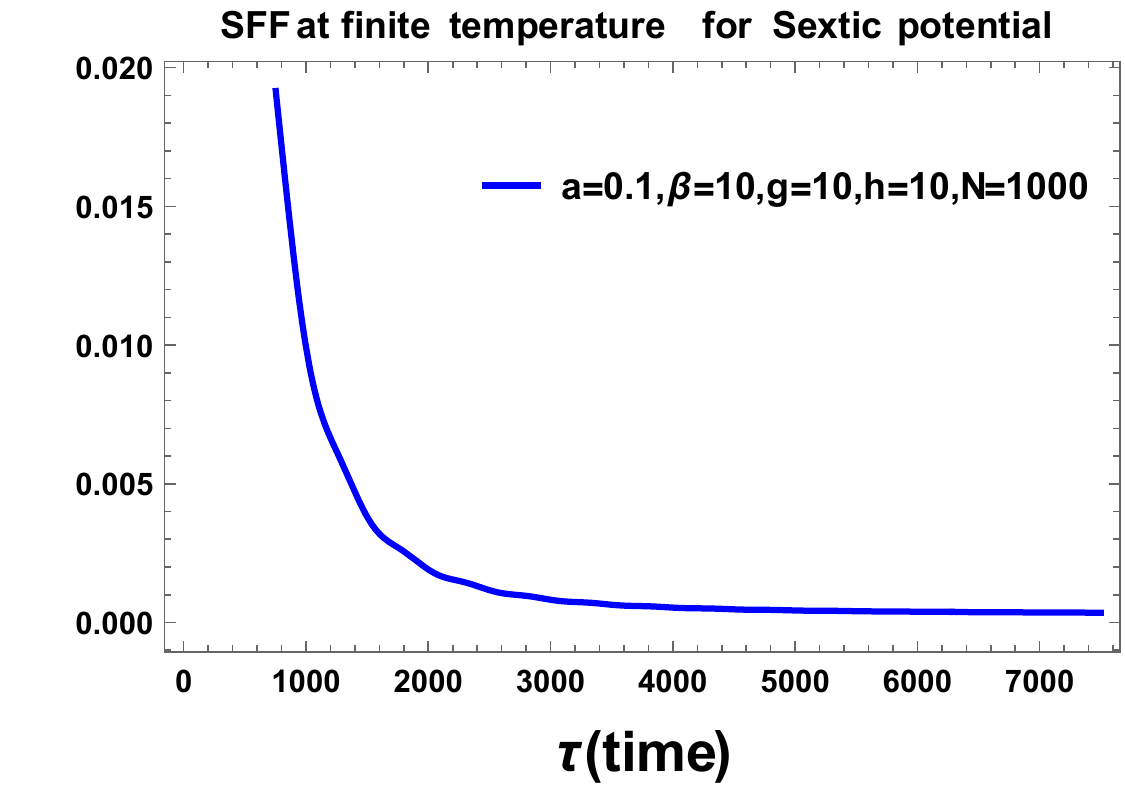}
    \label{ssua1x}
}
\subfigure[SFF for sextic potential at $\bg=100$.]{
    \includegraphics[width=7.8cm,height=8cm] {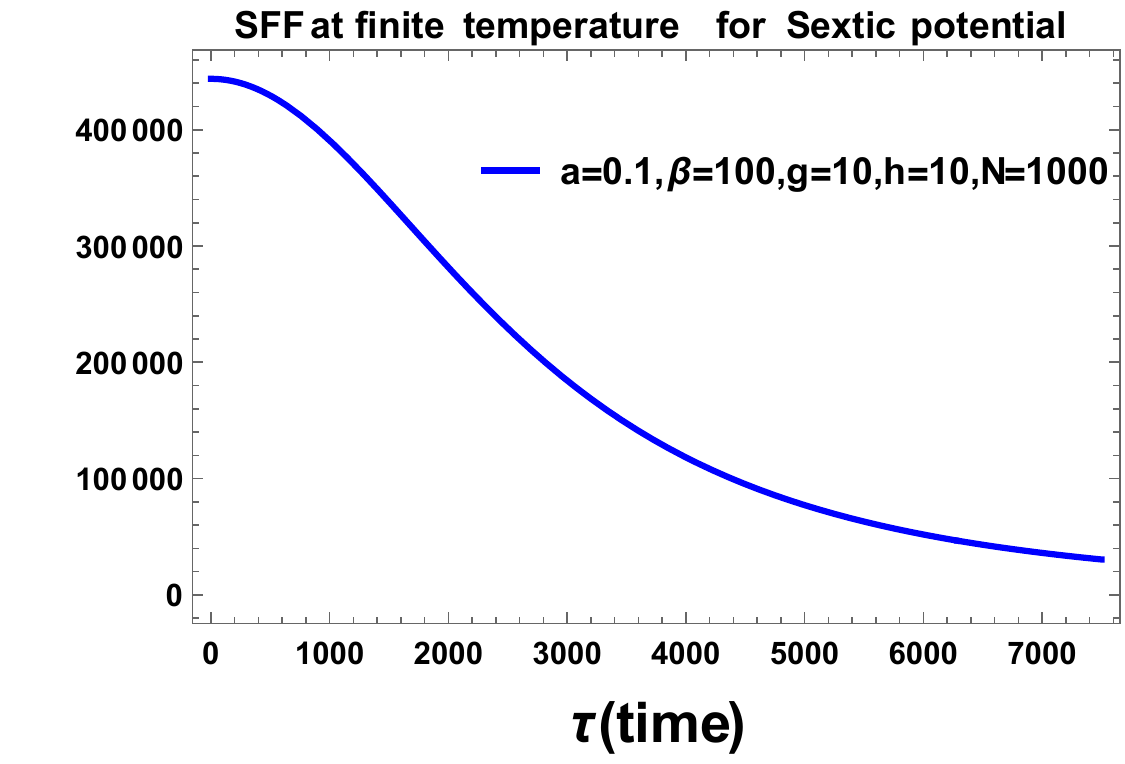}
    \label{ssua2x}
}
%\subfigure[SFF forsextic potential at $\bg=200$.]{
  %  \includegraphics[width=7.7cm,height=6cm] {SFF_S_B200.pdf}
  %  \label{ssua3x}
%}
%\subfigure[SFF for sextic potential at $\bg=1000$.]{
%    \includegraphics[width=7.7cm,height=6cm] {SFF_S_B1000.pdf}
%    \label{ssua4x}
%}
\caption{Spectral Form Factor for sextic  potential at different finite temperature[$\bg$] with $N=1000$ and $a=0.1$ }
\end{figure}
From fig.~\ref{ssua1x} and  fig.~\ref{ssua2x}, we see that SFF at finite temperature decays with increasing $\tau$ and reach zero. But with changing $\bg$ SFF values remains almost same initially (for higher $\bg$).

%\begin{figure}[H]
%\centering
%\subfigure[SFF for sextic potential at $N=10$.]{
 %   \includegraphics[width=7.8cm,height=7.8cm] {SFF_S_N10.pdf}
%    \label{xsua1x}
%}
%\subfigure[SFF for sextic potential at $N=100$.]{
%    \includegraphics[width=7.8cm,height=7.8cm] {SFF_S_N100.pdf}
 %   \label{xsua2}
%}
%\subfigure[SFF for sextic potential at $N=1000$.]{
 %   \includegraphics[width=7.8cm,height=7.8cm] {SFF_S_N1000.pdf}
 %   \label{xsua3}
%}
%\subfigure[SFF for sextic potential at $N=10000$.]{
%    \includegraphics[width=7.8cm,height=7.8cm] {SFF_S_N10000.pdf}
%    \label{xsua4}
%}
%\caption{Spectral Form Factor for quartic  potential varying with different N at finite temperature[$\bg=100 $] and $a=0.1$ }
%\end{figure}
%From fig.~\ref{xsua1x},  fig.~\ref{xsua1x}, fig.~\ref{xsua1x} and fig.~\ref{xsua1x},  we see that SFF at finite temperature decays with increasing $\tau$ and reach zero. But at different $N$ SFF  remains same initially. From both the plots we have shown that SFF decays to zero for finite temperature.
\begin{figure}[htb]
\centering
\subfigure[SFF for sextic for $a=.1,N=100$  with  $SFF|_{\tau=0}=0.0068169$ as origin.]{
    \includegraphics[width=7.8cm,height=8cm] {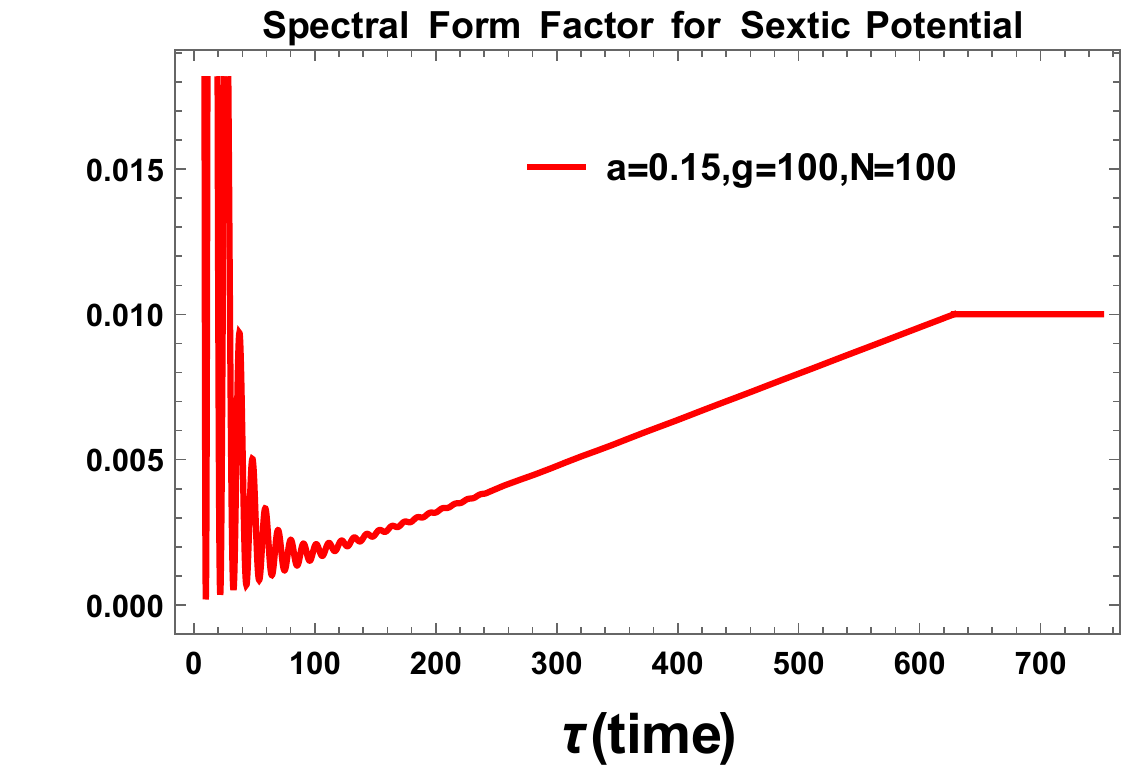}
    \label{SFFS21z}
}
\subfigure[SFF for sextic for $a=.1,N=1000$  with  $SFF|_{\tau=0}=0.00068169$ as origin]{
    \includegraphics[width=7.8cm,height=8cm] {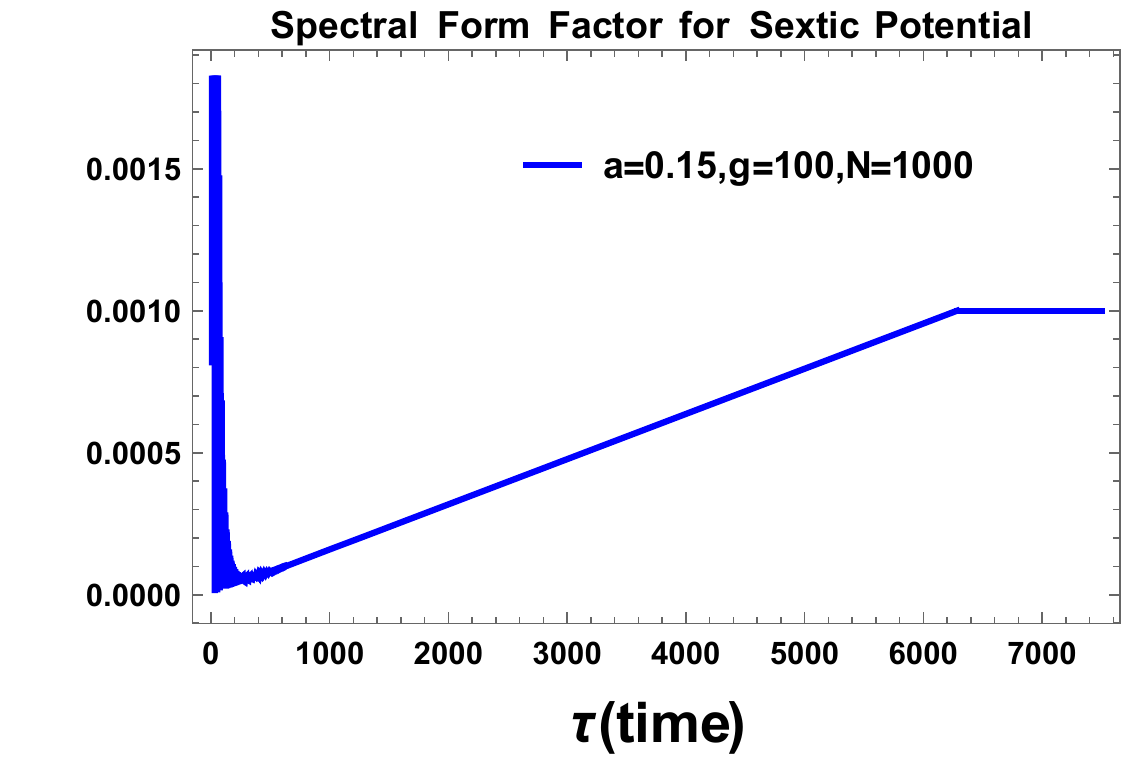}
    \label{SFFS23z}
}
\subfigure[SFF for sextic for $a=.1,N=10000$  with  $SFF|_{\tau=0}=0.000068169$ as origin]{
    \includegraphics[width=10.8cm,height=8cm] {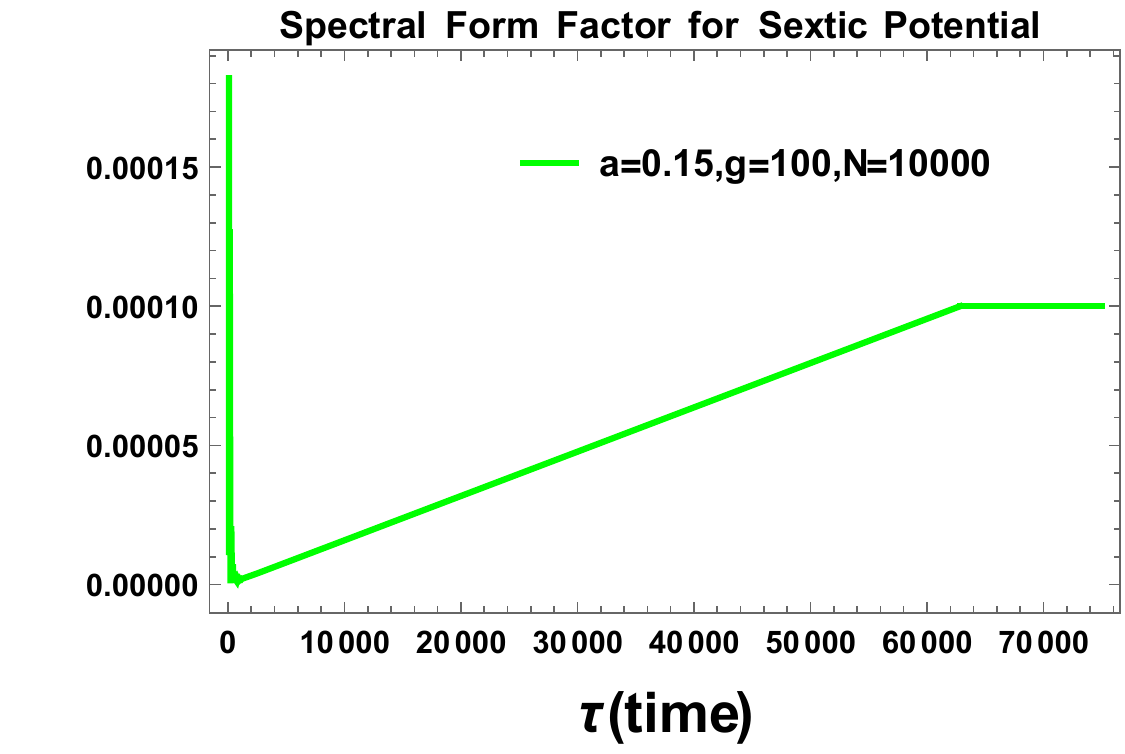}
    \label{SFFS22z}
}

\caption{Time variation of SFF for different N at $\bg=0$.Here we  shift  reference axis[SFF]  to $SFF|_{\tau=0}$ }
\end{figure} 
In fig.~\ref{SFFS21z},  fig.~\ref{SFFS23z}, fig.~\ref{SFFS22z},  it is observed that SFF with variation in $N$ get saturated at different value of $\tau$. But with increasing $N$ the value of the saturation point, will decrease.
Subtracting the change of axis[$SFF|_{\tau=0}$] we get the predicted bound of SFF.

%%%%%
%%%%%%
%%%%%%
%%%%%%%
%%%%%%%%
%%%%%%%%%%

\subsubsection{For Octa random potential}
Here we consider octa random potential, as given by the following expression:
\bea V(M)=\frac{1}{2}M^{2}+g M^{4}+h M^{6}+k M^{8}.
\eea
For a single interval ($n=1$) with end points -2a and 2a (semi-circle) we get the following expression for the density function in terms of the eigen value of the random matrix $M$, as given by:
\bea \rho(\lb)=\frac{1}{\pi}\sqrt{4 a^2-\lambda ^2} ~\left(a_3 \lambda ^6+a_2 \lambda ^4+a_1 \lambda ^2+a_0\right).
\eea
Then the function $\og(\lb+i0)$ can be expressed as:
\bea \og(\lb+i0)=\frac{1}{2} \left(4 g \lambda ^3+6 h \lambda ^5+8 k \lambda ^7+\lambda \right)+i \sqrt{4 a^2-\lambda ^2} \left(a_3 \lambda ^6+a_2 \lambda ^4+a_1 \lambda ^2+a_0\right).
\eea
Further Taylor expanding $\og(\lb+i0)$ near $\lb\rightarrow\infty$ we get:
\bea
\lambda ^5 \left(2 a_3 a^2-a_2+3 h\right)&&+\lambda ^3 \left(2 a_3 a^4+2 a_2 a^2-a_1+2 g\right)+\left(4 a_3 a^6+2 a_2 a^4+2 a_1 a^2-a_0+\frac{1}{2}\right) \lambda\nonumber\\
&&+\frac{10 a_3 a^8+4 a_2 a^6+2 a_1 a^4+2 a_0 a^2}{\lambda }+\lambda ^7 \left(4 k-a_3\right)+O\left(\left(\frac{1}{\lb}\right)^{3}\right)=\frac{1}{\lb}.~~~~~~~~~~~\eea
Therefore, equating both the sides of the above equation we get:
\bea a_{3}&=&4k,\\
a_{2}&=&3h+8a^{2}k,\\
a_{1}&=&24 a^4 k+6 a^2 h+2 g,\\
a_{0}&=&\frac{1}{2} \left(160 a^6 k+36 a^4 h+8 a^2 g+1\right),\eea
along with an additional constraint condition:
\be
 a^2 + 12 a^4 g + 60 a^6 h + 280 a^8 k=1
\ee
Solution of this constraint equation gives $ a^{2}$ in terms of $g$, $h$ and $k$. Since the solutions for $a^2$ are very complicated, we have not explicitly mentioned them here. Instead of writing full solution here we can check that putting $h=0$ the constraint condition reduces to the following simplified form:
\be
 12 g a^4+ a^{2}=1
\ee
and the solution of this equation is given by the following expression:
\be  \boxed{a^{2}=\frac{\sqrt{48 g+1}-1}{24 g}}~.\ee 
Here the critical value with $h=0$ and $k=0$ is given by the following expression:
\be \boxed{g_{c}=-\frac{1}{48}}~,\ee
which is exactly same result as obtained for quartic and sextic (with $h=0$) potential in the previous subsections.

Then, the final expression for the density function in terms of the eigen value of the random matrix $M$ can be written as:
\bea\label{ddf}
\rho(\lb)&=&\frac{1}{\pi}\sqrt{4 a^2-\lambda ^2}~\left(80 a^6 k+6 a^4 \left(3 h+4 k \lambda ^2\right)+a^2 \left(4 g+6 h \lambda ^2+8 k \lambda ^4\right)\right.\nonumber\\
&& \left.~~~~~~~~~~~~~~~~~~~~~~~~~~~~~~~~~~~~~~~~~~~ +2 g \lambda ^2+3 h \lambda ^4+4 k \lambda ^6+\frac{1}{2}\right).~~~~~~
\eea
\begin{figure}[htb]
\centering
\subfigure[$\rho(\lb)$ for octa potential for different g,h,k]{
    \includegraphics[width=7.8cm,height=7cm] {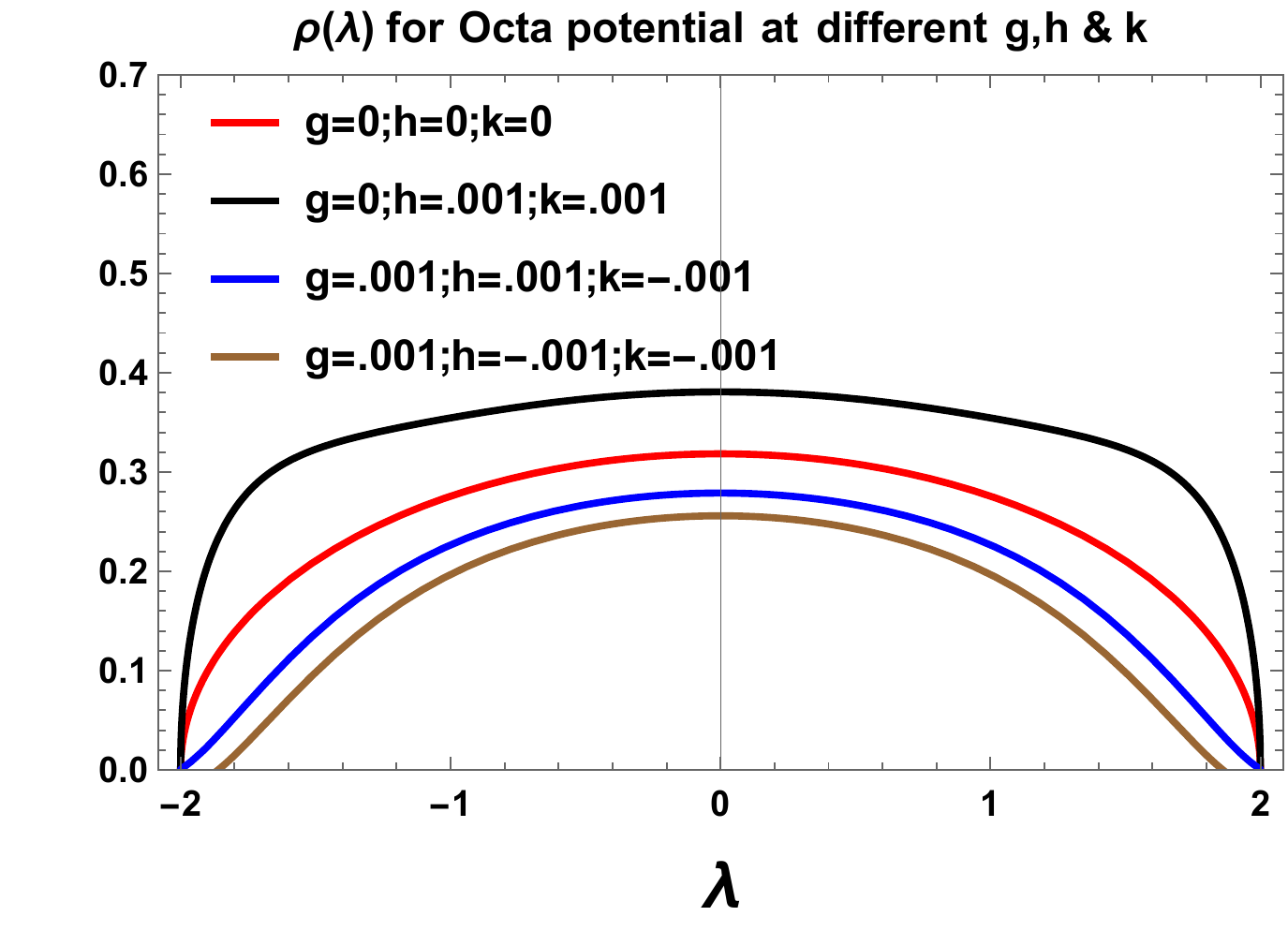}
    \label{Rhox41}
}
\subfigure[$\rho(\lb)$ for octa potential for different g,h,k]{
    \includegraphics[width=7.8cm,height=7cm] {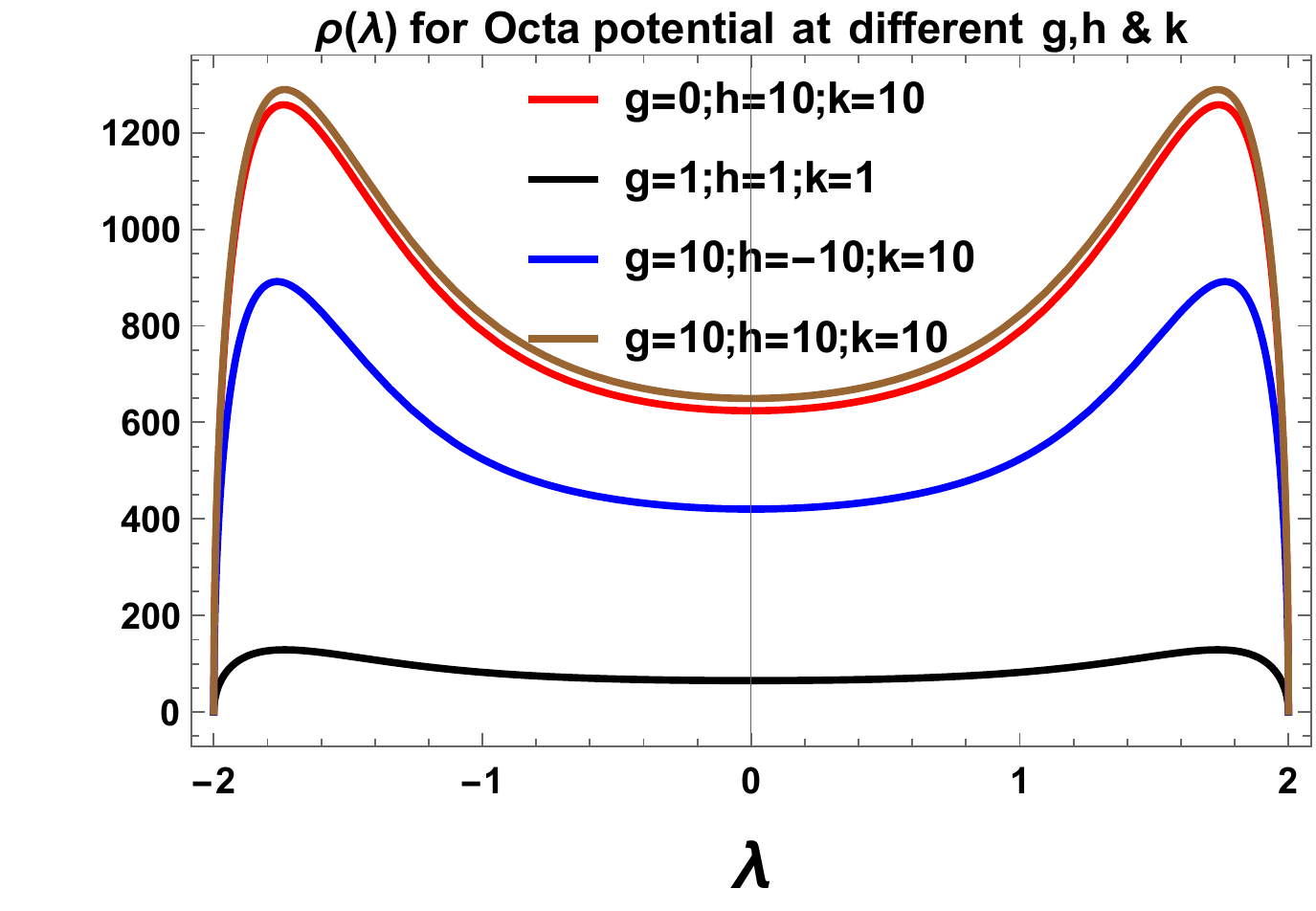}
    \label{Rhox42}
}
\caption{Eigen value distribution curve of density function for quartic and octa potential for different parameter values. Here we fix $a=1$.  }
\end{figure}
In fig.~\ref{Rhox41} and fig.~\ref{Rhox42} for octic potential behaviour of density function $\rho(\lb)$ is shown. The curve follows from Eq.~(\ref{ddf}). Again choosing $g=h=k=0$ will produce the {\it Wigner law}. Deviating $g$, $h$ and $k$ by small amount shows deviation from {\it Wigner semicircle law}. For $g=0,h>0,k>0$ the curve shows plateau region though merge with semicircle at end points. But choosing $g>0,h<0,k<0$ and $g>0,h>0,k<0$ show deviation from semicircle and don't converge even at end points. On the other hand, if we choose $g>0,h>0,k>0$ then we get a valley region lying between two peaks of the  maxima of the density distribution of eigen values of the random matrices under consideration. The same behaviour can be obtained by fixing $g>0,h<0,k>0$, $g=h=k=1$ and $g=0,h>>0,k>>0$. Only slight difference can be visualised  in the peak heights of the maxima and also in the spread in the valley region. But in all such cases in between it will not at all match with the {\it Wigner semicircle law}, but converge to the end points of the {\it Wigner semi-circle} , which is obtained by setting $g=h=k=0$.

Next, we compute the expression for the one point function of the partition function at finite temperature, which can be expressed as: 
\bea
 \displaystyle\langle Z(\bg\pm i\tau)\rangle&=&\frac{1}{\pi}\int_{-2 a}^{2 a}d\lb~ \sqrt{4 a^2-\lambda ^2}~\left(80 a^6 k+6 a^4 \left(3 h+4 k \lambda ^2\right)+a^2 \left(4 g+6 h \lambda ^2+8 k \lambda ^4\right)\right.\nonumber\\
&& \left.~~~~~~~~~~~~~~~~~~~~~~~~~~~~~+2 g \lambda ^2+3 h \lambda ^4+4 k \lambda ^6+\frac{1}{2}\right)~ e^{\mp i\tau \lambda}~e^{-\beta\lb}\nonumber\\
&=&\frac{1}{(\beta\pm i \tau)^6}\left[-24 a^2 I_2(2 (\beta\pm i \tau) \left| a\right| )\left(15 a^2 h \tau^4+\beta^4 \left(140 a^4 k+15 a^2 h+g\right)\right.\right.\nonumber\\ && \left.\left.~~ \pm 4 i \beta^3 \tau \left(140 a^4 k+15 a^2 h+g\right)-6\beta^2 \left(140 a^4 k \tau^2+5 h \left(3 a^2 \tau^2-1\right)-140 a^2 k+g \tau^2\right)\right.\right.\nonumber\\ && \left.\left.~~ \mp 4 i \beta \tau \left(140 a^4 k \tau^2+15 h \left(a^2 \tau^2-1\right)-420 a^2 k+g \tau^2\right)\right.\right.\nonumber\\ && \left.\left.~~~~~~~~~~~~~~~~~~~~ +140 k \left(a^4 \tau^4-6 a^2 \tau^2+12\right)+g \tau^4-30 h \tau^2\right)\right.\nonumber\\ && \left.~~~~~~~~~~~~~~~~~~~~ +\left| a\right|  (\beta\pm i \tau)^3 I_1(2 (\beta\pm i \tau) \left| a\right| )\left(-1120 a^6 k \tau^2+60 a^4 \left(112 k-3 h \tau^2\right)\right.\right.\nonumber\\ &&\left. \left.~~~~~~~~~~~~~~~~~~~~ +24 a^2 \left(15 h-g \tau^2\right)+\beta^2 \left(1120 a^6 k+180 a^4 h+24 a^2 g+1\right)\right.\right.\nonumber\\ && \left.\left. \pm 2 i \beta \tau \left(1120 a^6 k+180 a^4 h+24 a^2 g+1\right)-\tau^2\right)+20160 k \left| a\right| ^3 (\beta\pm i \tau) I_1(2 (\beta\pm i \tau) \left| a\right| )\right],\nonumber\\
  && \eea
where $ I_n(x)$ is the modified Bessel function of first kind with $n$ th order.

Further, considering the high temperature limiting situation we get the following simplified expression for the one point function of the partition function:
\bea
 \displaystyle\left[\langle Z(\bg\pm i\tau)\rangle\right]_{\beta=0}&=&\frac{1}{\pi}\int_{-2 a}^{2 a}d\lb~ \sqrt{4 a^2-\lambda ^2}~\left(80 a^6 k+6 a^4 \left(3 h+4 k \lambda ^2\right)+a^2 \left(4 g+6 h \lambda ^2+8 k \lambda ^4\right)\right.\nonumber\\
&& \left.~~~~~~~~~~~~~~~~~~~~~~~~~~~~~+2 g \lambda ^2+3 h \lambda ^4+4 k \lambda ^6+\frac{1}{2}\right)~ e^{\mp i\tau \lambda}\nonumber\\
&=&
\frac{a}{\tau^6} \left[\pm J_1(\pm 2 a \tau ) ( 1120 a^6 k \tau ^5+60 a^4 \tau ^3 (3 h \tau ^2-112 k)\right.\nonumber\\ && \left.+24 a^2 (g \tau ^5-15 h \tau ^3+840 k \tau)+\tau ^5)\right. \nonumber \\
&&\left.\displaystyle -24  J_2(\pm 2 a \tau ) (-30 \tau ^2 (28 a^2 k+h)+\tau ^4 (140 a^4 k+15 a^2 h+g)+1680 k)\right].~~~~~~~~~~~~~~
\eea
Next, simplifying the result for one point function in the limit ${\cal T}=\sqrt{N}\tau\rightarrow\infty$ we get:
\bea \label{tocta}
\left[\langle Z(\bg\pm i{\cal T})\rangle\right]_{\bg=0}&&=\sqrt{\frac{a}{\pi}}\frac{1}{(\pm {\cal T})^{\frac{3}{2}}}\left[\left(1120a^{6}k+180a^{4}h+2a^{2}g+1\right.\right.\nonumber\\
&&\left.\left.\displaystyle~~~~~~~~-\left(\frac{6720a^{4}k+360a^{2}h}{{\cal T}^{2}}\right)\right)\cos\left(\frac{\pi}{4}\pm 2a{\cal T}\right)\right.\nonumber\\
&&\left.\displaystyle~~~~~~~~\mp 24\left(\frac{140a^{4}k+15a^{2}h+g}{{\cal T}}-\left(\frac{840a^{2}k+30h}{{\cal T}^{3}}\right)\right)\sin\left(\frac{\pi}{4}\pm 2a{\cal T}\right)\right]\nonumber\\
&&~~~~~~~~~~~~~+O\left(\frac{1}{(\pm {\cal T})^{\frac{11}{2}}}\right).
\eea
Now for the octic random potential disconnected part of the Green's function can be computed at finite temperature as:
\bea G_{dc}(\beta,\tau)
&=&\frac{\langle Z(\bg+i\tau)\rangle \langle Z(\bg-i\tau)\rangle}{\langle Z(\bg)\rangle^{2}}=\frac{\beta^{12}}{(\beta^2+ \tau^2)^6}\left[-24 a^2 I_2(2 \beta \left| a\right| )\right.\nonumber\\ && \left. ~~~~~~~~~~~~~~~~~~~~\left(\beta^4 \left(140 a^4 k+15 a^2 h+g\right) +6\beta^2 \left(5 h+140 a^2 k\right)+1680 k \right)\right.\nonumber\\ && \left.~~~~~~~~~~~~~~~~~~~~ +\left| a\right|  \beta^3 I_1(2 (\beta) \left| a\right| )  \left(6720 a^4 k+360 a^2  h\right.\right.\nonumber\\ &&\left. \left.~~~~~~~~~~~~~~~~~~~~ +\beta^2 \left(1120 a^6 k+180 a^4 h+24 a^2 g+1\right)\right)+20160 k \left| a\right| ^3 \beta I_1(2 (\beta) \left| a\right| )\right]^{-2}\nonumber\\
&&~~~~~\times\left[-24 a^2 I_2(2 (\beta+ i \tau) \left| a\right| )\right.\nonumber\\ && \left. ~~~~~~~~~~~~~~~~~~~~\left(15 a^2 h \tau^4+\beta^4 \left(140 a^4 k+15 a^2 h+g\right)\right.\right.\nonumber\\ && \left.\left.~~~~~~~~~~~~~~~~~~~~ + 4 i \beta^3 \tau \left(140 a^4 k+15 a^2 h+g\right)\right.\right.\nonumber\\ && \left.\left.~~~~~~~~~~~~~~~~~~~~ -6\beta^2 \left(140 a^4 k \tau^2+5 h \left(3 a^2 \tau^2-1\right)-140 a^2 k+g \tau^2\right)\right.\right.\nonumber\\ && \left.\left.~~~~~~~~~~~~~~~~~~~~ - 4 i \beta \tau \left(140 a^4 k \tau^2+15 h \left(a^2 \tau^2-1\right)-420 a^2 k+g \tau^2\right)\right.\right.\nonumber\\ && \left.\left.~~~~~~~~~~~~~~~~~~~~ +140 k \left(a^4 \tau^4-6 a^2 \tau^2+12\right)+g \tau^4-30 h \tau^2\right)\right.\nonumber\\ && \left.~~~~~~~~~~~~~~~~~~~~ +\left| a\right|  (\beta+ i \tau)^3 I_1(2 (\beta+ i \tau) \left| a\right| ) \right.\nonumber\\ && \left.~~~~~~~~~~~~~~~~~~~~ \left(-1120 a^6 k \tau^2+60 a^4 \left(112 k-3 h \tau^2\right)\right.\right.\nonumber\\ &&\left. \left.~~~~~~~~~~~~~~~~~~~~ +24 a^2 \left(15 h-g \tau^2\right) +\beta^2 \left(1120 a^6 k+180 a^4 h+24 a^2 g+1\right)\right.\right.\nonumber\\ && \left.\left. ~~~~~~~~~~~~~~~~~~~~+ 2 i \beta \tau \left(1120 a^6 k+180 a^4 h+24 a^2 g+1\right)-\tau^2\right)\right.\nonumber\\ && \left.~~~~~~~~~~~~~~~~~~~~ +20160 k \left| a\right| ^3 (\beta+ i \tau) I_1(2 (\beta+ i \tau) \left| a\right| )\right]\nonumber\\
&&~~~~~~~~~~~~~~\times \left[-24 a^2 I_2(2 (\beta- i \tau) \left| a\right| )\right.\nonumber\\ && \left. ~~~~~~~~~~~~~~~~~~~~\left(15 a^2 h \tau^4+\beta^4 \left(140 a^4 k+15 a^2 h+g\right) - 4 i \beta^3 \tau \left(140 a^4 k+15 a^2 h+g\right)\right.\right.\nonumber\\ && \left.\left.~~~~~~~~~~~~~~~~~~~~ -6\beta^2 \left(140 a^4 k \tau^2+5 h \left(3 a^2 \tau^2-1\right)-140 a^2 k+g \tau^2\right)\right.\right.\nonumber\\ && \left.\left.~~~~~~~~~~~~~~~~~~~~ + 4 i \beta \tau \left(140 a^4 k \tau^2+15 h \left(a^2 \tau^2-1\right)-420 a^2 k+g \tau^2\right)\right.\right.\nonumber\\ && \left.\left.~~~~~~~~~~~~~~~~~~~~ +140 k \left(a^4 \tau^4-6 a^2 \tau^2+12\right)+g \tau^4-30 h \tau^2\right)\right.\nonumber\\ && \left.~~~~~~~~~~~~~~~~~~~~ +\left| a\right|  (\beta- i \tau)^3 I_1(2 (\beta- i \tau) \left| a\right| ) \right.\nonumber\\ && \left.~~~~~~~~~~~~~~~~~~~~ \left(-1120 a^6 k \tau^2+60 a^4 \left(112 k-3 h \tau^2\right)\right.\right.\nonumber\\ &&\left. \left.~~~~~~~~~~~~~~~~~~~~ -24 a^2 \left(15 h-g \tau^2\right)+\beta^2 \left(1120 a^6 k+180 a^4 h+24 a^2 g+1\right)\right.\right.\nonumber\\ && \left.\left. ~~~~~~~~~~~~~~~~~~~~- 2 i \beta \tau \left(1120 a^6 k+180 a^4 h+24 a^2 g+1\right)-\tau^2\right)\right.\nonumber\\ && \left.~~~~~~~~~~~~~~~~~~~~ +20160 k \left| a\right| ^3 (\beta - i \tau) I_1(2 (\beta- i \tau) \left| a\right| )\right]~~~~~~\eea
which can be further simplified in the high temperature limiting situation as:
\bea G_{dc}(\tau)&=&\left[\frac{\langle Z(\bg+i\tau)\rangle \langle Z(\bg-i\tau)\rangle}{\langle Z(\bg)\rangle^{2}}\right]_{\beta=0}\nonumber\\
&=&\frac{a^2}{N^2\tau^{12}}\nonumber\\
&&~~~~\times\ \left[ J_1( 2 a \tau ) ( 1120 a^6 k \tau ^5+60 a^4 \tau ^3 (3 h \tau ^2-112 k)\right.\nonumber\\
&&\left.\displaystyle~~~~~~~~+24 a^2 (g \tau ^5-15 h \tau ^3+840 k \tau)+\tau ^5)\right. \nonumber \\
&&\left.\displaystyle -24  J_2( 2 a \tau ) (-30 \tau ^2 (28 a^2 k+h)\right.\nonumber\\
&&\left.\displaystyle~~~~~~~~+\tau ^4 (140 a^4 k+15 a^2 h+g)+1680 k)\right]\nonumber\\
&&~~~~\times \left[- J_1(- 2 a \tau ) ( 1120 a^6 k \tau ^5+60 a^4 \tau ^3 (3 h \tau ^2-112 k)\right.\nonumber\\
&&\left.\displaystyle~~~~~~~~+24 a^2 (g \tau ^5-15 h \tau ^3+840 k \tau)+\tau ^5)\right. \nonumber \\
&&\left.\displaystyle -24  J_2(- 2 a \tau ) (-30 \tau ^2 (28 a^2 k+h)\right.\nonumber\\
&&\left.\displaystyle~~~~~~~~+\tau ^4 (140 a^4 k+15 a^2 h+g)+1680 k)\right].~~~~~~\eea
Further taking the limit ${\cal T}=\sqrt{N}\tau\rightarrow\infty$ we get the following simplified result:
\bea G_{dc}({\cal T})&=&\left[\frac{\langle Z(\bg+i{\cal T})\rangle \langle Z(\bg-i{\cal T})\rangle}{\langle Z(\bg)\rangle^{2}}\right]_{\beta=0}\nonumber\\&=&\frac{i}{{\cal T}^{3}}\frac{a}{N^2\pi}\nonumber\\
&&~~~~\times\left[\left(1120a^{6}k+180a^{4}h+2a^{2}g+1\right)\cos\left(\frac{\pi}{4}\pm 2a{\cal T}\right)\right.\nonumber\\
&&\left.\displaystyle~~~~~~~~\mp 24\left(\frac{140a^{4}k+15a^{2}h+g}{{\cal T}}\right)\sin\left(\frac{\pi}{4}\pm 2a{\cal T}\right)\right.\nonumber\\
&&\left.\displaystyle~~~~~~~~-\left(\frac{6720a^{4}k+360a^{2}h}{{\cal T}^{2}}\right)\cos\left(\frac{\pi}{4}\pm 2a{\cal T}\right)\right.\nonumber\\
&&\left.\displaystyle~~~~~~~~\pm 24\left(\frac{840a^{2}k+30h}{{\cal T}^{3}}\right)\sin\left(\frac{\pi}{4}\pm 2a{\cal T}\right)\right]\nonumber\\
&&~~~~\times \left[\left(1120a^{6}k+180a^{4}h+2a^{2}g+1\right)\cos\left(\frac{\pi}{4}\pm 2a{\cal T}\right)\right.\nonumber\\
&&\left.\displaystyle~~~~~~~~\mp 24\left(\frac{140a^{4}k+15a^{2}h+g}{{\cal T}}\right)\sin\left(\frac{\pi}{4}\pm 2a{\cal T}\right)\right.\nonumber\\
&&\left.\displaystyle~~~~~~~~-\left(\frac{6720a^{4}k+360a^{2}h}{{\cal T}^{2}}\right)\cos\left(\frac{\pi}{4}\pm 2a{\cal T}\right)\right.\nonumber\\
&&\left.\displaystyle~~~~~~~~\pm 24\left(\frac{840a^{2}k+30h}{{\cal T}^{3}}\right)\sin\left(\frac{\pi}{4}\pm 2a{\cal T}\right)\right].\nonumber\\
&&~~~~~~~~~~\eea
Now to compute  SFF we need to add both connected and disconnected part of the Green's function $G$($=G_{c}+G_{dc}$). Therefore, for octic polynomial potential we get finally the following expression for SFF at finite temp:
 \bea\boxed{\begin{array}{lll}\label{cqb1}
		  {\bf SFF}(\beta,\tau)&\equiv& \displaystyle \frac{\beta^{12}}{(\beta^2+ \tau^2)^6}\left[-24 a^2 I_2(2 \beta \left| a\right| )\left(\beta^4 \left(140 a^4 k+15 a^2 h+g\right)+6\beta^2 \left(5 h+140 a^2 k\right)+1680 k \right)\right.\\ && \left.~~~~ +\left| a\right|  \beta^3 I_1(2 (\beta) \left| a\right| )  \left(6720 a^4 k+360 a^2  h +\beta^2 \left(1120 a^6 k+180 a^4 h+24 a^2 g+1\right)\right)\right.\\ && \left.~~ +20160 k \left| a\right| ^3 \beta I_1(2 (\beta) \left| a\right| )\right]^{-2}\times\left[-24 a^2 I_2(2 (\beta+ i \tau) \left| a\right| )\right.r\\ && \left. ~~~\left(15 a^2 h \tau^4+\beta^4 \left(140 a^4 k+15 a^2 h+g\right) + 4 i \beta^3 \tau \left(140 a^4 k+15 a^2 h+g\right)\right.\right.\\ && \left.\left.~~ -6\beta^2 \left(140 a^4 k \tau^2+5 h \left(3 a^2 \tau^2-1\right)-140 a^2 k+g \tau^2\right)\right.\right.\\ && \left.\left.~~ - 4 i \beta \tau \left(140 a^4 k \tau^2+15 h \left(a^2 \tau^2-1\right)-420 a^2 k+g \tau^2\right)\right.\right.\\ && \left.\left.~~~~~~~~~~~~~~~~~~~~ +140 k \left(a^4 \tau^4-6 a^2 \tau^2+12\right)+g \tau^4-30 h \tau^2\right)\right.\\ && \left.~~+\left| a\right|  (\beta+ i \tau)^3 I_1(2 (\beta+ i \tau) \left| a\right| )  \left(-1120 a^6 k \tau^2+60 a^4 \left(112 k-3 h \tau^2\right)\right.\right.\\ &&\left. \left.~~~~~~~~~~~~~~~~~~~~ +24 a^2 \left(15 h-g \tau^2\right)+\beta^2 \left(1120 a^6 k+180 a^4 h+24 a^2 g+1\right)\right.\right.\\ && \left.\left. ~~~~~~~~~~~~~~~~~~~~+ 2 i \beta \tau \left(1120 a^6 k+180 a^4 h+24 a^2 g+1\right)-\tau^2\right)\right.\\ && \left.~~~~~~~~~~~~~~~~~~~~ +20160 k \left| a\right| ^3 (\beta+ i \tau) I_1(2 (\beta+ i \tau) \left| a\right| )\right]\\
&&~~~\times \left[-24 a^2 I_2(2 (\beta- i \tau) \left| a\right| )\left(15 a^2 h \tau^4+\beta^4 \left(140 a^4 k+15 a^2 h+g\right)\right.\right.\\ && \left.\left.~- 4 i \beta^3 \tau \left(140 a^4 k+15 a^2 h+g\right)-6\beta^2 \left(140 a^4 k \tau^2+5 h \left(3 a^2 \tau^2-1\right)-140 a^2 k+g \tau^2\right)\right.\right.\\ && \left.\left.~~~~~~~~~~~~~~~~~~~~ + 4 i \beta \tau \left(140 a^4 k \tau^2+15 h \left(a^2 \tau^2-1\right)-420 a^2 k+g \tau^2\right)\right.\right.\\ && \left.\left.~~~~~~~~~~~~~~~~~~~~ +140 k \left(a^4 \tau^4-6 a^2 \tau^2+12\right)+g \tau^4-30 h \tau^2\right)\right.\nonumber\\ && \left.~+\left| a\right|  (\beta- i \tau)^3 I_1(2 (\beta- i \tau) \left| a\right| ) \left(-1120 a^6 k \tau^2+60 a^4 \left(112 k-3 h \tau^2\right)\right.\right.\\ &&\left. \left.~~~~~~~~~~~~~~~~~~~~ -24 a^2 \left(15 h-g \tau^2\right)+\beta^2 \left(1120 a^6 k+180 a^4 h+24 a^2 g+1\right)\right.\right.\\ && \left.\left. ~~~~~~~~~~~~~~~~~~~~- 2 i \beta \tau \left(1120 a^6 k+180 a^4 h+24 a^2 g+1\right)-\tau^2\right)\right.\\ && \left.~~~~~~~~~~~~~~~~~~~~ +20160 k \left| a\right| ^3 (\beta - i \tau) I_1(2 (\beta- i \tau) \left| a\right| )\right]\\
&&~~+\left\{\begin{array}{lll}
			\displaystyle  
			\frac{\tau}{(2\pi N)^{2}}-\frac{1}{N}+\frac{1}{(\pi N)}\,,~~~~~~ &
			\mbox{\small  \textcolor{red}{\bf  {$\tau<2\pi N$ }}}  \\ \\
			\displaystyle  
			\frac{1}{\pi N}\,,~~~~~~ &
			\mbox{\small  \textcolor{red}{\bf  {$\tau>2\pi N$}}}  
		\end{array}
		\right.
	\end{array}}\nonumber\\
	&&\label{boundocta1}\eea
where ${\bf SFF}(\beta,\tau)$ is defined with proper normalization and in our prescription it gives the total Green's function as mentioned above. 

Further simplifying the result for high temperature limit we get the following expression for SFF, as given by:
\bea\boxed{\begin{array}{lll}\label{cqm2}
		\displaystyle   {\bf SFF}(\tau)&\equiv & \displaystyle \frac{a^2}{N^2\tau^{12}}\left[ J_1( 2 a \tau ) ( 1120 a^6 k \tau ^5+60 a^4 \tau ^3 (3 h \tau ^2-112 k)\right.\\
&&\left.\displaystyle~~~~~~~~+24 a^2 (g \tau ^5-15 h \tau ^3+840 k \tau)+\tau ^5)\right. \\
&&\left.\displaystyle -24  J_2( 2 a \tau ) (-30 \tau ^2 (28 a^2 k+h)+\tau ^4 (140 a^4 k+15 a^2 h+g)+1680 k)\right]\\
&&~~~~\times \left[- J_1(- 2 a \tau ) ( 1120 a^6 k \tau ^5+60 a^4 \tau ^3 (3 h \tau ^2-112 k)\right.\\
&&\left.\displaystyle~~~~~~~~+24 a^2 (g \tau ^5-15 h \tau ^3+840 k \tau)+\tau ^5)\right. \\
&&\left.\displaystyle -24  J_2(- 2 a \tau ) (-30 \tau ^2 (28 a^2 k+h)+\tau ^4 (140 a^4 k+15 a^2 h+g)+1680 k)\right]\\
&&+\left\{\begin{array}{lll}
			\displaystyle  
			\frac{\tau}{(2\pi N)^{2}}-\frac{1}{N}+\frac{1}{(\pi N)}\,,~~~~~~~~~~~~ &
			\mbox{\small  \textcolor{red}{\bf  {$\tau<2\pi N$ }}}  \\ 
			\displaystyle  
			\frac{1}{\pi N}\,,~~~~~~~~~~~~ &
			\mbox{\small  \textcolor{red}{\bf  {$\tau>2\pi N$}}}  
		\end{array}
		\right.
	\end{array}}~~~~~~~~~\label{boundocta2}\eea
	Further taking the limit ${\cal T}=\sqrt{N}\tau\rightarrow\infty$ we get the following simplified result for SFF:
\bea\boxed{\begin{array}{lll}\label{cqm3}
		\footnotesize\displaystyle   {\bf SFF}({\cal T})&\equiv&\displaystyle 
		\frac{i}{{\cal T}^{3}}\frac{a}{N^2\pi}\left[\left(1120a^{6}k+180a^{4}h+2a^{2}g+1\right)\cos\left(\frac{\pi}{4}\pm 2a{\cal T}\right)\right.\\
&&\left.\displaystyle~~~~~~~~\mp 24\left(\frac{140a^{4}k+15a^{2}h+g}{{\cal T}}\right)\sin\left(\frac{\pi}{4}\pm 2a{\cal T}\right)\right.\\
&&\left.\displaystyle~~~~~~~~-\left(\frac{6720a^{4}k+360a^{2}h}{{\cal T}^{2}}\right)\cos\left(\frac{\pi}{4}\pm 2a{\cal T}\right)\pm 24\left(\frac{840a^{2}k+30h}{{\cal T}^{3}}\right)\sin\left(\frac{\pi}{4}\pm 2a{\cal T}\right)\right]\\
&&~~~~\times \left[\left(1120a^{6}k+180a^{4}h+2a^{2}g+1\right)\cos\left(\frac{\pi}{4}\pm 2a{\cal T}\right)\right.\\
&&\left.\displaystyle~~~~~~~~\mp 24\left(\frac{140a^{4}k+15a^{2}h+g}{{\cal T}}\right)\sin\left(\frac{\pi}{4}\pm 2a{\cal T}\right)\right.\\
&&\left.\displaystyle~~~~~~~~-\left(\frac{6720a^{4}k+360a^{2}h}{{\cal T}^{2}}\right)\cos\left(\frac{\pi}{4}\pm 2a{\cal T}\right)\pm 24\left(\frac{840a^{2}k+30h}{{\cal T}^{3}}\right)\sin\left(\frac{\pi}{4}\pm 2a{\cal T}\right)\right]\\
&&+ \nonumber \left\{\begin{array}{lll}
			\displaystyle  
			\frac{{\cal T}}{(2\pi)^2 N^{5/2}}-\frac{1}{N}+\frac{1}{(\pi N)}\,,~~~~~~~~~~~~ &
			\mbox{\small  \textcolor{red}{\bf  {${\cal T}<2\pi N^{3/2}$ }}}  \\ 
			\displaystyle  
			\frac{1}{\pi N}\,,~~~~~~~~~~~~ &
			\mbox{\small  \textcolor{red}{\bf  {${\cal T}>2\pi N^{3/2}$}}}  
		\end{array}
		\right.
	\end{array}}~~.\\
	&&\label{boundocta3}\eea	
\begin{figure}[htb]
\centering
\subfigure[SFF for octa potential at $\bg=10$.]{
    \includegraphics[width=7.8cm,height=8.8cm] {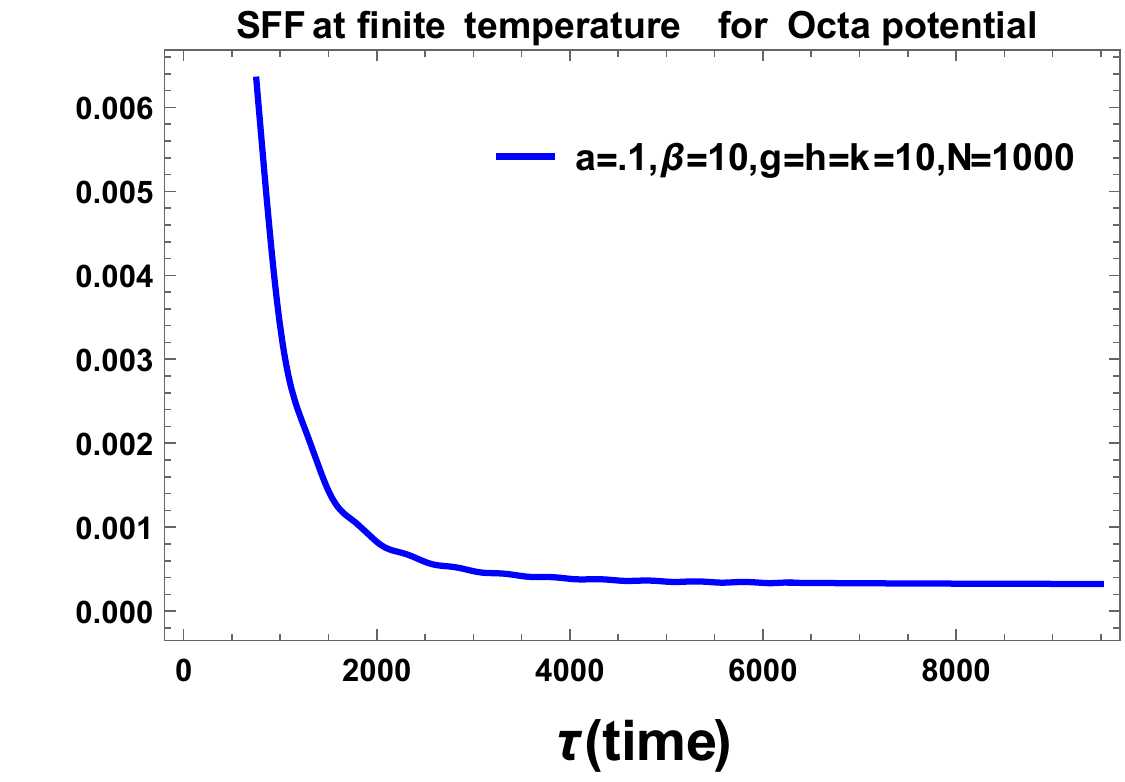}
    \label{osua1}
}
\subfigure[SFF for octa potential at $\bg=100$.]{
    \includegraphics[width=7.8cm,height=8.8cm] {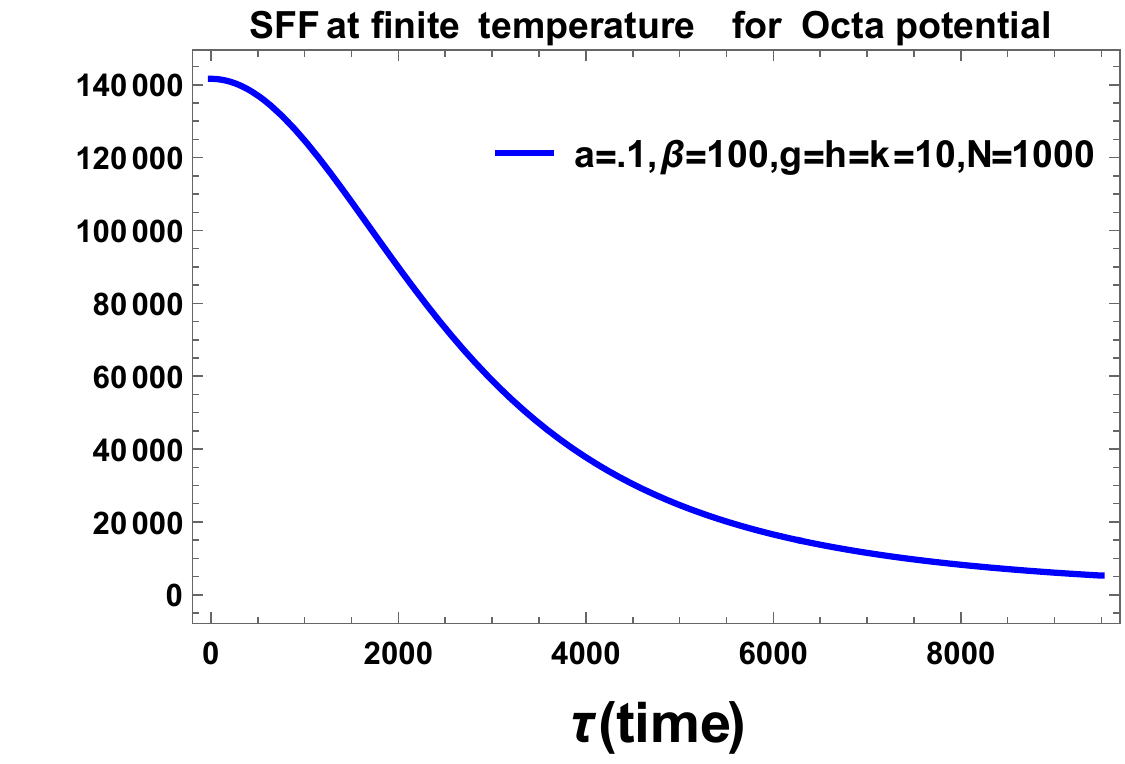}
    \label{osua2}
}
\subfigure[SFF for octa potential at $\bg=200$.]{
    \includegraphics[width=10.8cm,height=8.8cm] {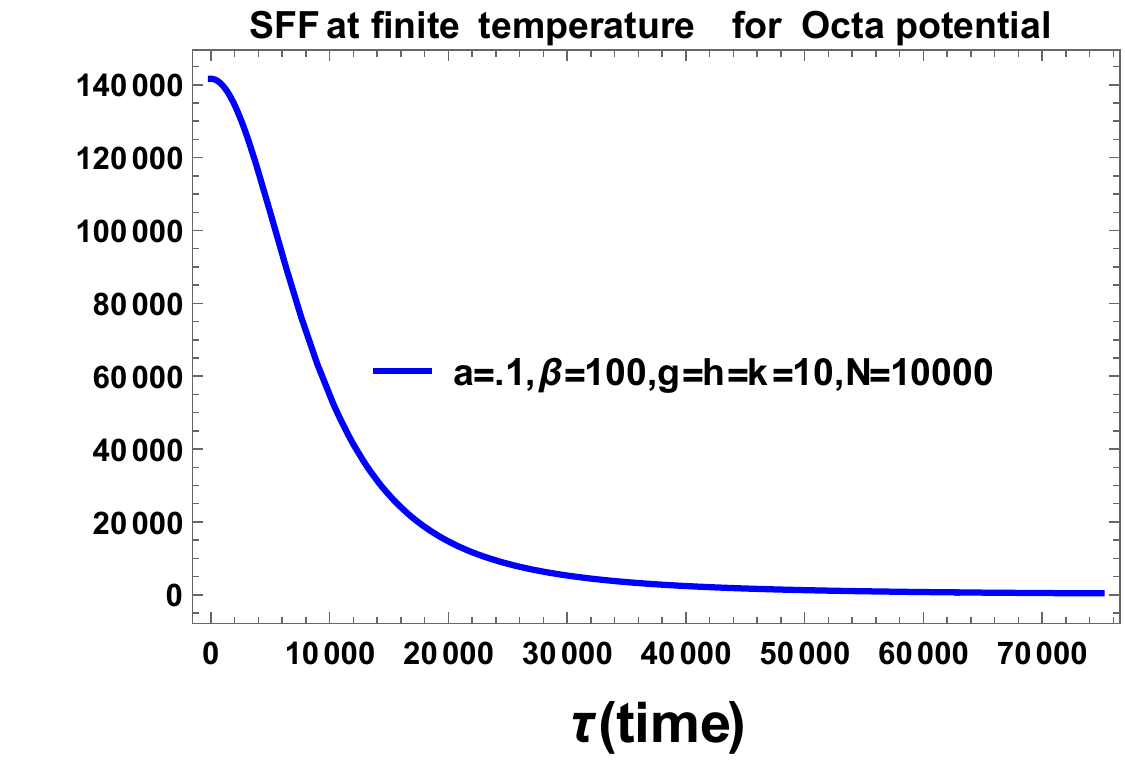}
    \label{osua3}
}
%\subfigure[SFF for octa potential at $\bg=1000$.]{
 %   \includegraphics[width=7.8cm,height=8.8cm] {SFF_O_B1000.pdf}
 %   \label{osua4}
%}
\caption{Spectral Form Factor for sextic  potential at different finite temperature[$\bg$] with $N=1000$ and $a=0.1$ }
\end{figure}
From fig.~\ref{osua1}, fig.~\ref{osua2} and fig.~\ref{osua3}, we see that SFF at finite temperature decays with increasing $\tau$ and reach zero. But with changing $\bg$ SFF values remains almost same initially (For higher $\bg$).

%\begin{figure}[H]
%\centering
%\subfigure[SFF for octa potential at $N=10$.]{
 %   \includegraphics[width=7.8cm,height=8.5cm] {SFF_O_N10.pdf}
%    \label{oua1x}
%}
%\subfigure[SFF for octa potential at $N=100$.]{
%    \includegraphics[width=7.8cm,height=8.5cm] {SFF_O_N100.pdf}
 %   \label{oua2x}
%}
%\subfigure[SFF for octa potential at $N=1000$.]{
 %   \includegraphics[width=7.8cm,height=8.5cm] {SFF_O_N1000.pdf}
 %   \label{oua3x}
%}
%\subfigure[SFF for octa potential at $N=10000$.]{
 %   \includegraphics[width=7.8cm,height=8.5cm] {SFF_O_N10000.pdf}
 %   \label{oua4x}
%}
%\caption{Spectral Form Factor for octa  potential varying with different N at finite temperature[$\bg=100 $] and $a=0.1$ }
%\end{figure}
%From fig.~\ref{oua1x}, fig.~\ref{oua2x}, fig.~\ref{oua3x} and fig.~\ref{oua4x}, we see that SFF at finite temperature decays with increasing $\tau$ and reach zero. But at different $N$ SFF values remains same initially. From both the figures we have shown that SFF decays to zero for finite temperature.
\begin{figure}[htb]
\centering
\subfigure[SFF for octa for $a=.1,N=100$  with  $SFF|_{\tau=0}=0.0068169$ as origin]{
    \includegraphics[width=7.8cm,height=8cm] {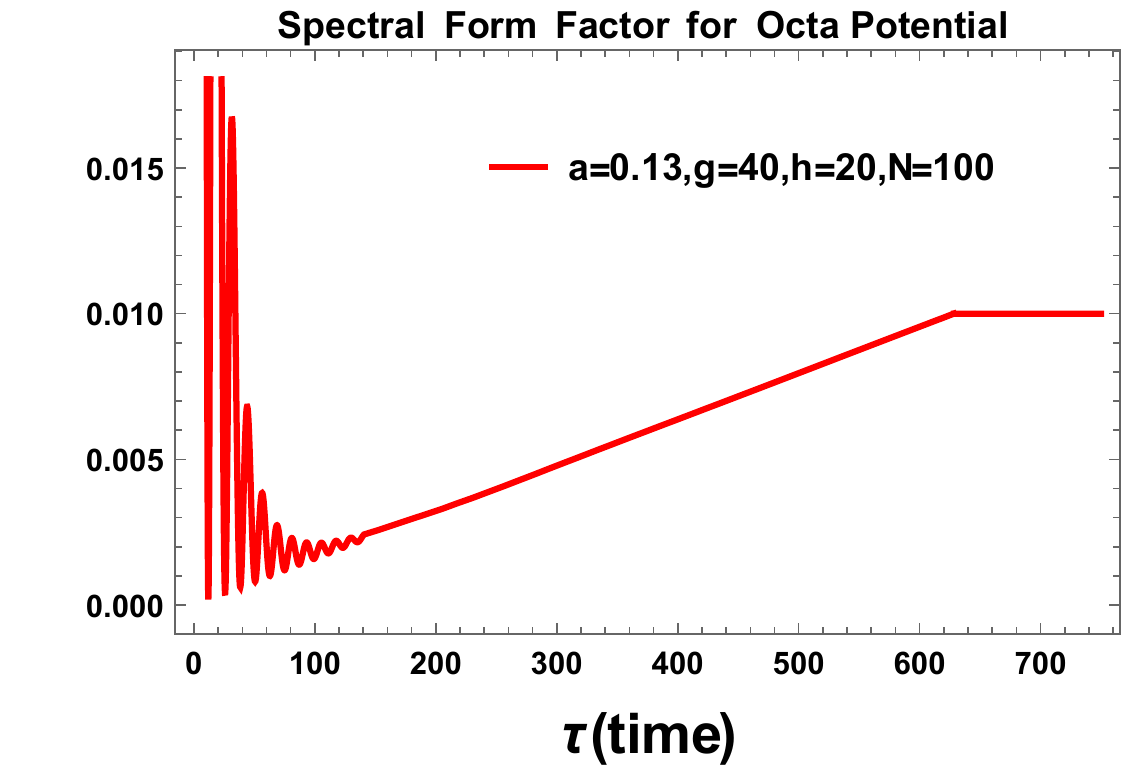}
    \label{SFFO21c}
}
\subfigure[SFF for octa for $a=.1,N=1000$ with  $SFF|_{\tau=0}=0.00068169$ as origin]{
    \includegraphics[width=7.8cm,height=8cm] {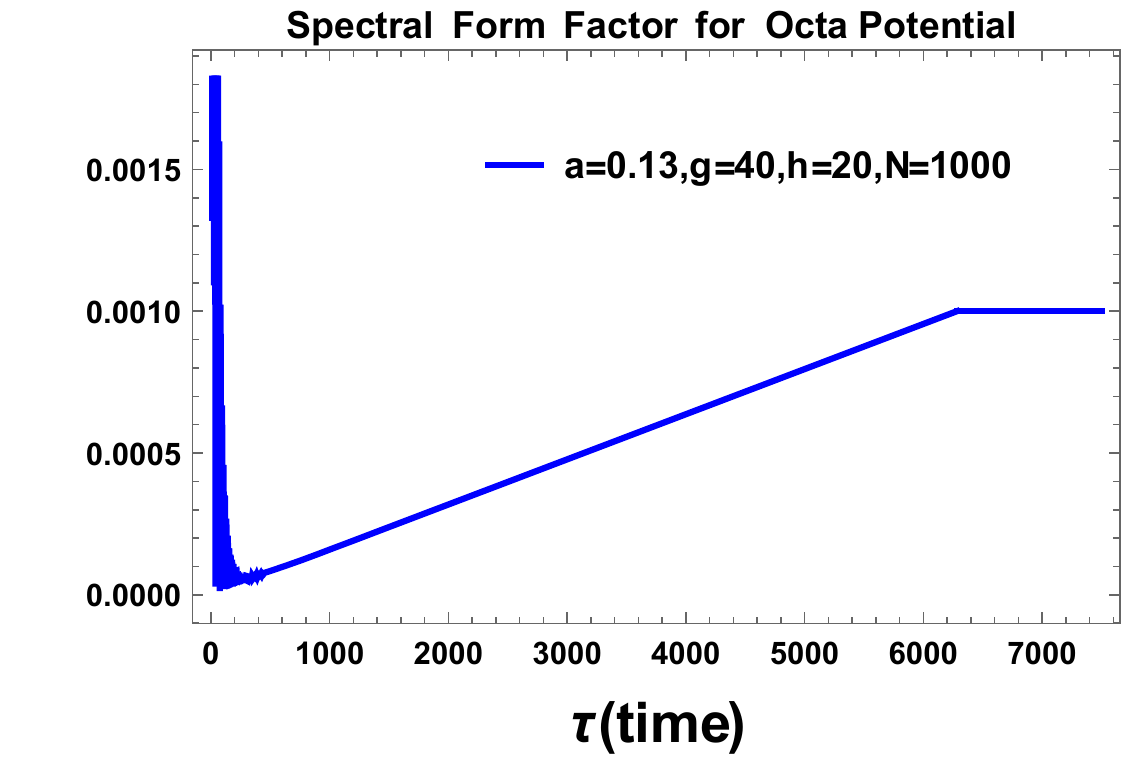}
    \label{SFFO23c}
}
\subfigure[SFF for octa for $a=.1,N=10000$  with  $SFF|_{\tau=0}=0.000068169$ as origin]{
    \includegraphics[width=10.8cm,height=8cm] {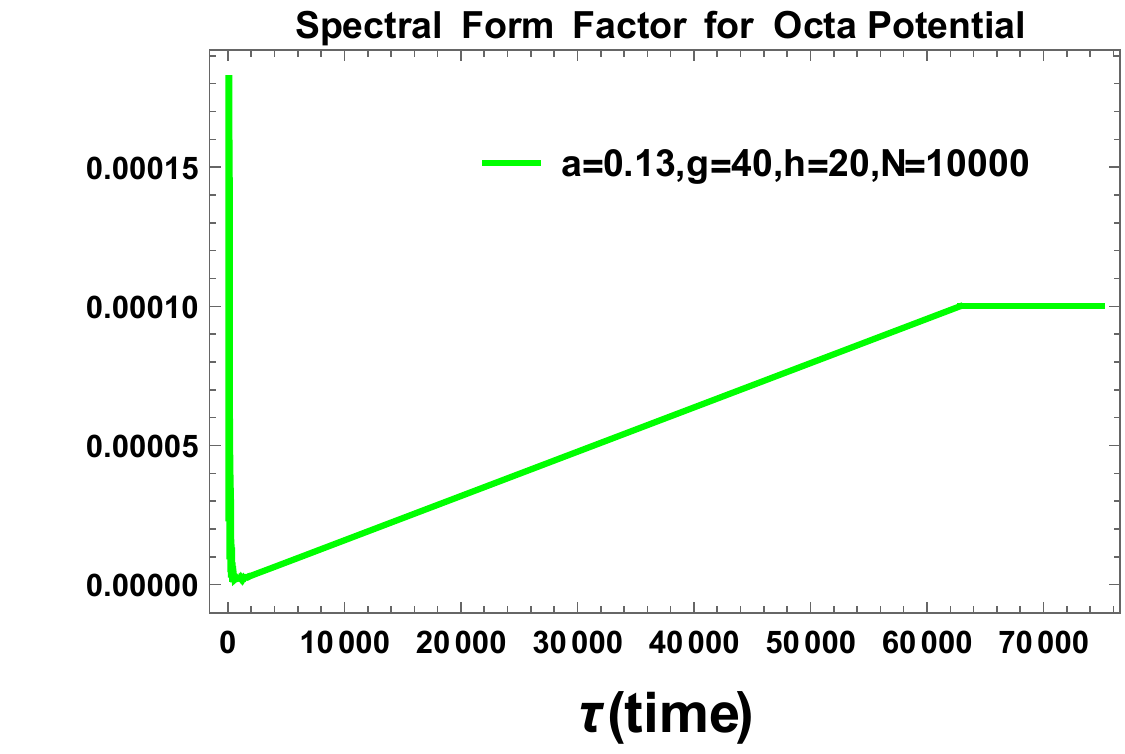}
    \label{SFFO22c}
}

\caption{Time variation of SFF for different N at $\bg=0$.Here we  shift  reference axis[SFF]  to $SFF|_{\tau=0}$  }
\end{figure} 
In fig~\ref{SFFO21c}, fig~\ref{SFFO23c} and fig~\ref{SFFO22c}, it is observed that SFF with variation in $N$ get saturated at different value of $\tau$. But with increasing $N$ the value of the saturation point, will decrease.Subtracting the change of axis[$SFF|_{\tau=0}$] we get the predicted bound of SFF. From these plots we can say that time variation of SFF follow oscillatory pattern initially but after certain time it has linear decaying amplitude for dominance of linear part. Then after $\tau>2\pi N$ region SFF abruptly saturated due to second part of the connected part of the total Green's function $G_{c}$. On the other hand, for $\tau<2\pi N$ region SFF is decaying in amplitude and increasing with time. After $\tau>2\pi N$ region the function will  be constant thereafter. 

Here it is important to note that, depending on the specific structure of the even polynomial random potential the upper bound on chaos very slightly changes (i.e. the amplitude for saturation of SFF is almost at the same order of magnitude for different even polynomial random potentials). But the late time behaviour for different random potentials are almost same as it shows complete saturation with respect to time. The saturation depends only on value of $N$.
Also it is import to note from the plots that, for each even polynomial potential sudden transition from the random oscillatory behaviour to the perfect saturation of SFF take place at the unique time, $\tau=2\pi N$. 

 \subsubsection{Estimation of dip-time scale from SFF}
 Now we introduce the concept of dip-time which denotes the change in fall-off behaviour of SFF near the critical points. It is estimated by comparing the initial fall-off behaviour with late time behaviour of the curve from which it starts the linear increase (ramp part).
 
 For different even polynomial random model (see Eq~(\ref{tgaussian} )~, Eq~(\ref{tquartic}) ~, Eq~(\ref{tsextic}), Eq~(\ref{tocta})), we see that the fall off behaviour varies with $\tau^{-\frac{N}{2}}$. Consequently, the disconnected part of the Green's function ($G_{dc}$) fall off as $\tau^{-N}$ and we tabulated different fall off behaviour with linear increase rate in table.~\ref{falloff}.  Physical time $t$, conformal time $\tau$ and newly define time scale are dined as~\footnote{Here to define the new time scale ${\cal T}$ we assume that the \underline{\textcolor{red}{\bf reheating approximation}} is perfectly valid. This implies that we can really neglect the expansion of the universe. This further pointing towards the fact the conformal time ($\tau$) and the physical time ($t$)                                                                                                                                                                                                                                                                                                                                                                                                                                                                                                                                                                                                                                                                                                                                                                                                                                                                                                                                                                   are related by the following expression:
 \bea \tau=t/a\propto t,\eea
 where during reheating we have assumed that the conformal and physical time are almost same and the proportionality constant is the inverse of the scale factor $a^{-1}$, which is independent of both the time scales discussed in this paper.  . }:
\bea {\cal T}=\sqrt{N}\tau\approx \sqrt{N}\frac{t}{a}\propto \sqrt{N}~t,\eea
 which specifically  depends on different values of $N$, where it represents order of the even polynomial used in our paper to compute {\bf SFF}.
 \bc
 \begin{table}[H]
\small\begin{tabular}{|||c||c||c||c||c||c|||}
\hline\hline\hline
\textcolor{blue}{\bf Potential} & \textcolor{blue}{\bf 1st critical point} & \textcolor{blue}{\bf 2nd Critical point} & \textcolor{blue}{\bf 3rd Critical point} & \textcolor{blue}{\bf 4th critical point} \\
\hline\hline
\textcolor{red}{\bf Gaussian}  & $\tau^{-3}=\frac{\tau}{N^{2}}$ & - & - & -\\
\hline\hline
\textcolor{red}{\bf Quartic} & $\tau^{-3}=\frac{\tau}{N^{2}}$ & $\tau^{-5}=\frac{\tau}{N^{2}}$ & - & -  \\
\hline\hline
\textcolor{red}{\bf Sextic} & $\tau^{-3}=\frac{\tau}{N^{2}}$ & $\tau^{-5}=\frac{\tau}{N^{2}}$ & $\tau^{-7}=\frac{\tau}{N^{2}}$ & - \\
\hline\hline
\textcolor{red}{\bf Octa} & $\tau^{-3}=\frac{\tau}{N^{2}}$ & $\tau^{-5}=\frac{\tau}{N^{2}}$ & $\tau^{-7}=\frac{\tau}{N^{2}}$ & $\tau^{-9}=\frac{\tau}{N^{2}}$ \\
\hline\hline\hline
\end{tabular}
\caption{Fall-off behaviour near critical points for different even polynomial random potential.}
\label{falloff}
\end{table}
\ec
\begin{table}[H]
\centering
\begin{tabular}{|||c||c||c||c|||}
\hline\hline\hline
\textcolor{blue}{\bf Equation of $\tau$ } & \textcolor{blue}{\bf$\tau$ in order of $N$} & \textcolor{blue}{\bf $t$ in order of $N$}  \\
\hline\hline
$\tau^{-3}=\frac{\tau}{N^{2}}$ & $\tau=O(\sqrt{N})$ & $t=O(1)$ \\
\hline\hline
$\tau^{-5}=\frac{\tau}{N^{2}}$ & $\tau=O(N^{\frac{1}{3}})$ & $t=O(N^{-\frac{1}{6}})$ \\
\hline\hline
$\tau^{-7}=\frac{\tau}{N^{2}}$ & $\tau=O(N^{\frac{1}{4}})$ & $t=O(N^{-\frac{1}{4}})$ \\
\hline\hline
$\tau^{-9}=\frac{\tau}{N^{2}}$ & $\tau=O(N^{\frac{1}{5}})$ & $t=O(N^{-\frac{3}{10}})$  \\
\hline\hline\hline
\end{tabular}
\caption{Order of magnitude estimation of conformal time ($\tau$) and physical time ($t$) in terms of the order of polynomial $N$}
\label{t_in_N}
\end{table}
 From table~\ref{falloff}, we get the proper estimation of dip time for different even order polynomial random potential at different critical point. 
 Further, in table~\ref{t_in_N}, we get an order of magnitude estimation of conformal time ($\tau$) and physical time ($t$) in terms of the order of polynomial $N$.

In the next section we will discuss about quantum correction of {\it Fokker-Planck equation} in the context of cosmological particle production. Here it is important to note that, particle production in cosmology can be treated as a chaotic random event and through our calculation we get quantum corrections due to the non-Gauaaian contribution in the probability distribution function. In this context the system can be treated as semiclassical. As a result, SFF shows a saturating behaviour on large time limit which implies that randomness in eigen value density has a upper bound though it is chaotic \cite{Gaikwad:2017odv}. This relate that particle production also can have an upper bound which also confirmed using the computation of {\it Lyapunov exponent}.
 
 \subsection{ Universal bound on quantum chaos from SFF and  its application to cosmology}
 In the previous subsection we have explicitly computed the analytical expression for SFF for generalized even polynomial random potential at finite temperature ($\beta=1/T$=finite) in Eq:.~(\ref{bound1}) and at very high temperature ($\beta=1/T\rightarrow 0$) in Eq.~(\ref{bound2}). Now in this subsection our prime objective is to compute the analytical bound on SFF at long time interval i.e. $\tau\rightarrow\infty$. To derive the bound on SFF  we first use the asymptotic behaviour of Hypergeomteric PFQ regularized function, which is given below:
\bea
\lim_{\tau\to\infty} \, _1\tilde{F}_2\left[-m+n+1;\frac{3}{2},-m+n+\frac{5}{2};a^2 (\beta\pm  i \tau)^2)\right] &=&0~~~\forall k=1,2,\cdots,n,\\
\lim_{\tau\to\infty} \, _1\tilde{F}_2\left[-m+n+\frac{1}{2};\frac{1}{2},-m+n+2;a^2 (\beta\pm  i \tau)^2)\right] &=&0~~~\forall k=1,2,\cdots,n.
\eea
This asymptotic behaviour of the Hypergeomteric PFQ regularized function remains same in the high temperature limit ($\beta=1/T\rightarrow 0$) also.

Consequently, the asymptotic behaviour of the disconnected part of the Green's function can be expressed at finite temperature as well as in the limit $\beta\to 0$ with finite $N$ as:
\be \lim_{\tau\to\infty} G_{dc}(\beta,\tau)=0~~~~\textcolor{red}{\bf\forall  \tau(\to \infty)>2\pi N}.\ee
\be \lim_{\tau\to\infty}  \lim_{\beta\to 0}G_{dc}(\beta,\tau)=0~~~~\textcolor{red}{\bf\forall  \tau(\to \infty)>2\pi N}.\ee
Similarly, the asymptotic behaviour of the connected part of the Green's function can be expressed at finite temperature as well as in the limit $\beta\to 0$ with finite $N$ as:
\be \lim_{\tau\to\infty} G_{c}(\beta,\tau)=\frac{1}{\pi N}~~~~\textcolor{red}{\bf\forall  \tau(\to \infty)>2\pi N}.\ee
\be \lim_{\tau\to\infty}  \lim_{\beta\to 0}G_{c}(\beta,\tau)=0~~~~\textcolor{red}{\bf\forall  \tau(\to \infty)>2\pi N}.\ee
Finally, adding the contribution from the disconnected and connected part of the Green's function in the asymptotic limit ($\tau\to \infty$) we get the following simplified expression for SFF  at finite $N$ as given by:
\be\boxed{ \boxed{{\bf SFF}(\beta,\tau\to \infty)=\lim_{\tau\to\infty}\left( G_{dc}(\beta,\tau)+G_{c}(\beta,\tau)\right)=\frac{1}{\pi N}>0~~~~\textcolor{red}{\bf\forall  \tau(\to \infty)>2\pi N~(Finite~N),~~\beta\leq\infty}}}~.\ee
Also in the high temperature limit with finite $N$ we get:
\be\boxed{ \boxed{{\bf SFF}(\beta\to 0,\tau\to \infty)=\lim_{\tau\to\infty} \lim_{\beta\to 0}\left( G_{dc}(\beta,\tau)+G_{c}(\beta,\tau)\right)=0~~~~\textcolor{red}{\bf\forall  \tau(\to \infty)>2\pi N~(Finite~N),~~\beta\geq0}}}~.\ee
Here we considered the part of SFF only after $\tau>2\pi N$ with finite $N$ as we are considering $\tau\to \infty$ asymptotic limit. The main obstruction of the 
taking $\tau\to \infty$ asymptotic limit in the $\tau<2\pi N$ with finite $N$ divergent contribution in the connected part of the total Green's function $G_c$ as given by:
\be \lim_{\tau\to\infty} G_{c}(\beta,\tau)= \lim_{\tau\to\infty} \left(\frac{\tau}{(2\pi N)^2}-\frac{1}{N}+\frac{1}{\pi N}\right)\to \infty~~~~\textcolor{red}{\bf\forall  \tau(\to \infty)<2\pi N~(Finite~N)}.\ee
On the other hand, for the disconnected part of the Green's function  we get the same result as obtained for $\tau(\to \infty)>2\pi N$ with finite $N$ case.

As a result, it gives divergent contribution to SFF at finite $N$ is given by:
\be \boxed{\boxed{{\bf SFF}(\beta,\tau\to\infty)\to \infty~~~~\textcolor{red}{\bf\forall  \tau(\to \infty)<2\pi N~(Finite~N)}}}~.\ee
For this reason, we will concentrate on the finite contribution on SFF coming from $\tau(\to \infty)>2\pi N$ with finite $N$ region.

Finally, adding both the contribution from connected and disconnected part of the total Green's function for the asymptotic region $\tau(\to \infty)>2\pi N~(Finite~N)$ with $0\leq \beta(=1/T)\leq \infty$ we get the following upper and lower bound on SFF, as given by:
\be\label{e1}  \boxed{\boxed{\textcolor{blue}{\underline{\bf Bound~on~SFF:}}~~~~~~~~0\leq {\bf SFF}\leq \frac{1}{\pi N}~~~~\textcolor{red}{\bf\forall  \tau(\to \infty)>2\pi N~(Finite~N),~~0\leq \beta(=1/T)\leq \infty}}}~.\ee

On the other hand, with large $N$ limit one can consider the $\tau<2\pi N$ region for the computation of the bound on SFF.  To justify this statement we take the $\tau\to \infty$ asymptotic limit in the $\tau<2\pi N$ with large $N$ gives finite contribution in the connected part of the total Green's function $G_c$ as given by:
\be \lim_{\tau\to\infty} G_{c}(\beta,\tau)= \lim_{\tau\to\infty} \left(\frac{\tau}{(2\pi N)^2}-\frac{1}{N}+\frac{1}{\pi N}\right)\simeq-\frac{1}{N}\left(1-\frac{1}{\pi }\right)<0~~~~\textcolor{red}{\bf\forall  \tau(\to \infty)<2\pi N~(Large~N)}.\ee
Similarly, for the disconnected part of the Green's function  we get the same result as obtained for $\tau(\to \infty)>2\pi N$ in previous case.

Finally, adding both the contribution from connected and disconnected part of the total Green's function for the asymptotic region $\tau(\to \infty)<2\pi N~(Large~N)$ with $0\leq \beta(=1/T)\leq \infty$ we get the following upper and lower bound on SFF, as given by:
\be \label{d1} \boxed{\boxed{\textcolor{blue}{\underline{\bf Bound~on~SFF:}}~ -\frac{1}{N}\left(1-\frac{1}{\pi }\right)\leq {\bf SFF}\leq 0~~\textcolor{red}{\bf\forall  \tau(\to \infty)<2\pi N~(Large~N),~~0\leq \beta(=1/T)\leq \infty}}}~.\ee
Combining Eq.~(\ref{e1}) and Eq.~(\ref{d1}) we can write for all range of $\tau$ the following bound on SFF:
\be \label{xc1} \boxed{\boxed{\textcolor{blue}{\underline{\bf Bound ~on~SFF~from~theory:}}~ -\frac{1}{N}\left(1-\frac{1}{\pi }\right)\leq {\bf SFF}\leq \frac{1}{\pi N}~~\textcolor{red}{\bf\forall  \tau,~~0\leq \beta(=1/T)\leq \infty}}}~.\ee
	\begin{figure}[htb]
\centering
\subfigure[For quartic $a=.1,N=1000,\bg=0.001$]{
    \includegraphics[width=7.8cm,height=9cm] {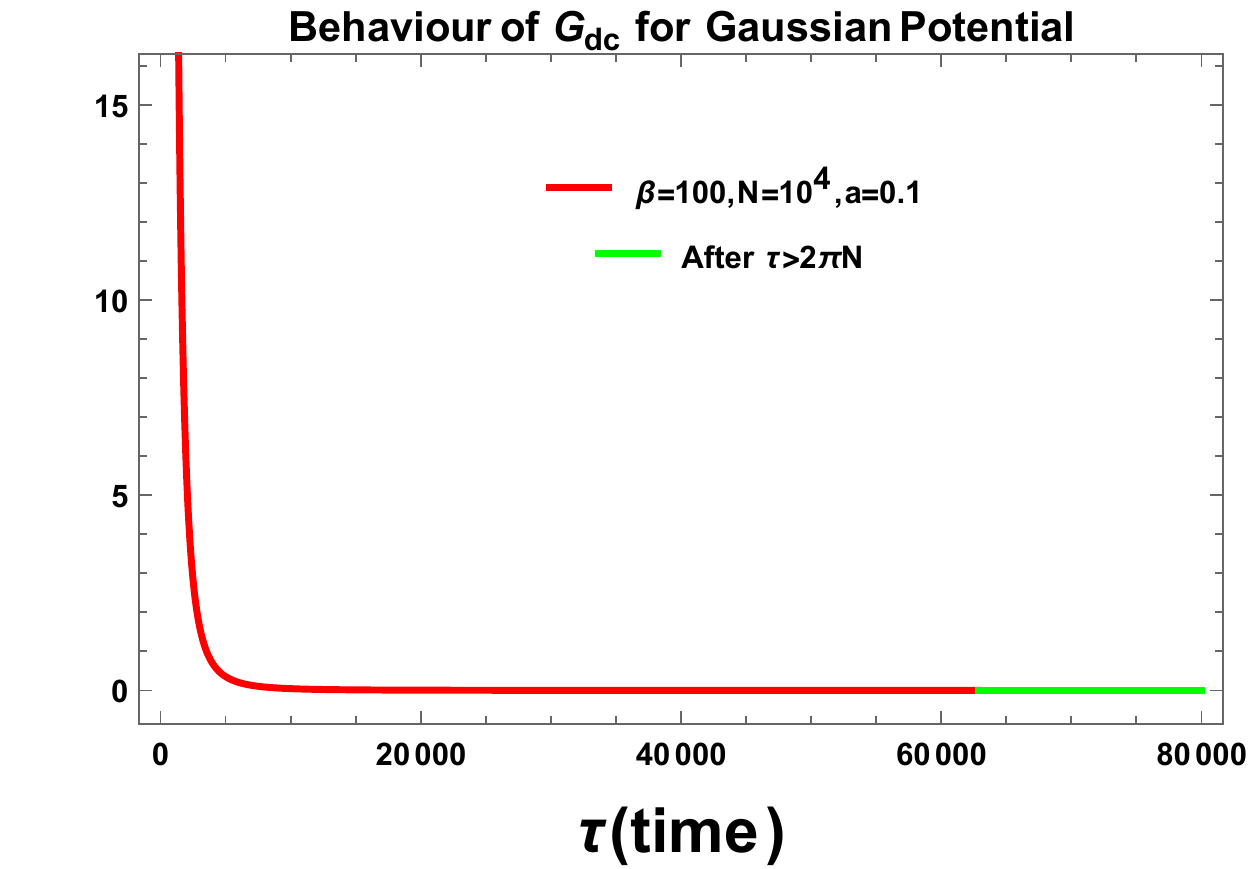}
    \label{GDC1}
}
\subfigure[For gaussian $a=.1,N=10^{3},\bg=1$]{
    \includegraphics[width=7.8cm,height=9cm] {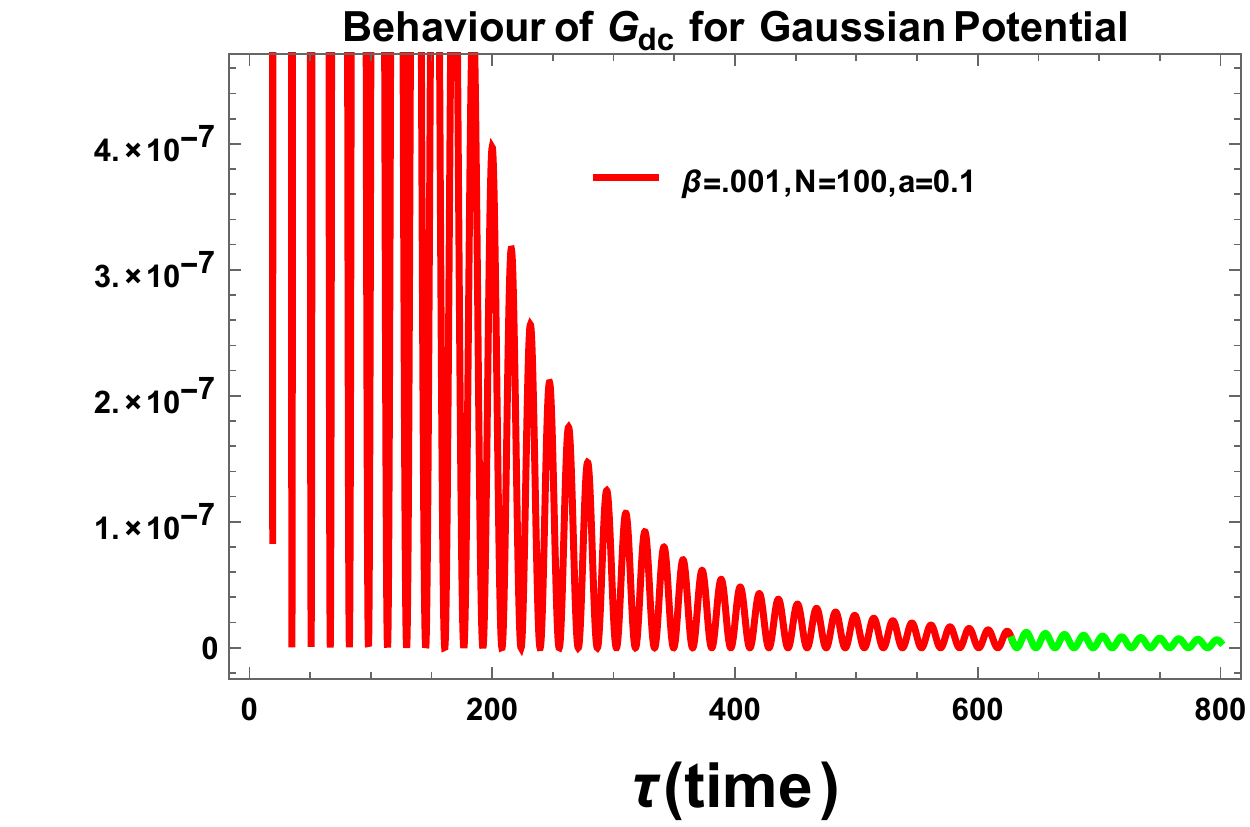}
    \label{GDC3}
}
\subfigure[For quartic $a=.1,N=10,\bg=10$]{
    \includegraphics[width=14.8cm,height=9cm] {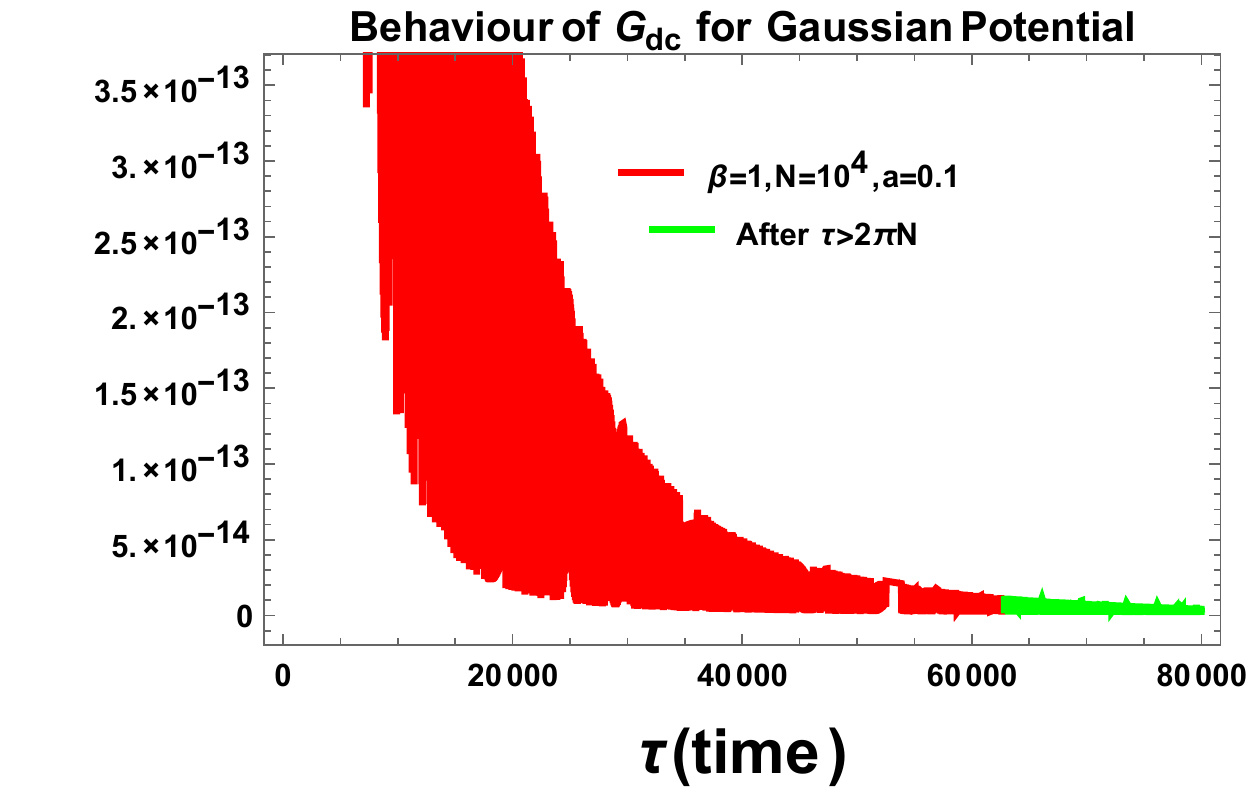}
    \label{GDC2}
}
\caption{$G_{dc}$ at different N and $\bg$}
\end{figure}
In Fig:-\ref{GDC1},\ref{GDC2},\ref{GDC3} we have shown the nature of $G_{dc}$ with different parameter. For all the cases $G_{dc}$ decays to zero at $\tau\rightarrow\infty$ which matches our analytical conclusion.Here we differentiate the $\tau<2\pi N$ and $\tau>2\pi N$ region with different color.
%\begin{figure}[H]
%\centering
%\subfigure[For gaussian $a=.1,N=1000,\bg=0.001$]{
%    \includegraphics[width=7.8cm,height=8.5cm] {SFF_HT_FN2.pdf}
 %   \label{SFFBOUND1}
%}
%\subfigure[For gaussian $a=.1,N=10,\bg=10$]{
%    \includegraphics[width=7.8cm,height=8.5cm] {SFF_MT_LN1.pdf}
 %   \label{SFFBOUND2}
%}
%\subfigure[For gaussian $a=.1,N=10^{3},\bg=1$]{
%    \includegraphics[width=14.8cm,height=8.5cm] {SFF_HT_FN1.pdf}
 %   \label{SFFBOUND3}
%}
%\caption{SFF at finite N and low $\bg$ shows an osscilatory increase with a sharp saturation. $\tau<2\pi N$ and $\tau>2\pi N$ region is marked using different colour.}
%\end{figure}
%\begin{figure}[H]
%\centering
%\subfigure[For gaussian $a=.1,N=1000,\bg=0.001$]{
    %\includegraphics[width=14cm,height=7cm] {SFF_HT_F_N.pdf}
    %\label{SFFsp1}
%}
%\subfigure[For gaussian $a=.1,N=1000,\bg=10$]{
 %   \includegraphics[width=7cm,height=5cm]{SFF_MT1_F_N.pdf}
 %   \label{SFFBOUND4}
%}
%\subfigure[For gaussian $a=.1,N=1000,\bg=1$]{
%    \includegraphics[width=7cm,height=5cm] {SFF_MT_F_N.pdf}
 %   \label{SFFBOUND5}
%}
%\end{figure}
\begin{figure}[htb]
\centering
\subfigure[For gaussian with $a=0.1,N=10^{12},\bg=10$]{
    \includegraphics[width=7.8cm,height=8cm] {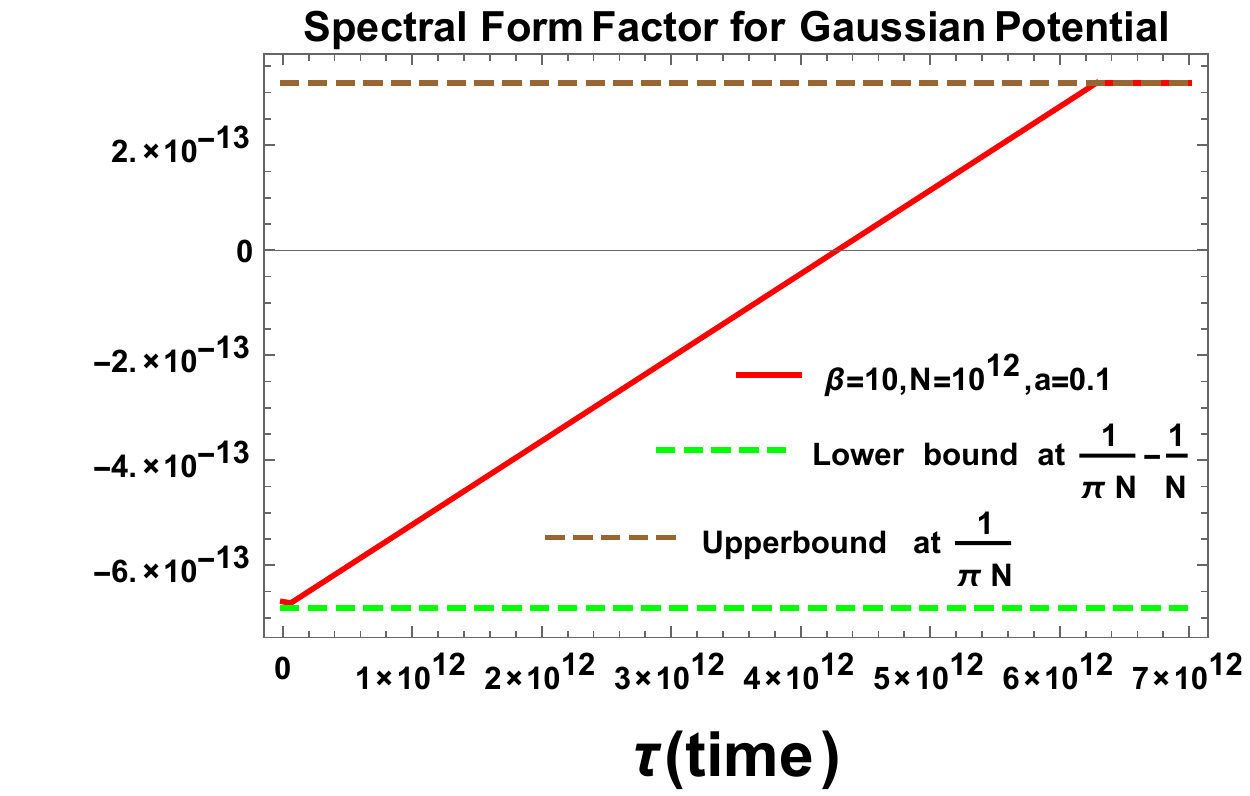}
    \label{SFFbound1}
}
\subfigure[For gaussian with $a=0.1,N=10000,\bg=0.001$]{
    \includegraphics[width=7.8cm,height=8cm] {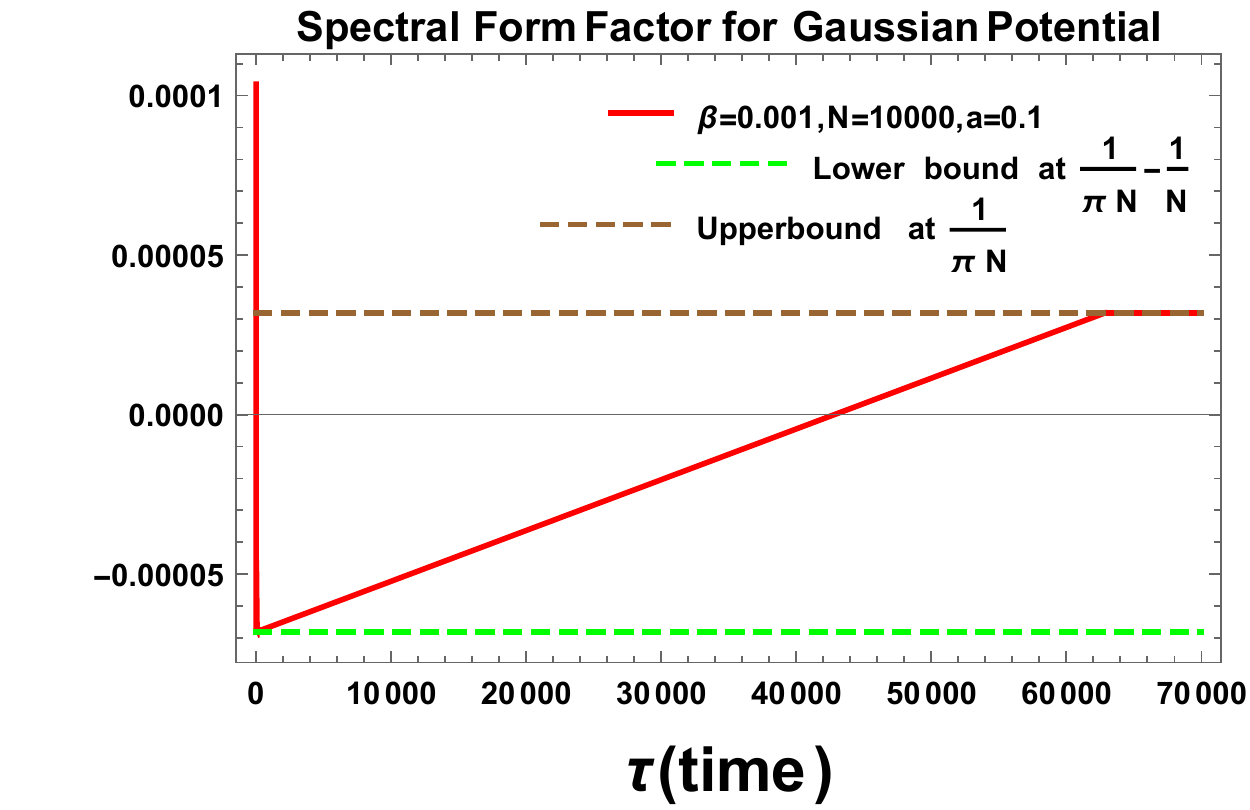}
    \label{SFFbound2}
}
\subfigure[For gaussian with $a=0.1,N=1000,\bg=10$]{
    \includegraphics[width=7.8cm,height=8cm] {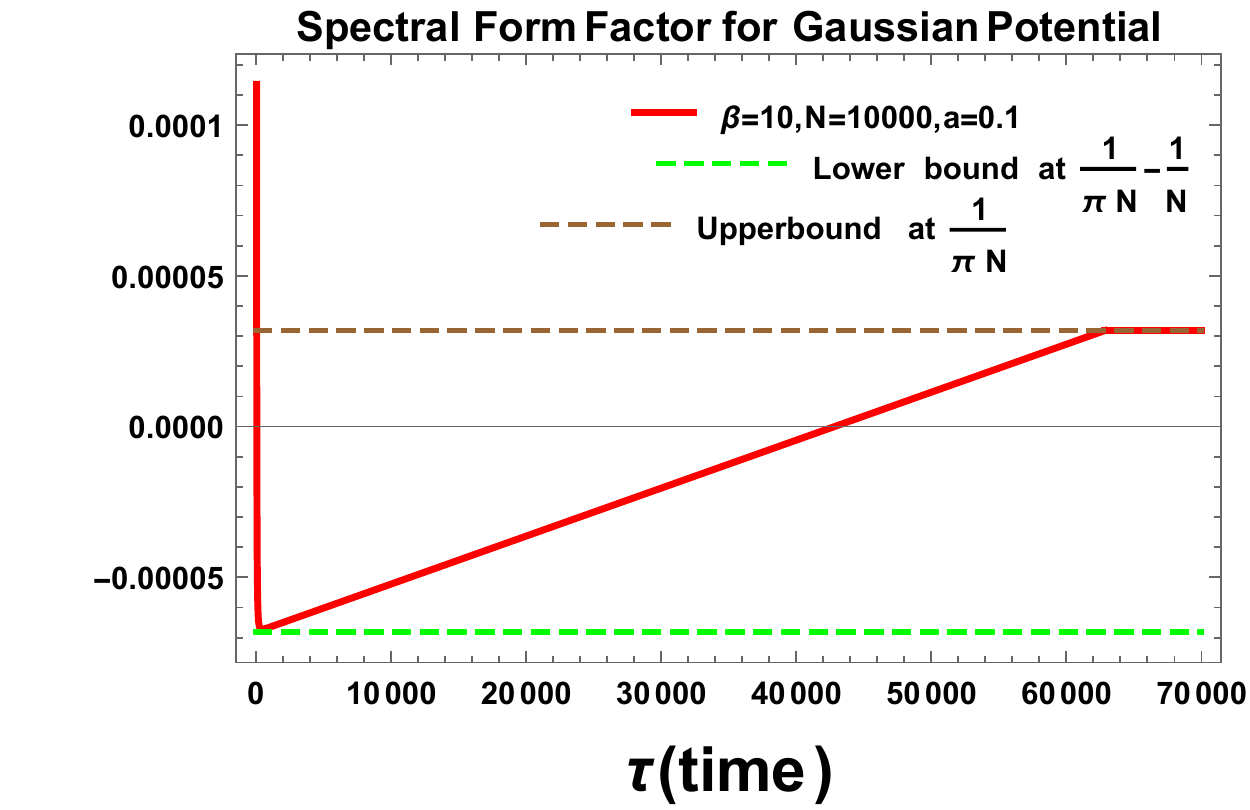}
    \label{SFFbound3}
}
\subfigure[For gaussian with $a=0.1,N=1000,\bg=100$]{
    \includegraphics[width=7.8cm,height=8cm] {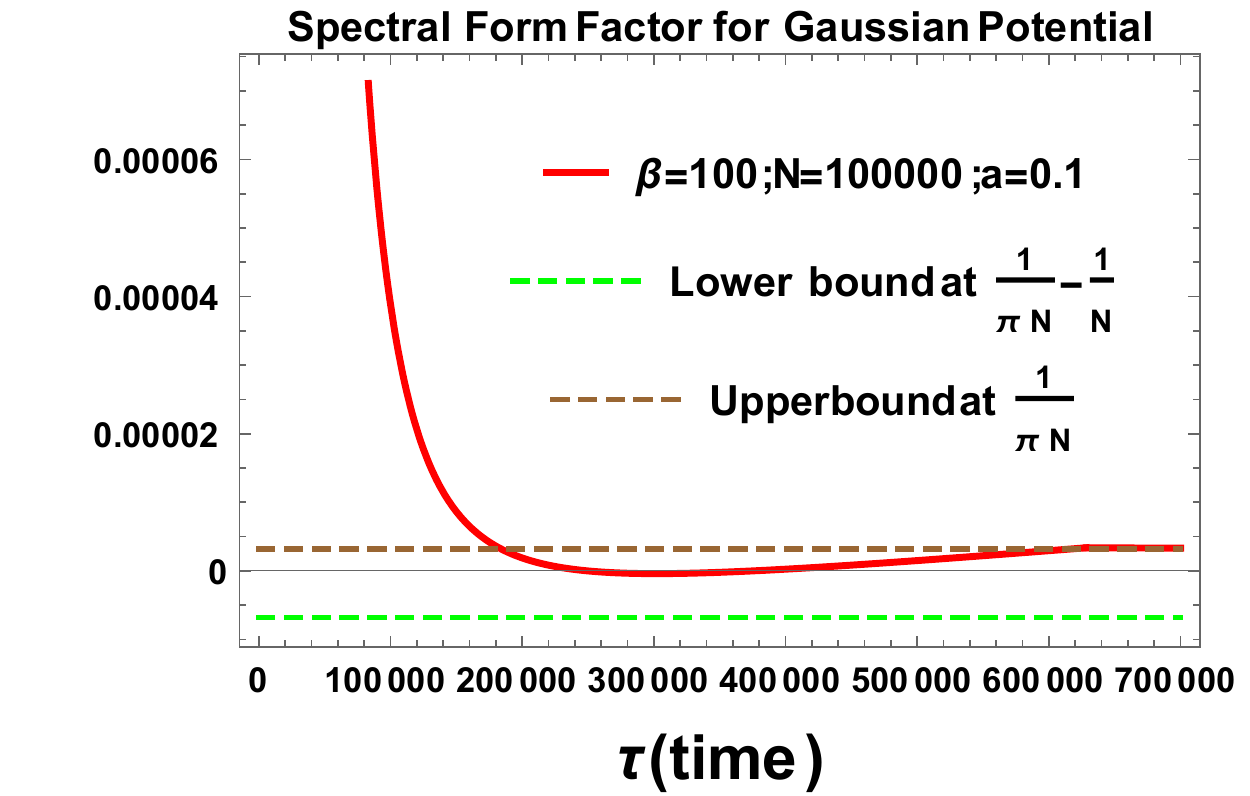}
    \label{SFFbound4}
}
\caption{Different nature of SFF at finite[$\bg\neq 0$] and Infinite temperature[$\bg=0$] }
\label{sffbound}
\end{figure} 
In fig.~\ref{sffbound} we show the behaviour of SFF with temperature and time for large and small N. At higher $\tau$ (Fig.~\ref{SFFBOUND4}),\ref{SFFBOUND5} SFF decays with $\tau$ and at last goes to $\frac{1}{\pi N}$. Else (Fig:-\ref{SFFbound1},\ref{SFFbound2},\ref{SFFbound3},\ref{SFFbound4}) SFF increases  with $\tau$ for $\tau < 2\pi N$ and at last saturate to $\frac{1}{\pi N}$. Now from the analytical solution we know that for large $\tau$ or large $\bg$ $G_{dc}$ doesn't contribute to the SFF as the $G_{c}$ has $\frac{\tau}{(2\pi N)^{2}}$ term.But for higher N and low $\tau$ there should be a minima($\frac{1}{\pi N}-\frac{1}{N}$) for this function within the range $\tau < 2\pi N$. As soon as the point $\tau=2\pi N$ is crossed the function change its form and get saturated. This way of calculation of bound conclude that nature of SFF at late $\tau$ shows same nature independent of type of potential. For infinite temperature SFF saturate at same level and at same $\tau$ value- for different potential with same $N$. For finite temperature it decays to zero irrespective of nature of potential. SFF is a measure of quantum chaos in a dynamical system. Bound on SFF prove that whatever be the interaction, every system at infinite temperature with same N saturated at same value. But at finite temperature randomness in the system decays to zero at late time and the system equilibriate within itself.

Here we consider Gaussian (Eq.~(\ref{cqf1})), quartic (Eq.~(\ref{cqs1}),Eq.~(\ref{cqs3}), Eq.~(\ref{cqs2})), sextic (Eq.~(\ref{cqv1}), Eq.~(\ref{cqw2}), Eq.~(\ref{cqw3})), octa (Eq.~(\ref{cqb1}), Eq.~(\ref{cqm2}), Eq.~(\ref{cqm3})) potential and applying same limit to get the SFF.

Here we have Bessel's Function of first kind ($I_{k}(2a\tau$) in which taking asymptotic time limit we get:
\be
\lim_{\tau\to\infty} \frac{I_{k}(2a\tau)}{\tau^{n}}=0~~~~~\forall ~k=1,2,\cdots, n.
\ee
Here $n$ is order of the polynomial random potential. 
As a result for finite $N$ and large $N$ we get the bound on SFF s mentioned earlier. Here it is important to note that, 
our prescribed bound on SFF is also the same as the saturation value of SFF for different potential for finite $N$ and large $N$.

\section{Randomness from higher order  Fokker-Planck equation: A probabilistic treatment in cosmology}
\label{QuamCorrected} 
\subsection{Cosmological scattering problem}
Here we discuss about the cosmological scattering problem due to the particle creation in the context of early universe physics (mostly during reheating epoch). For detailed derivation of the results see 
refs.~\cite{Amin:2015ftc,Mello1}, which we have followed in this discussion mostly. As we have already discussed in the first half of the paper that, the  Klein-Gordon equation, which is the dynamical master equation of the particles created during reheating can be solved in the same way as Schr{\"o}dinger problem by formulating it as scattering problem in presence of an impurity potential inside a conduction wire\cite{Battefeld:2011yj} and can be related to the phenomena of chaos
 \cite{Bassett:1997gb,Zanchin:1997gf,Green:2014xqa,Elec1}. In this section our prime objective is to establish this connection including the possible quantum effects (corrections) and we will try to develop a formalism to explain the quantum analogue of the chaos during cosmological particle creation.

To serve this purpose let us start with the solution of the Fourier mode of the field (created particle during reheating) after $j$- th non-adiabatic event, which can be expressed as: 
\bea \label{eq1}
\textcolor{blue}{\underline{\bf Fourier~mode~solution~after~j-th~event:}}~~~~x_{j}(\ta)= \frac{1}{\sqrt{2\pi}}[\bg_{j}e^{ik\ta}+\ag{j}e^{-ik\ta}],
\eea
where $\bg_{j}$  and  $ \ag_{j}$ are the Bogoliulov coefficients. For the
vacuum solution we set the initial condition as:
\bea \textcolor{blue}{\underline{\bf Vacuum~ initial~ condition:}} ~~~ \bg_{0} = 0 , ~\ag_{0} =e^{i\del}~~~~{\rm at ~\tau=0},\eea 
where $\delta$ is a phase factor. Now the vacuum initial condition implies:
\bea \label{eqw}
 x_{j}(\ta=0)=x_0= \frac{1}{\sqrt{2\pi}}[\bg_{0}+\ag_{0}].
\eea
In this context the Bogoliulov coefficients satisfy following normalization condition:
\be
\boxed{\textcolor{blue}{\underline{\bf Normalization:}}~~W[x_{j}x_{j}^{*}]=\left(x_j \frac{dx^{*}_j}{d\tau}-x^{*}_j \frac{dx_j}{d\tau}\right)=i \Longrightarrow  |\ag_{j}|^{2}-|\bg_{j}|^{2}=1~\forall j=1,2,\cdots,N}~ .\ee
This is analogous with scattering in presence of an impurity inside the conduction wire. Here we can relate Bogoliubov coefficients before and after the non-adiabatic event by transfer matrix as:
\be\label{eq2}
 \boxed{\underbrace{\left( {\begin{array}{c}
  \bg_{j}   \\
  \ag_{j}
  \end{array} } \right)}_{\textcolor{red}{\bf Co-efficient~matrix~for~j-th~event}}=
  \underbrace{\left( {\begin{array}{cc}
  M_{11} & M_{12}   \\
  M_{21} & M_{22}
  \end{array} } \right) }_{\textcolor{red}{\bf Transfer~matrix~\equiv M_j}}~~~
 \underbrace{\left({\begin{array}{c}
   \bg_{j-1}   \\
   \ag_{j-1}
  \end{array} } \right)}_{\textcolor{red}{\bf Co-efficient~matrix~for~(j-1)-th~event}}\forall j=1,\cdots,N}
\ee
When the wavelength of incoming mode is much larger than coherence interval of the non-adiabatic event, then the time dependent  mass profile evolution $m^2(\tau)$, can't be resolved in wave.
For this purpose, we take the following Dirac Delta profile of time dependent mass function:
\be \boxed{m^{2}(\tau)=\sum_{j=1}^{N} m_{j}\del_{D}(\tau-\tau_{j})}~,\ee
which is localized at time $\tau=\tau_j$. Here $j$ represents the number of non-adiabatic events and the total number of the events can be expressed as:
\be \sum_{j=1}^{N}1=N\ee
Now, it is important to note that, these solutions will satisfy the following two fold junction conditions:
\bea \label{eq4}
 \textcolor{red}{\underline{\bf Condition~ I:}}~~~~x_{j}(\tau_{j})&=& x_{j-1}(\tau_{j} )\\
 \label{eq5}
\textcolor{red}{\underline{\bf Condition~ II:}}~~~~ x_{j}^{'}(\tau_{j})&=&x_{j-1}^{'}(\tau_{j})-m_{j} x_{j-1}(\tau_{j})
\eea
In the present context, the transfer matrix $M_j$ can be expressed as:
\bea M_{j}=\left( {\begin{array}{cc}
  M_{11} & M_{12}   \\
  M_{21} & M_{22}
  \end{array} } \right)={\bf I}+i\lb_{j}
  \left({\begin{array}{cc}
  1 & e^{-2ik \tau_{j}}   \\
   -e^{2ik \tau_{j}}& -1
  \end{array} } \right),
 \eea
 where $\lb_j$ is defined as:
 \be  \lb_{j}=\frac{m_{j}}{2k}.\ee
Therefore in this computation the transmission and reflection co-efficient can be expressed as:
\bea T_j=|t_{j}|^{2}=\frac{1}{(1-i\lb_{j})(1+i\lb_{j})}=\frac{1}{1+\lb_{j}^{2}}=\frac{1}{1+\frac{m_{j}^{2}}{4k^2}}\\ R_j=|r_j|^2=\frac{-(i\lb_j)(i\lb_j)}{(1-i\lb_{j})(1+i\lb_{j})}=\frac{\lb^2_j}{1+\lb_{j}^{2}}=\frac{\frac{m_{j}^{2}}{4k^2}}{1+\frac{m_{j}^{2}}{4k^2}}.\eea
Further, the local change in occupation number $ n_{j}$ can be written 
in terms of transmission co-efficient as:
\bea n_{j}=T^{-1}_{j}-1=\lb_{j}^{2}=\frac{R_j}{T_j}=\frac{m_{j}^{2}}{4k^2}.~\eea
Further, assuming the local change of occupation number is large only for $ k << m_{j}$ i.e. $\lb_j>>1/2$ then the transfer matrix can be simplified to the following polar form as:
\bea \label{eq7}
M_j=
  \left( {\begin{array}{cc}
   e^{i \thg_j } \sqrt{1+n_j} & e^{i(2 \phi_j - \thg_j) } \sqrt{1+n_j}   \\
   e^{-i(2 \phi_j - \thg_j) } \sqrt{1+n_j}  & e^{-i \thg_j } \sqrt{1+n_j}
  \end{array} } \right)
\eea
where the phase factors $\theta_j$ and $\phi_j$ are defined as: 
\bea
 \thg_{j}&=&\tan^{-1}(\lb_{j})=\tan^{-1}\left(\frac{m_j}{2k}\right),\\
 \phi_{j}&=&tan^{-1}(\lb_{j}) -k\tau_{j}+\frac{\pi}{4}=\tan^{-1}\left(\frac{m_j}{2k}\right) -k\tau_{j}+\frac{\pi}{4}.\eea
 Further, using the transfer matrix in polar form
we define the transmission, reflection probability and the total occupation number as:
\bea
 t_j&=&e^{i \thg_j } \sqrt{(1+n_j)^{-1}}=\sqrt {T_j} ~e^{i\thg_j},\\
 r_j&=&-\sqrt{n_j(1+n_j)^{-1}}e^{2i(\thg_j-\phi_j)}=-\sqrt{1-T_j} e^{2i(\thg_j-\phi_j)}=-\sqrt{R_j} e^{2i(\thg_j-\phi_j)},\\
 n_j&=&T^{-1}_j-1=\frac{R_j}{T_j}.
\eea
Additionally, it is important to note that, the total occupation number before ($n(j)$) and after ($n(j-1)$) the $j$-th scattering are related by the following expression:
\bea\label{rec1x}
n(j)&=&n(j-1)+\lb_{j}^{2}\left[1+2n(j-1)+2\sqrt{n(j-1)[1+n(j-1)]}\right]\cos{\Del_{j}}\nonumber\\&&~~~~~~~~~~~~~~~~~~~~~~~~~~~~~~~~~~+2\lb_{j}\sqrt{n(j-1){(1+n(j-1))}}\sin{\Del_{j}}\,\eea
where $ \Del_{j}$ is the phase factor which is defined in terms of the Bogoliubov coefficients of the $(j-1)$~th events as:
\be  \Del_{j}\equiv=-arg[\ag_{j-1}]+arg[\bg_{j-1}]-2k\tau_{j}.\ee
Now the vacuum  initial condition demands that, \be  n(0)=0,\ee
which further implies the following equation for $n(-1)$:
\bea\label{sdxx1}
A~n^2(-1)
+B~n(-1)+C&=&0. ~~~~~~\eea
where we define $A$, $B$ and $C$ as:
\bea  A&=&\left[\left(1+2\lb_{0}^{2}\cos{\Del_{0}}\right)^2-4\lb^2_0 \left(\lb_{0}\cos{\Del_{0}}+\sin{\Del_{0}}\right)^2\right],\\
B&=&\left[2\lb_{0}^{2}\cos{\Del_{0}}\left(1+2\lb_{0}^{2}\cos{\Del_{0}}\right)-4\lb^2_0 \left(\lb_{0}\cos{\Del_{0}}+\sin{\Del_{0}}\right)^2\right],\\
C&=&\lb_{0}^{4}\cos^2{\Del_{0}}.\eea
Using Eq~(\ref{sdxx1}) the solution for $n(-1)$ can be written as:
\bea n(-1)=\frac{1}{2A}\left[-B\pm \sqrt{B^2-4AC}\right].\eea
Similarly using Eq~(\ref{rec1x}) recursively one can find out the expressions for occupation number for many scattering processes. 

Alternatively, using the concept of transfer matrices one can also compute the occupation number in the present context. To serve this purpose one can first write down the total transfer matrices for $N_s$ number of scatterer as:
\bea M(N_{s})=\prod^{N_s}_{i=1}M_{i}=M_{N_{s}}...M_{2}M_{1}.\eea
Using this the total occupation number can be expressed as:
\bea n(N_{s})=[M(N_{s})]_{11}^{*}M(N_{s})_{11}-1.
\eea
To model a phenomenological situation where width is finite, the scattering event is relevant and consider  "sech"  scatterers:
\bea m^{2}(\tau)=\sum_{j=1}^{N}\frac{ m_{j}}{2w_{j}}sech^{2}\left(\frac{\tau-\tau_{j}}{w_{j}}\right).
\eea
Now, if we take the limit $w_j\to \infty$ then we get back the Dirac Delta mass profile as given by:
\bea m^{2}(\tau)=\lim_{w_j\to \infty}~\sum_{j=1}^{N}\frac{ m_{j}}{2w_{j}}sech^{2}\left(\frac{\tau-\tau_{j}}{w_{j}}\right)=\sum_{j=1}^{N} m_{j}\del_{D}(\tau-\tau_{j}).\eea
Further, using the results obtained for transmission co-efficient we finally get the following simplified expression for the total occupation number:
\bea n_{j}=\frac{\cos^{2}(\frac{1}{2}\pi\sqrt{1+2m_{j}w_{j}})}{\sinh^{2}(\pi k w_{j})}.
\eea

\subsection{Fokker Planck Equation}
In this subsection our prime objective is to construct {\it Fokker Planck equation} from the basic principles. To serve this purpose we start with the concept of  probability density, which can be expressed in terms of {\it Smoluchowski equation}:
\bea &&\textcolor{blue}{\bf  \underline{Smoluchowski~Equation:}}~\nonumber\\
&&P(M;\tau+\del\tau)=\int_{-\infty}^{\infty} P(M_{1},\tau )P(M_{2},\del\tau) d M_{2}=\langle P(M_{1},\tau)\rangle_{M_{2}}
\eea
 It actually explain the probability density for particle position of Brownian motion in a random system. For a Markovian process one can further express this in terms of {\it Chapman-Kolmogorov equation} where the probability density is conditional. It is important to note that, {\it Smoluchowski equation} describes a two point conditional probability distribution satisfying the following criteria:
\be P_{2}(Y_{1},t_{1}|Y_{3},t_{3})=\int_{-\infty}^{\infty}dY_2~P_{2}(Y_{1},t_{1}|Y_{2},t_{2})P_{3}(Y_{1},t_{1};Y_{2},t_{2}|Y_{3},t_{3})~~~~~{\rm for}~t_1<t_2<t_3.\ee
where we have added a small interval $\del\tau$ to an existing interval $\tau$ to construct the probability density function. in such a situation the transfer matrix for the elongate interval can be expressed as:
\bea M=M_{\del\tau}M_{\tau}=M_{\tau+\del\tau}.
\eea
Here it is important to note that, to write this expression we have used the following set of rules:
\begin{enumerate}
\item First of all, we identify, $M_{1}=M_{\tau}$ and  $M_{2}=M_{\del\tau}$.
\item Then we apply the composition law \bea  M=M_{2}M_{1}.\eea
\item Finally, we write $M_{1}=M_{\tau}$ as: 
\bea M_{1}=M_{2}^{-1}M=M+\del M(M,M_{2}).\eea
This implies:
\bea \del M=(M^{-1}_2-1)M.\eea
\end{enumerate}
Then, the time evolution of the probability density function can be expressed as:
\bea\label{eq8}
\pl_{\tau} P(M,\tau)=\frac{\langle \del M\rangle_{M_{2}}}{\del\tau} \pl_{M} P(M,\tau) + \frac{\langle \del M \del M\rangle_{M_{2}}}{\del\tau} \pl_{M} \pl_{M} P(M,\tau)+...
\eea
This gives {\t Fokker Planck equation} upto different order for occupation number $n$ after appropriate parameterization of the transfer matrices and a marginalization over certain parameters.

Additionally it is important to note that, in order to reduce the complexity of the computation we suppress the wave number ($k$) dependence in our obtained results. As a consequence, the {\it Smoluchowski equation} as stated in Eq~(\ref{eq1}) can be simplified to the following form:
\bea\label{eq11}
P({n,\thg,\phi};\ta+\del\ta)=\int d n_{2} \frac{d \phi _{2}}{2\pi}\frac{d \thg _{2}}{2\pi} ~P({n_{1},\thg_{1},\phi_{1}};\ta) P({n_{2},\thg_{2},\phi_{2}};\del\ta)\,   
		=\langle P({n_{1},\thg_{1},\phi_{1}};\ta)\rangle_{\del\ta}.
\eea
Now, we Taylor expand both side after writing $ [n_{1},\thg_{1},\phi_{1}] $ in terms of $ [n,\thg.\phi] $.

Here one can write the occupation umber $n_1$ as:
\bea\label{eq13}
n_{1}&=&T^{-1}_1-1=[M_{1}]_{11}^{*}[M_{1}]_{11}-1 \equiv n+\del n,
\\ \label{eq14}
\thg_{1}&=&-\frac{i}{2} ln\left[\frac{[M_{1}]_{11}}{[M_{1}]_{11}^{*}}\right]\equiv \thg+\del\thg.
\eea
where the perturbed part of the occupation number $\delta n$ can be expressed as:
\bea\label{eq16}
\del n \equiv n_{2}(1+2n)-2 \sqrt{(1+n_{2})(1+n)n_{2}n} \cos{2(\phi _{2}-\thg)\equiv f(\phi _{2}-\thg)} .
\eea
In this context, the right hand side of the above equation represents the perturbed Hamiltonian. Here we use perturbation theory to find the eigenvalues of occupation number $n_1$ in terms of the eigenvalues of occupation number $n$ and the matrix elements of the perturbed part $\delta n$ in the preferred choice of basis which diagonalizes the matrix $n$.  Additionally, it is important to note that, the explicit expression for the perturbed angular parameter $\delta \theta$ is not very significant for our discussion. Instead of this, the angular difference $(\phi _{2}-\thg)$
play crucial role to quantify the perturbed contribution to the occupation number.

Now, the conditional probability of getting $Y$ at time $t+\tau$ in terms of probability of getting nearby to $Y-\xi $ at time $t$ and then to $Y$ in time $\tau$ is given by:
\bea
\label{eq15a}
P_{2}(Y_{0}|Y,t+\tau)=\int_{-\infty}^{\infty}d\xi ~P_{2}(Y_{0}|Y-\xi,t)P_{2}(Y-\xi|Y,\tau).\eea
On the other hand, using Taylor series expansion of $P_{2}(Y_{0}|Y,t+\tau)$ around $\tau=0$ we get:
\bea\label{eq15v} \boxed{P_{2}(Y_{0}|Y,t+\tau)\approx P_{2}(Y_{0}|Y,t)+\frac{\pl P_{2}(Y_{0}|Y,t)}{\pl t}\tau}~.
\eea
Here it is important to note that, in Taylor series expansion of the probability density function $P_{2}(Y_{0}|Y,t+\tau)$ we truncate the series by considering upto the second term in the series.

Further, comparing the right hand sides of Eq~(\ref{eq15a}) and Eq~(\ref{eq15v}), we finally get the second term of the Taylor expansion as given by:
\bea \label{xac11}\frac{\pl P_{2}(Y_{0}|Y,t)}{\pl t}\tau = -P_{2}(Y_{0}|Y,t)+\int_{-\infty}^{\infty}P_{2}(Y_{0}|Y-\xi,t)P_{2}(Y-\xi|Y,\tau) d \xi\eea
Next using Eq~(\ref{eq15a}) in Eq~(\ref{xac11}) we get the following simplified result for the second term of the Taylor expansion as given by:
\bea\label{scatta1} \frac{\pl P_{2}(Y_{0}|Y,t)}{\pl t}\tau &=&- \int_{-\infty}^{\infty}P_{2}(Y_{0}|Y,t)P_{2}(Y-|Y-\xi,\tau) d \xi  \nonumber\\ 
&&~~~~~~~~~+ \int_{-\infty}^{\infty}P_{2}(Y_{0}|Y-\xi,t)P_{2}(Y-\xi|Y,\tau) d \xi\eea
Now in this context the normalization condition for the probability density function can be written as:
\bea  \int_{-\infty}^{\infty}P_{2}(Y|Y-\xi,\tau) d \xi = 1.\eea
 From Eq~(\ref{scatta1}) we observe that it has {\it scattering out} and {\it scattering in} contributions respectively.

Now, we can determine $ P_{2} \equiv P({n_{2},\thg_{2},\phi _{2}};\del\tau)$  by the condition that it maximizes the Shannon entropy: 
\bea   &&\textcolor{blue}{\bf \underline{Shannon~Entropy:}}\nonumber\\
S&=&-\langle \ln P_{2}({n_{2},\thg_{2},\phi _{2}};\del\tau) \rangle_{\del\tau} -\underbrace{g_1\left[\langle 1 \rangle_{\delta\tau}-1\right]}_{\textcolor{red}{\bf Constraint~I}}-\underbrace{g_2\left[\langle n_2\rangle_{\delta\tau}-\mu\delta\tau\right]}_{\textcolor{red}{\bf Constraint~II}}+\underbrace{g_3\left[ \langle U(\theta_2) \rangle_{\delta\tau}-\alpha \delta\tau\right]}_{\textcolor{red}{\bf Constraint~III}}~.\nonumber\\ &&
\eea
using the principles of {\it maximum entropy ansatz}. In the above expression, $g_1, g_2$ and $g_3$ are the {\it Lagrange multipliers}. Here it is important to note that, $U(\theta_2)$ is an arbitrary function of $\theta_2$ which has an extremum at the location $\theta_2=0$ and can be explicitly determined by imposing 
additional constraint conditions i.e. symmetry arguments, consistency requirements and available knowledge of the microscopic sector of the system under consideration.

To apply the concept of {\it maximum entropy ansatz} we choose the following set of constraints, which are helpful to minimize {\it Shannon entropy} in the present computation:
\begin{enumerate}
\item  \textcolor{red}{\underline{\bf Constraint I:}}\newline 
First of all, we talk about the \underline{ Constraint I}, which will fix the normalization condition of the probability density distribution as given by:
\bea \langle 1 \rangle_{\delta\tau}=1.\eea
This is obtained by setting the co-efficient of the {\it Lagrange multiplier} $g_1$ to zero.
\item \textcolor{red}{\underline{\bf Constraint II:}}\newline 
Using the \underline{ Constraint II}, it is possible  to fix the local mean particle production rate , which is quantitatively defined as:
\bea \label{ra1}\frac{\langle n_{2}\rangle_{\del\tau}}{\del\tau} = \mu. \eea
This is obtained by setting the co-efficient of the {\it Lagrange multiplier} $g_2$ to zero.
\item  \textcolor{red}{\underline{\bf Constraint III:}}\newline
 Finally the \underline{ Constraint III} demands that:
 \bea \lim_{\delta\tau\to 0}M_{\tau+\del\tau} \rightarrow M_{\tau},\eea 
 which basically implies that the addition of infinitesimal interval can't correspond to a finite significant change in transfer matrix.
 
 To establish this statement we start with the following transfer matrix written in the polar form for $j=2$:
 \be \label{eqpol}
M_2=
  \left( {\begin{array}{cc}
   e^{i \thg_2 } \sqrt{1+n_2} & e^{i(2 \phi_2 - \thg_2) } \sqrt{1+n_2}   \\
   e^{-i(2 \phi_2 - \thg_2) } \sqrt{1+n_2}  & e^{-i \thg_2 } \sqrt{1+n_2}
  \end{array} } \right)
\ee
 Now in the limit $\delta\tau\to 0$ we have:
 \bea  \lim_{\delta\tau\to 0}n_2=0,~~~~~~\lim_{\delta\tau\to 0}e^{\pm i\theta_2}=1,~~~~~\lim_{\delta\tau\to 0}e^{\pm i(2\phi_2-\theta_2)}=0.\eea
 Consequently the transfer matrix can be simplified as:
 \bea \label{eqpol}
 \lim_{\delta\tau\to 0} M_2=
  \left( {\begin{array}{cc}
   1~~~ &~~~ 0   \\
  0~~~  &~~~ 1
  \end{array} } \right)={\bf I}.
\eea
To impose this specific non-trivial constraint we assume that the following condition is satisfied:
\bea \label{ra2}\langle U(\theta_2)\rangle_{\delta\tau}=\alpha\delta\tau\Longrightarrow \lim_{\delta\tau\to 0} \langle U(\theta_2)\rangle={\rm fixed},\eea
where $ U(\theta_2)$ is a real valued and positive definite arbitrary function. This is possible if the function $U(\theta_2)$ has an extremum at $\theta_2=0$ where $e^{i\theta_2}=1$. One can choose various types
of function which can satisfy these constraints. For an example, as a phenomenological choice one can consider the following functional form:
\bea  U(\theta_2)=\left[(e^{i\theta_2}-1)(e^{-i\theta_2}-1)\right]^{p}=|e^{i\theta_2}-1|^{2p}=4\sin^{2p}\frac{\theta_2}{2}~~\forall~p=1,2,3,\cdots.\eea 
As a result, the probability density function reaches its maximum at $\theta_2=0$ when the time interval $\delta\tau\to 0$. Further, extremizing the expression for the {\it Shannon entropy} we get the following expression for the probability density distribution function 
 $P({n_{2},\thg_{2},\phi _{2}};\del\tau)$ as given by:
 \bea P_2=P({n_{2},\thg_{2},\phi _{2}};\del\tau)=\left(\frac{1}{{\cal K}(g_2)}~e^{-g_2n_2}\right)\left(\frac{1}{{\cal K}(g_3)}~e^{-g_3 U(\theta_2)}\right),\eea
 where we introduce two new functions ${\cal K}(g_2)$ and ${\cal K}(g_3)$ which are defined as:
 \bea  {\cal K}(g_2)&\equiv&  \int^{\infty}_{0} dn_2~e^{-g_2n_2}=\frac{1}{g_2},\\
  {\cal K}(g_3)&\equiv&  \int\frac{d\theta_2}{2\pi}~e^{-g_3 U(\theta_2)}.
 \eea
 Further using Eq~(\ref{ra1}) and Eq~(\ref{ra2}), we get the following simplified expression for the probability density function:
  $P({n_{2},\thg_{2},\phi _{2}};\del\tau)$ as given by:
\bea  \textcolor{blue}{\bf \underline{Maximum~Entropy~Ansatz:}}~~~~~~~P_2&=&P({n_{2},\thg_{2},\phi _{2}};\del\tau)\nonumber\\
 &=&\left(\frac{1}{\mu\delta\tau}~e^{-\frac{n_2}{\mu\delta\tau}}\right)\left(\frac{1}{{\cal K}(\alpha \delta\tau)}~e^{-g_3 \alpha\delta\tau U(\theta_2)}\right)\nonumber\\
 &=&P(n_2;\delta\tau)P(\theta_2;\delta\tau)\nonumber\\
 &=&P(n_2,\theta_2;\delta\tau),\eea
which implies that the probability density function is independent of  $\phi_{2}$ after applying the {\it maximum entropy ansatz}. For weak scattering, this corresponds to scattering events being uniformly distributed. Now if we consider large number of scatterings, then applying Central Limit Theorem one can show that the final result is not sensitive to the probability density function $P_2$. In this discussion we have explicitly provided the mathematical form of the probability density function, which is not very important to derive the {\it Fokker-Planck equation}.
\end{enumerate}
Now if we use the fact that the probability density distribution function 
$P_2$ is completely independent of $\phi_2$ one can further express the {\it Smoluchowski equation} in the following simplified form:
\bea 
P(n,\theta,\phi;\tau+\del\tau)\equiv P(n,\theta;\tau+\del\tau)&=&\int P(n,\theta,\tau )P(dn+dn^{'},d\theta+d\theta^{'};\del\tau) dn^{'}d \theta^{'}\nonumber\\
&=&\langle P(n+\delta n,\theta+\delta\theta;\tau)\rangle_{\delta\tau}
\eea
Further, integrating both sides of the above equation with respect to the parameter $\thg$ we get the following simplified expression:
\bea \label{eq17}
P(n;\tau+\delta\tau)&=\int d\theta~P(n,\theta;\tau+\del\tau) \nonumber\\
&=&\int d\theta~ \langle P(n+\del n,\theta+\delta\theta ;\tau)\rangle_{\del\tau}\nonumber\\
&=&\langle P(n+\del n ;\tau)\rangle_{\del\tau}
\eea
where during performing the integration over $\theta$ we explicitly use the information that the infinitesimal change in $\theta$ i.e. $\delta\theta$ is not functionally dependent on $\theta$.

Now, using Taylor expansion of $\langle P(n+\del n ;\tau)\rangle_{\del\tau}$ with respect to the infinitesimal occupation number $\delta n$ we get:
\bea \label{asq1}  \langle P(n+\del n ;\tau)\rangle_{\del\tau}&=&\langle P(n;\tau)\rangle_{\del\tau}+                                                                                                                                                                                                                                                                                                                                                                                                                   \sum^{\infty}_{q=1}\frac{1}{q!}\frac{\partial^q P(n;\tau)}{\partial n^q}\langle (\delta n)^q\rangle_{\del\tau}\nonumber\\
&=&\langle P(n;\tau)\rangle_{\del\tau}+                                                                                                                                                                                                                                                                                                                                                                                                                    \frac{\partial P(n;\tau)}{\partial n}\langle \delta n\rangle_{\del\tau}+\frac{1}{2!}\frac{\partial^2 P(n;\tau)}{\partial n^2}\langle (\delta n)^2\rangle_{\del\tau}+\cdots\nonumber\\
&=&\langle P(n;\tau)\rangle_{\del\tau}+                                                                                                                                                                                                                                                                                                                                                                                                                    \left\{\frac{\partial P(n;\tau)}{\partial n}\frac{\partial\langle \delta n\rangle_{\del\tau}}{\partial\tau}+\frac{1}{2!}\frac{\partial^2 P(n;\tau)}{\partial n^2}\frac{\langle (\delta n)^2\rangle_{\del\tau}}{\delta\tau}\right\}\delta\tau+\cdots,~~~~~~~~ \eea
where in this context we have:
\bea  \langle P(n;\tau)\rangle_{\del\tau}=P(n,\tau).\eea
On the other hand, taking Taylor expansion of the probability density function $P(n;\tau+\delta\tau)$ with respect to the infinitesimal time interval $\delta\tau$ we get:
\bea \label{asq2} P(n;\tau+\delta\tau)&=&P(n;\tau)+                                                                                                                                                                                                                                                                                                                                                                                                                   \sum^{\infty}_{q=1}\frac{1}{q!}\frac{\partial^q P(n;\tau)}{\partial \tau^q} (\delta \tau)^q\nonumber\\
&=&P(n;\tau)+                                                                                                                                                                                                                                                                                                                                                                                                                    \frac{\partial P(n;\tau)}{\partial \tau}\delta \tau+\frac{1}{2!}\frac{\partial^2 P(n;\tau)}{\partial \tau^2}(\delta \tau)^2+\cdots, \eea
Further, substituting Eq~(\ref{asq1}) and Eq~(\ref{asq2}) in Eq~(\ref{eq17}) and equating both the sides we get:
\bea \label{eq18}
\frac{\pl P(n;\tau)}{\pl \tau}=\frac{\pl P(n;\tau)}{\pl n} \frac{\langle \del n\rangle_{\del\tau}}{\del\tau} + \frac{1}{2} \frac{\pl^{2} P(n;\tau)}{\pl n^{2}} \frac{\langle (\del n)^{2}\rangle_{\del\tau}}{\del\tau}+...\eea
Consequently, using Eq~(\ref{eq16}) one can define the following statistical moments:
\bea 
\langle \del n\rangle_{\del\tau}&=&(1+2n) \langle n_{2}\rangle=\mu \del\tau (1+2n) 
\\
\langle (\del n)^2\rangle_{\del\tau}&=& 2n(n+1)\langle n_{2}\rangle+ (1+6n+6n^{2})\langle n_{2}\rangle^{2}\nonumber\\
&= 2n(n+1)\mu \del\tau + (1+6n+6n^{2})(\mu \del\tau)^{2}.
\eea
Here it is important to note that, for proper truncation of the moments we assume that the particle production rate is small locally i.e. $\mu\delta\tau<1$. For this reason the second factor is ignored in $\langle (\del n)^2\rangle_{\del\tau}$ and finally we get:
\be \label{eq19}
\textcolor{blue}{\bf \underline{Fokker~Planck~Equation:}}~~~~~\frac{1}{\mu_{k}} \frac{\pl P(n;\tau)}{\partial\tau} =\underbrace{ (1+2n) \frac{\pl P(n;\tau)}{\pl n}}_{\textcolor{red}{\bf Drift~term }} +\underbrace{ n(1+n) \frac{\pl^{2} P(n;\tau)}{\pl n ^{2} }}_{\textcolor{red}{\bf Diffusion~term }},
\ee
where in the mean particle production rate (defined earlier) we have restored the Fourier mode dependence i.e. $\mu=\mu_k$. On the other hand, in the occupation number we have ignored the Fourier mode dependence. For more details see ref.~\cite{Mello2}. Additionally, it is important to note that 
 in presence of diffusion one can derive  the {\it Fokker-Planck equation} from {\it Langevin equation} of the following mathematical form:
 \bea \textcolor{blue}{\bf \underline{Langevin~Equation~I:}}~~~~~~~~~\frac{dn(\tau)}{d\tau}=a(n)+b(\tau),\eea
 where $a(n)$ is defined in terms of an external deterministic force $f(n)$ and contribution from frictional damping (which is characterised by the damping coefficient $\gamma$) as:
 \bea  a(n)=\frac{f(n)}{m\gamma}=0,\eea
 and $b(\tau)$ is the Gaussian random function (also known as Gaussian white noise), which satisfies the following criteria:
 \bea  \langle b(\tau)\rangle &=&0,\\
 \langle b(\tau) b(\tau^{'})\rangle &=& 2D(n)\delta(\tau-\tau^{'}).\eea
 In our prescription, the diffusion coefficient $D(n)$ is given by the following expression:
 \bea  \textcolor{blue}{\bf \underline{Diffusion~Coefficient~(Einstein's ~Relation):}}~~~~~~~~~D(n)=\frac{kT}{\eta}=n(1+n).\eea
Here it is important to note that, in a most generalised situation $b(\tau)$ has a finite microscopic autocorrelation time for which it is a coloured noise defined as:
\bea   \langle b(\tau) b(\tau^{'})\rangle &=g(\tau-\tau^{'}),\eea 
where $g(\tau-\tau^{'})$ is an arbitrary function of the time interval $\tau-\tau^{'}$. In such a situation it describes a non-Markovian process.

One can also recast the {\it Langevin equation} in the following alternative form:
 \bea \textcolor{blue}{\bf \underline{Langevin~Equation~II:}}~~~~~~~~~\frac{dn(\tau)}{d\tau}=a(n)+\sqrt{D(n_0)}~b(\tau),\eea
where, the Gaussian white noise satisfies the following criteria:
 \bea  \langle b(\tau)\rangle &=&0,\\
 \langle b(\tau) b(\tau^{'})\rangle &=&2\delta(\tau-\tau^{'}).\eea
 However, due to the presence of Dirac Delta function in the two point correlation function the white noise function $b(\tau)$ is singular and consequently the factor $\sqrt{D(n)}~b(\tau)$ is not defined in this context. This will finally lead to {\it It$\hat{o}$ vs. Stratonovitch dilemma}.
For an infinitesimal time interval $[\tau,\tau+\epsilon]$, in the present context the occupation number $n_0$, is defined as a function of a new parameter $Q$:
\bea  n_0&=&n(\tau)+\left(1-Q\right)\left[X(\tau+\epsilon)-x(\tau)\right]= y+\left(1-Q\right)\left[X(\tau+\epsilon)-y\right]~~~0\leq Q\leq 1.~~~~~\eea
Now we define:
\bea  {\cal B}_{\epsilon}&=&\int^{\tau+\epsilon}_{\tau}d\tau^{'}~b(\tau^{'}),\eea
using which we finally get the following results:
\bea  \label{zzv1}\langle {\cal B}_{\epsilon} \rangle &=& 0,\\
\label{zzv2} \langle {\cal B}_{\epsilon}{\cal B}_{\epsilon} \rangle &=& 2.\eea
\begin{enumerate}
\item \textcolor{red}{\bf \underline{It$\hat{o}$ prescription:}}\\
According to this prescription one can write:
\bea  n(\tau+\epsilon)&=&y+\epsilon a(y)+\sqrt{D(y)}\int^{\tau+\epsilon}_{\tau}d\tau^{'}~b(\tau^{'})\nonumber\\
&=&y+\epsilon a(y)+\sqrt{D(y)}~{\cal B}_{\epsilon}, \eea
which is true for $Q=1$. Now using the {\it Chapman-Kolmogorov equation} in the present context we get:
\bea  P(n,\tau+\epsilon|y,\tau)&=&\left\langle \delta\left(n-y-\epsilon a(y)-\sqrt{D(y)}~{\cal B}_{\epsilon}\right)\right\rangle\nonumber\\
&\simeq& \left(1-\epsilon \frac{\pl a(n)}{\pl n}-{\cal B}_{\epsilon}~\frac{\pl (\sqrt{D(n)})}{\pl n}+\frac{{\cal B}^2_{\epsilon}}{2}\frac{\pl^2 D(n)}{\pl n^2}\right)\nonumber\\
&&~~~~~~~~~~~~~~~~~\times\left\langle \delta\left(n-y-\epsilon a(y)-\sqrt{D(y)}~{\cal B}_{\epsilon}\right)\right\rangle. \eea
Here upto the order $\epsilon$ we use the following fact:
\bea  a(n)=a(y)+{\cal O}(\epsilon).\eea
Also we have used the following well known identity of Dirac Delta function, as given by:
\bea  \delta(f(y))=\frac{1}{|f^{'}(y)|}\delta(y-y_0),~~~~~~{\rm where}~~f(y_0)=0.\eea
Now, we expand the Dirac Delta function in the powers of $\epsilon$, as given by:
\bea  \delta\left(n-y-\epsilon a(y)-\sqrt{D(y)}~{\cal B}_{\epsilon}\right)&=&\delta(n-y)+\left[\epsilon a(n)+\sqrt{D(y)}~{\cal B}_{\epsilon}\right]\delta^{'}(x-y)\nonumber\\
&&~~~~+\frac{1}{2}\left[\epsilon a(n)+\sqrt{D(y)}~{\cal B}_{\epsilon}\right]^2\delta^{''}(x-y)+\cdots~~~~~~~~~~ \eea
where it is important to note that we have Taylor expanded the Dirac Delta function of the order of ${\cal B}^2_{\epsilon}$, this is because of the reason that ${\cal B}_{\epsilon}\sim {\cal O}(\sqrt{\epsilon})$. Hence we use this result in {\it Chapman-Kolmogorov equation} and we get the following simplified integral:
\bea  P(n;\tau+\epsilon|n_0)&=&\int dy~P(y,\tau|n_0)~\left\langle \left[\left(1-\epsilon \frac{\pl a(n)}{\pl n}\right.\right.\right.\nonumber\\
&& \left.\left. \left.-{\cal B}_{\epsilon}~\frac{\pl (\sqrt{D(n)})}{\pl n}+\frac{{\cal B}^2_{\epsilon}}{2}\frac{\pl^2 D(n)}{\pl n^2}\right)\delta(y-n)\right.\right.\nonumber\\
&& \left.\left. +\left(\epsilon a(n)+\sqrt{D(n)}{\cal B}_{\epsilon}\right)\delta^{'}(y-n)+\frac{D(n) {\cal B}^2_{\epsilon}}{2}\delta^{''}(y-n)\right]\right\rangle \eea
Further, performing the integration and using Eq~(\ref{zzv1}) and Eq~(\ref{zzv2}) we finally get the following simplified result of this integral:
\bea  P(n;\tau+\epsilon|n_0)&=&P(n;\tau|n_0)+\epsilon \frac{\pl P(n;\tau|n_0)}{\pl \tau}\nonumber\\
&=&P(n;\tau|n_0)+\epsilon \left\{ -\frac{\pl}{\pl n}\left(a(n)P(n,\tau|n_0)\right)\right.\nonumber\\
&&\left.~~~~~~~~~~~~~~~~~~~~~~~~~~+\frac{\pl^2}{\pl n^2}\left(D(n)P(n,\tau|n_0)\right)\right\}+{\cal O}(\epsilon^2).~~~~~~~~\eea
So for $Q=1$ the {\it Fokker-Planck equation} can be written starting from {\it Langevin equation II} in the following form:
\bea  &&\textcolor{blue}{\bf \underline{Fokker~Planck~Equation~(From~It\hat{o})}:}\nonumber\\
&&\frac{\pl P(n;\tau)}{\pl \tau}=-\frac{\pl }{\pl n}\left(a(n)P(n;\tau)\right)+\frac{\pl^2}{\pl n^2}\left(D(n)P(n;\tau)\right),\eea
where we have used the notation $P(n;\tau|n_0)\equiv P(n;\tau)$ for simplicity. 
\item \textcolor{red}{\bf \underline{Generalized It$\hat{o}$ prescription:}}\\
In this situation for general $Q$ one can write:
\bea  \langle n(\tau+\epsilon)-y\rangle =\epsilon\left[a(y)+(1-Q)D^{'}(y)\right].\eea
Further, we have to make the following substitution:
\bea  a(y)\longrightarrow a(y)+(1-Q)D^{'}(y).\eea
This will finally lead to the following {\it Fokker Planck equation}:
\bea  &&\textcolor{blue}{\bf \underline{Fokker~Planck~Equation~(For~Generalized~It\hat{o})}:}\nonumber\\
&&\frac{\pl P(n;\tau)}{\pl \tau}=-\frac{\pl }{\pl n}\left(a(n)P(n;\tau)\right)+\frac{\pl}{\pl n}\left((D(n))^{1-Q}\frac{\pl}{\pl n}\left((D(n))^Q P(n;\tau)\right)\right),~~~~~~\eea
where in the present situation the {\it Stratonovich prescription} corresponds to $Q=1/2$.  However, for our problem, we consider the simplest situation , where $a(n)=0$, $Q=0$ and $D(n)=n(1+n)$.
\end{enumerate}

Now integrating the {\it Langevin equation II} over a small time interval $\epsilon$ we get:
\bea  n(\tau+\epsilon)-n(\tau)=\epsilon a(n(\tau))+\int^{\tau+\epsilon}_{\tau} d\tau^{'}~\sqrt{D(n(\tau^{'})}~b(\tau^{'}).\eea
Now to deal with the product $\sqrt{D(n(\tau)}~b(\tau)$ one can use various prescriptions and that will finally lead to different form of {\it Fokker Planck equations}. One of the possibility is to apply {\it Stratonovich prescription}, using which
we can compute the integral as~\footnote{In case of It$\hat{o}$ prescription one can recast the integral ${\cal I}(n;\tau|\epsilon)$ in to the following form:
\bea   {\cal I}(n;\tau|\epsilon)=\int^{\tau+\epsilon}_{\tau} d\tau^{'}~\sqrt{D(n(\tau^{'}))}~b(\tau^{'})=~\sqrt{D\left(n(\tau)\right)}~\int^{\tau+\epsilon}_{\tau} d\tau^{'}~b(\tau^{'}).\eea}:
\bea  {\cal I}(n;\tau|\epsilon)=\int^{\tau+\epsilon}_{\tau} d\tau^{'}~\sqrt{D(n(\tau^{'}))}~b(\tau^{'})=~\sqrt{D\left(\frac{\left[n(\tau)+n(\tau+\epsilon)\right]}{2}\right)}~\int^{\tau+\epsilon}_{\tau} d\tau^{'}~b(\tau^{'}).\eea
This will correspond to the following form of {\it Fokker Planck equation}, as given by:
\bea  &&\textcolor{blue}{\bf \underline{Fokker~Planck~Equation~(From~Stratonovitch)}:}\nonumber\\
&&\frac{\pl P(n;\tau)}{\pl \tau}=-\frac{\pl }{\pl n}\left(a(n)P(n;\tau)\right)+\frac{\pl}{\pl n}\left(\sqrt{D(n)}\frac{\pl }{\pl n}\left(\sqrt{D(n)}P(n;\tau)\right)\right),\eea
Further, one can write the {\it Fokker-Planck equation} in terms of a {\it continuity equation}, given by the following expression:
\bea   \textcolor{blue}{\bf \underline{Continuity~Equation:}}~~~~~~~~~\frac{\pl P(n;\tau)}{\pl \tau}=-\frac{\pl J(n;\tau)}{\pl n},\eea
where the {\it Fokker-Planck current} is defined as:
\bea  &&\textcolor{blue}{\bf \underline{Fokker-Planck~current~(From~It\hat{o}):}}\nonumber\\
&&J(n;\tau)=\mu_k\left(a(n)-D(n)\frac{\pl}{\pl n}\right)P(n;\tau),\\
 &&\textcolor{blue}{\bf \underline{Fokker-Planck~current~(From~Stratonovitch):}}\nonumber\\
&&J(n;\tau)=\mu_k\left(a(n)P(n;\tau)-\sqrt{D(n)}\frac{\pl}{\pl n}\left(\sqrt{D(n)}P(n;\tau)\right)\right),\\
 &&\textcolor{blue}{\bf \underline{Fokker-Planck~current~(From~Our~Paper):}}\nonumber\\
&&J(n;\tau)=-\mu_k\left(n(1+n)\frac{\pl}{\pl n}\right)P(n;\tau).~~~~~~~\eea
Additionally, it is important to note that the {\it Fokker-Planck equation} explicitly mimics the role of a {\it Schr$\ddot{o}$dinger equation}, provided the real time should be replaced by imaginary time in the present context. and such an analogy usually used to describe the convergence to the equilibrium. To establish this statement let us start with the time dependent {\it Schr$\ddot{o}$dinger equation} for an electron moving in one dimension conduction wire in presence impurity potential $V(x)$, as given by:
\bea  \textcolor{blue}{\bf \underline{Schr\ddot{o}dinger~ Equation:}}~~~~~~~\left[-\frac{1}{2m}\frac{\pl^2}{\pl x^2}+V(x)\right]\psi(x,t)=H\psi(x,t)=i\frac{\pl \psi(x,t)}{\pl t}.~~~~~~~~~\eea
Now changing $t=-i\tau$, $x=n$, $\psi(x,\tau)=P(n;\tau)$ we get:
\bea ~~~~~~~\frac{\pl}{\pl n}\left(D(n)\frac{\pl P(n;\tau)}{\pl n}\right)=\frac{\pl P(n;\tau)}{\pl \tau}=-\frac{\pl J(n;\tau)}{\pl n}.\eea
Further taking, $V=0$ one can identify the above equation as diffusion equation and identify the diffusion coefficient as:
\bea \textcolor{blue}{\bf \underline{Diffusion~Coefficient:}}~~~~~D(n)=\frac{1}{2m}=\frac{kT}{\eta}=n(1+n).\eea
Now if we consider the contribution from the impurity potential is non vanishing then one can write:
\bea &&\textcolor{blue}{\bf \underline{Generalized~Fokker~Planck~Equation:}}\nonumber\\
&&\frac{\pl}{\pl n}\left(D(n)\left\{\frac{\pl P(n;\tau)}{\pl n}+\beta~ P(n;\tau)\frac{\pl V(n)}{\pl n}\right\}\right)=\frac{\pl P(n;\tau)}{\pl \tau}=-\frac{\pl J(n;\tau)}{\pl n}.\eea
 then for equilibrium we set the {\it Fokker-Planck current} is zero~\footnote{Here it is important to note that, in one dimension $J$=constant directly implies $J$=0.  But in the case of higher dimension one can get stationary out-of-equilibrium currents.}, for which we get finally the following result:
\bea \left\{\frac{\pl P(n)}{\pl n}+\beta~P(n)\frac{\pl V(n)}{\pl n}\right\}=0,\eea
from which we get the following Boltzmann probability distribution function for equilibrium:
\bea  P(n)=P_0~\exp(-\beta V(n)),\eea
where $P_0=P(n=0)$ is the normalization constant for the probability distribution.

Now in the situation where the {\it Fokker Planck current} is non-vanishing one can use the following solution ansatz to solve the most {\it Generalized Fokker Planck Equation}, as given by:
\bea \textcolor{blue}{\bf \underline{Solution~Ansatz:}}~~~~~~~~
P(n;\tau)=\exp\left(-\frac{\beta}{2} V(n)\right)W(n;\tau)~.\eea
Using this ansatz one can write the following expression:
\be  \left\{\frac{\pl P(n;\tau)}{\pl n}+\beta~ P(n;\tau)\frac{\pl V(n)}{\pl n}\right\}=\exp\left(-\frac{\beta}{2} V(n)\right) \left\{\frac{\pl W(n;\tau)}{\pl n}+\frac{\beta}{2}~ W(n;\tau)\frac{\pl V(n)}{\pl n}\right\}.~\ee
Further, substituting this result in the {\it Generalized~Fokker~Planck~Equation} we get the following partial differential equation for the unknown function $W(n;\tau)$, as given by:
\bea 
&&\frac{\pl}{\pl n}\left(D(n)\exp\left(-\frac{\beta}{2} V(n)\right) \left\{\frac{\pl W(n;\tau)}{\pl n}+\frac{\beta}{2}~ W(n;\tau)\frac{\pl V(n)}{\pl n}\right\}\right) \nonumber\\
&&~~~~~~~~~~~~~~~~~~~~~~~~=\exp\left(-\frac{\beta}{2} V(n)\right)\frac{\pl W(n;\tau)}{\pl \tau},\eea
which can be recast in to the following simplified form:
\bea\frac{\pl}{\pl n}\left(D(n)\frac{\pl W(n;\tau)}{\pl n}\right)-U(n)W(n;\tau)=\frac{\pl W(n;\tau)}{\pl \tau},\eea
where the effective potential $U(n)$ is defined as:
\bea &&\textcolor{blue}{\bf \underline{Effective ~Potential:}}\nonumber\\
&&U(n)=\left[\frac{\beta^2}{4}D(n)\left(\frac{\pl V(n)}{\pl n}\right)^2-\frac{\beta}{2}D(n)\left(\frac{\pl^2V(n)}{\pl n^2}\right)-\frac{\beta}{2}\left(\frac{\pl D(n)}{\pl n}\right)\left(\frac{\pl V(n)}{\pl n}\right)\right].\eea
Now let us only consider the time independent part of the {\it Generalized Fokker Planck equation} for which the wave function for the equilibrium is described by the following expression:
\bea\frac{\pl}{\pl n}\left(D(n)\frac{\pl \psi_0(n)}{\pl n}\right)=U(n)\psi_0(n),\eea
where the solution is given by:
\bea\psi_0(n)={\cal N}_{\rm eq}~\exp\left(-\frac{\beta}{2} V(n)\right),\eea
where ${\cal N}_{\rm eq}$ is the normalization constant which can be fixed by the following normalization condition of the equilibrium wave function:
\bea \int dn~|\psi_0(n)|^2=1~~~ \Longrightarrow~~~ |{\cal N}_{\rm eq}|=\frac{1}{\sqrt{\int dn~\exp(-\beta V(n))}},\eea
where $\psi_0(n)$ physically represents the ground state wave function with energy eigen value $E_0=0$. All the exicied state have energy eigen value $E_p>0$ (for $p>1$). To get the time dependence of the evolution equation we use the initial condition at time $\tau=0$ in terms of complete set of eigenfunctions $\psi_{p}(n)=\langle n| p \rangle$, which satisfy the following eigen value equation:
\bea -\frac{\pl}{\pl n}\left(D(n)\frac{\pl \psi_p(n)}{\pl n}\right)+U(n)\psi_p(n)=E_p\psi_{p}(n),\eea
which implies the following result:
\bea W(n,0|n_0)=\sum_{p}C_p \psi_p(n).\eea
Here the expansion coefficient of the basis is defined as:
\be C_p=\int dn~\psi_p(n)W(n,0|n_0)=\psi_p(n_0)\exp\left(\frac{\beta}{2}V(n_0)\right)=\langle n|n_0\rangle\exp\left(\frac{\beta}{2}V(n_0)\right),\ee
where we have used the following expression:
\bea W(n,0|n_0)=\exp\left(\frac{\beta}{2}V(n_0)\right)\delta(n-n_0). \eea
Now for the time dependent part the solution of the {\it Generalized Fokker Planck equation} can be written as:
\bea W(n,\tau|n_0)&=&\sum_{p}C_p \psi_p(n)~\exp\left(-E_p\tau\right)\nonumber\\
&=&\sum_p \exp\left(\frac{\beta}{2}V(n_0)\right) \langle n|p \rangle ~\exp\left(-E_p\tau\right)\langle p|n_0\rangle\\
&=&\exp\left(\frac{\beta}{2}V(n_0)\right) \langle n|\exp\left(-\left[-\frac{\pl}{\pl n}\left(D(n)\frac{\pl }{\pl n}\right)+U(n)\right]\tau\right)|n_0\rangle \nonumber.\eea
Then the probability distribution function can be expressed as:
\be P(n;\tau)=\exp\left(-\frac{\beta}{2}\left[V(n)-V(n_0)\right]\right) \langle n|\exp\left(-\left[-\frac{\pl}{\pl n}\left(D(n)\frac{\pl }{\pl n}\right)+U(n)\right]\tau\right)|n_0\rangle.\ee
In this paper we investigate the physical outcomes of the simplest possibility where $V(n)=V(n_0)=0$ and $U(n)=0$ for which one can write the following simplified expression:
\bea P(n;\tau)= \langle n|\exp\left(-\left[-\frac{\pl}{\pl n}\left(D(n)\frac{\pl }{\pl n}\right)\right]\tau\right)|n_0\rangle.\eea
Now from Eq~(\ref{eq19}), one can write down the {\it Fokker Planck equation} in terms of the following operator equation:
\be \hat{\cal D}_{\bf FP}P(n;\tau)=0,\ee
where $\hat{\cal D}_{\bf FP}$ is the {\it Fokker Planck operator} represented by:
\bea  \hat{\cal D}_{\bf FP}&\equiv &  \left[(1+2n) \frac{\pl }{\pl n} + n(1+n) \frac{\pl^{2} }{\pl n ^{2}}-\frac{1}{\mu_{k}} \frac{\pl }{\partial\tau}\right]\nonumber\\
&=&\left[\frac{\partial}{\partial n}\left(n(n+1)\frac{\partial}{\partial n}\right)-\frac{1}{\mu_{k}} \frac{\pl }{\partial\tau}\right]. \eea
Now, we consider a special case where $n>>1$, which gives the most simplest outcome in the present context. In such a situation one can approximately write down the following simplified form of the {\it Fokker Planck equation}, as given by:
\bea\label{eq19}
\frac{1}{\mu_{k}} \frac{\pl P(n;\tau)}{\partial\tau} = 2n \frac{\pl P(n;\tau)}{\pl n} +n^2 \frac{\pl^{2} P(n;\tau)}{\pl n ^{2} }=\frac{\partial}{\partial n}\left(n^2\frac{\partial P(n;\tau)}{\partial n}\right),
\eea
To solve this partial differential equation in the $n>>1$ limit we use method of separation of variable, using which we can write:
\bea \label{cvq1} P(n;\tau)=P_1(n)P_2(\tau).\eea
Further, using Eq~(\ref{cvq1}) we get the following two sets of independent differential equations, as given by:
\bea
\label{i1} \left[n^2 \frac{d^{2} }{d n ^{2} }+2n \frac{d }{d n} +q\right]P_1(n)&=&0,\\
\label{i2} \left[\frac{d}{d\tau}+q\right]P_2(\tau)&=0.\eea
Solution of Eq~(\ref{i1}) and Eq~(\ref{i2}) is given by:
\bea
P_1(n)&=&\left[A~ n^{-\frac{1}{2} \left(\sqrt{1-4q}+1 \right)}+B~ n^{\frac{1}{2} \left(\sqrt{1-4q}-1\right) }\right],\\
P_2(\tau)&=&C~e^{-q\tau},\eea
where we define a new constant:
\bea q=\mu_k Q^2.\eea 
Additionally,
 $A, ~B$ and $C$ are arbitrary constants, which can be determined after imposing appropriate boundary conditions.
 
Consequently, the most general total solution for the probability density function in the limit $n>>1$ can be expressed as:
\be\label{solu1} P(n;\tau)=\sum^{\infty}_{q=0}\left[A_1~ n^{-\frac{1}{2} \left(\sqrt{1-4q}+1 \right)}+B_1~ n^{\frac{1}{2} \left(\sqrt{1-4q}-1\right) }\right]~e^{-q \tau},\ee
where we define $A_1$ and $B_1$ as:
\bea A_1=AC,~~~~~~B_1=BC.\eea
Now, after imposing the boundary condition it can be shown that in the large $n$ limit ($n\to \infty$) the solution obtained in Eq~(\ref{solu1}) can be expressed in terms of the following log-normal distribution. To check that explicitly, let us write the probability distribution as the Fourier transformation with respect to the occupation number $n$, which is given by:
\bea\label{ft1} P(n;\tau|n^{'};\tau^{'})=\frac{1}{2\pi}\int dk~e^{ikn}\bar{P}(k;\tau|n^{'};\tau^{'}).\eea
Using this one can write down the {\it Fokker Planck equation} in the Fourier space as:
\bea\label{ss1c} \frac{\pl \bar{P}(k;\tau)}{\pl \tau}=\mu_k\left(2ink-k^2n^2\right)\bar{P}(k;\tau),\eea
which is obviously a simplest version of the {\it Fokker Planck equation} as it contains a single derivative with respect to time $\tau$. Now, one can choose the initial condition such that the probability distribution function at time $\tau^{'}$ is given by the following expression:
\bea P(n;\tau^{'}|n^{'};\tau^{'})=\delta(n-n^{'}).\eea
This is only true when the probability distribution function after Fourier transform at time $\tau=\tau^{'}$ can be written as:
\bea \bar{P}(k;\tau^{'}|n^{'};\tau^{'})=e^{-ikn^{'}}.\eea
Then solution of Eq~(\ref{ss1c}) can be expressed after imposing the initial condition as:
\bea \bar{P}(k;\tau|n^{'};\tau^{'})=e^{\mu_k(2ink-k^2n^2)(\tau-\tau^{'})-ikn^{'}}.\eea
Hence substituting back the above mentioned result into the definition of Fourier transformation and setting the initial condition $n^{'}=0$ and $\tau^{'}=0$ and further considering $n\rightarrow \infty$ limit we get the following result for the probability distribution function, as 
given by:
\bea P(n;\tau)=\frac{1}{n}\frac{1}{\sigma \sqrt{2\pi } }\exp\left[-\frac{(\ln n-\mu_k\tau)^2}{2\sigma^2}\right],\eea
which is precisely a log-normal distribution function and $\sigma$ is standard deviation of the log-normal distribution, which can be expressed as:
\bea \sigma=\sqrt{2\mu_k\tau}.\eea
\begin{figure}[htb]
	\includegraphics[width=16cm,height=8cm]{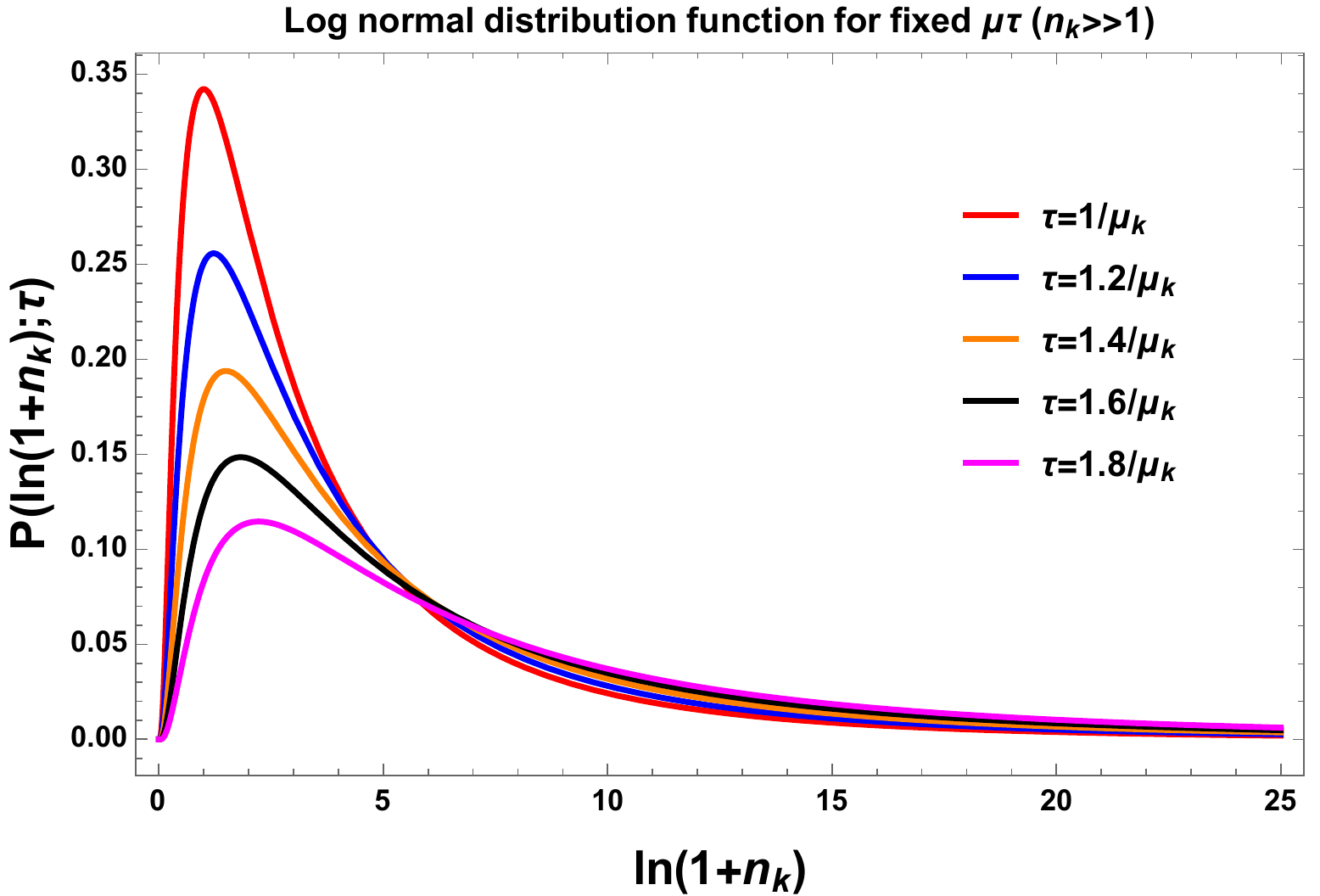}
	\caption{Evolution of the log normal probability density function with respect to the logarithm of the occupation number per mode $\ln(1+n_k)$, for a fixed time for $n_k>>1$.}
	\label{lognormal}
\end{figure}
One can explain the physics of this obtained result in the large $n$ limit . In the earlier section we have discussed that the averaging over the phase factor of $\ln n$  can be expressed in terms of the logarithms of the occupation number of particles produced in each scattering events.
Further using Central Limit Theorem, one can further interpret that $\ln n$ follows Gaussian profile (on the other hand, one can also say that in such a case $n$ follows a log-normal distribution). But this physical explanation is only valid for large $n$ limiting approximation. In figure~(\ref{lognormal}), we have shown the Evolution of the log normal probability density function with respect to the logarithm of the occupation number per mode $\ln(1+n_k)$, for a fixed time ($\mu_k\tau$=fixed). Additionally, it is important to note that from the plot that for very large value of the occupation number $n$ (large $n$ limit) the log normal profile shows Gaussian features perfectly, which indicates the initial assumption regarding large $n$ was consistent. 

Now, we consider another special case where $n<<1$ in the present context. In such a situation one can approximately write down the following simplified form of the {\it Fokker Planck equation}, as given by:
\bea\label{eqxz1}
\frac{1}{\mu_{k}} \frac{\pl P(n;\tau)}{\partial\tau} =  \frac{\pl P(n;\tau)}{\pl n} +n \frac{\pl^{2} P(n;\tau)}{\pl n ^{2} }=\frac{\pl }{\pl n}\left(n\frac{\pl P(n;\tau)}{\pl n}\right),
\eea
To solve this partial differential equation in the $n<<1$ limit we use method of separation of variable, using which we can write:
\bea \label{eqzx2} P(n;\tau)=P_1(n)P_2(\tau).\eea
Further, using Eq~(\ref{eqzx2}) we get the following two sets of independent differential equations, as given by:
\bea
\label{i1} \left[n \frac{d^{2} }{d n ^{2} }+ \frac{d }{d n} +w\right]P_1(n)&=0,\\
\label{i2} \left[\frac{d}{d\tau}+w\right]P_2(\tau)&=0.\eea
Solution of Eq~(\ref{i1}) and Eq~(\ref{i2}) is given by:
\bea
P_1(n)&=&\left[D~ \ln n+E-n w\right],\\
P_2(\tau)&=&F~e^{-w\tau},\eea
where we define a new constant:
\bea w=\mu_k W^2.\eea
Additionally,
 $D, ~E$ and $F$ are arbitrary constants, which can be determined after imposing appropriate boundary conditions.
 
Consequently, the most general total solution for the probability density function in the limit $n>>1$ can be expressed as:
\bea \label{solu1} P(n;\tau)=\sum^{\infty}_{w=0}\left[D_1~ \ln n+E_1-n w\right]~e^{-w \tau},\eea
where we define $D_1$ and $E_1$ as:
\bea D_1=DF,~~~~~~E_1=EF.\eea
Further using the  Fourier transformation with respect to the occupation number $n$ as mentioned in Eq~(\ref{ft1}), we get the following simplified expression for the {\it Fokker Planck equation} in $n<<1$ limit:
\bea\label{aasc} \frac{\pl \bar{P}(k;\tau)}{\pl \tau}=\mu_k\left(ik-k^2n\right)\bar{P}(k;\tau),\eea
which is obviously a simplest version of the {\it Fokker Planck equation} as it contains a single derivative with respect to time $\tau$. Further imposing the previously used boundary condition for the limit $n>>1$ in the present context we get the following result for the probability distribution function in the Fourier transformed space, as given by:
\bea \bar{P}(k;\tau|n^{'};\tau^{'})=e^{\mu_k(ik-k^2n)(\tau-\tau^{'})-ikn^{'}}.\eea
Hence substituting back the above mentioned result into the definition of Fourier transformation and setting the initial condition $n^{'}=0$ and $\tau^{'}=0$ and further considering $n\rightarrow 0$ limit we get the following result for the probability distribution function, as 
given by:
\bea P(n;\tau)=\frac{1}{2  \sqrt{\mu_k n \tau \pi}}~\exp\left[-\frac{(n+\mu_k \tau)^2}{4 \mu_k n \tau}\right],\eea
which is not a log normal distribution function in $n<<1$ limit.
\begin{figure}[htb]
	\includegraphics[width=16cm,height=8cm]{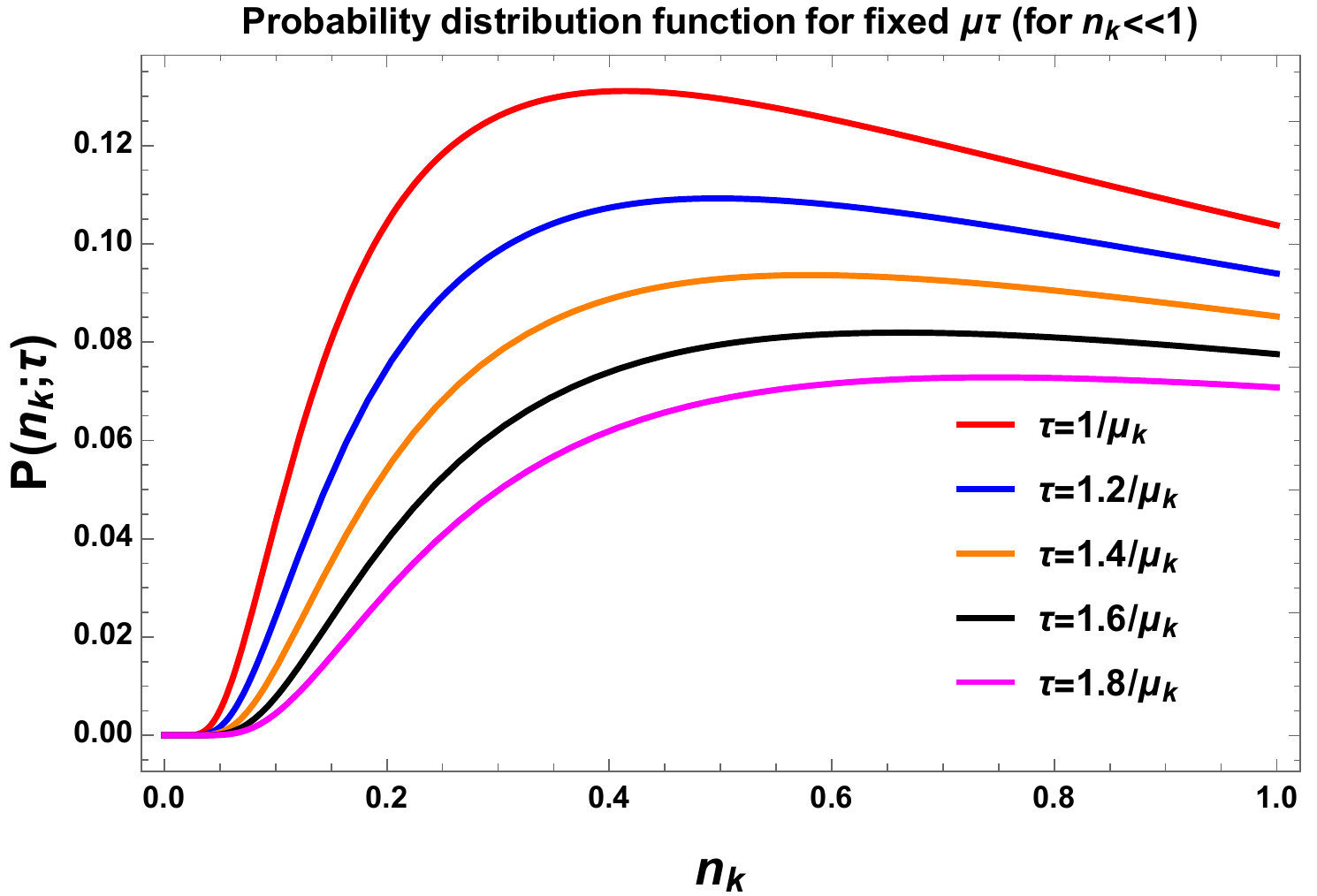}
	\caption{Evolution of the probability density function with respect to the the occupation number per mode $n_k$, for a fixed time in the limit $n_k<<1$.}
	\label{newn}
\end{figure}
One can explain the physics of this obtained result in the small $n$ limit .  In this situation one can observe deviations in the profile function. The prime reason for such deviations in small $n$ limit is appearing due to the fact that, the total transmission probability is bounded by unity. In other words, on can say that this is only possible when $n$ is bounded by zero in this context of discussion. In figure~(\ref{newn}), we have shown the evolution of the probability density function with respect to the occupation number per mode $n_k$, for a fixed time ($\mu_k\tau$=fixed). Additionally, it is important to note that for very small values of the parameter $n$ we have observed from the plot that the deviation from Gaussian feature is observed. In other words, one can interpret that the deviation from log normal probability distribution function corresponds to the significant non-Gaussian features at small values of $n$. Apart from this one can also comment on the quantum mechanical origin of higher order non-Gaussian contributions appearing in {\it Fokker Planck equation} which are more appropriate at small values of $n$.  In the next subsection we will discuss about the physical impacts of this additional higher order contributions in detail.

\subsection{Corrected probability distribution profiles: Quantum effects from non-Gaussianity}
In this subsection we get different order correction to the {\it Fokker Planck equation} that we have derived by Taylor expansion. As we already know that the Taylor expansion of the probability density distribution function is taken with respect to time $\tau$. On the other hand, using {\it Maximum entropy ansatz} we have considered the Taylor expansion of the ensemble average of the distribution function with respect to the occupation number $n$. After that we equate both the results and comparing the coefficient of $\del\tau $ from the both sides of the expansion (see previous Eq~(\ref{eq18}) for more details). Now without truncating both the sides of this expression one can get additional contributions in $\delta\tau$ and in its higher order. If we do the comparison including such additional contributions then it will give rise to corrected version of the {\it Fokker Planck equation} valid upto higher orders. Further solving these sets of differential equation order by order one can explicitly justify the validity of all such corrections in the {\it Fokker Planck equation}. In this paper we have investigated this possibility by considering the contributions upto fourth order. All such higher order correction terms are very useful to describe the non-Gaussian effects appearing during the process of cosmological particle production during reheating phase of early universe. On top of that, one can explain the origin of such higher order contributions in the quantum mechanical ground as it produces non vanishing significant effects in the expression for the higher order statistical moments directly originating from the various quantum mechanical correlations (one-point, two-point, three-point etc.) computed during cosmological particle production at the epoch of reheating of early universe. More precisely, the deviation from Gaussianity (in other words the deviation from log-normal distribution) in the present context can be directly linked with the quantum mechanical effects appearing during reheating epoch of early universe and for this reason one can interpret the higher order corrected version of the {\it Fokker Planck equation} as a {\it quantum corrected Fokker Planck equation}. Since in this paper we have provided the analytical correction upto the fourth order, one can say that in this derivation we have actually provided the {\it fourth order quantum corrected Fokker Planck equation}.
The details of this derivations are explicitly discussed in the following sub sections, where doing the analysis we justify order by order that how such specific corrections will modify the log-normal distribution and its impact in the quantum mechanical ground.

%%%%%%%%%%%
%%%%%%%%%
%%%%%%%
%%%%%
%%%
%%
%
%\textcolor{blue}{All calculation of this part can be found on [References$\rightarrow$Chaos$\rightarrow$all-calculation.nb]}
\subsubsection{First order contribution}
In this context, our prime objective is to find out the first order contribution to the {\it Fokker Planck equation} and to solve this equation analytically, which will help us to understand the background physics related to the present formalism. To serve this purpose we equate both the sides of Eq~(\ref{eq18}) after Taylor expansion and compare coefficient of $\del\tau$. Consequently, we get the following partial differential equation:
\bea\label{eq19cc}
&&\textcolor{blue}{\bf \underline{First~order~Fokker~Planck~Equation:}}~~~\nonumber\\
&&\frac{1}{\mu_{k}} \frac{\pl P(n;\tau)}{\del\tau} = (1+2n) \frac{\pl P(n;\tau)}{\pl n} + n(1+n) \frac{\pl^{2} P(n;\tau)}{\pl n ^{2}}.
\eea
Now to solve this partial differential equation we apply method of separation of variable, using which we can write the total solution in the following form:
\bea  \label{eq19ccc} P(n;\tau)=P_1(n)P_2(\tau).\eea
Further, using the solution ansatz stated in Eq~(\ref{eq19ccc}) we get the following sets of independent differential equations, as given by:
\bea\label{eq20a}
\left[n (n+1) \frac{d^{2}}{dn^{2}}+(2 n+1)\frac{d }{dn}+m_{1}\right] P_1(n)&=&0,\\ 
\label{eq20b} \frac{d P_2(\tau)}{d \tau}+m_{1} P_2(\tau)&=0.
\eea
Solution of Eq~(\ref{eq20a}) and Eq~(\ref{eq20b}) is given by:
\bea \label{eq21a} P_1(n)&=& C_{1} P_{\frac{1}{2}(-1+\sqrt{1-4m_{1}})}(1+2 n)+C_{2} Q_{\frac{1}{2} (-1+\sqrt{1-4m_{1}})}(1+2 n), \\
P_2(\tau)&=& C_3~e^{-\tau \mu_k m_{1}}.\eea
Here $C_1, C_2$ and $C_3$ are arbitrary integration constants which can be obtained by imposing appropriate boundary conditions.
Additionally, we introduce a constant $m_1$ which is defined as:
\bea m_1=m^2.\eea
which will follow certain constraints in the present context.

It is important to note that, to get real valued solution the constant $m_1$ satisfy the following condition:
\bea\label{c4} 
\frac{1}{2}[ -1+\sqrt{1-4m_{1}}]\equiv N\in \mathbb{Z} > 0\Longrightarrow \boxed{m_{1}=\frac{1}{4}\left[1-(2N+1)^{2}\right]}~,\eea
as Legendre polynomial has general form $P_{N}(x)$ with condition that $N$ should be an integer greater than zero. For different values of $N$ we get different $m_{1}$ following Eq~(\ref{c4}).

Consequently, the most general solution of probability distribution function $P(n;\tau)$ is given by the following expression:
\be\label{eq22}
 \textcolor{blue}{\bf\underline{ First~order~solution:}}P(n;\tau) =\sum_{N=0}^{\infty}\left [D_{1} P_{N} (1+2 n)+D_{2} Q_{N}(1+2 n)\right]
e^{-\frac{\tau}{4} \mu_k\left[1-(2N+1)^{2}\right]}, 
\ee
where we define two new constants, $D_1$ and $D_2$ by the following expressions:
\bea D_1= C_1C_3,~~~~~~~~D_2=C_2C_3.\eea
Here it is important to note that, this solution on limit n $\rightarrow \infty$ converge to log-normal distribution as we have discussed earlier.
In further section we will compare this result with obtained higher order calculations. From Eq~(\ref{c4}), we see that the quantization property of $m_{1}$  eventually help us to predict the quantum nature of the present set-up. In other words, one can interpret $N$ as a quantum number given by~\footnote{In the present context, $N$ actually mimics the role of principle quantum number. Also $m_1$ is another quantum number which is derived from $N$ using Eq~(\ref{c4}).}:
\bea\textcolor{blue}{\bf \underline{Quantum~number~ I:}}~~~~~N=0,1,2,\cdots,\infty \in \mathbb{Z}.\eea
Further using Eq~(\ref{c4}), one can further introduce another quantum number $m_1$ given by the following expression:
   \bea\textcolor{blue}{\bf \underline{Quantum~number~II:}}~~~~~m_1=0,-2,-6,\cdots,\infty.\eea
Further using the  Fourier transformation with respect to the occupation number $n$ as mentioned in Eq~(\ref{ft1}), we get the following simplified expression for the {\it Fokker Planck equation} in $n<<1$ limit:
\bea\label{xzx} \frac{\pl \bar{P}(k;\tau)}{\pl \tau}=\mu_k\left((2n+1)ik-k^2n(n+1)\right)\bar{P}(k;\tau),\eea
which is obviously a simplest version of the {\it Fokker Planck equation} as it contains a single derivative with respect to time $\tau$. Further imposing the previously used boundary condition used for the limit $n>>1$ and $n<<1$ in the present context we get the following result for the probability distribution function in the Fourier transformed space, as given by:
\bea\bar{P}(k;\tau|n^{'};\tau^{'})=e^{\mu_k((2n+1)ik-k^2n(n+1))(\tau-\tau^{'})-ikn^{'}}.\eea
Hence substituting back the above mentioned result into the definition of Fourier transformation and setting the initial condition $n^{'}=0$ and $\tau^{'}=0$ we get the following result for the probability distribution function, as 
given by:
\bea P(n;\tau)=\frac{1}{2 \sqrt{\mu_k n (n+1) \tau \pi }}~\exp\left[-n \left(\mu_k (n+1) \tau +\frac{1}{4 \mu_k  \tau(n +1) }+1\right)\right],\eea
 which is coming from the first order contribution in the {\it Fokker Planck equation}. This expression is actually equivalent to the result obtained in Eq~(\ref{eq22}).
 
\begin{figure}[H]
	\includegraphics[width=16cm,height=8cm]{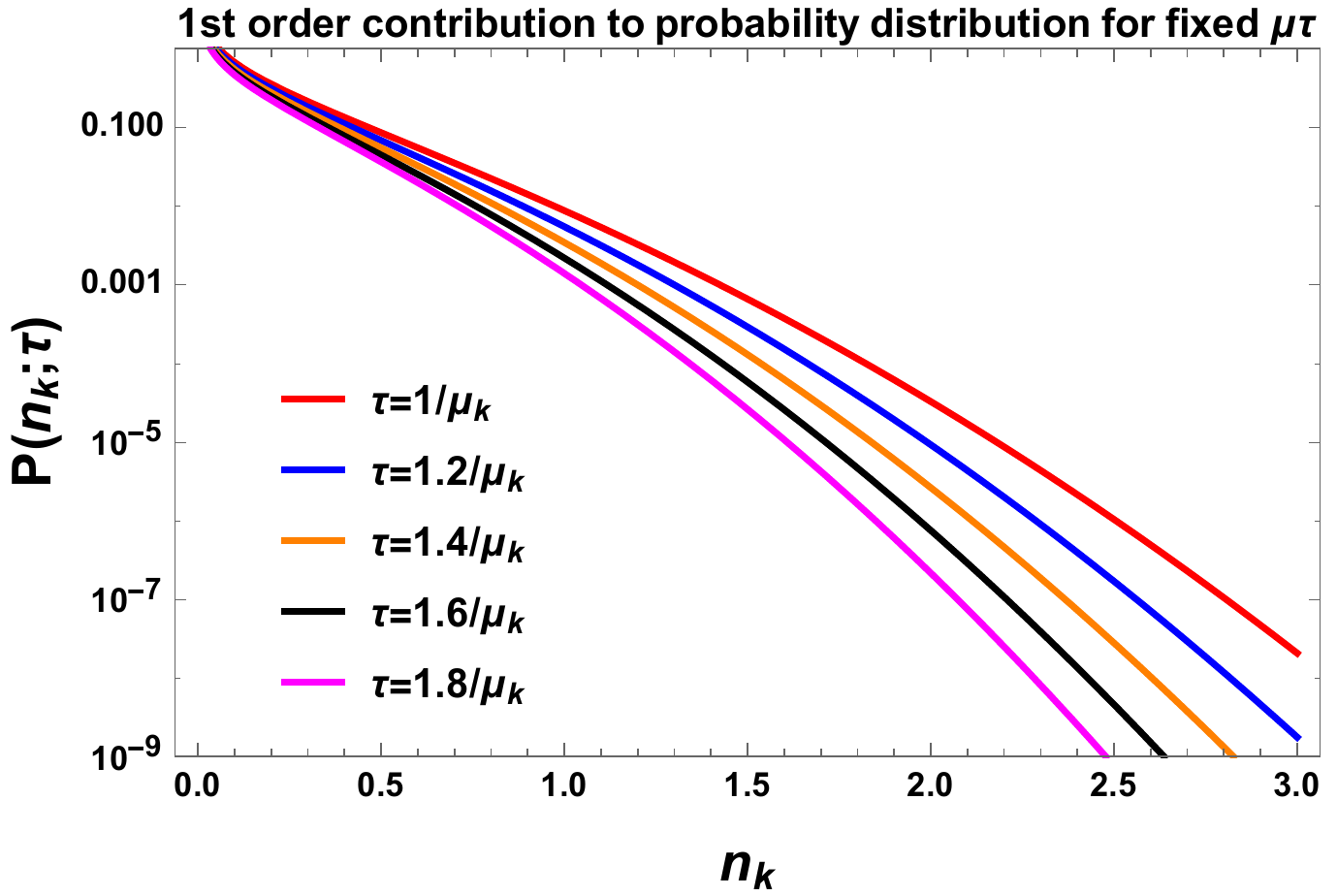}
	\caption{Evolution of the first order contribution to the probability density function with respect to the the occupation number per mode $n_k$, for a fixed time.}
	\label{1st}
\end{figure}

In figure~(\ref{1st}), we have shown the evolution of the probability density function with respect to the occupation number per mode, for a fixed time ($\mu_k\tau$=fixed). For very small values of the parameter $n$ we have observed from the plot that the deviation from Gaussian feature is observed. 

\subsubsection{Second order contribution}
In this context, our objective is to find out the contributions coming from second order in the {\it Fokker Planck equation} and to solve this equation
numerically~\footnote{Including the contributions from second order we will see that the {\it Fokker Planck equation} can not solvable analytically.}. To serve this purpose we equate both sides of Eq~(\ref{eq18}) after Taylor expansion and compare the coefficient of $\del\tau ^{2}$.
Consequently, we get the following partial differential equation:  
\bea\label{eq24}
&& \textcolor{blue}{\bf \underline{Second~order~Fokker~Planck~Equation:}}\nonumber\\
   &&\frac{n^2}{2}\left(1 + n\right)^{2}\frac{\pl^{4}P(n;\tau)}{\pl n^{4}}+2n \left(1 + 3 n + 2 n^2\right)\frac{\pl ^{3}P(n;\tau)}{\pl n^{3}}\nonumber\\
   &&+\left(1 + 6 n + 6 n^2\right) \frac{\pl ^{2}P(n;\tau)}{\pl n^{2}} = \frac{1}{\mu^{2}_k} \frac{\pl^{2} P(n;\tau)}{\pl \tau ^{2}}.
\eea
Now to solve this partial differential equation we apply method of separation of variable, using which we can write the total solution in the following form:
\bea \label{eq19cccas} P(n;\tau)=P_1(n)P_2(\tau).\eea
Further, using the solution ansatz stated in Eq~(\ref{eq19cccas}) we get the following sets of independent differential equations, as given by:
\bea\label{eq25}
&&\left[\frac{n^2}{2}\left(1 + n\right)^{2}\frac{d^{4}}{d n^{4}}+2n \left(1 + 3 n + 2 n^2\right)\frac{d ^{3}}{d n^{3}}+\left(1 + 6 n + 6 n^2\right) \frac{d^{2}}{d n^{2}}-m^2_{2}\right] P_1(n)=0,~~~\\
\label{eq25b} &&\left[\frac{d^{2} }{d \tau ^{2}}-m^2_2\mu^2_k\right]P_2(\tau)=0.
\eea
It is important to note that, the analytical solution of $P_1(n)$ is not possible for any arbitrary values of the constant $m_2$, except the special case $m_2=0$. For this reason we use numerical technique to solve Eq~(\ref{eq25}). On the other hand Eq~(\ref{eq25b}) is exactly solvable in the present context and the solution can be written as:
\bea P_2(\tau)=\left[C_3 e^{\tau  m_2\mu_k}+C_4 e^{-\tau m_2\mu_k}\right],\eea
where $C_3$ and $C_4$ are two arbitrary constants which can be fixed by choosing proper boundary conditions.

Now we  solve Eq~(\ref{eq25}) numerically for different values of $m_{2}$ along with given initial condition and also we consider the special case $m_{2} =0$ where we solve this equation analytically. Here it important to mention that, since arbitrary values of $m_2$ is allowed, one can consider integer as well as non integer values at the level of solution of differential equation. However, the only physically acceptable solution restrict us to only consider the integer values of $m_2$ because such second order corrected solution of the {\it Fokker Planck equation} is directly related to the quantum effects as we have mentioned earlier. As a result such integer values of $m_2$ can be interpreted as the (principal) quantum number i. e. 
\bea\textcolor{blue}{\bf \underline{Quantum ~Number~III:}}~~~~m_2=0,\pm 1,\pm 2,\cdots,\pm \infty \in \mathbb{Z}.\eea
For numerical solution we take the following assumptions:
\bea  \label{ew1} P_1(n=0.001) &=& 100, ~~~
\left[\frac{d P_1(n)}{d n}\right]_{n=0.001} = 100,~~~\left[\frac{d^{2} P_1(n)}{d n^{2}}\right]_{n=0.001} = 100,\nonumber\\
\left[ \frac{d^{3} P_1(n)}{d n^{3}}\right]_{n=0.001} &=& 100.\eea
Here we assume that the particle production rate at low $n$(= $0.001$) has a constant value and its derivatives also have same constant value for a given $m_{2}$. Getting the numerical solution we plot ($ P_{2}(n,\tau)$ vs  $n$) them for some particular range of $n$. 
%%%%%%%%
%%%%%%%%
%%%%%%
%%%%%
%%%%%%
%%%Add pictures for second order correction

\begin{figure}[htb]
\centering
%\subfigure[Second order solution for all values of $m_{2}$]{
%    \includegraphics[width=14cm,height=6.4cm] {allm2.pdf}
%    \label{x21}
%}

%\subfigure[Second order solution for all integer $m_{2}$]{
 %   \includegraphics[width=7cm,height=7cm] {m2_0-1-2.pdf}
 %   \label{x22}
%}
\subfigure[ Second order corrected distribution for   $m_{2}=0$]{
    \includegraphics[width=7.8cm,height=8cm] {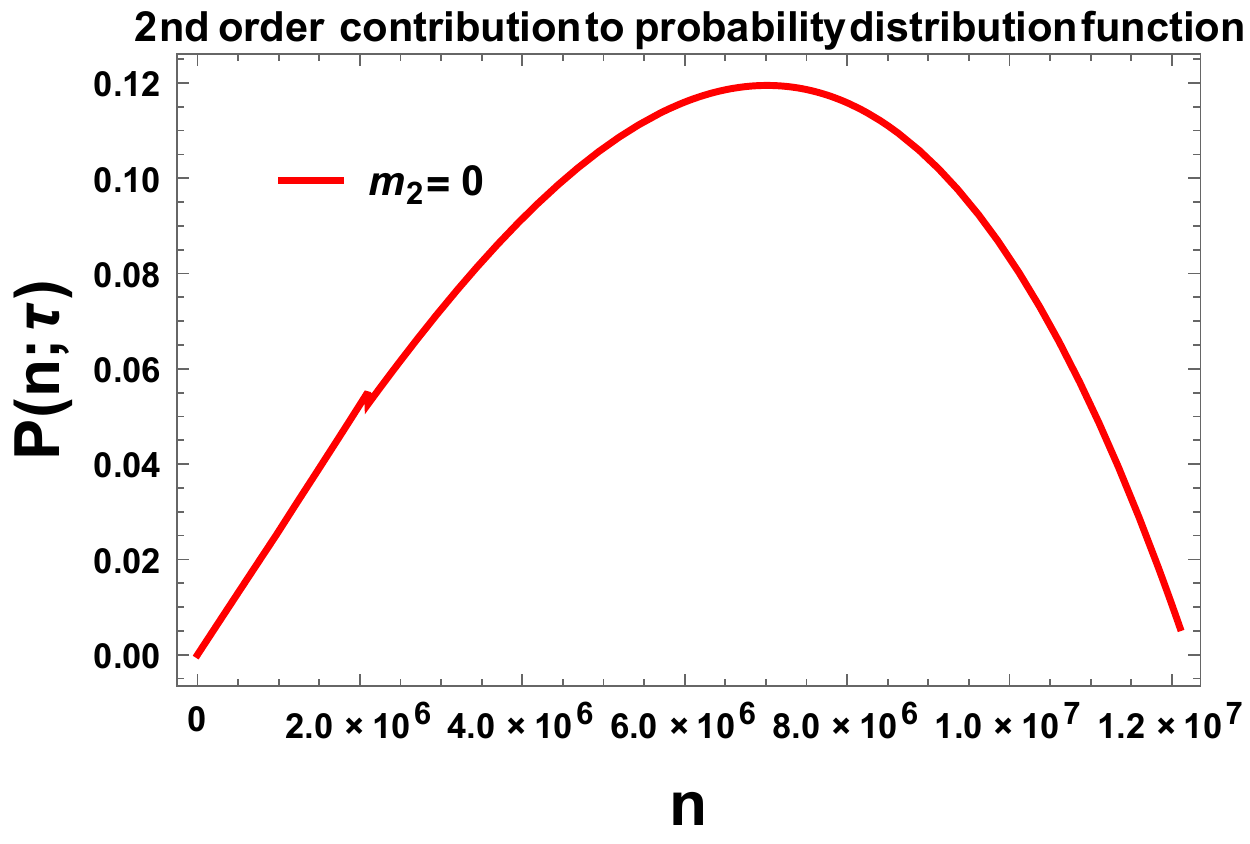}
    \label{x23}
}
\subfigure[Second order corrected distribution for   $m_{2}=-1$]{
    \includegraphics[width=7.8cm,height=8cm] {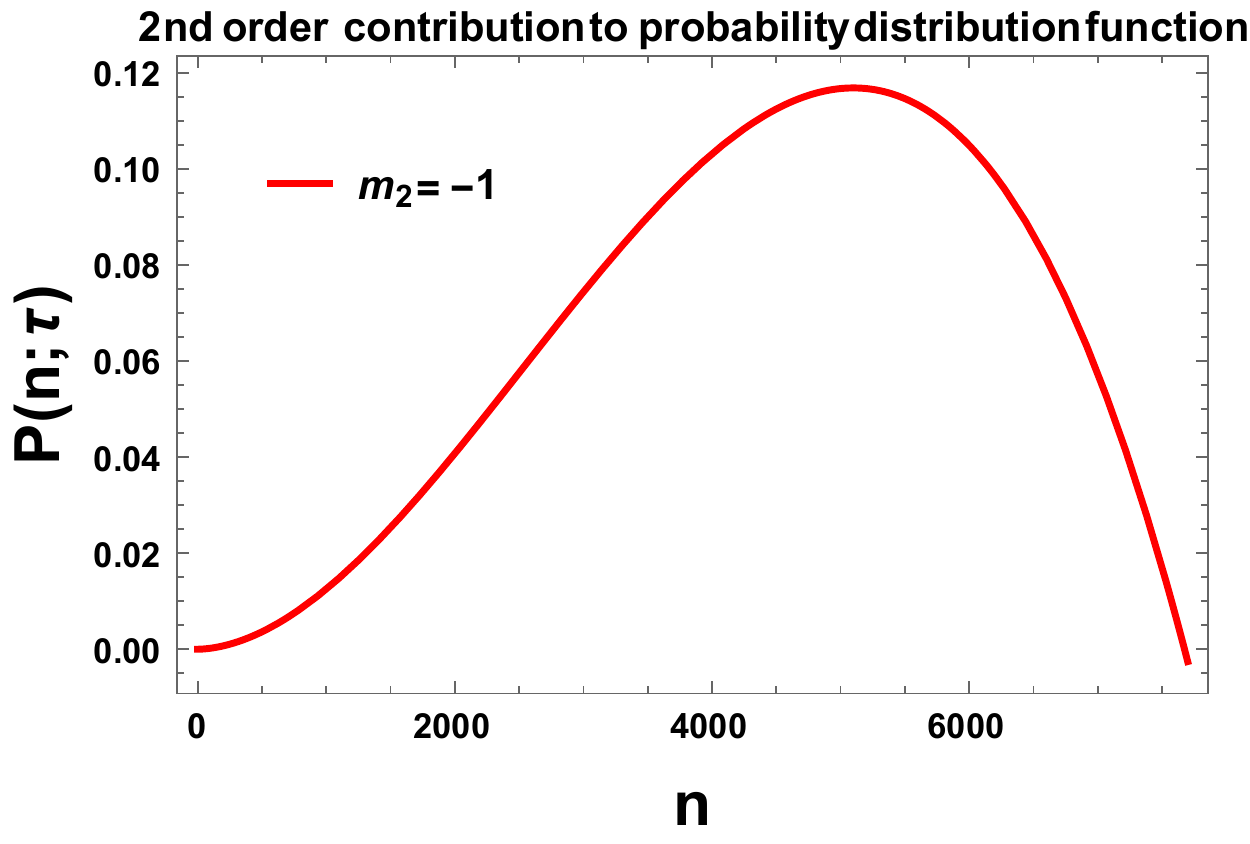}
    \label{x25}
}
\subfigure[Second order corrected distribution for   $m_{2}=1$]{
    \includegraphics[width=7.8cm,height=8cm] {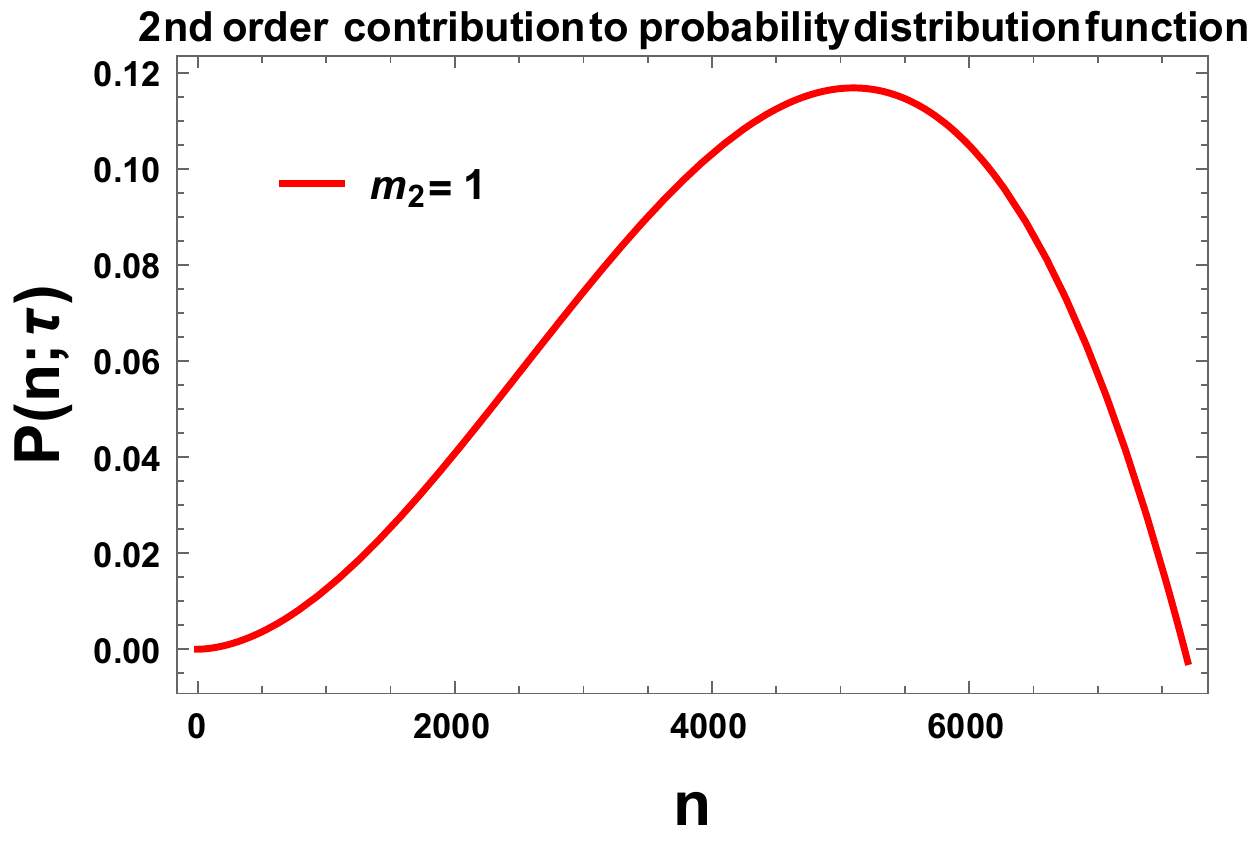}
    \label{x26}
}
\subfigure[Second order corrected distribution LogLog plot for $m_{2}=\pm2,\pm3$]{
    \includegraphics[width=7.8cm,height=8cm] {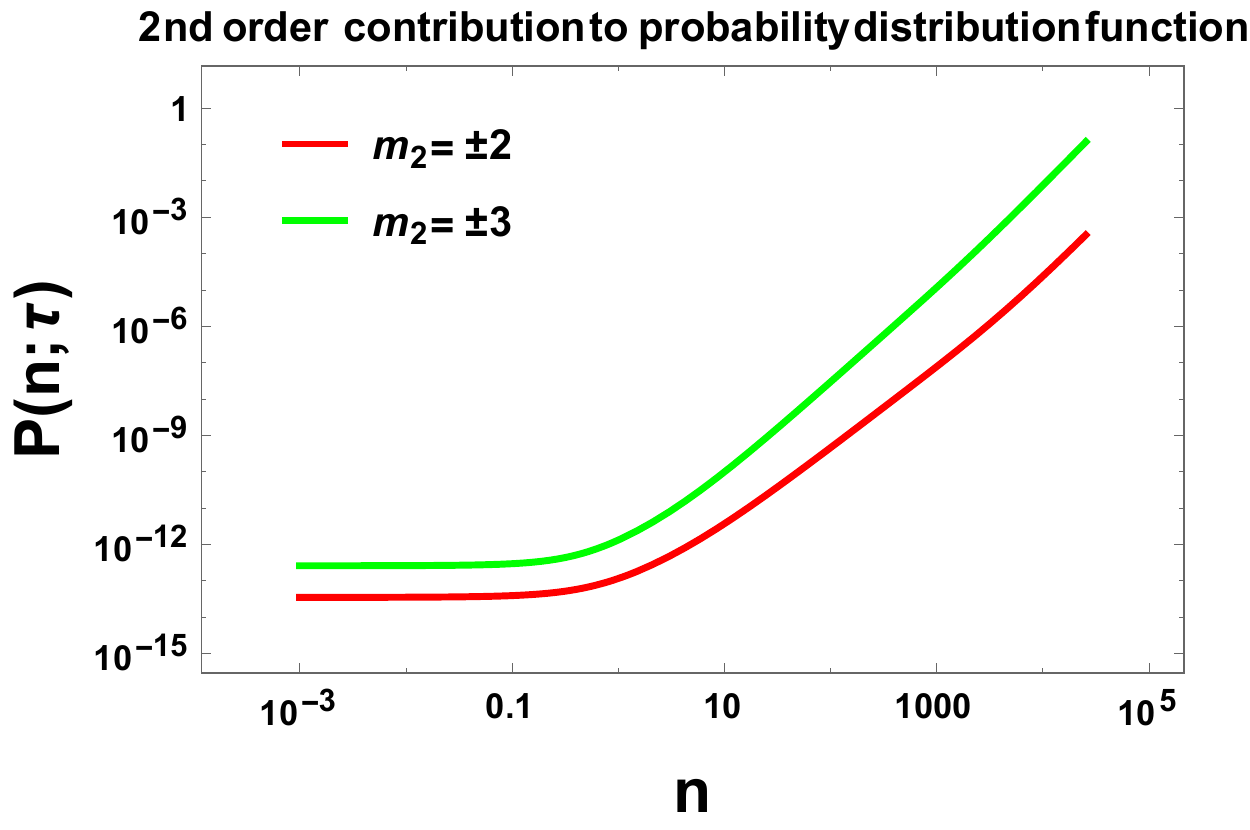}
    \label{x24}
}

\caption{ Evolution of probability distribution function obtained from the second order corrected Fokker-Planck equation with the occupation number $n$ for different value of $m_{2}$. Here we use the initial conditions as mentioned in Eq~(\ref{ew1}). }
\label{Fig2ndorder}
\end{figure}
%%%%%%
%%%%%%%
 From figure~\ref{Fig2ndorder} we can say that for $m_{2}=\pm 2,\pm 3$ second order corrected probability distribution function almost overlap at lower values of the occupation number $n$ but deviate significantly as $n$ increases to large number. As a consequence, for low value of $n$ particle production rate is independent on $m_{2}$ but as $n$ increases they significantly deviate. It also implies that for higher values of $n$ the integer $m_{2}$ constrains particle production rate.
 For $m_2=\pm 1$ we found that the second order corrected probability distribution function significantly deviates from log normal (Gaussian) distribution and both of them explicitly show the signature of non-Gaussianity is the second order corrected distribution function. Finally, we have shown that for $m_2=0$ the amount of deviation from log normal (Gaussian) distribution is small compared to results obtained from the other values of $m_2$.
 
Now we discuss about the analytical solution of Eq~(\ref{eq25}) for the special case when we fix $m_{2}=0$. In this situation one can recast the Eq~(\ref{eq25}) in to the following simplified form:
\bea\label{eq26}
\left[\frac{n^2}{2}\left(1 + n\right)^{2}\frac{d^{4}}{d n^{4}}+2n \left(1 + 3 n + 2 n^2\right)\frac{d ^{3}}{d n^{3}}+\left(1 + 6 n + 6 n^2\right) \frac{d^{2}}{d n^{2}}\right] P_1(n)=0
\eea
Then analytical solution of Eq~(\ref{eq26}) can be written as:
\bea
P_1(n)&=&-C_1\sum _{i=0}^{\infty } \frac{2 \Gamma\left(i-\frac{i \sqrt{7}}{2}+\frac{1}{2}\right) \Gamma \left(i-\frac{1}{2} i (\sqrt{47}+i)\right) \Gamma \left(i+\frac{1}{2} (\sqrt{47}+i) i+1\right) n^{i-\frac{i \sqrt{7}}{2}+\frac{1}{2}}}{(-2 i+\sqrt{7} i+1) i! \Gamma(i-i \sqrt{7}+1)\Gamma \left(-\frac{1}{2} i (\sqrt{47}+i)\right) \Gamma \left(1+\frac{1}{2} i(\sqrt{47}+i)\right)}\nonumber\\
&&~~~~~~~~~~~~~~~~~~~~~~\times \, _2\tilde{F}_1\left(\frac{3}{2}-\frac{i \sqrt{7}}{2},i-\frac{i \sqrt{7}}{2}-\frac{1}{2};i-\frac{i \sqrt{7}}{2}+\frac{3}{2};-n\right)\nonumber \\
&&+C_2 \sum _{i=0}^{\infty }\sum _{j=0}^{\infty }\sum _{m=0}^{\infty } \frac{\Gamma (c) 2^{u+1} \sqrt{\pi } 2^{-b-1} ((\sqrt{\pi } 2^{-b-1}) \Gamma (b+u+1)) }{(u-1)\Gamma (a) \Gamma (b_1) \Gamma (b_2)\Gamma\left(\frac{1}{2} (b+u+1)\right) \Gamma \left(\frac{1}{2} (b+u+2)\right)}\nonumber\\
&&\times \frac{2^j (-1)^{m+\frac{u}{2}+1} n^{i+m+\frac{u-1}{2}+1}\Gamma \left(\frac{1}{2} (b+u+1)+i\right) \Gamma\left(\frac{1}{2} (b+u+2)+i\right) }{i!j!m!\left(i+m+\frac{u-1}{2}+1\right) \Gamma\left(b+i+\frac{3}{2}\right) \Gamma (c+j+m)}\nonumber\\
&&~~~~~~~~~~~~~~~~~~~~~~~~~~~~~~~~~~~~~~~~~~~\times  \Gamma(b_1+m) \Gamma (b_2+n) \Gamma (a+j+m).
\label{2ndanalytical}
\eea
where $a$, $b$, $c$, $b_1$, $b_2$ and $u$ are all functions of $i$ which is summed over and we have introduced them to use shorthand notation. In this context the functional dependence of all of these $i$ dependent parameters are given by the following expressions:
\bea a=a(i)&=&\frac{1}{2} \left(-1+i \sqrt{7}\right),~~~
        b=b(i)=\frac{1}{2} i \left(\sqrt{47}+i\right),\nonumber \\
        c=c(i)&=&\frac{1}{2} \left(1+i \sqrt{7}\right),~~~
        b_1=b_1(i)=\frac{1}{2} \left(3-i \sqrt{7}\right),\nonumber \\
        b_2=b_2(i)&=&\frac{1}{2} \left(4 i+2 \sqrt{7} i+\sqrt{47} i+1\right),~~~
        u=u(i)=i \sqrt{7}.\eea
 The solution contains generalized Hypergeometric PFQ Regularized function. Also $C_{1}$ and $C_{2}$ are arbitrary constant of integration which can be evaluated by imposing appropriate initial conditions.

For $m_2=0$ the total probability distribution function can be expressed as:
\bea
P(n;\tau)&=&\left[-C_1\sum _{i=0}^{\infty } \frac{2 \Gamma\left(i-\frac{i \sqrt{7}}{2}+\frac{1}{2}\right) \Gamma \left(i-\frac{1}{2} i (\sqrt{47}+i)\right) \Gamma \left(i+\frac{1}{2} (\sqrt{47}+i) i+1\right) n^{i-\frac{i \sqrt{7}}{2}+\frac{1}{2}}}{(-2 i+\sqrt{7} i+1) i! \Gamma(i-i \sqrt{7}+1)\Gamma \left(-\frac{1}{2} i (\sqrt{47}+i)\right) \Gamma \left(1+\frac{1}{2} i(\sqrt{47}+i)\right)}\right.\nonumber\\
&&\left.~~~~~~~~~~~~~~~~~~~~~~\times \, _2\tilde{F}_1\left(\frac{3}{2}-\frac{i \sqrt{7}}{2},i-\frac{i \sqrt{7}}{2}-\frac{1}{2};i-\frac{i \sqrt{7}}{2}+\frac{3}{2};-n\right)\right.\nonumber \\
&&\left.+C_2 \sum _{i=0}^{\infty }\sum _{j=0}^{\infty }\sum _{m=0}^{\infty } \frac{\Gamma (c) 2^{u+1} \sqrt{\pi } 2^{-b-1} ((\sqrt{\pi } 2^{-b-1}) \Gamma (b+u+1)) }{(u-1)\Gamma (a) \Gamma (b_1) \Gamma (b_2)\Gamma\left(\frac{1}{2} (b+u+1)\right) \Gamma \left(\frac{1}{2} (b+u+2)\right)}\right.\nonumber\\
&&\left.\times \frac{2^j (-1)^{m+\frac{u}{2}+1} n^{i+m+\frac{u-1}{2}+1}\Gamma \left(\frac{1}{2} (b+u+1)+i\right) \Gamma\left(\frac{1}{2} (b+u+2)+i\right) }{i!j!m!\left(i+m+\frac{u-1}{2}+1\right) \Gamma\left(b+i+\frac{3}{2}\right) \Gamma (c+j+m)}\right.\nonumber\\
&&\left.~~~~~~~~~~~~~~~~~~~~~~~~~~~~~~~~~~~~~~~~~~~\times  \Gamma(b_1+m) \Gamma (b_2+n) \Gamma (a+j+m)\right]\nonumber\\
&&~~~~~~~~~~~~~~~~~~~~~~~~~~~~~~~~~~~~~~~~~\times\left[C_3 e^{\tau  m_2\mu_k}+C_4 e^{-\tau m_2\mu_k}\right].
\label{gjgjl}
\eea
For the special case $m_{2}=0$ and considering the large $n$ limit ($n \rightarrow \infty$) the  Eq~(\ref{eq26}) reduces to the following form:
\bea\label{eq27}
\left[\frac{n^4}{4}\frac{d^{4}}{d n^{4}}+2n^3\frac{d ^{3}}{d n^{3}}+3 n^2\frac{d^{2}}{d n^{2}}\right] P_1(n)=0
% (3 n^{2})\frac{d^{2}P[n]}{dn^{2}}]+(2 n^{3})\frac{d^{3}P[n]}{dn^{3}}+\frac{1}{4} n^{3} \frac{d^{4}P[n]}{dn^{4}}=0
\eea
After solving Eq~(\ref{eq17}) we get the following solution in the large $n$ limit:
\bea
\label{eq28}
 P_1(n)=\left[\frac{1}{6} \left(\frac{C_5}{n^2}+\frac{3 C_6}{n}\right)+C_7 n+C_8\right], \eea
where $C_{5}$, $C_{6}$, $C_{7}$ and $C_{8}$ are the arbitrary constants of integration which can be evaluated by imposing appropriate initial condition.

In the large $n$ limit with $m_2=0$ the total probability distribution function can be expressed as:
\bea P(n;\tau)&=&\left[\frac{1}{6} \left(\frac{C_5}{n^2}+\frac{3 C_6}{n}\right)+C_7 n+C_8\right]\left[C_3 e^{\tau  m_2\mu_k}+C_4 e^{-\tau m_2\mu_k}\right],\eea
which shows huge deviation from log normal (Gaussian) distribution. 

Further using the  Fourier transformation with respect to the occupation number $n$ as mentioned in Eq~(\ref{ft1}), we get the following simplified expression for the {\it Fokker Planck equation} at the second order:
\bea\label{xzx2} \frac{\pl^2 \bar{P}(k;\tau)}{\pl \tau^2}=\mu^2_k\left[\frac{n^2}{2}(1+n)^2k^4-2ink^3(1+3n+2n^2)-k^2(1+6n+6n^2)\right]\bar{P}(k;\tau),\eea
which is obviously a simplest version of the {\it Fokker Planck equation} as it contains only two derivative with respect to time $\tau$. 
In the present context we get the following result for the probability distribution function in the Fourier transformed space, as given by:
\bea \bar{P}(k;\tau|n^{'};\tau^{'})&=&C_1~\exp\left[{\cal G}(k;n^{'})~(\tau-\tau^{'})\right]+C_2~\exp\left[-{\cal G}(k;n^{'})~(\tau-\tau^{'})\right],\eea
where ${\cal G}(k;n^{'})$ is defined as:
\bea {\cal G}(k;n^{'})=\mu_k\sqrt{\left\{\frac{(n^{'})^2}{2}(1+n^{'})^2k^4-2in^{'}k^3(1+3n^{'}+2(n^{'})^2)-k^2(1+6n^{'}+6(n^{'})^2)\right\}}. \eea
Additionally, $C_1$ and $C_2$ are arbitrary constants which is fixed by the
following two fold boundary conditions, as given by:
\bea P(n;\tau|n^{'}=0;\tau^{'}=\tau)&=&\delta(n),\\
\left(\frac{\pl  P(n;\tau |n^{'};\tau^{'})}{\pl \tau}\right)_{n^{'}=0,\tau^{'}=\tau} &=&- \frac{\delta(n)}{n},\eea
which are necessary to solve the above mentioned second order differential equation.

As a result, we get the following set of constraints equations:
\bea C_1+C_2&=&1,\nonumber\\
C_1-C_2&=&-\frac{1}{i\mu_k k n},
\eea
Solving these equations we get:
\bea C_1&=&\frac{1}{2}\left(1-\frac{1}{i\mu_k k n}\right),\\
 C_2&=&\frac{1}{2}\left(1+\frac{1}{i\mu_k k n}\right).\eea
 Using this solution we get the following probability distribution function in Fourier space with $n^{'}=0$ and $\tau^{'}=0$, as given by:
\bea \bar{P}(k;\tau)=& \cos (\mu_k k \tau) -\frac{\sin (\mu_k k \tau)}{\mu_k k n}.\eea
Hence substituting back the above mentioned result into the definition of Fourier transformation and setting the initial condition $n^{'}=0$ and $\tau^{'}=0$ we get the following result for the probability distribution function, as 
given by:
\bea P(n;\tau)=\frac{1}{2\pi}\int^{\infty}_{-\infty}dk~\exp[ikn]\left( \cos (\mu_k k \tau) -\frac{\sin (\mu_k k \tau)}{\mu_k k n}\right).\eea
However this integral is not convergent within $-\infty<k<\infty$. For this reason we introduce a momentum cut-off $-\Lambda_{C}<k<\Lambda_{C}$. Consequently, we get the following regularised expression for the probability distribution function:
\bea P(n;\tau)&=&\frac{n \sin (\Lambda_C  n) \cos (\Lambda_C  \mu_k  \tau )-\mu_k  \tau  \cos (\Lambda_C  n) \sin (\Lambda_C  \mu_k  \tau )}{\pi  (n^2- \mu ^2_k \tau ^2)}\nonumber\\
&&~~-\frac{1}{4 \pi  \mu_k  n}\left[i (\text{Ci}(-\Lambda_C  (n+\mu_k  \tau ))-\text{Ci}(\Lambda_C  (n+\mu_k  \tau ))-\text{Ci}(\Lambda_C  \mu_k  \tau -n \Lambda_C )\right.\nonumber\\
&&\left.~~+\text{Ci}(\Lambda_C  (n-\mu_k  \tau ))
-2 i \text{Si}(\Lambda_C  (n+\mu_k  \tau ))+2 i \text{Si}(\Lambda_C  (n-\mu_k  \tau )))\right].
\label{ana_2}
\eea
 which is coming from the second contribution in the {\it Fokker Planck equation}.

\begin{figure}[H]
\centering
\includegraphics[width=16cm,height=8cm] {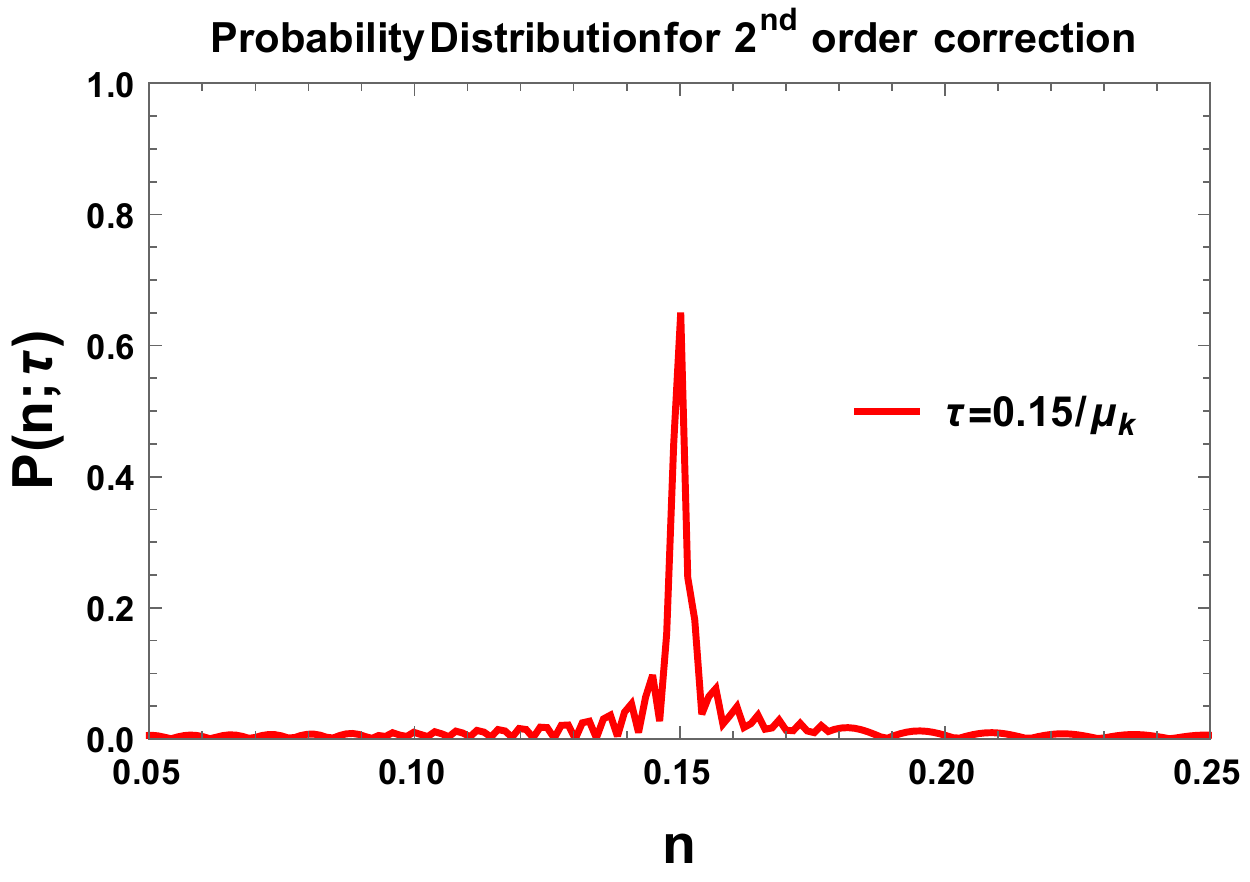}

\caption{ Second order contribution to the probability density function with respect to the the occupation number per mode $n$, for a fixed time from the analytical solution[\ref{ana_2}] of second order equation[\ref{eq24}].}
\label{Fig2ndana1}
\end{figure}

In figure~[\ref{Fig2ndana1}], we have shown the probability density  function for second order correction with respect to the occupation number per mode, for a fixed time ($\tau=\frac{0.15}{\mu_{k}}$). From this plot we have observed irregular oscillations with deviation from Gaussian feature.We have selected a high value for momentum cutoff[$\lambda_{c}$] for this particular plot.

%%%%%%%%%%%%%%%%%%5
%%%%%%%%%%%%%%%%%%
%%%%%%%%%%%%%%%%%%%
%%%%%%%%%%%%%%%%%
%%%%%%%%%%%%%%%%%
%%%%%%%%%%%%%%%%%%
%%%%%%%%%%%%%%%%%%%
%%%%%%%%%%%%%%%%%%%
\subsubsection{Third order correction} 
In this context, our objective is to find out the contributions coming from third order in the {\it Fokker Planck equation} and to solve this equation
numerically~\footnote{Including the contributions from third order we will see that the {\it Fokker Planck equation} can not solvable analytically.}. To serve this purpose we equate both sides of Eq~(\ref{eq18}) after Taylor expansion and compare the coefficient of $\del\tau ^{3}$.
Consequently, we get the following partial differential equation:  
\bea
&&\frac{ n^3}{6} (1 + n)^{3}\frac{\pl^{6}P(n;\tau)}{\pl n^{6}}\nonumber\\
&&+ \frac{3n^2}{2} (1 + n)^2 (1 + 2 n) \frac{\pl^{5}P(n;\tau)}{\pl n^{5}}\nonumber\\
&&+ 3 n (1 + n) (1 + 5 n + 5 n^2)\frac{\pl^{4}P(n;\tau)}{\pl n^{4}}\nonumber\\
&&+(1 + 2 n) (1 + 10 n + 10 n^2)\frac{\pl^{3}P(n;\tau)}{\pl n^{3}}=\frac{1}{\mu^3_k}\frac{\pl^3P(n;\tau)}{\pl \tau^3}   
\label{cxz}
\eea
%\be\begin{array}{lllll}
%\displaystyle m_{3}^{2} P[n]= (1 + 2 n) (1 + 10 n + 10 n^2)\frac{d^{3}P[n]}{dn^{3}} + 3 n (1 + n) (1 + 5 n + 5 n^2))\frac{d^{4}P[n]}{dn^{4}} \\
%\displaystyle \hspace{2cm}+ \frac{3n^2}{2} (1 + n)^2 (1 + 2 n) \frac{d^{5}P[n]}{dn^{5}} +\frac{ n^3}{6} (1 + n)^{3}\frac{d^{6}P[n]}{dn^{6}}
%\label{3rdodmain}
%\end{array}
%\ee
which can not able to solve analytically with any integer values of $m_3$.
We solve this equation for different values of $m_{3}$ numerically with assumed initial condition. Only for the special case, $m_3=0$ with large $n$ limit we can able to provide an analytical solution in the present context.

Now to solve this partial differential equation we apply method of separation of variable, using which we can write the total solution in the following form:
\bea  \label{eq33cv} P(n;\tau)=P_1(n)P_2(\tau).\eea
Further, using the solution ansatz stated in Eq~(\ref{eq19cccas}) we get the following sets of independent differential equations, as given by:
\bea
&&\frac{ n^3}{6} (1 + n)^{3}\frac{d^{6}P_1(n)}{d n^{6}}\nonumber\\
&&+ \frac{3n^2}{2} (1 + n)^2 (1 + 2 n) \frac{d^{5}P_1(n)}{d n^{5}}\nonumber\\
&&+ 3 n (1 + n) (1 + 5 n + 5 n^2)\frac{d^{4}P_1(n)}{d n^{4}}\nonumber\\
&&+(1 + 2 n) (1 + 10 n + 10 n^2)\frac{d^{3}P_1(n)}{d n^{3}}-m^2_3P_1(n)=0   
\label{cxz},\\
\label{eq25bccx} &&\left[\frac{d^{3} }{d \tau ^{3}}-m^2_3\mu^3_k\right]P_2(\tau)=0.
\eea
It is important to note that, the analytical solution of $P_1(n)$ is not possible for any arbitrary values of the constant $m_3$, except the special case $m_3=0$. For this reason we use numerical technique to solve Eq~(\ref{cxz}). On the other hand Eq~(\ref{eq25bccx}) is exactly solvable in the present context and the solution can be written as:
\bea P_2(\tau)=\left[C_7 e^{(-1)^{2/3} m_3^{2/3} \tau  \mu _k}+C_8 e^{-\sqrt[3]{-1} m_3^{2/3} \tau  \mu _k}+C_9 e^{m_3^{2/3} \tau  \mu _k}\right],\eea
where $C_7$, $C_8$ and $C_9$ are three arbitrary constants which can be fixed by choosing proper boundary conditions.

Now to  solve Eq~(\ref{cxz}) numerically for different values of $m_{3}$ along with given initial condition and we also prove the analytical solution for the special case $m_3=0$. Here it important to mention that, since arbitrary values of $m_3$ is allowed, one can consider integer as well as non integer values at the level of solution of differential equation. However, the only physically acceptable solution restrict us to only consider the integer values of $m_3$ because such third order corrected solution of the {\it Fokker Planck equation} is directly related to the quantum effects as we have mentioned earlier. As a result such integer values of $m_3$ can be interpreted as the quantum number i. e. 
\bea \textcolor{blue}{\bf \underline{Quantum ~Number~IV:}}~~~~m_3=0,\pm 1,\pm 2,\cdots,\pm \infty \in \mathbb{Z}.\eea
For numerical solution we take the following assumptions:
\bea \label{ddsa} P_1(n=0.0001)&=&100,~~~~~~
\left[\frac{d P_1(n)}{d n}\right]_{n=0.0001}=100,\nonumber\\
\left[\frac{d^{2} P_1(n)}{d n^{2}}\right]_{n=0.0001}&=&100,~~~~~~
\left[\frac{d^{3} P_1(n)}{d n^{3}}\right]_{n=0.0001}]=100,\nonumber\\
\left[\frac{d^{4} P_1(n)}{d n^{4}}\right]_{n=0.0001}&=&100,~~~~~~
\left[\frac{d^{5} P_{1}(n)}{d n^{5}}\right]_{n=0.0001}=100.\eea
Asper our assumption particle production probability has constant value at some particular small n[$0.0001$] or for small n $P[n,\tau]$ and all its derivative has constant values and all those values are same.

%%% Add pictures
\begin{figure}[htb]
\centering
\subfigure[Third order corrected distribution for   $m_{3}=0,1,2,3$]{
    \includegraphics[width=7.8cm,height=8cm] {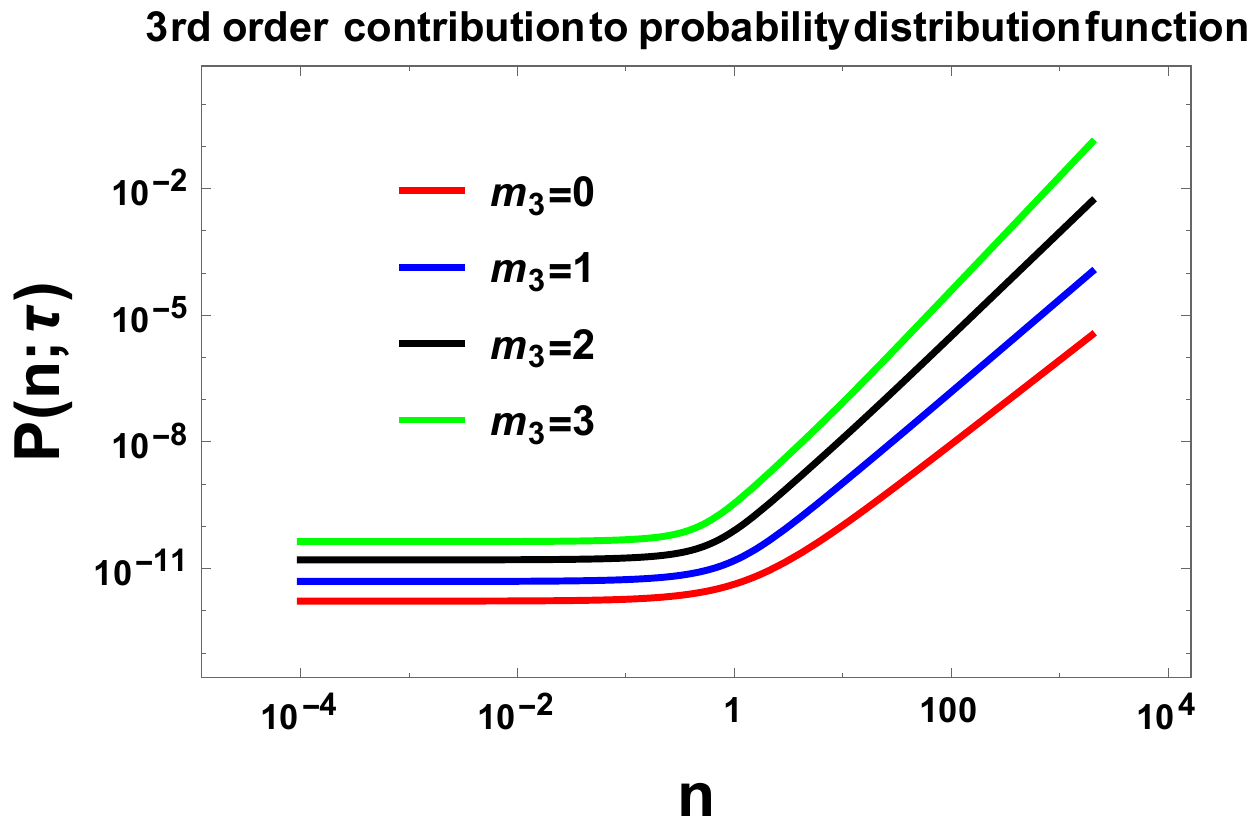}
    \label{x31}
}
\subfigure[Third order corrected distribution for   $m_{3}=0,-1,-2,-3$]{
    \includegraphics[width=7.8cm,height=8cm] {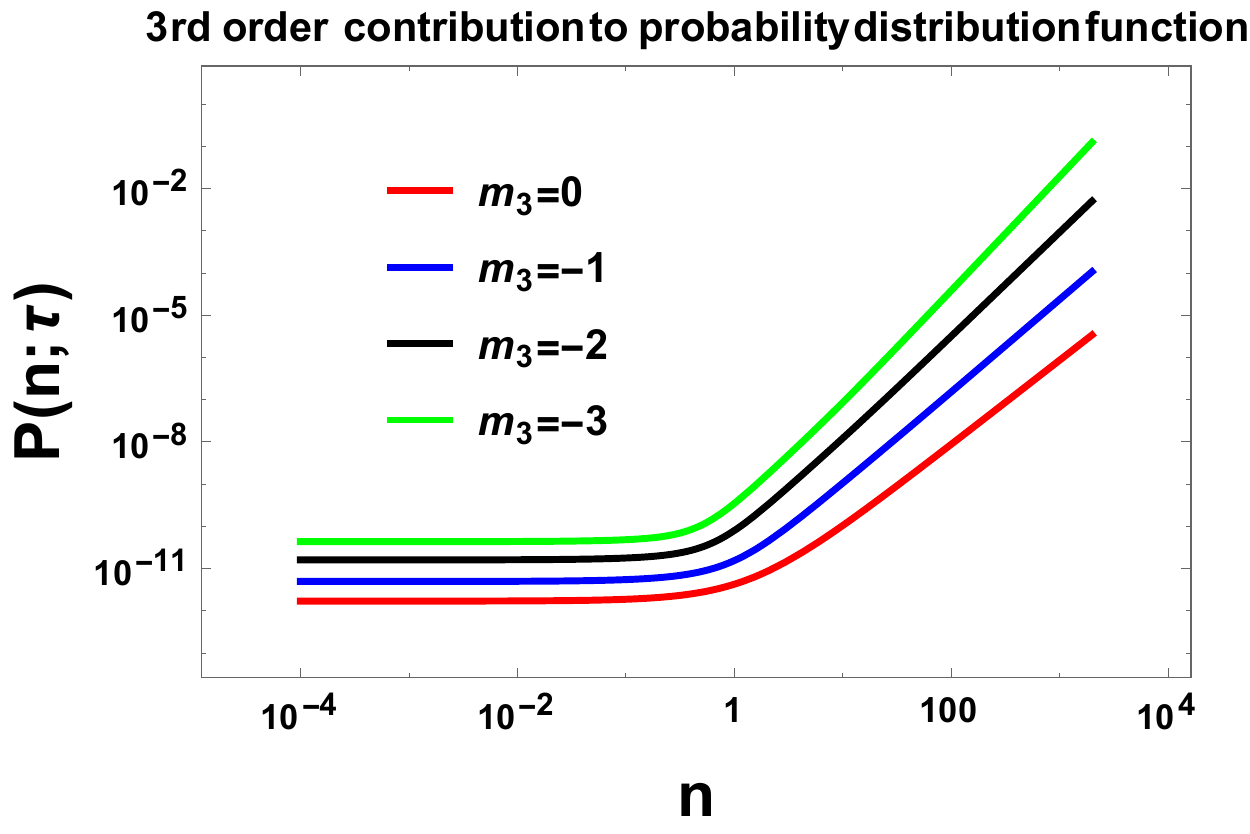}
    \label{x31}
}
\caption{Evolution of probability distribution function obtained from the third order corrected Fokker-Planck equation with the occupation number $n$ for different value of $m_{3}$. Here we use the initial conditions as mentioned in Eq~(\ref{ddsa}).}
\label{Fig3rdorder}
\end{figure}

From Fig \ref{Fig3rdorder} one can say that the third order corrected probability distribution function for different value of $m_{3}=0$ overlap at lower $n$ limit (n $\rightarrow 0$) though deviate significantly at large $n$ limit. At lower values of $n$, particle production probability is independent of $m_{3}$ and almost flat. But as soon as $n$ reaches values greater than unity the distribution increase exponentially 
and for different $m_{3}$ they differ from each other.

Now we discuss about the analytical solution of Eq~(\ref{cxz}) for the special case when we fix $m_{3}=0$. In this situation one can recast the Eq~(\ref{cxz}) in to the following simplified form:
\bea
&&\frac{ n^3}{6} (1 + n)^{3}\frac{d^{6}P_1(n)}{d n^{6}}\nonumber\\
&&+ \frac{3n^2}{2} (1 + n)^2 (1 + 2 n) \frac{d^{5}P_1(n)}{d n^{5}}\nonumber\\
&&+ 3 n (1 + n) (1 + 5 n + 5 n^2)\frac{d^{4}P_1(n)}{d n^{4}}\nonumber\\
&&+(1 + 2 n) (1 + 10 n + 10 n^2)\frac{d^{3}P_1(n)}{d n^{3}}=0.   
\label{cxzss}
\eea
To find the analytical solution of Eq~(\ref{cxzss}) one can further use the following simplification:
\bea
\label{eq29f}
&& \frac{n^3(1 + n)^{3}}{6}\frac{d^{3}Q_1(n)}{dn^{3}} \nonumber\\
&&+ \frac{3 n^{2} (1 + n)^{2} (1 + 2 n)}{2} \frac{d^{2}Q_1(n)}{dn^{2}}  \nonumber\\
&& + 3 n (1 + n) (1 + 5 n + 5 n^2)\frac{dQ_1(n)}{dn}\nonumber\\
&&+ (1 + 2 n) (1 + 10 n + 10 n^2) Q_1(n)= 0
\eea
where we introduce a new function $Q_1(n)$ which can be expressed in terms of $P_1(n)$ by following identification:
\bea  Q_1(n)=\frac{d^{3}P_1(n)}{dn^{3}}.\eea 
 On large $n$ limit ($n\rightarrow \infty$) the Eq~\ref{eq29f} reduces to following extremely simplified form:
\bea\label{eq30}
\frac{n^6}{6} \frac{d^{3}Q_1(n)}{dn^{3}}+3 n^{5}  \frac{d^{2}Q_1(n)}{dn^{2}}+15 n^4 \frac{dQ_1(n)}{dn} +20n^3 Q_1(n)=0.
\eea
The solution of this equation at large $n$ limit can be expressed as:
\bea\label{eq31}
Q_1(n)&=&\left[C_3-\frac{1}{148 n^{15/2}}\left\{\left(15 C_1-\sqrt{71} C_2\right) \sin \left(\frac{1}{2} \sqrt{71} \ln n\right)\right.\right.\nonumber\\
&& \left.\left.~~~~~~~~~~~~~+\left(15 C_2+\sqrt{71} C_1\right) \cos \left(\frac{1}{2} \sqrt{71}\ln n\right)+240 n^{15/2}\ln n\right\}\right],~~~
\eea
where $C_{1}$,$C_{2}$ and $C_{3}$ are arbitrary constant of integration and can be evaluated by imposing appropriate initial condition.

Here, in the large $n$ limit the solution for $P_1(n)$ can be written as:
\bea P_1(n)&=&\left[\frac{1}{674880 n^{9/2}}\left\{3040 n^{15/2} \left(37 C_3-60 \ln n+110\right)\right.\right.\nonumber\\&& \left.\left.~~~~~~~~~~~~~~~~~~~~~-\left(189 C_1+17 \sqrt{71} C_2\right) \sin \left(\frac{1}{2} \sqrt{71} \ln n\right)\right.\right.\nonumber\\&& \left.\left.~~~~~~~~~~~~~~~~~~~~~+\left(17 \sqrt{71} C_1-189 C_2\right) \cos \left(\frac{1}{2} \sqrt{71} \ln n\right)\right\}\right.\nonumber\\&& \left.~~~~~~~~~~~~~~~~~~~~~~~~~~~~~~~~~~~~~~~~~~~~~~~~~~~~+ C_6 n^2+C_5 n+C_4\right], \eea
where $C_{4}$, $C_{5}$ and $C_{6}$ are arbitrary constant of integration and can be evaluated by imposing appropriate initial condition.

Finally, in the large $n$ limit the total probability distribution function can be expressed as:
\bea P(n;\tau)&=&\left[\frac{1}{674880 n^{9/2}}\left\{3040 n^{15/2} \left(37 C_3-60 \ln n+110\right)\right.\right.\nonumber\\&& \left.\left.~~~~~~~~~~~~~~~~~~~~~-\left(189 C_1+17 \sqrt{71} C_2\right) \sin \left(\frac{1}{2} \sqrt{71} \ln n\right)\right.\right.\nonumber\\&& \left.\left.~~~~~~~~~~~~~~~~~~~~~+\left(17 \sqrt{71} C_1-189 C_2\right) \cos \left(\frac{1}{2} \sqrt{71} \ln n\right)\right\}\right.\nonumber\\&& \left.~~~~~~~~~~~~~~~~~~~~~~~~~~~~~~~~~~~~~~~~~~~~~~~~~~~~+ C_6 n^2+C_5 n+C_4\right]\nonumber\\
&&\times \left[C_7 e^{(-1)^{2/3} m_3^{2/3} \tau  \mu _k}+C_8 e^{-\sqrt[3]{-1} m_3^{2/3} \tau  \mu _k}+C_9 e^{m_3^{2/3} \tau  \mu _k}\right],\eea
which shows large deviation from log normal (Gaussian) distribution. 

Further using the  Fourier transformation with respect to the occupation number $n$ as mentioned in Eq~(\ref{ft1}), we get the following simplified expression for the {\it Fokker Planck equation} at the third order:
\bea\label{xzazx2} \frac{\pl^3 \bar{P}(k;\tau)}{\pl \tau^3}&=&\mu^3_k\left[-\frac{n^3}{6}(1+n)^3k^6+\frac{3n^2i}{2}(1+n)^2(1+2n)k^5\right.\nonumber\\ &&
\left.+3n(1+n)(1+5n+5n^2)k^4-ik^3(1+2n)(1+10n+10n^2)\right]\bar{P}(k;\tau),~~~~~~~~~~\eea
which is obviously a simplest version of the {\it Fokker Planck equation} as it contains only three derivative with respect to time $\tau$. 
In the present context we get the following result for the probability distribution function in the Fourier transformed space, as given by:
\bea \bar{P}(k;\tau|n^{'};\tau^{'})&=&C_1 ~\exp\left[(-1)^{2/3}\sqrt[3]{{\cal O}(k;n^{'})}~(\tau-\tau^{'}) \right]\nonumber\\
&&~~~+C_2 ~\exp\left[(-1)^{1/3}\sqrt[3]{{\cal O}(k;n^{'})}~(\tau-\tau^{'}) \right]\nonumber\\
&&~~~~~~~~~~+C_3~\exp\left[\sqrt[3]{{\cal O}(k;n^{'})}~(\tau-\tau^{'}) \right],\eea
where ${\cal O}(k;n^{'})$ is defined as:
\bea {\cal O}(k;n^{'})&=&\mu^3_k\left[-\frac{(n^{'})^3}{6}(1+n^{'})^3k^6+\frac{3(n^{'})^2i}{2}(1+n^{'})^2(1+2n^{'})k^5\right.\nonumber\\ &&
\left. +3n^{'}(1+n^{'})(1+5n^{'}+5(n^{'})^2)k^4-ik^3(1+2n^{'})(1+10n^{'}+10(n^{'})^2)\right].~~~~~~ \eea
Additionally, $C_1$ , $C_2$ and $C_3$ are arbitrary constants which is fixed by the 
following three fold boundary conditions, as given by:
\bea P(n;\tau|n^{'}=0;\tau^{'}=\tau)&=&\delta(n),\\
\left(\frac{\pl  P(n;\tau |n^{'};\tau^{'})}{\pl \tau}\right)_{n^{'}=0,\tau=\tau^{'}} &=&-\frac{\delta(n)}{n},\\
\left(\frac{\pl^2  P(n;\tau |n^{'};\tau^{'})}{\pl \tau^{2}}\right)_{n^{'}=0,\tau=\tau^{'}} &=&\frac{2~\delta(n)}{n^2}.\eea
which are necessary to solve the above mentioned third order differential equation.

As a result, we get the following set of constraints equations:
\bea &&C_1 +C_2 +C_3=1,\\
&&C_1 -(-1)^{2/3}~C_2 -(-1)^{1/3}~C_3=\frac{1}{i^{1/3}\mu_k k n},~~~~\\
&&C_1-(-1)^{4/3}C_2 -(-1)^{2/3}C_3=-\frac{2}{i^{2/3}\mu^2_k k^2 n^2},~~~~
\eea
Solving these equations we get:
\bea C_1&=&\frac{(-1)^{2/3} \left(-\sqrt[6]{-1} k n \mu _k-i k n \mu _k+2 \sqrt[3]{-1}-2\right)}{\left(3 \sqrt[3]{-1}-1\right) k^2 n^2 \mu _k^2},\\
 C_2&=&-\frac{\sqrt[3]{-1} \left(-k n \mu _k-(-1)^{2/3} k n \mu _k-4 i+2 \sqrt[6]{-1}+6 (-1)^{5/6}\right)}{\left(\sqrt[6]{-1}-i\right) \left(3 \sqrt[3]{-1}-1\right) k^2 n^2 \mu _k^2},\\
 C_3&=&-\frac{\sqrt[3]{-1} k n \mu _k-(-1)^{2/3} k n \mu _k-8 i+10 \sqrt[6]{-1}+10 (-1)^{5/6}}{\left(\sqrt[6]{-1}-i\right) \left(3 \sqrt[3]{-1}-1\right) k^2 n^2 \mu _k^2}.\eea
 Using this solution get the following probability distribution function in Fourier space, as given by:
\bea \bar{P}(k;\tau)&=&\frac{(-1)^{2/3} \left(-\sqrt[6]{-1} k n \mu _k-i k n \mu _k+2 \sqrt[3]{-1}-2\right)}{\left(3 \sqrt[3]{-1}-1\right) k^2 n^2 \mu _k^2} ~\exp\left[-i^{1/3}\mu_k k~\tau\right]\nonumber\\
&&-\frac{\sqrt[3]{-1} \left(-k n \mu _k-(-1)^{2/3} k n \mu _k-4 i+2 \sqrt[6]{-1}+6 (-1)^{5/6}\right)}{\left(\sqrt[6]{-1}-i\right) \left(3 \sqrt[3]{-1}-1\right) k^2 n^2 \mu _k^2} ~\exp\left[(-1)^{2/3}i^{1/3}\mu_k k~\tau \right]\nonumber\\
&&+-\frac{\sqrt[3]{-1} k n \mu _k-(-1)^{2/3} k n \mu _k-8 i+10 \sqrt[6]{-1}+10 (-1)^{5/6}}{\left(\sqrt[6]{-1}-i\right) \left(3 \sqrt[3]{-1}-1\right) k^2 n^2 \mu _k^2}~\exp\left[(-1)^{1/3}i^{1/3}\mu_k k~\tau \right].\nonumber\\
&& \eea
Hence substituting back into the definition of Fourier transformation and setting the initial condition $n^{'}=0$ and $\tau^{'}=0$ we get the following result for the probability distribution function, as 
given by:
\bea P(n;\tau)
&=&\frac{ \left(\left(\sqrt{3}+3 i\right) \mu_k +2 \left(\sqrt{3}+i\right)\right) n^3}{4 \left(\sqrt{3}+2 i\right) \mu ^2_k n^2 \left((-1)^{2/3} \mu_k  \tau +n\right) \sqrt{\left((-1)^{2/3} \mu_k  \tau +n\right)^2}}\nonumber\\
&&+2 i n^2 \left(2 i \sqrt{3} \mu ^2_k \tau +\mu_k  \left(\sqrt{-\sqrt[3]{-1} \mu ^2_k \tau ^2+n^2+2 (-1)^{2/3} \mu_k  n \tau }+3 i \sqrt{3} \tau +3 \tau \right)\right.\nonumber\\
&&\left.~~-2 \sqrt{-\sqrt[3]{-1} \mu ^2_k \tau ^2+n^2+2 (-1)^{2/3} \mu_k  n \tau }\right)-\mu_k  n \tau  \left(\left(-\left(\sqrt{3}-3 i\right)\right) \mu ^2_k \tau\right.\nonumber\\
&&\left.~~~~~~~~~~~~~~~ +2 \sqrt[6]{-1} \mu  \left(\sqrt{-\sqrt[3]{-1} \mu ^2_k \tau ^2+n^2+2 (-1)^{2/3} \mu_k  n \tau }+3 i \sqrt{3} \tau +3 \tau \right)\right.\nonumber\\
&&\left.~~~~~~~~~~~~~~~-2 \left(\sqrt{3}-i\right) \sqrt{-\sqrt[3]{-1} \mu ^2_k \tau ^2+n^2+2 (-1)^{2/3} \mu_k  n \tau }\right)\nonumber\\
&&~~+\left(\sqrt{3}+i\right) \mu ^2_k \tau ^2 \left(2 \mu_k  \tau +\sqrt{-2 i \left(\sqrt{3}-i\right) \mu ^2_k \tau ^2+4 n^2+4 i \left(\sqrt{3}+i\right) \mu_k  n \tau }\right),\nonumber\\
&&
\label{ana_3}
\eea
 which is coming from the third contribution in the {\it Fokker Planck equation}. 
 
 \begin{figure}[H]
\centering
\includegraphics[width=16cm,height=8cm] {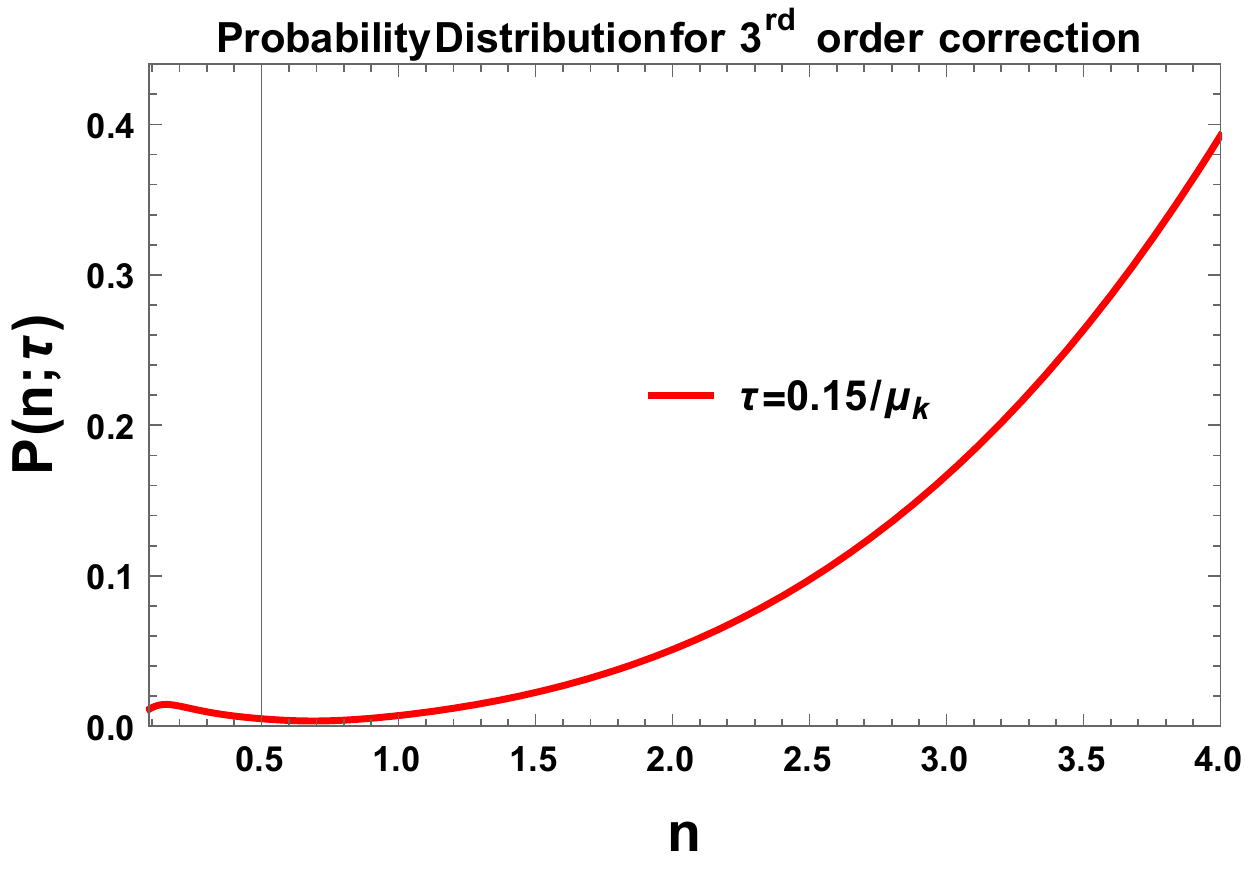}
\caption{ Third order contribution to the probability density function with respect to the the occupation number per mode $n$, for a fixed time from the analytical solution[Eq:-\ref{ana_3}] of third order correction equation[Eq:-\ref{cxz}]}
\label{Figana3rd}
\end{figure}

In Fig:-\ref{Figana3rd}, we have shown the third order correction of probability density function with respect to the occupation number per mode, for a fixed time ($\mu_k\tau$=fixed). From this plot we have observed primary gaussian feature followed by an exponential type increase.We consider the perturbative expansion to be valid. So the higher order correction contribute less than the previous order.To normalize the analytical solution we have used this. In latter part we can see the exponential type increase contribute in longer tail effect [ higher kurtosis]. 
\subsubsection{Fourth order correction}
In this context, our objective is to find out the contributions coming from fourth order in the {\it Fokker Planck equation} and to solve this equation
numerically~\footnote{Including the contributions from fourth order we will see that the {\it Fokker Planck equation} can not solvable analytically.}. To serve this purpose we equate both sides of Eq~(\ref{eq18}) after Taylor expansion and compare the coefficient of $\del\tau ^{4}$.
Consequently, we get the following partial differential equation:  
\bea\label{ss12}
&&70 n^4 (1+n)^4\frac{\pl^{8}P(n;\tau)}{\pl n^{8}}+140 n^3 (1+2 n) \frac{\pl^{7}P(n;\tau)}{\pl n^{7}}\nonumber\\
&&+30 n^2 (1+n)^2 (3+14 n+14 n^2)\frac{\pl^{6}P(n;\tau)}{\pl n^{6}}\nonumber\\
&&+20 n (1+n) (1+2 n) (1+7 n+7 n^2)\frac{\pl ^{5}P(n;\tau)}{\pl n^{5}}\nonumber\\
&&+(1+20 n+90 n^2+140 n^3+70 n^4)\frac{\pl^{4}P(n;\tau)}{\pl n^{4}}=\frac{1}{\mu^4_k}\frac{\pl^4P(n;\tau)}{\pl \tau^4}.
\eea
which can not able to solve analytically with any integer values of $m_4$.
We solve this equation for different values of $m_{4}$ numerically with assumed initial condition. Only for the special case, $m_4=0$ with large $n$ limit we can able to provide an analytical solution in the present context.

Now to solve this partial differential equation we apply method of separation of variable, using which we can write the total solution in the following form:
\bea  \label{eq33cv} P(n;\tau)=P_1(n)P_2(\tau).\eea
Further, using the solution ansatz stated in Eq~(\ref{ss12}) we get the following sets of independent differential equations, as given by:
\bea\label{ss12a}
&&70 n^4 (1+n)^4\frac{d^{8}P_1(n)}{d n^{8}}+140 n^3 (1+2 n) \frac{d^{7}P_1(n)}{d n^{7}}\nonumber\\
&&+30 n^2 (1+n)^2 (3+14 n+14 n^2)\frac{d^{6}P_1(n)}{d n^{6}}\nonumber\\
&&+20 n (1+n) (1+2 n) (1+7 n+7 n^2)\frac{d^{5}P_1(n)}{d n^{5}}\nonumber\\
&&+(1+20 n+90 n^2+140 n^3+70 n^4)\frac{d^{4}P_1(n)}{d n^{4}}-m^2_4P_1(n)=0,\\
\label{ss12b} &&\left[\frac{d^{4} }{d \tau ^{4}}-m^2_4\mu^4_k\right]P_2(\tau)=0.
\eea
It is important to note that, the analytical solution of $P_1(n)$ is not possible for any arbitrary values of the constant $m_4$. For this reason we use numerical technique to solve Eq~(\ref{ss12a}). Also considering the large $n$ limit we have checked that Eq~(\ref{ss12a}) is not analytically solvable. On the other hand Eq~(\ref{ss12b}) is exactly solvable in the present context and the solution can be written as:
\bea P_2(\tau)&=&\left[C_9 e^{-\sqrt{m_4}\tau\mu _k}+C_{10} e^{\sqrt{m_4} \tau  \mu _k}+C_{11} \sin \left(\sqrt{m_4} \tau  \mu _k\right)+C_{12} \cos \left(\sqrt{m_4} \tau  \mu _k\right)\right],~~~~\eea
where $C_9$, $C_{10}$, $C_{11}$ and $C_{12}$ are three arbitrary constants which can be fixed by choosing proper boundary conditions.

Now to  solve Eq~(\ref{ss12a}) numerically for different values of $m_{4}$ along with given initial condition. Here it important to mention that, since arbitrary values of $m_4$ is allowed, one can consider integer as well as non integer values at the level of solution of differential equation. However, the only physically acceptable solution restrict us to only consider the integer values of $m_4$ because such third order corrected solution of the {\it Fokker Planck equation} is directly related to the quantum effects as we have mentioned earlier. As a result such integer values of $m_4$ can be interpreted as the quantum number i. e. 
\bea\textcolor{blue}{\bf \underline{Quantum ~Number~V:}}~~~~m_4=0,\pm 1,\pm 2,\cdots,\pm \infty \in \mathbb{Z}.\eea
For numerical solution we take the following assumptions:
\bea \label{ddsax} P_1(n=0.0001)&=&100,\nonumber\\
\left[\frac{d P_1(n)}{d n}\right]_{n=0.0001}&=& 100,\nonumber\\
\left[\frac{d^{2} P_1(n)}{d n^{2}}\right]_{n=0.0001}&=&100,\nonumber\\
\left[\frac{d^{3} P_1(n)}{d n^{3}}\right]_{n=0.0001}&=&100,\nonumber\\
\left[\frac{d^{4} P_1(n)}{d n^{4}}\right]_{n=0.0001}&=&100,\nonumber\\
\left[\frac{d^{5} P_{1}(n)}{d n^{5}}\right]_{n=0.0001}&=&100,\nonumber\\
\left[\frac{d^{6} P_1(n)}{d n^{6}}\right]_{n=0.0001}&=&100,\nonumber\\
\left[\frac{d^{7} P_{1}(n)}{d n^{7}}\right]_{n=0.0001}&=&100.\eea
Accoding to our assumption particle production probability has constant value at some particular small $n$ value ($n=0.0001$) and all its derivative has constant values and all those values are same.
%%%%%%%% ADD pictures
\begin{figure}[htb]
\centering
\subfigure[Fourth order corrected distribution for $m_{4} = 0$]{
    \includegraphics[width=7.8cm,height=8cm] {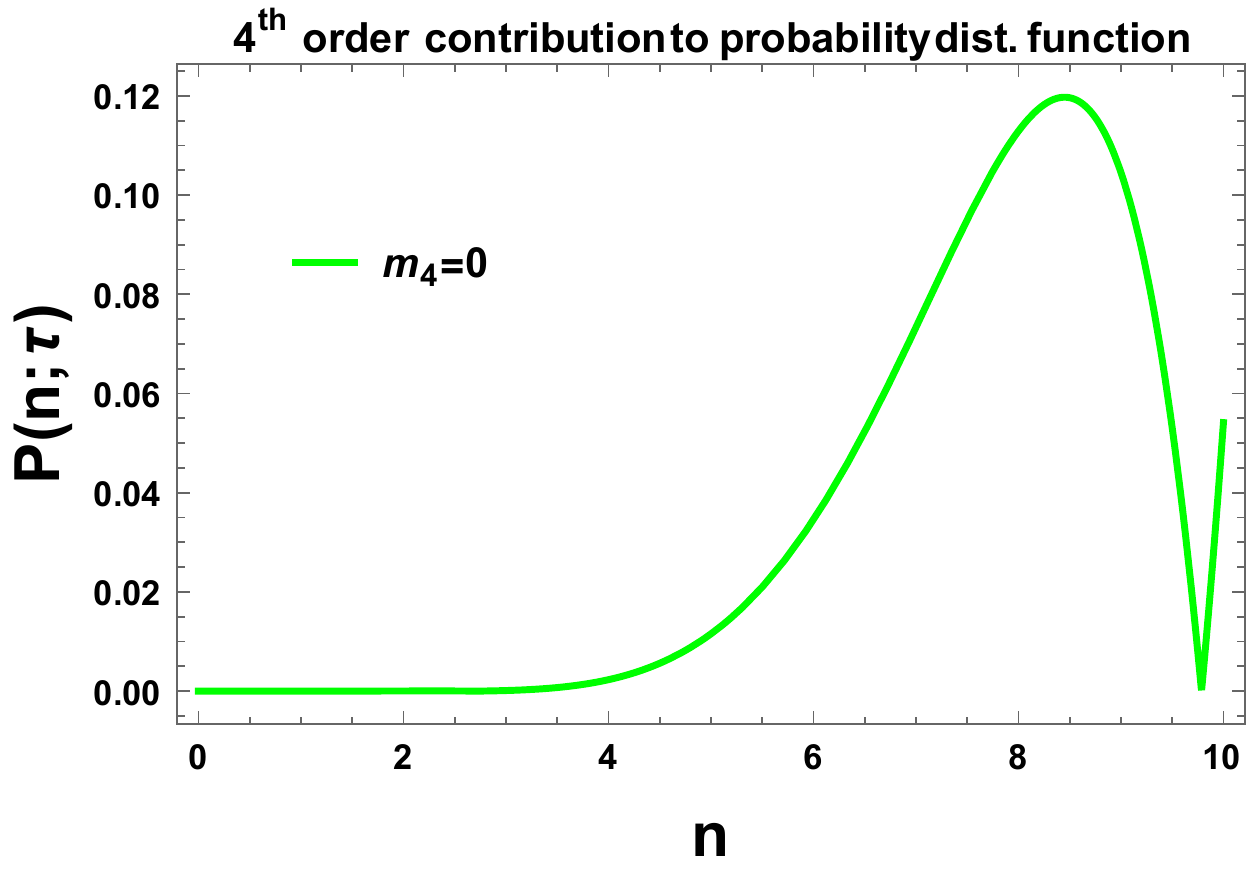}
    \label{x41}
}
\subfigure[Fourth order corrected distribution for $m_{4} = \pm 1$]{
    \includegraphics[width=7.8cm,height=8cm] {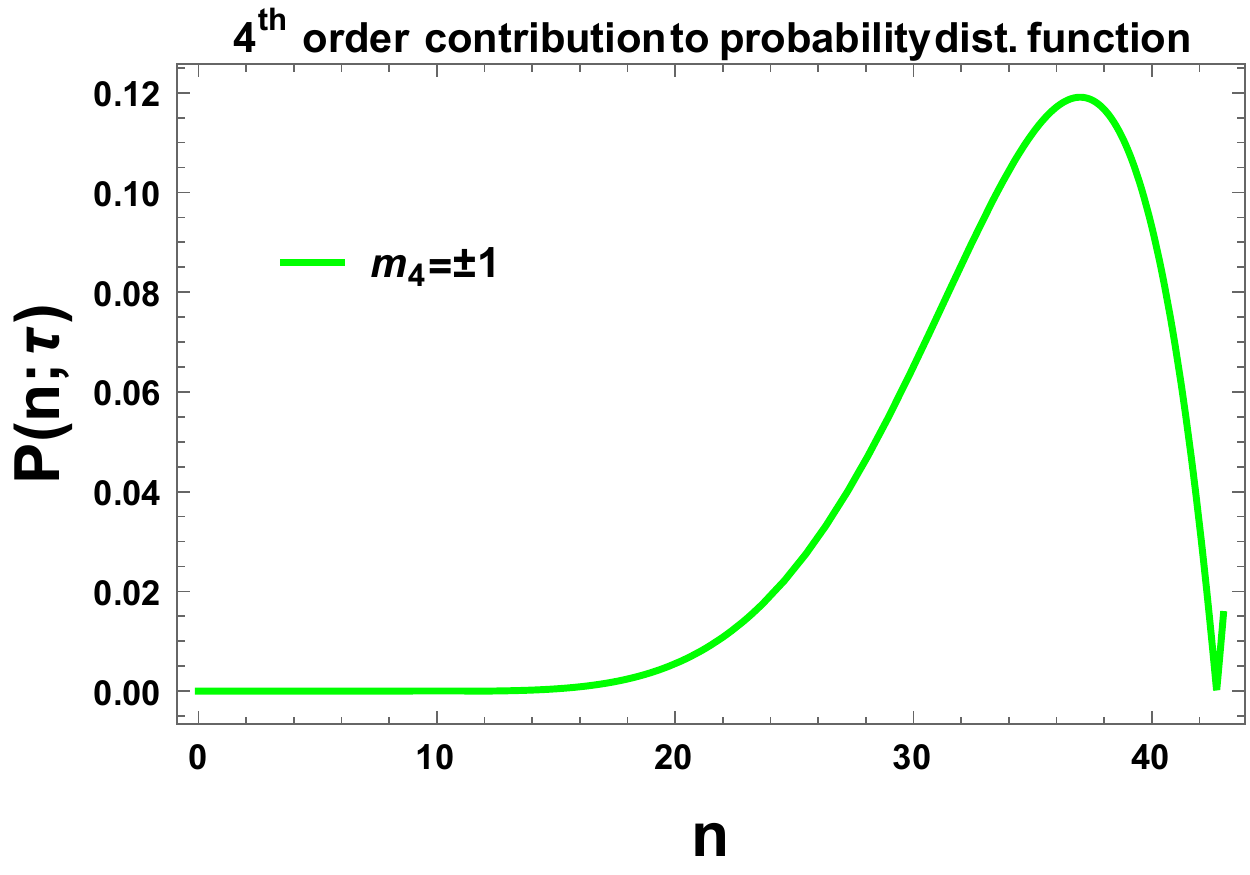}
    \label{x42}
}
\subfigure[Fourth order corrected distribution for $m_{4}=\pm 2$ ]{
    \includegraphics[width=7.8cm,height=8cm] {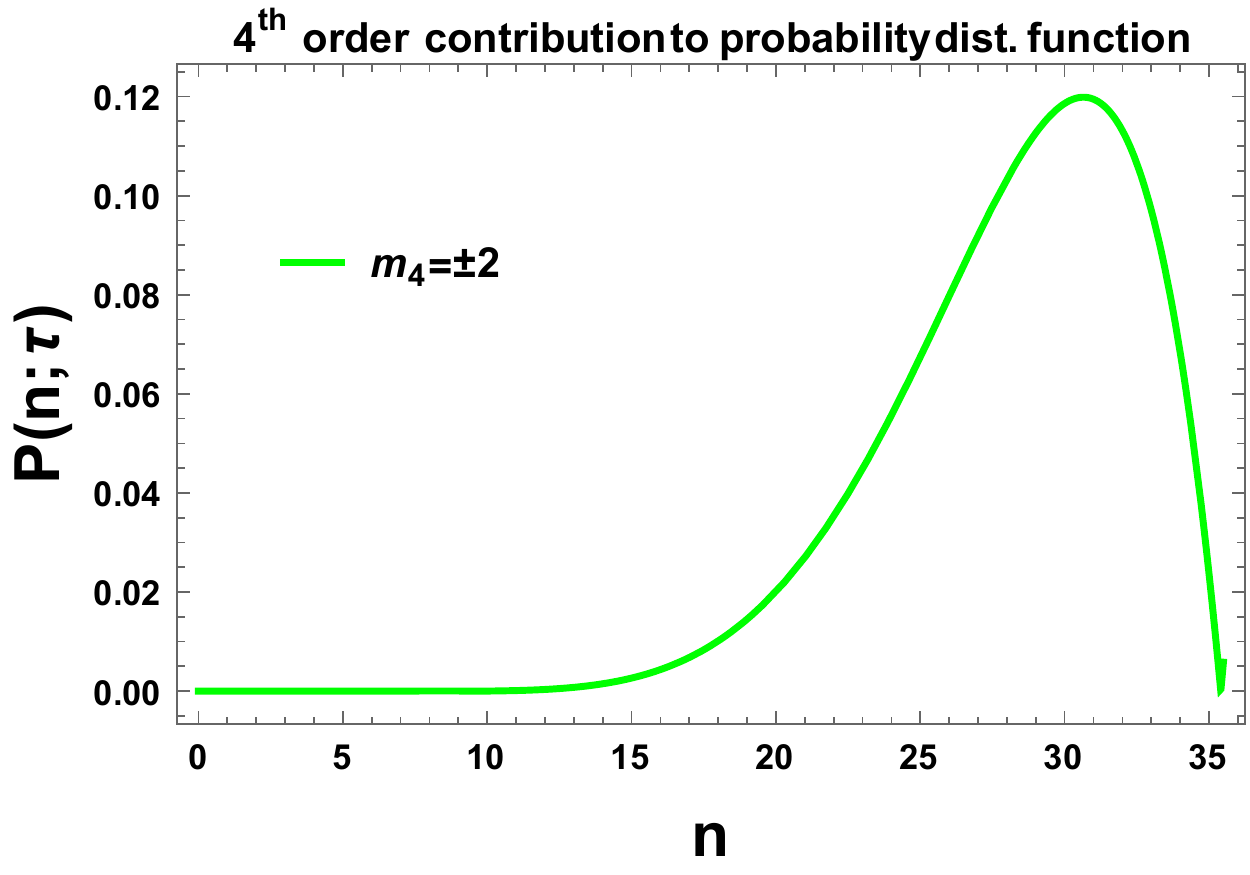}
    \label{x43}
}
\subfigure[Fourth order corrected distribution for $m_{4} =\pm 3 $  ]{
    \includegraphics[width=7.8cm,height=8cm] {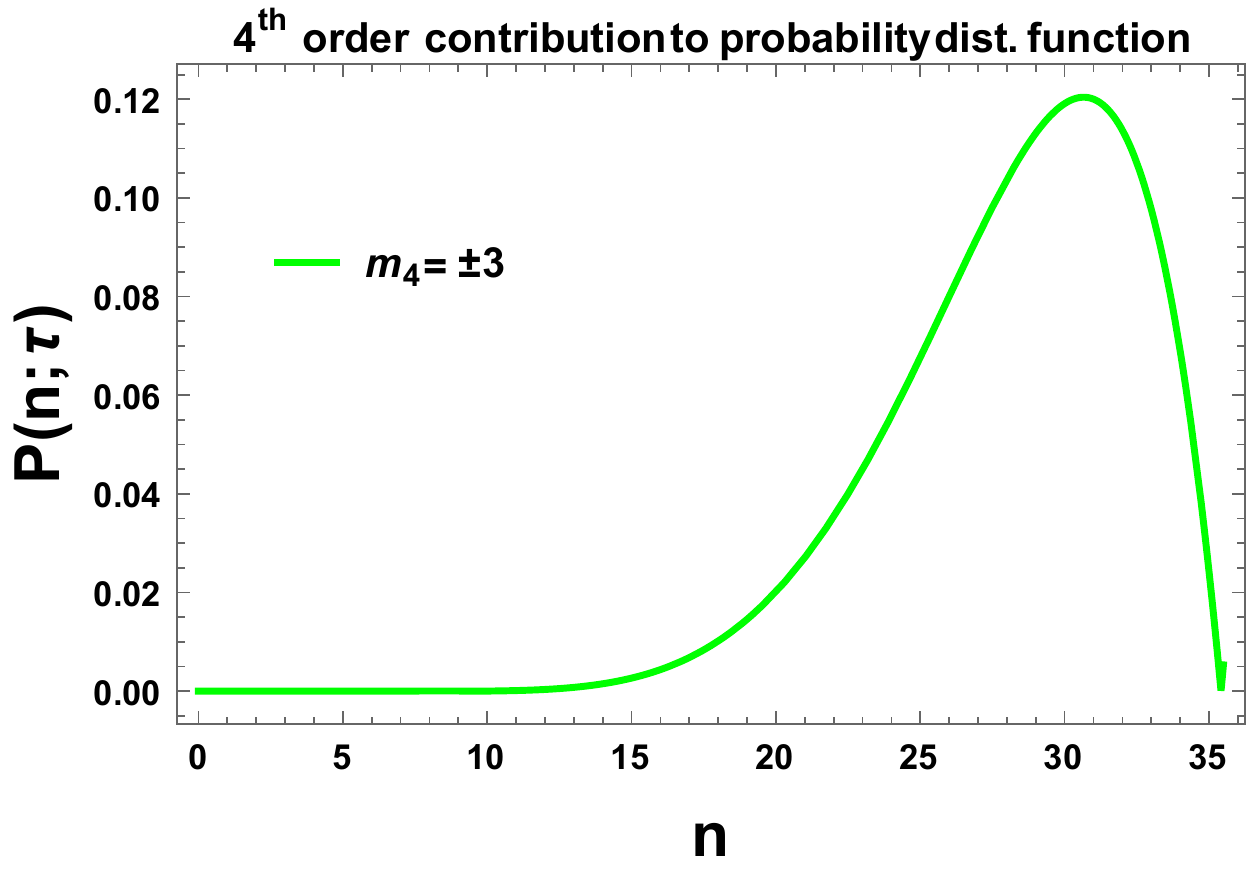}
    \label{x44}
}
\caption{Evolution of probability distribution function obtained from the fourth order corrected Fokker-Planck equation with the occupation number $n$ for different value of $m_{4}$. Here we use the initial conditions as mentioned in Eq~(\ref{ddsax}).}
\label{Fig4thorder}
\end{figure}

During the analysis we assume that the particle production probability and all its derivative has a constant same value for $n=0.0001$, which is very very helpful for us to deal with the initial conditions during performing numerical techniques to solve the Eq~(\ref{ss12a}).

From fig.~\ref{Fig4thorder} we observe that the fourth order corrected probability distribution function for different $m_{4}$ is almost flat upto $n=1$ and after that the distribution function suddenly increases. Additionally, we observe that the fourth order correction has deviation from Gaussianity at small values of the occupation number. On the other hand, for large values of the occupation number we get a Gaussian like feature and that is shown explicitly in the mentioned plot.

Further using the  Fourier transformation with respect to the occupation number $n$ as mentioned in Eq~(\ref{ft1}), we get the following simplified expression for the {\it Fokker Planck equation} at the fourth order:
\bea\label{xzazx24} \frac{\pl^4 \bar{P}(k;\tau)}{\pl \tau^4}&=&\mu^4_k\left[70n^4(1+n)^4k^8-140in^3(1+2n)k^7\right.\nonumber\\
&&\left.-30n^2(1+n)^2(3+14n+14n^2)k^6\right.\nonumber\\
&&\left.+20ni(1+n)(1+2n)(1+7n+7n^2)k^5\right.\nonumber\\
&&\left.+(1+20n+90n^2+140n^3+70n^4)k^4\right]\bar{P}(k;\tau),~~~~~~~~~~\eea
which is obviously a simplest version of the {\it Fokker Planck equation} as it contains only four derivative with respect to time $\tau$. 
In the present context we get the following result for the probability distribution function in the Fourier transformed space, as given by:
\bea \bar{P}(k;\tau|n^{'};\tau^{'})&=&C_1 \exp\left[-\sqrt[4]{{\cal J}(k;n^{'})} \left(\tau-\tau^{'}\right)\right]+C_2 \exp\left[\sqrt[4]{{\cal J}(k;n^{'})} \left(\tau-\tau^{'}\right)\right]\nonumber\\
&&+C_3 \sin \left(\sqrt[4]{{\cal J}(k;n^{'})} \left(\tau-\tau^{'}\right)\right)+C_4 \cos \left(\sqrt[4]{{\cal J}(k;n^{'})} \left(\tau-\tau^{'}\right)\right),~~~~~~~~\eea
where ${\cal J}(k;n^{'})$ is defined as:
\bea {\cal J}(k;n^{'})&=&\mu^4_k\left[70n^{'4}(1+n^{'})^4k^8-140in^{'3}(1+2n^{'})k^7\right.\nonumber\\
&&\left.-30n^2(1+n^{'})^2(3+14n^{'}+14n^{'2})k^6\right.\nonumber\\
&&\left.+20ni(1+n^{'})(1+2n^{'})(1+7n^{'}+7n^{'2})k^5\right.\nonumber\\
&&\left.+(1+20n^{'}+90n^{'2}+140n^{'3}+70n^{'4})k^4\right].~~~~~~ \eea
Additionally, $C_1$, $C_2$, $C_3$ and $C_4$ are arbitrary constants which is fixed by the 
following three fold boundary conditions, as given by:
\bea P(n;\tau|n^{'}=0;\tau^{'}=\tau)&=&\delta(n),\\
\left(\frac{\pl  P(n;\tau |n^{'};\tau^{'})}{\pl \tau}\right)_{n^{'}=0,\tau=\tau^{'}} &=&-\frac{\delta(n)}{n},\\
\left(\frac{\pl^2  P(n;\tau |n^{'};\tau^{'})}{\pl \tau^{2}}\right)_{n^{'}=0,\tau=\tau^{'}} &=&\frac{2~\delta(n)}{n^2}\\
\left(\frac{\pl^3  P(n;\tau |n^{'};\tau^{'})}{\pl \tau^{3}}\right)_{n^{'}=0,\tau=\tau^{'}} &=&-\frac{6~\delta(n)}{n^3}.\eea
which are necessary to solve the above mentioned fourth order differential equation.

As a result, we get the following set of constraints equations:
\bea &&C_1 +C_2 +C_4=1,\\
&&C_1 -C_2 -C_3=\frac{1}{\mu_k k n},~~~~\\
&&C_1+C_2-C_3=\frac{2}{\mu^2_k k^2 n^2},~~~~\\
&&C_1-C_2 +C_3=\frac{6}{\mu^3_k k^3 n^3}.
\eea
Solving these equations we get:
\bea C_1&=&\frac{-k^2 n^2 \mu _k^2-2 k n \mu _k-6}{4 k^3 n^3 \mu _k^3},~~~~~~
 C_2=-\frac{k^2 n^2 \mu _k^2-2 k n \mu _k+6}{4 k^3 n^3 \mu _k^3},\\
 C_3&=&-\frac{k^2 n^2 \mu _k^2-6}{2 k^3 n^3 \mu _k^3},~~~~~~
 C_4=-\frac{1}{k^2 n^2 \mu _k^2}.\eea
 Using this solution get the following probability distribution function in Fourier space, as given by:
\bea \bar{P}(k;\tau)&=&\frac{-k^2 n^2 \mu _k^2-2 k n \mu _k-6}{4 k^3 n^3 \mu _k^3}~ \exp\left[-\mu_k k \tau\right]-\frac{k^2 n^2 \mu _k^2-2 k n \mu _k+6}{4 k^3 n^3 \mu _k^3}~ \exp\left[\mu_k \tau\right]\nonumber\\
&&-\frac{k^2 n^2 \mu _k^2-6}{2 k^3 n^3 \mu _k^3} \sin \left(\mu_k k \tau\right)-\frac{1}{k^2 n^2 \mu _k^2} \cos \left(\mu_k k \tau\right),~~~~~~~~\eea
Hence substituting back into the definition of Fourier transformation and setting the initial condition $n^{'}=0$ and $\tau^{'}=0$ we get the following result for the probability distribution function, as 
given by:
\bea P(n;\tau)&=&\frac{1}{2\pi}\int^{\infty}_{-\infty}dk~\exp\left[ikn\right]\left\{\frac{-k^2 n^2 \mu _k^2-2 k n \mu _k-6}{4 k^3 n^3 \mu _k^3}~ \exp\left[-\mu_k k \tau\right]\right.\nonumber\\
&&\left.~~~~~~~~~~~~~~~~~~-\frac{k^2 n^2 \mu _k^2-2 k n \mu _k+6}{4 k^3 n^3 \mu _k^3}~ \exp\left[\mu_k \tau\right]\right.\nonumber\\
&&\left.~~~~~~~~~~~~~~~~~~-\frac{k^2 n^2 \mu _k^2-6}{2 k^3 n^3 \mu _k^3} \sin \left(\mu_k k \tau\right)-\frac{1}{k^2 n^2 \mu _k^2} \cos \left(\mu_k k \tau\right)\right\}
\label{ana_4}
\eea
 which is divergent within the interval $-\infty<k<\infty$. After introducing an IR and UV regulators, $Q<k<L$ we can get a finite result, which we have not presented in this paper for its huge length. The origin of such corrections are the fourth order contribution in the {\it Fokker Planck equation}. 
\begin{figure}[H]
\centering
    \includegraphics[width=14.8cm,height=8cm] {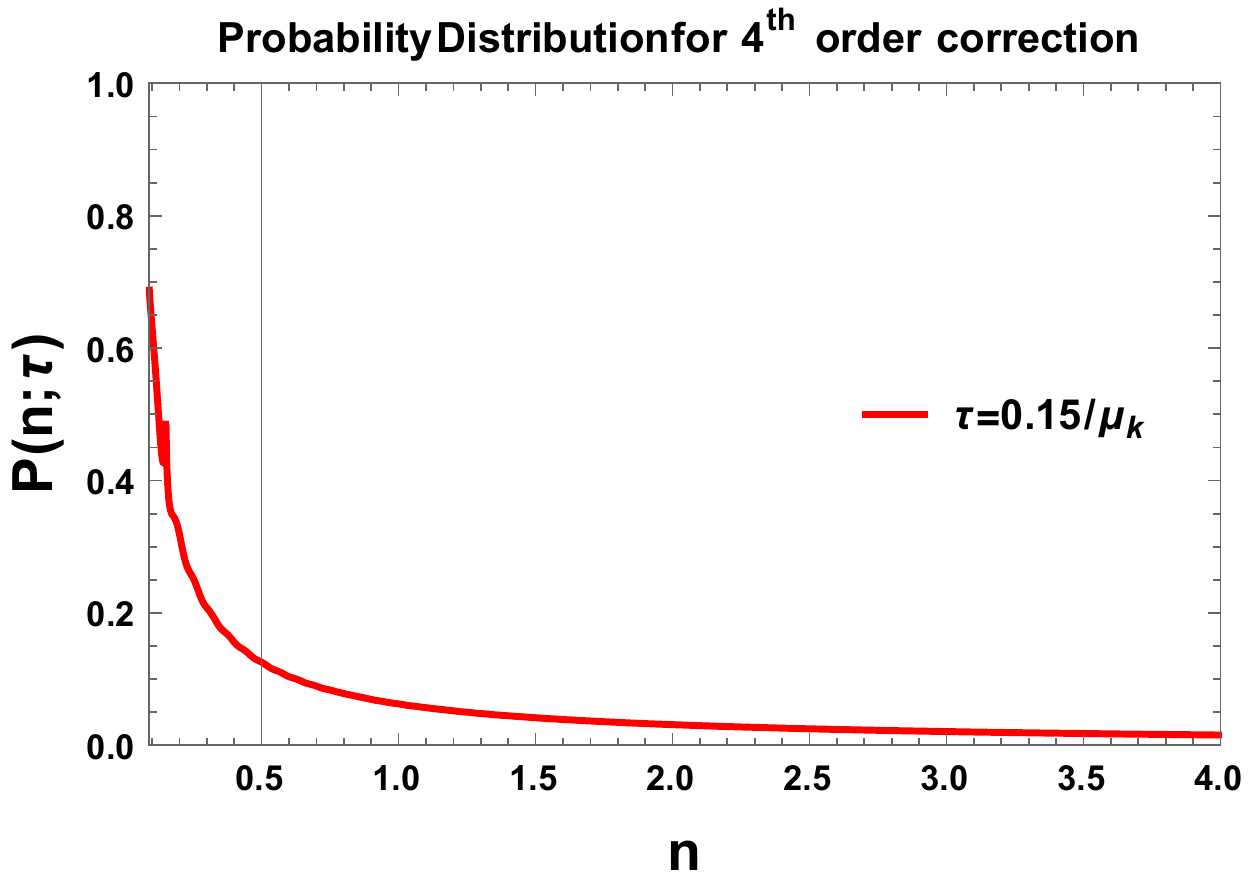}
\caption{ Fourth order contribution to the probability density function with respect to the the occupation number per mode $n$, for a fixed time from the analytical solution [Eq:-\ref{ana_4}] of the fourth order correction distribution function[Eq:-\ref{ss12}].}
\label{Figana4th}
\end{figure}

In Fig:-\ref{Figana4th}, we have shown the fourth order correction to  the probability density function with respect to the occupation number per mode, for a fixed time ($\mu_k\tau$=fixed) flow the intial gaussian then exponential decay type distribution.This decreasing feature suggest very low effect from fourth order correction which is also supported by perturbative expansion assumption.

\subsubsection {Total solution considering different order correction}
Now we plot total solutions with different order of correction with main solution.Then we check validation of different order of correction altogether and how they merge with each other at what limit.\\
\textcolor{red}{\textsf{\bf \underline{A. Upto second order correction}}}:-\\
Further we consider upto second order corrected solution with first order contribution at $m_{1}=2$ for different $m_{2}$. 
\begin{figure}[H]
\centering
\subfigure[Upto second order corrected distribution for $m_{2} = 1 $ ]{
    \includegraphics[width=7.8cm,height=8cm] {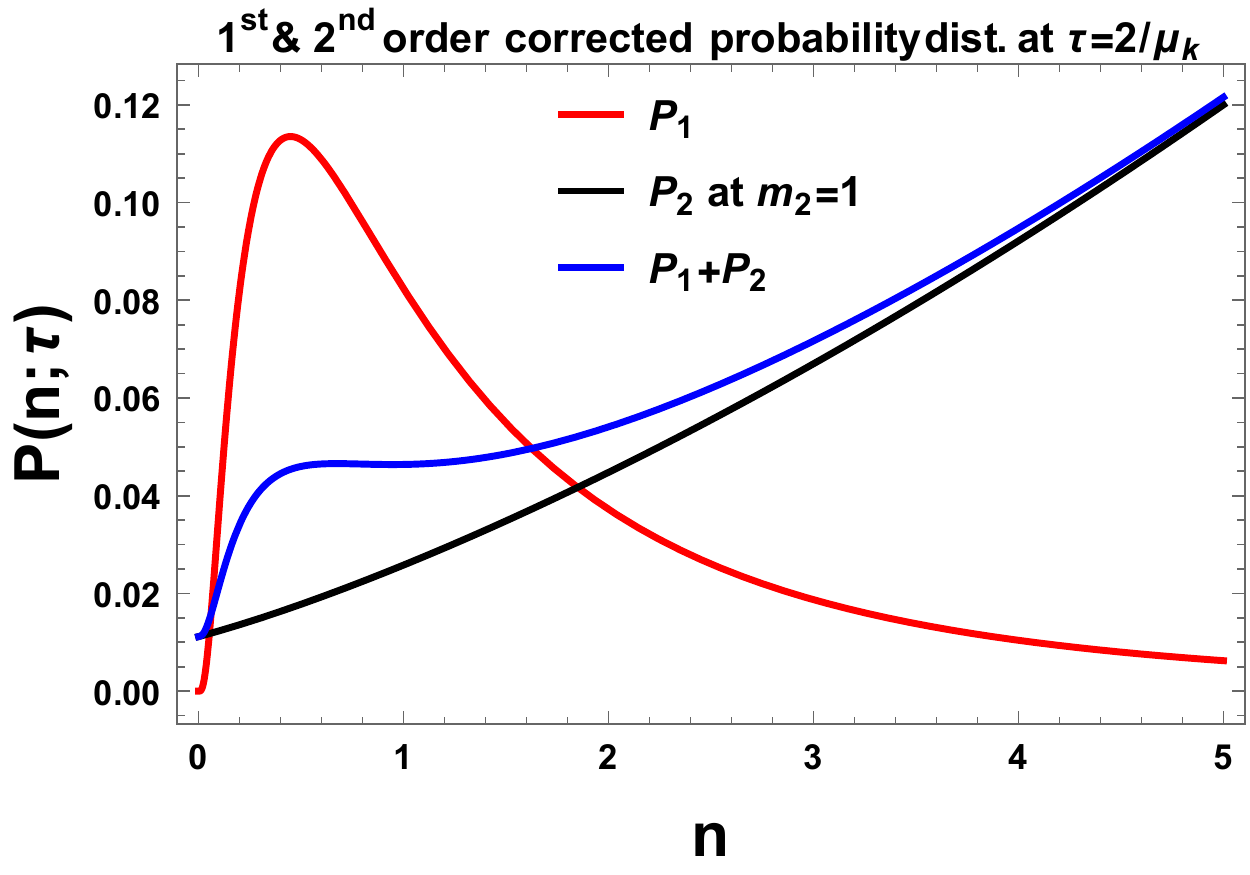}
    \label{xt21}
}
\subfigure[Upto second order corrected distribution for $m_{2} = 0 $ ]{
    \includegraphics[width=7.8cm,height=8cm] {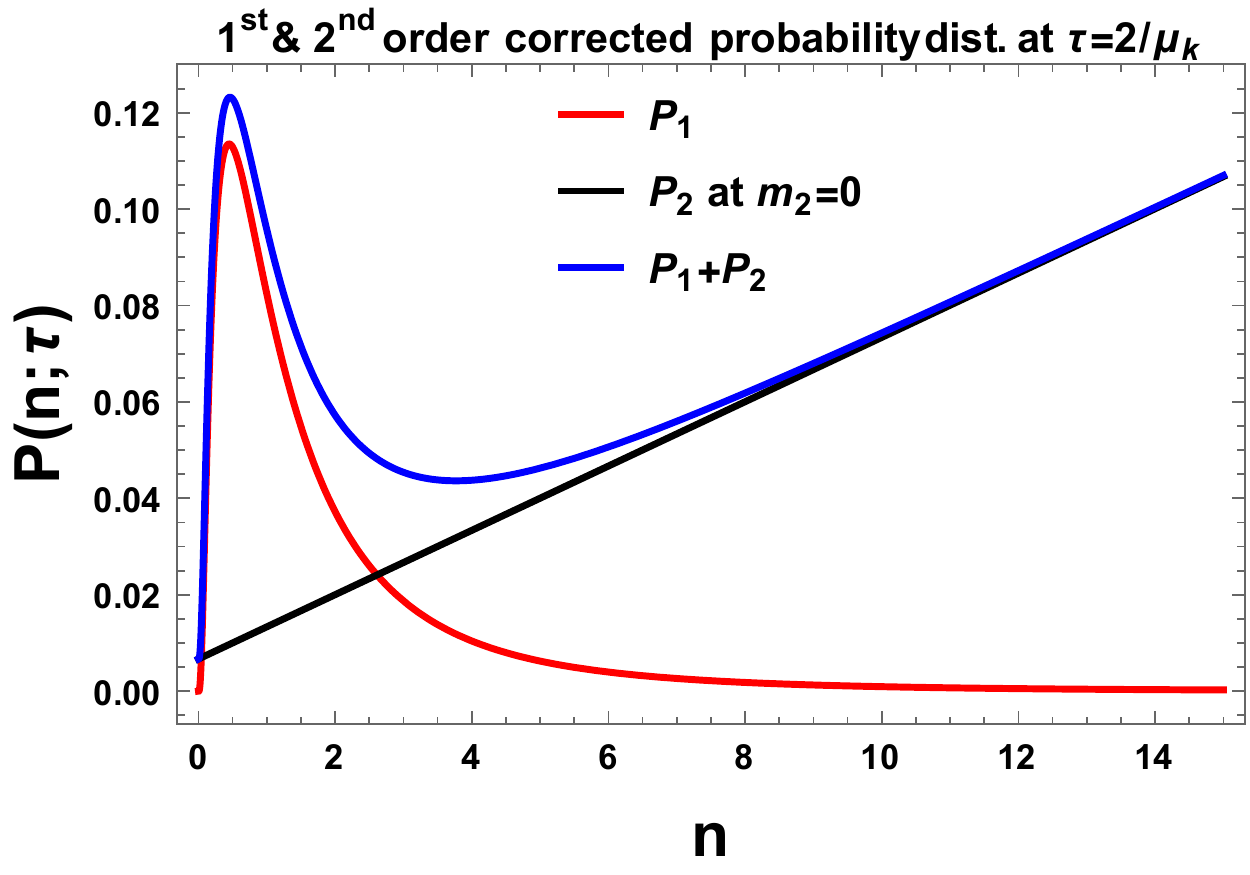}
    \label{xt22}
}
\subfigure[second order analytic solution]{
  \includegraphics[width=7.8cm,height=8cm] {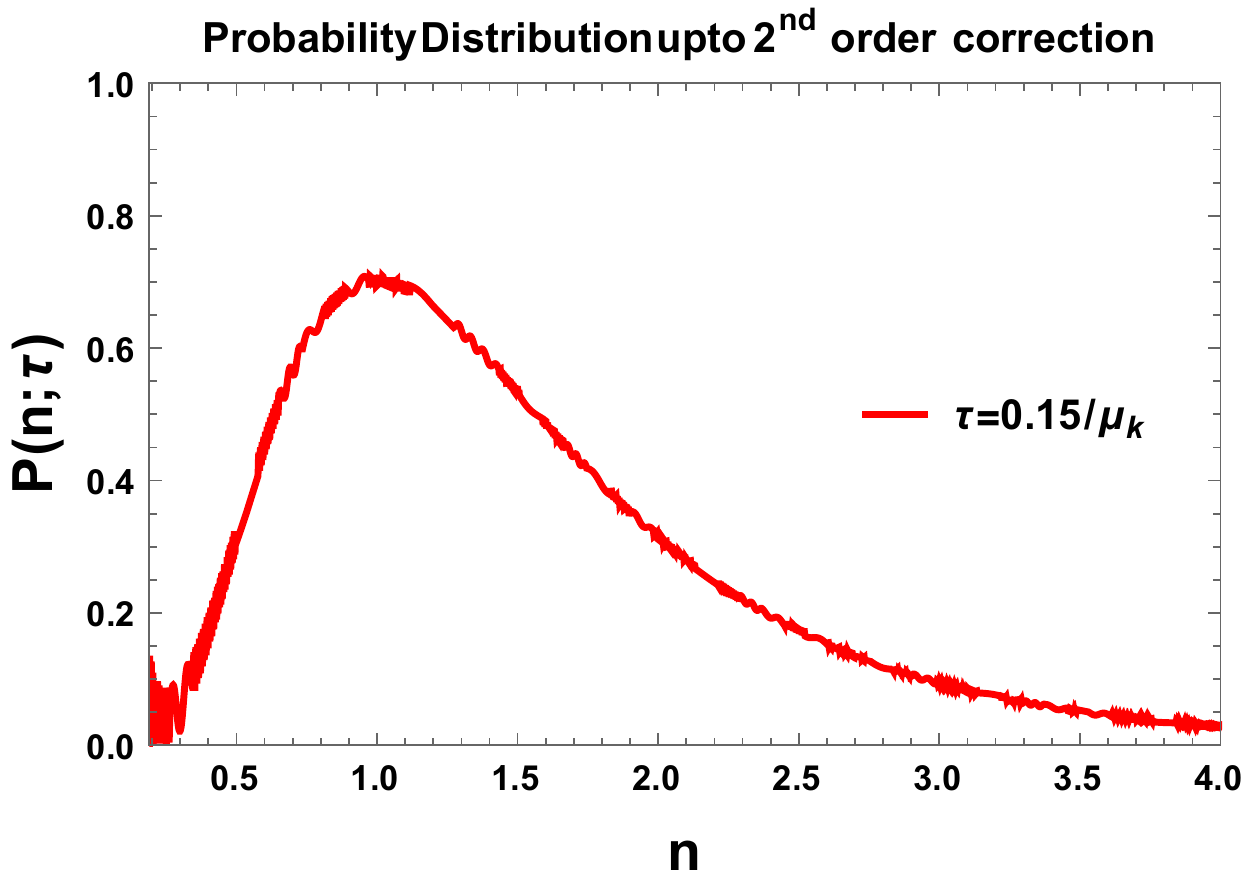}
    \label{Fig1998}
}
\caption{Second order corrected probability distribution profile for different $m_{2}$  with previously mentioned initial conditions. Here we fix $m=2$. Here the subscript $1$ and $2$ stands for the corrected order in the distribution.}
%\label{Fig2nd1st}
\end{figure}
From Fig:-\ref{xt21},\ref{xt22} we observe that at low values of $n$,  $P_{1}+P_{2}$ and $P_{2}$ are significantly different, but as we increase $n$ they overlapped and $P_{2}$ effect is more over $P_{1}+P_{2}$ so the second order solution dominate over the first order solution.On the other hand Fig:-\ref{Fig1998} shows how the second order contribution effect the primary gaussian curve. Distinct oscillation effect on gaussian curve shows the effect of second order correction which is not available from the numerical solution.
\\ \\
\textcolor{red}{\textsf{\bf \underline{B. Upto third order correction}}}:-\\ \\
Here we add all the three previously derived contributions to produce the total probability distribution corrected upto third order.
\begin{figure}[H]
\centering
\subfigure[upto third order corrected distribution for $m_{3}=0$,$m_{2} = 0 $ ]{
    \includegraphics[width=7.8cm,height=7cm] {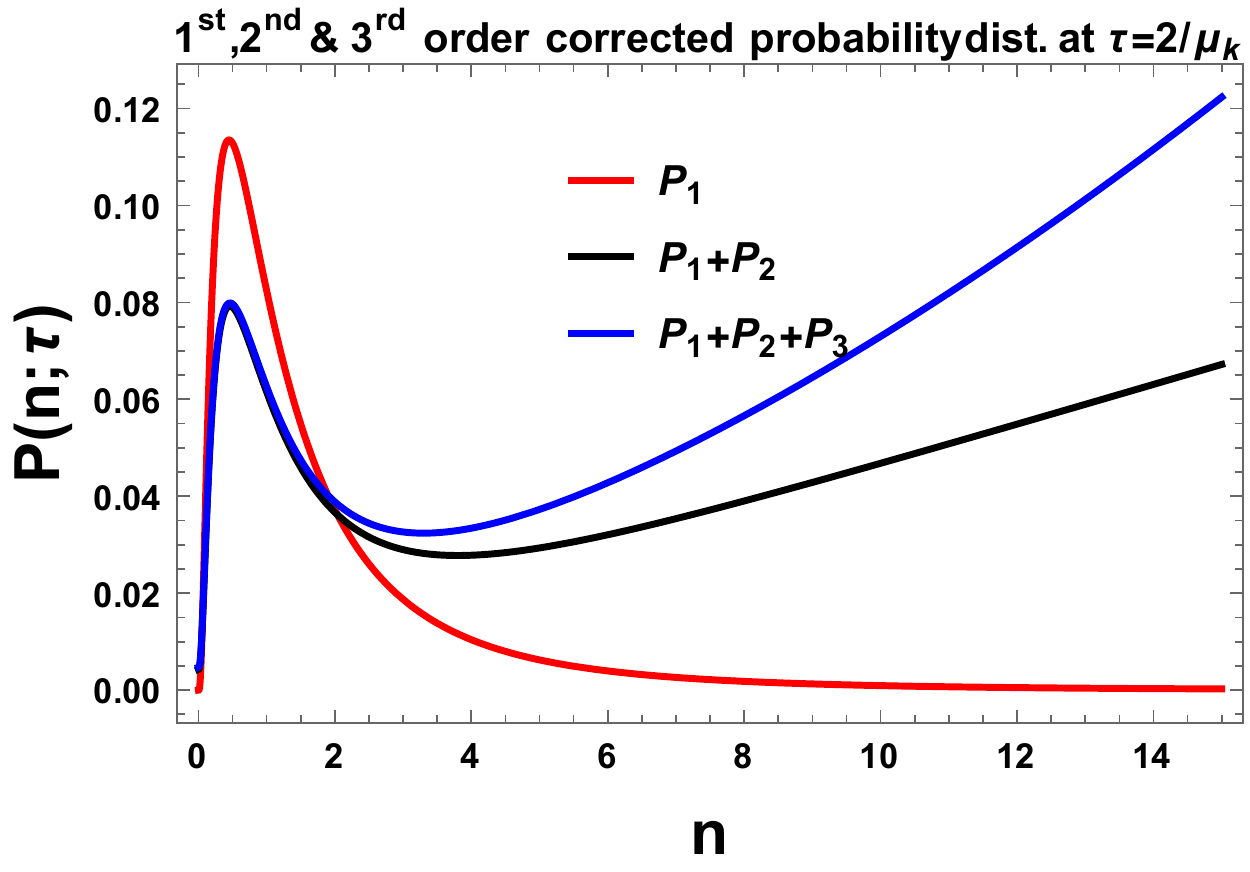}
    \label{xt321}
}
\subfigure[Upto third order corrected distribution for $m_{2} = 0 $ and $m_{3}=1$. ]{
    \includegraphics[width=7.8cm,height=7cm] {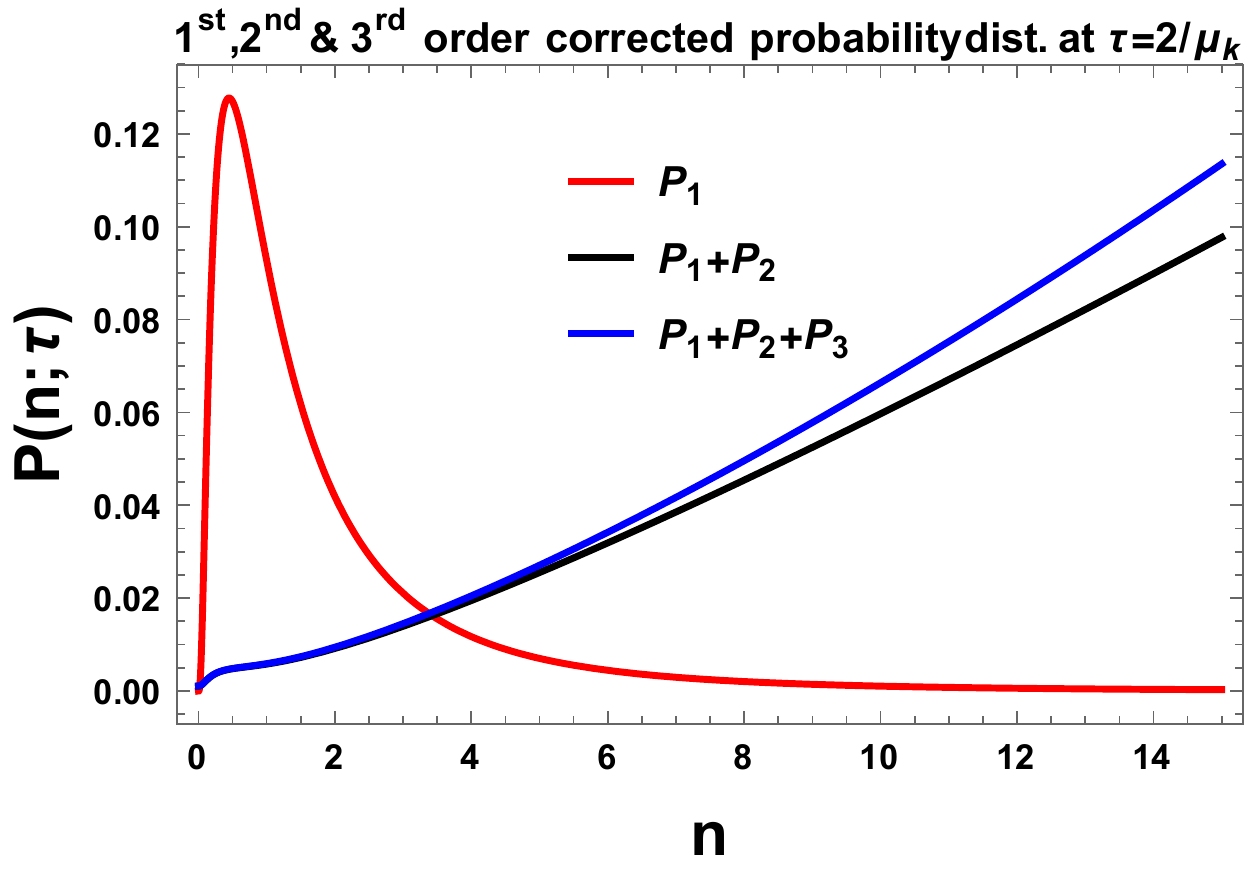}
    \label{xt322}
}
\subfigure[upto third order corrected analytical solution  ]{
    \includegraphics[width=7.8cm,height=7cm] {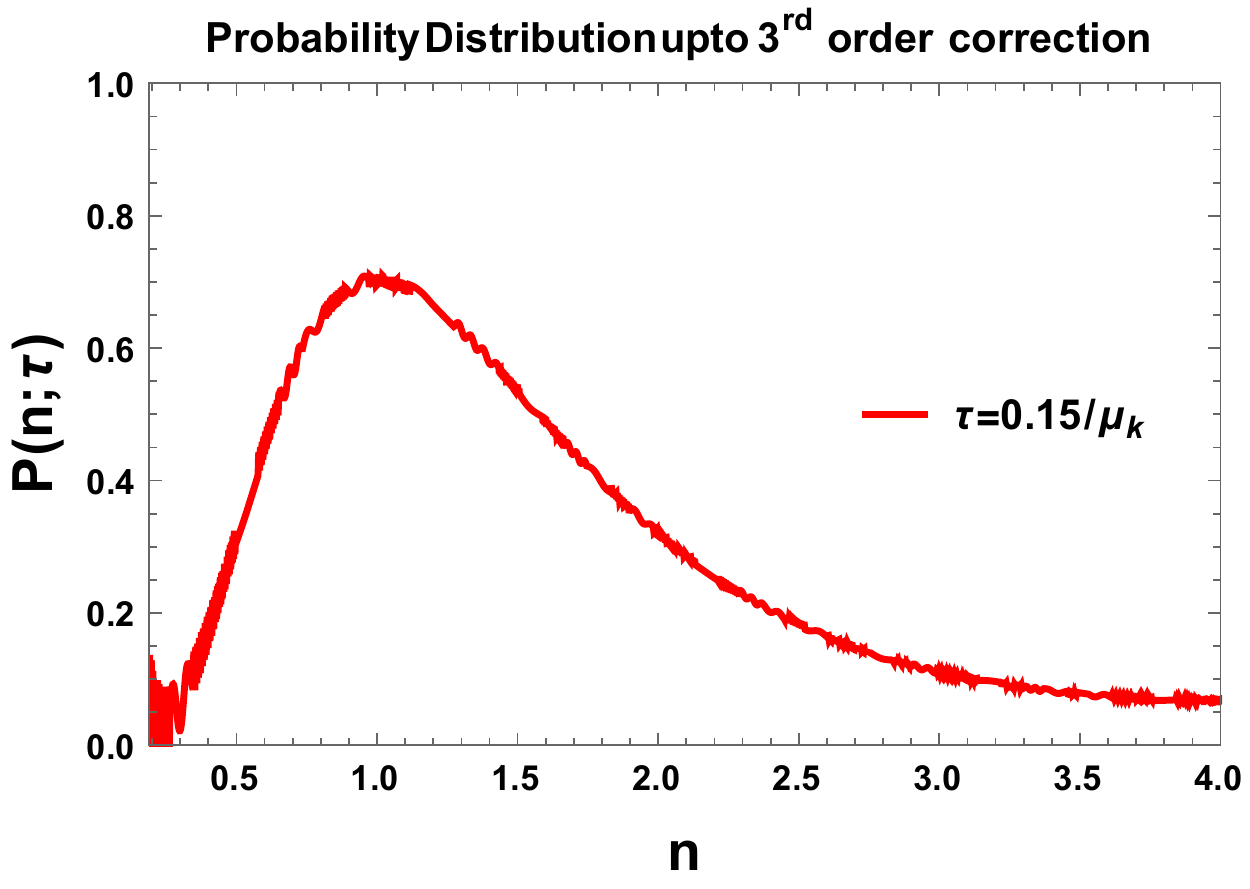}
    \label{xt323}
}
\subfigure[comparison between 2nd order and third order effect]{
    \includegraphics[width=7.8cm,height=7cm] {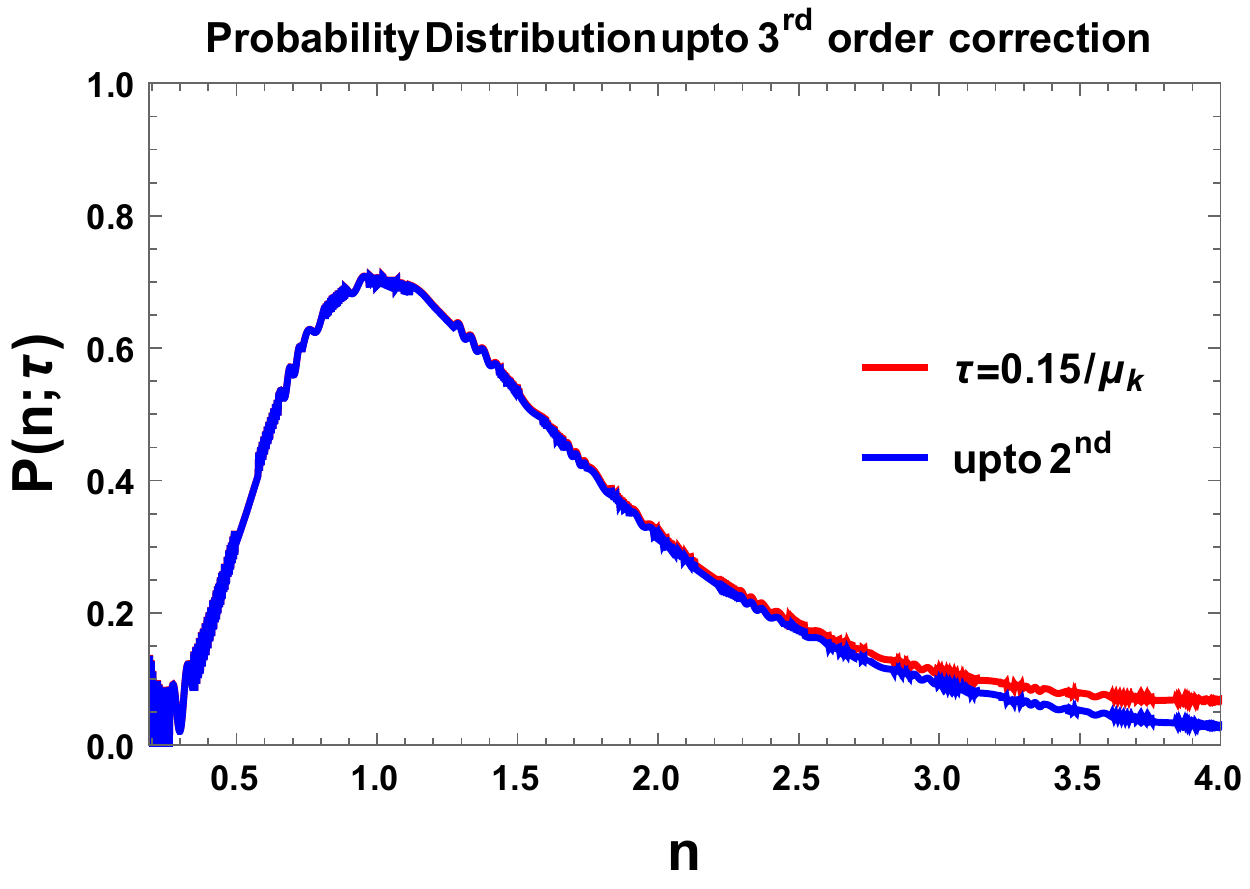}
    \label{xt443n}
}
\caption{Upto third order corrected probability distribution profile for different $m_{3}$,$m_{2}$ and $m_{1}=2$ with previously mentioned initial conditions and Exact analytical solution for probability distribution with previous correction included}
\label{Fig3rd2nd1st}
\end{figure}
From fig.~\ref{xt321},\ref{xt322} we observe that $P_{1}+P_{2}+P_{3}$, $P_{2}+P_{3}$ and $P_{3}$ overlap at higher $n$ limit though separated. At low $n$ limit and $P_{1}+P_{2}$ and $P_{2}$ overlap with each other but remain separated from $P_{1}+P_{2}+P_{3}$ for the complete range. This implies that third order contribution is dominant over the other two
due to the non-linearities in the differential equation and behaves like a non-perturbative quantum effect at the level of solution.
From fig~\ref{xt323} we show the third order correction affects the tail of the gaussian and shift it bit higher.It is clear from fig~\ref{xt443n}.\\
\textcolor{red}{\textsf{\bf \underline{C. Upto fourth order correction}}}:-\\
Here we add all the four previously derived contributions to produce the total probability distribution corrected upto fourth order.
\begin{figure}[H]
\centering
\subfigure[upto fourth order corrected solution for different $m_{1}$,$m_{2}$,$m_{3}$  and $m_{4}$]{
    \includegraphics[width=7.8cm,height=6cm] {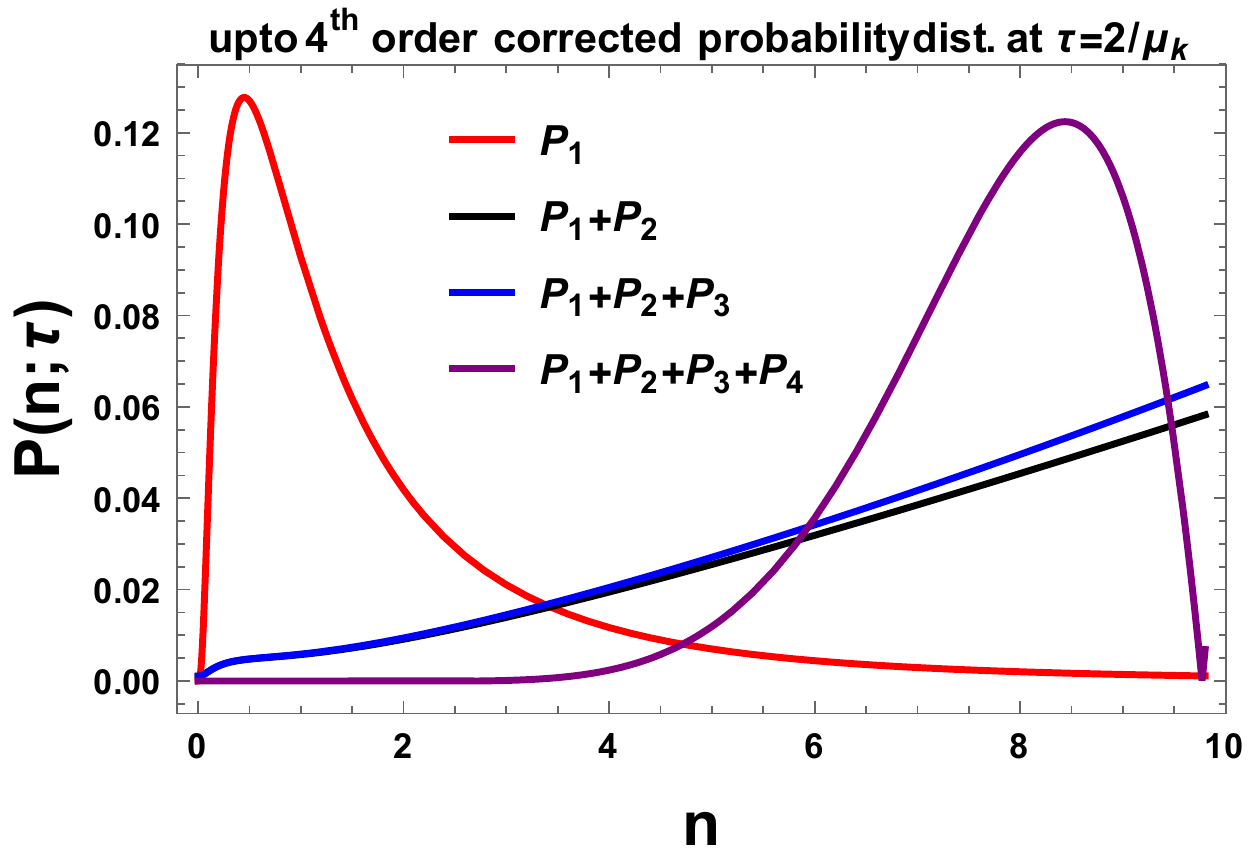}
    \label{xt4322}
}
\subfigure[Upto 4th order correction]{
    \includegraphics[width=7.8cm,height=6cm] {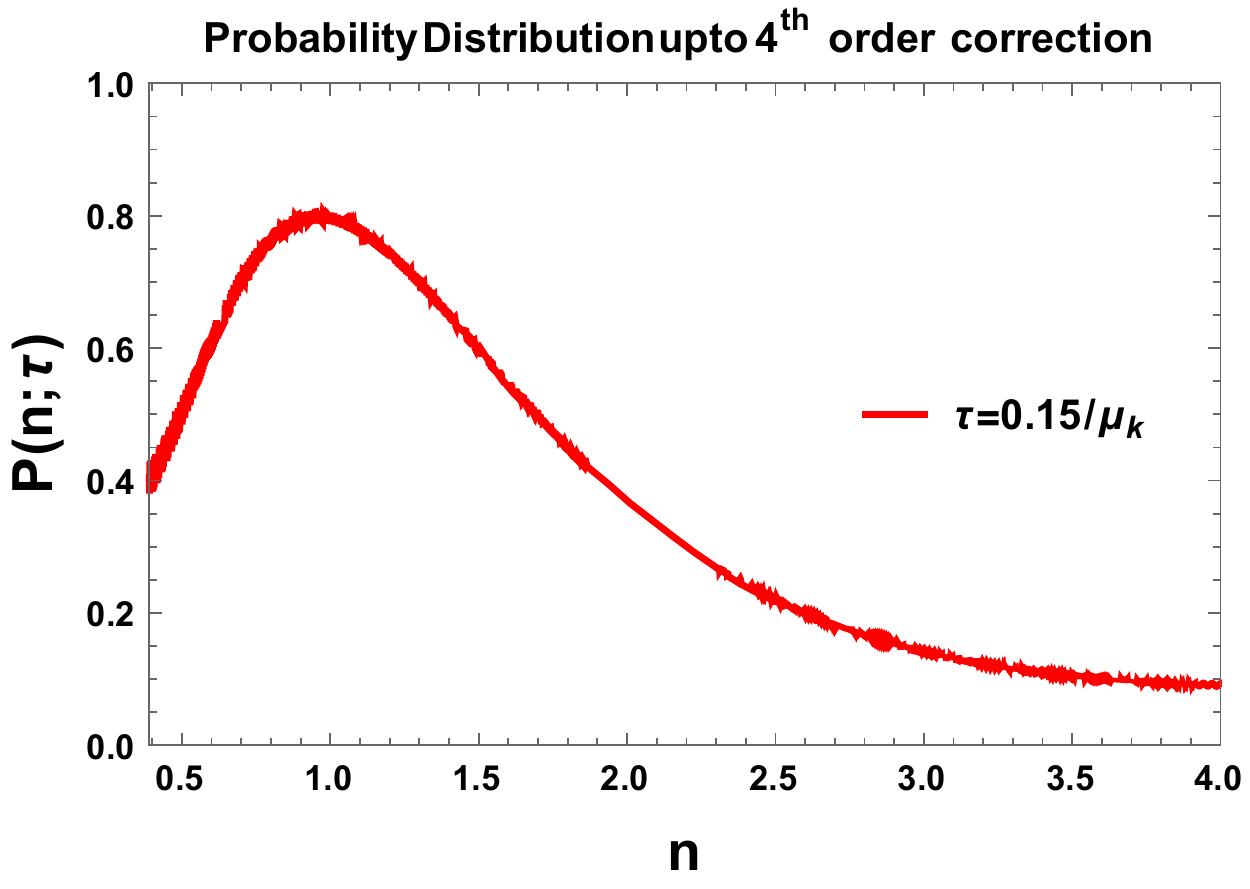}
    \label{xt41n}
}
\subfigure[Upto 4th order correction ]{
    \includegraphics[width=7.8cm,height=6cm] {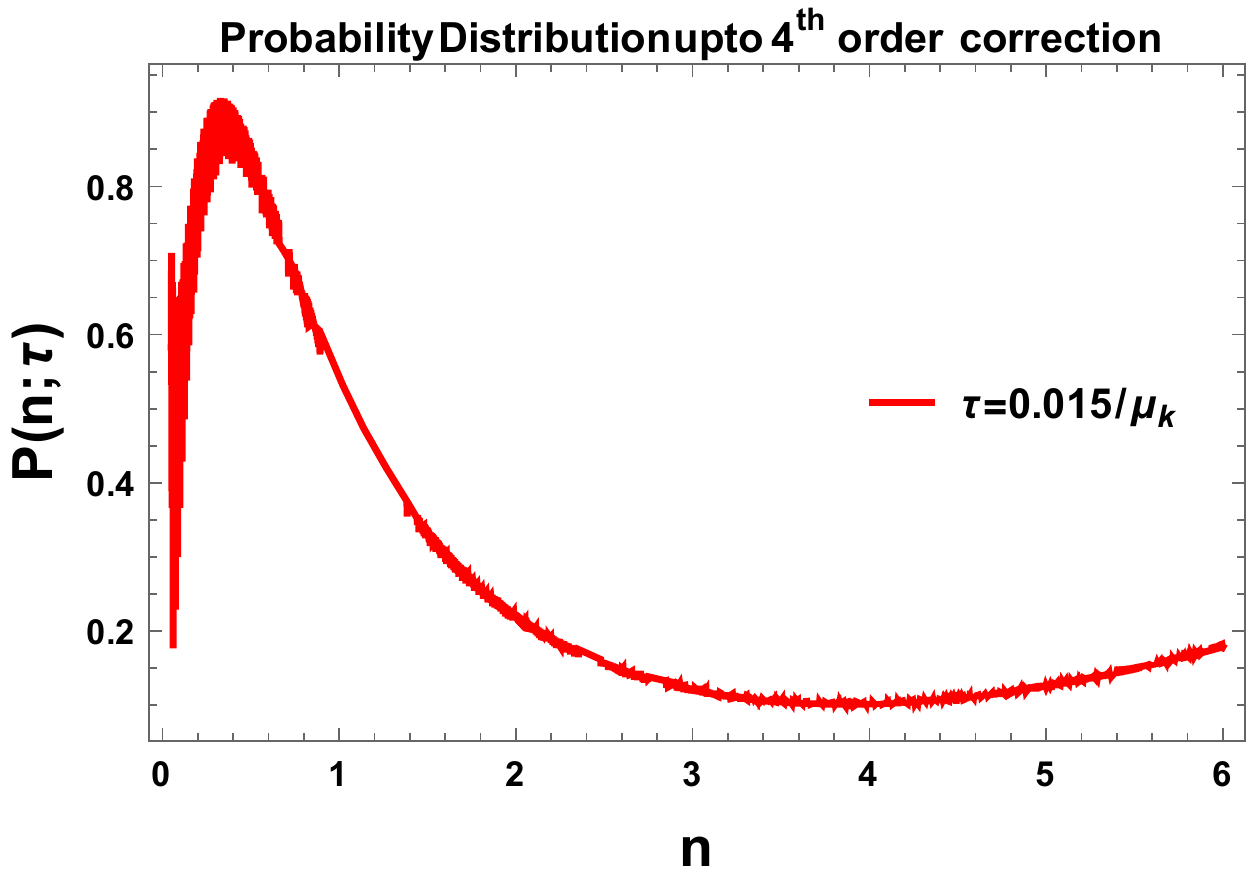}
    \label{xt42n}
}
\subfigure[Upto 4th order correction comparison between different correction]{
    \includegraphics[width=7.8cm,height=6cm] {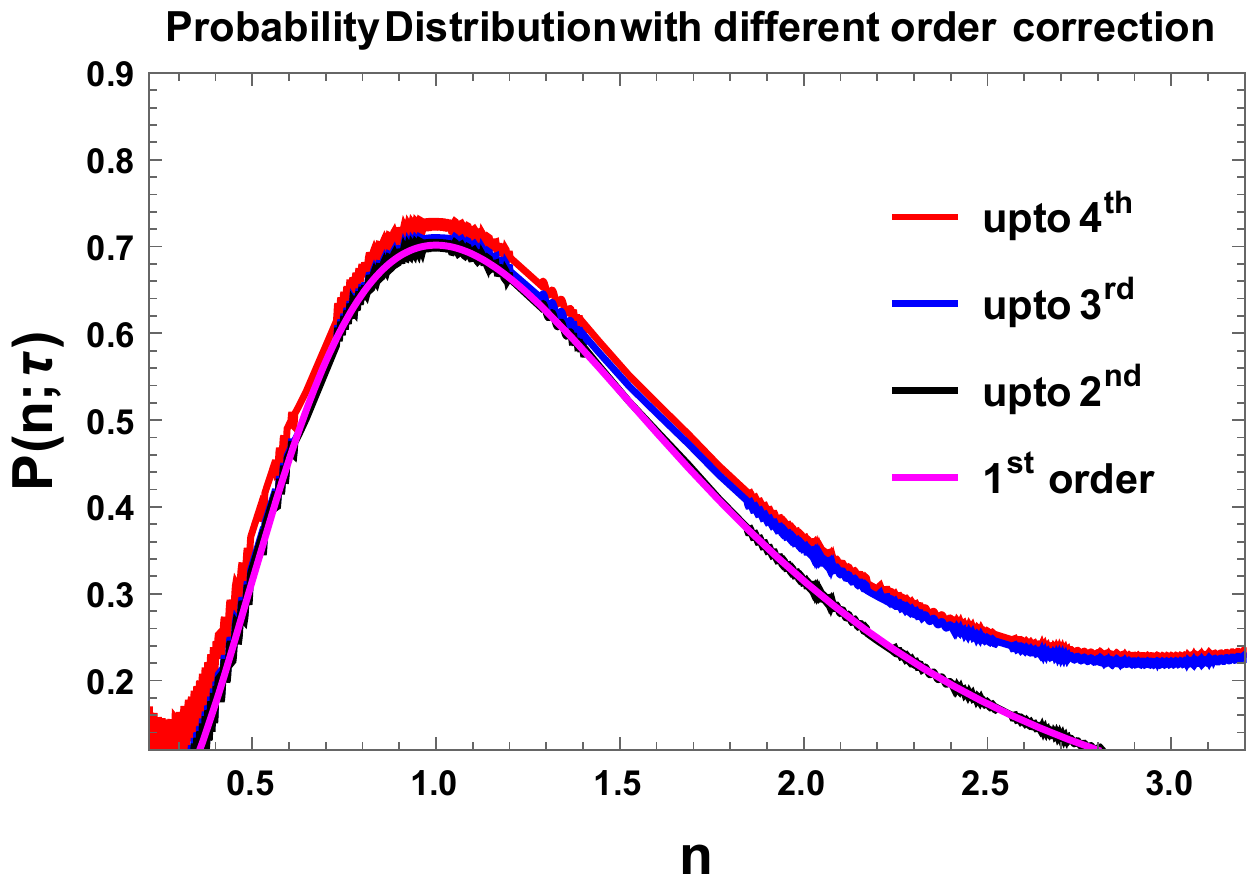}
    \label{xt43n}
}

\caption{Fourth order corrected probability distribution for different $m_{4}$,$m_{3}$,$m_{2}$ and $m_{1}=2$ with previously mentioned initial conditions.}
\label{Fig4th3rd2nd1st}
\end{figure}
From fig.~\ref{xt4322} we observe the final curve represented by  $P_{1}+P_{2}+P_{3}+P_{4}$ shifted the mean of the gaussian from its value at $P_{1}$.Initial gaussian feature for $P_{1}+P_{2}+P_{3}$ and $P_{1}+P_{2}$ are deviated at high $n$ limit . Here $P_{1}+P_{2}+P_{3}+P_{4}$ and $P_{1}$ follow exact gaussian nature but they are mirror image of one another. $P_{1}+P_{2}+P_{3}$  and $P_{1}+P_{2}$ don't have this gaussian nature due to their divergence property as n increases.
In fig~\ref{xt41n} and fig~\ref{xt42n} the effect of all order correction can be observed but effect of different order correction become evident from fig~\ref{xt43n}.Second order correction introduce the oscillating feature whereas third order correction increase the tail and responsible for higher kurtosis. Fourth order correction add a small positive effect to the previous correction without any shape change.

Previously it is shown that in ref.~\cite{Amin:2015ftc} the distribution will be log-normal at large $n$, considering the lowest order contribution coming from the solution of {\it Fokker-Planck} equation. We extend this result upto fourth order and shown the effect of different order correction.The oscillating nature and long tail can be specific feature of stochastic particle production in inflation epoch.This may be the effect of background field or noise at that time.The numerical solutions give different quantum numbers which support the quantum nature. In the next subsection we will calculate various statistical moments and from that we can discuss about the role of quantum effects and non-Gaussinanity from the Probability distribution profile for particle production in the context of early universe cosmology (mostly during reheating).
%%%%%%%
%%%%%%%%%
%%%%%%%%%%%%
%%%%%%%%%%%%%
%%%%%%%%%%%%%%%
%%%%%%%%%%%%%%%
%%%%%%%%%%%%%%%%
\subsection{Calculation of statistical moments (or quantum correlation functions) from corrected probability distribution function}    
Here our prime objective is to compute the different statistical moments from the quantum corrected probability distribution function as obtain by Taylor expanding in order by order from Eq~(\ref{eq18}). From is corrected probability distribution function we compute the expression for $\langle n\rangle $,$\langle n^{2}\rangle$,$\langle n^{3}\rangle$ and $\langle n^{4}\rangle$ and then calculate standard deviation, skewness and kurtosis for a given time. We have explicitly shown that the non vanishing contributions of skewness and kurtosis carries the signature of significant effect of non-Gaussianity. In this analysis the values of these moments are compared with predicted results obtained from log-normal (Gaussian) distribution and non zero values of kurtosis and skewness define the deviation from that. 

To compute the moments we start with the following sets of master evolution equations valid in different orders, as given by:
\bea
&&\textcolor{blue}{\bf \underline{First~Order~Master~Evolution~Equation:}}~~~~~~~~~\nonumber\\
&&~~~~~~~~~~~~~~~~~~~~~~~~~~~~\frac{1}{\mu_{k}} \frac{\pl \langle F\rangle}{\pl \tau} = \left\langle (1+2n) \frac{\pl F}{\pl n}+n(n+1)\frac{\pl^{2} F}{\pl n^{2}}\right\rangle,
\\
&&\textcolor{blue}{\bf \underline{Second~Order~Master~Evolution~Equation:}}~~~~~~~~~\nonumber\\
&&~~~~~~\frac{1}{\mu^2_{k}} \frac{\pl^2 \langle F\rangle}{\pl \tau^2} = \left\langle \frac{n^2}{2}\left(1 + n\right)^{2}\frac{\pl^{4}F}{\pl n^{4}}+2n \left(1 + 3 n + 2 n^2\right)\frac{\pl ^{3}F}{\pl n^{3}}+\left(1 + 6 n + 6 n^2\right) \frac{\pl ^{2}F}{\pl n^{2}}\right\rangle,~~~~~~~~~~~~
\\
&&\textcolor{blue}{\bf \underline{Third~Order~Master~Evolution~Equation:}}~~~~~~~~~\nonumber\\
&&~~~~~~\frac{1}{\mu^3_{k}} \frac{\pl^3 \langle F\rangle}{\pl \tau^3} =\left\langle \frac{ n^3}{6} (1 + n)^{3}\frac{\pl^{6}F}{\pl n^{6}}+ \frac{3n^2}{2} (1 + n)^2 (1 + 2 n) \frac{\pl^{5}F}{\pl n^{5}}\nonumber\right. \\
&&\left.~~~~~~~~~~~~~~~~~~~~+ 3 n (1 + n) (1 + 5 n + 5 n^2)\frac{\pl^{4}F}{\pl n^{4}}+(1 + 2 n) (1 + 10 n + 10 n^2)\frac{\pl^{3}F}{\pl n^{3}}\right\rangle,~~~~~~~~~~~~
\eea \bea
&&\textcolor{blue}{\bf \underline{Fourth~Order~Master~Evolution~Equation:}}~~~~~~~~~\nonumber\\
&&~~~~~~\frac{1}{\mu^4_{k}} \frac{\pl^4 \langle F\rangle}{\pl \tau^4} =\left\langle 70 n^4 (1+n)^4\frac{\pl^{8}F}{\pl n^{8}}+140 n^3 (1+2 n) \frac{\pl^{7}F}{\pl n^{7}}\nonumber\right.\\
&&~~~~~~~~~~~~~~~~~~~~~~~~~\left.+30 n^2 (1+n)^2 (3+14 n+14 n^2)\frac{\pl^{6}F}{\pl n^{6}}\nonumber\right.\\
&&~~~~~~~~~~~~~~~~~~~~~~~~~\left.+20 n (1+n) (1+2 n) (1+7 n+7 n^2)\frac{\pl ^{5}F}{\pl n^{5}}\nonumber\right.\\
&&~~~~~~~~~~~~~~~~~~~~~~~~~\left.+(1+20 n+90 n^2+140 n^3+70 n^4)\frac{\pl^{4}F}{\pl n^{4}}\right\rangle,~~~~~~~~~~~~.\eea
where the first moment or the expectation value of the observable $F$
is define as:      
 \be\label{gew} \textcolor{blue}{\bf \underline{First~Moment:}}~~~~~~\langle F(n)\rangle(\tau) \equiv \int ~dn ~ F(n)P(n;\tau)~. \ee                     
 Here $F(n)$ is the physical observable in which we are interested in and $P(n;\tau)$ is the corrected probability distribution function which is not necessarily log-normal (Gaussian) in nature. In the present context of discussion, Eq~(\ref{gew}) plays the role of generating function, which is commonly used in calculating various mathematical special functions. In our discussion, Eq~(\ref{gew}) represents the {\it statistical moment generating function} in presence of quantum corrected probability distribution function. In terms of quantum mechanical language, Eq~(\ref{gew}) signify the one point quantum correlation function and it is exactly equal to the statistical first moment in this discussion.
 
 Now, we explicitly compute the expressions for one point function (or first moment) of the occupation number i.e. $\langle n\rangle $, two point function (or second moment) of the occupation number i.e. $\langle n^2\rangle $,  three point function (or third moment) of the occupation number i.e. $\langle n^3\rangle $ and  four point function (or fourth moment) of the occupation number i.e. $\langle n^2\rangle $ using the previously mentioned first, second, third and fourth order master equations. The detailed steps of the computations are appended bellow:
 \begin{enumerate}
 \item \textcolor{red}{\bf \underline{Step~I:}}\\
 First of all, we use the first order master evolution equation. Then we replace the function $F$ by the occupation number $n$. Consequently, we get the following time evolution equation of the first moment or one point function $\langle n\rangle $, given by:
  \bea\frac{1}{\mu_{k}}  \frac{\pl \langle n\rangle}{\pl\tau} =\langle (1+2n)\rangle= 1 + 2 \langle n\rangle~.\eea
 \item \textcolor{red}{\bf \underline{Step~II:}}\\
 Secondly, we want to compute the expression for $\langle n^2\rangle $. To compute this we consider here the first and second order master equations, as mentioned earlier. Considering only the first order master equation we get the following analytical expression:
 \bea\frac{1}{\mu_{k}}  \frac{\pl \langle n^2\rangle}{\pl\tau} =\langle 2n(1+2n)+2n(1+n)\rangle=  \langle 4n + 6n^2\rangle=4\langle n\rangle+6\langle n^2 \rangle~.\eea
On the other hand, using the second order master equation we get the following analytical expression for the time evolution of the second moment or two point correlation:
\be\frac{1}{\mu_{k}^{2}}\frac{\pl^{2} \langle n^2\rangle}{\pl \tau^{2}}=\langle 2(1+6n+6n^2)\rangle=12 \langle n\rangle+12 \langle n^2\rangle+2
~.
\ee

\item  \textcolor{red}{\bf \underline{Step~III:}}\\
  Next, we want to compute the expression for $\langle n^3\rangle $. To compute this we consider here the first, second and third order master equations, as mentioned earlier. Considering only the first order master equation we get the following analytical expression:
\be\frac{1}{\mu_{k}}  \frac{\pl \langle n^3\rangle}{\pl\tau} =\langle 3n^2(1+2n)+6n(1+n)\rangle=  \langle 6n + 9n^2 + 6n^3\rangle=6\langle n\rangle+9\langle n^2 \rangle+6\langle n^3 \rangle~.
\ee
On the other hand, using the second order master equation we get the following analytical expression for the time evolution of the third moment or three point correlation:
\be\frac{1}{\mu_{k}^{2}}\frac{\pl^{2} \langle n^3\rangle}{\pl \tau^{2}}=\langle 12n(1+3n+3n^2)+6n(1+6n+6n^2)\rangle=18 \langle n\rangle+72 \langle n^2\rangle+60 \langle n^3\rangle
~.
\ee
Finally, using the third order master equation we get the following analytical expression for the time evolution of the third moment or three point correlation:
\be\frac{1}{\mu_{k}^{3}}\frac{\pl^{3} \langle n^3\rangle}{\pl \tau^{3}}=\langle 6(1+2n)(1+10n+10n^2)\rangle=72 \langle n\rangle+180 \langle n^2\rangle+120\langle n^3\rangle+6
~.
\ee
 \item \textcolor{red}{\bf \underline{Step~IV:}}\\
 Next, we want to compute the expression for $\langle n^4\rangle $. To compute this we consider here the first, second, third and fourth order master equations, as mentioned earlier. Considering only the first order master equation we get the following analytical expression:
\bea\frac{1}{\mu_{k}}  \frac{\pl \langle n^4\rangle}{\pl\tau} &=&\langle 4n^3(1+2n)+12n^2(1+n)\rangle=  \langle 16n^3 + 20n^4 \rangle=16\langle n^3\rangle+20\langle n^4 \rangle~.~~~~~~
\eea
On the other hand, using the second order master equation we get the following analytical expression for the time evolution of the fourth moment or four point correlation:
\bea\frac{1}{\mu_{k}^{2}}\frac{\pl^{2} \langle n^4\rangle}{\pl \tau^{2}}&=&\langle 12n^2(1+n)^2+48n^2(1+3n+2n^2)+12n^2(1+6n+6n^2)\rangle\nonumber\\&=&72 \langle n^2\rangle+240 \langle n^3\rangle+180 \langle n^4\rangle
~.\eea
Then, using the third order master equation we get the following analytical expression for the time evolution of the fourth moment or fourth point correlation:
\bea\frac{1}{\mu_{k}^{3}}\frac{\pl^{3} \langle n^4\rangle}{\pl \tau^{3}}&=&\langle 72n(1+n)(1+5n+5n^2)+24n(1+2n)(1+10n+10n^2)\rangle\nonumber\\
&=&96 \langle n\rangle+720 \langle n^2\rangle+1440\langle n^3\rangle+840\langle n^4\rangle
~.\eea
Finally, using the third order master equation we get the following analytical expression for the time evolution of the fourth moment or fourth point correlation:
\bea\frac{1}{\mu_{k}^{4}}\frac{\pl^{4} \langle n^4\rangle}{\pl \tau^{4}}&=&\langle 24(1+20n+90n^2+140n^3+70n^4)\rangle\nonumber\\
&=&480 \langle n\rangle+2160 \langle n^2\rangle+3360\langle n^3\rangle+1680\langle n^4\rangle+24
~.\eea
 \item \textcolor{red}{\bf \underline{Step~V:}}\\
 Further we apply the  boundary conditions, i.e.
$\langle n\rangle$, $\langle n^{2}\rangle$, $\langle n^{3}\rangle$, $\langle n^{4}\rangle$, $\frac{d\langle n^{2}\rangle}{d \tau^2}$, $\frac{d\langle n^{3}\rangle}{d \tau^3}$ and , $\frac{d\langle n^{4}\rangle}{d \tau^4}$ are vanishingly small at $\tau =0$. Using these conditions 
we get expressions for $\langle n\rangle$, $\langle n^{2}\rangle$, $\langle n^{3}\rangle$ and $\langle n^{4}\rangle$.
\item \textcolor{red}{\bf \underline{Step~VI:}}\\ 
 Using the result obtained in \textcolor{red}{\bf Step~I} and using the previously mentioned boundary condition we get the following expression for the one point function~\footnote{Here it is important to note that as far as quantum mechanical computation is concerned, it produces same result for the one point function and first moment of the occupation number and both of them is equal to the expectation or average value of the occupation number in this context.} (or first moment) of occupation number:
\bea
\textcolor{blue}{\bf \underline{First~Moment~(First~Order):}}~~~~~~~~~\langle n\rangle_{\textcolor{red}{\bf I}}=\frac{1}{2} (e^{2 \tau \mu_{k}}-1)
~.\eea
which is further used to compute all the higher order moments from master evolution equation considering higher order Taylor expansion.
Additionally, it is important to note that if we use higher order equations for the first moment then after imposing the boundary conditions we get the following results:
\bea
\textcolor{blue}{\bf \underline{First~Moment~(Second~Order):}}~~~~~~~~~\langle n\rangle_{\textcolor{red}{\bf II}}=0,\\
\textcolor{blue}{\bf \underline{First~Moment~(Third~Order):}}~~~~~~~~~\langle n\rangle_{\textcolor{red}{\bf III}}=0,\\
\textcolor{blue}{\bf \underline{First~Moment~(Fourth~Order):}}~~~~~~~~~\langle n\rangle_{\textcolor{red}{\bf IV}}=0
~.\eea
Consequently, the total first moment can be written as:
\bea\label{1stdg}
\textcolor{blue}{\bf \underline{Total~First~Moment:}}~~~~\langle n\rangle &=&\langle n\rangle_{\textcolor{red}{\bf I}}+\langle n\rangle_{\textcolor{red}{\bf II}}+\langle n\rangle_{\textcolor{red}{\bf III}}+\langle n\rangle_{\textcolor{red}{\bf IV}}\nonumber\\
~~~~~~~~~~~~~~~~~~~~&=&\langle n\rangle_{\textcolor{red}{\bf I}}=\frac{1}{2} (e^{2 \tau \mu_{k}}-1)
~.\eea
\begin{figure}[htb]
\centering
{
    \includegraphics[width=14.8cm,height=8cm] {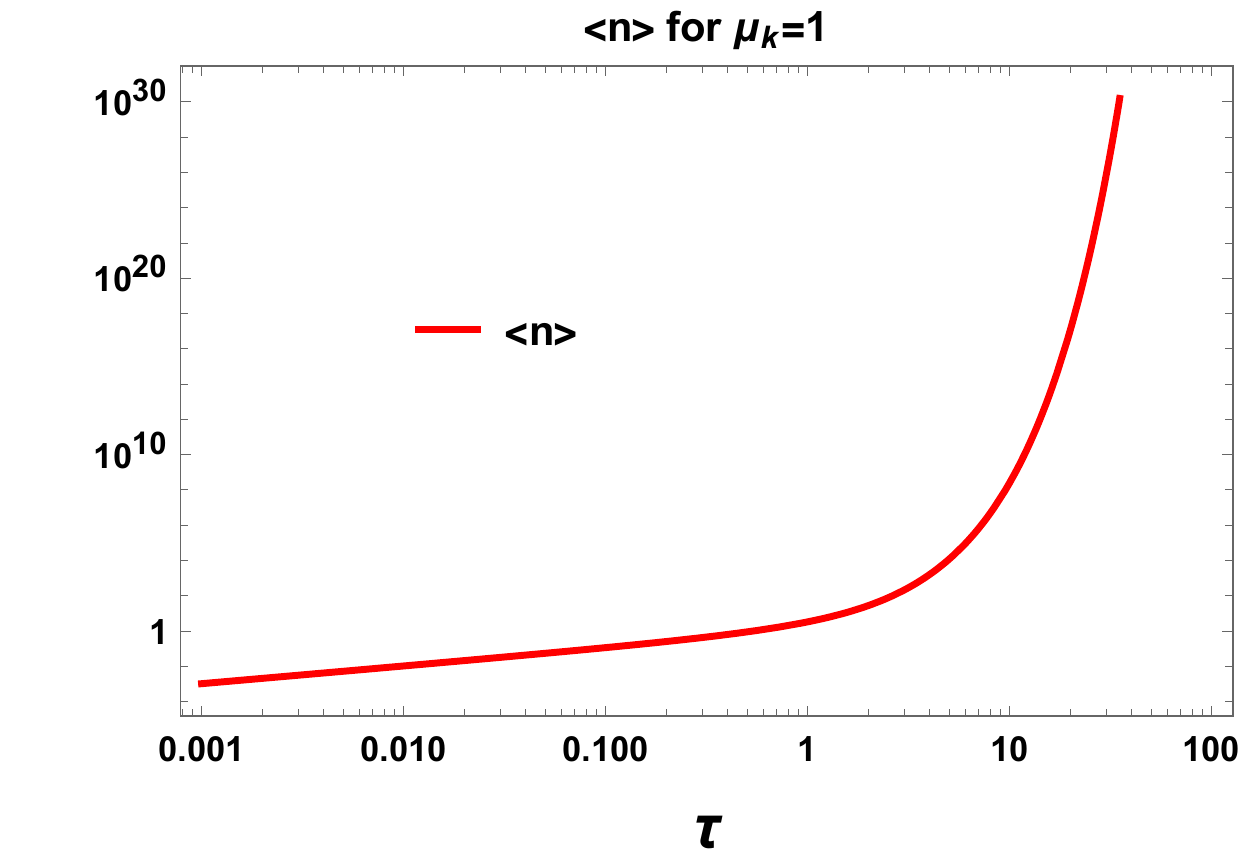}
}
\caption{Time dependent behaviour of $\langle n\rangle$ for $\mu_k=1$.}
\label{NNM1}
\end{figure}
In fig.~(\ref{NNM1}), we have explicitly shown the time dependent behaviour of first moment or one point function of the occupation number $\langle n\rangle$. As there is no contributions are coming from the second, third and fourth order moment generating master evolution equation for $\langle n\rangle$, the only contribution is coming from the first order master evolution equation. From this plot we see that for a fixed value of the parameter $\mu_k=1$, at the lower values of the time the first moment or the one point function of the occupation number initially increase with time very very slowly. Then after a certain time when $\tau>>1$ it shows suddenly huge increment in the  behaviour.  Most importantly, this plot shows the first moment or one point function of the occupation number is not zero. This shows the first signature of the non-Gaussianity as we know for Gaussian probability distribution profile this is exactly zero. 

\item \textcolor{red}{\bf \underline{Step~VII:}}\\ 
 Using the results obtained in \textcolor{red}{\bf Step~II} and using the previously mentioned boundary condition we get the following expression for the two point function~\footnote{It is important to note that, as far as quantum mechanical computation is concerned, it produces not exactly same result for the one point function and first moment of the occupation number. For two point function we actually get the following result:
 \be \langle n(\tau) n(\tau^{'}) \rangle = A(\tau)\delta(\tau+\tau^{'}),\ee
 where $A(\tau)$ is the amplitude of the two point function as given by the following expression:
 \be A(\tau)=\langle n^{2}\rangle=\langle n^{2}\rangle_{\textcolor{red}{\bf I}}+\langle n^{2}\rangle_{\textcolor{red}{\bf II}}.\ee
 This implies that the amplitude part is exactly matches with the  second moment  of the occupation number in this context.}  (or second moment) of occupation number:
\bea\label{2nd1storder}
&&\textcolor{blue}{\bf \underline{Second~Moment~(First~Order):}}\nonumber\\
&&~~~~~~~~~~~~~~~~~~\langle n^{2}\rangle_{\textcolor{red}{\bf I}}=\frac{1}{6} e^{6 \tau \mu_{k}} + \frac{1}{6} (2 - 3 e^{2 \tau \mu_{k}})
~.\eea
\bea\label{2nd2ndorder}
&&\textcolor{blue}{\bf \underline{Second~Moment~(Second~Order):}}\nonumber\\
&&\langle n^{2}\rangle_{\textcolor{red}{\bf II}}=\frac{1}{24} e^{-2 \sqrt{3} \mu_k  \tau } \left[8 e^{2 \sqrt{3} \mu_k  \tau }-18 e^{2 \left(\sqrt{3}+1\right) \mu_k  \tau }+\left(3 \sqrt{3}+5\right) e^{4 \sqrt{3} \mu_k  \tau }-3 \sqrt{3}+5\right]
~.\eea
\begin{figure}[htb]
\centering
%\subfigure[Different $\langle n^{2}\rangle$ for lower values $\mu_k$.  ]{
%    \includegraphics[width=7.8cm,height=8.2cm] {N2_u1.pdf}
%    \label{NNN1}
%}
%\subfigure[Different $\langle n^{2}\rangle$  for medium values $\mu_k$.]{
  %  \includegraphics[width=7.8cm,height=8.2cm] {N2_u10.pdf}
 %   \label{NNN2}
%}
%\subfigure[Different $\langle n^{2}\rangle$  for higher values $\mu_k$.]{
%    \includegraphics[width=7.8cm,height=8.2cm] {N2_u100.pdf}
 %   \label{NNN3}
%}
%\subfigure[Different $\langle n^{2}\rangle$ in different order for higher values $\mu_k$.]
{
    \includegraphics[width=14.8cm,height=8cm] {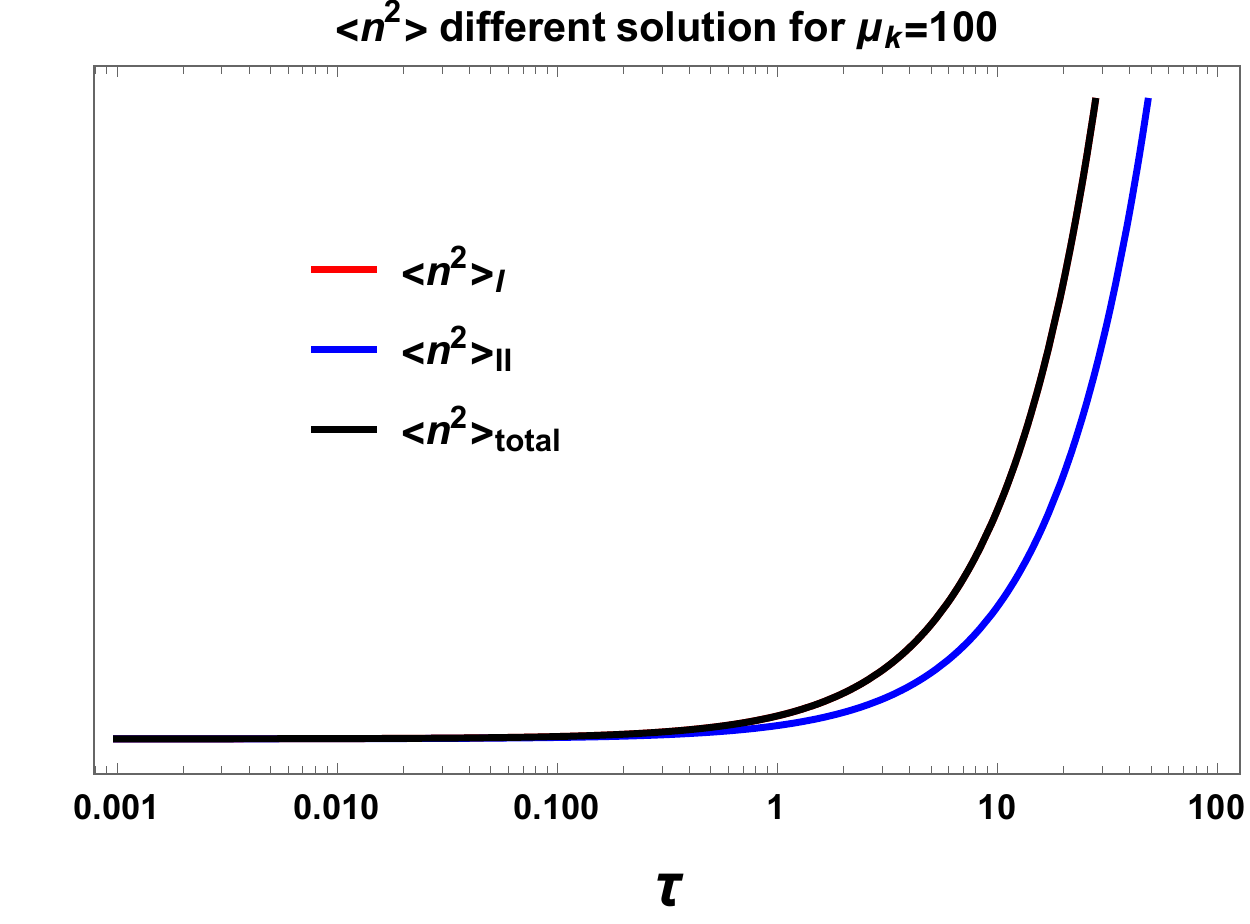}
   % \label{NNN3}
}
\caption{Time dependent behaviour of different $\langle n^{2}\rangle$   at different values of the parameter $\mu_k$.}
\label{NNN1a}
\end{figure}
which is further used to compute all the higher order moments from master evolution equation considering higher order Taylor expansion.
Additionally, it is important to note that if we use higher order equations for the second moment then after imposing the boundary conditions we get the following results:
\bea
\textcolor{blue}{\bf \underline{Second~Moment~(Third~Order):}}~~~~~~~~~\langle n^2\rangle_{\textcolor{red}{\bf III}}&=&0,\\
\textcolor{blue}{\bf \underline{Second~Moment~(Fourth~Order):}}~~~~~~~~~\langle n^2\rangle_{\textcolor{red}{\bf IV}}&=&0
~.\eea
Consequently, the total second moment can be written as:
\bea\label{2ndg}
\textcolor{blue}{\bf \underline{Total~Second~Moment:}}~~~~\langle n^{2}\rangle &=&\langle n^{2}\rangle_{\textcolor{red}{\bf I}}+\langle n^{2}\rangle_{\textcolor{red}{\bf II}}+\langle n^{2}\rangle_{\textcolor{red}{\bf III}}+\langle n^{2}\rangle_{\textcolor{red}{\bf IV}}\nonumber\\
&=&\langle n^{2}\rangle_{\textcolor{red}{\bf I}}+\langle n^{2}\rangle_{\textcolor{red}{\bf II}}
~.\eea

In fig.~(\ref{NNN1a}), we have explicitly shown the time dependent behaviour of second moment or amplitude of the two point function of the occupation number $\langle n^2\rangle$. As there is no contributions are coming from the third and fourth order moment generating master evolution equation for $\langle n^2\rangle$, the only contribution is coming from the first and second order master evolution equation. From this plot we see that for a fixed value of the parameter $\mu_k=1, 10, 100$, at the lower values of the time the second moment or the amplitude of the two point function of the occupation number initially increase with time very very slowly. Then after a certain time when $\tau>>1$ it shows suddenly huge increment in the  behaviour.  
\item \textcolor{red}{\bf \underline{Step~VIII:}}\\ 
 Using the results obtained in \textcolor{red}{\bf Step~III} and using the previously mentioned boundary condition we get the following expression for the three point function~\footnote{It is important to note that, as far as quantum mechanical computation is concerned, it produces not exactly same result for the three point function and third moment of the occupation number. For three point function we actually get the following result:
 \be \langle n(\tau) n(\tau^{'})n(\tau^{''}) \rangle = B(\tau,\tau^{'},\tau^{"})\delta(\tau+\tau^{'}+\tau^{"}),\ee
 where $ B(\tau,\tau^{'},\tau^{"})$ is the amplitude of the three point function. If we fix $\tau=\tau^{'}=\tau^{"}$ (equal time) then we get the following expression:
 \be B(\tau,\tau,\tau)=\langle n^{3}\rangle=\langle n^{3}\rangle_{\textcolor{red}{\bf I}}+\langle n^{3}\rangle_{\textcolor{red}{\bf II}}+\langle n^{3}\rangle_{\textcolor{red}{\bf III}}.\ee
 This implies that the equal time amplitude part is exactly matches with the  third moment  of the occupation number in this context.}  (or third moment) of occupation number:
\bea\label{3rd1storder}
&&\textcolor{blue}{\bf \underline{Third~Moment~(First~Order):}}\nonumber\\
&&~~~~~~~~~~~~~~~~~~\langle n^{3}\rangle_{\textcolor{red}{\bf I}}=\frac{1}{8 \mu_k }\left[e^{6 \mu_k  \tau } \left(12 \mu ^2_k \tau +2 \mu_k -5\right)+(9-6 \mu_k ) e^{2 \mu_k  \tau }+4 (\mu_k -1)\right]
~.\eea
\bea\label{3rd2ndorder}
&&\textcolor{blue}{\bf \underline{Third~Moment~(Second~Order):}}\nonumber\\
&&~~~~~~~~~\langle n^{3}\rangle_{\textcolor{red}{\bf II}}=\frac{1}{560} \left[35 \left(3 \sqrt{3}-5\right) e^{-2 \sqrt{3} \mu_k  \tau }+450 e^{2 \mu_k  \tau }\right.\nonumber\\ && \left.~~~~~~~~~~~~~~~~~~~~~~~~~~~~-35 \left(3 \sqrt{3}+5\right) e^{2 \sqrt{3} \mu_k  \tau }+\left(6 \sqrt{15}+20\right) e^{2 \sqrt{15} \mu_k  \tau }\right.\nonumber\\ && \left.~~~~~~~~~~~~~~~~~~~~~~~~~~~~~~~~+\left(20-6 \sqrt{15}\right) e^{-2 \sqrt{15} \mu_k  \tau }-140\right]
~.\eea
\bea\label{3rd3rdorder}\footnotesize
&&\textcolor{blue}{\bf \underline{Third~Moment~(Third~Order):}}\nonumber\\
&&\langle n^{3}\rangle_{\textcolor{red}{\bf III}}=\frac{1}{36960}\left[\frac{2 }{\sqrt{3}-3 i}\left(-6135 i+2045 \sqrt{3}-1654\ 3^{5/6} \sqrt[3]{5}+1011 \sqrt[6]{3} 5^{2/3}\right.\right.\nonumber\\
&& \left.\left.~~~~~~~~~~+1011 i 15^{2/3}\right) e^{-\sqrt[3]{15} \left(1+i \sqrt{3}\right) \mu_{k}  \tau }+32670 e^{2 \mu_{k}  \tau }-1050 \left(10 \sqrt{3}+17\right) e^{2 \sqrt{3} \mu_{k}  \tau }\right.\nonumber\\
&& \left.~~~~~~+\left(4090+827 i 3^{5/6} \sqrt[3]{5}-1011 i \sqrt[6]{3} 5^{2/3}-827 \sqrt[3]{15}-337\ 15^{2/3}\right) e^{i \sqrt[3]{15} \left(\sqrt{3}+i\right) \mu_{k}  \tau }\right.\nonumber\\
&& \left.~~~~~~+1050 \left(10 \sqrt{3}-17\right) e^{-2 \sqrt{3} \mu_{k}  \tau }-9240+\frac{2 }{\sqrt{3}-3 i}\left(-6135 i+2045 \sqrt{3}\right.\right.\nonumber\\
&& \left.\left.~~~~~~+827\ 3^{5/6} \sqrt[3]{5}+1011 \sqrt[6]{3} 5^{2/3}-2481 i \sqrt[3]{15}-1011 i 15^{2/3}\right) e^{2 \sqrt[3]{15} \mu_{k}  \tau }\right]
~.\eea
which is further used to compute all the higher order moments from master evolution equation considering higher order Taylor expansion.
Additionally, it is important to note that if we use higher order equations for the third moment then after imposing the boundary conditions we get the following results:
\bea
\textcolor{blue}{\bf \underline{Third~Moment~(Fourth~Order):}}~~~~~~~~~\langle n^2\rangle_{\textcolor{red}{\bf IV}}&=&0
~.\eea
Consequently, the total third moment can be written as:
\bea\label{4rntg}
\textcolor{blue}{\bf \underline{Total~Third~Moment:}}~~~~\langle n^{3}\rangle &=&\langle n^{3}\rangle_{\textcolor{red}{\bf I}}+\langle n^{3}\rangle_{\textcolor{red}{\bf II}}+\langle n^{3}\rangle_{\textcolor{red}{\bf III}}+\langle n^{3}\rangle_{\textcolor{red}{\bf IV}}\nonumber\\&
=&\langle n^{3}\rangle_{\textcolor{red}{\bf I}}+\langle n^{3}\rangle_{\textcolor{red}{\bf II}}+\langle n^{3}\rangle_{\textcolor{red}{\bf III}}~.\eea

%\subfigure[Different $\langle n^{3}\rangle$ for lower values $\mu_k$.  ]{
 %   \includegraphics[width=7.8cm,height=9.2cm] {N3_u1.pdf}
 %   \label{NN1}
%}
%\subfigure[Different $\langle n^{3}\rangle$  for medium values $\mu_k$.]{
%    \includegraphics[width=7.8cm,height=9.2cm] {N3_u10.pdf}
 %   \label{NN2}
%}
%\subfigure[Different $\langle n^{3}\rangle$ for higher values $\mu_k$.]{
%    \includegraphics[width=7.8cm,height=9.2cm] {N3_u100.pdf}
 %   \label{NN3}
%}
%\subfigure[Different $\langle n^{3}\rangle$ with the total solution for higher values $\mu_k$.]
\begin{figure}[htb]
\centering
{
    \includegraphics[width=14.8cm,height=8cm] {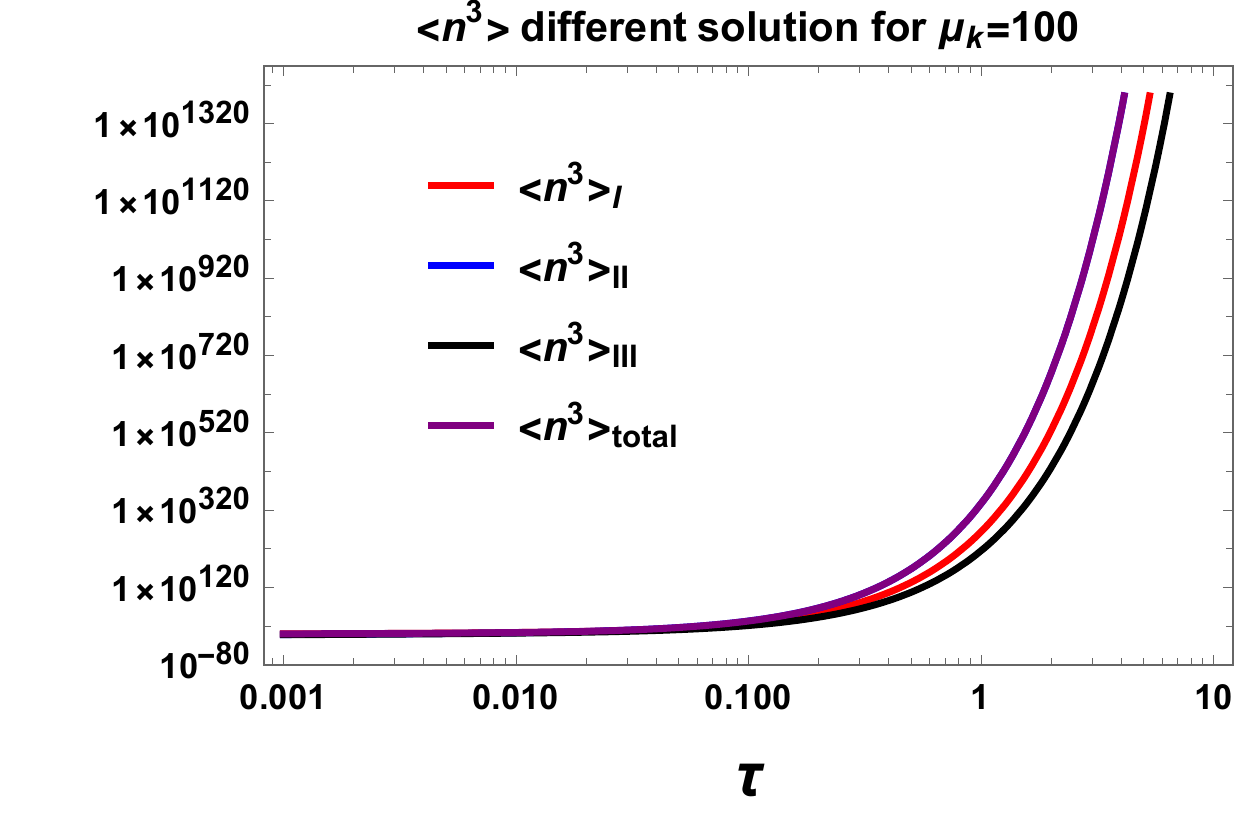}
}
\caption{Time dependent behaviour of the third moment $\langle n^{3}\rangle$ for different fixed values of the parameter $\mu_k$.}
\label{NN1}
\end{figure}
In fig.~(\ref{NN1}), we have explicitly shown the time dependent behaviour of third moment or three point function of the occupation number $\langle n^3\rangle$. As there is no contributions are coming from the fourth order moment generating master evolution equation for $\langle n^3\rangle$, the only contribution is coming from the first, second and third order master evolution equation. From this plot we see that for a fixed value of the parameter $\mu_k=1,10,100$, at the lower values of the time the third moment or the equal time amplitude of the three point function of the occupation number initially increase with time very slowly. Then after a certain time it shows suddenly huge increment in the  behaviour.  Most importantly, this plot shows the third moment or equal time amplitude of the three point function of the occupation number is not zero. This shows the second signature of the non-Gaussianity as we know for Gaussian probability distribution profile this is exactly zero. 
\item \textcolor{red}{\bf \underline{Step~IX:}}\\ 
 Using the results obtained in \textcolor{red}{\bf Step~IV} and using the previously mentioned boundary condition we get the following expression for the four point function~\footnote{It is important to note that, as far as quantum mechanical computation is concerned, it produces not exactly same result for the three point function and third moment of the occupation number. For three point function we actually get the following result:
 \be \langle n(\tau) n(\tau^{'})n(\tau^{''})n(\tau^{'''})  \rangle = C(\tau,\tau^{'},\tau^{"},\tau^{'''})\delta(\tau+\tau^{'}+\tau^{"}+\tau^{"'}),\ee
 where $C(\tau,\tau^{'},\tau^{"},\tau^{'''})$ is the amplitude of the four point function. If we fix $\tau=\tau^{'}=\tau^{"}=\tau^{'''}$ (equal time) then we get the following expression:
 \be C(\tau,\tau,\tau,\tau)=\langle n^{4}\rangle=\langle n^{4}\rangle_{\textcolor{red}{\bf I}}+\langle n^{4}\rangle_{\textcolor{red}{\bf II}}+\langle n^{4}\rangle_{\textcolor{red}{\bf III}}+\langle n^{4}\rangle_{\textcolor{red}{\bf IV}}.\ee
 This implies that the equal time amplitude part is exactly matches with the  third moment  of the occupation number in this context.} (or fourth moment) of occupation number:
\bea\label{4rth1storder}
&&\textcolor{blue}{\bf \underline{Fourth~Moment~(First~Order):}}\nonumber\\
&&\langle n^{4}\rangle_{\textcolor{red}{\bf I}}=\frac{6 (\mu -1) e^{2 \mu  \tau }+3 (5 \mu -2) e^{8 \mu  \tau }-2 e^{6 \mu  \tau } (3 \mu  (4 \mu  \tau +3)-5)-3 \mu +2}{2 \mu }
~.\eea
\bea\label{4rth2ndorder}
&&\textcolor{blue}{\bf \underline{Fourth~Moment~(Second~Order):}}\nonumber\\
&&\langle n^{4}\rangle_{\textcolor{red}{\bf II}}=\frac{\left(67-45 \sqrt{5}\right) e^{-6 \sqrt{5} \mu  \tau }}{10080}+\frac{\left(45 \sqrt{5}+67\right) e^{6 \sqrt{5} \mu  \tau }}{10080}+\frac{1}{5040}(-5265 \sinh (2 \mu  \tau )\nonumber\\
&&~~~~~~~~~~~~-216 \sqrt{15} \sinh \left(2 \sqrt{15} \mu  \tau \right)+2610 \sqrt{3} \sinh \left(2 \sqrt{3} \mu  \tau \right)-5265 \cosh (2 \mu  \tau )\nonumber\\
&&~~~~~~~~~~~~~~~~~+4350 \cosh \left(2 \sqrt{3} \mu  \tau \right)-720 \cosh \left(2 \sqrt{15} \mu  \tau \right)+1568)
\eea
The third and the fourth order corrected version of the fourth order moment equations are not exactly solvable analytically. For this reason we have applied numerical techniques to solve these differential equations.

Consequently, the total fourth moment can be written as:
\bea\label{4rddntg}
\textcolor{blue}{\bf \underline{Total~Fourth~Moment:}}~~~~\langle n^{4}\rangle &=&\underbrace{\langle n^{4}\rangle_{\textcolor{red}{\bf I}}+\langle n^{4}\rangle_{\textcolor{red}{\bf II}}}_{\textcolor{red}{\bf Analytical}} +\underbrace{\langle n^{4}\rangle_{\textcolor{red}{\bf III}}+\langle n^{4}\rangle_{\textcolor{red}{\bf IV}}}_{\textcolor{red}{Numerical}}~.\eea

\begin{figure}[H]
\centering
\subfigure[Different $\langle n^{4}\rangle$ for lower values $\tau $.  ]{
    \includegraphics[width=7.8cm,height=7cm] {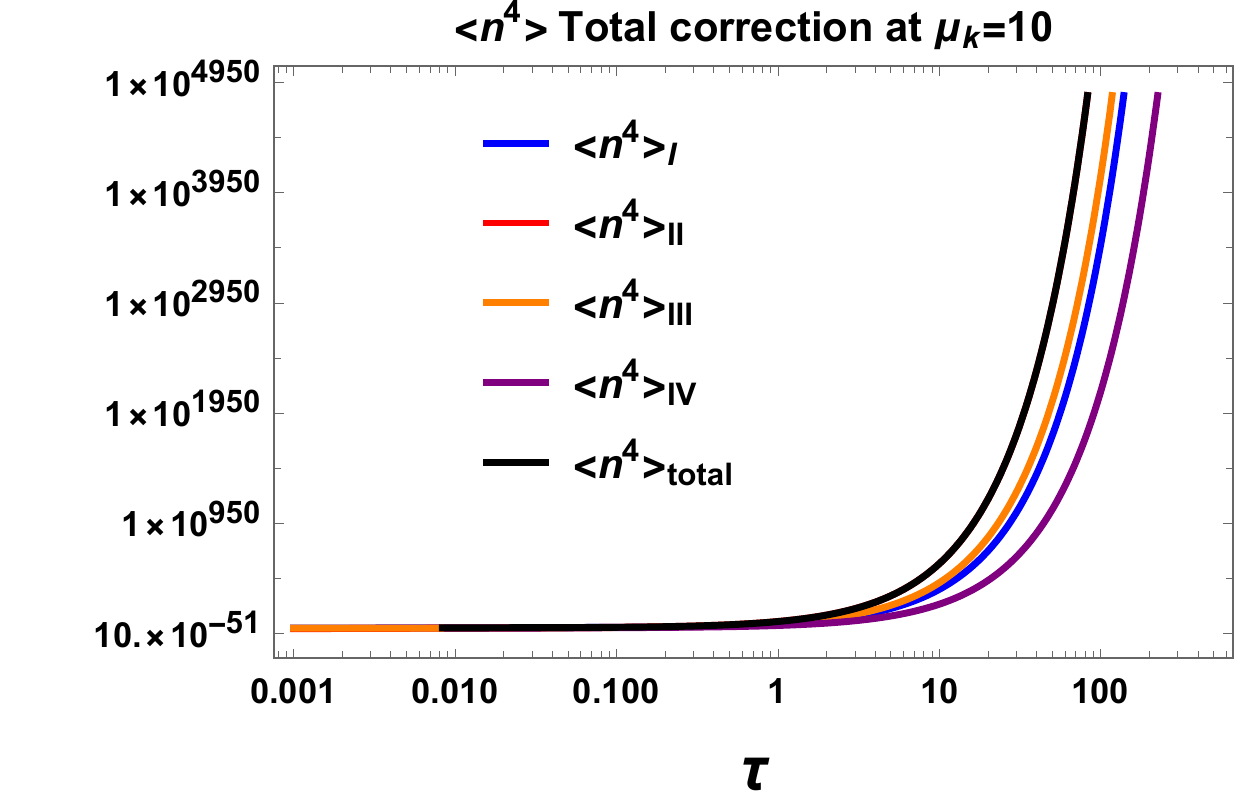}
    \label{N41a}
}
%\subfigure[Different $\langle n^{4}\rangle$ for higher values $\tau$.]{
    %\includegraphics[width=7.8cm,height=6cm] {N4_u100_full.pdf}
    %\label{N43}
%}
\subfigure[Different $\langle n^{4}\rangle$ for higher values $\tau$.]{
    \includegraphics[width=7.8cm,height=7cm] {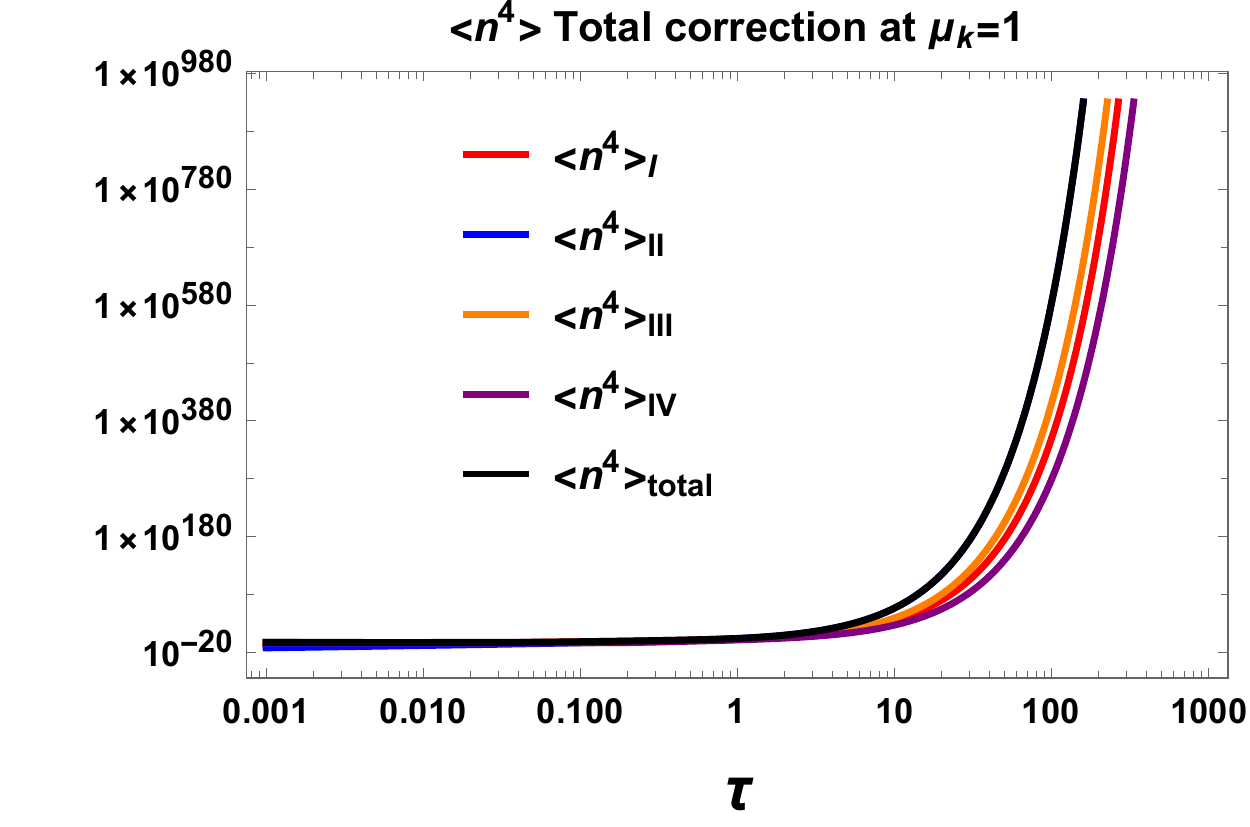}
    \label{N43b}
}
%\subfigure[Different $\langle n^{4}\rangle$  for medium values $\tau$.]{
%    \includegraphics[width=8.8cm,height=7cm] {N4_u100_vlow.pdf}
%    \label{N42c}
%}
\caption{Log plot of Time dependent behaviour of the Fourth moment $\langle n^{4}\rangle$ for different fixed values of the parameter $\mu_k$.}
\label{N41}
\end{figure}
In fig.~(\ref{N41}), we have explicitly shown the time dependent behaviour of fourth moment or amplitude of the four point function of the occupation number $\langle n^4\rangle$. From this plot we see that for a fixed value of the parameter $\mu_k=1,10,100$, at the lower values of the time the third moment or the equal time amplitude of the four point function of the occupation number initially increase with time very slowly. Then after a certain time it shows suddenly huge increment in the  behaviour.  
\end{enumerate}
\subsubsection{Standard Deviation}
Further, using the results obtained in the context of second moment or two point correlation function, in this subsection our prime objective is compute the expression for the {\it Standard Deviation} from the corrected version of the probability distribution function. In the present context of discussion {\it Standard Deviation} actually gives the spread of the peak of the corrected  probability distribution function. Therefore, {\it Standard Deviation} considering upto first order is given by the following expression:
\bea\label{uncorrected-s-d}
{\bf S.D._{uc}}&=&\sqrt{\langle n^2 \rangle_{\textcolor{red}{\bf I}}-\left(\langle n \rangle_{\textcolor{red}{\bf I}}\right)^2}=\frac{\sqrt{2e^{6\tau \mu_k}-3e^{4\tau\mu_k}+1}}{2\sqrt{3}},
\eea
where the subscript "uc" stands for uncorrected.

On the other hand, after including the result from the second order the corrected expression for the {\it Standard Deviation} can be expressed as:
\bea\label{corrected-s-d}
 {\bf S.D._{c}}&=&\sqrt{\left(\langle n^2 \rangle_{\textcolor{red}{\bf I}}+\langle n^2 \rangle_{\textcolor{red}{\bf II}}\right)-\left(\langle n \rangle_{\textcolor{red}{\bf I}}\right)^2}\nonumber\\
&=&\frac{\sqrt{\left(3 \sqrt{3}+5\right) e^{2 \sqrt{3} \mu_k  \tau }-18 e^{2 \mu_k  \tau }-6 e^{4 \mu_k  \tau }+4 e^{6 \mu_k  \tau }+\left(5-3 \sqrt{3}\right) e^{-2 \sqrt{3} \mu_k  \tau }+10}}{2 \sqrt{6}}.~~~~~~~~~~~~~
\eea

%%%%%%%%%%%%%%%%% add pictures
\begin{figure}[H]
\centering
\subfigure[Variance for lower values $\mu_{k}$  ]{
    \includegraphics[width=7.8cm,height=8cm] {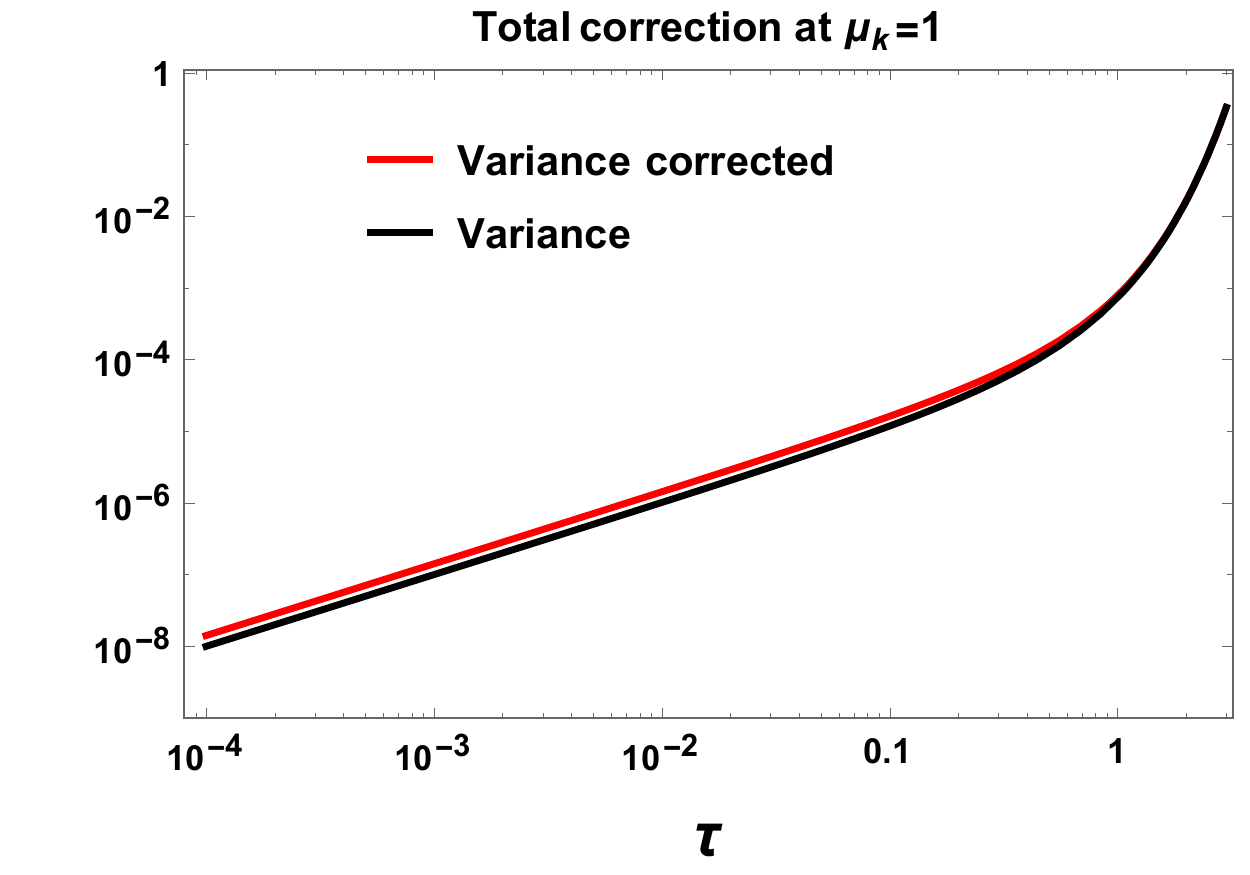}
    \label{sd1}
}
\subfigure[Variance for higher values $\mu_{k}$]{
    \includegraphics[width=7.8cm,height=8cm] {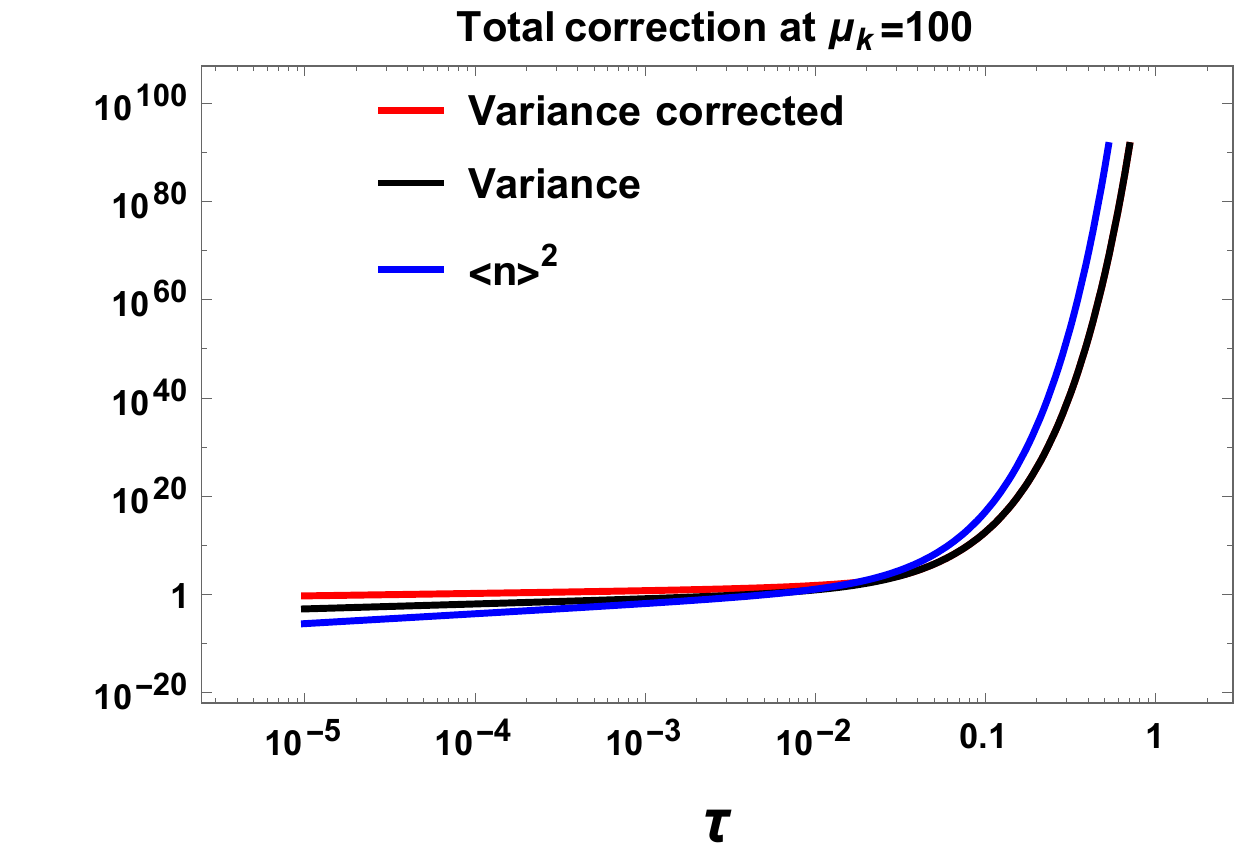}
    \label{sd3}
}
\caption{Time dependent behaviour of variance without second order correction and with second order correction and $\langle n\rangle^{2}$ are shown for different $\mu_{k}$.}
\label{SD1}
\end{figure}

From the fig.-~\ref{SD1}, we can see that the uncorrected {\it Standard Deviation} (first order) and corrected  {\it Standard Deviation} (second order) has significant difference in low $\mu_{k} \tau $ limit and second order overlapped as they approach higher $\mu_{k} \tau$ . So for lower limit this second order correction is significant and for this reason during the computation of {\it Kurtosis} and {\it Skewness} we use total solution of standard deviation over the uncorrected one alone.In fig~\ref{sd1} the variance is with in the value 1 but in fig~\ref{sd2} variance has a enormous value.It is mere effect of tuning.The plots or values of variance can always be tuned to be within 1 using proper prefactor.

\subsubsection{Skewness}
In this subsection, our prime objective is to computed the expression for the {\it Skewness} from the corrected probability distribution function.
{\it Skewness} actually measure the asymmetry of the probability distribution function of a real-valued random variable about its mean value. This measure can be positive or negative, or undefined. From positive skewness (for unimodal distribution) we can say normal curve has longer right tail. 

Therefore, skewness without correction can be expressed as:
\bea\label{skewnessunc}
{\bf Skewness_{uc}}&=&\frac{\langle n^3 \rangle_{\textcolor{red}{\bf I}}}{\left[\langle n^2 \rangle_{\textcolor{red}{\bf I}}-\left(\langle n \rangle_{\textcolor{red}{\bf I}}\right)^2\right]^{3/2}}=\frac{\langle n^3 \rangle_{\textcolor{red}{\bf I}}}{\left[{\bf S.D._{uc}}\right]^{3}}\nonumber\\
&=&\frac{3 \sqrt{3} \left(e^{6 \mu  \tau } \left(12 \mu ^2 \tau +2 \mu -5\right)+(9-6 \mu ) e^{2 \mu  \tau }+4 (\mu -1)\right)}{\mu  \left(-3 e^{4 \mu  \tau }+2 e^{6 \mu  \tau }+1\right)^{3/2}}.
\eea
On the other hand, the third order corrected value of the {\it Skewness} can be expressed as:
\bea\label{skewnessunc}
{\bf Skewness_{uc}}&=&\frac{\langle n^3 \rangle_{\textcolor{red}{\bf I}}+\langle n^3 \rangle_{\textcolor{red}{\bf II}}+\langle n^3 \rangle_{\textcolor{red}{\bf III}}}{\left[\left(\langle n^2 \rangle_{\textcolor{red}{\bf I}}+\langle n^2 \rangle_{\textcolor{red}{\bf II}}\right)-\left(\langle n \rangle_{\textcolor{red}{\bf I}}\right)^2\right]^{3/2}}=\frac{\langle n^3 \rangle_{\textcolor{red}{\bf I}}}{\left[{\bf S.D._{c}}\right]^{3}}.
\eea
The explicit detail of the corrected version of {\it Skewness} is not written
to avoid writing  the unnecessary lengthy expression.

%%%%%%%ADD DIAGRAMS
\begin{figure}[H]
\centering
\subfigure[Skewness for lower values $\mu_{k}\tau$  ]{
    \includegraphics[width=7.8cm,height=8cm] {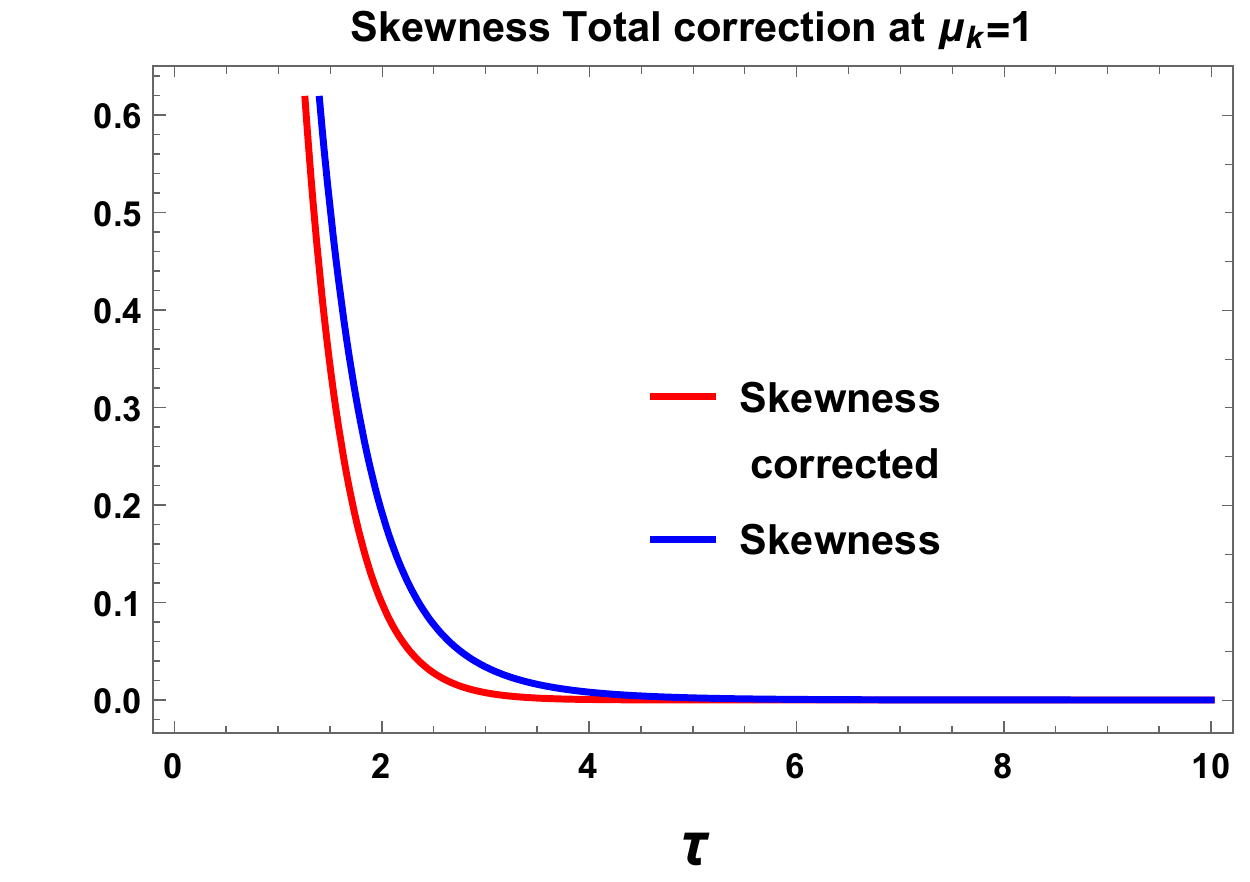}
    \label{sk1}
}
\subfigure[Skewness for all values $\mu_{k}\tau$ ]{
    \includegraphics[width=7.8cm,height=8cm] {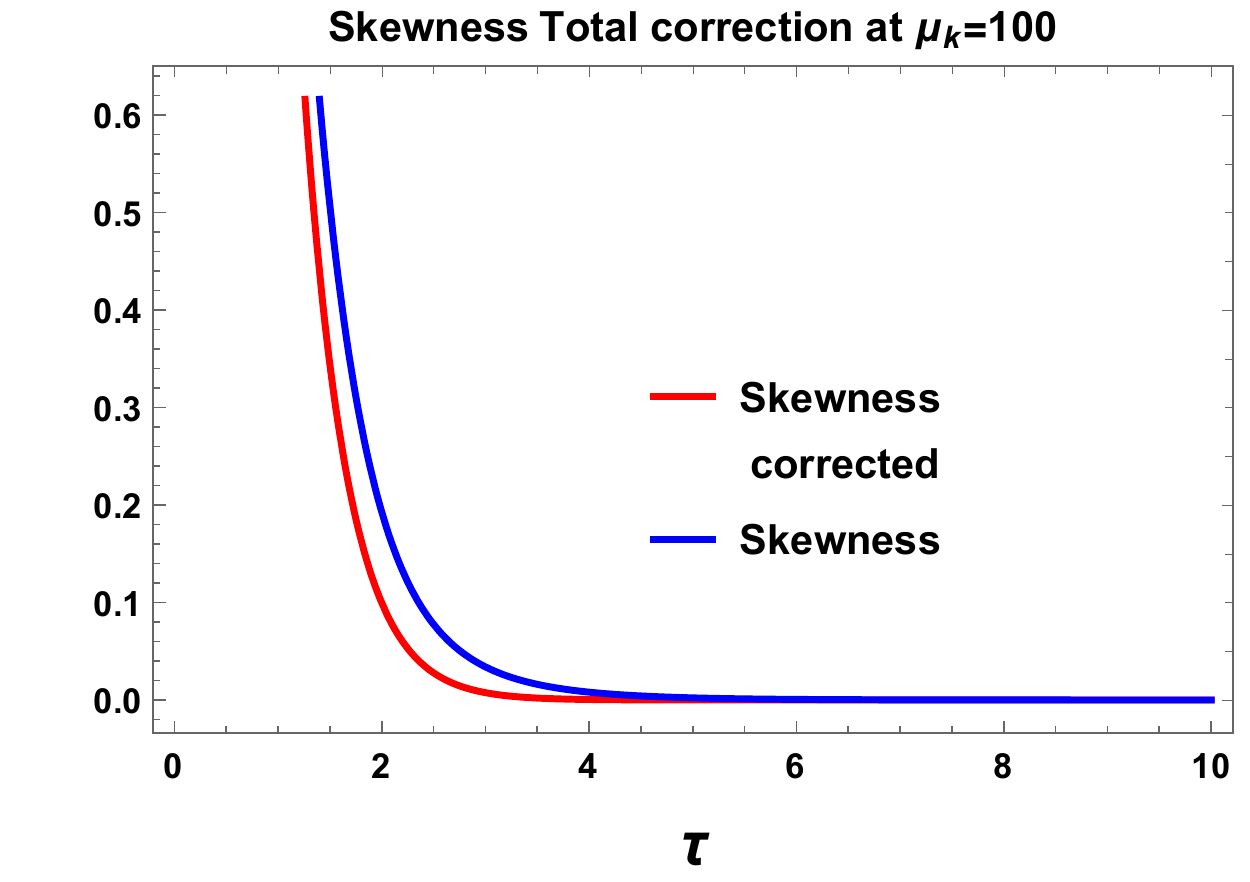}
    \label{sk3}
}
\caption{Time dependent behaviour of {\it  Skewness} without correction and with correction at different fixed range of $\mu_{k}$.}
\label{SKRR1}
\end{figure}
Now from fig.~\ref{SKRR1}, we can say that the corrected  {\it Skewness} deviate significantly from the uncorrected one at low $\mu_{k} \tau$ limit. But we can see that at higher limit they overlap. 
Also {\it Skewness} is positive for the whole range which implies that the normal distribution curve has longer right tail. Moreover, there is a discontinuity of third order corrected {\it Skewness} in between the range $0.1<\tau<1$ and for the rest of the whole range of time {\it Skewness} decreased upto unity and then it is increased. 

\subsubsection{Kurtosis}
{\it Kurtosis} is a measure of the {\it tailedness} of the probability distribution of a real-valued random variable. This is actually a descriptor of the shape of a probability distribution function and there are specific ways of quantifying it for a theoretical probability distribution and corresponding ways of estimating it from a sample from a population.
It is important to note that, the {\it Kurtosis} of any univariate normal distribution is $3$. For practical purposes it is common practice to compare the expression for {\it Kurtosis} of a probability distribution function to $3$. Probability distributions with {\it Kurtosis} less than the value $3$ are identified as {\it platykurtic}, although this information does not imply the distribution is {\it flat-topped} in nature. Rather, it implies that the probability distribution produces fewer and less extreme outliers than does the normal probability distribution. Probability distributions with {\it Kurtosis} greater than the value $3$ are said to be {\it leptokurtic}.  It is also common practice to use, the {\it excess Kurtosis}, which is the {\it Kurtosis} minus $3$, to provide the comparison to the normal probability distribution profile. Like {\it Skewness }here also we calculate kurtosis from different distribution and get it at different order of correction. 

Therefore, {\it Kurtosis} without correction can be expressed as:
\bea
{\bf Kurtosis_{uc}}&=&\frac{\langle n^{4}\rangle_{\textcolor{red}{\bf I}}}{\left[\langle n^2 \rangle_{\textcolor{red}{\bf I}}-\left(\langle n \rangle_{\textcolor{red}{\bf I}}\right)^2\right]^2}\nonumber\\
&=&\frac{\langle n^{4}\rangle_{\textcolor{red}{\bf I}}}{\left[{\bf S.D._{uc}}\right]^2}\nonumber\\
&=&\frac{6}{\mu  \left(-3 e^{4 \mu_k  \tau }+2 e^{6 \mu_k  \tau }+1\right)}\nonumber\\
&& ~\times\left[6 (\mu_k -1) e^{2 \mu_k  \tau }+3 (5 \mu_k -2) e^{8 \mu_k  \tau }\right.\nonumber\\&& \left.~~~~~~~~~~~~~~ -2 e^{6 \mu_k  \tau } (3 \mu  (4 \mu_k  \tau +3)-5)-3 \mu_k +2\right].
\eea
On the other hand, the fourth order corrected value of the {\it Kurtosis} can be expressed as:
\bea
{\bf Kurtosis_{c}}&=&\frac{\langle n^{4}\rangle_{\textcolor{red}{\bf I}}+\langle n^{4}\rangle_{\textcolor{red}{\bf II}}+\langle n^{4}\rangle_{\textcolor{red}{\bf III}}+\langle n^{4}\rangle_{\textcolor{red}{\bf IV}}}{\left[\langle n^2 \rangle_{\textcolor{red}{\bf I}}+\langle n^2 \rangle_{\textcolor{red}{\bf I}}-\left(\langle n \rangle_{\textcolor{red}{\bf I}}\right)^2\right]^2}\nonumber\\
&=&\frac{\langle n^{4}\rangle_{\textcolor{red}{\bf I}}+\langle n^{4}\rangle_{\textcolor{red}{\bf II}}+\langle n^{4}\rangle_{\textcolor{red}{\bf III}}+\langle n^{4}\rangle_{\textcolor{red}{\bf IV}}}{\left[{\bf S.D.}_{c}\right]^2}\eea

%%%%ADD pic
\begin{figure}[H]
\centering
\subfigure[Corrected {\it Kurtosis}  for large values $\mu_{k}\tau$  ]{
    \includegraphics[width=7.8cm,height=8cm] {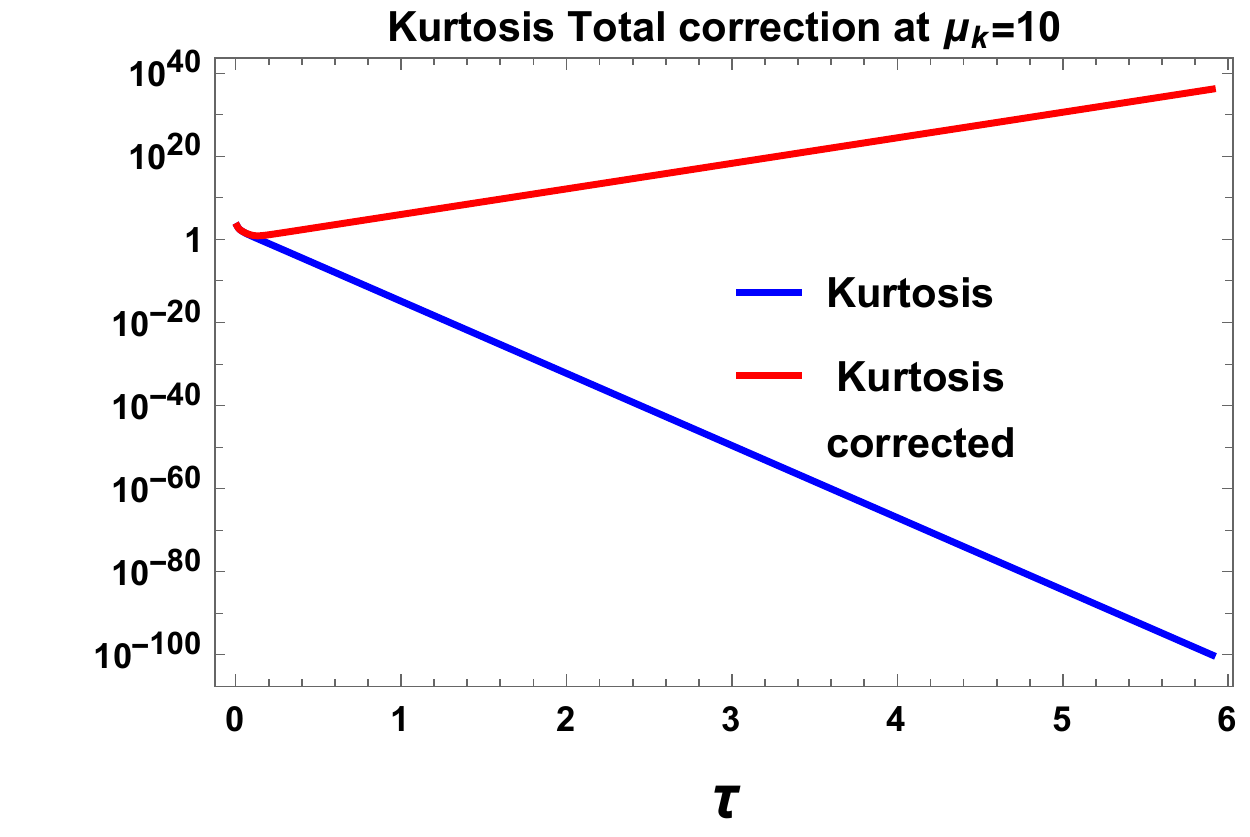}
    \label{kw2}
}
\subfigure[Corrected {\it Kurtosis} for small values $\tau$ ]{
	\includegraphics[width=7.8cm,height=8cm] {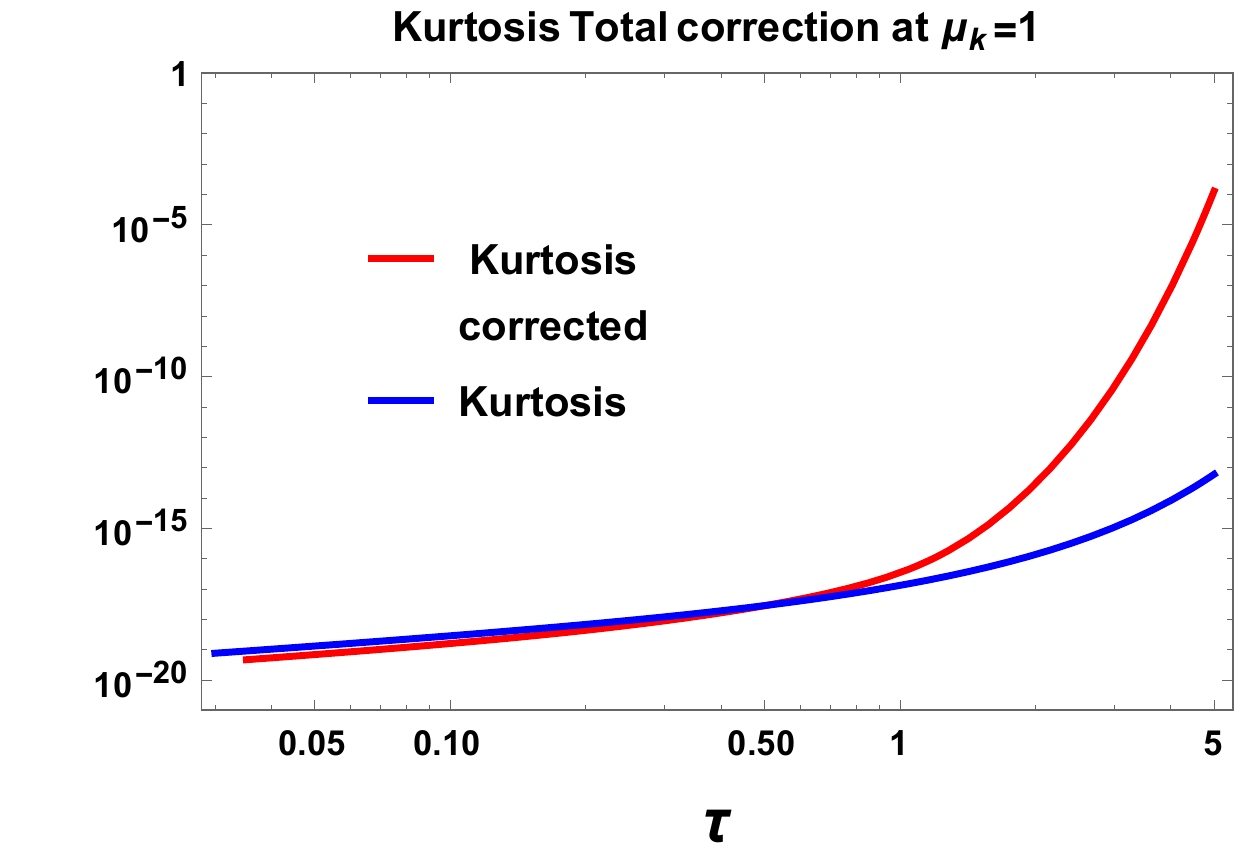}
	\label{kw4}
}
\caption{Time dependent behaviour  of {\it Kurtosis} without correction and with correction in the probability density distribution function at different range of  $\tau$  considering different $\mu_{k}$ }
\label{k2}
\end{figure}

From fig.~(\ref{k2}), we can say that uncorrected {\it kurtosis} deviate from corrected one in lower $\tau$ regime though overlapped in higher order [$\tau>1$]. So the contribution from the correction factor is important in lower regime. Also it is important to note that, {\it Kurtosis} is greater than the value $3$ for the whole time regime, so the distribution is Leptokurtic and have fatter tails.Here fig~\ref{kw2} has a huge value for corrected kurtosis which can be bounded within a particular value using tuning parameter.

From the calculation of the higher-order statistical moments (or equivalently the amplitude of the quantum mechanical correlation functions) we get the following overall features to analyze  the nature and physical outcomes of the corrected probability density function derived in this paper.
\begin{enumerate}
\item 
The {\bf Standard Deviation} is significantly large for higher $\mu_{k} \tau$, though very small for lower regime.
\item 
 {\bf Skewness} is positive throughout the time regime, though becomes vanishingly small at a specific time  interval ($0.1<\tau<0.7$).
\item 
{\bf Kurtosis } is greater than $3$ for the whole time regime.
\item The predicted Log-Normal Gaussian Distribution shows deviations at significant levels. Effects of the non-Gaussianity in the distribution function is clearly visualized. 

\item  The probability distribution has longer trailing ends and the trails go broad higher.
\item The probability distribution has a very low spread at lower $\mu_{k} \tau$ limit though highly spread out in larger limit.
\end{enumerate}

Conequently, from the previously predicted result of \cite{Amin:2015ftc} we show that distribution deviates from a Log-Normal distribution and the predicted distribution function appears to be {\it  leptokurtic} and has a broad right trailing end.All of this portion have a simmilar connection with ref~ \cite{Choudhury:2018bcf} 

\section{Conclusion}
\label{conc}

In this paper we have addressed the issues which are 
appended below:
\begin{itemize}
\item In this paper, we have provided the analogy between particle creation in primordial cosmology and scattering problem inside a conduction wire in presence of impurities. Such impurities are characterized by effective potential in the context of quantum mechanical description. On the other hand, in the context of primordial cosmology time dependent mass profile of created particles (couplings) mimics the same role.

\item Specific time dependence of mass profile actually restricts the structure of the scattering effective potential. To establish the analogy between two theoretical frameworks we have further computed various characteristic features of conduction wire i.e. resistance, conductance (electrical properties), {\it Lyapunov exponent} (dynamical property), reflection and transmission coefficients (optical properties), occupation number and energy density (energetics) using the expression for Bogoliubov coefficients for different mass profiles which connects the ingoing and outgoing solution of the mode functions obtained in the context of particle creation process in cosmology.  

\item We have solved this particle creation problem using the following crucial steps:
\begin{enumerate}
\item First of all assuming that the interactions are well known we have studied the one to one correspondence between the particle creation problem in early universe cosmology with the scattering problem inside a conduction wire. Here we have additionally neglected the effect of the expansion of our universe and this is perfectly justifiable during the epoch of reheating. For this reason we call it as \textcolor{red}{\bf \underline{Reheating Approximation}}.

\item Secondly we have studied the same problem where the particle interactions are not known at all at the level of action. In such a situation, assuming the gravitational background is classical in nature and also assuming the previously mentioned \textcolor{red}{\bf \underline{Reheating Approximation}} we have demonstrated the problem with the help of {\it Random Matrix Theory}. 

\item Further we have solved the dynamics of the particle creation problem by studying the higher order corrections in the {\it Fokker Planck equation} for previously mentioned random system where the interactions are not easily quantifiable at the level of action. We have constructed the fourth order corrected {\it Fokker Planck equation} from which we have provided the solution of the random probability distribution function. Such distributions are very very useful to study the dynamical systems when particle interactions are not well known. In our analysis we have identified all of these modifications as the quantum correction to the {\it Fokker-Planck equation}, the physical implications of which we have studied in detail in this paper.
\end{enumerate}

\item In this work, we have shown that the {\it Lyapunov exponent} varies inversely with the number of scatterers. Therefore, with an increase in the number of scatterers the {\it Lyapunov exponent} also reduces thereby reducing the amount of randomness in the system. This may be a hint to the fact that the {\it Lyapunov exponent} has a dependence on the momenta values of the incoming wave-function of the scalar field. Additionally, it is important to note that the upper bound of {\it Lyuapunov exponent} is restricted by the constraint $\lambda\leq 2\pi/\beta$ (where $\beta=1/T$), which is a generic bound on chaos obtained in the context of quantum field theory. As a consequence, one can find restriction on the upper bound on the reheating temperature for the different quenched mass profiles for which the chaos bound saturates. This is obviously a remarkable result in the present context as it can able to provide the explicit expression for the reheating temperature for a specified momentum scale, which was not predicted earlier in the detailed study of reheating. Most importantly, the bound on quantum chaos in terms of {\it Lyapunov exponent} directly restrict the value of reheating temperature
without explicitly knowing the details of the particle interactions as appearing in the action. Just the knowledge of time dependence of the quenched mass profiles (in other words the knowledge of effective impurity potential as appearing inside the conduction wire) is sufficient enough to restrict the upper bound of reheating temperature due to quantum chaos.

\item In this context we have also provided the expression for the two point quantum correlation function, which is known as {\it Spectral Form Factor} (SFF) for both in finite and zero temperature. {\it Spectral Form Factor} is actually a more strong measure to find chaotic behaviour of a dynamical system compared to {\it Lyapunov exponent}. We get saturating behaviour of SFF at late time scale, which indicates that it has an upper-bound. We can relate SFF for any potential (Even Polynomial Potential in this case). In the calculation of the Lyapunov Exponent for the specific time dependent mass profiles, we choose three different quenched protocols for mass profiles. Potential functions which can be represented by polynomial potential (Even only in our case) can be used to get the SFF-saturation. In this connection ,we have provided a model independent upper and lower bound of SFF, which is treated as the significant bound of quantum chaos ($-1/N\left(1-1/\pi\right)\leq {\bf SFF} \leq 1/\pi N$) in the context of particle production event in cosmology. This is obviously a remarkable result which we have explicitly computed in this paper.In \cite{Choudhury:2018lcb} this same has been calculated for general polynomial with GUE.

\item We have also presented the computation of quantum corrected {\it Fokker- Planck equation} which corresponds to the delta-scatterers. From this computation we have derived the corrected statistical distribution of the 
particle production events in cosmology. The distribution which has been predicted in \cite{Amin:2015ftc} to be 
Gaussian doesn't retain its form when more correction terms are taken into account. This may be treated as a signature of
 non-Gaussian in particle production events during reheating (in cosmology). 
 
\end{itemize}

%\bc
%\begin{figure}[htb]
	%\includegraphics[width=19cm,height=18cm]{chaoscosmo.pdf}
	%\caption{The above figure represents a flow chart describing the Open Quantum System approach to the universe}
	%\label{chaos5}
%\end{figure}
%\ec
The future discussions of the present work are mentioned in the following:
\begin{itemize}
\item In this paper for our study of quantum chaos in the context of cosmology we have used a closed quantum system. As we have mentioned that the present computation has been performed for a massless scalar field which interacts with the
heavy fields (which acts like scatterers inside the conduction wire). The entire calculation is being done for the set up when there is only a single massless scalar field that interacts with the scatterer. One may repeat the calculation for a large number of these scalar fields interacting with the scatterers which needs the introduction of the random matrix approach in a more generalized fashion.

\item The system we have studied in this paper have no interactions with the background as the definition of the background in this set-up is itself an ill-defined one during reheating. To treat the entire system having being interacted with a background one needs to have a detailed description of the nature of background in the cosmological scenario. Then it will be possible to introduce the other non-linear and dissipative effects into the system introduced by the background itself. Such a treatment will be studied within the framework of an open quantum system interacting with the defined background set-up. One then needs to consider the entire system having being interacted with the background under a weak coupling limit. One such model as has been studied in \cite{Shandera:2017qkg}.

\item We have calculated the {\it Lyapunov exponent} and {\it Spectral Form Factor} in this paper which is a measure of chaos or non-linearity into the system. With the system prescribed in this work being treated as an open quantum system one may study the effects of dissipation being introduced into such
a system which renders the system to be a stochastic one. With this, one may be able to study the effects of the non-linearity
being introduced into the system which may well be a good study to look for the behaviour of {\it Lyapunov exponent} and {\it Spectral Form Factor}.

\item During the study of quantum correction in the {\it Fokker Planck equation} and the deviation from log normal distribution we have followed a specific approach in which we have considered the following possibilities:
\begin{enumerate}

\item We have neglected the contribution from the damping term in the {\it Fokker Planck equation}. One can include such effect and study its role in the context of cosmology (specifically during reheating).

\item During the computation we have followed a specific approach in which we have also neglected the effect of impurity potential at very high temperature during reheating. This will give rise to a simplest form of the 
{\it Fokker Planck equation} where only diffusion and drift contributions are appearing explicitly. But if we include the effect of impurity potential in presence of finite temperature then it will surely effect the final solution of the probability distribution function. One can include such additional effects and study its impact during reheating epoch of the early universe.

\item Furthermore, during the construction of the {\it Fokker Planck equation} from the basic principles we have followed a special approach in which the effect of diffusion and drift is appearing in a very simplified manner. However, in the study of statistical field theory {\it It$\hat{o}$} and {\it Stratnovitch} or more {\it generalized} prescriptions are used commonly to construct the {\it Fokker Planck equation}. Here it is important to note that in each case it will give rise to different mathematical structure of {\it Fokker Planck equations}. In the present context of discussion, one can follow such well known prescriptions to
see its physical outcomes to solve the probability distribution function for the particle production and compare the results to check the appropriateness of these approaches during reheating.

\end{enumerate}

\end{itemize}

%%%%%%%%%%%%%%%%%%%%%%%%%%%%%%%%%%%%%%%%%%%%%%%%%%%%%%%%%%%%%%%%%%%%%%%%%%%%%%%%%%%%%%%%%%%%%%%%%%%%%%%%%%%%%%%%%%%%%%%%%%%%%%%%%%%%%%%%%%%%%%%%%%%%%%%%%%%%%%%%%%%%%%%%%%%%%%%%%%%%%%%%%%%%%%%%%%%%%%%%
	%\newpage
	\section*{\textcolor{blue}{Acknowledgments}}
	SC would like to thank Quantum Gravity and Unified Theory and Theoretical Cosmology Group, Max Planck Institute for Gravitational Physics, Albert Einstein Institute for providing the Post-Doctoral
	Research Fellowship. SC would like to thank IUCAA, Pune, India where the problem was formulated and part of the work has been done during post doctoral tenure. SC take this opportunity to thank sincerely to  Jean-Luc Lehners, Shiraz Minwalla, Sudhakar Panda and Varun Sahni for their constant support
	and inspiration. SC thank the organisers of Summer School on Cosmology 2018, ICTP, Trieste, 15 th Marcel Grossman Meeting, Rome, The European Einstein Toolkit meeting 2018, Centra, Instituto Superior Tecnico, Lisbon, The Universe as a Quantum Lab, APC, Paris, LMU Quantum Gravity School, Munich, Nordic String Meeting 2019, AEI, Summer meeting on String Cosmology, DESY, Zeuthen,  Kavli Asian WInter School on Strings, Particles and Cosmology 2018 
	for providing the local hospitality during the work. SC also thank DTP, TIFR, Mumbai, ICTS, TIFR, Bengaluru, IOP, Bhubaneswar, CMI, Chennai, SINP, Kolkata and 
IACS, Kolkata for
	providing the academic visit during the work. AM, PC and SB are thankful to IUCAA, Pune for the visit during the work for winter project. Last but not the least, We would all like to acknowledge our
	debt to the people of India for their generous and steady support for research in natural sciences, especially
	for theoretical high energy physics, string theory and cosmology.
%%%%%%%%%%%%%%%%%%

\appendix
 \section{It$\hat{o}$ solution of Fokker Planck equation }
 \begin{figure}[htb]
	\includegraphics[width=16cm,height=8cm]{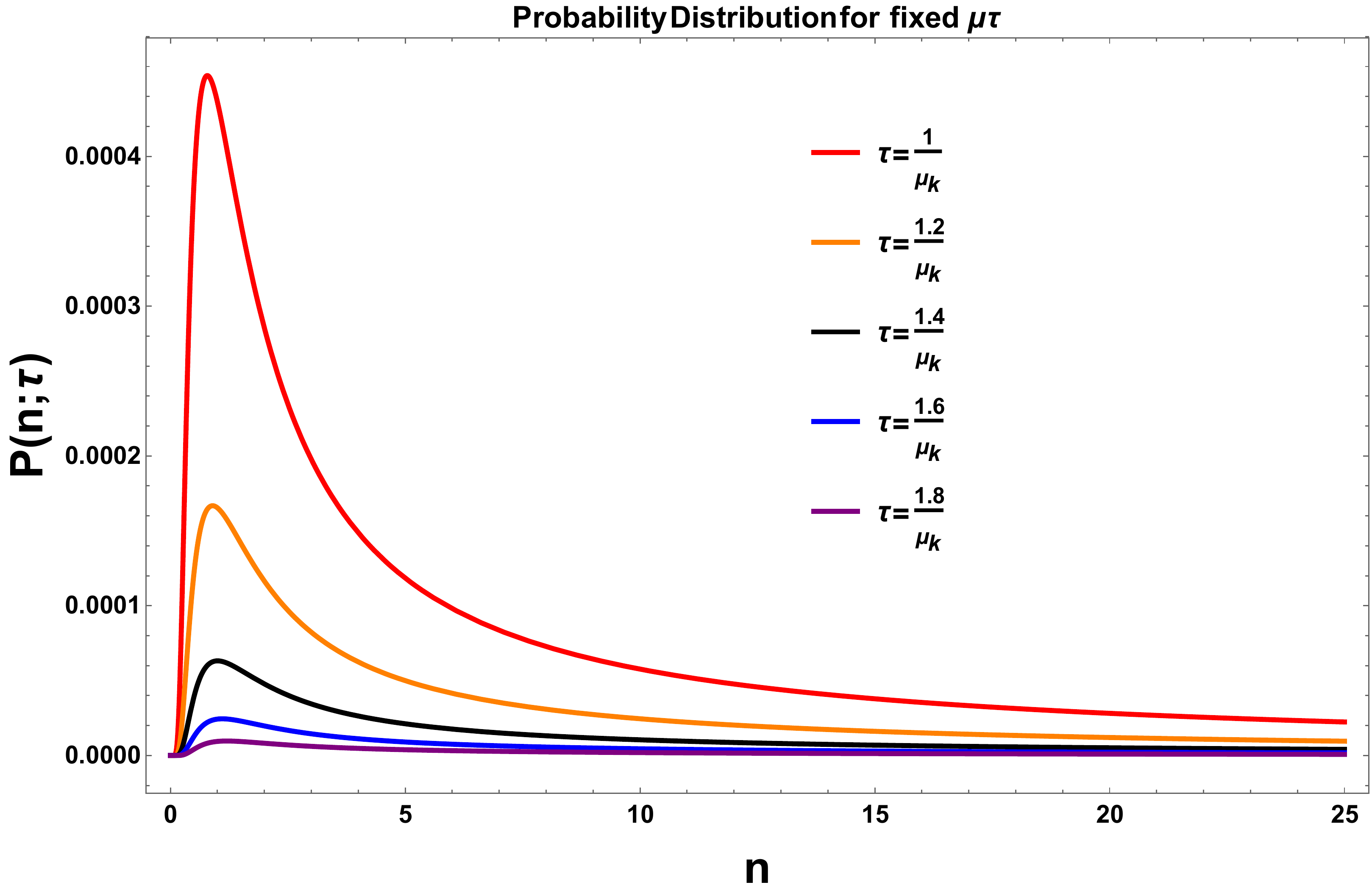}
	\caption{Evolution of the probability density function for It$\hat o$ prescription with respect to the the occupation number per mode $n$, for a fixed time.}
	\label{ito1}
\end{figure}
 From It$\hat{o}$ perspective the Fokker Planck equation can be expressed as:
 \bea  &&\textcolor{blue}{\bf \underline{Fokker~Planck~Equation~(From~It\hat{o})}:}\nonumber\\
\frac{\pl P(n;\tau)}{\pl \tau}&=&-\frac{\pl }{\pl n}\left(a(n)P(n;\tau)\right)+\frac{\pl^2}{\pl n^2}\left(D(n)P(n;\tau)\right),\eea
Here we take $a(n)=0$ and $D(n)=n(n+1)$. Using this we get the following solution of probability distribution:
\bea P(n,\tau)&=&\frac{1}{2 \sqrt{\pi } \sqrt{n (n+1) \tau  \mu _k}}\exp\left[-\frac{((4 n+2) \tau  \mu _k+n){}^2 }{4 n (n+1) \tau  \mu _k}\right]
\eea
In fig.~(\ref{ito1}) we have shown the probability distribution function obtained from the It$\hat{o}$ solution of the {\it Fokker Planck equation}. This solution is similar to the log normal distribution obtained from the present computation. From the plot we observe that for large value of occupation number $n$ ($n>>1$) the distribution function decays to a finite saturation value. On the other hand for small occupation number $n$ ($n<<1$) we get peak in the distribution function for different values of $\mu_k\tau$. 
\section{ Stratonovitch solution of Fokker Planck equation}
\begin{figure}[htb]
	\includegraphics[width=16cm,height=8cm]{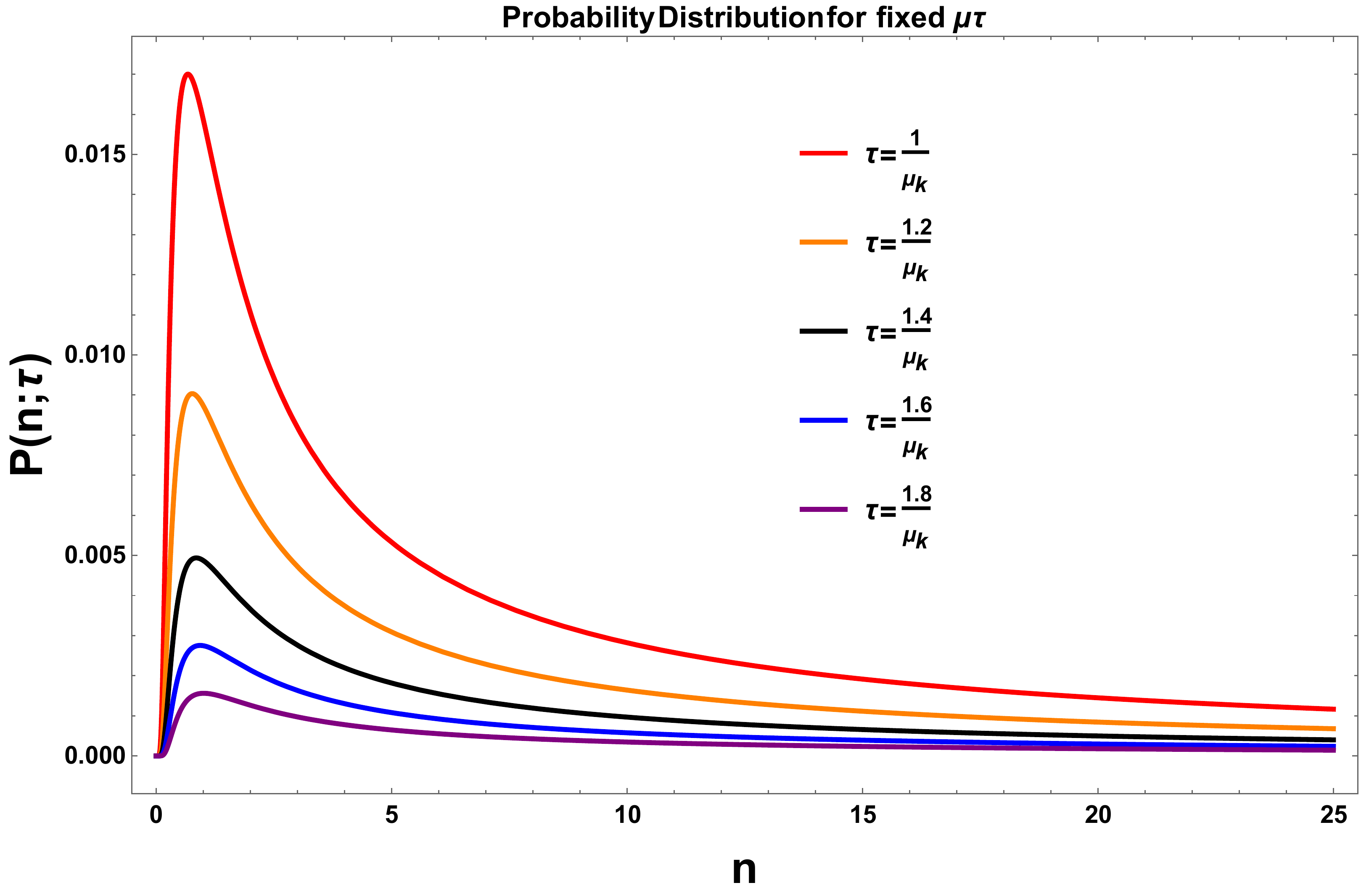}
	\caption{Evolution of the probability density function for Stratonovitch prescription with respect to the the occupation number per mode $n$, for a fixed time in the limit $n<<1$.}
	\label{STAT1}
\end{figure}
 From Stratonovitch perspective the Fokker Planck equation can be expressed as:
\bea  &&\textcolor{blue}{\bf \underline{Fokker~Planck~Equation~(From~Stratonovitch)}:}\nonumber\\
\frac{\pl P(n;\tau)}{\pl \tau}&=&-\frac{\pl }{\pl n}\left(a(n)P(n;\tau)\right)+\frac{\pl}{\pl n}\left(\sqrt{D(n)}\frac{\pl }{\pl n}\left(\sqrt{D(n)}P(n;\tau)\right)\right),\eea
Here we take $a(n)=0$ and $D(n)=n(n+1)$. Using this we get the following solution of probability distribution:
\bea P(n,\tau)&=&\frac{1}{2 \sqrt{\pi } \sqrt{n (n+1) \tau  \mu _k}}\exp\left[-\frac{9 (2 n+1)^2 \tau  \mu _k}{16 n (n+1)}\right]
\eea
In fig.~(\ref{STAT1}) we have shown the probability distribution function obtained from the Stratonovitch solution of the {\it Fokker Planck equation}. This solution is similar to the log normal distribution obtained from the present computation. From the plot we observe that for large value of occupation number $n$ ($n>>1$) the distribution function decays to a finite saturation value. On the other hand for small occupation number $n$ ($n<<1$) we get peak in the distribution function for different values of $\mu_k\tau$. 
\section{ Generalized solution of Fokker Planck equation at infinite temperature}
\begin{figure}[htb]
\centering
\subfigure[Probability distribution for $Q=1/4$.  ]{
    \includegraphics[width=7.8cm,height=8cm] {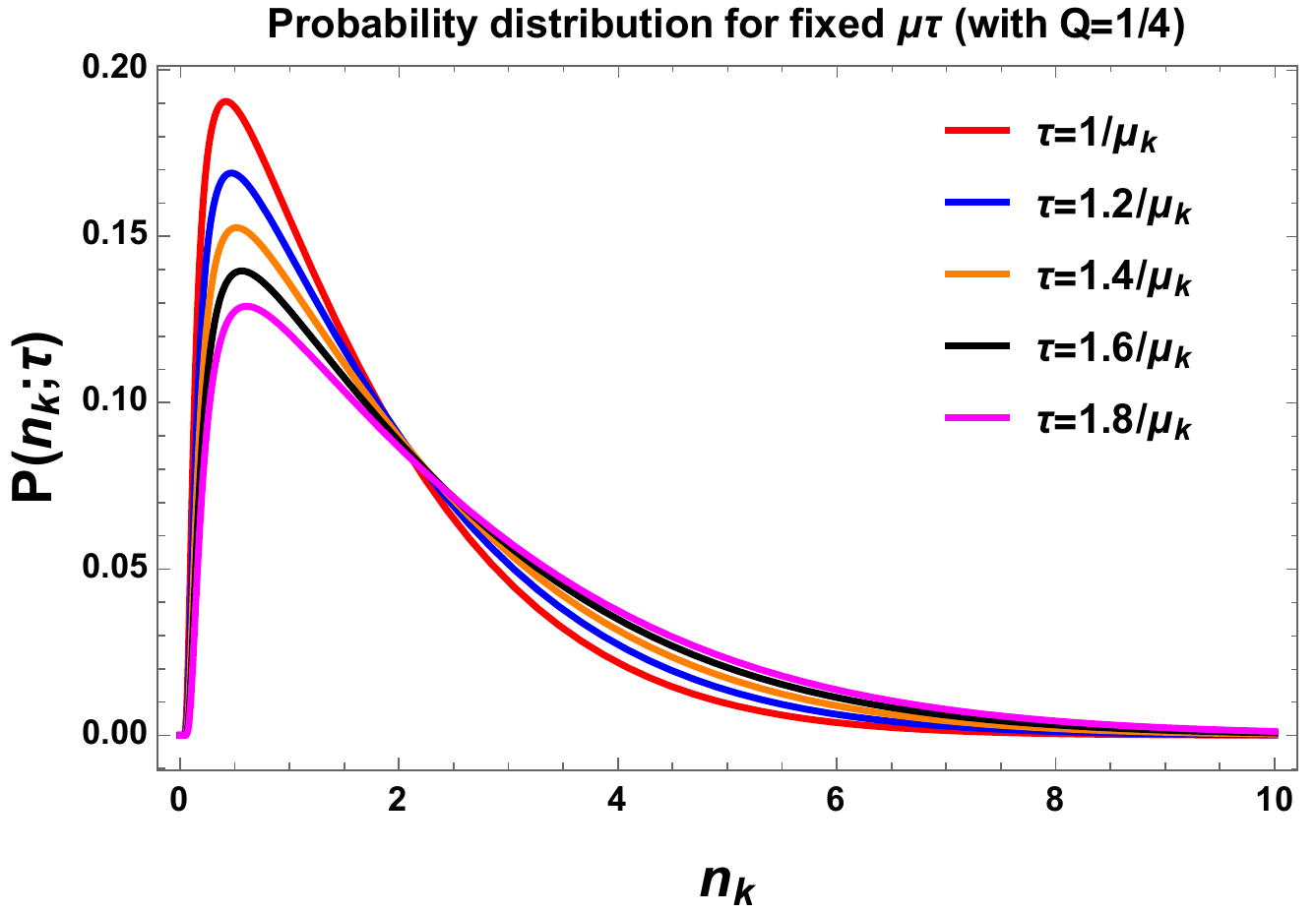}
    \label{G1a}
}
%\subfigure[Different $\langle n^{4}\rangle$ for higher values $\tau$.]{
    %\includegraphics[width=7.8cm,height=6cm] {N4_u100_full.pdf}
    %\label{N43}
%}
\subfigure[Probability distribution for $Q=1/8$.  ]{
    \includegraphics[width=7.8cm,height=8cm] {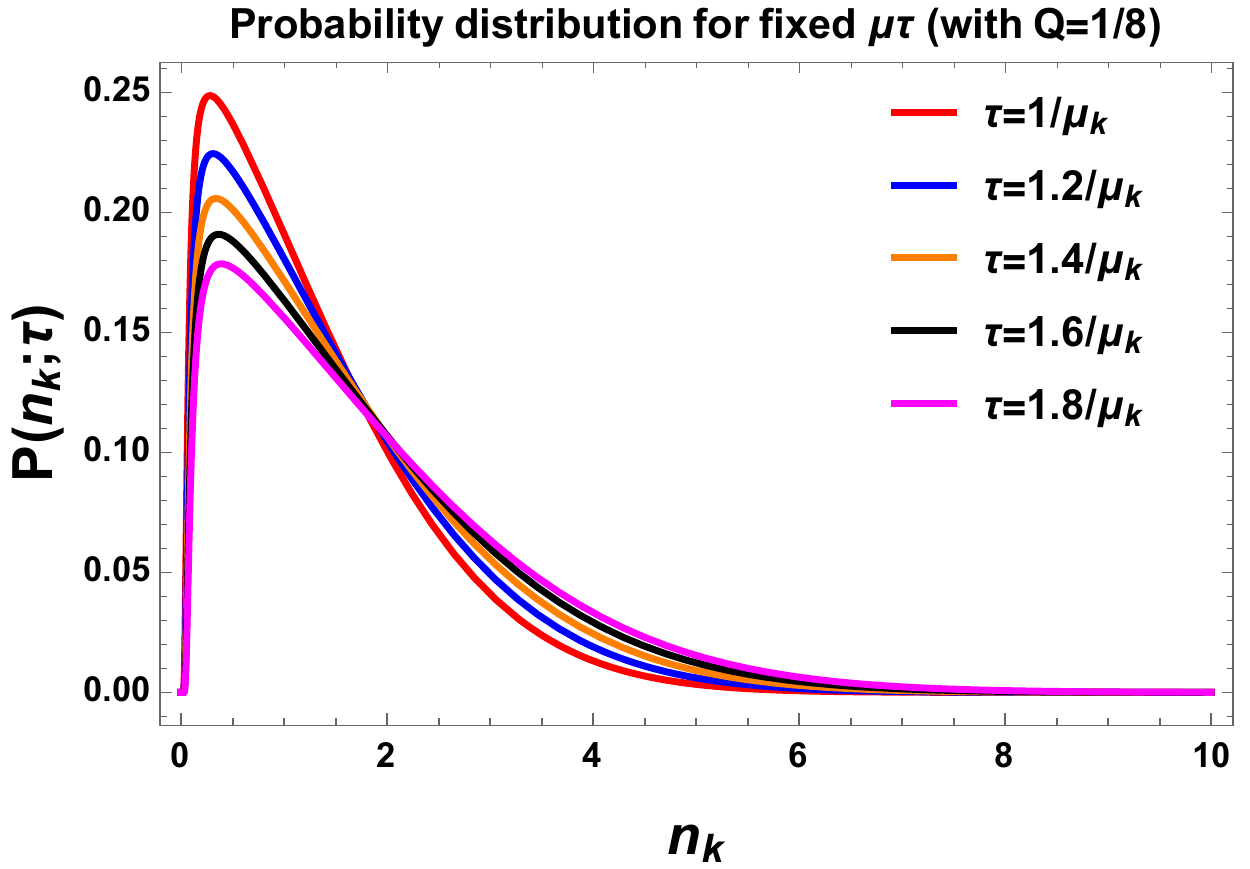}
    \label{G1b}
}
	\caption{Evolution of the probability density function for generalized prescription with respect to the the occupation number per mode $n$, for a fixed time.}
	\label{gQ1}
\end{figure}
 From General perspective the Fokker Planck equation can be expressed as:
\bea  &&\textcolor{blue}{\bf \underline{Fokker~Planck~Equation~(For~Generalized~It\hat{o})}:}\nonumber\\
\frac{\pl P(n;\tau)}{\pl \tau}&=&-\frac{\pl }{\pl n}\left(a(n)P(n;\tau)\right)+\frac{\pl}{\pl n}\left((D(n))^{1-Q}\frac{\pl}{\pl n}\left((D(n))^Q P(n;\tau)\right)\right),~~~~~~\eea
Here we take $a(n)=0$ and $D(n)=n(n+1)$. Using this we get the following solution of probability distribution:
\bea P(n,\tau)&=&\frac{1}{2 \sqrt{\mu_k \pi  \tau (n (n+1))^Q}}\nonumber\\
&&~~~~~~~~~\times\exp \left[-\frac{\left(n^2 (n+1)+\mu_k \tau (2 n+1) Q (Q+1) (n (n+1))^Q\right)^2}{4 \mu_k \tau (n (n+1))^{Q+2}}\right]~~~~~~
\eea
In fig.~(\ref{gQ1}) we have shown the probability distribution function obtained from the generalized solution of the {\it Fokker Planck equation} without dissipation in very very large temperature. From the plot we observe that for large value of occupation number $n$ ($n>>1$) the distribution function decays to a finite saturation value. On the other hand for small occupation number $n$ ($n<<1$) we get peak in the distribution function for different values of $\mu_k\tau$. 
%%%%%%%%%%%%%%%%%
%%%%
%%%
%%%
%%%%%
%%%%%%%
%%%%%%%
%%%%%%%
\section{ Generalized solution of Fokker Planck equation at finite temperature}
 From General perspective the Fokker Planck equation with effect of potential can be expressed as:
\bea &&\textcolor{blue}{\bf \underline{Effective ~Potential:}}\nonumber\\
U(n)&=&\left[\frac{\beta^2}{4}D(n)\left(\frac{\pl V(n)}{\pl n}\right)^2-\frac{\beta}{2}D(n)\left(\frac{\pl^2V(n)}{\pl n^2}\right)-\frac{\beta}{2}\left(\frac{\pl D(n)}{\pl n}\right)\left(\frac{\pl V(n)}{\pl n}\right)\right].\eea
\bea\frac{\pl}{\pl n}\left(D(n)\frac{\pl W(n;\tau)}{\pl n}\right)-U(n)W(n;\tau)&=&\frac{\pl W(n;\tau)}{\pl \tau},\eea
\bea P(n;\tau)&=&\exp\left(-\frac{\beta}{2} V(n)\right)W(n;\tau)~.\eea

Here we take $a(n)=0$,$V[n]=n^{2}$and $D(n)=n(n+1)$. Using this we get the following solution of probability distribution:
\bea~P(n,\tau)&=&\frac{1}{2 \sqrt{\pi } \sqrt{n (n+1) \tau  \mu _k}}\nonumber\\
&&~~~\times\exp\left[-\frac{(n-\mu _k (2 n \tau +\tau )){}^2}{4 n (n+1) \tau  \mu _k}-\frac{\beta  n^2}{2}-\beta  n \left\{n (\beta  n (n+1)-3)-2\right\}\right]~~~~
\eea
\begin{figure}[htb]
\centering
\subfigure[Probability distribution for $\beta=0.01$.  ]{
    \includegraphics[width=7.8cm,height=8cm] {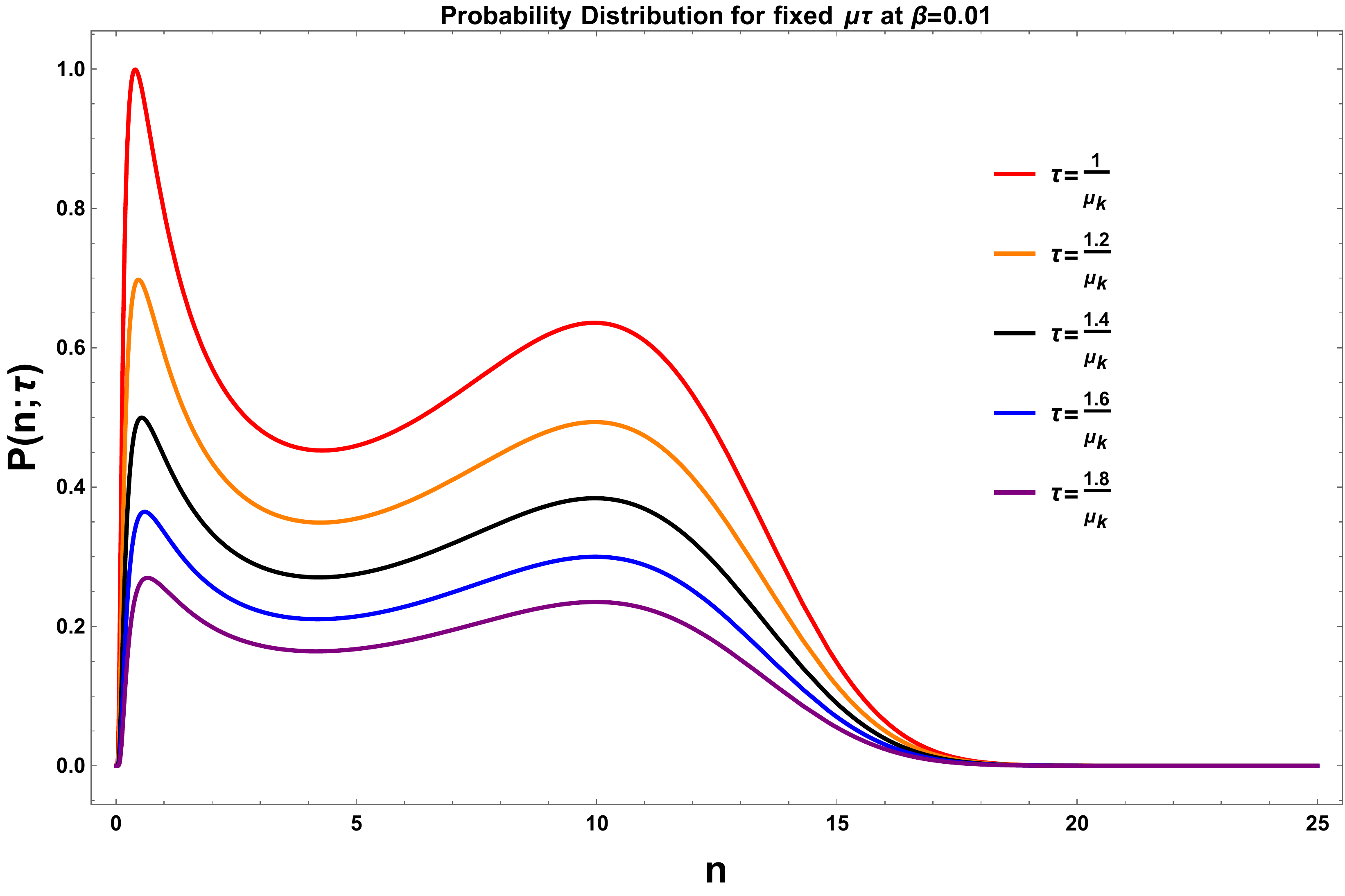}
    \label{4a}
}
\subfigure[Probability distribution for $\beta=1$.  ]{
    \includegraphics[width=7.8cm,height=8cm] {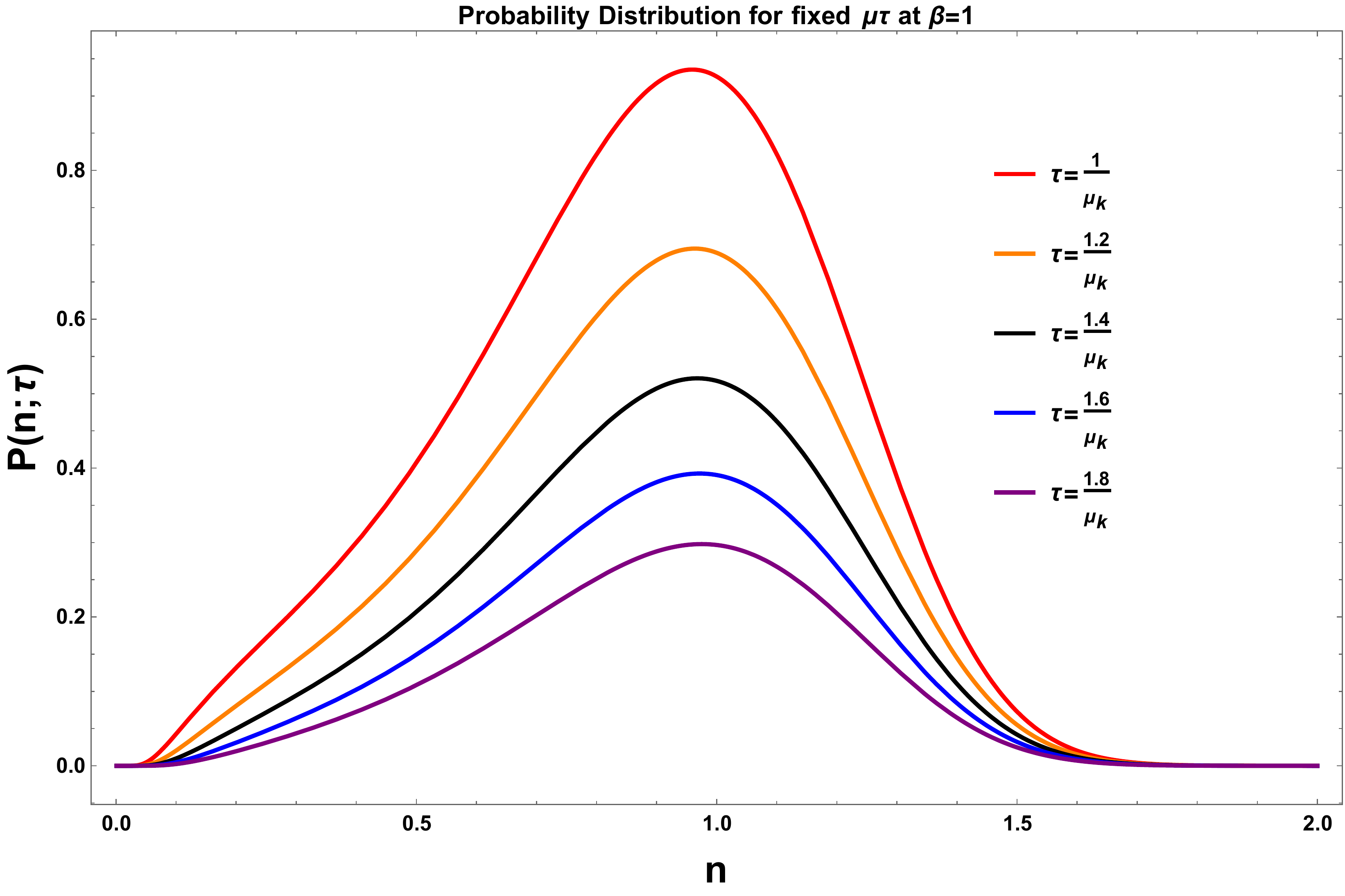}
    \label{4b}
}
\subfigure[Probability distribution for $\beta=10$. ]{
    \includegraphics[width=7.8cm,height=8cm] {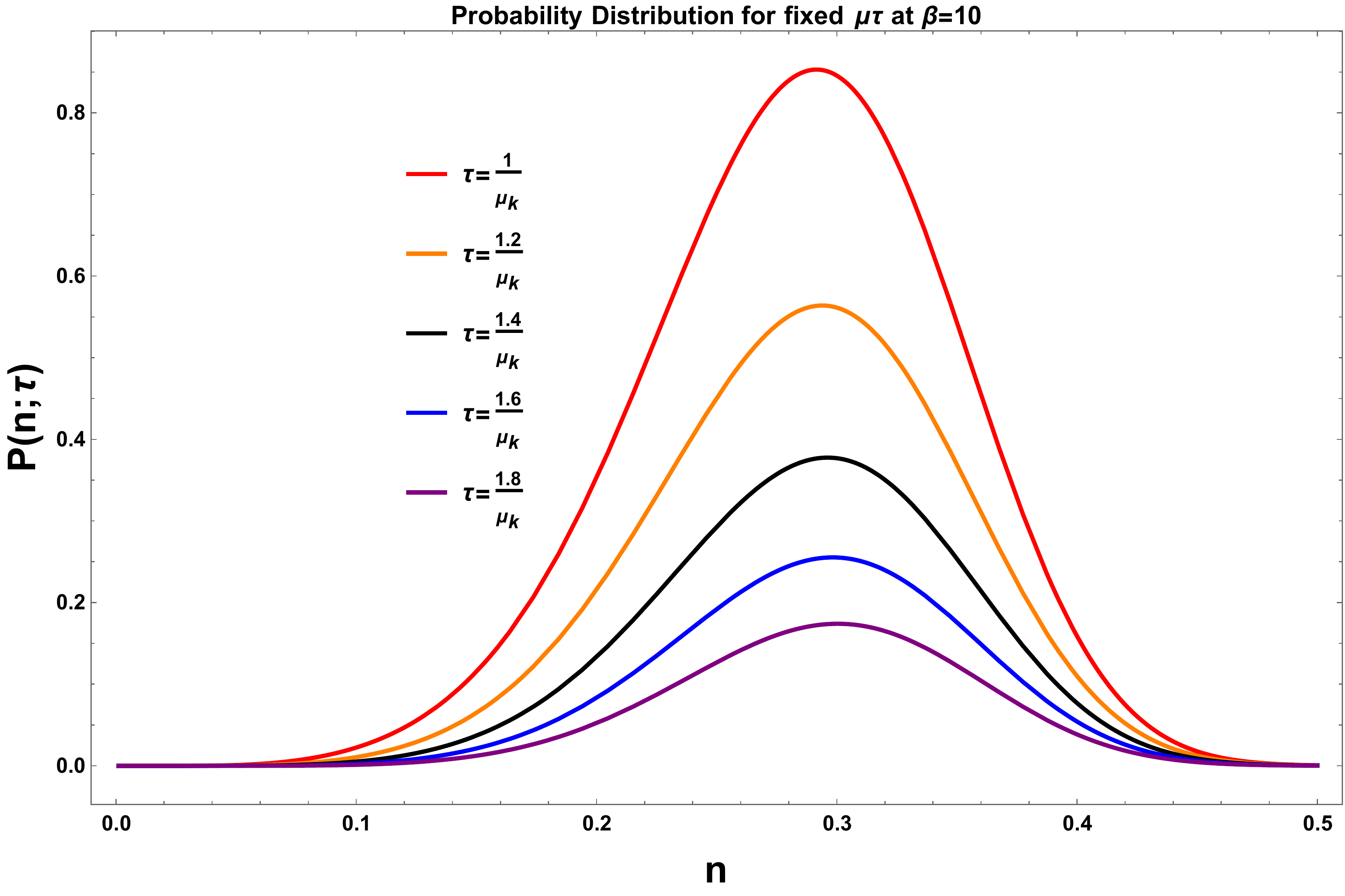}
    \label{4c}
}
\subfigure[Probability distribution for $\beta=100$.   ]{
    \includegraphics[width=7.8cm,height=8cm] {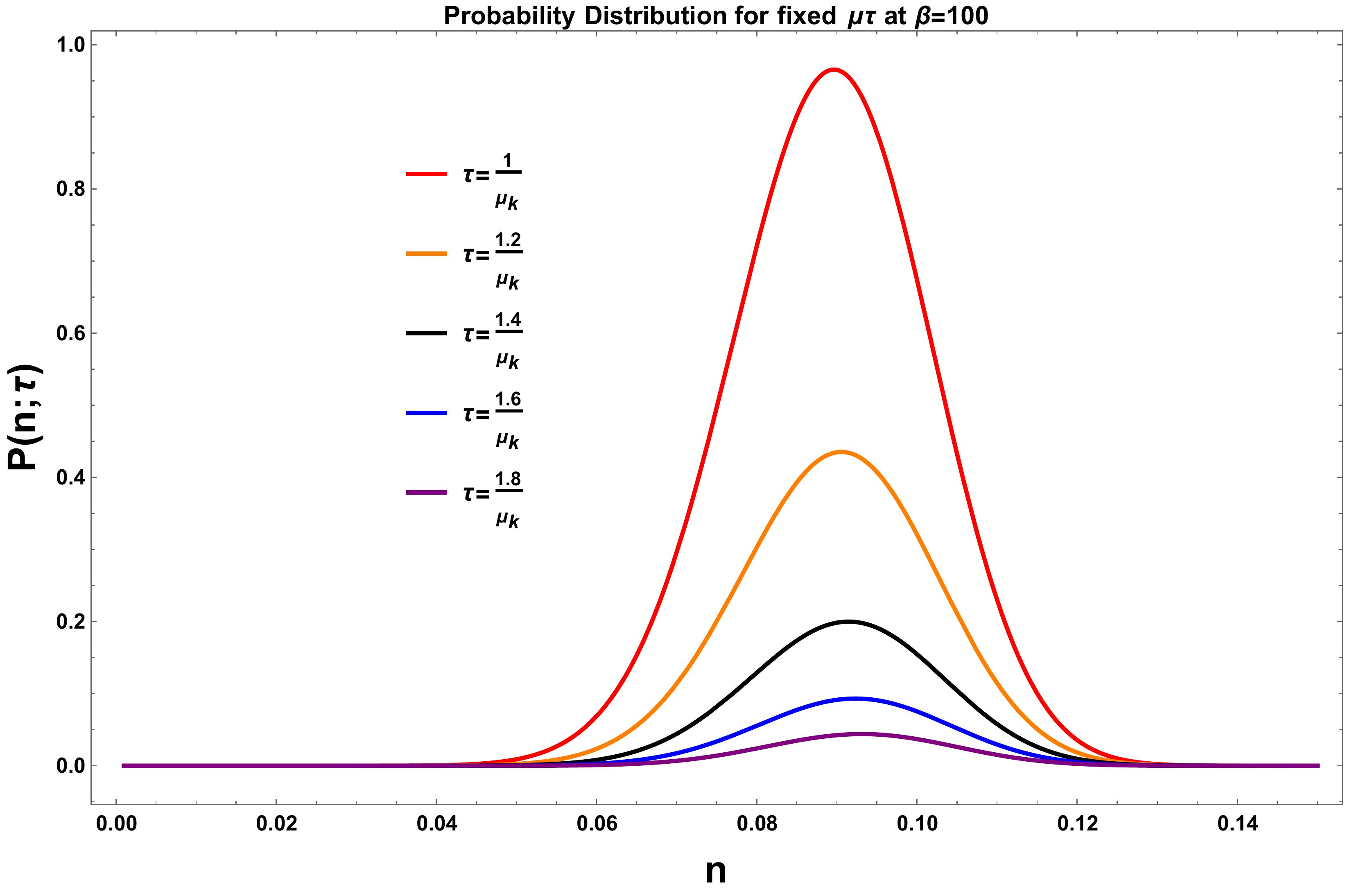}
    \label{4d}
}
	\caption{Evolution of the probability density function for generalized prescription with respect to the the occupation number per mode $n$, for a fixed time and $\beta$.}
	\label{r4e}
\end{figure}
In fig.~(\ref{r4e}) we have shown the probability distribution function obtained from the solution of of the {\it Fokker Planck equation} derived in presence of finite temperature effective potential solution. From the plot we observe that for large value of occupation number $n$ ($n>>1$) the distribution function decays to a finite saturation value. On the other hand for small occupation number $n$ ($n<<1$) we get peak in the distribution function for different values of $\mu_k\tau$.

\end{document}